\documentclass[prx,twocolumn,notitlepage,showpacs, superscriptaddress,nofootinbib,amsmath,amsthm,amssymb,UFT8]{revtex4-2}
\usepackage[utf8]{inputenc}
\usepackage{algorithm}
\usepackage{algorithmicx}
\usepackage{algpseudocode}
\usepackage{amsmath}
\usepackage{amssymb}
\usepackage{amsthm}
\usepackage{bm}
\usepackage{braket}
\usepackage{cancel}
\usepackage{color}
\usepackage{dsfont}
\usepackage{bm}
\usepackage{parskip}
\usepackage{subcaption}
\usepackage{graphicx}
\usepackage{xurl}
\usepackage[colorlinks=true,linkcolor=red,urlcolor=blue, citecolor=blue, bookmarks,breaklinks]{hyperref}
\usepackage{empheq}
\usepackage{comment}
\usepackage{ulem}
\usepackage{tikz}
\usetikzlibrary{positioning}

\makeatletter
\newtheorem*{rep@theorem}{\rep@title}
\newcommand{\newreptheorem}[2]{%
\newenvironment{rep#1}[1]{%
 \def\rep@title{#2 \ref{##1} (restated)}%
 \begin{rep@theorem}}%
 {\end{rep@theorem}}}
\makeatother

\newtheorem{theorem}{Theorem}
\newtheorem{assumption}{Assumption}

\newtheorem{corollary}[theorem]{Corollary}
\newtheorem{definition}[theorem]{Definition}
\newtheorem{lemma}[theorem]{Lemma}

\newtheorem{proposition}[theorem]{Proposition}
\newtheorem{remark}[theorem]{Remark}

\newreptheorem{lemma}{Lemma}
\newreptheorem{proposition}{Proposition}
\newreptheorem{theorem}{Theorem}
\numberwithin{theorem}{section}

\definecolor{darkgreen}{rgb}{0,0.6,0}
\definecolor{darkred}{rgb}{.7,0.0,.5}
\newcommand{\sbcomment}[1]{{}}
\newcommand{\sbedit}[1]{{#1}}

\newcommand{\jscomment}[1]{{}}
\newcommand{\todo}[1]{{}}

\newcommand{\of}[1]{\left( #1 \right)} %

\makeatletter
\newcommand{\vast}{\bBigg@{3}}
\newcommand{\Vast}{\bBigg@{4}}
\newcommand{\VAST}{\bBigg@{6}}
\makeatother

\allowdisplaybreaks

\begin{document}

\title{Quantum Approximate Optimization Algorithm in Finite Size and Large Depth and Equivalence to Quantum Annealing} %

\author{Sami~Boulebnane}
\email{sami.boulebnane@jpmchase.com}
\author{James Sud}
\author{Ruslan Shaydulin}
\email{ruslan.shaydulin@jpmchase.com}
\author{Marco Pistoia}
\affiliation{Global Technology Applied Research, JPMorganChase, New York, NY 10001, USA}

\begin{abstract}
The quantum approximate optimization algorithm (QAOA) and quantum annealing are two of the most popular quantum optimization heuristics. While QAOA is known to be able to approximate quantum annealing, the approximation requires QAOA angles to vanish with the problem size $n$, whereas optimized QAOA angles are observed to be size-independent for small $n$ and constant in the infinite-size limit. This fact led to a folklore belief that QAOA has a mechanism that is fundamentally different from quantum annealing. In this work, we provide evidence against this by analytically showing that QAOA energy approximates that of quantum annealing under two conditions, namely that angles vary smoothly from one layer to the next and that the sum is bounded by a constant. These conditions are known to hold for near-optimal QAOA angles empirically. Our proof relies on a series expansion of QAOA energy in sum of angles, which we show converges to quantum annealing limit as QAOA depth grows for constant sum of angles even if angles do not vanish with problem size $n$. A corollary of our results is a quadratic improvement for the bound on depth required to compile Trotterized quantum annealing of the SK model in the average case.
\end{abstract}

\maketitle

\section{Introduction}

The Quantum Approximate Optimization Algorithm (QAOA) \cite{Hogg2000,Hogg2000search,farhi_qaoa} and quantum annealing are among the most popular optimization heuristics. For the purposes of this paper, we will use ``quantum annealing'' to refer to a heuristic version of the quantum adiabatic algorithm~\cite{farhi_adiabatic} wherein the total evolution time is shorter than the (typically, exponential) value prescribed by the adiabatic theorem~\cite{Albash2018}. Part of the appeal of these algorithms resides in the simplicity of implementation, making them a viable target for nearer-term quantum computers~\cite{ICCAD_qaoapara} and enabling small-scale demonstrations on quantum devices available today~\cite{Shaydulin2023npgeq,Pelofske2023,Pelofske2024,2409.12104,Tasseff2024}. 

The performance of both QAOA and quantum annealing depends on the choice of schedule. In the context of QAOA, the schedule is specified by hyperparameters that are commonly referred to as \textit{angles} and can be interpreted as Hamiltonian evolution times. 
QAOA with small angles (order $1/n$ where $n$ is the problem {instance} size) can provably recover the quantum adiabatic algorithm \cite{farhi_adiabatic}. However, optimal QAOA performance requires angles that are much larger than prescribed by this adiabatic regime.
For instance, if the number of circuit layers is fixed and the problem instance size tends to infinity, optimal QAOA angles were observed to be of constant order for a broad family of optimization problems \cite{qaoa_maxcut_high_girth,qaoa_ksat,qaoa_labs}\footnote{Here, ``constant" assumes a conventional normalization of the problem's cost function such that its optimum is of order the instance size $n$. We note that this is different from the standard normalization used in Hamiltonian simulation community, wherein the cost Hamiltonian is normalized to have norm one. This inconsistency is why the equivalence we establish is between QAOA with \textit{constant} angles and quantum annealing with \textit{linear} time.}. 
Correspondingly, quantum annealing is typically executed with a short time limited by the coherence time of hardware, far from the adiabatic regime.

Recently, substantial progress has been made in understanding QAOA  with a constant number of layers (depth) in infinite size limit~\cite{qaoa_sk,qaoa_maxcut_high_girth,qaoa_spin_glass_models,qaoa_ksat,qaoa_spiked_tensor,qaoa_quantum_problems_zhou,qaoa_quantum_problems_sud}. From a practical perspective, these methods allow to search for good QAOA angles on a classical computer, removing the need for optimizing the parameters of a quantum circuit on a quantum computer. Unfortunately, for all these methods, the time and memory complexity scale exponentially with the depth, rendering their numerical implementation impractical beyond small constant depth ---in practice, up to 20 \cite{qaoa_maxcut_high_girth}. Moreover, to the best of our knowledge, there is currently no sound mathematical understanding on the infinite-depth %
limit, unlike in the quantum adiabatic algorithm case. This is despite optimal QAOA schedules showing stark resemblances to those used in quantum annealing \cite{qaoa_performance_mechanism_implementation,qaoa_beyond_low_depth}, such as smoothly varying parameters between consecutive layers and special boundary values.

In this work, we establish an equivalence between infinite-depth QAOA with constant angles and linear-time quantum annealing. Specifically, we prove that if the angles are smaller than an absolute constant independent of the instance size, QAOA achieves energy density arbitrarily close to that of quantum annealing with corresponding schedule for sufficiently large depth (Section~\ref{sec:approximation_constant_time_annealing_constant_time_qaoa} and Theorem~\ref{th:approximation_continuous_time_annealing_qaoa} therein). More precisely, we prove this equivalence result for the disorder-average of the energy density achieved by QAOA on the Sherrington-Kirkpatrick \sbedit{(SK)} model. \sbedit{Furthermore, for our choice of QAOA schedule, we show that the SK-QAOA expected cost function concentrates across the disorder, and does so uniformly in $p$ in the infinite-size limit; we also prove a similar result for the cost output by linear-time quantum annealing. From these concentration properties, the approximation result between QAOA and quantum annealing applies not only to the disorder-averaged expected costs, but also instance-wise with high probability in the infinite-size limit.} Note that this approximation result is nontrivial, since with angles independent of $n$, one cannot invoke usual results for products of non-commutative operator exponentials, e.g. the Baker-Campbell-Hausdorff formula, to constant order (Section~\ref{sec:trotter_analysis_qaoa_vs_qaa}). We nonetheless note that qualitative and numerical arguments for the validity of this expansion in the constant angles regime were provided in \cite{Wurtz2022}.

We show numerically that our convergence results apply to QAOA with angles smaller than but close to the the optimized infinite-size-limit angles of Ref.~\cite{qaoa_maxcut_high_girth} (Section~\ref{sec:numerical_results_equivalence_qaoa_qaa}). While we do not observe arbitrarily close approximation at exactly optimized angles, we nonetheless see a strong correlation between QAOA and quantum annealing in this regime. Remarkably, for angles even slightly larger than optimal, we observe that QAOA energy \textit{diverges} from that of quantum annealing. We leave to future work a deeper investigation of why optimized QAOA angles correspond to the inflection point separating the regime of convergence to quantum annealing and divergence from it.

On a practical level, our results suggest that {linear-time} (and, more speculatively, beyond {linear-time}) quantum annealing can be compiled with a more aggressive Trotter step than naive generic bounds suggest. 
Specifically, when compiled to quantum gates, the running time of Trotterized Hamiltonian evolution for a sum of two fast-forwardable Hamiltonians is not proportional to the physical total Hamiltonian evolution time, but rather to the number of Trotter layers, with at most a logarithmic dependence on each layer's Hamiltonian evolution time. In this context, a more aggressive Trotter step, allowing for less layers, is desirable. For a fixed total evolution time, our results allow for a Trotter step of size $\mathcal{O}(n)$, or equivalently QAOA angles of size $\mathcal{O}(1)$, hence constant number of Trotter layers, while standard Trotter error bounds require a constant Trotter step (QAOA angles of size $1/n$) and $n$ layers.
Consequently, our results imply a reduction in gate count and logical circuit running time by a linear factor $n$. 

On a more fundamental level, our results show that QAOA at small (but $n$-independent) angles can be understood by the perhaps more tractable model of linear-time analog quantum annealing. This raises the question whether QAOA at constant (but not necessarily small) angles, covering infinite-size optimized angles \cite{qaoa_maxcut_high_girth}, can be understood by another proxy analog Hamiltonian evolution model. The identification of such a model would greatly improve the understanding of the QAOA mechanism at large depth in the infinite size limit, which at the moment remains unsolved.

Our central technical contribution is an expansion of QAOA energy around the non-interacting limit (phase operator angles $\bm\gamma=0$) in sum of angles. In the non-interacting limit, QAOA circuit includes only the mixer operator (i.e., $e^{-i\beta \sum_jX_j}$) and has a natural continuous equivalent since all QAOA layers commute and there is no discretization error. Surprisingly, we find that this expansion converges to the \textit{same continuous limit} when the angles are constant with system size as it does in the ``Trotter approximation regime'' when the angles vanish as $1/n$. We remark that this expansion holds whether QAOA is used as an approximate optimizer (see e.g. \cite{qaoa_sk,qaoa_maxcut_high_girth,qaoa_spin_glass_models}) or an exact solver (see e.g. \cite{qaoa_labs,qaoa_ksat,qaoa_near_symmetric_optimization_problems}).

\subsection{Summary of technical results}
\label{sec:summary_technical_results}

We now give a brief overview of our proof techniques. The object of interest is the energy produced by fixed-angles QAOA applied to a random instance of the Sherrington-Kirkpatrick (SK) model (referred to as SK-QAOA below). 
As shown by earlier work~\cite{qaoa_sk}, running QAOA with fixed, instance-independent angles on a random SK model instance produces a non-trivial energy in the average-instance case and after taking the thermodynamic ($n \to \infty$) limit. The ratio between QAOA energy and the optimal energy is further empirically conjectured to reach 1 as the number of QAOA layers $p$ goes to infinity~\cite{qaoa_maxcut_high_girth}. 

In this study, rather than considering the $n \to \infty$ limit, we work at finite $n$. 
In this setting, we give a representation of the instance-averaged SK-QAOA energy as a Quadratic Generalized Multinomial Sum (QGMS) in Appendix Section~\ref{sec:sk_qaoa_qgms}. Generalized Multinomial Sums were considered in the context of QAOA in earlier works \cite{qaoa_spin_glass_models,qaoa_ksat,qaoa_spiked_tensor}, with only the quadratic case being relevant to this study; basic technical definitions are provided in Appendix Section~\ref{sec:qgms_background}, with more advanced ones deferred to Appendix Section~\ref{sec:qgms_saddle_point_correlations_tensors}. We express the QAOA energy in terms of a specific family of objects associated to a QGMS, called \textit{QGMS moments}. We then perform a convergent power series expansion of the QGMS moments as a function of the interaction parameter. In the case of SK-QAOA, the interaction parameter corresponds to the maximum magnitude $\gamma_{\mathrm{max}}$ of $\bm{\gamma}$ angles. However, the method can be described in a more general setting, by abstracting the QGMS as well as the definition of the interaction parameter from SK-QAOA. The expansion of QGMS moments in this abstracted setting is detailed in Appendix Section~\ref{sec:pqgms_series_expansion_noninteracting_limit}. More precisely, this expansion consists of two nested expansions. The first level of expansion, described in Section~\ref{sec:qgms_moments_series_expansion} and embodied by Proposition~\ref{prop:qgms_integral_series_expansion}, expresses the 
QGMS moments in terms of thermodynamic limit quantities called \textit{correlation tensors}. These may be interpreted as multi-point functions in the standard quantum many-body physics sense, and can (in principle) be computed elementarily from $n \to \infty$ SK-QAOA objects introduced in \cite{qaoa_sk,qaoa_maxcut_high_girth}. The second level of expansion, described in Section~\ref{sec:pqgms_saddle_point_expansion}, expands each correlation tensor as a series in \textit{noninteracting correlation tensors}. 
Loosely speaking, noninteracting correlation tensors are the values assumed by correlation tensors in the $\bm\gamma = \bm{0}$ limit of QAOA angles; in this case, the QAOA circuit collapses to unitaries acting independently over qubits, hence the ``noninteracting" qualifier. 

With this expansion in hand, we assume a specific ansatz for the QAOA angles, based on the discretization of smooth functions. At a high level, this ansatz formalizes the idea that angles should be individually small, vary mildly from one layer to another, and that their sum should be upper-bounded by a constant. Under this ansatz, we show that the noninteracting correlation tensors converge to a well-defined continuous limit, leading to Theorem~\ref{th:approximation_continuous_time_annealing_qaoa}. Since the QAOA energy was ultimately expressed in terms of noninteracting correlation tensors by the previously derived nested expansion, a sequence of elementary arguments leads to identifying a continuous limit for the QAOA energy. Surprisingly, the same limit is attained regardless of whether the angles vanish with $n$. Concretely, the continuous limit when angles are $O(1/n)$ is equal to the limit for when angles are smaller than an absolute $n$-independent constant. This equivalence of convergence is the main technical insight of our work.

\section{Background and related work}
\label{sec:background}

In this work, we analyze QAOA and quantum annealing in the approximate optimization setting, with the goal of producing approximate minima of the Sherrington-Kirkpatrick model. The following provides a precise definition of this random optimization problem:

\begin{definition}[Sherrington-Kirkpatrick (SK) model]
\label{def:sk_model}
The Sherrington-Kirkpatrick (SK) model \cite{sk_original_paper} at size $n$ is a random optimization problem over $n$ spin variables $\bm{\sigma} = \left(\sigma_j\right)_{1 \leq j \leq n} \in \{1, -1\}^n$, defined by cost function:
\begin{align}
    C\left(\bm\sigma\right) & := \frac{1}{\sqrt{n}}\sum_{1 \leq j < k \leq n}J_{j,\,k}\sigma_j\sigma_k, \quad \forall \bm{\sigma} = \left(\sigma_j\right)_{1 \leq j \leq n},
\end{align}
parametrized by random i.i.d normal variables $\bm{J} := \left(J_{j,\,k}\right)_{1 \leq j < k\leq n}$, $J_{j,\,k} \sim \mathcal{N}\left(0, 1\right)$. SK model is encoded on qubits by a cost Hamiltonian
\begin{align}
    C & = \frac{1}{\sqrt{n}}\sum_{1 \leq j < k \leq n}J_{j,\,k}Z_jZ_k.
\end{align}
\end{definition}

Efficient classical algorithms exist for obtaining $\epsilon$-approximate solutions of SK model if $\epsilon$ is independent of $n$~\cite{sk_solve_amp}. However, no efficient classical or quantum algorithms are known for solving SK to arbitrary precision (e.g. $\epsilon \approx 1/n$).

\subsection{The quantum approximate optimization algorithm}

The Quantum Approximate Optimization Algorithm (QAOA)~\cite{farhi_qaoa,Hogg2000,Hogg2000search} solves optimization problems by preparing a quantum state alternating Hamiltonian evolution under the cost Hamiltonian $C$ and the mixing Hamiltonian $B$:
\begin{align}
    \ket{\bm\gamma, \bm\beta} & = \overleftarrow{\prod_{t = 1}^p}e^{-i\beta_tB}e^{-i\gamma_tC}\ket{+}^{\otimes n},\label{eq:def_qaoa_state}
\end{align}
where $B := \sum_{1 \leq j \leq n}X_j$ and $X_j$ is the Pauli $X$ matrix acting on qubit $j$. 

The exponentiated cost Hamiltonian $e^{-i\gamma C}$ and exponentiated mixer Hamiltonian $e^{-i\beta B}$ are referred to as the \textit{cost unitary} and \textit{mixer unitary} respectively. The parameter $p$ ---the number of cost or mixer unitaries--- is called the \textit{number of layers}, or the \textit{depth} of QAOA. After being prepared, the state is measured in the computational basis, hopefully producing a low-cost bitstring. The parameters $\bm\gamma, \bm\beta$ are commonly referred to as \textit{QAOA angles} and are not specified in the description of the algorithm. We define the following quantities, which are important to our analysis.

\begin{definition}[{Total $\gamma$ and $\beta$ angles}]
\label{def:total_gamma_beta_evolution_time}
Given a QAOA schedule with $p$ layers given by angles $\bm\gamma = \left(\gamma_1, \ldots, \gamma_p\right)$ and $\bm\beta = \left(\beta_1, \ldots, \beta_p\right)$, we define the \textnormal{{total $\gamma$ angle}} as:
\begin{align}
    \gamma_{\mathrm{tot}} & := \sum_{1 \leq t \leq p}\left|\gamma_t\right|\label{eq:total_gamma_evolution_time_definition}
\end{align}
and the \textnormal{{total $\beta$ angle}} as:
\begin{align}
    \beta_{\mathrm{tot}} & := \sum_{1 \leq t \leq p}\left|\beta_t\right|.\label{eq:total_beta_evolution_time_definition}
\end{align}
\end{definition}

The total $\gamma$ and $\beta$ angles introduced in Definition \ref{def:total_gamma_beta_evolution_time} can be understood as Hamiltonian evolution times. For instance, $n\gamma_t$ is the physical Hamiltonian evolution time under Hamiltonian $C$ when applying unitary
\begin{align}
    \exp\left(-i\gamma_tC\right) = \exp\left(-in\gamma_t\left(C/n\right)\right).
\end{align}
As a result, the total evolution time under Hamiltonian $C$ is $n\gamma_{\mathrm{tot}}$. The $n$ normalization factor comes from the fact that in the QAOA literature, cost Hamiltonians are commonly normalized so their operator norm $\left\lVert \cdot \right\rVert_{\infty}$ is of order the instance size $n$, whereas in the Hamiltonian simulation literature, Hamiltonians are normalized to a size-independent constant. We choose to adopt the latter convention, hence the need for the $n$ normalization besides the QAOA angles.
This definition of evolution time is furthermore natural from a quantum computing perspective, in the sense that implementing $\exp\left(-i\tau H\right)$ for a generic Hamiltonian $H$, $\left\lVert H \right\rVert_{\infty} = 1$ requires physical time $\Omega(\tau)$ (no fast-forwarding theorem, e.g. \cite[Theorem 3]{berry_efficient_quantum_algorithms_sparse_hamiltonians}, \cite[Theorem 2]{Haah2021}). 

It is common practice to optimize the QAOA parameters with respect to the average cost of a sampled bitstring $\braket{\bm\gamma, \bm\beta|C|\bm\gamma, \bm\beta}$ or probability of measuring an optimal bitstring $\braket{\bm{\gamma}, \bm{\beta}|\Pi_{\ker C}|\bm{\gamma}, \bm{\beta}}$, where $\Pi_{\ker C}$ the the projector onto the kernel of $C$. This has motivated the development of classically computable formulae to evaluate the relevant objective function without a quantum computer~\cite{quadratic_single_layer_qaoa_expectations,qaoa_maxcut_high_girth,qaoa_spin_glass_models,qaoa_ksat}, though they are restricted to the infinite size $n \to \infty$ and their evaluation typically scales exponentially in the number of layers $p$, making it prohibitively expensive to optimize parameters for large depth. Optimizing these formulae gives parameters that are of constant order and that do not vanish with problem size, highlighting the difference between the regime where QAOA is performant and the regime where QAOA trivially approximates the adiabatic evolution. Importantly, the total angles grow with depth $p$. For problems classes where such formulae are not readily available, optimizing the objective for a few small instances gives parameters that generalize well for most instances~\cite{qaoa_labs,qaoa_weighted_maxcut,2502.04277}. 

\subsection{Quantum adiabatic algorithm and quantum annealing}
The Quantum Adiabatic Algorithm (QAA) aims at producing a minimum cost bitstring of $C$ by performing {the following time-dependent Hamiltonian evolution:}
\begin{align}
    \ket{\Psi\left(0\right)} & := \ket{+}^{\otimes n},\label{eq:quantum_adiabatic_initial_state}\\
    i\frac{\mathrm{d}\ket{\Psi\left(u\right)}}{\mathrm{d}u} & = H(u)\ket{\Psi(u)}, \qquad u \in [0, 1],\label{eq:quantum_adiabatic_evolution}\\
    H\left(u\right) & := T\widetilde{H}\left(u\right),\\
    \widetilde{H}\left(u\right) & := \left(1 - s(u)\right)(-B) + s(u)C.\label{eq:quantum_adiabatic_hamiltonian}
\end{align}
In the above equations, $s: [0, 1] \to [0, 1]$ is a smooth function taking boundary values $s(0) = 0, s(1) = 1$. Hence, from $u = 0$ to $u = 1$, time-dependent Hamiltonian $H(u)$ smoothly interpolates between $-B$ and $C$, {with evolution parameter $T$ controlling the speed of the interpolation. Note we took the time parameter in the Schrodinger equation as a dimensionless parameter here, absorbing $T$ in the definition of the interpolating Hamiltonian $H\left(u\right)$ instead; this choice will prove more convenient when relating annealing to QAOA and describing numerical experiments in Section \ref{sec:numerical_results_equivalence_qaoa_qaa}.} Let us define the minimum spectral gap of all Hamiltonians $\widetilde{H}(u)$:
\begin{align}
    \Delta_{\min} & := \min_{u \in [0, 1]}\left(\mathrm{spectral\,gap\,of\,} \widetilde{H}(u)\right)\\
    & = \min_{u \in [0, 1]}\left(\lambda_1\left(\widetilde{H}(u)\right) - \lambda_0\left(\widetilde{H}(u)\right)\right).
\end{align}
Then, the adiabatic theorem states {that} for
\begin{align}
    T \gtrsim \frac{1}{\Delta_{\mathrm{min}}},\label{eq:adiabatic_condition}
\end{align}
the final state is close to the computational basis state $\ket{\bm{x}^*}$ minimizing classical cost function $C$:
\begin{align}
    \left|\braket{\bm{x}^*|\Psi\left(1\right)}\right| & \gtrsim 1.
\end{align}
In this informal statement, we assumed for simplicity non-vanishing of the minimum spectral gap, as well as unicity of the minimizer $\bm{x}^*$ of $C$. In this work, we focus on the regime where the evolution time may be shorter than required by condition \eqref{eq:adiabatic_condition}. In this regime, we will refer to QAA as quantum annealing.

\subsection{Connection between QAOA and quantum annealing}

The seminal paper of Farhi et al.~\cite{farhi_qaoa} already notes the connection between QAOA and quantum annealing. Since both QAOA and Trotterized quantum annealing involve alternation between evolution with phase and mixing operators, appropriate choice of QAOA angles enables matching the Trotterization of quantum annealing and, with sufficient number of steps, approximate the continuous quantum annealing. This proof is formalized in Ref.~\cite{Binkowski2024}. 
However, standard Trotter error analysis along these lines does not capture QAOA with optimized angles. Specifically, as we show in Sec.~\ref{sec:trotter_analysis_qaoa_vs_qaa}, such analysis requires angles to vanish with problem size as $1/n$ to control the approximation error. At the same time, many theoretical and empirical studies (see e.g. \cite{qaoa_sk,qaoa_maxcut_low_girth,qaoa_spin_glass_models,qaoa_ksat,qaoa_labs,qaoa_spiked_tensor,qaoa_maxcut_low_girth,qaoa_hundreds_qubits,qaoa_beyond_low_depth,qaoa_phase_diagram_quantum_chemistry} for recent examples) observe that QAOA angles must remain constant as problem size grows to achieve good performance. Therefore, the simple Trotter analysis does not capture the regime in which QAOA is performant, leaving open the question of the mechanism by which QAOA solves optimization problems and prompting further investigation.

Despite the theoretical challenges of connecting the two algorithms, several works have tried to analyze QAOA through the lens of the QAA. Ref.~\cite{qaoa_performance_mechanism_implementation} considered the dynamics of the quantum state produced by QAOA applied to diverse instances of the MAXCUT problem. For some instances, QAOA was found to behave adiabatically, while for others, non-adiabatic features were clearly observed. In Ref.~\cite{qaoa_alignment_mixer_initial_state}, moving beyond the present work's setting of unconstrained optimization over bitstrings, the authors observed that alignment between QAOA mixer and initial state improved the QAOA success probability, as should be expected with the QAA. It is also folklore knowledge (see e.g. QAOA angles plots from Refs.~\cite{qaoa_maxcut_high_girth,qaoa_labs}) that the parameters controlling the Hamiltonian evolution in QAOA and the QAA obey similar boundary conditions and are ``smoothly varying in time".
Using this insight, Ref.~\cite{quantum_annealing_initialization_qaoa} proposed to guess QAOA parameters based on predicted good quantum adiabatic parameters. On a more fundamental level, Ref.~\cite{qaoa_beyond_low_depth} proposes a semi-rigorous analysis of the QAOA state dynamics beyond low depth assuming smooth variation of parameters, relying on insights from the unitary adiabatic theorem (see e.g. Ref.~\cite{unitary_adiabatic_theorem} for an up-to-date presentation and proof). Similar qualitative observations on the performance of QAOA with varying magnitude of angles were made in Ref.~\cite{Wurtz2022}. While these works produce empirically correct predictions, they may not be considered fully rigorous when the QAOA angles have magnitude independent of the problem instance size.

\section{Results}
\label{sec:results}

\subsection{Equivalence between {linear}-time annealing and QAOA {with constant total angle}}
\label{sec:approximation_constant_time_annealing_constant_time_qaoa}

In this Section, we informally state our main theoretical result and illustrate it numerically, with full proofs being deferred to the appendices. Our result focuses on QAOA applied to the approximate optimization of the Sherrington-Kirkpatrick model for simplicity; we nonetheless believe they may extend to a broader variety of problems, including in the exact solver setting.

Our main technical result connects QAOA with constant order ($n$-independent) magnitude angles to {linear}-time annealing. To establish this connection, we 
need to define a correspondence between the parameters of QAOA and those of the annealing schedule. 

\begin{definition}[QAOA angles derived from continuous annealing schedule]
\label{def:continuous_schedule_informal}
Consider quantum annealing with schedules $\gamma^{\mathrm{cont}}: [0, 1] \longrightarrow \mathbf{R}$, $\beta^{\mathrm{cont}}: [0, 1] \longrightarrow \mathbf{R}$ defined by the time-dependent Hamiltonian
\begin{align}
    H\left(u\right) & := \gamma^{\mathrm{cont}}\left(u\right)C + \beta^{\mathrm{cont}}\left(u\right)B \qquad \forall u \in [0, 1].\label{eq:continuous_time_annealing_schedule_informal}
\end{align}

Then we define QAOA angles corresponding to the discretization of this annealing schedule as 
\begin{align}
    \gamma_t & := \frac{1}{p+1}\gamma^{\mathrm{cont}}\left(\frac{t - 1}{p + 1/2}\right) & \forall 1 \leq t \leq p,\label{eq:gamma_from_continuum_main}\\
    \beta_t & := \int_{(t - 1)/(p + 1/2)}^{t/(p + 1/2)}\mathrm{d}x\,\beta^{\mathrm{cont}}\left(x\right) & \forall 1 \leq t \leq p.\label{eq:beta_from_continuum_main}
\end{align}
\end{definition}

We note that we use slightly different discretizations for parameters $\gamma$ (left end of the interval) and $\beta$ (average over the interval) due to the particularities of the proof techniques. However, both discretizations give approximately the same values, namely 
\begin{equation}\label{eq:discretzation_approximate}
\gamma_t \approx \frac{1}{p}\gamma^{\mathrm{cont}}\left(\frac{t}{p}\right), \;\;\; \beta_t \approx \frac{1}{p}\beta^{\mathrm{cont}}\left(\frac{t}{p}\right).
\end{equation}

{The total $\gamma$, $\beta$ angles may be defined for a continuous schedule analogously to a discrete one} (Definition \ref{def:total_gamma_beta_evolution_time}):

\begin{definition}[{Total evolution times for continuous schedule}]
\label{eq:total_gamma_beta_evolution_time_continuous}
For a continuous schedule $\gamma^{\mathrm{cont}}, \beta^{\mathrm{cont}}$ the {total $\gamma$ angle} is defined as:
\begin{align}
    {\gamma_{\mathrm{tot}}^{\mathrm{cont}}} & := \int_0^1\!\mathrm{d}s\,\left|\gamma^{\mathrm{cont}}\left(s\right)\right|.
\end{align}
Likewise, the {total $\beta$ angle} is defined as:
\begin{align}
    {\beta^{\mathrm{cont}}_{\mathrm{tot}}} & := \int_0^1\!\mathrm{d}s\,\left|\beta^{\mathrm{cont}}\left(s\right)\right|.
\end{align}
\end{definition}

The discrete total evolution times (Definition \ref{def:total_gamma_beta_evolution_time}) approach the continuous ones $\gamma_{\mathrm{tot}} \rightarrow \gamma^{\mathrm{cont}}_{\mathrm{tot}}, \beta_{\mathrm{tot}} \rightarrow \beta^{\mathrm{cont}}_{\mathrm{tot}}$ in the fixed $n$, $p \to \infty$ limit. \sbedit{We note the following elementary bounds on the total angles and total evolution times:
\begin{align}
    \gamma^{\mathrm{cont}}_{\mathrm{tot}}, \gamma_{\mathrm{tot}} \leq \gamma_{\mathrm{max}}, && \beta^{\mathrm{cont}}_{\mathrm{tot}}, \beta_{\mathrm{tot}} \leq \beta_{\mathrm{max}},\label{eq:total_angle_maximum_angle_bound}
\end{align}
where
\begin{align}
    \gamma_{\mathrm{max}} := \max_{s \in [0, 1]}\left|\gamma^{\mathrm{cont}}\left(s\right)\right|, && \beta_{\mathrm{max}} := \max_{s \in [0, 1]}\left|\beta^{\mathrm{cont}}\left(s\right)\right|
\end{align}
are the maxima of the continuous schedules. The bounds in the discrete case ($\gamma_{\mathrm{tot}}, \beta_{\mathrm{tot}}$) hold uniformly in $p$ for our choice of finite $p$ schedules Definition~\ref{def:continuous_schedule_informal}.} We remark that while there are many continuous schedules corresponding to a given QAOA angle sequence, our results apply to any of them provided the conditions of Theorem~\ref{th:approximation_continuous_time_annealing_qaoa} are satisfied, i.e. the schedule is bounded and Lipschitz-continuous.

In this work, we consider the instance-averaged energy produced by QAOA and continuous-time quantum annealing, parametrized by $\gamma^{\mathrm{cont}}, \beta^{\mathrm{cont}}$ (Equation \ref{eq:continuous_time_annealing_schedule_informal}), and QAOA at finite number of layers, where the QAOA angles at any $p$ are understood to be defined from a continuous schedule (Equations \ref{eq:gamma_from_continuum_main}, \ref{eq:beta_from_continuum_main}). We introduce specific notations for the energy produced in both these cases:

\begin{definition}[QAOA and quantum annealing energy]
Consider a continuous schedule $\gamma^{\mathrm{cont}}, \beta^{\mathrm{cont}}$ and corresponding quantum annealing and QAOA schedules given by Def.~\ref{def:continuous_schedule_informal}. 

Consider QAOA applied to the SK model defined in Def.~\ref{def:sk_model}. Denote by $\ket{\Psi_{p,\,n}}$ state produced by QAOA with $p$ layers on $n$ qubits defined by Eq.~\ref{eq:def_qaoa_state}. Then the instance-averaged energy output by SK-QAOA at size $n$ is defined by
\begin{align}
    \nu_{p,\,n} & := \mathbb{E}\braket{\Psi_{p,\,n}|C_n/n|\Psi_{p,\,n}},\label{eq:qaoa_energy_definition}
\end{align}
with the expectation taken over random SK couplings.

Likewise, denote by $\ket{\Psi_{\infty,\,n}}$ the state produced by quantum annealing (Eqs.~\ref{eq:quantum_adiabatic_initial_state},~\ref{eq:quantum_adiabatic_evolution},~\ref{eq:quantum_adiabatic_hamiltonian}) with time-dependent Hamiltonian $H(u)$ defined by Equation \ref{eq:continuous_time_annealing_schedule_informal}.
Then we define the instance-averaged energy output by continuous-time quantum annealing at size $n$ by
\begin{align}
    \nu_{\infty,\,n} & = \mathbb{E}\braket{\Psi_{\infty,\,n}|C_n/n|\Psi_{\infty,\,n}}.\label{eq:qaoa_annealing_energy_definition}
\end{align}
\end{definition}

We remark that by Euler discretization of ordinary differential equations, it holds 
\begin{align}
    \ket{\Psi_{p,\,n}} \xrightarrow[p \to \infty]{} \ket{\Psi_{\infty,\,n}}.
\end{align}
\sbedit{
From there, by a simple dominated convergence argument (deferred to Lemma~\ref{lemma:qa_disorder_averages} from Appendix~\ref{sec:other_moments_concentration}),
\begin{align}
    \nu_{p,\,n} & \xrightarrow[p \to \infty]{} \nu_{\infty,\,n}.
\end{align}
}
However, this simple analysis requires $p$ to grow with $n$ to achieve a fixed convergence error, as the norm of the Hamiltonians grows with $n$ (see Section~\ref{sec:trotter_analysis_qaoa_vs_qaa} for a detailed discussion). To capture the regime in which QAOA is typically used and in which it performs well (i.e. with constant, size-independent angles), we prove the following result:

\begin{figure*}[t]
    \centering
    \includegraphics[width=\textwidth]{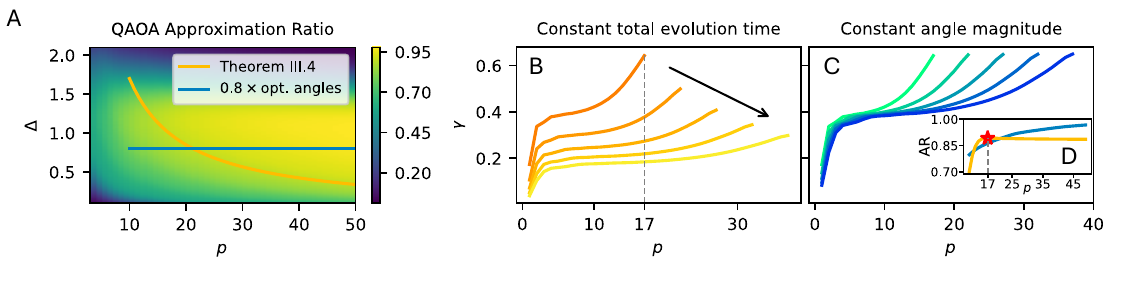}
    \caption{Summary of the numerical experiments. \textbf{A} Two numerically evaluated regimes overlaid onto an example QAOA performance diagram for a 20-spin SK model. Points along the orange line correspond to constant total evolution time and satisfy the conditions of Theorem~\ref{th:approximation_continuous_time_annealing_qaoa}, with example schedules for varying $p$ plotted in \textbf{B}. Points along the blue line correspond approximately to the conjectured optimal angle behavior, with example schedules in \textbf{C}. When conjectured optimal angles are used, QAOA approximation ratio approaches 1 (\textbf{D}), whereas for angles with constant total evolution time QAOA approximation ratio is flat. The point marked with red star corresponds exactly to infinite-size-limit optimized parameters for $p=17$ of Ref.~\cite{qaoa_maxcut_high_girth}. For lines in \textbf{D}, use the legend from \textbf{A}.} %
    \label{fig:numerical_intuition}
\end{figure*}

\begin{theorem}[Equivalence between quantum annealing and QAOA with $n$-independent angles]
\label{th:approximation_continuous_time_annealing_qaoa}
Consider QAOA applied to the SK model, with angles schedules $\bm\gamma, \bm\beta$ arising from the discretization of a fixed continuous schedule $\gamma^{\mathrm{cont}}, \beta^{\mathrm{cont}}$ as specified in Definition~\ref{def:continuous_schedule_informal}. Assume the continuous schedule is
bounded:
\begin{align}
    \left|\beta^{\mathrm{cont}}\left(s\right)\right| & \leq \beta_{\mathrm{max}},\\
    \left|\gamma^{\mathrm{cont}}\left(s\right)\right| & \leq \gamma_{\mathrm{max}},
\end{align}
with $\gamma^{\mathrm{cont}}$ being furthermore $M_{\gamma}$-Lipschitz:
\begin{align}\label{eq:lipschitz}
    \left|\gamma^{\mathrm{cont}}\left(s\right) - \gamma^{\mathrm{cont}}\left(s'\right)\right| & \leq M_{\gamma}\left|s - s'\right| \qquad \forall s, s' \in [0, 1].
\end{align}
Then, there exists an absolute constant $\gamma_{\mathrm{max}}^*$ such that whenever,
\begin{align}
    \gamma_{\mathrm{max}} < \gamma^*_{\mathrm{max}},\label{eq:main_theorem_max_angle_bound_condition}
\end{align}
the following additive error bound holds for the energy density output by QAOA run with discretized schedules and quantum annealing:
\begin{align}
    \left|\nu_{p,\,n} - \nu_{\infty,\,n}\right| & \leq \frac{c\left(\beta_{\mathrm{max}}, \gamma_{\mathrm{max}}, M_{\gamma}/\gamma_{\mathrm{max}}\right)}{p + 1}..\label{eq:approximation_continuous_time_annealing_qaoa}
\end{align}
In the above inequality, $c\left(\beta_{\mathrm{max}}, \gamma_{\mathrm{max}}, M_{\gamma}/\gamma_{\mathrm{max}}\right)$ is a function of the continuous schedules magnitudes $\beta_{\mathrm{max}}, \gamma_{\mathrm{max}}$ and the relative Lipschitz constant $M_{\gamma}/\gamma_{\mathrm{max}}$ of $\gamma^{\mathrm{cont}}$; this constant further remains bounded as these 3 parameters are. As a consequence of this error bound, for {constant} and sufficiently small maximum continuous $\gamma$ angle $\gamma_{\mathrm{max}}$, the difference can be made arbitrarily small (in additive terms) provided $p$ is chosen sufficiently large, uniformly in the instance size $n$. \sbedit{Besides, by concentration of the expected QAOA cost $\bra{\Psi_{p,\,n}}C_n/n\ket{\Psi_{p,\,n}}$ across the SK Gaussian disorder (Proposition~\ref{prop:uniform_concentration_qaoa} from Appendix~\ref{sec:other_moments_concentration}), and similarly for the expected quantum annealing cost (Proposition~\ref{prop:concentration_linear_time_qa}), the approximation result holds with high probability in the infinite size limit $n \to \infty$.}
\end{theorem}

\sbedit{Note that by inequalities \eqref{eq:total_angle_maximum_angle_bound} condition \eqref{eq:main_theorem_max_angle_bound_condition} of the Theorem constrains the total angle to satisfy an absolute bound:
\begin{align}
    {\gamma_{\mathrm{tot}}}, {\gamma_{\mathrm{tot}}^{\mathrm{cont}}} & < {\gamma^*_{\mathrm{max}}}. \label{eq:main_theorem_total_angle_bound_condition}
\end{align}
As discussed earlier, this restriction means that for sufficiently large $p$, the Theorem fails to capture the optimal angles regime, where it is conjectured $\gamma_{\mathrm{tot}} = \Theta(p)$. However, numerical experiments (Section~\ref{sec:numerical_results_equivalence_qaoa_qaa}) empirically suggest the result may extend to $\gamma_{\mathrm{max}} = \Theta(p)$, though the implicit constant cannot be taken sufficiently large to reach the optimal angles regime. Said differently, the error bounds of the theorem are presumably loose in terms of $\gamma_{\mathrm{max}}$; on the other hand, we numerically conjecture the $1/p$ dependence in $p$ to be tight. Finally, the result only shows closeness of the QAOA energy density to the quantum annealing one, and makes no claim about the closeness of states.
}

The proof of this theorem is deferred to Appendix~\ref{sec:sk_qaoa_energy_continuum_limit}. While Theorem~\ref{th:approximation_continuous_time_annealing_qaoa} focuses on QAOA with angles arising from a discretization of the annealing schedule, it is also possible to go in the opposite direction, i.e. start with some fixed QAOA angles and show equivalence to quantum annealing with a corresponding schedule. We do so in the numerical results in Sec.~\ref{sec:numerical_results_equivalence_qaoa_qaa}.

We highlight the crucial part in the theorem's claim, namely that the inequality \eqref{eq:approximation_continuous_time_annealing_qaoa} holds uniformly in $p \geq 1$. This is in sharp contrast to $p = \Omega(n)$ and $\gamma = \mathcal{O}(1/n)$ required by the standard analysis sketched in Sec.~\ref{sec:trotter_analysis_qaoa_vs_qaa}. Unfortunately, due to the absolute constant bound $\gamma^*_{\mathrm{tot}}$ on the total $\gamma$ angle, the conditions of Theorem~\ref{th:approximation_continuous_time_annealing_qaoa} are not satisfied for infinite-size optimal angles for arbitrary $p$ under a widely believed conjecture on the structure of the latter. Indeed, \cite{qaoa_maxcut_high_girth} provided numerical evidence that in this infinite size regime, optimal $\bm\gamma$ angles (as well as $\bm\beta$ angles) have magnitude of constant order, independent of $p$. Hence, their sum grows linearly with $p$ and will therefore exceed constant bound $\gamma^*_{\mathrm{tot}}$ for sufficiently large $p$. However, for the largest value considered in Ref.~\cite{qaoa_maxcut_high_girth}, i.e. $p=17$, our results still apply as we show in the following Section. 
Moreover, the central conclusion of the Theorem~\ref{th:approximation_continuous_time_annealing_qaoa}, namely that QAOA is equivalent to annealing, is observed to hold numerically even as the sum of angles grows as long as the magnitude at each step remains constant, matching the conjectured optimal behavior in large $p$ limit.

\subsection{QAOA with optimized angles is~equivalent to quantum annealing}
\label{sec:numerical_results_equivalence_qaoa_qaa}

We now provide numerical evidence that the equivalence between QAOA and quantum annealing shown analytically in Theorem~\ref{th:approximation_continuous_time_annealing_qaoa} holds in the regime where QAOA is most performant, that is for QAOA with optimized angles or angles closed to optimized. 
Optimized QAOA angles for a wide range of problems have the property that the angles vary gradually from one layer to the next~\cite{qaoa_sk,qaoa_labs,qaoa_maxcut_high_girth,qaoa_weighted_maxcut,2502.04277,qaoa_performance_mechanism_implementation}. In this regime, QAOA parameters for a small number of layers can be naturally extrapolated by converting them to a continuous annealing schedule and using a discretization of this continuous schedule at a larger depth. The specific procedure we use for constructing the equivalent annealing schedule and extrapolating the parameters in the numerical experiments in this Section is inspired by the Fourier extrapolation of Zhou et al.~\cite{qaoa_performance_mechanism_implementation} and is described in Appendix~\ref{sec:optimal_angles_continuous_schedules}. 

The performance of QAOA with gradually changing parameters can be summarized by a ``performance diagram''~\cite{qaoa_phase_diagram_quantum_chemistry,qaoa_beyond_low_depth}, an example of which for an SK model on 20 spins is given in Figure~\ref{fig:numerical_intuition}A. Starting from $p=17$ parameters optimized with respect to the infinite-size limit QAOA energy $\nu_{p,\infty}$~\cite{qaoa_maxcut_high_girth}, the performance diagram is drawn by evaluating QAOA performance with parameters extrapolated to larger $p$ and rescaled by some constant $\Delta$. The contour line of $\Delta=17/p$ corresponds to keeping the sum of angles fixed, with example schedules shown in Figure~\ref{fig:numerical_intuition}B. The contour line of $\Delta=1$ corresponds to the conjectured~\cite{qaoa_maxcut_high_girth} behavior of QAOA with optimal angles as $p$ grows (example schedules in Figure~\ref{fig:numerical_intuition}C). As we show later in this Section, for angles set exactly to their conjectured optimal values, we do not see an arbitrarily close convergence of QAOA to quantum annealing as predicted by Theorem~\ref{th:approximation_continuous_time_annealing_qaoa}. Therefore we instead highlight $\Delta=0.8$, which corresponds to angles that are close to but slightly smaller than their conjectured optimal values and for which the predictions of Theorem~\ref{th:approximation_continuous_time_annealing_qaoa} hold numerically.
Despite this suboptimal angles choice, as Figure~\ref{fig:numerical_intuition}D shows, the approximation ratio of QAOA approaches 1 (blue line). In contrast, for the constant total evolution time approximation ratio is flat with $p$ ($\Delta=17/p$, orange line). 

The formal procedure for obtaining the schedules is specified in Appendix~\ref{sec:optimal_angles_continuous_schedules}. We now summarize it. First, the continuous annealing schedule $\gamma^{\mathrm{cont}}_{\mathrm{reference}}$, $\beta^{\mathrm{cont}}_\mathrm{reference}$ is obtained from optimized QAOA angles. Second, the QAOA angles for a given value of $p$ and an equivalent annealing schedule are obtained by rescaling
\begin{align}
    \gamma^{\mathrm{cont}} & = \Delta\cdot p \cdot \gamma^{\mathrm{cont}}_{\mathrm{reference}}, \\
    \beta^{\mathrm{cont}} & = \Delta\cdot p \cdot \beta^{\mathrm{cont}}_{\mathrm{reference}},
\end{align}
and discretizing following Definition~\ref{def:continuous_schedule_informal}. 
We remark that, by construction, the discretization of \eqref{eq:gamma_from_continuum_main},~\eqref{eq:beta_from_continuum_main} ensures that the sum of QAOA angles (Def.~\ref{def:total_gamma_beta_evolution_time}) is approximately equal to the total annealing time (Def.~\ref{eq:total_gamma_beta_evolution_time_continuous}). Specifically, $\gamma_{\mathrm{tot}}\simeq 0.43 \cdot \Delta\cdot p$ and $\beta_{\mathrm{tot}}\simeq 0.32\cdot \Delta\cdot p$, where the difference between constants $0.43$, $0.32$ arises from relative magnitudes of parameters $\bm\gamma$, $\bm\beta$. 

\begin{figure*}[t]
    \centering
    \includegraphics[width=\textwidth]{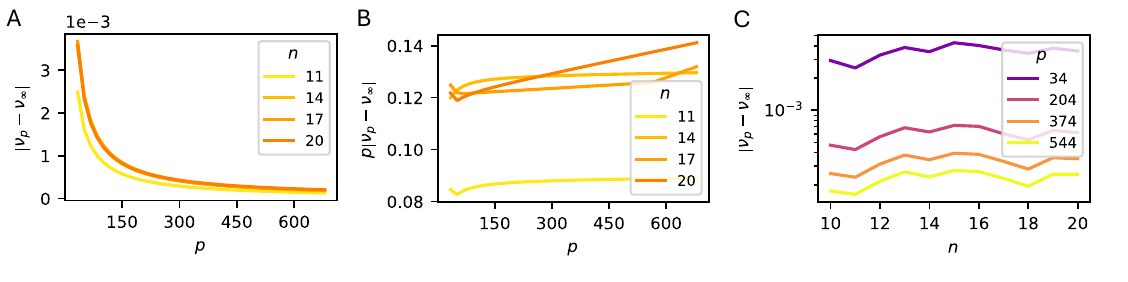}
    \caption{Equivalence between QAOA and quantum annealing for constant total evolution time. \textbf{A} Difference between energy achieved by QAOA and quantum annealing decays with $p$. The rate of decay is $1/p$ (\textbf{B}) and is independent of $n$ (\textbf{C}), as predicted by Theorem~\ref{th:approximation_continuous_time_annealing_qaoa}.}
    \label{fig:constant_evolution_time}
\end{figure*}

We begin by evaluating numerically the regime where QAOA depth is increased while the total evolution time (sum of QAOA angles) remains fixed, i.e. $\Delta \sim 1/p$. The condition \sbedit{
\eqref{eq:main_theorem_max_angle_bound_condition} of Theorem~\ref{th:approximation_continuous_time_annealing_qaoa}} is satisfied since $\gamma_{\mathrm{max}}\approx 0.65\cdot \Delta\cdot p$. An example for $\Delta = 17/p$ is shown by the line ``Theorem~\ref{th:approximation_continuous_time_annealing_qaoa}'' line in Figure~\ref{fig:numerical_intuition}A. The results for $\Delta = 17/p$ are shown in Figure~\ref{fig:constant_evolution_time}. We observe that as $p$ grows, the difference between QAOA and annealing energy $\left|\nu_{p,\,n} - \nu_{\infty,\,n}\right|$ goes down (Fig.~\ref{fig:constant_evolution_time}A), with the $\mathcal{O}(1/p)$ decay matching the prediction of Eq.~\ref{eq:approximation_continuous_time_annealing_qaoa} (Fig.~\ref{fig:constant_evolution_time}B). As predicted by Eq.~\ref{eq:approximation_continuous_time_annealing_qaoa}, there is no dependency on $n$ (Fig.~\ref{fig:constant_evolution_time}C). We present numerical results for other values of total evolution time in Appendix~\ref{sec:appendix_extra_numerics}.

\begin{figure*}[t]
    \centering
    \includegraphics[width=\textwidth]{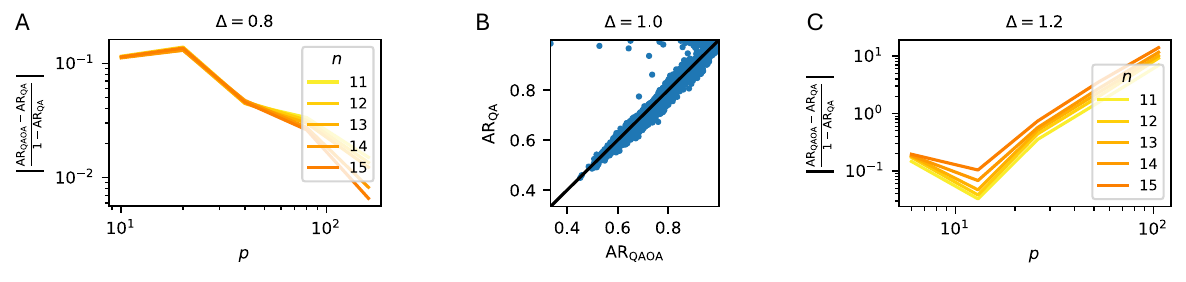}
    \caption{Equivalence between QAOA and quantum annealing for constant angle magnitude. \textbf{A} For angle magnitude close to but smaller than those corresponding to conjectured optimal QAOA parameters ($\Delta=0.8$), relative residual approximation ratio decays rapidly with $p$. \textbf{B} At conjectured optimal QAOA angles ($\Delta = 1$), a correlation is still observed between QAOA and annealing approximation ratios, despite the error between the two not vanishing in the large $p$ limit (see complementary numerical results in appendix \ref{sec:appendix_extra_numerics}). \textbf{C} For large $\Delta$, QAOA and quantum annealing energies diverge.}
    \label{fig:constant_angle_magnitude}
\end{figure*}

We next explore the regime which corresponds to optimized QAOA parameters and which is conjectured to correspond to optimal QAOA angles for large $p$, namely constant $\Delta$ and total evolution time growing as $\Delta\cdot p$. An example for $\Delta=0.8$ is shown by the line ``$0.8 \times$ optimal angles'' in Figure~\ref{fig:numerical_intuition}. The conditions of Theorem~\ref{th:approximation_continuous_time_annealing_qaoa} are no longer satisfied since the total evolution time is not bounded. We remark that for fixed $\Delta$, both QAOA and quantum annealing achieve approximation ratio that approaches 1 as $p$ grows (Fig~\ref{fig:numerical_intuition}D). Consequently, it may be the case that the difference in energy between two algorithms goes to zero simply by virtue of both of them solving the problem exactly. 

To mitigate this issue, we plot relative residual approximation ratio, defined as $(\mathrm{AR}_{\mathrm{QAOA}}-\mathrm{AR}_{\mathrm{QA}})/(1-\mathrm{AR}_{\mathrm{QA}})$, where $\mathrm{AR}_{\mathrm{QAOA}}$ ($\mathrm{AR}_{\mathrm{QA}})$ is the ratio between QAOA (quantum annealing) energy and the optimal energy. For this metric, we still observe that QAOA approximates quantum annealing well if angles are close to but smaller than the conjectured optimal ones, with the error decaying rapidly with $p$ (Figure~\ref{fig:constant_angle_magnitude}A  for $\Delta=0.8$, additional results in Appendix~\ref{sec:appendix_extra_numerics}). If parameters are set exactly to their conjectured optimal value ($\Delta=1$), the approximation error no longer vanishes (see Appendix~\ref{sec:appendix_extra_numerics}); however, there is still close correlation between the QAOA and annealing approximation ratios, as shown in Figure~\ref{fig:constant_angle_magnitude}B. Finally, if the angle magnitude is slightly larger than conjectured optimal, QAOA rapidly diverges from quantum annealing ($\Delta = 1.2$, Figure~\ref{fig:constant_angle_magnitude}C). The fact that the transition from approximation error vanishing with $p$ to error exploding with $p$ takes place at exactly the optimal QAOA angle magnitudes ($\Delta=1$) is not explained by our theory and merits further investigation.

\subsection{Technical overview}
\label{sec:small_total_gamma_angle_expansion_results}

\begin{figure*}
\begin{tikzpicture}[auto]

\node[draw, rectangle, text width=3cm, align=center, font=\small] (box1) {SK-QAOA energy \\ $\nu_{p, n}$};
\node[draw, rectangle, text width=2.5cm, align=center, right=0.3cm of box1, font=\small] (box2) {QGMS moments \\ $\frac{\partial S_n\left(\bm{\mu}\right)}{\partial\mu_{\alpha}}$, \\ $\frac{\partial^2S_n\left(\bm{\mu}\right)}{\partial\mu_{\alpha}^2}$};
\node[draw, rectangle, text width=3cm, align=center, right=0.3cm of box2, font=\small] (box3) {Correlations tensors \\ $\bm{C}^{(d)}$ of \\ $n$-independent order $d$};
\node[draw, rectangle, text width=2.5cm, align=center, right=0.3cm of box3, font=\small] (box4) {Noninteracting correlations \\ tensors $\bm{\overline{C}}^{(d)}$};
\node[draw, rectangle, text width=3cm, align=center, right=0.3cm of box4, font=\small] (box5) {Continuum limit of \\ noninteracting correlations \\ tensors $\overline{C}^{(d),\,\mathrm{cont}}$};

\draw[->, bend left] ([xshift=0.5cm]box1.north) to node[midway, above, text width=3cm, align=center, font=\small] {Representation of SK-QAOA energy as sum of QGMS moments (Appendix~\ref{sec:sk_qaoa_qgms})} ([xshift=-0.5cm]box2.north);
\draw[->, bend left] ([xshift=0.5cm]box2.north) to node[midway, above, text width=3.5cm, align=center, font=\small] {QGMS moments are sum of tensor networks of correlations tensors (Proposition~\ref{prop:qgms_integral_series_expansion})} ([xshift=-0.5cm]box3.north);
\draw[->, bend left] ([xshift=0.5cm]box3.north) to node[midway, above, text width=3cm, align=center, font=\small] {Expansion of correlations tensors around the noninteracting limit (Proposition~\ref{propSaddlePointEquationAnalyticity})} ([xshift=-0.5cm]box4.north);
\draw[->, bend left] ([xshift=0.5cm]box4.north) to node[midway, above, text width=3cm, align=center, font=\small] {Explicit calculation (Appendix~\ref{sec:noninteracting_correlations_tensors_continuum_limit})} ([xshift=-0.5cm]box5.north);

\end{tikzpicture}
\caption{Outline of general method to take the $p \to \infty$ limit in the constant total angle regime. The central insight is that the order of correlation tensors $\bm{C}^{(d)}$ can be shown to be independent of $n$. Expanding correlation tensors in terms of noninteracting correlation tensors which have a natural continuum limit, this give uniform in $n$ convergence of SK-QAOA energy to its continuum limit.}
\label{fig:proof_structure_outline}
\end{figure*}

In this Section, we review the main ideas behind the proof of our main theoretical result Theorem~\ref{th:approximation_continuous_time_annealing_qaoa}. The main line of proof is deferred to Appendix~\ref{sec:sk_qaoa_energy_continuum_limit}, with the core argument laid down in Section~\ref{sec:main_theorem_derivation}. For reader's convenience, we outline the flow of the argument in Fig.~\ref{fig:proof_structure_outline}.

\subsubsection{SK-QAOA energy from Quadratic Generalized Multinomial Sums (QGMS)}
\label{sec:qgms_background_main_text}

In this Section, we informally review elements of background on the main technical tool allowing the analysis of QAOA expectation: Quadratic Generalized Multinomial Sums (QGMS). Formal definitions and proofs are deferred to Appendix Section~\ref{sec:qgms_background} and \ref{sec:qgms_saddle_point_correlations_tensors}. A Quadratic Generalized Multinomial Sums is a sequence of functions $\left(S_n\right)_{n \geq 1}$, indexed by a non-negative integer $n$, of the following form:

\begin{align}
    S_n\left(\bm{\mu}\right) & = \sum_{\substack{\bm{n} = \left(n_{\bm{a}}\right)_{\bm{a} \in \mathcal{S}}\\\sum\limits_{\bm{a} \in \mathcal{S}}n_{\bm{a}} = n}}\binom{n}{\bm{n}}e^{\bm{n}^T\bm{L}^T\bm{L}\bm{n}/(2n) + \bm{\mu}^T\bm{L}\bm{n}/n}\prod_{\bm{a} \in \mathcal{S}}Q_{\bm{a}}^{n_{\bm{a}}},\label{eq:qgms_mgf_main_text}\\
    \bm{\mu} & = \left(\mu_{\alpha}\right)_{\alpha \in \mathcal{A}} \in \mathbf{C}^{\mathcal{A}}.
\end{align}
In the above equations, $\mathcal{A}, \mathcal{S}$ are finite index sets.  Function $S_n$ is defined over vectors $\bm{\mu} = \left(\mu_{\alpha}\right)_{\alpha \in \mathcal{A}}$ indexed by set $\mathcal{A}$. The sum is over $\mathcal{S}$-indexed tuples $\bm{n} = \left(n_{\bm{a}}\right)_{\bm{a} \in \mathcal{S}}$ of non-negative integers $n_{\bm{a}}$ summing to $n$; by finiteness of set $\mathcal{S}$, the sum has a finite number of terms. The multinomial coefficient, generalizing the binomial coefficient, is defined as:
\begin{align}
    \binom{n}{\bm{n}} & := \binom{n}{\left(n_{\bm{a}}\right)_{\bm{a} \in \mathcal{S}}} := \frac{n!}{\prod\limits_{\bm{a} \in \mathcal{S}}n_{\bm{a}}!}.
\end{align}
Besides,
\begin{align}
    \bm{Q} & := \left(Q_{\bm{a}}\right)_{\bm{a} \in \mathcal{S}}
\end{align}
is a vector indexed by $\mathcal{S}$, and
\begin{align}
    \bm{L} & := \left(L_{\alpha,\,\bm{a}}\right)_{\alpha \in \mathcal{A},\,\bm{a} \in \mathcal{S}}
\end{align}
is a matrix with rows indexed by $\mathcal{A}$ and columns indexed by $\mathcal{S}$. In Appendix Section~\ref{sec:sk_qaoa_qgms}, we will express the SK-QAOA instance-averaged energy from a specific QGMS. While the current overview does not depend on detailed understanding of this Appendix, a few facts about the QGMS describing the SK-QAOA energy will be relevant. First, for $p$-layers QAOA, finite index sets $\mathcal{A}, \mathcal{S}$ 
are defined as:
\begin{align}
    \mathcal{A} & := \mathcal{I}^2 = \left\{(r, s)\,:\,r, s \in \mathcal{I}\right\},\\
    \mathcal{S} & := \{1, -1\}^{\mathcal{I}} = \left\{\left(a_t\right)_{t \in \mathcal{I}}\,:\,\forall t \in \mathcal{I},\,a_t \in \{1, -1\}\right\},
\end{align}
where
\begin{align}
    \mathcal{I} & := \left\{0, 1, \ldots, 2p, 2p + 1\right\},
\end{align}
so that
\begin{align}
    \left|\mathcal{A}\right| & = \left(2p + 2\right)^2,\label{eq:sk_qaoa_energy_qgms_A_index_set_size_main_text}\\
    \left|\mathcal{S}\right| & = 2^{2p + 2}.
\end{align}
Second, it holds:
\begin{align}
    S_n\left(\bm{0}\right) & = \sum_{\substack{\bm{n} = \left(n_{\bm{a}}\right)_{\bm{a} \in \mathcal{S}}\\\sum\limits_{\bm{a} \in \mathcal{S}}n_{\bm{a}} = n}}\binom{n}{\bm{n}}\exp\left(\frac{1}{2n}\bm{n}^T\bm{L}^T\bm{L}\bm{n}\right)\prod_{\bm{a} \in \mathcal{S}}Q_{\bm{a}}^{n_{\bm{a}}}\nonumber\\
    & = 1.
\end{align}
This identity suggests to regard the family of complex numbers
\begin{align}
    \exp\left(\frac{1}{2n}\bm{n}^T\bm{L}^T\bm{L}\bm{n}\right)\prod_{\bm{a} \in \mathcal{S}}Q_{\bm{a}}^{n_{\bm{a}}},\label{eq:pseudo_probability_weights_configuration_numbers_main_text}
\end{align}
indexed by tuple $\bm{n}$, as a quasiprobability distribution over such tuples. From this interpretation, one may consider the ``pseudo-moments" of vector $\bm{n}$ under this pseudo-probability distribution. For instance, the generic second-order pseudo-moment, indexed by $\bm{b}, \bm{c} \in \mathcal{S}$, can be defined as:
\begin{align}
    \sum_{\substack{\bm{n} = \left(n_{\bm{a}}\right)_{\bm{a} \in \mathcal{S}}\\\sum\limits_{\bm{a} \in \mathcal{S}}n_{\bm{a}} = n}}n_{\bm{b}}n_{\bm{c}}\binom{n}{\bm{n}}\exp\left(\frac{1}{2n}\bm{n}^T\bm{L}^T\bm{L}\bm{n}\right)\prod_{\bm{a} \in \mathcal{S}}Q_{\bm{a}}^{n_{\bm{a}}}\label{eq:qgms_general_order_2_moment}
\end{align}
The derivatives of $S_n\left(\bm{\mu}\right)$ evaluated at $\bm{\mu} = \bm{0}$ generate certain linear combinations of these moments. For instance, still considering the example of order 2 moments:
\begin{align}
    & \frac{\partial^2S_n\left(\bm{\mu}\right)}{\partial\mu_{\alpha}\partial\mu_{\beta}}\Bigg|_{\bm{\mu} = \bm{0}}\nonumber\\
    & = \sum_{\substack{\bm{n} = \left(n_{\bm{a}}\right)_{\bm{a} \in \mathcal{S}}\\\sum\limits_{\bm{a} \in \mathcal{S}}n_{\bm{a}} = n}}\binom{n}{\bm{n}}\exp\left(\frac{1}{2n}\bm{n}^T\bm{L}^T\bm{L}\bm{n}\right)\prod_{\bm{a} \in \mathcal{S}}Q_{\bm{a}}^{n_{\bm{a}}}\nonumber\\[-20px]
    & \hspace*{60px} \times \left[\bm{L}\bm{n}\right]_{\alpha}\left[\bm{L}\bm{n}\right]_{\beta}\\
    & = \sum_{\substack{\bm{n} = \left(n_{\bm{a}}\right)_{\bm{a} \in \mathcal{S}}\\\sum\limits_{\bm{a} \in \mathcal{S}}n_{\bm{a}} = n}}\binom{n}{\bm{n}}\exp\left(\frac{1}{2n}\bm{n}^T\bm{L}^T\bm{L}\bm{n}\right)\prod_{\bm{a} \in \mathcal{S}}Q_{\bm{a}}^{n_{\bm{a}}}\nonumber\\[-20px]
    & \hspace*{60px} \times \left(\sum_{\bm{b} \in \mathcal{S}}L_{\alpha,\,\bm{b}}n_{\bm{b}}\right)\left(\sum_{\bm{c} \in \mathcal{S}}L_{\beta,\,\bm{c}}n_{\bm{c}}\right)\label{eq:qgms_l_rowspace_order_2_moment}
\end{align}
The latter quantity is a linear combination of the generic order 2 moments defined in Eq.~\ref{eq:qgms_general_order_2_moment}, with coefficients $L_{\alpha,\,\bm{b}}L_{\beta,\,\bm{c}}$. In fact, as we show in Appendix~\ref{sec:sk_qaoa_qgms}, those moments defined by the column span of $\bm{L}$ are sufficient to express the instance-averaged SK-QAOA energy. More specifically, Appendix Proposition~\ref{prop:qgms-mgf_formulation_sk_qaoa_energy} derives the following representation of $\nu_{p, n}$:
\begin{align}
    \nu_{p,\,n} & := -\frac{i}{\gamma_{p + 1}}\sum_{r \in [p]}\frac{\partial^2 S_n\left(\bm{\mu}\right)}{\partial\mu_{(r,\,p + 1)}^2}\Bigg|_{\bm{\mu} = \bm{0}}.\label{eq:sk_qaoa_energy_from_qgms_moments_main_text}
\end{align}
In the following, we shall simply refer to the derivatives the $S_n\left(\bm{\mu}\right)$ as \textit{QGMS moments}.

\subsubsection{Analysis of QGMS: saddle point and correlations tensors}
\label{sec:qgms_saddle_point_correlations_tensors_main_text}

\begin{figure}[!htbp]
    \centering
    \includegraphics[width=0.6\columnwidth]{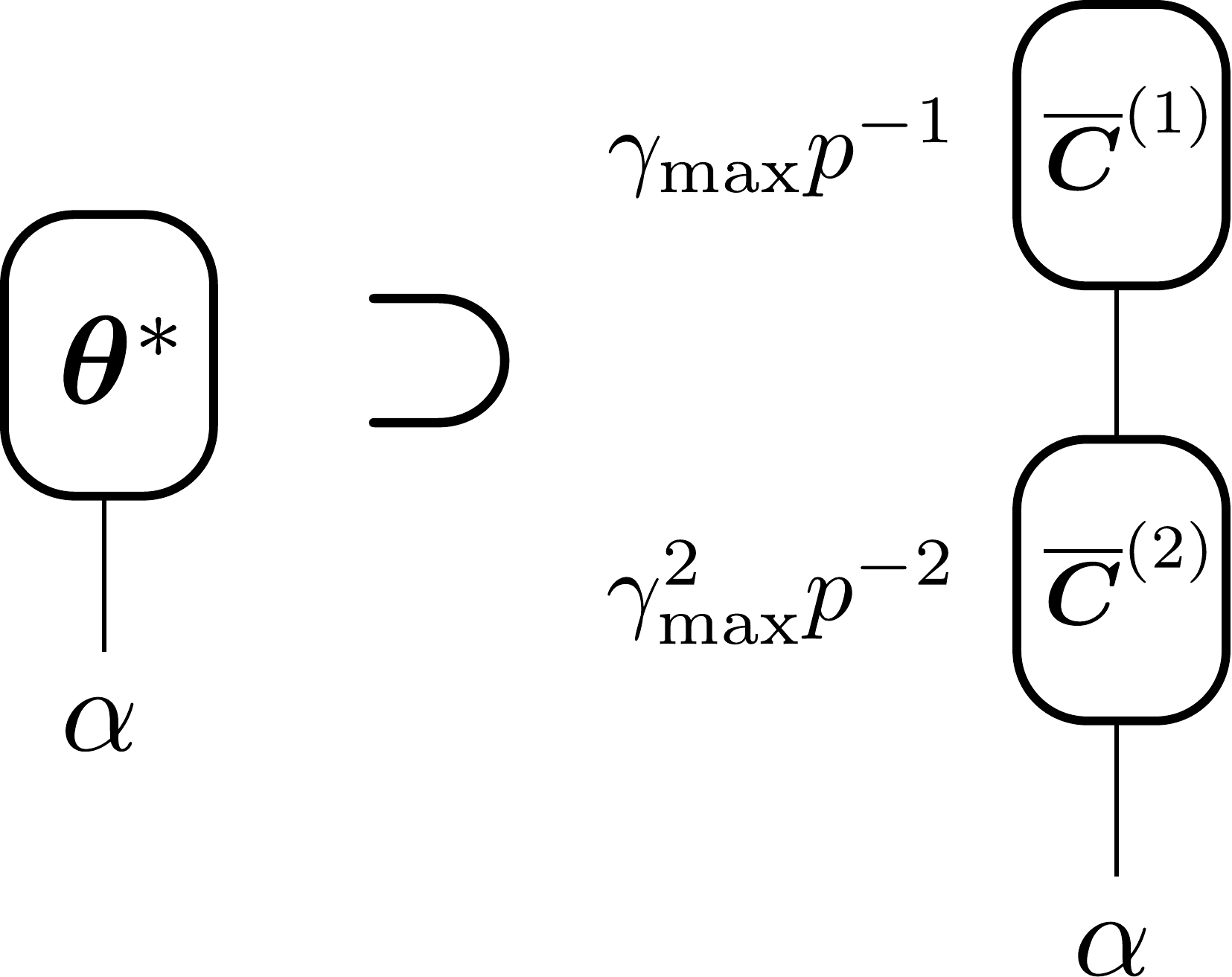}
    \caption{An example term in the expansion of $\bm{\theta}^*\left(\lambda\right)$ as a series in $\lambda$. Each noninteracting tensor block $\bm{\overline{C}}^{(d)}$ contributes a factor $\lambda^d$, where $\lambda = \gamma_{\mathrm{max}}p^{-1}$. Hence, the represented term is of order $\lambda^3$. Besides this explicit prefactor, $p$ also appears implicitly in the bond dimensions, since tensors are indexed by $\mathcal{A}$ and $\left|\mathcal{A}\right| = \left(2p + 2\right)^2$ (Eq.~\ref{eq:sk_qaoa_energy_qgms_A_index_set_size_main_text}). Symbol $\supset$ signals that the right-hand-side is a single additive contribution to the left-hand side.}
    \label{fig:saddle_point_noninteracting_contribution_example}
\end{figure}

\begin{figure}[!htbp]
    \centering
    \includegraphics[width=0.8\columnwidth]{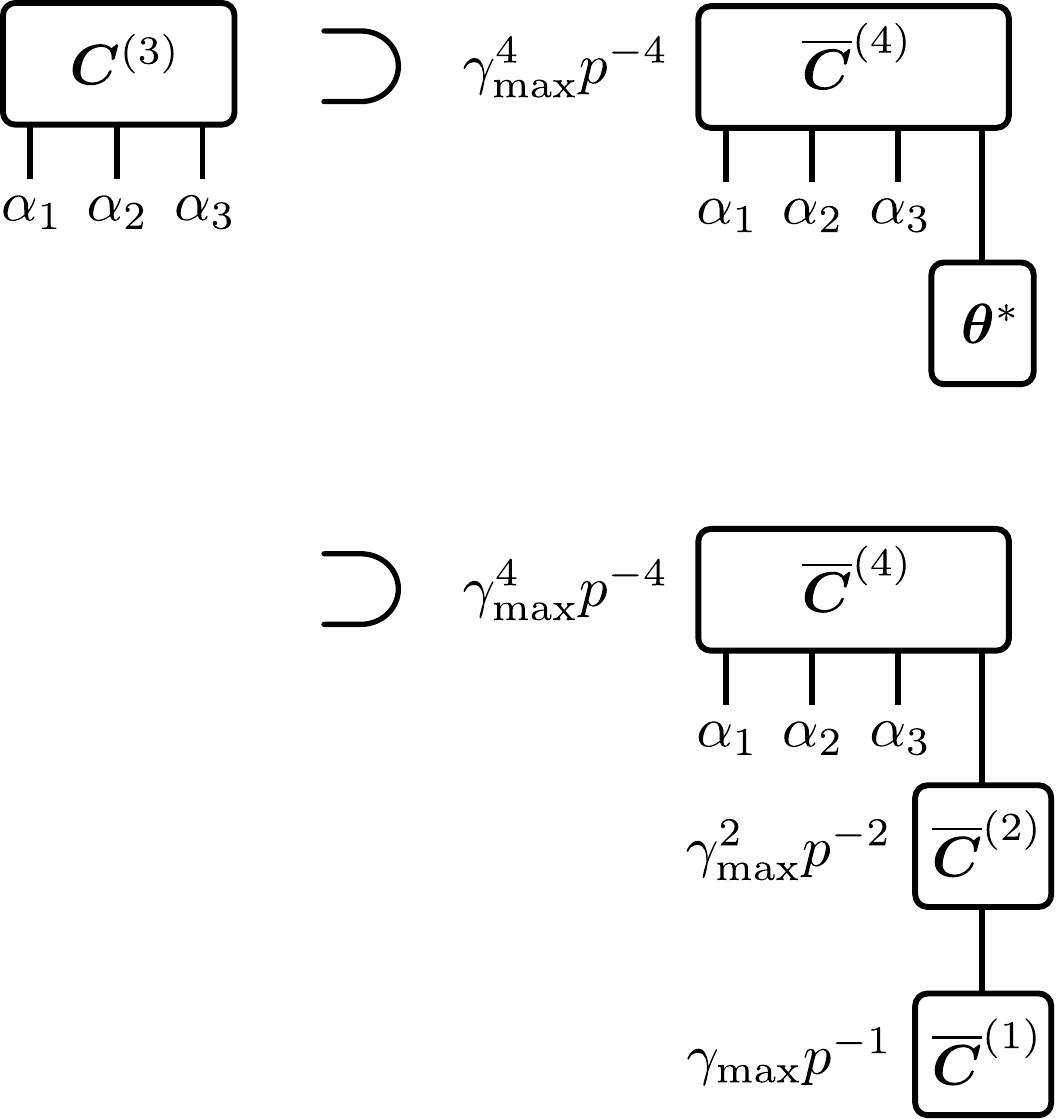}
    \caption{An example term in the expansion of $\bm{C}^{(3)} = C^{(3)}\left(\lambda\right)$ as a series in $\lambda$. The first calculation step results from expressing $\bm{C}^{(3)}$ in terms of noninteracting correlations tensors $\bm{\overline{C}}^{(d)}$ and $\bm{\theta}^*$, using Eq.~\ref{eq:correlations_tensor_definition_main_text} defining $\bm{\overline{C}}^{(3)}$ from $\bm{\theta}^*$. The second calculation step results from picking one term (the same as in figure \ref{fig:saddle_point_noninteracting_contribution_example}) in the series expansion of $\bm{\theta}^*$. In both steps, each noninteracting correlations tensor $\bm{\overline{C}^{(d)}}$ contributes a factor $\lambda^d = \gamma_{\mathrm{max}}^dp^{-d}$. This yields an additive contribution to $\bm{C}^{(3)}$ of order $\lambda^7 = \gamma_{\mathrm{max}}^7p^{-7}$.}
    \label{fig:correlations_tensor_noninteracting_contribution_example}
\end{figure}

We now come back to a fully general QGMS, and sketch how its pseudo-moments can be analyzed. A central object associated to any QGMS (see Appendix Section~\ref{sec:qgms_saddle_point_correlations_tensors} for details) is the \textit{saddle point} $\bm{\theta}^*$, defined by the fixed-point equation:
\begin{align}
    \theta^*_{\alpha} & = \frac{\sum\limits_{\bm{a} \in \mathcal{S}}Q_{\bm{a}}\exp\left(\sum\limits_{\beta \in \mathcal{A}}\theta^*_{\beta}L_{\beta,\,\bm{a}}\right)L_{\alpha,\,\bm{a}}}{\sum\limits_{\bm{a} \in \mathcal{S}}Q_{\bm{a}}\exp\left(\sum\limits_{\beta \in \mathcal{A}}\theta^*_{\beta}L_{\beta,\,\bm{a}}\right)}.\label{eq:saddle_point_equation_main_text}
\end{align}
Existence and uniqueness of a solution to this equation is a priori unclear, and addressed in Appendix Section~\ref{sec:pqgms_saddle_point_expansion} under some assumptions ---in particular, the sufficiently small $\gamma_{\mathrm{max}}$ condition from main Theorem~\ref{th:approximation_continuous_time_annealing_qaoa}. Given the saddle point $\bm{\theta}^*$ associated to this QGMS, the \textit{correlations tensor} of order $d$ associated to the same QGMS, denoted by $\bm{C}^{(d)}$, is defined by:
\begin{align}
    \bm{C}^{(d)} & := \frac{\sum\limits_{\bm{a} \in \mathcal{S}}Q_{\bm{a}}\exp\left(\bm{\theta}^*\bm{L}_{:,\,\bm{a}}\right)\bm{L}_{:,\,\bm{a}}^{\otimes d}}{\sum\limits_{\bm{a} \in \mathcal{S}}Q_{\bm{a}}\exp\left(\bm{\theta}^{*T}\bm{L}_{:,\,\bm{a}}\right)}.\label{eq:correlations_tensor_definition_main_text}
\end{align}
More explicitly, in indexed notation:
\begin{align}
    \bm{C}^{(d)} & := \left(C^{(d)}_{\alpha_1,\,\ldots,\,\alpha_d}\right)_{\alpha_1,\,\ldots,\,\alpha_d \in \mathcal{A}} \in \left(\mathbf{C}^{\mathcal{A}}\right)^{\otimes d},\\
    C^{(d)}_{\alpha_1,\,\ldots,\,\alpha_d} & := \frac{\sum\limits_{\bm{a} \in \mathcal{S}}Q_{\bm{a}}\exp\left(\sum\limits_{\beta \in \mathcal{A}}\theta^*_{\beta}L_{\beta,\,\bm{a}}\right)L_{\alpha_1,\,\bm{a}}\ldots L_{\alpha_d,\,\bm{a}}}{\sum\limits_{\bm{a} \in \mathcal{S}}Q_{\bm{a}}\exp\left(\sum\limits_{\beta \in \mathcal{A}}\theta^*_{\beta}L_{\beta,\,\bm{a}}\right)}.
\end{align}
From this definition, the correlations tensor is manifestly symmetric. Combining Eqns.~\ref{eq:saddle_point_equation_main_text}, \ref{eq:correlations_tensor_definition_main_text} gives:
\begin{align}
    \bm{C}^{(1)} & = \bm{\theta}^*.
\end{align}
Rephrasing, the correlations tensor of order 1, also known as the saddle point, is obtained by solving fixed point equation \ref{eq:saddle_point_equation_main_text}, and higher-order correlations tensors can then be explicitly computed by Eq.~\ref{eq:correlations_tensor_definition_main_text} as functions of $\bm{C}^{(1)}$ and the QGMS parameters $\bm{Q}, \bm{L}$. The resolution of saddle point equation \ref{eq:saddle_point_equation_main_text} and the the corresponding formulae for correlations tensors are developed in Appendix Section~\ref{sec:pqgms_saddle_point_expansion}. It will be helpful to sketch the formulae here as they will provide valuable intuition for the existence of the $p \to \infty$ limit. The main idea of Appendix Section~\ref{sec:pqgms_saddle_point_expansion} is to solve Eq.~\ref{eq:saddle_point_equation_main_text} in the form of a series expansion. For that purpose, we assume matrix $\bm{L} \in \mathbf{C}^{\mathcal{A} \times \mathcal{S}}$ defining the QGMS to be of the form:
\begin{align}
    \bm{L}\left(\lambda\right) & := \lambda\bm{\overline{L}},\label{eq:qgms_parametrization_main_text}
\end{align}
with $\lambda \in \mathbf{C}$ a scalar and $\bm{\overline{L}}$ a matrix of same shape as $\bm{L}$, regarded as independent of $\lambda$. We refer to the family of $\lambda$-indexed QGMS thereby defined as a \textit{parametrized QGMS}. In practice, for the QGMS representing the SK-QAOA instance-average energy (via Eq.~\ref{eq:sk_qaoa_energy_from_qgms_moments_main_text}), we will be able to take (Appendix Section~\ref{sec:error_bounds_discrete_continuum_iterations}):
\begin{align}
    \lambda & = \frac{2^{-1/2}\gamma_{\mathrm{max}}}{p + 1} \simeq \frac{\gamma_{\mathrm{max}}}{p},\label{eq:sk_qaoa_energy_pqgms_lambda_main_text}
\end{align}
with
\begin{align}
    \gamma_{\mathrm{max}} & := \max_{s \in [0, 1]}\left|\gamma^{\mathrm{cont}}\left(s\right)\right| \leq \gamma^*_{\mathrm{max}}.
\end{align}
We then express the saddle point, then correlations tensors, as a series expansion in $\lambda \in \mathbf{C}$. Each term of the series involves contractions of \textit{noninteracting correlations tensors}. In simple terms, noninteracting correlations tensors are the limits of correlations tensors $\bm{C}^{(d)}\left(\lambda\right)$, computed for $\bm{L}\left(\lambda\right)$, as $\lambda \to 0$. Explicitly, the noninteracting correlations tensor $\bm{\overline{C}}^{(d)}$ is defined by:
\begin{align}
    \overline{C}^{(d)}_{\alpha_1,\,\ldots,\,\alpha_d} & := \frac{\sum\limits_{\bm{a} \in \mathcal{S}}Q_{\bm{a}}L_{\alpha_1,\,\bm{a}} \ldots L_{\alpha_d,\,\bm{a}}}{\sum\limits_{\bm{a} \in \mathcal{S}}Q_{\bm{a}}},\label{eq:noninteracting_correlations_tensor_definition_main_text}
\end{align}
indeed satisfying limit:
\begin{align}
    \bm{\overline{C}}^{(d)} & = \lim_{\lambda \to 0}\left(\lambda^{-d}\bm{C}^{(d)}\left(\lambda\right)\right).
\end{align}
To further stress the distinction between noninteracting correlations tensors $\bm{\overline{C}}^{(d)}$ and standard correlations tensors $\bm{C}^{(d)}$, we will frequently refer to $\bm{C}^{(d)}$ as \textit{interacting correlations tensors}. Note that unlike saddle point equation \ref{eq:saddle_point_equation_main_text} defining $\bm{\theta}^*$ implicitly, and related equation \ref{eq:correlations_tensor_definition_main_text} defining $\bm{C}^{(d)}$ from $\bm{\theta}^*$, Eq.~\ref{eq:noninteracting_correlations_tensor_definition_main_text} is fully explicit in the QGMS parameters $\bm{Q}, \bm{L}$. Appendix Section~\ref{sec:pqgms_saddle_point_expansion} develops an expansion of correlations tensors $\bm{C}^{(d)}\left(\lambda\right)$ as a power series in $\lambda$, with coefficients given by tensor networks in the noninteracting correlations tensors $\bm{\overline{C}}^{\left(m\right)}$. More specifically, the saddle point equation \ref{eq:saddle_point_equation_main_text} is first solved as a series in $\lambda$. Figure \ref{fig:saddle_point_noninteracting_contribution_example} gives an example series term in the form of a tensor network diagram. 
The series representation of $\bm{\theta}^*\left(\lambda\right)$ thereby obtained can then be plugged into Eq.~\ref{eq:correlations_tensor_definition_main_text} defining correlations tensors to obtain a series representation of the latters. An example series term obtained by this method is represented on figure \ref{fig:correlations_tensor_noninteracting_contribution_example}.

\subsubsection{Analysis of QGMS: the continuum limit of correlations tensors}
\label{sec:qgms_correlations_tensors_continuum_limit_main_text}

\begin{figure*}[!htbp]
    \includegraphics[width=\textwidth]{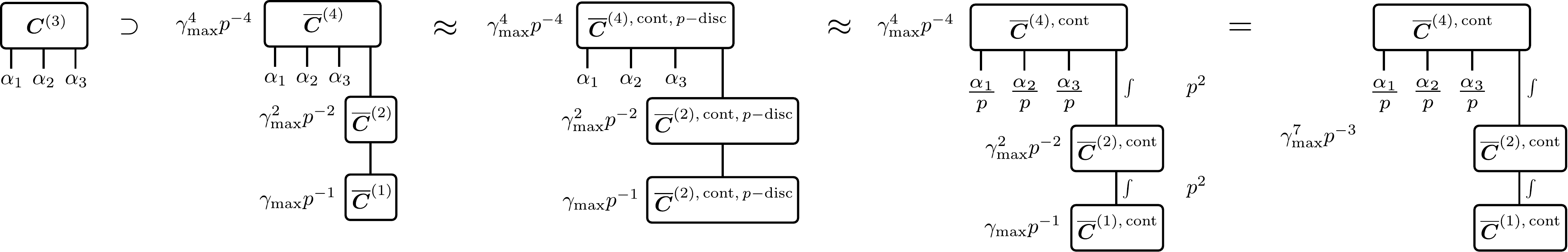}
    \caption{An example term in the series expansion of $\bm{C}^{(3)}$ and its convergence to a continuum limit as $p \to \infty$. The diagrams in the first 3 steps are standard tensor network diagrams: boxes represent discrete tensors with a finite number of indices assuming a finite number of valued, and edges between boxes denote finite summation over the corresponding indices. The diagrams in the final two steps are formally similar to tensor network diagrams: boxes now represent multivariate functions of real variables, each variable living in $[0, 2]^2$; edges denote integration each such continuous variables, as reminded by the integral sign against the edges.}
    \label{fig:higher_order_correlations_tensor_continuum_limit_example}
\end{figure*}

After sketching the procedure to obtain interacting correlation tensors $\bm{C}^{(d)}$ from noninteracting ones $\bm{\overline{C}}^{(d)}$, we are now ready to explain the intuition behind their convergence to a continuum limit. To reach this result, the first step is to establish a continuum limit for noninteracting correlations tensors. This is done in Appendix Section~\ref{sec:noninteracting_correlations_tensors_continuum_limit}, proving tensors $\bm{\overline{C}}^{(d)}$ are \textit{exact discretizations} of continuous, $p$-independent functions $\overline{C}^{(d),\,\mathrm{cont}}$. Explicitly:
\begin{align}
    \overline{C}^{(d)}_{\alpha_1,\,\ldots,\,\alpha_d} & = \overline{C}^{(d),\,\mathrm{cont},\,p-\mathrm{disc}}_{\alpha_1,\,\ldots,\,\alpha_d}.\label{eq:noninteracting_correlations_tensors_discretization_main_text}
\end{align}
with tensor $\bm{C}^{(d),\,\mathrm{cont},\,p-\mathrm{disc}} \in \left(\mathbf{C}^{\mathcal{A}}\right)^{\otimes d}$ defined as the discretization of the continuous function $\overline{C}^{(d),\,\mathrm{cont}}$ over a linear grid of spacing (approximately) $1/p$:
\begin{align}
    \overline{C}^{(d),\,\mathrm{cont},\,p-\mathrm{disc}} & := \overline{C}^{(d),\,\mathrm{cont}}\left(\frac{\alpha_1}{p + 1/2}, \ldots, \frac{\alpha_d}{p + 1/2}\right).
\end{align}
We emphasize that on the left-hand side,
\begin{align}
    \bm{\overline{C}}^{(d),\,\mathrm{cont},\,p-\mathrm{disc}} & = \left(\overline{C}^{(d),\,\mathrm{cont},\,p-\mathrm{disc}}_{\alpha_1,\,\ldots,\,\alpha_d}\right)_{\alpha_1,\,\ldots,\,\alpha_d \in \mathcal{A}} \in \left(\mathbf{C}^{\mathcal{A}}\right)^{\otimes d},\\
    \mathcal{A} & := \left\{0, 1, \ldots, 2p, 2p + 1\right\},
\end{align}
is a discrete tensor, while on the right-hand side,
\begin{align}
    \overline{C}^{(d),\,\mathrm{cont}}: \left\{\begin{array}{ccc}
         \left([0, 2]^2\right)^d & \longrightarrow & \mathbf{C}\\
         \left(\xi_1, \ldots, \xi_d\right) & \longmapsto & \overline{C}^{(d),\,\mathrm{cont}}\left(\xi_1, \ldots, \xi_d\right) 
    \end{array}\right.
\end{align}
is a continuous function. Rephrasing Eq.~\ref{eq:noninteracting_correlations_tensors_discretization_main_text}, discrete tensor $\bm{\overline{C}}^{(d)}$, whose definition depends on $p$, is given by a discretization of a $p$-independent function $\overline{C}^{(d),\,\mathrm{cont}}$ over a linear grid of spacing approximately $1/p$. Admitting discretization identity Eq.~\ref{eq:noninteracting_correlations_tensors_discretization_main_text} for non-interacting correlations tensors of arbitary degree, let us now examine its consequence on interacting correlations tensors. For definiteness, we focus again on the example of the interacting correlations tensor of order 3 introduced in Section~\ref{sec:qgms_saddle_point_correlations_tensors_main_text}. More specifically, we consider the term of the series expansion of $\bm{C}^{(3)}$ (in terms of noninteracting correlations tensors) depicted on Fig.~\ref{fig:correlations_tensor_noninteracting_contribution_example}. Writing this term explicitly:
\begin{widetext}
\begin{align}
    C^{(3)}_{\alpha_1,\,\alpha_2,\,\alpha_3} & \supset \gamma_{\mathrm{max}}^7p^{-7}\sum_{\alpha_4,\,\alpha_5 \in \mathcal{S}}\overline{C}^{(4)}_{\alpha_1,\,\alpha_2,\,\alpha_3,\,\alpha_4}\overline{C}^{(2)}_{\alpha_4,\,\alpha_5}\overline{C}^{(1)}_{\alpha_5},\nonumber\\
    & = \gamma_{\mathrm{max}}^7p^{-7}\sum_{\alpha_4,\,\alpha_5 \in \mathcal{A}}\overline{C}^{(4),\,\mathrm{cont},\,p-\mathrm{disc}}_{\alpha_1,\,\alpha_2,\,\alpha_3,\,\alpha_4}\overline{C}^{(2),\,\mathrm{cont},\,p-\mathrm{disc}}_{\alpha_4,\,\alpha_5}\overline{C}^{(1),\,\mathrm{cont},\,p-\mathrm{disc}}_{\alpha_5}\nonumber\\
    & = \gamma_{\mathrm{max}}^7p^{-7}\sum_{\alpha_4,\,\alpha_5 \in \mathcal{A}}\overline{C}^{(4),\,\mathrm{cont}}\left(\frac{\alpha_1}{p + 1/2}, \frac{\alpha_2}{p + 1/2}, \frac{\alpha_3}{p + 1/2}, \frac{\alpha_4}{p + 1/2}\right)\overline{C}^{(2),\,\mathrm{cont}}\left(\frac{\alpha_4}{p + 1/2}, \frac{\alpha_5}{p + 1/2}\right)\overline{C}^{(1),\,\mathrm{cont}}\left(\frac{\alpha_5}{p + 1/2}\right)\nonumber\\
    & \approx \gamma_{\mathrm{max}}^4p^{-3}\int_{[0, 2]^2 \times [0, 2]^2}\!\mathrm{d}\xi_4\,\mathrm{d}\xi_5\,\overline{C}^{(4),\,\mathrm{cont}}\left(\frac{\alpha_1}{p + 1/2}, \frac{\alpha_2}{p + 1/2}, \frac{\alpha_3}{p + 1/2}, \xi_4\right)\overline{C}^{(2),\,\mathrm{cont}}\left(\xi_4, \xi_5\right)\overline{C}^{(1),\,\mathrm{cont}}\left(\xi_5\right).\label{eq:higher_order_correlations_tensor_continuum_limit_example}
\end{align}
\end{widetext}
In the second and third lines, we plugged in discretization identity Eq.~\ref{eq:noninteracting_correlations_tensors_discretization_main_text} for noninteracting correlations tensors $\bm{\overline{C}}^{(d)}$. In the fourth line, we replaced discrete sum over $\mathcal{A} = \left\{0, 1, \ldots, 2p, 2p + 1\right\}$ by integrals over $[0, 2]^2$. This follows from Riemann sum approximation of integrals, whereby sum over a single variable is related to the integral by prefactor:
\begin{align}
    \left(\frac{2p + 1}{2}\right)^2 \sim p^2.
\end{align}
Fig.~\ref{fig:higher_order_correlations_tensor_continuum_limit_example} summarizes the reasoning. 
Eq.~\ref{eq:higher_order_correlations_tensor_continuum_limit_example} asserts that in the limit $p \to \infty$, the additive contribution to $\bm{C}^{(3)}$ considered here consists of a prefactor $\gamma_{\mathrm{max}}^7p^{-3}$, times the discretization over a linear grid of spacing approximately $1/p$ of a continuous function independent of $p$:
\begin{align}
    \begin{array}{ccl}
         \left([0, 2]^2\right)^3 & \longrightarrow & \mathbf{C}\\
         \left(\xi_1, \xi_2, \xi_3\right) & \longmapsto & \int\limits_{[0, 2]^2 \times [0, 2]^2}\mathrm{d}\xi_4\,\mathrm{d}\xi_5\overline{C}^{(4),\,\mathrm{cont}}\left(\xi_1, \xi_2, \xi_3, \xi_4\right)\nonumber\\
         & & \hspace*{40px} \times \overline{C}^{(2),\,\mathrm{cont}}\left(\xi_4, \xi_5\right)\overline{C}^{(1),\,\mathrm{cont}}\left(\xi_5\right).
    \end{array}
\end{align}
A crucial point in the application of the Riemann sum approximation in this context is the independence of the integral's dimension: 2 in this case, on the problem instance size $n$. It follows that the $p$ required to achieve a target additive approximation error is \textit{independent of $n$}. This holds for a single term of the series representing $\bm{C}^{(3)}$ in terms of non-interacting correlations tensors. The tedious part of the proof, deferred to Appendix~\ref{sec:sk_qaoa_continuum_limit}, is to show these insights hold for all other terms of the series individually, to precisely quantify the errors and bound their sum over all series terms.

\subsubsection{Analysis of QGMS: series expansion of QGMS moments and the continuum limit of the SK-QAOA energy}
\label{sec:qmgs_moments_series_expansion_main_text}

After introducing two fundamental objects in the analysis of QGMS: the saddle point and correlations tensors, we focus our attention back on QGMS moments, which, according to Eq.~\ref{eq:sk_qaoa_energy_from_qgms_moments_main_text}, encode the instance-averaged SK-QAOA energy. Proposition~\ref{prop:qgms_integral_series_expansion} from Appendix~\ref{sec:pqgms_series_expansion_noninteracting_limit} provides a series expansion of the second order diagonal moments:
\begin{align}
    \frac{\partial^2S_n\left(\bm{\mu}\right)}{\partial\mu_{\alpha}^2}\Bigg|_{\bm{\mu} = \bm{0}}
\end{align}
in terms of interacting correlations tensors. This series is distinct from the series expressing interacting correlations tensors discussed in Section~\ref{sec:qgms_saddle_point_correlations_tensors_main_text}, and may be regarded as a different level of series expansion. More specifically, Appendix Proposition~\ref{prop:qgms_integral_series_expansion} provides an expansion of the moments as a series where each term is a tensor network of interacting correlations tensors; within this each term of this summation, interacting correlations tensors may then be expanded from noninteracting correlations tensors according to the distinct series expansion introduced in Section~\ref{sec:qgms_saddle_point_correlations_tensors_main_text}.  In this Section, we sketch the structure of the series expansion of QGMS moments proven in Proposition~\ref{prop:qgms_integral_series_expansion}. For simplicity, we defer general definitions underlying the Proposition to appendices, and focus on a single term for which all objects will admit simple explicit definition.

Proposition~\ref{prop:qgms_integral_series_expansion} identifies 3 additive contributions to the second-order diagonal QGMS moment of index $\alpha$, and focuses on exactly one of these contributions, denoted $\nu_{\alpha}$ ---the others being similarly treatable:
\begin{align}
    \frac{\partial^2S_n\left(\bm{\mu}\right)}{\partial\mu_{\alpha}^2}\Bigg|_{\bm{\mu} = \bm{0}} & \supset \nu_{\alpha}.\label{eq:qgms_moments_nu_contribution_main_text}
\end{align}
Proposition~\ref{prop:qgms_integral_series_expansion} then asserts the following series expansion of $\nu_{\alpha}$:
\begin{align}
    \nu_{\alpha} & = \frac{2}{\sqrt{n}}\sum_{\left(n_d\right)_{d \geq 2}}\binom{n}{\left(n_d\right)_{d \geq 2}}\frac{n^{-\sum\limits_{d \geq 0}dn_d/2}}{\prod\limits_{d \geq 2}d!^{n_d}}\theta^*_{\alpha}I^{\left(n_d\right)_{d \geq 2}}_{\alpha},\label{eq:qgms_integral_series_expansion_main_text}\\
    I^{\left(n_d\right)_{d \geq 2}}_{\alpha} & := \left\langle \bm{\mathcal{I}}_{\alpha}^{\left(1 + \sum\limits_{d \geq 2}dn_d\right)}, \bigotimes_{d \geq 2}\bm{\delta C}^{(d)\otimes n_d} \right\rangle.\label{eq:qgms_integral_series_expansion_dot_product_main_text}
\end{align}
The sum in equation \ref{eq:qgms_integral_series_expansion_main_text} is over infinite sequence of non-negative integers indexed by $d \geq 2$:
\begin{align}
    \left(n_d\right)_{d \geq 2} & = \left(n_2, n_3, n_4, \ldots\right) \in \mathbf{N}^{[[2, +\infty[[}.
\end{align}
The multinomial coefficient is defined as:
\begin{align}
    \binom{n}{\left(n_d\right)_{d \geq 2}} & := \frac{n!}{\left(n - \sum\limits_{d \geq 2}n_d\right)!\prod\limits_{d \geq 2}n_d!}.
\end{align}
Note that we do not enforce constraint $\sum_{d \geq 2}n_d = n$ in the summation, hence the first factor in the denominator of the multinomial coefficient. We note that for any finite $n$ this first factor restricts $\sum_{d \geq 2}n_d$ to $[0, n]$; hence, for any finite $n$, sequences $\left(n_d\right)_{d \geq 2}$ can be restricted to those with at most $n$ nonzero elements. This means that uniformly in $n$, the summation may be restricted to sequences with a finite number of nonzero elements. It follows that all algebraic quantities defined from $\left(n_d\right)_{d \geq 2}$ and occurring in equation \ref{eq:qgms_integral_series_expansion_main_text}, such as
\begin{align}
    n^{-\sum\limits_{d \geq 2}dn_d},
\end{align}
as well as tensor product
\begin{align}
    \bigotimes_{d \geq 2}\bm{\delta C}^{(d)\otimes n_d},
\end{align}
are well-defined and finite. Next, the brackets $\left\langle \cdot, \cdot\right\rangle$ in Eq.~\ref{eq:qgms_integral_series_expansion_dot_product_main_text} denote the Euclidean dot product between its argument tensors. Finally, $\bm{\mathcal{I}}^{\left(1 + D\right)}$ and $\bm{\delta C}^{(d)}$ occurring inside this dot product are respectively the matching tensor of order $D$ and the centered correlations tensor of order $d$. We defer general definitions of these tensors to Appendix Section~\ref{sec:qgms_moments_series_expansion}, and focus for now on a specific series terms, writing all tensor definitions that can be written explicitly and concisely. Namely, we consider term indexed by:
\begin{align}
    \left(n_d\right)_{d \geq 2} & = \left(n_2, n_3, n_4, n_5, \ldots\right) := \left(0, 1, 0, 0, \ldots\right).
\end{align}
Then,
\begin{align}
    D := \sum_{d \geq 2}dn_d = 3,
\end{align}
and the tensor product in defining equation~\ref{eq:qgms_integral_series_expansion_dot_product_main_text} for $I_{\left(n_d\right)_{d \geq 2}}$ reduces to:
\begin{align}
    \bigotimes_{d \geq 2}\bm{\delta C}^{(d)\otimes n_d} & = \bm{\delta C}^{(3)}.
\end{align}
The centered correlation tensor of order 3: $\bm{\delta C}^{(3)}$ admits the following explicit expression:
\begin{align}
    \bm{\delta C}^{(3)} & = \left(\delta C^{(3)}_{\alpha_1,\,\alpha_2,\,\alpha_3}\right)_{\alpha_1,\,\alpha_2,\,\alpha_3 \in \mathcal{A}} \in \left(\mathbf{C}^{\mathcal{A}}\right)^{\otimes 3},\\
    \delta C^{(3)}_{\alpha_1,\,\alpha_2,\,\alpha_3} & := C^{(3)}_{\alpha_1,\,\alpha_2,\,\alpha_3} - C^{(2)}_{\alpha_1,\,\alpha_2}C^{(1)}_{\alpha_3} - C^{(2)}_{\alpha_1,\,\alpha_3}C^{(1)}_{\alpha_2}\nonumber\\
    & \hspace*{15px} - C^{(2)}_{\alpha_2,\,\alpha_3}C^{(1)}_{\alpha_1} + 2C^{(1)}_{\alpha_1}C^{(1)}_{\alpha_2}C^{(1)}_{\alpha_3}.\label{eq:qgms_moment_series_example_term_centered_correlations_tensor_definition}
\end{align}
Note the first additive contribution of $\bm{\delta C}^{(3)}$ is simply $\bm{C}^{(3)}$. As for $\bm{\mathcal{I}}^{(1 + D)}_{\alpha}$ ---in the general vocabulary of Appendix Section~\ref{sec:qgms_moments_series_expansion}, slice $\alpha$ of the matching tensor of order $(D + 1)$, it is explicitly given by:
\begin{align}
    \bm{\mathcal{I}}^{(4)}_{\alpha} & := \left(\mathcal{I}^{(4)}_{\alpha,\,\alpha_1,\,\alpha_2,\,\alpha_3}\right)_{\alpha_1,\,\alpha_2,\,\alpha_3 \in \mathcal{A}} \in \left(\mathbf{C}^{\mathcal{A}}\right)^{\otimes 3},\\
    \mathcal{I}^{(4)}_{\alpha,\,\alpha_1,\,\alpha_2,\,\alpha_3} & := \delta_{\alpha,\,\alpha_1}\delta_{\alpha_2,\,\alpha_3} + \delta_{\alpha,\,\alpha_2}\delta_{\alpha_1,\,\alpha_3}\nonumber\\
    & \hspace*{15px} + \delta_{\alpha,\,\alpha_3}\delta_{\alpha_1,\,\alpha_2}.\label{eq:qgms_moment_series_example_term_matching_tensor_definition}
\end{align}
Combining Eqs.~\ref{eq:qgms_moment_series_example_term_centered_correlations_tensor_definition}, \ref{eq:qgms_moment_series_example_term_matching_tensor_definition}, we may now compute dot product $I^{\left(n_d\right)_{d \geq 2}}$ generally defined by Eq.~\ref{eq:qgms_integral_series_expansion_dot_product_main_text}. For conciseness, we only consider a single additive contribution to this dot product, resulting from picking the first term: $C^{(3)}_{\alpha_1,\,\alpha_2,\,\alpha_3}$ in the additive decomposition of $\delta C^{(3)}_{\alpha_1,\,\alpha_2,\,\alpha_3}$ given in equation \ref{eq:qgms_moment_series_example_term_centered_correlations_tensor_definition}.
\begin{align}
    I^{\left(n_d\right)_{d \geq 2}}_{\alpha} & = \left\langle \bm{\mathcal{I}}^{\left(n_d\right)_{d \geq 2}}_{\alpha}, \bigotimes_{d \geq 2}\bm{\delta C}^{(d)\otimes n_d} \right\rangle\nonumber\\
    & = \left\langle \bm{\mathcal{I}}^{(4)}_{\alpha}, \bm{\delta C}^{(3)} \right\rangle\nonumber\\
    & \supset \left\langle \bm{\mathcal{I}}^{(4)}_{\alpha}, \bm{C}^{(3)} \right\rangle\nonumber\\
    & = \sum_{\alpha_1,\,\alpha_2,\,\alpha_3 \in \mathcal{A}}\mathcal{I}^{(4)}_{\alpha,\,\alpha_1,\,\alpha_2,\,\alpha_3}C^{(3)}_{\alpha_1,\,\alpha_2,\,\alpha_3}\nonumber\\
    & = 3\sum_{\beta \in \mathcal{A}}C^{(3)}_{\alpha,\,\beta,\,\beta},\label{eq:qgms_moments_series_expansion_term_example_dot_product}
\end{align}
where to perform the calculation in the final step, we recalled the symmetry of $\bm{C}^{(3)}$ (following from its defining equation~\ref{eq:correlations_tensor_definition_main_text}, see Section~\ref{sec:qgms_saddle_point_correlations_tensors_main_text}). This concludes the (partial) computation of dot product $I^{\left(n_d\right)_{d \geq 2}}_{\alpha}$ in Eq.~\ref{eq:qgms_integral_series_expansion_main_text}. The numerical prefactor against this dot product is:
\begin{align}
    \frac{2}{\sqrt{n}}\binom{n}{\left(n_d\right)_{d \geq 2}}\frac{n^{-\sum\limits_{d \geq 2}dn_d/2}}{\prod\limits_{d \geq 2}d!^{n_d}} & = \frac{2}{\sqrt{n}}\frac{n!}{n_3!(n - n_3)!}\frac{n^{-3n_3/2}}{3!^{n_3}}\nonumber\\
    & = \frac{1}{3n}.\label{eq:qgms_moments_series_expansion_term_example_combinatorial_factor}
\end{align}
We note this combinatorial factor is uniformly bounded in $n$, i.e. it is less than a constant for any integer $n$. 
We show in Appendix that this insight generalizes to other terms in the series. Indeed:
\begin{align}
    \frac{2}{\sqrt{n}}\binom{n}{\left(n_d\right)_{d \geq 2}}\frac{n^{-\sum\limits_{d \geq 2}dn_d/2}}{\prod\limits_{d \geq 2}d!^{n_d}} & \leq \frac{2}{\sqrt{n}}\frac{n^{\sum\limits_{d \geq 2}n_d}}{\prod\limits_{d \geq 2}n_d!}\frac{n^{-\sum\limits_{d \geq 2}dn_d/2}}{\prod\limits_{d \geq 2}d!^{n_d}}\nonumber\\
    & =  \frac{2}{\sqrt{n}}\frac{n^{\sum\limits_{d \geq 2}n_d}}{\prod\limits_{d \geq 2}n_d!}\frac{n^{-\sum\limits_{d \geq 2}2n_d/2}}{\prod\limits_{d \geq 2}d!^{n_d}}\nonumber\\
    & = \frac{2}{\sqrt{n}}\frac{1}{\prod\limits_{d \geq 2}d!^{n_d}n_d!}.
\end{align}
As will be shown in detail in Appendix Section~\ref{sec:qgms_moments_series_expansion_sk_qaoa}, this uniform bound in $n$ implies the series expansion of the QGMS moments (Eq.~\ref{eq:qgms_integral_series_expansion_main_text}) can be truncated to a finite, $n$-independent set of terms to achieve an arbitrary constant error on the QGMS moments, hence on the SK-QAOA instance-averaged energy density. Since the order of truncation of series \ref{eq:qgms_integral_series_expansion_main_text} bounds the degrees of correlations tensors involved in dot products $I^{\left(n_d\right)_{d \geq 2}}$, this means the energy density approximately \textit{only depends on correlations tensors of constant, $n$-independent order}. It remains to understand the potential dependence of these series terms on $p$. For that purpose, we combine Eqns.~\ref{eq:qgms_moments_series_expansion_term_example_dot_product}, \ref{eq:qgms_moments_series_expansion_term_example_combinatorial_factor} to complete the (partial) computation of the single series term:
\begin{align}
    \nu_{\alpha} & \supset \frac{2}{\sqrt{n}}\binom{n}{\left(n_d\right)_{d \geq 2}}\frac{n^{-\sum\limits_{d \geq 2}dn_d/2}}{\prod\limits_{d \geq 2}d!^{n_d}}\theta^*_{\alpha}I^{\left(n_d\right)_{d \geq 2}}_{\alpha}\nonumber\\
    & \supset \frac{1}{n}\sum_{\beta \in \mathcal{A}}\theta^*_{\alpha}C^{(3)}_{\alpha,\,\beta,\,\beta}\nonumber\\
    & = \frac{1}{n}\sum_{\beta \in \mathcal{A}}C^{(1)}_{\alpha}C^{(3)}_{\alpha,\,\beta,\,\beta}
\end{align}
From Eqns.~\ref{eq:qgms_moments_nu_contribution_main_text}, \ref{eq:sk_qaoa_energy_from_qgms_moments_main_text}, we deduce the following additive contribution to the instance-averaged SK-QAOA energy:
\begin{align}
    \nu_{p,\,n} & = -\frac{i}{\gamma_{p + 1}}\sum_{t \in \mathcal{I}}\frac{\partial^2S_n\left(\bm{\mu}\right)}{\partial\mu_{(p + 1,\,t)}^2}\Bigg|_{\bm{\mu} = \bm{0}}\nonumber\\
    & \supset -\frac{i}{\gamma_{p + 1}}\sum_{t \in \mathcal{I}}\nu_{(p + 1,\,t)}\nonumber\\
    & \supset -\frac{i}{n\gamma_{p + 1}}\sum_{t \in \mathcal{I}}\sum_{\beta \in \mathcal{A}}C^{(1)}_{(p + 1,\,t)}C^{(3)}_{(p + 1,\,t),\,\beta,\,\beta}\nonumber\\
    & \supset -\frac{i}{n\gamma_{p + 1}}\sum_{t, u, v \in \mathcal{I}}C^{(1)}_{(p + 1,\,t)}C^{(3)}_{(p + 1,\,t),\,(u,\,v),\,(u,\,v)}\label{eq:sk_qaoa_energy_contribution_example_step_1}
\end{align}
We then recall the result from Section~\ref{sec:qgms_correlations_tensors_continuum_limit_main_text} stating the existence of a continuum limit for correlations tensors; applied to correlations tensors $\bm{C}^{(1)}, \bm{C}^{(3)}$ involved in the last equation, it states:
\begin{align}
    C^{(1)}_{\alpha_1} & \approx p^{-1}C^{(1),\,\mathrm{cont}}\left(\frac{\alpha_1}{p + 1/2}\right),\\
    C^{(3)}_{\alpha_1,\,\alpha_2,\,\alpha_3} & \approx p^{-3}C^{(3),\,\mathrm{cont}}\left(\frac{\alpha_1}{p + 1/2},\,\frac{\alpha_2}{p + 1/2},\,\frac{\alpha_3}{p + 1/2}\right),
\end{align}
with $C^{(1),\,\mathrm{cont}}, C^{(3),\,\mathrm{cont}}$ being continuous functions independent of $p$. Besides, by construction of the $\bm{\gamma}$ schedule at finite $p$ (Definition~\ref{def:continuous_schedule_informal}),
\begin{align}
    \gamma_{p + 1} & \sim \frac{1}{p}\gamma^{\mathrm{cont}}(1) \qquad \textrm{as } p \to \infty.
\end{align}
Plugging these estimates into Eq.~\ref{eq:sk_qaoa_energy_contribution_example_step_1} gives the following estimate for our single additive contribution to the SK-QAOA energy:
\begin{widetext}
    \begin{align}
        \nu_{p,\,n} & \supset -\frac{i}{n\gamma_{p + 1}}\sum_{t,\,u,\,v \in \mathcal{I}}C^{(1)}_{(p + 1,\,t)}C^{(3)}_{(p + 1,\,t),\,(u,\,v),\,(u,\,v)}\nonumber\\
        & \approx -\frac{ip}{n\gamma^{\mathrm{cont}}(1)}\sum_{t,\,u,\,v \in \mathcal{I}}p^{-1}C^{(1),\,\mathrm{cont}}\left(2, \frac{t}{p + 1/2}\right)p^{-3}C^{(3),\,\mathrm{cont}}\left(2, \frac{t}{p + 1/2}, \frac{u}{p + 1/2}, \frac{v}{p + 1/2}, \frac{u}{p + 1/2}, \frac{v}{p + 1/2}\right)\nonumber\\
        & \approx -\frac{ip}{n\gamma^{\mathrm{cont}}(1)}p^3\int\limits_{[0, 2]^3}\!\mathrm{d}\tau\,\mathrm{d}\upsilon\,\mathrm{d}\omega\,p^{-1}C^{(1),\,\mathrm{cont}}\left(2, \tau\right)p^{-3}C^{(3),\,\mathrm{cont}}\left(2, \tau, \upsilon, \omega, \upsilon, \omega\right)\nonumber\\
        & = -\frac{i}{n\gamma^{\mathrm{cont}}(1)}\int_{[0, 2]^3}\!\mathrm{d}\tau\,\mathrm{d}\upsilon\,\mathrm{d}\omega\,C^{(1),\,\mathrm{cont}}\left(2, \tau\right)C^{(3),\,\mathrm{cont}}\left(2, \tau, \upsilon, \omega, \upsilon, \omega\right).
    \end{align}
\end{widetext}
The quantity in the final line is manifestly independent of $p$ and expressed as an integral of continuous multivariate functions. Importantly, similarly to the continuum limits of correlations tensors discussed in Section~\ref{sec:qgms_correlations_tensors_continuum_limit_main_text}, the integral's dimension is independent of $n$, so the $p$ achieving a given constant target error also is. Again similar to the conclusions of Section~\ref{sec:qgms_correlations_tensors_continuum_limit_main_text}, the tedious part of the full proof, developed in Appendix Section~\ref{sec:main_theorem_derivation}, is to precisely bound the Riemann sum error and combine the errors contributed by all terms in series \ref{eq:qgms_integral_series_expansion_main_text}. This concludes the outline of the central argument of the proof of Theorem~\ref{th:approximation_continuous_time_annealing_qaoa} which states that QAOA energy $\nu_{p,n}$ converges to its continuum limit uniformly in $n$ and $p$, provided total $\gamma$ angle is bounded by a small but $n$-independent constant.

\subsection{Standard Trotter analysis does not show equivalence between QAOA and annealing}\label{sec:trotter_analysis_qaoa_vs_qaa}

The QAA relies on Hamiltonian evolution smoothly interpolating between Hamiltonians $-B$ and $C$, unlike the QAOA which proceeds by discretely switching between two Hamiltonians $C$ and $B$. For a specific angles regime, QAOA can nonetheless approximate the QAA via Trotterization. Trotterization (see \cite{theory_trotter_error} for an extensive discussion of the theory) consists to approximate the exponential of a sum of Hermitian operators $A, B$: $e^{i(A + B})$, by a product of exponentials of scaled $A$ and $B$, i.e. a product with factors of the form $e^{itA}$ or $e^{itB}$, with $t \in \mathbf{R}$. The simplest example of Trotterization identity is
\begin{align}
    e^{it\left(A + B\right)} & = e^{itA}e^{itB} + \mathcal{O}_{\left\lVert \cdot \right\rVert}\left(t^2\right),
\end{align}
assuming operator norms bounded by $1$: $\left\lVert A \right\rVert, \left\lVert B \right\lVert \leq 1$, and where the error term is understood in operator norm. Let us sketch how Trotterization can be applied to discretize the quantum adiabatic evolution in Equation \ref{eq:quantum_adiabatic_evolution}. For simplicity, we may start by considering the $N$-steps Euler discretization of this ODE, whereby the final state may be approximated as:
\begin{align}
    \ket{\Psi\left(T\right)} & \approx \overleftarrow{\prod_{j = 0}^{N - 1}}e^{-i\tau H\left(j\tau\right)}\ket{+}^{\otimes n},
\end{align}
where the {discretization step} is
\begin{align}
    \tau & := \frac{T}{N}.
\end{align}
We may then consider applying Trotterization to each term of this product individually. The terms of the sum to exponentiate are
\begin{align}
    & -\tau\left(1 - s\left(j\tau\right)\right)B,\\
    & \tau s\left(j\tau\right)C.
\end{align}
Observe that $s, 1 - s$ are of order unity while $B$ is of {operator norm} $n$; in fact, the {operator norm} of $C$ will also always be of order $n$ by convention. As a result, to satisfy the norm bound assumption in Trotter's formula, we are forced to assume:
\begin{align}
    \tau = \frac{T}{N} \lesssim \frac{1}{n},
\end{align}
i.e.
\begin{align}
    N \gtrsim nT.\label{eq:trotterization_number_steps_constraint}
\end{align}
In particular, the number of Euler discretization steps needs to scale at least linearly with the problem size to allow for Trotterization. With this condition satisfied, we may now write
\begin{align}
    e^{-i\tau H(j\tau)} & = e^{i\tau\left(1 - s\left(j\tau\right)\right)B}e^{-i\tau s\left(j\tau\right)C} + \mathcal{O}_{\left\lVert \cdot \right\rVert}\left(\tau^2\right)\\
    & = e^{i\tau\left(1 - s\left(j\tau\right)\right)B}e^{-i\tau s\left(j\tau\right)C} + \mathcal{O}_{\left\lVert \cdot \right\rVert}\left(\frac{T^2}{N^2}\right).
\end{align}
Compounding the error over the $N$ discretized Euler steps yields a global approximation
\begin{align}
    \ket{\Psi\left(T\right)} & \approx \overleftarrow{\prod_{j = 0}^{N - 1}}e^{i\tau\left(1 - s\left(j\tau\right)\right)B}e^{-i\tau s\left(j\tau\right) C}\ket{+}^{\otimes n} + \mathcal{O}\left(\frac{T^2}{N}\right).
\end{align}
The product on the right-hand side corresponds to a QAOA state with $p := N$ layers and angles
\begin{align}
    \gamma_t & := \tau s\left((t - 1)\tau\right),\\
    \beta_t & := -\tau \left(1 - s\left((t - 1)\tau\right)\right).\label{eq:trotterized_qaa_qaoa_angles}
\end{align}
for $1 \leq t \leq p$. Making the Trotterization error negligible requires
\begin{align}
    N \gtrsim T^2.
\end{align}
This constraint could be made looser (decreasing the exponent 2 of $T$) by increasing the order of the Trotter scheme. However, we would still need to satisfy constraint \ref{eq:trotterization_number_steps_constraint} to apply any Trotter scheme at all, ultimately lower-bounding the number of Trotter step by $n$ for constant $T$, and upper-bounding the QAOA angles (Equation \ref{eq:trotterized_qaa_qaoa_angles}) by $\mathcal{O}\left(1/n\right)$. All in all, the standard analysis of Trotter error, aimed at approximating the Hamiltonian evolution operator in operator norm, may only be applied to QAOA for angles of order $1/n$. In contrast, optimal angles in the constant $p$, infinite-$n$ limit are known to be of constant order for several optimization problems \cite{qaoa_sk,qaoa_maxcut_high_girth,qaoa_ksat}, including the SK model considered in this work. Hence, the standard analysis of Trotter does not allow to interpret QAOA as a Trotterization of the adiabatic algorithm, or even as a Trotterized Hamiltonian evolution in general, in the $n \to \infty$ limit. In contrast, our analysis allows to interpret QAOA as a Trotterized form of quantum annealing in the regime where angles are 
independent of the size $n$. 

\subsection{Improved error analysis of compiled Trotterized quantum annealing}

In Section \ref{sec:trotter_analysis_qaoa_vs_qaa}, we recalled why the standard analysis of Trotter error only allowed to establish equivalence between annealing and QAOA in the regime of QAOA angles upper-bounded by $\mathcal{O}(1/n)$. Under this assumption, QAOA can then be regarded as a circuit compilation of quantum annealing. In the convention of the Hamiltonian simulation literature (see e.g. \cite{theory_trotter_error}), where Hamiltonian are normalized to infinite norm $\mathcal{O}(1)$ ---instead of the typical $\mathcal{O}(n)$ in the QAOA literature---, this angle scaling corresponds to a constant Trotter step. The main result of this study, Theorem \ref{th:approximation_continuous_time_annealing_qaoa}, implies that a Trotter step increasing of linear order $n$ in the instance size is possible. This linear improvement factor in the allowed step size in turn corresponds to a linear factor reduction in the required number of Trotterization layers to approximation annealing to a fixed error. This improvement holds assuming a linear total Hamiltonian evolution time (corresponding to a constant total QAOA angle), and considering the large size $n$ limit. In particular, it is theoretically possible that the proportionality constant in the linearly growing Trotter step size must decrease with the total evolution time, rendering the improvement less favorable for large total evolution time. Numerical results nonetheless suggest that the proportionality constant can in fact be allowed to increase with total evolution time. Hence, a more aggressive size-independent Trotter step size empirically appears possible as the total evolution time increases. For the special case of the SK model, the linear factor reduction in the allowed number of Trotter layers translates to a quadratic depth improvement in the Trotterization of quantum annealing, a fact stated formally in the following corollary of Theorem \ref{th:approximation_continuous_time_annealing_qaoa} for convenience:

\begin{corollary}[Quadratic depth reduction in Trotterization of SK model annealing]
\label{cor:sk_compilation_quadratic_depth_improvement}
Let a quantum annealing schedule for the (transverse) SK model be defined by controls $\gamma^{\mathrm{cont}}$, $\beta^{\mathrm{cont}}$ as in equation \ref{eq:continuous_time_annealing_schedule_informal}. Assume, as in the statement of Theorem \ref{th:approximation_continuous_time_annealing_qaoa} (equation \ref{eq:main_theorem_max_angle_bound_condition}), that the maximum of $\gamma^{\mathrm{cont}}$ and the total continuous $\gamma$ angles are bounded by absolute constants. Then, for all $\varepsilon > 0$, there exists a circuit of depth $\mathcal{O}\left(n/\varepsilon\right)$ outputting an energy $\varepsilon$-close to the energy of the annealing evolution with high probability over the random choice of SK instance. 
With respect to $n$, this is a quadratic improvement compared to the $\Omega\left(n^2\right)$ 
required by the standard analysis of Trotterized Hamiltonian evolution.
\begin{proof}
The result follows from the fact that in the standard analysis of Trotter error, $n$ QAOA layers would be required to compile quantum annealing to a constant fidelity, while according to the analysis of theorem \ref{th:approximation_continuous_time_annealing_qaoa}, $\mathcal{O}\left(1/\varepsilon\right)$ layers (independent of $n$) suffice to approximate the energy to error $\varepsilon$. We then recall that each layer of QAOA for the SK model requires depth $n$ 
due to the all-to-all connected nature of this quadratic Hamiltonian. Consequently, the standard analysis of Trotter error and theorem \ref{th:approximation_continuous_time_annealing_qaoa} respectively require a circuit of depth $n^2$ and $n/\varepsilon$.
\end{proof}
\end{corollary}

An interesting future research direction is characterizing the equivalence between Hamiltonian ansatz and analog Hamiltonian evolution for a greater variety of Hamiltonians. Establishing this equivalence may translate to improved bounds on circuit complexity of Hamiltonian evolution, as shown in the very special case of the SK model in Corollary \ref{cor:sk_compilation_quadratic_depth_improvement}.

\section{Discussion}
\label{sec:discussion}

In this work, we show the equivalence between QAOA with gradually varying angles and quantum annealing with an appropriately constructed continuous schedule in the regime where the total evolution time is bounded by a constant. While our proof techniques are restricted to the Sherrington-Kirkpatrick model and constant total evolution time, we expect the results to generalize broadly. We show numerically that the results generalize to QAOA with large $p$ and parameters conjectured to be optimal, which corresponds to linear total evolution time. 
Our results take a step towards disproving the folklore belief that QAOA mechanism is different from that of quantum annealing for large QAOA angles and are in sharp contrast with prior works which conjecture different QAOA mechanisms for constant total evolution time (``small-angle'') and constant angle magnitude regimes~\cite{qaoa_phase_diagram_quantum_chemistry,qaoa_beyond_low_depth}. 

An important limitation of our findings is their restriction to gradually varying QAOA angles. If this restriction is relaxed, exponential separations exist between the performance of QAOA and quantum annealing~\cite{bapat2019bangbang}. A limitation of these separations is that they are obtained using symmetric problems, which are not representative of hard optimization and constrained-satisfaction problems targeted by classical solvers or arising in practice. In contrast, for problems that are well-studied in classical literature like LABS~\cite{qaoa_labs} or $k$-SAT~\cite{qaoa_ksat}, QAOA with gradually varying angles appears empirically to work best and our results are likely to apply. An interesting direction for future work is understanding for what problems good QAOA performance leads to QAOA approximating quantum annealing.

Our central technical result is a novel analysis of the expansion of expected QAOA energy in total $\gamma$ angle, with expectation taken over random instance choice. Unlike prior results~\cite{qaoa_sk,qaoa_maxcut_high_girth,qaoa_spin_glass_models,qaoa_ksat,qaoa_spiked_tensor,qaoa_quantum_problems_zhou,qaoa_quantum_problems_sud}, our techniques work in finite size and cover both the setting where QAOA is used as an exact as well as approximate solver. 

A practical consequence of our approximation result is that linear-time annealing {(approximated by constant total angle QAOA)} may be compiled with a Trotter step size of order $\mathcal{O}(n)$ {(equivalently, QAOA angles of constant order)} rather than $\mathcal{O}(1)$ {(equivalently, rather than QAOA angles of order $1/n$). This in turn results in a reduction of the number of Trotter layers by a factor $n$, implying a saving in gate count by a similar factor for fault-tolerant compilation. Applying this to the SK model, where each Trotter layer requires depth $\widetilde{\mathcal{O}}\left(n\right)$, the total depth is reduced from $\widetilde{\mathcal{O}}\left(n^2\right)$ to $\widetilde{\mathcal{O}}\left(n\right)$ ---a quadratic improvement.} Unfortunately, our theoretical results fall short of predicting how the Trotter step size varies as a function of the total evolution time, though numerical results seem to show it can actually be increased with total evolution time.

\section*{Acknowledgements}

The authors thank Anuj Apte, Abid Khan, Jacob Watkins, Shouvanik Chakrabarti, and Brandon Augustino for helpful discussions and feedback on the manuscript. The authors thank their colleagues at the Global Technology Applied Research center of JPMorganChase for support.

\section*{Data Availability}

The full data presented in this work is available at  \url{https://doi.org/10.5281/zenodo.15013474}.

\bibliography{bibliography}

\section*{Disclaimer}
This paper was prepared for informational purposes by the Global Technology Applied Research center of JPMorganChase. This paper is not a product of the Research Department of JPMorganChase or its affiliates. Neither JPMorganChase nor any of its affiliates makes any explicit or implied representation or warranty and none of them accept any liability in connection with this position paper, including, without limitation, with respect to the completeness, accuracy, or reliability of the information contained herein and the potential legal, compliance, tax, or accounting effects thereof. This document is not intended as investment research or investment advice, or as a recommendation, offer, or solicitation for the purchase or sale of any security, financial instrument, financial product or service, or to be used in any way for evaluating the merits of participating in any transaction.

\appendix
\onecolumngrid

\tableofcontents

\newpage
\section{Complements to numerical experiments}
This appendix provides additional details about numerical experiments omitted from the main text.

\subsection{Estimation of continuous schedules corresponding to optimal angles}
\label{sec:optimal_angles_continuous_schedules}

We start by explaining how the continuous $\gamma$ and $\beta$ schedules corresponding to optimal SK-QAOA angles were estimated. The method relies on Fourier extrapolation, which was observed to efficiently predict QAOA angles at larger $p$ from optimized angles at smaller $p$ \cite{qaoa_performance_mechanism_implementation}. We precisely state the Fourier extrapolation formulae we used in the experiments. Let be given $\gamma, \beta$ angles optimized for $p_0$-layers SK-QAOA; we denote these by $\bm{\gamma}^{(p_0)}, \bm{\beta}^{(p_0)}$. These angle sequences are then decomposed in appropriate discrete Fourier bases:
\begin{align}
    \gamma^{(p_0)}_t & =: \frac{1}{p_0}\sum_{1 \leq k \leq p_0}\widehat{\gamma}^{(p_0)}_k\sin\left(\pi\left(k - 1/2\right)\frac{t - 1/2}{p_0}\right), & 1 \leq t \leq p_0,\\
    \beta^{(p_0)}_t & =: \frac{1}{p_0}\sum_{1 \leq k \leq p_0}\widehat{\beta}^{(p_0)}_k\cos\left(\pi\left(k - 1/2\right)\frac{t - 1/2}{p_0}\right), & 1 \leq t \leq p_0,
\end{align}
where $\bm{\widehat{\gamma}^{(p_0)}}, \bm{\widehat{\beta}}^{(p_0)}$ are the Fourier components of $\bm{\gamma}^{(p_0)}$, $\bm{\beta}^{(p_0)}$ respectively. Technically, $\bm\gamma$ is applied the \textit{real odd Fourier transform of type IV}, and $\bm\beta$ the \textit{real even Fourier transform of type IV}, as implemented in the FFTW package \cite{FFTW05}; these choices of Fourier transforms were dictated by the performance at extrapolation. From these Fourier components, extrapolated angles $\bm{\gamma}^{(p),\,\mathrm{extrap}}, \bm{\beta}^{(p),\,\mathrm{extrap}}$ can be computed at any number of layers $p$ by formulae:
\begin{align}
    \gamma^{(p),\,\mathrm{extrap}}_t & := \frac{1}{p}\sum_{1 \leq k \leq p_0}\widehat{\gamma}^{(p_0)}_k\sin\left(\pi\left(k - 1/2\right)\frac{t - 1/2}{p}\right), & 1 \leq t \leq p\label{eq:gammas_extrapolated}\\
    \beta^{(p),\,\mathrm{extrap}}_t & := \frac{1}{p}\sum_{1 \leq k \leq p_0}\widehat{\beta}^{(p_0)}_k\cos\left(\pi\left(k - 1/2\right)\frac{t - 1/2}{p}\right), & 1 \leq t \leq p.\label{eq:betas_extrapolated}
\end{align}
In accordance with the previous formulae, we define the continuous extrapolated schedules as:
\begin{align}
    \gamma^{\mathrm{cont}}\left(s\right) & := \frac{1}{p}\sum_{1 \leq k \leq p_0}\widehat{\gamma}^{(p_0)}_k\sin\left(\pi\left(k - 1/2\right)s\right), & 1 \leq t \leq p,\label{eq:gammas_extrapolated_continuous}\\
    \beta^{\mathrm{cont}}\left(s\right) & := \frac{1}{p}\sum_{1 \leq k \leq p_0}\widehat{\beta}^{(p_0)}_k\cos\left(\pi\left(k - 1/2\right)s\right), & 1 \leq t \leq p.\label{eq:betas_extrapolated_continuous}
\end{align}

We then define the following discretization rule for generating a QAOA schedule $\left(\bm{\gamma}^{(p)}, \bm{\beta}^{(p)}\right)$ at finite $p$:
\begin{align}
    \gamma^{(p)}_t & := \frac{1}{p}\gamma^{\mathrm{cont}}\left(\frac{t - 1/2}{p}\right),\label{eq:gamma_schedule_from_continuous_numerics}\\
    \beta^{(p)}_t & := \frac{1}{p}\beta^{\mathrm{cont}}\left(\frac{t - 1/2}{p}\right).\label{eq:beta_schedule_from_continuous_numerics}
\end{align}

This definition is consistent with the main text's 
definition of discrete schedules from continuous ones (Eq.~\ref{eq:discretzation_approximate}).
In practice, in numerical experiments we set $p_0 := 17$, the largest number of layers where QAOA angles were exactly optimized in the infinite size limit to the best of our knowledge \cite{qaoa_maxcut_high_girth}. The parameters we use are reproduced in Table~\ref{table:optimal_params}.

\begin{table}[H]
\centering
\begin{tabular}{|c|c|}
\hline
$\bm\gamma$ &  $\bm\beta$ \\
\hline \hline
0.1735, 0.3376, 0.3562, 0.3789, 0.3844, &  0.6375, 0.5197, 0.4697, 0.4499, 0.4255, \\
0.3907, 0.3946, 0.4016, 0.4099, 0.4217, &  0.4054, 0.3832, 0.3603, 0.3358, 0.3092, \\
0.4370, 0.4565, 0.4816, 0.5138, 0.5530, &  0.2807, 0.2501, 0.2171, 0.1816, 0.1426, \\
0.5962, 0.6429 &  0.1001, 0.0536 \\
 \hline
\end{tabular}
\caption{Optimized values of $\bm\gamma$ and $\bm\beta$ for $p=17$, reproduced from Ref.~\cite{qaoa_maxcut_high_girth}.
}
\label{table:optimal_params}
\end{table}

\begin{figure*}[t]
    \centering
    \begin{subfigure}{0.45\textwidth}
        \centering
        \includegraphics[width=\textwidth]{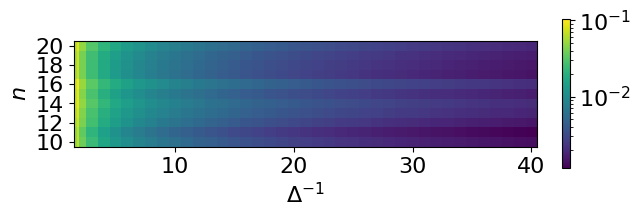}
        \caption{$T = 4$}
    \end{subfigure}
    \begin{subfigure}{0.45\textwidth}
        \centering
        \includegraphics[width=\textwidth]{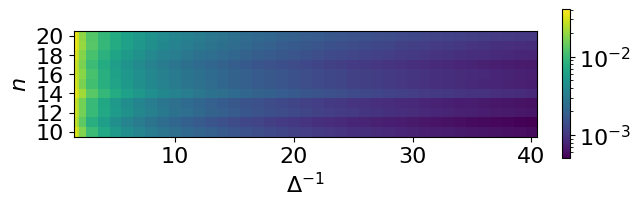}
        \caption{$T = 8$}
    \end{subfigure}\\
    \begin{subfigure}{0.45\textwidth}
        \centering
        \includegraphics[width=\textwidth]{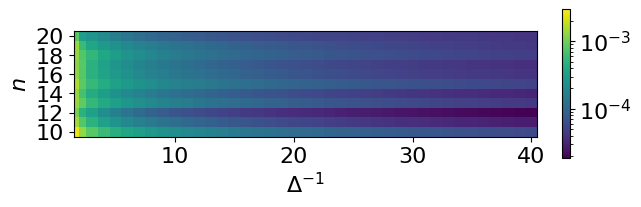}
        \caption{$T = 32$}
    \end{subfigure}
    \begin{subfigure}{0.45\textwidth}
        \centering
        \includegraphics[width=\textwidth]{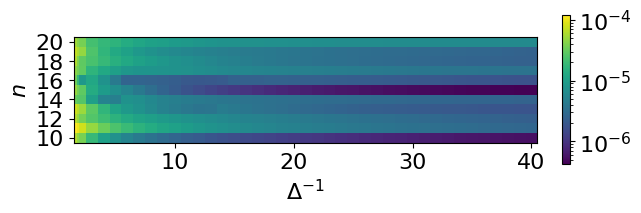}
        \caption{$T = 128$}
    \end{subfigure}
    \caption{Relative energy error between QAOA and analog annealing as a function of $\Delta \in [1/40, 1/2]$ and $n \in [10, 20]$, for schedules of increasing total angles. On this range, the error uniformly vanishes as $T$ grows. 
    }
    \label{fig:varying_trotter_step_extrapolated_optimal}
\end{figure*}
\begin{figure}[!htbp]
    \centering
    \begin{subfigure}{0.45\textwidth}
        \centering
        \includegraphics[width=\textwidth]{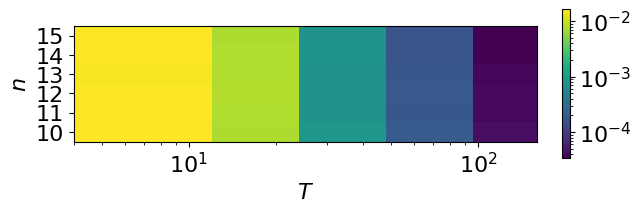}
        \caption{$\Delta = 0.5$}
    \end{subfigure}
    \begin{subfigure}{0.45\textwidth}
        \centering
        \includegraphics[width=\textwidth]{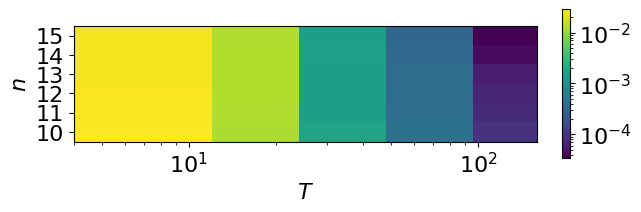}
        \caption{$\Delta = 0.8$}
    \end{subfigure}\\
    \begin{subfigure}{0.45\textwidth}
        \centering
        \includegraphics[width=\textwidth]{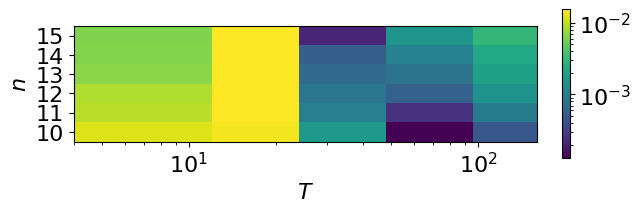}
        \caption{$\Delta = 1.0$}
    \end{subfigure}
    \begin{subfigure}{0.45\textwidth}
        \centering
        \includegraphics[width=\textwidth]{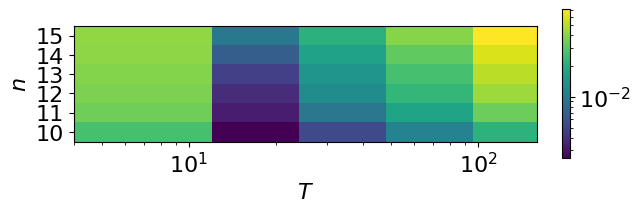}
        \caption{$\Delta = 1.2$}
    \end{subfigure}
    \caption{Relative energy error between QAOA and analog annealing as a function of $T$ and $n$, for schedules with different $\Delta$ parameters. The choice of $\Delta = 1$ approximately sets QAOA angles to their optimal values; in this case, the behavior of the error as the total angle parameter $T$ increases is unclear. For $Delta < 1$, approximation error vanishes and for $\Delta > 1$, the approximation error explodes with total time.
    }
    \label{fig:varying_total_time_fixed_delta}
\end{figure}

\begin{figure}[!htbp]
    \centering
    \begin{subfigure}{0.4\textwidth}
        \centering
        \includegraphics[width=\textwidth]{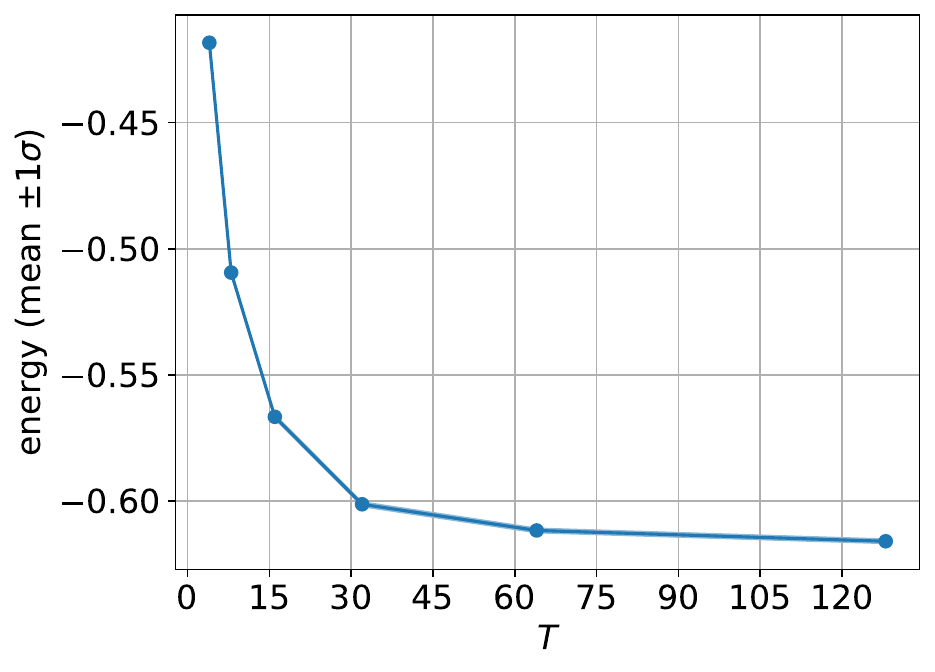}
        \caption{With energy as figure of merit}
    \end{subfigure}
    \begin{subfigure}{0.4\textwidth}
        \centering
        \includegraphics[width=\textwidth]{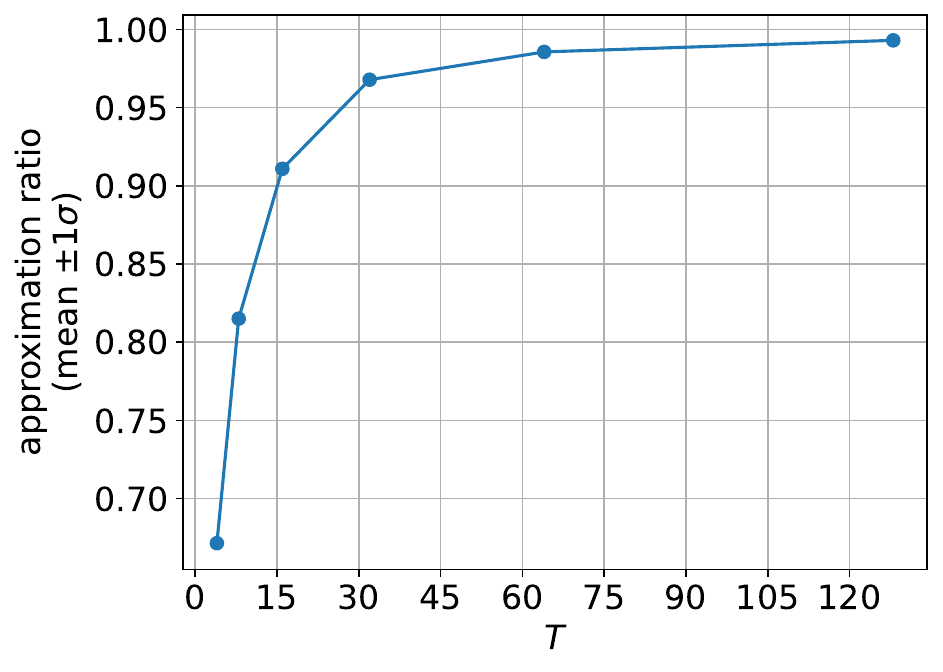}
        \caption{With approximation ratio as figure of merit}
    \end{subfigure}
    \caption{Performance of analog quantum annealing as a function of $T$. Two figures of merit are considered for the performance: the energy and approximation ratio of the energy to the optimal one. The energy is normalized to be a constant in the $n \to \infty$ limit. Figures of merit are aggregated over all SK instances of all sizes.}
    \label{fig:continuous_time_annealing_performance}
\end{figure}

\begin{figure}[!htbp]
    \centering
    \begin{subfigure}{0.45\textwidth}
        \centering
        \includegraphics[width=\textwidth]{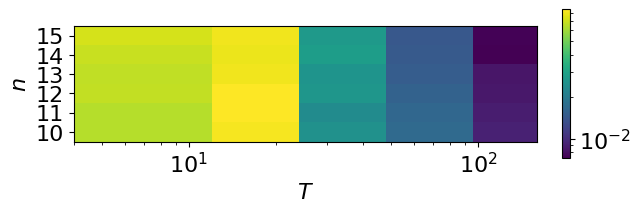}
        \caption{$\Delta = 0.4$}
    \end{subfigure}
    \begin{subfigure}{0.45\textwidth}
        \centering
        \includegraphics[width=\textwidth]{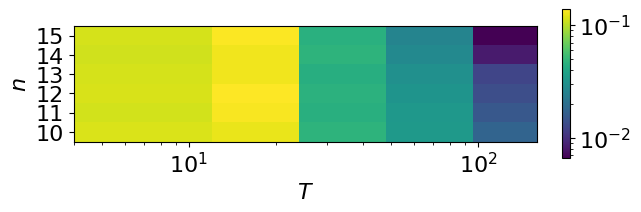}
        \caption{$\Delta = 0.8$}
    \end{subfigure}\\
    \begin{subfigure}{0.45\textwidth}
        \centering
        \includegraphics[width=\textwidth]{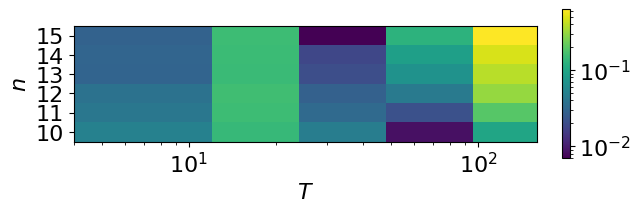}
        \caption{$\Delta = 1.0$}
    \end{subfigure}
    \begin{subfigure}{0.45\textwidth}
        \centering
        \includegraphics[width=\textwidth]{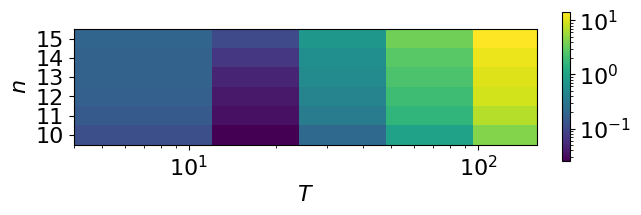}
        \caption{$\Delta = 1.2$}
    \end{subfigure}
    \caption{Relative error in residual approximation ratio between QAOA and analog annealing as a function of $T$ and $n$, for schedules with different $\Delta$ parameters. $\Delta = 1$ approximately sets QAOA angles to their optimal values. Clear convergence is observed for $\Delta < 1$ and clear divergence is observed for $\Delta > 1$.}
    \label{fig:varying_total_time_fixed_delta_residual_approximation_ratio}
\end{figure}

\begin{figure}[htbp]
    \captionsetup[subfigure]{justification=centering}
    \centering
    \begin{subfigure}{0.33\textwidth}
        \centering
        \includegraphics[width=\textwidth]{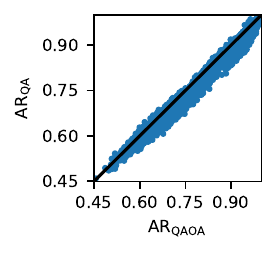}
        \caption{$\Delta = 0.8$}
    \end{subfigure}
    \begin{subfigure}{0.32\textwidth}
        \centering
        \includegraphics[width=\textwidth]{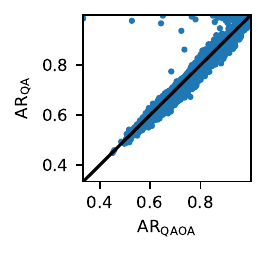}
        \caption{$\Delta = 1.0$}
    \end{subfigure}
    \begin{subfigure}{0.33\textwidth}
        \centering
        \includegraphics[width=\textwidth]{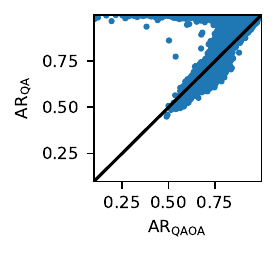}
        \caption{$\Delta = 1.2$}
    \end{subfigure}
    \caption{Comparison of QAOA and quantum annealing approximation ratios (ARs) for $3$ different values of $\Delta$. Points corresponding to all instances (no aggregation) are represented. At $\Delta = 0.8$, vanishing of the QAOA-QA error with increasing total evolution time is observed (Figure \ref{fig:constant_angle_magnitude} from main text), which is reflected in the strong correlation between the two approximation ratios. At $\Delta = 1.0$, while the behavior of the error between QAOA AR and QA AR is unclear (see Figure \ref{fig:varying_total_time_fixed_delta_residual_approximation_ratio} from present appendix), a strong correlation persists between the two quantities. In contrast, at $\Delta = 1.2$, where the error between QAOA and QA empirically manifestly explodes, the correlation significantly weakens.}
    \label{fig:fixed_Delta_ar_qaoa_vs_ar_qa}
\end{figure}

\subsection{Additional numerical results}
\label{sec:appendix_extra_numerics}

We now present additional numerical results. We denote total evolution time by $T=\Delta\cdot p$ and report additional results for constant $T$ (Fig.~\ref{fig:varying_trotter_step_extrapolated_optimal}) and for constant angle magnitude ($T$ growing linearly with $p$, Figs.~\ref{fig:varying_total_time_fixed_delta},~\ref{fig:varying_total_time_fixed_delta_residual_approximation_ratio},~\ref{fig:fixed_Delta_ar_qaoa_vs_ar_qa}). We additionally show that in the latter regime, approximation ration for quantum annealing approaches 1 (Fig.~\ref{fig:continuous_time_annealing_performance}; see Fig.~\ref{fig:numerical_intuition}D for corresponding plot for QAOA).

\section{Tensor notations and background on Quadratic Generalized Multinomial Sums (QGMS)}
\label{sec:notations}

Quadratic Generalized Multinomial Sums (QGMS) are the main mathematical objects allowing the analysis of QAOA in this work. In this section, we provide elementary background on QGMS, with more technical definitions deferred to section \ref{sec:qgms_saddle_point_correlations_tensors}. Since the definition and analysis of QGMS involves tensors, we start by introducing a variety of tensor notations.

\subsection{Tensor notations}

In this section, we introduce notations that will be used throughout the technical appendices to manipulate tensors.

We start with notations for multi-indices, which will be needed to index tensors. Consider a degree $d$ tensor
\begin{align}
    \bm{T} & = \left(T_{\alpha_1,\,\ldots,\,\alpha_d}\right)_{\alpha_1,\,\ldots,\,\alpha_d \in \mathcal{A}} \in \left(\mathbf{C}^{\mathcal{A}}\right)^{\otimes d} \simeq \mathbf{C}^{\mathcal{A}^d},
\end{align}
indexed by $d$ indices taking values in $\mathcal{A}$. We will then frequently denote
\begin{align}
    \bm{\alpha}_{1:d} & := \left(\alpha_1,\,\ldots,\,\alpha_d\right) \in \mathcal{A}^d
\end{align}
for the $d$-tuple of these indices. Said differently, a vector notation indexed by a colon range indicates a tuple with elements named as the vectors, and indices iterating in the specified range. Using notation
\begin{align}
    T_{\bm{\alpha}_{1:d}} & := T_{\alpha_1,\,\ldots,\,\alpha_d}
\end{align}
will occasionally prove convenient to avoid writing many indices. Similarly, if the $d$ indices of $\bm{T}$ can be obtained by concatenating a $d'$ tuple of indices and a $d''$-tuple of indices:
\begin{align}
    \bm{\alpha'} & := \bm{\alpha'}_{1:d'} := \left(\alpha'_{1},\,\ldots,\,\alpha'_{d'}\right) \in \mathcal{A}^{d'},\\
    \bm{\alpha''} & := \bm{\alpha''}_{1:d''} := \left(\alpha''_{1},\,\ldots,\,\alpha''_{d''}\right) \in \mathcal{A}^{d''},\\
    d & = d' + d'',
\end{align}
we denote
\begin{align}
    T_{\bm{\alpha'},\,\bm{\alpha''}} := T_{\bm{\alpha'}_{1:d'},\,\bm{\alpha''}_{1:d''}} := T_{\alpha'_1,\,\ldots,\,\alpha'_{d'},\,\alpha''_1,\,\ldots,\,\alpha''_{d''}}.
\end{align}
Using similar notations, we denote by
\begin{align}
    \bm{T}_{\bm{\alpha'}}
\end{align}
the degree $d''$ tensor defined by setting the first $d'$ indices of $\bm{T}$ to $\bm{\alpha}$. That is:
\begin{align}
    \left[\bm{T}_{\bm{\alpha'}}\right]_{\bm{\alpha''}} & := T_{\bm{\alpha'},\,\bm{\alpha''}}, \qquad \bm{\alpha'} \in \mathcal{A}^{d'},\,\bm{\alpha''} \in \mathcal{A}^{d''},\,d' + d'' = d.
\end{align}
where indexed square brackets around a tensor denote the entry of this tensor identified by the index. We also use notation
\begin{align}
    \bm{T}_{\alpha'}
\end{align}
for $\alpha' \in \mathcal{A}$ in the special case $d' = 1$, i.e. $\bm{\alpha'} = \left(\alpha\right)$. We define the Euclidean dot product between two tensors $\bm{S}, \bm{T} \in \left(\mathbf{C}^{\mathcal{A}}\right)^{\otimes d}$ by:
\begin{align}
    \left\langle \bm{S}, \bm{T} \right\rangle & := \sum_{\bm{\alpha} \in \mathcal{A}^d}S_{\bm{\alpha}}T_{\bm{\alpha}}.
\end{align}
Note that there is no Hermitian conjugation even if tensors are complex-valued (as a result, this is not really a dot product defining a Hilbert space, but it does not impact the derivations). We also extend the definition of the dot products to tensors of different degrees. Let us assume for definiteness that $\bm{S}$ has degree $d' + d''$ and $\bm{T}$ degree $d''$ (the converse is defined by symmetry). Then, the dot product of $\bm{S}$ and $\bm{T}$ is a tensor of degree $d'$
\begin{align}
    \bm{U} & := \left\langle \bm{S}, \bm{T} \right\rangle \in \left(\mathbf{C}^{\mathcal{A}}\right)^{\otimes d'}
\end{align}
with entries defined by:
\begin{align}
    U_{\bm{\alpha'}} & := \sum_{\bm{\alpha''} \in \mathcal{A}^{d''}}S_{\alpha',\,\alpha''}T_{\bm{\alpha''}}.\label{eq:tensor_dot_product_different_dimensions}
\end{align}

\subsection{Background on Quadratic Generalized Multinomial Sums (QGMS)}
\label{sec:qgms_background}

In this Section, we precisely introduce quadratic generalized multinomial sums (QGMS), whose definition was sketched in Section \ref{sec:small_total_gamma_angle_expansion_results}. We also introduce some further definitions and concepts, motivating them by the $n \to \infty$ asymptotic analysis of QGMS. In particular, the saddle point $\bm{\theta}^*$, introduced by its defining Equation \ref{eq:saddle_point_equation_main_text}, plays an important role in this asymptotic analysis. In the present work, we will always reason at finite instance size $n$, but the saddle point and related objects will still play an important role.

We start with a general definition of a quadratic generalized multinomial sum (QGMS). A QGMS is specified by two finite index sets $\mathcal{A}, \mathcal{S}$, a vector $\bm{Q} = \left(Q_{\bm{a}}\right)_{\bm{a} \in \mathcal{S}}$ indexed by $\mathcal{S}$ and a matrix $\bm{L} = \left(L_{\alpha,\,\bm{a}}\right)_{\alpha \in \mathcal{A},\,\bm{a} \in \mathcal{S}}$ indexed by $\mathcal{A} \times \mathcal{S}$. The QGMS associated to these objects is the sequence $\left(S_n\right)_{n \geq 1}$ indexed by integer $n$:
\begin{align}
    S_n & = \sum_{\bm{n} \in \mathcal{P}(n)}\binom{n}{\bm{n}}\exp\left(\frac{1}{2n}\bm{n}^T\bm{L}^T\bm{L}\bm{n}\right)\prod_{\bm{a} \in \mathcal{S}}Q_{\bm{a}}^{n_{\bm{a}}}, \label{eq:qgms_basic_definition_restated}
\end{align}
where we denoted by
\begin{align}
    \mathcal{P}(n) & := \left\{\bm{n} = \left(n_{\bm{a}}\right)_{\bm{a} \in \mathcal{S}}\,:\,\sum_{\bm{a} \in \mathcal{S}}n_{\bm{a}} = n\right\}
\end{align}
the set of $\mathcal{S}$-indexed tuples summing up to $n$. The standard form of a QGMS, as restated in Equation \ref{eq:qgms_basic_definition_restated}, is not sufficient to express QAOA observables of several cases, in particular the SK-QAOA energy. However, these can often be expressed by a QGMS variant with general term amended by a polynomial in $\bm{n}$, namely:
\begin{align}
    \sum_{\bm{n} \in \mathcal{P}\left(n\right)}\binom{n}{\bm{n}}\exp\left(\frac{1}{2n}\bm{n}^T\bm{L}^T\bm{L}\bm{n}\right)n_{\bm{a}^{(1)}}n_{\bm{a}^{(2)}}\ldots n_{\bm{a}^{(d - 1)}}n_{\bm{a}^{(d)}}\prod_{\bm{a} \in \mathcal{S}}Q_{\bm{a}}^{n_{\bm{a}}},\label{eq:qgms_pseudo_moment}
\end{align}
for specific choices of $\bm{a}^{(1)}, \bm{a}^{(2)}, \ldots, \bm{a}^{(d - 1)}, \bm{a}^{(d)} \in \mathcal{S}$. These quantities can be informally regarded as moments due to the following observation. Indeed, consider a random variable $\bm{N} = \left(N_1, \ldots, N_k\right)$, distributed according to a multinomial distribution of probabilities $\left(p_1, \ldots, p_k\right)$ with $n$ trials. The joint distribution of $\bm{N}$ is given by
\begin{align}\label{eq:multinomial_distribution}
    \mathbb{P}\left[\bm{N} = \left(n_1, \ldots, n_k\right)\right] & = \binom{n}{\left(n_j\right)_{1 \leq j \leq k}}\prod_{1 \leq j \leq k}p_j^{n_j}.
\end{align}

The moment of $\bm{N}$ of order $d$ and coordinates $\left(j^{(1)}, j^{(2)}, \ldots, j^{(d - 1)}, j^{(d)}\right)$ is given by
\begin{align}
    \sum_{\substack{\bm{n} = \left(n_j\right)_{1 \leq j \leq k}\\\sum\limits_{1 \leq j \leq k}n_j = n}}\binom{n}{\bm{n}}n_{j^{(1)}}n_{j^{(2)}}\ldots n_{j^{(d - 1)}}n_{j^{(d)}}\prod_{1 \leq j \leq k}p_j^{n_j}.
\end{align}
This expression can be formally likened to Equation \ref{eq:qgms_pseudo_moment} when $\bm{L} := \bm{0}$; namely, complex numbers $\left(Q_{\bm{a}}\right)_{\bm{a} \in \mathcal{S}}$ play the role of $\left(p_j\right)_{1 \leq j \leq k}$ in the multinomial distribution of Eq.~\ref{eq:multinomial_distribution}. While deriving the QGMS representation of the SK-QAOA energy (Section \ref{sec:sk_qaoa_qgms}), we will see that not all moments in Equation \ref{eq:qgms_pseudo_moment} occur in this expression, but rather only the moments of monomials in the linear span of $\bm{L}$, i.e.:
\begin{align}
    \sum_{\bm{n} \in \mathcal{P}\left(n\right)}\binom{n}{\bm{n}}\exp\left(\frac{1}{2n}\bm{n}^T\bm{L}^T\bm{L}\bm{n}\right)\frac{\left(\bm{L}\bm{n}\right)_{\alpha_1}}{n}\frac{\left(\bm{L}\bm{n}\right)_{\alpha_2}}{n}\ldots \frac{\left(\bm{L}\bm{n}\right)_{\alpha_{d - 1}}}{n}\frac{\left(\bm{L}\bm{n}\right)_{\alpha_d}}{n}\prod_{\bm{a} \in \mathcal{S}}Q_{\bm{a}}^{n_{\bm{a}}}.\label{eq:qgms_pseudo_moment_restricted}
\end{align}
The additional $1/n$ normalization in the above equation will emerge naturally when deriving the SK-QAOA QGMS. Equation \ref{eq:qgms_pseudo_moment_restricted} for the ``restricted" pseudo-moments of the QGMS motivates the introduction of a generating function for these pseudo-moments:

\begin{definition}[Pseudo-Moment Generating Function of Quadratic Generalized Multinomial Sums (QGMS-MGF)] We define a pseudo-moment generating function of a quadratic generalized multinomial sum (QGMS-MGF) as the following sum:

\label{def:qgms-mgf}
\begin{align}
    S_n\of{\bm\mu} & = \sum_{\bm{n} = \of{n_{\bm{a}}}_{\bm{a} \in \mathcal{S}} \in \mathcal{P}(n)}\binom{n}{\bm{n}}\exp\of{\frac{1}{2n}\bm{n}\bm{L}^T\bm{L}\bm{n} + \bm{\mu}^T\bm{L}\frac{\bm{n}}{n}}\prod_{\bm{a} \in \mathcal{S}}Q_{\bm{a}}^{n_{\bm{a}}},\label{eq:qgms_mgf}
\end{align}

where

\begin{align}
    \bm{L} = \left(L_{\alpha,\,\bm{a}}\right)_{\alpha \in \mathcal{A},\,\bm{a} \in \mathcal{S}} \in \mathbf{C}^{\mathcal{A} \times \mathcal{S}},
\end{align}

is an arbitrary complex matrix with columns indexed by indices $\bm{a}\in \mathcal{S}$ (``bitstring index") and rows indexed by indices $\alpha \in \mathcal{A}$, $\mathcal{P}(n)$ denotes the set of all partitions of $n$ elements in to sets, restricting $\sum_{\bm{a}} n_{\bm{a}}=n$, and $\binom{n}{\bm{n}}$ denotes the standard multinomial coefficient
\begin{align}
\binom{n}{\{n_{\bm{a}}\}} & := \frac{n!}{\prod_{\bm{a} \in \mathcal{S}}n_{\bm{a}}!}
\end{align}
\end{definition}

We now introduce the pseudo moment-generating function of a parametrized QGMS, similar to the moment-generating function of a standard MGF in Definition \ref{def:qgms-mgf}. First, observe that whenever $\bm{L} = \bm{0}$, the QMGS-MGF is trivial to compute as it collapse to a standard multinomial sum:
\begin{align}
    S_n\left(\bm\mu\right) & = \sum_{\bm{n} \in \mathcal{P}(n)}\binom{n}{\bm{n}}\exp\left(\frac{1}{n}\bm{\mu}^T\bm{n}\right)\prod_{\bm{a} \in \mathcal{S}}Q_{\bm{a}}^{n_{\bm{a}}} \qquad \left(\bm{L} = \bm{0}\right)\nonumber\\
    & = \left(\sum_{\bm{a} \in \mathcal{S}}Q_{\bm{a}}\exp\left(\frac{1}{n}\bm{\mu}^T\bm{L}_{:,\,\bm{a}}\right)\right)^n.
\end{align}
The following definition now introduces the pseudo-moment-generating function of a paramerized QGMS, which recovers the above trivial limit when parameter $\lambda$ is set to $0$:

\begin{definition}[Pseudo-Moment Generating Function of Parametrized Quadratic Generalized Multinomial Sums (QGMS-MGF)]
Let parameters $\mathcal{A}, \mathcal{S}, \bm{Q}, \bm{L}$ be given as in Definition \ref{def:pqgms-mgf}. We define a pseudo-moment generating function of a parametrized quadratic generalized multinomial sum (PQGMS-MGF) as the following sum:

\label{def:pqgms-mgf}
\begin{align}
    S_n\of{\lambda, \bm\mu} & = \sum_{\bm{n} = \of{n_{\bm{a}}}_{\bm{a} \in \mathcal{S}} \in \mathcal{P}(n)}\binom{n}{\bm{n}}\exp\of{\frac{\lambda^2}{2n}\bm{n}\bm{L}^T\bm{L}\bm{n} + \lambda\bm{\mu}^T\bm{L}\frac{\bm{n}}{n}}\prod_{\bm{a} \in \mathcal{S}}Q_{\bm{a}}^{n_{\bm{a}}},\label{eq:pqgms_mgf}
\end{align}

\end{definition}

\section{QAOA expectations from moment generating functions of Quadratic Generalized Multinomial Sums}
\label{sec:sk_qaoa_qgms}

After giving elements of background on Quadratic Generalized Multinomial Sums (QGMS) in section \ref{sec:qgms_background}, we establish QGMS representations of the expected cost function and and cost function squared in SK-QAOA. These were previously derived in Ref.~\cite{qaoa_sk}, but for completeness and consistency, we offer explicit derivations adapted to our notations here.

In this Section, we formulate the computation of relevant observables in SK-QAOA in terms of Quadratic Generalized Multinomial Sums (QGMS). More precisely, we consider the disorder average of the expected cost function $\mathbb{E}\bra{\bm{\gamma}, \bm{\beta}}C/n\ket{\bm{\gamma}, \bm{\beta}}$ (Section~\ref{sec:sk_qaoa_first_order_moment_qgms}), and the disorder average of the expected squared cost function $\mathbb{E}\bra{\bm{\gamma}, \bm{\beta}}\left(C/n\right)^2\ket{\bm{\gamma}, \bm{\beta}}$ (Section~\ref{sec:sk_qaoa_second_order_moment_qgms}). The continuum limit ($p \to \infty$ with angles schedules prescribed by Definition~\ref{def:angles_from_continuum}) of the first order moment will constitute the main object of study of Appendix~\ref{sec:sk_qaoa_energy_continuum_limit}, leading to the proof of main Theorem~\ref{th:approximation_continuous_time_annealing_qaoa}. The second order moment will serve to establish concentration of the expected cost across disorder for large size $n$ uniformly in this limit; Appendix Section~\ref{sec:other_moments_concentration} is dedicated to this reasoning.

\subsection{The first order moment}
\label{sec:sk_qaoa_first_order_moment_qgms}

In this Section, we derive a QGMS representation (more specifically, a representation in terms of QGMS moments, see Definition \ref{def:qgms-mgf}) for the instance-averaged SK-QAOA energy $\mathbb{E}\bra{\bm{\gamma}, \bm{\beta}}C/n\ket{\bm{\gamma}, \bm{\beta}}$. This representation of the SK-QAOA energy was for the first time explicitly derived in \cite{qaoa_sk}. For completeness, we 
reprove it here using a slighly modified version of this original work's arguments, as well as adapted notations. Recalling Definition \ref{def:sk_model} of the SK model, the problem's cost function (expressed here as a diagonal Hamiltonian) depends on a random upper matrix triangle $\bm{J} = \left(J_{j,\,k}\right)_{1 \leq j < k \leq n}$:
\begin{align}
    C & := \frac{1}{\sqrt{n}}\sum_{1 \leq j < k \leq n}J_{j,\,k}Z_jZ_k.
\end{align}
The $p$-layers QAOA state:
\begin{align}
    \ket{\bm\gamma, \bm\beta} & = \overleftarrow{\prod_{t = 1}^p}\exp\left(-i\beta_tB\right)\exp\left(-i\gamma_tC\right)\ket{+}^{\otimes n}
\end{align}
then implicitly depends on randomness $\bm{J}$. In this context, we take as figure of merit the expected energy of a string produced by QAOA, averaged over randomness:
\begin{align}
    \nu_{p,\,n}\left(\bm\gamma, \bm\beta\right) & := \mathbb{E}\braket{\bm\gamma, \bm\beta|C|\bm\gamma, \bm\beta},
\end{align}
also referred to as \textit{SK-QAOA energy} for short. This metric was shown (Ref.~\cite{qaoa_sk}) to concentrate over randomness $\bm{J}$ in the infinite size limit $n \to \infty$. The QGMS representation of the SK-QAOA energy is summarized in the following proposition:

\begin{proposition}[SK-QAOA energy from generalized multinomial sum]
\label{prop:qgms-mgf_formulation_sk_qaoa_energy}
Consider SK-QAOA at finite size $n$ and number of layers $p+1$, with angles $\bm\gamma = \left(\gamma_1, \ldots, \gamma_{p}\right)$ and $\bm\beta = \left(\beta_1, \ldots, \beta_{p}\right)$. Then, the instance-averaged energy achieved by QAOA at size $n$ after layer $t$ ($1 \leq t \leq p - 1$) can be expressed as
\begin{align}
    \nu_{p,\,n,\,t}\of{\bm{\gamma},\bm{\beta}} & = -\frac{i}{\Gamma_{t + 1}}\sum_{r \in \{\pm 1,\,\ldots,\,\pm p\}}\frac{\partial^2S_n\left(\bm\mu\right)}{\partial\mu_{\left(r, t + 1\right)}^2}\Bigg|_{\bm\mu = \bm{0}}.\label{eq:layer_energy_from_restricted_mgf}
\end{align}
In the above equation, $S_n\left(\bm\mu\right)$ is the generalized multinomial sum defined in Equation \ref{eq:qgms_mgf}. The parameters $\mathcal{S}, \mathcal{A}, \bm{L}, \bm{Q}$ of the generalized multinomial sum are as follows:
\begin{align}
    \mathcal{S} & := \left\{\bm{a} = \left(a_1, a_2, \ldots, a_{p - 1}, a_p, a_0, a_{-p}, a_{-(p - 1)}, \ldots, a_{-2}, a_{-1}\right) \in \{1, -1\}^{2p + 1}\right\},\\
    \mathcal{A} & := \{\pm 1,\,\ldots,\,\pm p\}^2,\\
    \bm{L} & := \left(L_{\bm{\alpha},\,\bm{a}}\right)_{\substack{\alpha \in \mathcal{A}\\\bm{a} \in \mathcal{S}}},\quad L_{(r, s),\,\bm{a}} := \sqrt{-\frac{\Gamma_r\Gamma_s}{2} + i\varepsilon}\;a_ra_s,\\
    \bm{Q} & := \left(Q_{\bm{a}}\right)_{\bm{a} \in \mathcal{S}},\quad
    Q_{\bm{a}} := \frac{1}{2}\braket{a_p|e^{i\beta_pX}|a_0}\braket{a_0|e^{-i\beta_pX}|a_{-p}}\hspace{-6px}\prod_{1 \leq l \leq p - 1}\hspace{-6px}\braket{a_l|e^{i\beta_lX}|a_{l + 1}}\braket{a_{-l - 1}|e^{-i\beta_lX}|a_{-l}}.
\end{align}
where
\begin{align}
    \bm\Gamma & := \left(\Gamma_1, \Gamma_2, \ldots, \Gamma_p, \Gamma_0, \Gamma_{-p}, \ldots, \Gamma_{-2}, \Gamma_{-1}\right)\\
    & = \left(\gamma_1, \gamma_2, \ldots, \gamma_{p}, 0, -\gamma_{p}, \ldots, -\gamma_2, -\gamma_1\right).
\end{align}
\end{proposition}

Consider the energy achieved by $p$-layers SK-QAOA at any intermediate layer $t$. Note one may simply do that by cancelling all layers beyond $t$ by unitarity. Then, one would simply need to use the configuration basis $(2t + 1)$-bits bitstrings, with bits indexed by
\begin{align}
    1,\,2,\,\ldots,\,t - 1,\,t,\,0,\,-t,\,-(t - 1),\,\ldots,\,-2,\,-1.
\end{align}
However, in this Section, we show how to perform the same calculation using configuration-basis numbers indexed by $(2p + 1)$-bits bitstrings, with bits indexed as
\begin{align}
    1,\,2,\,\ldots,\,p - 1,\,p,\,0,\,-p,\,\ldots,\,-2,\,-1.
\end{align}
Of course, both calculations should give the same result since they are computing the same thing.

To unify the expressions of the energy computed after an intermediate layer $t$ with that computed after final layer $p$, it will be convenient to use a slightly different indexing for bitstrings. Namely, for a $p$-layer ansatz, we work with $(2p + 2)$-bits bitstrings, with bits indexed as
\begin{align}
    1,\,2,\,\ldots,\,p - 1,\,p,\,p + 1,\,-(p + 1),\,-p,\,-(p - 1),\,\ldots,\,-2,\,-1.
\end{align}
The QAOA state is
\begin{align}
    \ket{\bm\gamma, \bm\beta} & = \overleftarrow{\prod_{l = 1}^p}\exp\left(-i\beta_lB\right)\exp\left(-i\gamma_lC\right)\ket{+}^{\otimes n}
\end{align}
We wish to compute the energy achieved by this state at layer $t$, after averaging over instances
\begin{align}
    \nu_{p,\,n,\,t}\of{\bm{\gamma},\bm{\beta}} & = \mathbb{E}_{\bm{J}}\left\{\bra{+}^{\otimes n}\left(\overrightarrow{\prod_{l = 1}^t}\exp\left(i\gamma_lC\right)\exp\left(i\beta_lB\right)\right)C\left(\overleftarrow{\prod_{l = 1}^t}\exp\left(-i\beta_lB\right)\exp\left(-i\gamma_lC\right)\right)\ket{+}^{\otimes n}\right\}\label{eq:sk_qaoa_expectation_intermediate_layer}
\end{align}
However, as hinted previously, while it may appear awkward, we will rewrite this expectation explicitly keeping the $p$ ansatz layers
\begin{align}
    \nu_{p,\,n,\,t}\of{\bm{\gamma},\bm{\beta}} & = \mathbb{E}_{\bm{J}}\left\{\bra{+}^{\otimes n}\left(\overrightarrow{\prod_{l = 1}^p}\exp\left(i\gamma_lC\right)\exp\left(i\beta_lB\right)\right)\left(\overleftarrow{\prod_{l = t + 1}^p}\exp\left(-i\beta_lB\right)\exp\left(-i\gamma_lC\right)\right)\right.\nonumber\\
    & \left. \hspace*{60px} \times C\left(\overleftarrow{\prod_{l = 1}^t}\exp\left(-i\beta_lB\right)\exp\left(-i\gamma_lC\right)\right)\ket{+}^{\otimes n}\right\}\label{eq:sk_qaoa_intermediate_layer_expanded_unitary}
\end{align}
We now express this expectation by introducing a path integral for the computational basis state of each qubit $j$ after each layer (indexing layers from the unitary with negative indices, and layers from the inverse unitary with positive indices). The computational basis states of qubit $j$ across these layers is denoted by $(2p + 2)$ dimensional vector
\begin{align}
    \bm{z}_j & = \left(z_j^{[1]}, z_j^{[2]}, \ldots, z_j^{[p]}, z_j^{[p + 1]}, z_j^{[-(p + 1)]}, z_j^{[-p]}, \ldots, z_j^{[-2]}, z_j^{[-1]}\right) \in \{1, -1\}^{2p + 2}\label{eq:bit_matrix_qubit_slice}
\end{align}
We also introduce a vector notation for the computational basis states of all qubits at layer $t$
\begin{align}
    \bm{z}^{[l]} & := \left(z^{[l]}_1, z^{[l]}_2, \ldots, z^{[l]}_{n - 1}, z^{[l]}_n\right) \in \{1, -1\}^n
\end{align}
Finally, we collect computational basis states of all qubits after all layers in a matrix
\begin{align}
    \bm{z} & = \begin{pmatrix}
        \bm{z}^{[1]}\\
        \vdots\\
        \bm{z}^{[p + 1]}\\
        \bm{z}^{[-(p + 1)]}\\
        \vdots\\
        \bm{z}^{[-1]}
    \end{pmatrix} = \begin{pmatrix}
        \bm{z}_1 & \bm{z}_2 & \ldots & \bm{z}_{n - 1} & \bm{z}_n
    \end{pmatrix} \in \{1, -1\}^{(2p + 2) \times n}.\label{eq:bit_matrix_layer_slice}
\end{align}
Under this parametrization, expectation \ref{eq:sk_qaoa_intermediate_layer_expanded_unitary} can be expressed:
\begin{align}
    &\nu_{p,\,n,\,t}\of{\bm{\gamma},\bm{\beta}} = \mathbb{E}_{\bm{J}}\sum_{\bm{z} \in \{1, -1\}^{(2p + 2) \times n}}\hspace*{-20px}\braket{+|\bm{z}^{[1]}}\left(\prod_{l = 1}^p\exp\left(i\gamma_lC\left(\bm{z}^{[l]}\right)\right)\braket{\bm{z}^{[l]}|\exp\left(i\beta_lB\right)|\bm{z}^{[l + 1]}}\right)C\left(\bm{z}^{[t + 1]}\right)\nonumber\\
    &\hspace{10px} \times \mathbf{1}\left[\bm{z}^{[p + 1]} = \bm{z}^{[-(p + 1)]}\right]\left(\prod_{l = 1}^p\braket{\bm{z}^{[-l - 1]}|\exp\left(-i\beta_lB\right)|\bm{z}^{[-l]}}\exp\left(-i\gamma_lC\left(\bm{z}^{[-l]}\right)\right)\right)\braket{\bm{z}^{[-1]}|+}\label{eq:sk_qaoa_expectation_intermediate_layer_expanded_unitary_path_integral}
\end{align}
We now introduce the configuration basis numbers of bits $\bm{z}$
\begin{align}
    \bm{n} & = \left(n_{\bm{a}}\right)_{\bm{a} \in \{1, -1\}^{2p + 2}},\\
    n_{\bm{a}} & = \left|\left\{j \in [n]\,:\,\bm{z}_j = \bm{a}\right\}\right|,\\ 
    \sum_{\bm{a} \in \{1, -1\}^{2p + 2}}n_{\bm{a}} & = n.
\end{align}
We now separately consider the multiplicative contributions in the summand of Equation \ref{eq:sk_qaoa_expectation_intermediate_layer_expanded_unitary_path_integral}.
For the contributions coming from mixer unitaries and the overlap with the initial plus state
\begin{align}
    &\braket{+|\bm{z}^{[1]}}\braket{\bm{z}^{[-1]}|+}\prod_{l = 1}^p\braket{\bm{z}^{{[l]}}|\exp\left(i\beta_lB\right)|\bm{z}^{[l + 1]}}\braket{\bm{z}^{[-l - 1]}|\exp\left(-i\beta_lB\right)|\bm{z}^{[-l]}}\mathbf{1}\left[\bm{z}^{[p + 1]} = \bm{z}^{[-(p + 1)]}\right] \nonumber\\ 
    &\hspace{30px}=\prod_{\bm{a}}Q_{\bm{a}}^{n_{\bm{a}}}, \label{eq:sk_qaoa_expectation_intermediate_layer_expanded_unitary_mixer_contribution}
\end{align}
where
\begin{align}
    Q_{\bm{a}} & = \frac{1}{2}\prod_{l = 1}^p\braket{a_l|\exp\left(i\beta_lX\right)|a_{l + 1}}\braket{a_{-l - 1}|\exp\left(-i\beta_lX\right)|a_{-l}}\mathbf{1}\left[\bm{a}_{[p + 1]} = \bm{a}_{[-(p + 1)]}\right].\label{eq:sk_qaoa_expectation_intermediate_layer_expanded_unitary_q_definition}
\end{align}
In particular, this contribution can be expressed directly in terms of configuration basis numbers, which comes from permutation invariance of the mixer Hamiltonian. We now look at the contribution from the cost function, i.e.
\begin{align}
    & C\left(\bm{z}^{[t + 1]}\right)\prod_{l = 1}^p\exp\left(i\gamma_lC\left(\bm{z}^{[l]}\right)\right)\exp\left(-i\gamma_lC\left(\bm{z}^{[-l]}\right)\right)\nonumber\\
    & = C\left(\bm{z}^{[t + 1]}\right)\exp\left(\sum_{1 \leq l \leq p}i\gamma_lC\left(\bm{z}^{[l]}\right) - \sum_{1 \leq l \leq p}i\gamma_lC\left(\bm{z}^{[-l]}\right)\right)\nonumber\\
    & = C\left(\bm{z}^{[t + 1]}\right)\exp\left(\sum_{l \in \pm [p]}i\Gamma_lC\left(\bm{z}^{[l]}\right)\right)\nonumber\\
    & = \frac{1}{\sqrt{n}}\left(\sum_{1 \leq j < k \leq n}J_{j, k}z^{[t + 1]}_jz^{[t + 1]}_k\right)\exp\left(\frac{1}{\sqrt{n}}\sum_{l \in \pm [p]}i\Gamma_l\sum_{1 \leq j < k \leq n}J_{j, k}z^{[l]}_jz^{[l]}_k\right)\label{eq:sk_qaoa_expectation_intermediate_layer_expanded_unitary_nonaveraged_cost_contributions}
\end{align}
where in the third line, we introduced
\begin{align}
    \Gamma_l & = \left\{\begin{array}{cc}
         \gamma_l & \textrm{if } l > 0\\
         -\gamma_{-l} & \textrm{if } l < 0 
    \end{array}\right.
\end{align}
Before taking the explicit average over instances in Equation \ref{eq:sk_qaoa_expectation_intermediate_layer_expanded_unitary_nonaveraged_cost_contributions}, it will help to ``remove" the $\bm{J}$ ``downstairs" (not inside the exponential) using Gaussian integration by parts (proposition \ref{prop:gaussian_integration_by_parts}). We start by separating the contribution of each pair $\{j, k\}$ to the energy by linearity
\begin{align}
    & \mathbb{E}_{\bm{J}}\left\{\frac{1}{\sqrt{n}}\left(\sum_{1 \leq j < k \leq n}J_{j, k}z^{[t + 1]}_jz^{[t + 1]}_k\right)\exp\left(\frac{1}{\sqrt{n}}\sum_{l \in \pm [p]}i\Gamma_l\sum_{1 \leq j' < k' \leq n}J_{j', k'}z^{[l]}_{j'}z^{[l]}_{k'}\right)\right\}\nonumber\\
    & = \frac{1}{\sqrt{n}}\sum_{1 \leq j < k \leq n}\mathbb{E}_{\bm{J}}\left\{J_{j, k}z^{[t + 1]}_jz^{[t + 1]}_k\exp\left(\frac{1}{\sqrt{n}}\sum_{l \in \pm [p]}i\Gamma_l\sum_{1 \leq j' < k' \leq n}J_{j', k'}z^{[l]}_{j'}z^{[l]}_{k'}\right)\right\}
\end{align}
Then, using Gaussian integration by parts, the contribution of edge $\{j, k\}$ can be re-expressed
\begin{align}
    & \mathbb{E}_{\bm{J}}\left\{\frac{1}{\sqrt{n}}z^{[t + 1]}_jz^{[t + 1]}_kJ_{j, k}\exp\left(\frac{1}{\sqrt{n}}\sum_{l \in \pm [p]}i\Gamma_l\sum_{1 \leq j' < k' \leq n}J_{j', k'}z^{[l]}_{j'}z^{[l]}_{k'}\right)\right\}\nonumber\\
    & = \mathbb{E}_{\bm{J}}\left\{\frac{1}{\sqrt{n}}z^{[t + 1]}_jz^{[t + 1]}_k\sum_{1 \leq j'' < k'' \leq n}\mathbb{E}_{\bm{J}}\left[J_{j, k}J_{j'', k''}\right]\frac{\partial}{\partial J_{j'', k''}}\exp\left(\frac{1}{\sqrt{n}}\sum_{l \in \pm [p]}i\Gamma_l\sum_{1 \leq j' < k' \leq n}J_{j', k'}z^{[l]}_{j'}z^{[l]}_{k'}\right)\right\}\nonumber\\
    & = \mathbb{E}\left\{\frac{1}{\sqrt{n}}z^{[t + 1]}_jz^{[t + 1]}_k\sum_{1 \leq j'' < k'' \leq n}\mathbb{E}_{\bm{J}}\left[J_{j, k}J_{j'', k''}\right]\left(\frac{1}{\sqrt{n}}\sum_{l \in \pm [p]}i\Gamma_lz^{[l]}_{j''}z^{[l]}_{k''}\right)\right.\nonumber\\
    &\hspace{30px} \times \left. \exp\left(\frac{1}{\sqrt{n}}\sum_{l \in \pm [p]}i\Gamma_l\sum_{1 \leq j' < k' \leq n}J_{j', k'}z^{[l]}_{j'}z^{[l]}_{k'}\right)\right\} \nonumber\\
    & = \mathbb{E}_{\bm{J}}\left\{\frac{1}{n}z^{[t + 1]}_jz^{[t + 1]}_k\left(\sum_{l \in \pm [p]}i\Gamma_lz^{[l]}_{j}z^{[l]}_{k}\right)\exp\left(\frac{1}{\sqrt{n}}\sum_{l \in \pm [p]}i\Gamma_l\sum_{1 \leq j' < k' \leq n}J_{j', k'}z^{[l]}_{j'}z^{[l]}_{k'}\right)\right\}
\end{align}
Summing this over $1 \leq j < k \leq n$, we obtain
\begin{align}
    & \mathbb{E}_{\bm{J}}\left\{C\left(\bm{z}^{[t + 1]}\right)\exp\left(\frac{1}{\sqrt{n}}\sum_{l \in \pm [p]}i\Gamma_lC\left(\bm{z}^{[l]}\right)\right)\right\}\nonumber\\
    & = \mathbb{E}_{\bm{J}}\left\{\frac{1}{n}\left(\sum_{1 \leq j < k \leq n}J_{j, k}z^{[t + 1]}_jz^{[t + 1]}_k\right)\exp\left(\frac{1}{\sqrt{n}}\sum_{l \in \pm [p]}i\Gamma_l\sum_{1 \leq j < k \leq n}J_{j, k}z^{[l]}_jz^{[l]}_k\right)\right\}\nonumber\\
    & = \mathbb{E}_{\bm{J}}\left\{\frac{1}{n}\sum_{l \in \pm [p]}i\Gamma_l\left(\sum_{1 \leq j < k \leq n}z^{[t + 1]}_jz^{[t + 1]}_kz^{[l]}_jz^{[l]}_k\right)\exp\left(\frac{1}{\sqrt{n}}\sum_{l \in \pm [p]}i\Gamma_l\sum_{1 \leq j' < k' \leq n}J_{j', k'}z^{[l]}_{j'}z^{[l]}_{k'}\right)\right\}\nonumber\\
    & = \mathbb{E}_{\bm{J}}\left\{\frac{1}{2n}\sum_{l \in \pm [p]}i\Gamma_l\left(\sum_{\substack{1 \leq j, k \leq n\\j \neq k}}z^{[t + 1]}_jz^{[t + 1]}_kz^{[l]}_jz^{[l]}_k\right)\exp\left(\frac{1}{\sqrt{n}}\sum_{l \in \pm [p]}i\Gamma_l\sum_{1 \leq j' < k' \leq n}J_{j', k'}z^{[l]}_{j'}z^{[l]}_{k'}\right)\right\}
\end{align}
In fact, in the sum above, we may remove constraint $j \neq k$, since for $j = k$
\begin{align}
    & \sum_{l \in \pm [p]}i\Gamma_lz^{[t + 1]}_jz^{[t + 1]}_kz^{[l]}_jz^{[l]}_k\exp\left(\frac{1}{\sqrt{n}}\sum_{l \in \pm [p]}i\Gamma_l\sum_{1 \leq j' < k' \leq n}J_{j', k'}z^{[l]}_{j'}z^{[l]}_{k'}\right)\nonumber\\
    & =  \sum_{l \in \pm [p]}i\Gamma_l\exp\left(\frac{1}{\sqrt{n}}\sum_{l \in \pm [p]}i\Gamma_l\sum_{1 \leq j' < k' \leq n}J_{j', k'}z^{[l]}_{j'}z^{[l]}_{k'}\right)\nonumber\\
    & = \left(\sum_{l \in \pm [p]}i\Gamma_l\right)\exp\left(\frac{1}{\sqrt{n}}\sum_{l \in \pm [p]}i\Gamma_l\sum_{1 \leq j' < k' \leq n}J_{j', k'}z^{[l]}_{j'}z^{[l]}_{k'}\right)\nonumber\\
    & = 0.
\end{align}
As a provisional summary, we have re-expressed the (averaged) contribution from the cost function as follows
\begin{align}
    & \mathbb{E}_{\bm{J}}\left\{C\left(\bm{z}^{[t + 1]}\right)\exp\left(\frac{1}{\sqrt{n}}\sum_{l \in \pm [p]}i\Gamma_lC\left(\bm{z}^{[l]}\right)\right)\right\}\nonumber\\
    & = \mathbb{E}_{\bm{J}}\left\{\frac{1}{2n}\sum_{l \in \pm [p]}i\Gamma_l\left(\sum_{\substack{1 \leq j, k \leq n}}z^{[t + 1]}_jz^{[t + 1]}_kz^{[l]}_jz^{[l]}_k\right)\exp\left(\frac{1}{\sqrt{n}}\sum_{l \in \pm [p]}i\Gamma_l\sum_{1 \leq j' < k' \leq n}J_{j', k'}z^{[l]}_{j'}z^{[l]}_{k'}\right)\right\}
\end{align}
We may then average independently over each $J_{j, k}$ ($1 \leq j < k \leq n$) using Gaussian integration formula
\begin{align}
    \mathbb{E}_{X \sim \mathcal{N}\left(0, 1\right)}\left[e^{\alpha X}\right] & = e^{\alpha^2/2} \qquad \forall \alpha \in \mathbf{C}.
\end{align}
This gives us:
\begin{align}
    & \mathbb{E}_{\bm{J}}\left\{C\left(\bm{z}^{[t + 1]}\right)\exp\left(\frac{1}{\sqrt{n}}\sum_{l \in \pm [p]}i\Gamma_lC\left(\bm{z}^{[l]}\right)\right)\right\}\nonumber\\
    & = \frac{1}{2n}\sum_{l \in \pm [p]}i\Gamma_l\left(\sum_{\substack{1 \leq j, k \leq n}}z^{[t + 1]}_jz^{[t + 1]}_kz^{[l]}_jz^{[l]}_k\right)\exp\left(-\frac{1}{2n}\sum_{1 \leq j' < k' \leq n}\left(\sum_{l \in \pm [p]}\Gamma_lz^{[l]}_{j'}z^{[l]}_{k'}\right)^2\right)\nonumber\\
    & = \frac{1}{2n}\sum_{l \in \pm [p]}i\Gamma_l\left(\sum_{\substack{1 \leq j, k \leq n}}z^{[t + 1]}_jz^{[t + 1]}_kz^{[l]}_jz^{[l]}_k\right)\exp\left(-\frac{1}{4n}\sum_{\substack{1 \leq j', k' \leq n\\j' \neq k'}}\left(\sum_{l \in \pm [p]}\Gamma_lz^{[l]}_{j'}z^{[l]}_{k'}\right)^2\right)\nonumber\\
    & = \frac{1}{2n}\sum_{l \in \pm [p]}i\Gamma_l\left(\sum_{\substack{1 \leq j, k \leq n}}z^{[t + 1]}_jz^{[t + 1]}_kz^{[l]}_jz^{[l]}_k\right)\exp\left(-\frac{1}{4n}\sum_{\substack{1 \leq j', k' \leq n}}\left(\sum_{l \in \pm [p]}\Gamma_lz^{[l]}_{j'}z^{[l]}_{k'}\right)^2\right))\nonumber\\
    & = \frac{1}{2n}\sum_{l \in \pm [p]}i\Gamma_l\left(\sum_{\substack{1 \leq j, k \leq n}}z^{[t + 1]}_jz^{[t + 1]}_kz^{[l]}_jz^{[l]}_k\right)\exp\left(-\frac{1}{4n}\sum_{\substack{1 \leq j, k \leq n}}\sum_{r, s \in \pm [p]}\Gamma_r\Gamma_sz^{[r]}_{j}z^{[s]}_{j}z^{[r]}_{k}z^{[s]}_{k}\right)
\end{align}
We now decompose the summation over ordered pairs $(j, k)$ according to the configuration of bits $j, k$, i.e. the values of $(2p+ 2)$-bitstrings $\bm{z}_j$ and $\bm{z}_k$. By definition of configuration basis numbers, for each $(2p + 2)$-bitstring $\bm{a} \in \{1, -1\}^{2p + 2}$, there are $n_{\bm{a}}$ indices $j \in [n]$ such that $\bm{z}_j = \bm{a}$. The sums over $j, k$ can then be re-expressed as
\begin{align}
    \sum_{1 \leq j, k \leq n}z_j^{[t + 1]}z_k^{[t + 1]}z_j^{[l]}z_k^{[l]} & = \left(\sum_{1 \leq j \leq n}z^{[t + 1]}_jz^{[l]}_j\right)\left(\sum_{1 \leq k \leq n}z^{[t]}_kz^{[l]}_k\right)\nonumber\\
    & = \left(\sum_{\bm{a} \in \{1, -1\}^{2p + 2}}a_{t + 1}a_ln_{\bm{a}}\right)\left(\sum_{\bm{b} \in \{1, -1\}^{2p + 2}}b_{t + 1}b_ln_{\bm{a}}\right)\\
    \sum_{1 \leq j, k \leq n}z_j^{[r]}z_k^{[r]}z_j^{[s]}z_k^{[s]} & = \sum_{\bm{a},\,\bm{b} \in \{1, -1\}^{2p + 2}}a_rb_ra_sb_sn_{\bm{a}}n_{\bm{b}}
\end{align}
Plugging this into the last expression for the instance-average of the cost contributions, we obtain
\begin{align}
     & \mathbb{E}_{\bm{J}}\left\{C\left(\bm{z}^{[t]}\right)\exp\left(\frac{1}{\sqrt{n}}\sum_{l \in \pm [p]}i\Gamma_lC\left(\bm{z}^{[l]}\right)\right)\right\}\nonumber\\
     & = \frac{i}{2n}\sum_{l \in \pm [p]}\Gamma_l\left(\sum_{\bm{a} \in \{1, -1\}^{2p + 2}}a_{t + 1}a_ln_{\bm{a}}\right)\left(\sum_{\bm{b} \in \{1, -1\}^{2p + 2}}b_{t + 1}b_ln_{\bm{a}}\right)\exp\left(-\frac{1}{4n}\sum_{\bm{a}, \bm{b}}\Phi^2_{\bm{a}\bm{b}}n_{\bm{a}}n_{\bm{b}}\right),\label{eq:sk_qaoa_expectation_intermediate_layer_expanded_unitary_cost_contribution}
\end{align}
where we introduced
\begin{align}
    \Phi_{\bm{c}} & := \sum_{l \in \pm [p]}\Gamma_lc_l \qquad \forall \bm{c} \in \{1, -1\}^{2p + 2}\label{sk_qaoa_expectation_intermediate_layer_expanded_unitary_phi_definition}
\end{align}
and $\bm{a}\bm{c}$ denotes the dotwise product of bitstrings $\bm{a}$ and $\bm{b}$. Plugging Equations \ref{eq:sk_qaoa_expectation_intermediate_layer_expanded_unitary_mixer_contribution} and \ref{eq:sk_qaoa_expectation_intermediate_layer_expanded_unitary_cost_contribution} into path integral representation \ref{eq:sk_qaoa_expectation_intermediate_layer_expanded_unitary_path_integral} for the instance-averaged cost of SK-QAOA, we obtain the following expression for this cost:
\begin{align}
    & \frac{1}{n}\mathbb{E}_{\bm{J{}}}\braket{+|\left(\overrightarrow{\prod_{l = 1}^t}\exp\left(i\gamma_lC\right)\exp\left(i\beta_lB\right)\right)C\left(\overleftarrow{\prod_{l = 1}^t}\exp\left(-i\beta_lB\right)\exp\left(-i\gamma_lC\right)\right)|+}\\
    & = \frac{i}{2}\sum_{\bm{n}}\binom{n}{\bm{n}}\sum_{l \in \pm [p]}\Gamma_l\left(\sum_{\bm{a} \in \{1, -1\}^{2p + 2}}\hspace*{-10px}a_{t + 1}a_l\frac{n_{\bm{a}}}{n}\right)\left(\sum_{\bm{b} \in \{1, -1\}^{2p + 2}}\hspace*{-10px}b_{t + 1}b_l\frac{n_{\bm{b}}}{n}\right)\exp\left(-\frac{1}{4n}\sum_{\bm{a},\,\bm{b} \in \{1, -1\}^{2p + 2}}\hspace*{-15px}\Phi^2_{\bm{a}\bm{b}}n_{\bm{a}}n_{\bm{b}}\right)\nonumber\\
    & \hspace*{70px} \times \prod_{\bm{a} \in \{1, -1\}^{2p + 2}}\hspace*{-15px}Q_{\bm{a}}^{n_{\bm{a}}} \; \times \mathbf{1}\left[\forall \bm{a} \in \{1, -1\}, a_{p + 1} \neq a_{-(p + 1)} \implies n_{\bm{a}} = 0\right] \label{eq:instance_averaged_sk_intermediate_layer_energy}.
\end{align}
Note the extra factor of $1/n$ on the right-hand side of the expression --- this is because we re-normalized the left-hand size by $n$ so that the energy converges to a constant as $n \longrightarrow \infty$. We can now remove the constraint
\begin{align}
     \mathbf{1}\left[\forall \bm{a} \in \{1, -1\}, a_{p + 1} \neq a_{-(p + 1)} \implies n_{\bm{a}} = 0\right]
\end{align}
In other words, bits $p+1$ and $(p+1)$ assume a common value $a_0$. This is equivalent to summing over $(2p + 1)$-bit bitstrings
\begin{align}
    \bm{a} & = \left(a_1, a_2, \ldots, a_{p - 1}, a_p, a_0, a_{-p}, a_{-(p - 1)}, \ldots, a_{-2}, a_{-1}\right) \in \{1, -1\}^{2p + 1},
\end{align}
with the following amended definition for $Q_{\bm{a}}$
\begin{align}
    Q_{\bm{a}} & := \frac{1}{2}\left(\prod_{l = 1}^{p - 1}\braket{a_l|\exp\left(i\beta_lX\right)|a_{l + 1}}\braket{a_{-l - 1}|\exp\left(-i\beta_lX\right)|a_{l}}\right) \\
    &\times \braket{a_p|\exp\left(i\beta_pX\right)|a_0}\braket{a_0|\exp\left(-i\beta_pX\right)|a_{-p}}.\label{sk_qaoa_expectation_intermediate_layer_expanded_unitary_q_definition_amended}
\end{align}
The expression Eq.~\eqref{eq:instance_averaged_sk_intermediate_layer_energy} can then be factorized
\begin{align}
    \nu_{p,\,n,\,t}\of{\bm{\gamma},\bm{\beta}} & = \frac{i}{2}\sum_{r \in \{\pm 1,\,\pm 2,\,\ldots,\,\pm t\}}\Gamma_r\sum_{\{n_{\bm{a}}\}}\binom{n}{\{n_{\bm{a}}\}}\exp\left(-\frac{1}{4n}\sum_{\bm{a}, \bm{b}}\Phi^2_{\bm{a}\bm{b}}n_{\bm{a}}n_{\bm{b}}\right)\left(\prod_{\bm{a}}Q_{\bm{a}}^{n_{\bm{a}}}\right)\nonumber\\
    &\hspace*{135px}\times\left(\sum_{\bm{u}}\Gamma_ru_ru_{t + 1}\frac{n_{\bm{u}}}{n}\right)\left(\sum_{\bm{v}}\Gamma_rv_rv_{t + 1}\frac{n_{\bm{v}}}{n}\right)\nonumber\\
    & = -\frac{i}{\Gamma_{t + 1}}\sum_{r \in \{\pm 1,\,\pm 2,\,\ldots,\,\pm t\}}\sum_{\{n_{\bm{a}}\}}\binom{n}{\{n_{\bm{a}}\}}\exp\left(-\frac{1}{4n}\sum_{\bm{a}, \bm{b}}\Phi^2_{\bm{a}\bm{b}}n_{\bm{a}}n_{\bm{b}}\right)\left(\prod_{\bm{a}}Q_{\bm{a}}^{n_{\bm{a}}}\right)\nonumber\\
    & \hspace*{145px} \times \left(\sum_{\bm{u}}\sqrt{-\frac{\Gamma_r\Gamma_{t + 1}}{2} + i\varepsilon}\,u_ru_{t + 1}\frac{n_{\bm{u}}}{n}\right)\left(\sum_{\bm{v}}\sqrt{-\frac{\Gamma_r\Gamma_{t + 1}}{2} + i\varepsilon}\,v_rv_{t + 1}\frac{n_{\bm{v}}}{n}\right)\nonumber\\
    & = -\frac{i}{\gamma_{t + 1}}\sum_{r \in \{\pm 1,\,\pm 2,\,\ldots,\,\pm t\}}\sum_{\{n_{\bm{a}}\}}\binom{n}{\{n_{\bm{a}}\}}\exp\left(-\frac{1}{4n}\sum_{\bm{a}, \bm{b}}\Phi^2_{\bm{a}\bm{b}}n_{\bm{a}}n_{\bm{b}}\right)\left(\prod_{\bm{a}}Q_{\bm{a}}^{n_{\bm{a}}}\right)\nonumber\\
    &\hspace*{145px} \times \left(\sum_{\bm{u}}L_{\left(r,\,t + 1\right),\,\bm{u}}\frac{n_{\bm{u}}}{n}\right)\left(\sum_{\bm{v}}L_{\left(r,\,t + 1\right),\,\bm{v}}\frac{n_{\bm{v}}}{n}\right)\nonumber\\
    & =  -\frac{i}{\gamma_{t + 1}}\sum_{r \in \{\pm 1,\,\pm 2,\,\ldots,\,\pm t\}}\sum_{\{n_{\bm{a}}\}}\binom{n}{\{n_{\bm{a}}\}}\exp\left(-\frac{1}{4n}\sum_{\bm{a}, \bm{b}}\Phi^2_{\bm{a}\bm{b}}n_{\bm{a}}n_{\bm{b}}\right)\left(\prod_{\bm{a}}Q_{\bm{a}}^{n_{\bm{a}}}\right)\nonumber\\
    & = -\frac{i}{\gamma_{t + 1}}\sum_{r \in \{\pm 1, \pm 2, \ldots, \pm p\}}\sum_{\bm{n} \in \mathcal{P}(n)}\binom{\bm{n}}{n}\exp\left(\frac{1}{2n}\bm{n}^T\bm{L}^T\bm{L}\bm{n}\right)\left(\prod_{\bm{a}}Q_{\bm{a}}^{n_{\bm{a}}}\right)\frac{\left(\bm{L}\bm{n}\right)_{\left(r,\,t + 1\right)}}{n}\frac{\left(\bm{L}\bm{n}\right)_{\left(r,\,t + 1\right)}}{n}\nonumber\\
    & =  -\frac{i}{\gamma_{t + 1}}\left(\sum_{r \in \{\pm 1, \pm 2, \ldots, \pm p\}}\frac{\partial^2}{\partial\mu_{\left(r, t + 1\right)}\partial\mu_{\left(r, t + 1\right)}}\right)S\left(\bm\mu\right)\Bigg|_{\bm\mu = 0}.
\end{align}
as desired.

\subsection{The second order moment}
\label{sec:sk_qaoa_second_order_moment_qgms}

With the intention of studying the concentration of the SK-QAOA expected cost across disorder, we now express the second-order moment of the QAOA cost function:
\begin{align}
    \mathbb{E}\bra{\bm{\gamma}, \bm{\beta}}\left(C/n\right)^2\ket{\bm{\gamma}, \bm{\beta}}
\end{align}
in the QGMS formalism. This is not to be confused with the second-order moments of the QGMS
\begin{align}
    \frac{\partial^2}{\partial\mu_{\alpha}\partial\mu_{\beta}}, \qquad \alpha, \beta \in \mathcal{A}.
\end{align}
In fact, we will need QGMS moments up to order 4 to express the second-order moment of the cost function.  Compared to the first order moment calculation, we only need to evaluated the disorder-average of the cost function contributions to the path integral measure, which is now:
\begin{align}
    \frac{1}{n^2}C\left(\bm{z}^{[t + 1]}\right)^2\exp\left(\sum_{l \in \pm [p]}i\Gamma_lC\left(\bm{z}^{[l]}\right)\right).
\end{align}
Expanding this:
\begin{align}
    & \frac{1}{n^2}C\left(\bm{z}^{[t + 1]}\right)^2\exp\left(\sum_{l \in \pm [p]}i\Gamma_lC\left(\bm{z}^{[l]}\right)\right)\nonumber\\
    & = \frac{1}{n^2}\left(\frac{1}{\sqrt{n}}\sum_{1 \leq j < k \leq n}J_{j, k}z^{[t + 1]}_jz^{[t + 1]}_k\right)^2\exp\left(\sum_{l \in \pm [p]}i\Gamma_l\sum_{1 \leq j'' < k'' \leq n}J_{j'', k''}z^{[l]}_{j''}z^{[l]}_{k''}\right)\nonumber\\
    & = \frac{1}{n^3}\sum_{\substack{1 \leq j < k \leq n\\1 \leq j' < k' \leq n}}J_{j, k}J_{j', k'}z^{[t + 1]}_jz^{[t + 1]}_kz^{[t + 1]}_{j'}z^{[t + 1]}_{k'}\exp\left(\sum_{l \in \pm [p]}i\Gamma_l\sum_{1 \leq j'' < k'' \leq n}J_{j'', k''}z^{[l]}_{j''}z^{[l]}_{k''}\right).
\end{align}
We now average the above using Gaussian integration by parts for a quadratic monomial (Corollary~\ref{cor:gaussian_integration_by_parts_quadratic}), giving:
\begin{align}
    & \mathbb{E}\left[\frac{1}{n^2}C\left(\bm{z}^{[t + 1]}\right)^2\exp\left(\sum_{l \in \pm [p]}i\Gamma_lC\left(\bm{z}^{[l]}\right)\right)\right]\nonumber\\
    & = \frac{1}{n^3}\mathbb{E}\sum_{\substack{1 \leq j < k \leq n\\1 \leq j' < k' \leq n}}z^{[t + 1]}_jz^{[t + 1]}_kz^{[t + 1]}_{j'}z^{[t + 1]}_{k'}\left(\mathbb{E}\left[J_{j, k}J_{j', k'}\right] + \frac{\partial^2}{\partial J_{j, k}\partial J_{j', k'}}\right)\exp\left(\sum_{l \in \pm [p]}\frac{i\Gamma_l}{\sqrt{n}}\sum_{1 \leq j'' < k'' \leq n}J_{j'', k''}z^{[l]}_{j''}z^{[l]}_{k''}\right)\nonumber\\
    & = \frac{1}{n^3}\mathbb{E}\sum_{\substack{1 \leq j < k \leq n\\1 \leq j' < k' \leq n}}z^{[t + 1]}_jz^{[t + 1]}_kz^{[t + 1]}_{j'}z^{[t + 1]}_{k'}\left(\delta_{jj'}\delta_{kk'} - \frac{1}{n}\sum_{l, m \in \pm [p]}\Gamma_l\Gamma_mz^{[l]}_jz^{[l]}_kz^{[m]}_{j'}z^{[m]}_{k'}\right)\exp\left(\sum_{l \in \pm [p]}i\Gamma_lC\left(\bm{z}^{[l]}\right)\right)\nonumber\\
    & = \frac{1}{n^3}\mathbb{E}\sum_{\substack{1 \leq j < k \leq n}}\underbrace{z^{[t + 1]}_jz^{[t + 1]}_kz^{[t + 1]}_{j}z^{[t + 1]}_{k}}_{= 1}\exp\left(\sum_{l \in \pm [p]}i\Gamma_lC\left(\bm{z}^{[l]}\right)\right)\nonumber\\
    & \hspace*{10px} - \frac{1}{n^4}\mathbb{E}\left\{\left(\sum_{l \in \pm [p]}\Gamma_l\sum_{1 \leq j <  k\leq n}z^{[t + 1]}_jz^{[t + 1]}_kz^{[l]}_jz^{[l]}_k\right)\left(\sum_{m \in \pm [p]}\Gamma_m\sum_{1 \leq j' < k' \leq n}z^{[t + 1]}_{j'}z^{[t + 1]}_{k'}z^{[m]}_{j'}z^{[m]}_{k'}\right)\exp\left(\sum_{l \in \pm [p]}i\Gamma_lC\left(\bm{z}^{[l]}\right)\right)\right\}\nonumber\\
    & = \left\{\frac{1}{n^3}\binom{n}{2} - \frac{1}{n^4}\left(\sum_{m \in \pm [p]}\Gamma_m\sum_{1 \leq j' < k' \leq n}z^{[t + 1]}_{j'}z^{[t + 1]}_{k'}z^{[m]}_{j'}z^{[m]}_{k'}\right)^2\right\}\mathbb{E}\exp\left(\sum_{l \in \pm [p]}i\Gamma_lC\left(\bm{z}^{[l]}\right)\right)\label{eq:second_order_cost_function_moment_computation_step_1}
\end{align}
Similar to the computation of the cost function's first moment, we may exploit the vanishing of the $\bm{\Gamma}$ sum:
\begin{align}
    \sum_{m \in \pm [p]}\Gamma_m & = 0,
\end{align}
to extend summations $1 \leq j' < k' \leq n$ in the last line's parentheses to $1 \leq j', k' \leq n$:
\begin{align}
    \sum_{m \in \pm [p]}\Gamma_l\sum_{1 \leq j' < k' \leq n}z^{[t + 1]}_{j'}z^{[t + 1]}_{k'}z^{[m]}_{j'}z^{[m]}_{k'} & = \frac{1}{2}\sum_{l \in \pm [p]}\Gamma_l\sum_{\substack{1 \leq j', k' \leq n\\j' \neq k'}}z^{[t + 1]}_{j'}z^{[t + 1]}_{k'}z^{[l]}_{j'}z^{[l]}_{k'}\nonumber\\
    & = \frac{1}{2}\sum_{m \in \pm [p]}\Gamma_m\sum_{\substack{1 \leq j', k' \leq n\\j' \neq k'}}z^{[t + 1]}_{j'}z^{[t + 1]}_{k'}z^{[m]}_{j'}z^{[m]}_{k'}\nonumber\\
    & \hspace*{20px} + \frac{1}{2}\sum_{m \in \pm [p]}\Gamma_m\sum_{1 \leq j' \leq n}\underbrace{z^{[t + 1]}_{j'}z^{[t + 1]}_{j'}z^{[m]}_{j'}z^{[m]}_{k'}}_{= 1}\nonumber\\
    & = \frac{1}{2}\sum_{m \in \pm [p]}\Gamma_m\sum_{1 \leq j', k' \leq n}z^{[t + 1]}_{j'}z^{[t + 1]}_{k'}z^{[m]}_{j'}z^{[m]}_{k'}\nonumber\\
    & = \frac{1}{2}\sum_{m \in \pm [p]}\Gamma_m\left(\sum_{j' \in [n]}z^{[t + 1]}_{j'}z^{[m]}_{j'}\right)^2
\end{align}
Plugging this back into equation \ref{eq:second_order_cost_function_moment_computation_step_1}, we obtain the following expression for the disorder-average of the cost contribution to the path integral weight:
\begin{align}
    & \mathbb{E}\left[\frac{1}{n^2}C\left(\bm{z}^{[t + 1]}\right)^2\exp\left(\sum_{l \in \pm [p]}i\Gamma_lC\left(\bm{z}^{[l]}\right)\right)\right]\nonumber\\
    & = \left\{\frac{1}{n^3}\binom{n}{2} - \frac{1}{4n^4}\left(\sum_{m \in \pm [p]}\Gamma_m\left(\sum_{j' \in [n]}z^{[t + 1]}_{j'}z^{[m]}_{j'}\right)^2\right)^2\right\}\mathbb{E}\exp\left(\sum_{l \in \pm [p]}i\Gamma_lC\left(\bm{z}^{[l]}\right)\right)\label{eq:second_order_cost_function_moment_computation_step_2}.
\end{align}
This can be reexpressed in terms of configuration basis numbers as follows:
\begin{align}
    & \mathbb{E}\left[\frac{1}{n^2}C\left(\bm{z}^{[t + 1]}\right)^2\exp\left(\sum_{l \in \pm [p]}i\Gamma_lC\left(\bm{z}^{[l]}\right)\right)\right]\nonumber\\
    & = \left\{\frac{1}{n^3}\binom{n}{2} - \frac{1}{4n^4}\left(\sum_{m \in \pm [p]}\Gamma_m\left(\sum_{\bm{a} \in \mathcal{S}}a_{t + 1}a_mn_{\bm{a}}\right)^2\right)^2\right\}\mathbb{E}\exp\left(\sum_{l \in \pm [p]}i\Gamma_lC\left(\bm{z}^{[l]}\right)\right)\label{eq:second_order_cost_function_moment_computation_step_3}.
\end{align}
To obtain $\mathbb{E}\bra{\bm{\gamma}, \bm{\beta}}|\left(C/n\right)^2\ket{\bm{\gamma}, \bm{\beta}}$ from this quantity, we need to sum it over configuration basis numbers $\bm{n} = \left(n_{\bm{a}}\right)_{\bm{a} \in \mathcal{S}}$, $\sum_{\bm{a} \in \mathcal{S}}n_{\bm{a}} = n$, and against weight
\begin{align}
    \prod_{\bm{a} \in \mathcal{S}}Q_{\bm{a}}^{n_{\bm{a}}}.
\end{align}
This gives:
\begin{align}
    & \mathbb{E}\bra{\bm{\gamma}, \bm{\beta}}\left(C/n\right)^2\ket{\bm{\gamma}, \bm{\beta}}\nonumber\\
    & = \sum_{\substack{\bm{n} = \left(n_{\bm{a}}\right)_{\bm{a} \in \mathcal{S}}\\\sum_{\bm{a} \in \mathcal{S}}n_{\bm{a}} = n}}\binom{n}{\bm{n}}\left(\prod_{\bm{a} \in \mathcal{S}}Q_{\bm{a}}^{n_{\bm{a}}}\right)\left\{\frac{1}{n^3}\binom{n}{2} - \frac{1}{4n^4}\left(\sum_{m \in \pm [p]}\Gamma_m\left(\sum_{\bm{a} \in \mathcal{S}}a_{t + 1}a_mn_{\bm{a}}\right)^2\right)^2\right\}\mathbb{E}\exp\left(\sum_{l \in \pm [p]}i\Gamma_lC\left(\bm{z}^{[l]}\right)\right).
\end{align}
The first additive contribution inside the brace:
\begin{align}
    \frac{1}{n^3}\binom{n}{2}
\end{align}
gives a contribution:
\begin{align}
    \bra{\bm{\gamma}, \bm{\beta}}\ket{\bm{\gamma}, \bm{\beta}} & \supset \mathbb{E}\bra{\bm{\gamma}, \bm{\beta}}\frac{1}{n^3}\binom{n}{2}\ket{\bm{\gamma}, \bm{\beta}}\nonumber\\
    &= \frac{1}{n^3}\binom{n}{2}
\end{align}
to the second order cost moment; this can be seen by expanding this expectation as a path integral. As for the remaining term inside the braces, its contribution to the cost function's second order moment can be reexpressed:
\begin{align}
    \mathbb{E}\bra{\bm{\gamma}, \bm{\beta}}\left(C/n\right)^2\ket{\bm{\gamma}, \bm{\beta}} & \supset -\frac{1}{4}\sum_{\substack{\bm{n} = \left(n_{\bm{a}}\right)_{\bm{a} \in \mathcal{S}}\\\sum_{\bm{a} \in \mathcal{S}}n_{\bm{a}} = n}}\binom{n}{\bm{n}}\left(\prod_{\bm{a} \in \mathcal{S}}Q_{\bm{a}}^{n_{\bm{a}}}\right)\left(\sum_{m \in \pm [p]}\Gamma_m\left(\sum_{\bm{a} \in \mathcal{S}}a_{t + 1}a_m\frac{n_{\bm{a}}}{n}\right)^2\right)^2\mathbb{E}\exp\left(\sum_{l \in \pm [p]}i\Gamma_lC\left(\bm{z}^{[l]}\right)\right)\nonumber\\
    & = -\frac{1}{\Gamma_{t + 1}^2}\sum_{l, m \in \pm [p]}\frac{\partial^4S_n\left(\bm{\mu}\right)}{\partial\mu_{(t + 1,\,l)}^2\partial\mu_{(t + 1,\,m)}^2}\Bigg|_{\bm{\mu} = \bm{0}}
\end{align}
All in all, we established the following representation of the second order moment of the cost function in terms of QGMS moments:
\begin{align}
    \mathbb{E}\bra{\bm{\gamma}, \bm{\beta}}\left(C/n\right)^2\ket{\bm{\gamma}, \bm{\beta}} & = \frac{1}{n^3}\binom{n}{2} - \frac{1}{\Gamma_{t + 1}^2}\sum_{l, m \in \pm [p]}\frac{\partial^4S_n\left(\bm{\mu}\right)}{\partial\mu_{(t + 1,\,l)}^2\partial\mu_{(t + 1,\,m)}^2}\Bigg|_{\bm{\mu} = \bm{0}}.\label{eq:second_order_cost_function_moment}
\end{align}
Anticipating on Definition~\ref{def:qgms_integral_moments_tensor}, we express the above QGMS moments in terms of QGMS integral moments. As explained in the discussion motivating this Definition, the linear relation between QGMS moments and integral QGMS moments can be obtained by taking derivative:
\begin{align}
    \frac{\partial^4}{\partial\mu_{\alpha}^2\partial\mu_{\beta}^2}\exp\left(\bm{\mu}^T\left(\bm{\theta}^* + \frac{1}{\sqrt{n}}\bm{\chi}\right) - \frac{1}{2n}\bm{\mu}^T\bm{\mu}\right)\Bigg|_{\bm{\mu} = \bm{0}}.
\end{align}
After a slightly tedious calculation, this gives:
\begin{align}
    & \frac{\partial^4}{\partial\mu_{\alpha}^2\partial\mu_{\beta}^2}\exp\left(\bm{\mu}^T\left(\bm{\theta}^* + \frac{1}{\sqrt{n}}\bm{\chi}\right) - \frac{1}{2n}\bm{\mu}^T\bm{\mu}\right)\Bigg|_{\bm{\mu} = \bm{0}}\nonumber\\
    & = \frac{1 + 2\delta_{\alpha\beta}}{n^2} - \frac{4\delta_{\alpha\beta}}{n}\left(\theta^*_{\alpha}+ \frac{1}{\sqrt{n}}\chi_{\alpha}\right)^2 - \frac{1}{n}\left(\theta^*_{\alpha} + \frac{1}{\sqrt{n}}\chi_{\alpha}\right)^2 - \frac{1}{n}\left(\theta^*_{\beta} + \frac{1}{\sqrt{n}}\chi_{\beta}\right)^2 + \left(\theta^*_{\alpha} + \frac{1}{\sqrt{n}}\chi_{\alpha}\right)^2\left(\theta^*_{\beta} + \frac{1}{\sqrt{n}}\chi_{\beta}\right)^2\nonumber\\
    & = \frac{1}{n^2}\left(1 + 2\delta_{\alpha\beta} - 4\delta_{\alpha\beta}\chi_{\alpha}^2 - \chi_{\alpha}^2 - \chi_{\beta}^2 + \chi_{\alpha}^2\chi_{\beta}^2\right)\nonumber\\
    & \hspace*{20px} + \frac{1}{n^{3/2}}\left(-8\delta_{\alpha\beta}\theta^*_{\alpha}\chi_{\alpha} - 2\theta^*_{\alpha}\chi_{\alpha} - 2\theta^*_{\beta}\chi_{\beta} + 2\theta^*_{\alpha}\chi_{\alpha}\chi_{\beta}^2 + 2\theta^*_{\beta}\chi_{\beta}\chi_{\alpha}^2\right)\nonumber\\
    & \hspace*{20px} + \frac{1}{n}\left(-4\delta_{\alpha\beta}\left(\theta^*_{\alpha}\right)^2 -\left(\theta^*_{\alpha}\right)^2 - \left(\theta^*_{\beta}\right)^2 + 4\theta^*_{\alpha}\theta^*_{\beta}\chi_{\alpha}\chi_{\beta} + \left(\theta^*_{\alpha}\right)^2\chi_{\alpha}^2 + \left(\theta^*_{\beta}\right)^2\chi_{\beta}^2\right)\nonumber\\
    & \hspace*{20px} + \frac{1}{\sqrt{n}}\left(2\left(\theta^*_{\beta}\right)^2\theta^*_{\alpha}\chi_{\alpha} + 2\left(\theta^*_{\alpha}\right)^2\theta^*_{\beta}\chi_{\beta}\right)\nonumber\\
    & \hspace*{20px} + \left(\theta^*_{\alpha}\right)^2\left(\theta^*_{\beta}\right)^2.
\end{align}

This yields the corresponding identity for the QGMS moment in terms of QGMS integral moments:
\begin{align}
    \frac{\partial^4S_n\left(\bm{\mu}\right)}{\partial\mu_{\alpha}^2\partial\mu_{\beta}^2}\Bigg|_{\bm{\mu} = \bm{0}} & = \frac{1}{n^2}\left(1 + 2\delta_{\alpha\beta} - 4\delta_{\alpha\beta}\left[\bm{S}^{(2)}_n\right]_{\alpha,\,\alpha} - \left[\bm{S}^{(2)}_n\right]_{\alpha,\,\alpha} - \left[\bm{S}^{(2)}_n\right]_{\beta,\,\beta} + \left[\bm{S}^{(4)}_n\right]_{\alpha,\,\alpha,\,\beta,\,\beta}\right)\nonumber\\
    & \hspace*{20px} + \frac{1}{n^{3/2}}\left(-8\delta_{\alpha\beta}\theta^*_{\alpha}\left[\bm{S}^{(1)}_n\right]_{\alpha} - 2\theta^*_{\alpha}\left[\bm{S}^{(1)}_n\right]_{\alpha} - 2\theta^*_{\beta}\left[\bm{S}^{(1)}_n\right]_{\beta} + 2\theta^*_{\alpha}\left[\bm{S}^{(3)}_n\right]_{\alpha,\,\beta,\,\beta} + 2\theta^*_{\beta}\left[\bm{S}^{(3)}_n\right]_{\beta,\,\alpha,\,\alpha}\right)\nonumber\\
    & \hspace*{20px} + \frac{1}{n}\left(-4\delta_{\alpha\beta}\left(\theta^*_{\alpha}\right)^2 - \left(\theta^*_{\alpha}\right)^2 - \left(\theta^*_{\beta}\right)^2 + 4\theta^*_{\alpha}\theta^*_{\beta}\left[\bm{S}^{(2)}_n\right]_{\alpha,\,\beta} + \left(\theta^*_{\alpha}\right)^2\left[\bm{S}^{(2)}_n\right]_{\beta,\,\beta} + \left(\theta^*_{\beta}\right)^2\left[\bm{S}^{(2)}_n\right]_{\alpha,\,\alpha}\right)\nonumber\\
    & \hspace*{20px} + \frac{1}{\sqrt{n}}\left(2\left(\theta^*_{\beta}\right)^2\theta^*_{\alpha}\left[\bm{S}^{(1)}_n\right]_{\alpha} + 2\left(\theta^*_{\alpha}\right)^2\theta^*_{\beta}\left[\bm{S}^{(1)}_n\right]_{\beta}\right)\nonumber\\
    & \hspace*{20px} + \left(\theta^*_{\alpha}\right)^2\left(\theta^*_{\beta}\right)^2\nonumber\\
    & =: \frac{1}{n^2}\nu^{(7)}_{\alpha,\,\beta} + \frac{1}{n^{3/2}}\nu^{(6)}_{\alpha,\,\beta} + \frac{1}{n}\nu^{(5)}_{\alpha,\,\beta} + \frac{1}{\sqrt{n}}\nu^{(4)}_{\alpha,\,\beta} + \nu^{(3)}_{\alpha,\,\beta},\label{eq:second_order_cost_function_moment_qgms_moments}
\end{align}
where in the last line, we grouped terms by explicit power of $n$. Note there is still an implicit $n$ dependence of each $\nu^{(k)}_{\alpha,\,\beta}$ via that of the QGMS integral moments tensors $\bm{S}_n^{(k)}$.

\section{Series expansion of parametrized QGMS around the noninteracting limit ($\lambda = 0$)}
\label{sec:pqgms_series_expansion_noninteracting_limit}

After introducting Quadratic Generalized Multinomial Sums (QGMS) in Appendix~\ref{sec:notations}, Appendix~\ref{sec:sk_qaoa_qgms} established a representation of the object of interest of main theorem \ref{th:approximation_continuous_time_annealing_qaoa}: the instance-averaged SK-QAOA energy, as a QGMS. In this Section, we take a first technical step in the analysis of QGMS, underlying the proof of the main theorem in subsequent Appendix~\ref{sec:sk_qaoa_energy_continuum_limit}. In Section~\ref{sec:qgms_saddle_point_correlations_tensors}, we start by introducing further technical background on QGMS, leading to the central concept of the saddle point and correlations tensors associated to a QGMS. The saddle point has no ``simple" analytic expressions in terms of the QGMS data $\bm{Q} \in \mathbf{C}^{\mathcal{S}}, \bm{L} \in \mathbf{C}^{\mathcal{A} \times \mathcal{S}}$; the same statement holds for correlations tensors, whose definition depends on the saddle point. In fact, even existence and unicity of the saddle point are not guaranteed by the definitions introduced in Section~\ref{sec:qgms_saddle_point_correlations_tensors} alone. However, reasoning in the framework of parametrized QGMS (PQGMS, see Definition~\ref{def:pqgms-mgf}), one may consider a perturbative series expansion of the saddle point in the PQGMS parameter $\lambda$. $\lambda$ can intuitively be regarded as tuning an interaction strength; as we will see when specializing to the QGMS describing the SK-QAOA energy, it is proportional to the phase separator angles $\bm{\gamma}$ in this case, with $\gamma = 0$ corresponding to noninteracting QAOA. The expansion of the QGMS saddle point and correlations tensors as a series in $\lambda$ is rigorously analyzed in Section \ref{sec:pqgms_saddle_point_expansion} of this Appendix, establishing the existence of these objects for sufficiently small interaction parameter $\lambda$. With this existence result in hand, Section~\ref{sec:qgms_moments_series_expansion} develops an expansion of PQGMS moments (generated by the pseudo-moment-generating function introduced in Definition~\ref{def:pqgms-mgf}) as a series in $\lambda$, where each term is a contracted tensor network involving correlations tensors and the saddle point. While this series assumes existence of the saddle point and correlations tensors established by the results of earlier Section~\ref{sec:pqgms_saddle_point_expansion}, it does not explicitly require the series expansions for the saddle point and correlations tensors established there. Said differently, it can be regarded as a different level of series expansion in variable $\lambda$, and Section~\ref{sec:qgms_moments_series_expansion}, \ref{sec:pqgms_saddle_point_expansion} can be read independently. The contents of this Appendix apply to any Parametrized Quadratic Generalized Multinomial Sum (PQGMS) and are not restricted to the special case of the SK-QAOA energy. It would be interesting future research to further generalize these methods to Generalized Multinomial Sums (Refs.~\cite{qaoa_spin_glass_models,qaoa_spiked_tensor,qaoa_ksat}) beyond quadratic order.

\subsection{Further background on Quadratic Generalized Multinomial Sums: saddle point and correlations tensors}
\label{sec:qgms_saddle_point_correlations_tensors}

After defining Quadratic Generalized Multinomial Sums (QGMS) in section \ref{sec:qgms_background}, the current section provides more advanced background on (parametrized) QGMS in preparation of the series expansion developed in this appendix. To contextualize these developments, recall from Appendix \ref{sec:sk_qaoa_qgms} that expected values of QAOA can be expressed from pseudo-moments, generated by a pseudo-moment generating function (Definitions~\ref{def:qgms-mgf}, \ref{def:pqgms-mgf}). The starting point of the noninteracting series expansion is an integral representation of the pseudo-moment generating function, which we now derive.

Using Gaussian integration identity

\begin{align}
    \exp\left(\frac{1}{2n}\bm{n}^T\bm{L}^T\bm{L}\bm{n}\right) = \left(2\pi\right)^{-\left|\mathcal{A}\right|/2}\int\limits_{\mathbf{R}^{\mathcal{A}}}\!\mathrm{d}\bm{\theta}\,\exp\left(-\frac{1}{2}\bm{\theta}^T\bm{\theta}\right)\exp\left(\frac{1}{\sqrt{n}}\bm{\theta}^T\bm{L}\bm{n}\right),
\end{align}

the pseudo-moment generating function of any QGMS can be expressed:

\begin{align}
    S_n\left(\bm\mu\right) & = \sum_{\bm{n} \in \mathcal{P}(n)}\binom{n}{\bm{n}}\exp\left(\frac{1}{2n}\bm{n}^T\bm{L}^T\bm{L}\bm{n} + \frac{1}{n}\bm{\mu}^T\bm{L}\bm{n}\right)\prod_{\bm{a} \in \mathcal{S}}Q_{\bm{a}}^{n_{\bm{a}}}\nonumber\\
    & = \sum_{\bm{n} \in \mathcal{P}(n)}\binom{n}{\bm{n}}\left(2\pi\right)^{-\left|\mathcal{A}\right|/2}\int_{\mathbf{R}^{\mathcal{A}}}\!\mathrm{d}\bm{\theta}\,\exp\left(-\frac{1}{2}\bm{\theta}^T\bm{\theta}\right)\exp\left(\frac{1}{\sqrt{n}}\bm{\theta}^T\bm{L}\bm{n} + \frac{1}{n}\bm{\mu}^T\bm{L}\bm{n}\right)\prod_{\bm{a} \in \mathcal{S}}Q_{\bm{a}}^{n_{\bm{a}}}\nonumber\\
    & = \left(2\pi\right)^{-\left|\mathcal{A}\right|/2}\int_{\mathbf{R}^{\mathcal{A}}}\!\mathrm{d}\bm{\theta}\,\exp\left(-\frac{1}{2}\bm{\theta}^T\bm{\theta}\right)\left(\sum_{\bm{a} \in \mathcal{S}}Q_{\bm{a}}\exp\left(\frac{1}{\sqrt{n}}\bm{\theta}\bm{L}_{:,\,\bm{a}} + \frac{1}{n}\bm{\mu}^T\bm{L}_{:,\,\bm{a}}\right)\right)^n\nonumber\\
    & = \left(\frac{n}{2\pi}\right)^{|\mathcal{A}|/2}\int_{\mathbf{R}^{\mathcal{A}}}\!\mathrm{d}\bm\theta\,\exp\left(-\frac{n}{2}\bm{\theta}^T\bm{\theta}\right)\left(\sum_{\bm{a} \in \mathcal{S}}Q_{\bm{a}}\exp\left(\bm{\theta}^T\bm{L}_{:,\,\bm{a}} + \frac{1}{n}\bm{\mu}^T\bm{L}_{:,\,\bm{a}}\right)\right)^n\nonumber\\
    & = \left(\frac{n}{2\pi}\right)^{|\mathcal{A}|/2}\int_{\mathbf{R}^{\mathcal{A}}}\!\mathrm{d}\bm\theta\,\exp\left(-\frac{n}{2}\bm{\theta}^T\bm{\theta}\right)\left(\sum_{\bm{a} \in \mathcal{S}}Q_{\bm{a}}\exp\left(\left(\bm\theta + \frac{1}{n}\bm{\mu}\right)^T\bm{L}_{:,\,\bm{a}}\right)\right)^n\nonumber\\
    & = \left(\frac{n}{2\pi}\right)^{|\mathcal{A}|/2}\int_{\mathbf{R}^{\mathcal{A}}}\!\mathrm{d}\bm\theta\,\exp\left(-\frac{1}{2n}\bm{\mu}^T\bm{\mu} + \bm{\mu}^T\bm{\theta} - \frac{n}{2}\bm{\theta}^T\bm{\theta}\right)\left(\sum_{\bm{a} \in \mathcal{S}}Q_{\bm{a}}\exp\left(\bm{\theta}^T\bm{L}_{:,\,\bm{a}}\right)\right)^n, \label{eq:mgf_integral_form_after_gaussian_trick}
\end{align}

where in the last line we performed a translation change of variables $\bm{\theta} + \frac{1}{n} \bm{\mu} \rightarrow \bm{\theta}$. At $\bm{\mu}=\bm{0}$, this integral can be cast to the form required of the saddle-point method:

\begin{align}
     S_n\left(\bm{0}\right) & = \left(\frac{n}{2\pi}\right)^{|A|/2}\int_{\mathbf{R}^{\mathcal{A}}}\!\mathrm{d}\bm{\theta}\,\exp\left(-\frac{n}{2}\bm{\theta}^T\bm{\theta}\right)\left(\sum_{\bm{a} \in \mathcal{S}}Q_{\bm{a}}\exp\left(\bm{\theta}^T\bm{L}_{:,\,\bm{a}}\right)\right)^n.
\end{align}

Besides, the integrand can be rewritten as a single exponential, which will allow us to introduce the relevant definitions

\begin{align}
    \exp\left(-\frac{n}{2}\bm{\theta}^T\bm{\theta}\right)\left(\sum_{\bm{a} \in \mathcal{S}}Q_{\bm{a}}\exp\left(\bm{\theta}^T\bm{L}_{:,\,\bm{a}}\right)\right)^n & = \exp\left(n\left(-\frac{1}{2}\bm{\theta}^T\bm{\theta} + \log\sum_{\bm{a} \in \mathcal{S}}Q_{\bm{a}}\exp\left(\bm{\theta}^T\bm{L}_{:,\,\bm{a}}\right)\right)\right)\\
    & =: \exp\left(n\left(-\frac{1}{2}\bm{\theta}^T\bm{\theta} + \log\mathcal{Z}\left(\bm{\theta}\right)\right)\right)\\
    & =: \exp\left(n\Phi(\bm{\theta})\right),
\end{align}

where we introduced the \emph{pseudo-partition function} at (vector) temperature $\bm\theta$

\begin{align}
    \mathcal{Z}\left(\bm\theta\right) & := \sum_{\bm{a} \in \mathcal{S}}Q_{\bm{a}}\exp\left(\bm{\theta}^T\bm{L}_{:,\,\bm{a}}\right),\label{eq:pseudo_partition_function}
\end{align}

whose logarithm enters the \emph{phase function}

\begin{align}
    \Phi\left(\bm{\theta}\right) & := -\frac{1}{2}\bm{\theta}^T\bm{\theta} + \log\sum_{\bm{a} \in \mathcal{S}}Q_{\bm{a}}\exp\left(\bm{\theta}^T\bm{L}_{:,\,\bm{a}}\right).\label{eq:phase_function}
\end{align}

The saddle point equation, where we denote the saddle point as $\bm{\theta}^*$, then reads

\begin{align}
    \nabla \Phi\left(\bm{\theta}^*\right) = \bm{0} \implies \bm{\theta}^* & = \frac{\sum\limits_{\bm{a} \in \mathcal{S}}Q_{\bm{a}}\exp\left(\bm{\theta}^{*T}\bm{L}_{:,\,\bm{a}}\right)\bm{L}_{:,\,\bm{a}}}{\sum\limits_{\bm{a} \in \mathcal{S}}Q_{\bm{a}}\exp\left(\bm{\theta}^{*T}\bm{L}_{:,\,\bm{a}}\right)},\label{eq:saddle_point_equation}
\end{align}

or, using index rather than vector notations

\begin{align}
    \theta^{*}_{\alpha} & = \frac{\sum\limits_{\bm{a} \in \mathcal{S}}Q_{\bm{a}}\exp\left(\sum\limits_{\beta \in \mathcal{A}}\theta^{*}_{\beta}L_{\beta,\,\bm{a}}\right)L_{\alpha,\,\bm{a}}}{\sum\limits_{\bm{a} \in \mathcal{S}}Q_{\bm{a}}\exp\left(\sum\limits_{\beta \in \mathcal{A}}\theta^{*}_{\beta}L_{\beta,\,\bm{a}}\right)} \qquad \forall \alpha \in \mathcal{A}.
\end{align}

We currently set aside the question of existence and unicity of the saddle-point; this will be addressed in a perturbative setting in Appendix Section~\ref{sec:pqgms_saddle_point_expansion}. We then define notation:

\begin{definition}[Pseudo-Gibbs expectation]\label{def:gibbs_expectation}

    We refer to the pseudo-expectation over $\bm{a} \in \mathcal{S}$ according to the pseudo-Gibbs measure at vector temperature $\bm\theta$, given explicitly by 
    \begin{align}
    \left\langle h\left(\bm{a}\right) \right\rangle_{\bm\theta,\,\bm{a}} & := \frac{1}{\mathcal{Z}\left(\bm{\theta}\right)}\sum\limits_{\bm{a} \in \mathcal{S}}Q_{\bm{a}}\exp\left(\bm{\theta}^{T}\bm{L}_{:,\,\bm{a}}\right)h\left(\bm{a}\right) \label{eq:pseudo_gibbs_expectation_defn}\\
    & = \frac{\sum\limits_{\bm{a} \in \mathcal{S}}Q_{\bm{a}}\exp\left(\bm{\theta}^{T}\bm{L}_{:,\,\bm{a}}\right)h\left(\bm{a}\right)}{\sum\limits_{\bm{a} \in \mathcal{S}}Q_{\bm{a}}\exp\left(\bm{\theta}^{T}\bm{L}_{:,\,\bm{a}}\right)},
    \end{align}

    as the Pseudo-Gibbs expectation of $f$ at vector temperature $\bm\theta$. When $\bm\theta = \bm{\theta}^*$, we also define shorthand
    
    \begin{align}
        \left\langle h\left(\bm{a}\right) \right\rangle_{\bm{a}} & := \left\langle h\left(\bm{a}\right) \right\rangle_{\bm{\theta}^*,\,\bm{a}}.
    \end{align}
\end{definition}

The following family of pseudo-expectations will be particularly useful: \medskip

\begin{definition}[Correlations tensor]
\label{def:c_tensor}
Let $d \geq 1$ an integer. The correlations tensor of order $d$ is the symmetric tensor indexed by $d$ indices $\alpha_1, \ldots, \alpha_d$

\begin{align}
    \bm{C}^{\left(d\right)} & = \left(C^{(d)}_{\alpha_1,\,\alpha_2,\,\ldots,\,\alpha_d}\right),
\end{align}

with entries given by:

\begin{align}
    C^{(d)}_{\alpha_1,\,\alpha_2,\,\ldots,\,\alpha_d} & := \left\langle L_{\alpha_1,\,\bm{a}}L_{\alpha_2,\,\bm{a}}\ldots L_{\alpha_d,\,\bm{a}} \right\rangle_{\bm\theta,\,\bm{a}} \label{eq:c_defn}\\
    & = \frac{1}{\mathcal{Z}\left(\bm\theta\right)}\sum\limits_{\bm{a} \in \mathcal{S}}Q_{\bm{a}}\exp\left(\bm{\theta}^{T}\bm{L}_{:,\,\bm{a}}\right)L_{\alpha_1,\,\bm{a}}L_{\alpha_2,\,\bm{a}}\ldots L_{\alpha_d,\,\bm{a}} \\
    &= \frac{1}{\mathcal{Z}\left(\bm\theta\right)} \frac{\partial^d \mathcal{Z}\of{\bm\theta}}{\partial\theta_{\alpha_1} \cdots \partial\theta_{\alpha_d}} \label{eq:c_defn_in_terms_of_Z}. 
\end{align}

In this text, $\bm\theta$ is implicit in the notation of $\bm{C}^{(d)}$ and should be assumed to be $\bm\theta^*$ unless otherwise stated. We note that by definition (Eq.~\eqref{eq:c_defn}) the correlation tensors are symmetric under permutation of indices. By the saddle-point equation, the following relation holds at $\bm\theta = \bm\theta^*$:

\begin{align}
    \bm{C}^{(1)} & := \bm\theta^*.
\end{align}
\end{definition}

This concludes the definition of the QGMS saddle point and correlations tensor. We now adapt these definitions to the case of \textit{parametrized} multinomial sums, characterized (Definition~\ref{def:pqgms-mgf}) by a pseudo-moment generating function of the form:
\begin{align}
    S_n\left(\lambda, \bm{\mu}\right) & := \sum_{\bm{n} \in \mathcal{P}(n)}\binom{n}{\bm{n}}\exp\left(\frac{\lambda^2}{2n}\bm{n}^T\bm{L}^T\bm{L}\bm{n} + \frac{\lambda}{n}\bm{\mu}^T\bm{L}\bm{n}\right).
\end{align}
For each value of parameter $\lambda$, the above can be seen as the moment-generating function of a QGMS, of generic form:
\begin{align}
    S_n\left(\bm{\mu}\right) & := \sum_{\bm{n} \in \mathcal{P}(n)}\binom{n}{\bm{n}}\exp\left(\frac{1}{2n}\bm{n}^T\bm{L}^T\bm{L}\bm{n} + \frac{1}{n}\bm{\mu}^T\bm{L}\bm{n}\right),
\end{align}
after substitution $\bm{L} \longrightarrow \lambda\bm{L}$. Hence, for each value of parameter $\lambda$, a PQGMS-MGF defines a standard QGMS-MGF (Definition \ref{def:qgms-mgf}), with corresponding saddle point $\bm{\theta}^*$ and correlation tensors $\bm{C}^{(d)}$. More explicitly, the pseudo-partition function (Eq.~\ref{eq:pseudo_partition_function}) can now be regarded as a function of both $\bm{\theta}$ and $\lambda$:
\begin{align}
    \mathcal{Z}\left(\lambda, \bm{\theta}\right) & := \sum_{\bm{a} \in \mathcal{S}}Q_{\bm{a}}\exp\left(\lambda\bm{\theta}^T\bm{L}_{:,\,\bm{a}}\right).
\end{align}
The saddle point is still defined as extremizing the phase function defined in Eq.~\ref{eq:phase_function}, which nows also depends on $\lambda$:
\begin{align}
    \Phi\left(\lambda, \bm{\theta}\right) & := -\frac{1}{2}\bm{\theta}^T\bm{\theta} + \log\mathcal{Z}\left(\lambda, \bm{\theta}\right).
\end{align}
This leads to a $\lambda$-dependent saddle point $\bm{\theta}^*\left(\lambda\right)$, obeying equation:
\begin{align}
    \bm{\theta}^*\left(\lambda\right) & = \frac{\sum\limits_{\bm{a} \in \mathcal{S}}Q_{\bm{a}}\exp\left(\lambda\bm{\theta}^*\left(\lambda\right)^{T}\bm{L}_{:,\,\bm{a}}\right)\lambda\bm{L}_{:,\,\bm{a}}}{\mathcal{Z}\left(\lambda, \bm{\theta}^*\left(\lambda\right)\right)} = \frac{\sum\limits_{\bm{a} \in \mathcal{S}}Q_{\bm{a}}\exp\left(\lambda\bm{\theta}^*\left(\lambda\right)^{T}\bm{L}_{:,\,\bm{a}}\right)\lambda\bm{L}_{:,\,\bm{a}}}{\sum\limits_{\bm{a} \in \mathcal{S}}Q_{\bm{a}}\exp\left(\lambda\bm{\theta}^*\left(\lambda\right)^T\bm{L}_{:,\,\bm{a}}\right)}.
\end{align}
The definition of correlation tensors in terms of the pseudo-partition function and the saddle point (Eq.~\ref{eq:c_defn_in_terms_of_Z}) remains unchanged, but now leads to $\lambda$-dependent correlations tensors:
\begin{align}
    \bm{C}^{(d)}\left(\lambda\right) & = \frac{1}{\mathcal{Z}\left(\lambda, \bm{\theta}^*\left(\lambda\right)\right)}\sum_{\bm{a} \in \mathcal{S}}Q_{\bm{a}}\exp\left(\lambda\bm{\theta}^*\left(\lambda\right)^T\bm{L}_{:,\,\bm{a}}\right)\left(\lambda\bm{L}_{:,\,\bm{a}}\right)^{\otimes d}.\label{eq:c_defn_in_terms_of_Z_pqgms}
\end{align}
In the above equation, we specialized Eq.~\ref{eq:c_defn_in_terms_of_Z} to case $\bm{\theta} = \bm{\theta}^*$, which will be the only relevant case when considering correlations tensors in the rest of this manuscript. As a final simplification, when manipulating parametrized QGMS, we may often drop the explicit mention of $\lambda$ dependence, writing $\bm{\theta}^*$ for $\bm{\theta}^*\left(\lambda\right)$, $\bm{C}^{(d)}$ for $\bm{C}^{(d)}\left(\lambda\right)$, and $\mathcal{Z}^*$ for $\mathcal{Z}^*\left(\lambda\right) = \mathcal{Z}\left(\lambda, \bm{\theta}^*\left(\lambda\right)\right)$. The goal of Appendix Section~\ref{sec:pqgms_saddle_point_expansion} will be to prove the existence of the saddle point of a parametrized QGMS for sufficiently small $\lambda$, and produce a systematic expansion as a series in $\lambda$ for correlation tensors.

As we will see when deriving the QGMS representing the SK-QAOA energy, $\lambda$ can be viewed as an interaction tuning parameter; namely, we will observe that transformation $\bm{L} \longrightarrow \lambda\bm{L}$ in the SK-QAOA energy QGMS is equivalent to transformation $\bm{\gamma} \longrightarrow \lambda\bm{\gamma}$ on the QAOA angles. In particular, setting $\lambda = 0$ is equivalent to letting $\bm{\gamma} = \bm{0}$, where the circuit becomes non-interacting. This motivates the notion of \emph{noninteracting correlation tensors}, which will play an important role in the series expansion of the saddle point derived in Appendix Section~\ref{sec:pqgms_saddle_point_expansion}.

\begin{definition}[Noninteracting correlations tensors]
\label{def:noninteracting_c_tensor}
Let a PQGMS-MGF (Definition~\ref{eq:pqgms_mgf}) be given by parameter $\bm{Q} = \left(Q_{\bm{a}}\right)_{\bm{a} \in \mathcal{S}}$ and $\bm{L} = \left(L_{\alpha,\,\bm{a}}\right)_{\alpha \in \mathcal{A},\,\bm{a} \in \mathcal{S}}$. The noninteracting correlations tensor of order $d$ associated to this PQGMS is defined as:
\begin{align}
    \bm{\overline{C}}^{(d)} & = \left(\overline{C}^{(d)}_{\alpha_1,\,\ldots,\,\alpha_d}\right)_{\alpha_1,\,\ldots,\,\alpha_d\in \mathcal{A}} \in \left(\mathbf{C}^{\mathcal{A}}\right)^{\otimes d}\\
    \overline{C}^{(d)}_{\alpha_1,\,\ldots,\,\alpha_d} & = \frac{1}{\overline{\mathcal{Z}}^*}\sum_{\bm{a} \in \mathcal{S}}Q_{\bm{a}}\bm{L}_{:,\,\bm{a}}^{\otimes d},
\end{align}
where the noninteracting pseudo-partition function is defined as:
\begin{align}
    \overline{\mathcal{Z}}^* & := \sum_{\bm{a} \in \mathcal{S}}Q_{\bm{a}}.
\end{align}
\end{definition}

Note that unlike (interacting) correlations tensors introduced in Definition~\ref{def:c_tensor}, noninteracting correlations tensor do not depend on the saddle point, but can be ``elementarily" computed from QGMS parameters $\bm{Q}, \bm{L}$. In subsequent derivations, it will be useful to expand correlations tensors as a series involving noninteracting correlations tensor and the saddle point, a result we collect in the following proposition for conveninence:

\begin{proposition}[Series expansion of interacting correlations tensors in terms of noninteracting correlations tensors]
\label{prop:c_tensors_noninteracting_c_tensors_series_expansion}
Let be given a PQGMS-MGF as in Definition~\ref{def:pqgms-mgf}. The following series expansion of the correlations tensor of order $d$ (Definition~\ref{def:c_tensor}) in terms of noninteracting correlations tensors (Definition~\ref{def:noninteracting_c_tensor}) holds:
\begin{align}
    \bm{C}^{(d)}\left(\lambda\right) & = \frac{\overline{\mathcal{Z}}^*}{\mathcal{Z}^*\left(\lambda\right)}\sum_{m \geq 0}\frac{\lambda^{d + m}}{m!}\left\langle \bm{\theta}^*\left(\lambda\right)^{\otimes m}, \bm{\overline{C}}^{\left(d + m\right)} \right\rangle,
     \end{align}
where, consistent with the dot product notation (Eq.~\ref{eq:tensor_dot_product_different_dimensions}) introduced in Appendix~\ref{sec:notations},
\begin{align}
    \left\langle \bm{\theta}^*\left(\lambda\right)^{\otimes m}, \bm{\overline{C}}^{\left(d + m\right)} \right\rangle
\end{align}
is a tensor of degree $d$ with entries given by
\begin{align}
    \left[\bm{\theta}^*\left(\lambda\right)^{\otimes m}, \bm{\overline{C}}^{\left(d + m\right)}\right]_{\alpha_1,\,\ldots,\,\alpha_d} & := \sum_{\alpha_{d + 1},\,\ldots,\,\alpha_{d + m} \in \mathcal{A}}\theta^*_{\alpha_{d + 1}}\left(\lambda\right)\ldots\theta^*_{\alpha_{d + m}}\left(\lambda\right)\overline{C}^{(d + m)}_{\alpha_1,\,\ldots,\,\alpha_d,\,\alpha_{d + 1},\,\ldots,\,\alpha_{d + m}},\\
    \alpha_1, \ldots, \alpha_d & \in \mathcal{A}.
\end{align}
\begin{proof}
This results from direct computation, starting from the definition of correlations tensor in the context of a PQGMS (Eq.~\ref{eq:c_defn_in_terms_of_Z_pqgms}) Taylor-expanding the exponential and recalling the definition of noninteracting correlations tensors (Definition~\ref{def:noninteracting_c_tensor}):
\begin{align}
    \bm{C}^{(d)}\left(\lambda\right) & = \frac{1}{\mathcal{Z}^*\left(\lambda, \bm{\theta}^*\left(\lambda\right)\right)}\sum_{\bm{a} \in \mathcal{S}}Q_{\bm{a}}\exp\left(\lambda\bm{\theta}^*\left(\lambda\right)^T\bm{L}_{:,\,\bm{a}}\right)\left(\lambda\bm{L}_{:,\,\bm{a}}\right)^{\otimes d},\\
    & = \frac{1}{\mathcal{Z}^*\left(\lambda\right)}\sum_{\bm{a} \in \mathcal{S}}Q_{\bm{a}}\sum_{m \geq 0}\frac{1}{m!}\left\langle \bm{\theta}^*\left(\lambda\right), \lambda\bm{L}_{:,\,\bm{a}} \right\rangle^m\left(\lambda\bm{L}_{:,\,\bm{a}}\right)^{\otimes d}\nonumber\\
    & = \frac{1}{\mathcal{Z}^*\left(\lambda\right)}\sum_{\bm{a} \in \mathcal{S}}Q_{\bm{a}}\sum_{m \geq 0}\frac{1}{m!}\left\langle \bm{\theta}^*\left(\lambda\right)^{\otimes m}, \left(\lambda\bm{L}_{:,\,\bm{a}}\right)^{\otimes m} \right\rangle \left(\lambda\bm{L}_{:,\,\bm{a}}\right)^{\otimes d}\nonumber\\
    & = \frac{1}{\mathcal{Z}^*\left(\lambda\right)}\sum_{m \geq 0}\frac{\lambda^{d + m}}{m!}\left\langle \bm{\theta}^*\left(\lambda\right)^{\otimes m}, \sum_{\bm{a} \in \mathcal{S}}Q_{\bm{a}}\bm{L}_{:,\,\bm{a}}^{\otimes \left(d + m\right)} \right\rangle\nonumber\\
    & = \frac{1}{\mathcal{Z}^*\left(\lambda\right)}\sum_{m \geq 0}\frac{\lambda^{d + m}}{m!}\left\langle \bm{\theta}^*\left(\lambda\right)^{\otimes m}, \overline{\mathcal{Z}}^*\bm{\overline{C}}^{(d + m)} \right\rangle.
\end{align}
\end{proof}
\end{proposition}

We have now motivated the Definition of the saddle point $\bm{\theta}^*$ by considering the QGMS-MGF evaluated at $\bm{\mu} = \bm{0}$. From the saddle point, we also defined correlations tensors, which will play a central role in the series expansion of QGMS moments derived in Section~\ref{sec:qgms_moments_series_expansion}. As a starting point to this expansion, it will be convenient to rewrite the QGMS-MGF at a general parameter $\bm{\mu} \in \mathbf{C}^{\mathcal{A}}$ an an integral centered about the saddle point. We collect this important result in the following Proposition.

\begin{proposition}[Integral representation of QGMS-MGF (saddle-point-centered version)]
\label{prop:qgms_mgf_saddle_point_centered_integral_representation}
Consider a QGMS-MGF defined by parameters $\mathcal{A}, \mathcal{S}, \bm{Q}, \bm{L}$:
\begin{align}
    S_n\left(\bm{\mu}\right) & := \sum_{\substack{\bm{n} = \left(n_{\bm{a}}\right)_{\bm{a} \in \mathcal{S}}\\\sum\limits_{\bm{a} \in \mathcal{S}}n_{\bm{a}} = n}}\binom{n}{\bm{n}}\exp\left(\frac{1}{2n}\bm{n}^T\bm{L}^T\bm{L}\bm{n} + \frac{1}{n}\bm{\mu}^T\bm{L}\bm{n}\right)\prod_{\bm{a} \in \mathcal{S}}Q_{\bm{a}}^{n_{\bm{a}}},\\
    \bm{\mu} & \in \mathbf{C}^{\mathcal{A}}.
\end{align}
Assume existence (not necessarily uniqueness) of a saddle point $\bm{\theta}^*$, i.e. a vector $\bm{\theta}^*$ satisfying Eq.~\ref{eq:saddle_point_equation}. Then, the following integral representation holds for the QGMS-MGF:
\begin{align}
    S_n\left(\bm{\mu}\right) & = \frac{e^{-n\bm{\theta}^{*T}\bm{\theta}}}{\left(2\pi\right)^{|\mathcal{A}|/2}}\int_{\mathbf{R}^{\mathcal{A}}}\!\mathrm{d}\bm{\theta}\,\exp\left(-\frac{1}{2}\bm{\chi}^T\bm{\chi} + \bm{\mu}^T\left(\bm{\theta}^* + \frac{1}{\sqrt{n}}\bm{\chi}\right) - \frac{1}{2n}\bm{\mu}^T\bm{\mu}\right)\left(\sum_{\bm{a} \in \mathcal{S}}Q_{\bm{a}}\exp\left(\bm{\theta}^{*T}\bm{L}_{:,\,\bm{a}} + \frac{1}{\sqrt{n}}\bm{\chi}^T\left(\bm{L}_{:,\,\bm{a}} - \bm{\theta}^*\right)\right)\right)^n.\label{eq:qgms_mgf_saddle_point_centered_integral_representation}
\end{align}
\begin{proof}
We start from the integral representation of $S_n\left(\bm{\mu}\right)$ derived in Eq.~\ref{eq:mgf_integral_form_after_gaussian_trick} (not assuming knowledge of a saddle point):
\begin{align}
    S_n\left(\bm{\mu}\right) & = \left(\frac{n}{2\pi}\right)^{\left|\mathcal{A}\right|/2}\int_{\mathbf{R}^{\mathcal{A}}}\!\mathrm{d}\bm{\theta}\,\exp\left(-\frac{1}{2}\bm{\mu}^T\bm{\mu} + \bm{\mu}^T\bm{\theta} - \frac{n}{2}\bm{\theta}^T\bm{\theta}\right)\left(\sum_{\bm{a} \in \mathcal{S}}Q_{\bm{a}}\exp\left(\bm{\theta}^T\bm{L}_{:,\,\bm{a}}\right)\right)^n.
\end{align}
We perform a translation change of variables in this integral representation:
\begin{align}
    \bm{\theta} & \longleftarrow \bm{\theta}^* + \frac{1}{\sqrt{n}}\bm{\chi}.
\end{align}
This complex-valued translation change of variables is licit, since the integrand (expanding the quantity raised to the power $n$) is a finite linear combination of complex Gaussians, for which such a change of variables is valid. The result then follows from expressing the integrand in variable $\bm{\chi}$.
\end{proof}
\end{proposition}

The interest of the integral representation of the QGMS-MGF established in Prop.~\ref{prop:qgms_mgf_saddle_point_centered_integral_representation} is, the dependence in $\mu$ is concentrated in exponential factor
\begin{align}
    \exp\left(\bm{\mu}^T\left(\bm{\theta}^* + \frac{1}{\sqrt{n}}\bm{\chi}\right) - \frac{1}{2n}\bm{\mu}^T\bm{\mu}\right).
\end{align}
The QGMS moments are by definition obtained by differentiating $S_n\left(\bm{\mu}\right)$ with respect to $\bm{\mu}$ at $\bm{\mu} = \bm{0}$. By the last observation, this operation brings down a polynomial in $\bm{\chi}$ inside the integral. For instance,
\begin{align}
    \frac{\partial^2S_n\left(\bm{\mu}\right)}{\partial\mu_{\alpha}\partial\mu_{\beta}}\Bigg|_{\bm{\mu} = \bm{0}} & = \frac{e^{-n\bm{\theta}^{*T}\bm{\theta}/2}}{\left(2\pi\right)^{\left|\mathcal{A}\right|/2}}\int_{\mathbf{R}^{\mathcal{A}}}\!\mathrm{d}\bm{\theta}\,\frac{\partial}{\partial\mu_{\alpha}}\left(e^{-\bm{\chi}^T\bm{\chi}/2 + \bm{\mu}^T\left(\bm{\theta}^* + n^{-1/2}\bm{\chi}\right) - \bm{\mu}^T\bm{\mu}/(2n)}\right)\Bigg|_{\bm{\mu} = \bm{0}}\left(\sum_{\bm{a} \in \mathcal{S}}Q_{\bm{a}}\exp\left(\bm{\theta}^{*T}\bm{L}_{:,\,\bm{a}} + \frac{\bm{\chi}^T}{\sqrt{n}}\left(\bm{L}_{:,\,\bm{a}} - \bm{\theta}^*\right)\right)\right)^n\nonumber\\
    & = \frac{e^{-n\bm{\theta}^{*T}\bm{\theta}/2}}{\left(2\pi\right)^{\left|\mathcal{A}\right|/2}}\int_{\mathbf{R}^{\mathcal{A}}}\!\mathrm{d}\bm{\chi}\,\left(\left(\theta^*_{\alpha} + \frac{\chi_{\alpha}}{\sqrt{n}}\right)\left(\theta^*_{\beta} + \frac{\chi_{\beta}}{\sqrt{n}}\right) - \frac{\delta_{\alpha\beta}}{n}\right)e^{-\bm{\chi}^T\bm{\chi}/2}\left(\sum_{\bm{a} \in \mathcal{S}}Q_{\bm{a}}\exp\left(\bm{\theta}^{*T}\bm{L}_{:,\,\bm{a}} + \frac{\bm{\chi}^T}{\sqrt{n}}\left(\bm{L}_{:,\,\bm{a}} - \bm{\theta}^*\right)\right)\right)^n\nonumber\\
    & = \left(\theta^*_{\alpha}\theta^*_{\beta} - \frac{1}{n}\delta_{\alpha\beta}\right)S^{(0)}_n + \frac{1}{\sqrt{n}}\theta^*_{\alpha}\left[\bm{S}^{(1)}_n\right]_{\beta} + \frac{1}{\sqrt{n}}\theta^*_{\beta}\left[\bm{S}^{(1)}_n\right]_{\alpha} + \frac{1}{n}\left[\bm{S}^{(2)}_n\right]_{\alpha,\,\beta},\label{eq:qgms_second_order_moment_from_qgms_integral_moments}
\end{align}

where in the final line, we introduced the QGMS integral moments tensors, specified in the following Definition:

\begin{definition}[QGMS integral moments tensor]
\label{def:qgms_integral_moments_tensor}
Let a QGMS be defined by parameters $\mathcal{A}, \mathcal{S}, \bm{Q}, \bm{L}$ and assume the QGMS admits a saddle point $\bm{\theta}^*$. With reference to the integral representation of the QGMS-MGF stated in Proposition~\ref{prop:qgms_mgf_saddle_point_centered_integral_representation}, the integral QGMS moments tensor of order $k$ is defined for all integer $k \geq 0$ by:
\begin{align}
    \bm{S}_n^{(k)} & := e^{-n\bm{\theta}^{*T}\bm{\theta}^*/2}\int_{\mathbf{R}^{\mathcal{A}}}\!\mathrm{d}\bm{\chi}\,\bm{\chi}^{\otimes k}\frac{e^{-\bm{\chi}^T\bm{\chi}/2}}{\left(2\pi\right)^{|\mathcal{A}|/2}}\left(\sum_{\bm{a} \in \mathcal{S}}Q_{\bm{a}}\exp\left(\bm{\theta}^{*T}\bm{L}_{:,\,\bm{a}} + \frac{1}{\sqrt{n}}\bm{\chi}^T\left(\bm{L}_{:,\,\bm{a}} - \bm{\theta}^*\right)\right)\right)^n \in \left(\mathbf{C}^{\mathcal{A}}\right)^{\otimes k}.
\end{align}
\end{definition}
In particular, recalling the integral representation for QGMS moments stated in Proposition~\ref{prop:qgms_mgf_saddle_point_centered_integral_representation},
\begin{align}
    S_n^{(0)} & = e^{-n\bm{\theta}^*\bm{\theta}^*/2}\int_{\mathbf{R}^{\mathcal{A}}}\!\mathrm{d}\bm{\chi}\,\frac{e^{-\bm{\chi}^T\bm{\chi}/2}}{\left(2\pi\right)^{|\mathcal{A}|/2}}\left(\sum_{\bm{a} \in \mathcal{S}}Q_{\bm{a}}\exp\left(\bm{\theta}^{*T}\bm{L}_{:,\,\bm{a}} + \frac{1}{\sqrt{n}}\bm{\chi}^T\left(\bm{L}_{:,\,\bm{a}} - \bm{\theta}^*\right)\right)\right)^n\nonumber\\
    & = S_n\left(\bm{0}_{\mathcal{A}}\right).
\end{align}

\subsection{Expansion of parametrized QGMS saddle point around the noninteracting limit}
\label{sec:pqgms_saddle_point_expansion}

In this Section, we develop the expansion of parametrized quadratic generalized multinomial sums. We consider a generic parametrized QGMS as introduced in Definition \ref{def:pqgms-mgf}. Given $\mathcal{A}, \mathcal{S}, \bm{Q}, \bm{L}$ defining the parametrized QGMS, the pseudo-moment-generating function is:
\begin{align}
    S_n\left(\lambda, \bm{\mu}\right) & = \sum_{\bm{n} = \of{n_{\bm{a}}}_{\bm{a} \in \mathcal{S}} \in \mathcal{P}(n)}\binom{n}{\bm{n}}\exp\of{\frac{\lambda^2}{2n}\bm{n}\bm{L}^T\bm{L}\bm{n} + \bm{\mu}^T\bm{L}\frac{\bm{n}}{n}}\prod_{\bm{a} \in \mathcal{S}}Q_{\bm{a}}^{n_{\bm{a}}}.
\end{align}
The associated saddle point equation is obtained by replacing $\bm{L} \longrightarrow \lambda\bm{L}$ in Equation \ref{eq:saddle_point_equation}:
\begin{align}
    \bm{\theta}^*\left(\lambda\right) & = \frac{\sum\limits_{\bm{a} \in \mathcal{S}}Q_{\bm{a}}\exp\left(\bm{\theta}\left(\lambda\right)^{*T}\lambda\bm{L}_{:,\,\bm{a}}\right)\lambda\bm{L}_{:,\,\bm{a}}}{\sum\limits_{\bm{a} \in \mathcal{S}}Q_{\bm{a}}\exp\left(\bm{\theta}\left(\lambda\right)^{*T}\lambda\bm{L}_{:,\,\bm{a}}\right)}\label{eq:saddle_point_equation_pqgms}
\end{align}
We note that to order $0$ is $\lambda$, this equation has trivial solution $\bm{\theta}^*\left(0\right) = \bm{0}_{\mathcal{A}}$. The order $1$ in $\lambda$ (assuming existence of a $\lambda$ power series expansion) is also easily obtained as
\begin{align}
    \bm{\theta}^*\left(\lambda\right) & = \frac{\sum\limits_{\bm{a} \in \mathcal{S}}Q_{\bm{a}}\lambda\bm{L}_{:,\,\bm{a}}}{\sum\limits_{\bm{a} \in \mathcal{S}}Q_{\bm{a}}} + \mathcal{O}\left(\lambda^2\right).
\end{align}
The goal of this Section is to prove the existence of a series expansion for $\lambda$ in a sufficiently small (complex) neighborhood of $0$, and systematize the expansion to arbitrary order.

We start (Section~\ref{sec:heuristic_pqgms_expansion}) by providing a heuristic derivation of the series expansion of $\bm{\theta}^*$ as an analytic function of $\lambda$. The convergence of this expansion is then established in Section~\ref{sec:convergence_saddle_point_expansion}, leading to the following main Proposition:
\begin{proposition}[Analyticity of candidate solution to saddle-point equation]
\label{propSaddlePointEquationAnalyticity}
Assume correlation tensors to satisfy bound in assumption \ref{assp:gamma_zero_correlations_boundedness}, depending on constants $c_{\mathrm{min}}, c_{\mathrm{max}}$. From these constants, assume the following bound on parameter $\lambda$ and index set $\mathcal{A}$ of the PQGMS:
\begin{align}
    \left|\lambda\right|^2\left|\mathcal{A}\right| & \leq \max\left(\frac{1}{2c^2}, \frac{c_{\mathrm{min}}}{2c_{\mathrm{max}}}\right),\label{eq:lambda_analyticity_domain_initial_statement}
\end{align}
where $c > 0$ is a constant introduced in Proposition~\ref{prop:t_block_entrywise_bound} and depending only on $c_{\mathrm{min}}, c_{\mathrm{max}}$ from Assumption~\ref{assp:gamma_zero_correlations_boundedness}. Then, the following candidate solution to the saddle point equation:
\begin{align}
    \bm{\theta}^*\left(\lambda\right) & := \sum_{m \geq 0}\bm{\theta}^{*,\,m}\left(\lambda\right)\nonumber\\
    \bm{\theta}^{*,\,m}\left(\lambda\right) & := \sum_{d^{(2)},\,\ldots,\,d^{(m + 1)} \geq 1}\bm{\theta}^{*,\,\left(d^{(2)},\,\ldots,\,d^{(m + 1)}\right)}\left(\lambda\right)
\end{align}
defines an absolutely convergent series, which is further an analytic function of $\lambda$ on the domain specified by Eq.~\ref{eq:lambda_analyticity_domain_initial_statement}. The absolute convergence holds at the innermost level of summation, i.e.:
\begin{align}
    \sum_{m \geq 0}\hspace*{5px}\sum_{d^{(2)},\,\ldots,\,d^{(m + 1)} \geq 1}\left\lVert \bm{\theta}^{*,\,\left(d^{(2)},\,\ldots,\,d^{(m + 1)}\right)} \right\rVert_{\infty} & < \infty.
\end{align}
In the above formulae,
\begin{align}
    \bm{\theta}^*\left(\lambda\right), \quad \bm{\theta}^{*,\,m}\left(\lambda\right), \quad \bm{\theta}^{*,\,\left(d^{(2)},\,\ldots,\,d^{(m + 1)}\right)}\left(\lambda\right) \in \mathbf{C}^{\mathcal{A}}.
\end{align}
For all $m \geq 0$ and $d^{(2)}, \ldots, d^{(m + 1)} \geq 1$,
\begin{align}
    \bm{\theta}^{*,\,\left(d^{(2)},\,\ldots,\,d^{(m + 1)}\right)}\left(\lambda\right) & := \bm{T}_{1,\,d^{(2)}}\bm{T}_{d^{(2)},\,d^{(3)}}\bm{T}_{d^{(3)},\,d^{(4)}} \ldots \bm{T}_{d^{(m - 1)},\,d^{(m)}}\bm{T}_{d^{(m)},\,d^{(m + 1)}}\left(\lambda\bm{\overline{C}}^{(1)}\right)^{\otimes d^{(m + 1)}}\\
    & = \overrightarrow{\prod_{r = 1}^{m}}\bm{T}_{d^{(r)},\,d^{(r + 1)}}\left(\lambda\bm{\overline{C}}^{(1)}\right)^{\otimes d^{(m + 1)}},
\end{align}
where in the last equation, the product with right arrow signals an ordering of factors from left to right, and we let $d^{(1)} := 1$ in factor $r = 1$ of the product. Under this definition, for $m = 0$, one may interpret $\left(d^{(2)},\,\ldots,\,d^{(m + 1)}\right) = \varnothing$ as the empty tuple, and the above equation becomes:
\begin{align}
    \bm{\theta}^{*,\,\varnothing}\left(\lambda\right) & = \lambda\bm{\overline{C}}^{(1)}.
\end{align}
For all $q, d \geq 1$, $\bm{T}_{q,\,d}$ is a linear operator:
\begin{align}
    \bm{T}_{q,\,d}: \left(\mathbf{C}^{\mathcal{A}}\right)^{\otimes d} & \longrightarrow \left(\mathbf{C}^{\mathcal{A}}\right)^{\otimes q}
\end{align}
which can be decomposed as a sum over unordered partitions and tuple partitions:
\begin{align}
    \bm{T}_{q,\,d} & := \sum_{\substack{\left(\mu_l\right)_{l \geq 1}\\l_1,\,\ldots,\,l_q \geq 0\\l_1 + \ldots + l_q + \sum_ll\mu_l = d}}\bm{T}_{q,\,d\,;\,\left(\mu_l\right)_{l \geq 1},\,l_1,\,\ldots,\,l_q}.
\end{align}
For each tuple partition $\left(l_1, \ldots, l_q\right)$ and unordered partition $\left(\mu_l\right)_{l \geq 1}$, row $\bm{\alpha}_{1:q} = \left(\alpha_1,\,\ldots,\,\alpha_q\right) \in \mathcal{A}^{q}$ of $\bm{T}_{q,\,d\,;\,\left(\mu_l\right)_{l \geq 1},\,l_1,\,\ldots,\,l_q}$ is defined as:
\begin{align}
    \left[\bm{T}_{q,\,d\,;\,\left(\mu_l\right)_{l \geq 1},\,l_1,\,\ldots,\,\l_q}\right]_{\bm{\alpha}_{1:q},\,:} & := \lambda^{q + d}\frac{\left(q - 1 + \sum\limits_{l \geq 1}\mu_l\right)}{\left(q - 1\right)!}\frac{(-1)^{\sum_l\mu_l}}{l_1!\ldots l_q!\prod\limits_{l \geq 1}\mu_l!l!^{\mu_l}}\bigotimes_{l \geq 1}\bm{\overline{C}}^{(l)\otimes \mu_l} \otimes \bigotimes_{1 \leq r \leq q}\bm{\overline{C}}^{\left(l_r + 1\right)}_{\alpha_r},
\end{align}
where we recall from Section~\ref{sec:notations} that $\bm{\overline{C}}^{\left(l_r + 1\right)}_{\alpha_r} \in \left(\mathbf{C}^{\mathcal{A}}\right)^{\otimes (l_r + 1)}$ is the slice of tensor $\bm{\overline{C}}^{\left(l_r + 1\right)}$ obtained by setting first index to $\alpha_r$ (in fact, any index, given the tensor is symmetric).
\end{proposition}
Despite the established convergence, this result still does not prove $\bm{\theta}^*\left(\lambda\right)$ solves saddle point equation \ref{eq:saddle_point_equation_pqgms}. Section~\ref{sec:correctness_saddle_point_expansion} then establishes this final point, showing correctness of the heuristic expansion derived in Section~\ref{sec:heuristic_pqgms_expansion}.

\begin{proposition}[Solution of saddle-point equation]
\label{propSolutionSaddlePointEquation}
Under conditions of applicability of Proposition~\ref{propSaddlePointEquationAnalyticity}, candidate solution $\bm{\theta}^*\left(\lambda\right) \in \mathbf{C}^{\mathcal{A}}$ defined in this Proposition effectively solves the saddle-point equation, i.e.:
\begin{align}
    \bm{\theta}^*\left(\lambda\right) & = \frac{\sum\limits_{\bm{a} \in \mathcal{S}}Q_{\bm{a}}\exp\left(\lambda\bm{\theta}^{*}\left(\lambda\right)^T\bm{L}_{:,\,\bm{a}}\right)\lambda\bm{L}_{:,\,\bm{a}}}{\sum\limits_{\bm{a} \in \mathcal{S}}Q_{\bm{a}}\exp\left(\lambda\bm{\theta}^{*}\left(\lambda\right)^T\bm{L}_{:,\,\bm{a}}\right)}.
\end{align}
\end{proposition}

After proving correctness of the saddle point expansion, Section~\ref{sec:bounds_pseudo_partition_function_correlations} proceeds to establish simple bounds on the pseudo-partition function and correlation tensors implied by this expansion. Section~\ref{sec:simplifying_bounds_pseudo_partition_function_correlations} further simplifies these bounds for future use in Section~\ref{sec:qgms_moments_series_expansion}.

\subsubsection{Heuristic derivation of the expansion}
\label{sec:heuristic_pqgms_expansion}

We start by giving a heuristic and rather short description of the series expansion of saddle point $\bm{\theta}^*\left(\lambda\right)$. The account is heuristic in that it presumes existence of the series and legitimacy of several series compositions (for instance, ignoring all possibility of singularity in the denominator of the right-hand side of saddle point Equation \ref{eq:saddle_point_equation_pqgms}).

The general idea of expressing solution $\bm{\theta}^*\left(\lambda\right)$ to Equation \ref{eq:saddle_point_equation_pqgms} is to convert this nonlinear equation into a linear one over a new (infinite-dimensional) vector $\bm{\Theta}^*$, ``collecting all tensor powers of sought solution $\bm{\theta}^*\left(\lambda\right)$":
\begin{align}
    \bm{\Theta}^*\left(\lambda\right) & := \begin{pmatrix}
        \bm{\theta}^*\left(\lambda\right)\\
        \bm{\theta}^{*\otimes 2}\left(\lambda\right)\\
        \bm{\theta}^{*\otimes 3}\left(\lambda\right)\\
        \vdots
    \end{pmatrix}
\end{align}
To slightly lighten the notation, we will abstain from signaling the $\lambda$ dependence of $\bm{\theta}^*$ in the rest of the paragraph. To express Equation \ref{eq:saddle_point_equation_pqgms} as a linear equation in $\bm{\Theta}^*$, we express the right-hand side as a series involving tensor powers of $\bm{\theta}^*$; this series is obtained by Taylor-expanding the numerator and denominator separately, and finally putting both these expansions together.\\

\paragraph{Expanding the numerator}\mbox{}
\label{sec:pqgms_expansion_heuristic_numerator}

Let us then start with expanding the numerator.

\begin{align}
    \sum_{\bm{a} \in \mathcal{S}}Q_{\bm{a}}\exp\left(\lambda\bm{\theta}^{*T}\bm{L}_{:,\,\bm{a}}\right)\lambda \bm{L}_{:,\,\bm{a}} & = \sum_{\bm{a} \in \mathcal{S}}Q_{\bm{a}}\sum_{d \geq 0}\frac{\lambda^{l + 1}}{l!}\left(\bm{\theta}^{*T}\bm{L}_{:,\,\bm{a}}\right)^l\bm{L}_{:,\,\bm{a}}\nonumber\\
    & = \sum_{\bm{a} \in \mathcal{S}}Q_{\bm{a}}\sum_{l \geq 0}\frac{\lambda^{l + 1}}{l!}\left\langle \bm{\theta}^{*\otimes l}, \bm{L}_{:,\,\bm{a}}^{\otimes \left(l + 1\right)} \right\rangle\nonumber\\
    & = \overline{\mathcal{Z}^*}\sum_{d \geq 0}\frac{\lambda^{l + 1}}{l!}\left\langle \bm{\theta}^{*\otimes l}, \bm{\overline{C}}^{(l + 1)} \right\rangle\\
    & = \overline{\mathcal{Z}^*}\left(\lambda \bm{\overline{C}}^{(1)} + \sum_{l \geq 1}\frac{\lambda^{l + 1}}{l!}\left\langle \bm{\theta}^{*\otimes l}, \bm{\overline{C}}^{(l + 1)} \right\rangle\right).\label{eq:numerator_series_expansion}
\end{align}
From the third line, we introduced the \textit{noninteracting correlations tensor} of order $d$, defined by:
\begin{align}
    \bm{\overline{C}}^{(d)} & := \left(\overline{C}^{(d)}_{\bm{\alpha}_{1:d}}\right)_{\bm{\alpha}_{1:d} \in \mathcal{A}^d} = \left(\overline{C}^{(d)}_{\alpha_1,\,\ldots,\,\alpha_d}\right)_{\alpha_1,\,\ldots,\,\alpha_d \in \mathcal{A}},\\
    \overline{C}^{(d)}_{\alpha_1,\,\alpha_2,\,\ldots,\,\alpha_{d - 1},\,\alpha_d} & := \frac{1}{\overline{\mathcal{Z}^*}}\sum_{\bm{a}}Q_{\bm{a}}L_{\alpha_1,\,\bm{a}}L_{\alpha_2,\,\bm{a}}\ldots L_{\alpha_{d - 1},\,\bm{a}}L_{\alpha_d,\,\bm{a}},\label{eq:noninteracting_correlations_tensor}
\end{align}
where we introduced the \textit{noninteracting partition function}
\begin{align}
    \overline{\mathcal{Z}^*} & := \sum_{\bm{a} \in \mathcal{S}}Q_{\bm{a}}.\label{eq:noninteracting_partition_function}
\end{align}
Similar to correlations tensors, noninteracting correlations tensors can be regarded as pseudo-expectations under a quasiprobability measure, except the quasiprobability measure does not depend on $\bm{\theta}^*$.

The $q$ tensor power of the above can be expressed:
\begin{align}
    & \left(\sum_{\bm{a} \in \mathcal{S}}Q_{\bm{a}}\exp\left(\lambda\bm{\theta}^{*T}\bm{L}_{:,\,\bm{a}}\right)\lambda \bm{L}_{:,\,\bm{a}}\right)^{\otimes q}\nonumber\\
    & = \overline{\mathcal{Z}^*}^q\sum_{l_1,\,\ldots,\,l_q \geq 0}\frac{\lambda^{\left(l_1 + 1\right) + \ldots + \left(l_q + 1\right)}}{l_1!\ldots l_q!}\left\langle \bm{\theta}^{*\otimes \left(l_1 + \ldots + l_q\right)}, \bm{\overline{C}}^{(l_1 + 1)} \otimes \ldots \otimes \bm{\overline{C}}^{(l_q + 1)}\right\rangle\nonumber\\
    & = \overline{\mathcal{Z}^*}^q\sum_{d \geq 0}\lambda^{d + q}\left\langle \bm{\theta}^{*\otimes d}, \sum_{\substack{l_1,\,\ldots,\,l_q\\l_1 + \ldots + l_q = d}}\frac{1}{l_1!\ldots l_q!}\bm{\overline{C}}^{(l_1 + 1)} \otimes \ldots \otimes \bm{\overline{C}}^{(l_q + 1)}\right\rangle\label{eq:numerator_tensor_power_series_expansion}\\
    & = \overline{\mathcal{Z}^*}^q\lambda^q\bm{\overline{C}}^{(1)\otimes q} + \sum_{d \geq 1}\lambda^{d + q}\left\langle \bm{\theta}^{*\otimes d}, \sum_{\substack{l_1,\,\ldots,\,l_q\\l_1 + \ldots + l_q = d}}\frac{1}{l_1!\ldots l_q!}\bm{\overline{C}}^{(l_1 + 1)} \otimes \ldots \otimes \bm{\overline{C}}^{(l_q + 1)}\right\rangle,\label{eq:numerator_tensor_power_series_expansion_order_zero_singled_out}
\end{align}
where in the final line, we defined the dot product between a $d$-indices tensor and a $(d + q)$-indices one by:
\begin{align}
    \left\langle \bm{U}, \bm{V} \right\rangle_{\alpha_{d + 1},\,\ldots,\,\alpha_{d + q}} & := \sum_{\alpha_1,\,\ldots,\,\alpha_d}U_{\alpha_1,\,\ldots,\,\alpha_d}V_{\alpha_1,\,\ldots,\,\alpha_d,\,\alpha_{d + 1},\,\ldots,\,\alpha_{d + q}},\\
    \bm{U} & = \left(U_{\alpha_1,\,\ldots,\,\alpha_d}\right)_{\alpha_1,\,\ldots,\,\alpha_d \in \mathcal{A}},\\
    \bm{V} & = \left(V_{\alpha_1,\,\ldots,\,\alpha_d,\,\alpha_{d + 1},\,\ldots,\,\alpha_{d + q}}\right)_{\alpha_1,\,\ldots,\,\alpha_d,\,\alpha_{d + 1},\,\ldots,\,\alpha_{d + q} \in \mathcal{A}}.
\end{align}
The right-hand sides of Equations \ref{eq:numerator_series_expansion} and \ref{eq:numerator_tensor_power_series_expansion_order_zero_singled_out} are nonlinear in $\bm\theta^{*}$. They can however be regarded as affine in the collection of tensor powers of $\bm\theta^*$. That is, defining vector
\begin{align}
    \bm{\Theta}^* & := \begin{pmatrix}
        \bm{\theta}^*\\
        \bm{\theta}^{*\otimes 2}\\
        \bm{\theta}^{*\otimes 3}\\
        \vdots
    \end{pmatrix},
\end{align}
From this $\lambda$ series expansion of the numerator, a $\lambda$ series representation of the saddle point can be derived making additional assumption:
\begin{align}
    \mathcal{Z}^*\left(\lambda\right) = 1 \quad \forall \lambda \qquad \textrm{(special case)},
\end{align}
which implies in particular (setting $\lambda = 0$) $\overline{\mathcal{Z}^*} = 1$; this holds for the SK-QAOA energy sum in particular. From this additional assumption, the denominator in the saddle point equation reduces to $1$. Then, the right-hand side of Equation \ref{eq:numerator_series_expansion} reads
\begin{align}
    \lambda\bm{\overline{C}}^{(1)} + \bm{T}_{1,\,:}\bm{\Theta}^*,
\end{align}
and the right-hand side of Equation \ref{eq:numerator_tensor_power_series_expansion_order_zero_singled_out} reads
\begin{align}
    \lambda^q\bm{\overline{C}}^{(1)\otimes q} + \bm{T}_{q,\,:}\bm{\Theta}^*.\label{eq:tree_expansion_matrix_block_domain}
\end{align}
In the last two equations, we defined linear operators:
\begin{align}
    \bm{T}_{q,\,d}: \left(\mathbf{C}^{\mathcal{A}}\right)^{\otimes d} \longrightarrow \left(\mathbf{C}^{\mathcal{A}}\right)^{\otimes q},\label{eq:tree_expansion_matrix_block_coefficients}
\end{align}
with matrix coefficients given by
\begin{align}
    & \left[\bm{T}_{q,\,d}\right]_{\left(\alpha_{d + 1},\,\ldots,\,\alpha_{d + q}\right),\,\left(\alpha_1,\,\ldots,\,\alpha_d\right)}\nonumber\\
    & := \sum_{\substack{l_1,\,\ldots,\,l_q\\l_1 + \ldots + l_q = d}}\frac{\lambda^{d + q}}{l_1!\ldots l_q!}\overline{C}^{(l_1 + 1)}_{\alpha_{d + 1},\,\alpha_1,\,\ldots,\,\alpha_{l_1}}\overline{C}^{(l_2 + 1)}_{\alpha_{d + 2},\,\alpha_{l_1 + 1},\,\ldots,\,\alpha_{l_1 + l_2}}\ldots \overline{C}^{(l_q + 1)}_{\alpha_{d + q},\,\alpha_{l_1 + \ldots + l_{q - 1} + 1},\,\ldots,\,\alpha_{l_1 + \ldots + l_q}}.
\end{align}
Note the definition of these operators does not feature $\bm\theta^*$, but only the noninteracting correlations $\bm{\overline{C}}^{(d)}$, which are assumed easy to compute\footnote{In this context, ``easy to compute" means it is efficient to output an entry of such a tensor given its indices. This does not solve the problem of the dimension of $\bm{C}^{(d)}$, containing all entries, scales exponentially in $d$.}. Next, stacking matrices $\bm{T}_{q,\,d}$ defined in Equation \ref{eq:tree_expansion_matrix_block_domain}, \ref{eq:tree_expansion_matrix_block_coefficients} into a large matrix:
\begin{align}
    \bm{T} & = \begin{pmatrix}
        \bm{T}_{1,\,1} & \bm{T}_{1,\,2} & \bm{T}_{1,\,3} & \ldots\\
        \bm{T}_{2,\,1} & \bm{T}_{2,\,2} & \bm{T}_{2,\,3} & \ldots\\
        \bm{T}_{3,\,1} & \bm{T}_{3,\,2} & \bm{T}_{3,\,3} & \ldots\\
        \vdots & \vdots & \vdots & \ddots,
    \end{pmatrix},
\end{align}
and the tensor powers of the noninteracting order $1$ correlations into a block vector:
\begin{align}
    \overline{\bm{\Theta}^*} & := \begin{pmatrix}
        \lambda\overline{\bm{C}}^{(1)}\\
        \lambda^2\overline{\bm{C}}^{(1)\otimes 2}\\
        \lambda^3\overline{\bm{C}}^{(1)\otimes 3}\\
        \vdots
    \end{pmatrix}
\end{align}
the saddle point Equation \ref{eq:saddle_point_equation_pqgms} can be synthetically written
\begin{align}
    \bm{\Theta}^* & = \overline{\bm{\Theta}^*} + \bm{T}\bm{\Theta}^*,
\end{align}
where again $\bm{T}$ does not depend on $\bm{\theta}^*$. Hence, the right-hand side is affine in $\bm{\Theta}^*$. Assuming invertibility operator $\bm{I} - \bm{T}$ (for the QGMS related to the SK-QAOA energy, this can be shown to follow from nilpotence of $\bm{T}$), the above equation can formally be solved as:
\begin{align}
    \bm\Theta^* & = \left(\bm{I} - \bm{T}\right)^{-1}\overline{\bm{\Theta}^*}.
\end{align}
While simplified due to assumption $\mathcal{Z}^*\left(\lambda\right) = 1$ for all $\lambda$, these considerations already give a faithful idea of the general approach.\\

\paragraph{Expanding the denominator}\mbox{}
\label{sec:pqgms_expansion_heuristic_denominator}

To generalize the method to any parametrized QGMS, we have yet to expand the denominator. The expansion is:

\begin{align}
    \sum_{\bm{a} \in \mathcal{S}}Q_{\bm{a}}\exp\left(\lambda\bm{\theta}^{*T}\bm{L}_{:,\,\bm{a}}\right) & = \sum_{\bm{a} \in \mathcal{S}}Q_{\bm{a}}\sum_{l \geq 0}\frac{\lambda^l}{l!}\left(\bm{\theta}^{*T}\bm{L}_{:,\,\bm{a}}\right)^l\\
    & = \sum_{\bm{a} \in \mathcal{S}}Q_{\bm{a}}\sum_{l \geq 0}\frac{\lambda^l}{l!}\left\langle \bm{\theta}^{*\otimes l}, \bm{L}_{:,\,\bm{a}}^{\otimes l} \right\rangle\\
    & = \sum_{l \geq 0}\frac{\lambda^l}{l!}\overline{\mathcal{Z}^*}\left\langle \bm{\theta}^{*\otimes d}, \overline{\bm{C}}^{(d)} \right\rangle\\
    & = \overline{\mathcal{Z}^*}\left(1 + \sum_{l \geq 1}\frac{\lambda^l}{l!}\left\langle \bm{\theta}^{*\otimes l}, \overline{\bm{C}}^{(l)} \right\rangle\right),
\end{align}

In order to write the saddle point equation in the space of tensor powers, we will need to raise this to power $-q$. This expansion gives:

\begin{align}
    \left(\sum_{\bm{a} \in \mathcal{S}}Q_{\bm{a}}\exp\left(\lambda \bm{\theta}^{*T}\bm{L}_{:,\,\bm{a}}\right)\right)^{-q} & = \left(\overline{\mathcal{Z}^*}\right)^{-q}\left(1 + \sum_{l \geq 1}\frac{\lambda^l}{l!}\left\langle \bm{\theta}^{*\otimes l}, \overline{\bm{C}}^{(l)} \right\rangle\right)^{-q}\nonumber\\
    & = \left(\overline{\mathcal{Z}^*}\right)^{-q}\left(1 + \sum_{l \geq 1}\frac{\lambda^l}{l!}\left\langle \bm{\theta}^{*\otimes l}, \overline{\bm{C}}^{(l)} \right\rangle\right)^{-q}\nonumber\\
    & = \left(\overline{\mathcal{Z}^*}\right)^{-q}\sum_{\mu \geq 0}\binom{-q}{\mu}\left(\sum_{l \geq 1}\frac{\lambda^l}{l!}\left\langle \bm{\theta}^{*\otimes l}, \overline{\bm{C}}^{(l)} \right\rangle\right)^{\mu}\nonumber\\
    & = \left(\overline{\mathcal{Z}^*}\right)^{-q}\sum_{\mu \geq 0}\binom{-q}{\mu}\sum_{\substack{\left(\mu_l\right)_{l \geq 1}\\\sum_l\mu_l = \mu}}\binom{\mu}{\left(\mu_l\right)_{l \geq 1}}\prod_{l \geq 1}\left(\frac{\lambda^l}{l!}\left\langle \bm{\theta}^{*\otimes d}, \bm{\overline{C}}^{(l)} \right\rangle\right)^{\mu_l}\nonumber\\
    & = \left(\overline{\mathcal{Z}^*}\right)^{-q}\sum_{\substack{\left(\mu_l\right)_{l \geq 1}}}\frac{\left(q + \sum\limits_{l \geq 1}\mu_l - 1\right)!}{\left(q - 1\right)!}\prod_{l \geq 1}\frac{1}{\mu_l!}\left(-\frac{\lambda^l}{l!}\left\langle \bm{\theta}^{*\otimes l}, \overline{\bm{C}}^{(l)} \right\rangle\right)^{\mu_l}\label{eq:denominator_tensor_power_series_expansion}\\
    & = \left(\overline{\mathcal{Z}^*}\right)^{-q}\left(1 + \sum_{\substack{\left(\mu_l\right)_{l \geq 1}\\\sum_l\mu_l \geq 1}}\frac{\left(q + \sum\limits_{l \geq 1}\mu_l - 1\right)!}{\left(q - 1\right)!}\prod_{l \geq 1}\frac{1}{\mu_l!}\left(-\frac{\lambda^l}{l!}\left\langle \bm{\theta}^{*\otimes l}, \overline{\bm{C}}^{(l)} \right\rangle\right)^{\mu_l}\right)\label{eq:denominator_tensor_power_series_expansion_order_zero_singled_out}.
\end{align}

\paragraph{Putting numerator and denominator together}\mbox{}
\label{sec:pqgms_expansion_heuristic_putting_everything_together}

We are now ready to put together the series expansions of the numerator (paragraph \ref{sec:pqgms_expansion_heuristic_numerator}) and denominator (paragraph \ref{sec:pqgms_expansion_heuristic_denominator}) to phrase the saddle point equation as a linear equation in the ``vector of tensor powers" $\bm{\Theta}^*$. This linearization was previously sketched at the end of paragraph \ref{sec:pqgms_expansion_heuristic_numerator}, assuming a unit denominator for all $\lambda$.

Combining equations \ref{eq:numerator_tensor_power_series_expansion} and \ref{eq:denominator_tensor_power_series_expansion} for the (tensor) power $q$ series expansion of the numerator and denominator of the right-hand side of the saddle point equation \ref{eq:saddle_point_equation}, we obtain:
\begin{align}
    \left(\frac{\sum\limits_{\bm{a} \in \mathcal{S}}Q_{\bm{a}}\exp\left(\lambda\bm{\theta}^{*T}\bm{L}_{:,\,\bm{a}}\right)\lambda\bm{L}_{:,\,\bm{a}}}{\sum\limits_{\bm{a} \in \mathcal{S}}Q_{\bm{a}}\exp\left(\lambda\bm{\theta}^{*T}\bm{L}_{:,\,\bm{a}}\right)}\right)^{\otimes q} & = \left(\sum\limits_{\bm{a} \in \mathcal{S}}Q_{\bm{a}}\exp\left(\lambda\bm{\theta}^{*T}\bm{L}_{:,\,\bm{a}}\right)\right)^{-q}\left(\sum\limits_{\bm{a} \in \mathcal{S}}Q_{\bm{a}}\exp\left(\lambda\bm{\theta}^{*T}\bm{L}_{:,\,\bm{a}}\right)\lambda\bm{L}_{:,\,\bm{a}}\right)^{\otimes q}\nonumber\\
    & = \sum_{\substack{\left(\mu_l\right)_{l \geq 1}\\l_1,\,\ldots,\,l_q}}\lambda^{q + l_1 + \ldots + l_q + \sum\limits_{l \geq 1}l\mu_l}\frac{\left(q + \sum\limits_{l \geq 1}\mu_l - 1\right)!}{\left(q - 1\right)!}\frac{\left(-1\right)^{\sum\limits_{l \geq 1}\mu_l}}{l_1!\ldots l_q!\prod\limits_{l \geq 1}\mu_l!l!^{\mu_l}}\nonumber\\
    & \hspace*{50px} \times \left\langle \left(\bm{\theta}^{*}\right)^{\otimes\left(l_1 + \ldots + l_q + \sum\limits_{l \geq 1}l\mu_l\right)}, \bigotimes_{l \geq 1}\overline{\bm{C}}^{(l)\otimes \mu_l} \otimes \overline{\bm{C}}^{\left(l_1 + 1\right)} \otimes \ldots \otimes \overline{\bm{C}}^{\left(l_q + 1\right)} \right\rangle\label{eq:saddle_point_equation_rhs_power_q_expansion}
\end{align}
In the above sum, we can single out the term where both $\left(\mu_l\right)_{l \geq 1}$ and $\left(l_1, \ldots, l_q\right)$ are zero. This term is the term of order $q$ in $\lambda$, and evaluates as expected to:
\begin{align}
    \left\langle \bm{\theta}^{*\otimes 0}, \lambda\overline{\bm{C}}^{\left(l_1 + 1\right)} \otimes \ldots \otimes \lambda\overline{\bm{C}}^{\left(l_q + 1\right)}\right\rangle & = \overline{\bm{C}}^{\left(l_1 + 1\right)} \otimes \ldots \otimes \overline{\bm{C}}^{\left(l_q + 1\right)}\nonumber\\
    & = \left(\lambda\overline{\bm{C}}^{\left(1\right)}\right)^{\otimes q},\label{eq:tensor_powers_vector}
\end{align}
that is the $\overline{\bm{C}}$ tensor at lowest nontrivial order in $\lambda$. Similar to the simpler case where the denominator was trivial ($\mathcal{Z}^*\left(\lambda\right) = 1$ for all $\lambda$), we can then phrase the saddle point equation in the tensor powers space and in terms of an operator $\bm{T}$ acting on this space. Namely, introducing the block vector of tensor powers:
\begin{align}
    \bm{\Theta}^* & = \begin{pmatrix}
        \bm{\theta}^*\\
        \bm{\theta}^{*\otimes 2}\\
        \bm{\theta}^{*\otimes 3}\\
        \vdots
    \end{pmatrix},
\end{align}
and defining $\bm{T}$ by blocks:
\begin{align}
    \bm{T} & := \begin{pmatrix}
        \bm{T}_{1, 1} & \bm{T}_{1, 2} & \bm{T}_{1, 3} & \ldots\\
        \bm{T}_{2, 1} & \bm{T}_{2, 2} & \bm{T}_{2, 3} & \ldots\\
        \bm{T}_{3, 1} & \bm{T}_{3, 2} & \bm{T}_{3, 3} & \ldots\\
        \vdots & \vdots & \vdots & \ddots
    \end{pmatrix},\label{eq:saddle_point_equation_tensor_interpretation_t}
    \end{align}
with block $\bm{T}_{q,\,d}$ given by:
\begin{align}
    & \left[\bm{T}_{q,\,d}\right]_{\left(\alpha_{d + 1},\,\ldots,\,\alpha_{d + q}\right),\,\left(\alpha_1,\,\ldots,\,\alpha_d\right)}\nonumber\\
    & := \lambda^{q + d}\sum_{\substack{\left(\mu_{l}\right)_{l \geq 1}\\l_1,\,\ldots,\,l_q\\\sum_ll\mu_l + l_1 + \ldots + l_q = d}}\frac{\left(q - 1 + \sum\limits_{l \geq 1}\mu_l\right)!}{\left(q - 1\right)!}\frac{(-1)^{\sum\limits_{l \geq 1}\mu_l}}{l_1!\ldots l_q!\prod\limits_{l \geq 1}\mu_l!l!^{\mu_l}}\nonumber\\
    & \hspace*{120px} \times \overline{C}^{(1)}_{\alpha_1}\ldots \overline{C}^{(1)}_{\alpha_{q_1}}\overline{C}^{(2)}_{\alpha_{\mu_1 + 1},\,\alpha_{\mu_1 + 2}}\ldots \overline{C}^{(2)}_{\alpha_{\mu_1 + 2\mu_2 - 1,\,\alpha_{\mu_1 + 2\mu_2}}}\ldots\nonumber\\
    & \hspace*{120px} \times \overline{C}^{\left(l_1 + 1\right)}_{\alpha_{d + 1},\,\alpha_{\sum_{l}l\mu_l + 1},\,\ldots,\,\alpha_{\sum_{l}l\mu_l + l_1}}\ldots \overline{C}^{\left(l_q + 1\right)}_{\alpha_{d + q},\,\alpha_{\sum_{l}l\mu_l + l_1 + \ldots + l_{q - 1} + 1},\,\ldots,\,\alpha_{\sum_ll\mu_l + l_1 + \ldots + l_q}},\label{eq:saddle_point_equation_tensor_interpretation_t_block}
\end{align}
the saddle point equation reads:
\begin{align}
    \bm{\Theta}^* & = \bm{T}\bm{\Theta}^* + \overline{\bm{\Theta}^*},
\end{align}
where $\overline{\bm{\Theta}^*}$ is now defined as:
\begin{align}
    \overline{\bm{\Theta}^*} & = \begin{pmatrix}
        \lambda\overline{\bm{C}}^{(1)}\\
        \lambda^2\overline{\bm{C}}^{(1)\otimes 2}\\
        \lambda^3\overline{\bm{C}}^{(1)\otimes 3}\\
        \vdots
    \end{pmatrix}
\end{align}
For convenience, we parse the definition of block $T_{q,\,d}$ in equation \ref{eq:saddle_point_equation_tensor_interpretation_t_block}. The second line contains $\mu_1$ occurrences of $\overline{C}^{\left(1\right)}$, $\mu_2$ occurrences of $\overline{C}^{\left(2\right)}$, and more generally $\mu_l$ occurrences of $\overline{C}^{\left(l\right)}$ for all $l \geq 1$. This is still a finite number of tensors as there can only be a finite number of nonzero $\mu_l$ due to constraint
\begin{align}
    \sum_{l \geq 1}\mu_l \leq \sum_{l \geq 1}l\mu_l \leq \sum_{l \geq 1}l\mu_l + l_1 + \ldots + l_q = d.
\end{align}
These occurrences are indexed by indices $\alpha_1, \ldots, \alpha_{\sum_ll\mu_l}$. Since $\sum_ll\mu_l \leq d$, only column indices of block $\bm{T}_{q,\,d}$ are used in the second line of equation \ref{eq:saddle_point_equation_tensor_interpretation_t_block}. Next, in the third equation line, one finds an occurrence of $\overline{C}^{\left(l_1 + 1\right)}$, an occurrence of $\overline{C}^{\left(l_2 + 1\right)}$, \ldots, an occurrence of $\overline{C}^{\left(l_q + 1\right)}$. The first index in these occurrences is respectively $\alpha_{d + 1},\,\alpha_{d + 2},\,\ldots,\,\alpha_{d + q}$ ---these are all the rows indices of the $\bm{T}_{q,\,d}$. The remaining indices for these tensor occurrences are the remaining column indices of $\bm{T}_{q,\,d}$ (not used in the second line of equation \ref{eq:saddle_point_equation_tensor_interpretation_t_block}), i.e. $\alpha_{\sum_ll\mu + 1},\,\alpha_{\sum_ll\mu + 2},\,\ldots,\,\alpha_d$. Writing a single row of block $\bm{T}_{q,\,d}$ in equation \ref{eq:saddle_point_equation_tensor_interpretation_t_block} may lead to a more readable formula:
\begin{align}
    \left[\bm{T}_{q,\,d}\right]_{\left(\alpha_{d + 1},\,\ldots,\,\alpha_{d + q}\right),\,:}  & = \lambda^{q + d}\sum_{\substack{\left(\mu_l\right)_{l \geq 1}\\l_1,\,\ldots,\,l_q\\\sum_ll\mu_l + l_1 + \ldots + l_q = d}}\hspace*{-20px}\frac{\left(q - 1 + \sum\limits_{l \geq 1}\mu_l\right)!}{\left(q - 1\right)!}\frac{\left(-1\right)^{\sum\limits_{l \geq 1}\mu_l}}{l_1!\ldots l_q!\prod\limits_{l \geq 1}\mu_l!l!^{\mu_l}}\nonumber\\
    & \hspace*{120px} \times \bigotimes_{l \geq 1}\overline{\bm{C}}^{\left(l\right)\otimes \mu_l} \otimes \overline{\bm{C}}^{\left(l_1 + 1\right)}_{\alpha_{d + 1}} \otimes \ldots \otimes \overline{\bm{C}}^{\left(l_1 + q\right)}_{\alpha_{d + q}}
\end{align}

Assuming invertibility of $\bm{I} - \bm{T}$, this is uniquely solved by:
\begin{align}
    \bm{\Theta}^* & = \left(\bm{I} - \bm{T}\right)^{-1}\overline{\bm{\Theta}^*}.\label{eq:saddle_point_equation_solution_non_rigorous}
\end{align}
Further assuming the inverse can be expanded as a power series:
\begin{align}
\label{eq:fixed_point_solution}
    \left(\bm{I} - \bm{T}\right)^{-1} & = \bm{I} + \bm{T} + \bm{T}^2 + \bm{T}^3 + \ldots,
\end{align}
the solution can also be expressed as a series:
\begin{align}
    \bm{\Theta}^*& = \overline{\bm{\Theta}^*} + \bm{T}\overline{\bm{\Theta}^*} + \bm{T}^2\overline{\bm{\Theta}^*} + \bm{T}^3\overline{\bm{\Theta}^*} + \ldots\label{eq:saddle_point_equation_tensor_interpretation_series_solution}.
\end{align}
The latter power series sketches a method for solving the saddle-point equation to a given order in $\lambda$. To deduce an approximate solution up to a certain order in $\lambda$, it will help to introduce an intuitive interpretation of matrix blocks $\bm{T}_{q, d}$, including a tensor network representation. We first consider the $\lambda$ orders of the series terms \ref{eq:saddle_point_equation_tensor_interpretation_series_solution}. For that purpose, let us consider a single power of the $\bm{T}$ and express it in terms of the blocks $\bm{T}_{q',\,d'}$. More specifically, consider block $\left(q, d\right)$ of the $m$-th power of $\bm{T}$:
\begin{align}
    \left[\bm{T}^m\right]_{q,\,d} & = \sum_{d^{(1)},\,d^{(2)},\,\ldots,\,d^{(m - 1)} \geq 1}\bm{T}_{q,\,d^{(1)}}\bm{T}_{d^{(1)},\,d^{(2)}}\bm{T}_{d^{(2)},\,d^{(3)}}\ldots\bm{T}_{d^{(m - 3)},\,d^{(m - 2)}}\bm{T}_{d^{(m - 2)},\,d^{(m - 1)}}\bm{T}_{d^{(m - 1)},\,d}\label{eq:saddle_point_equation_tensor_interpretation_t_power_from_blocks}
\end{align}

\begin{figure}[!htbp]
    \centering
    \includegraphics[width=0.6\linewidth]{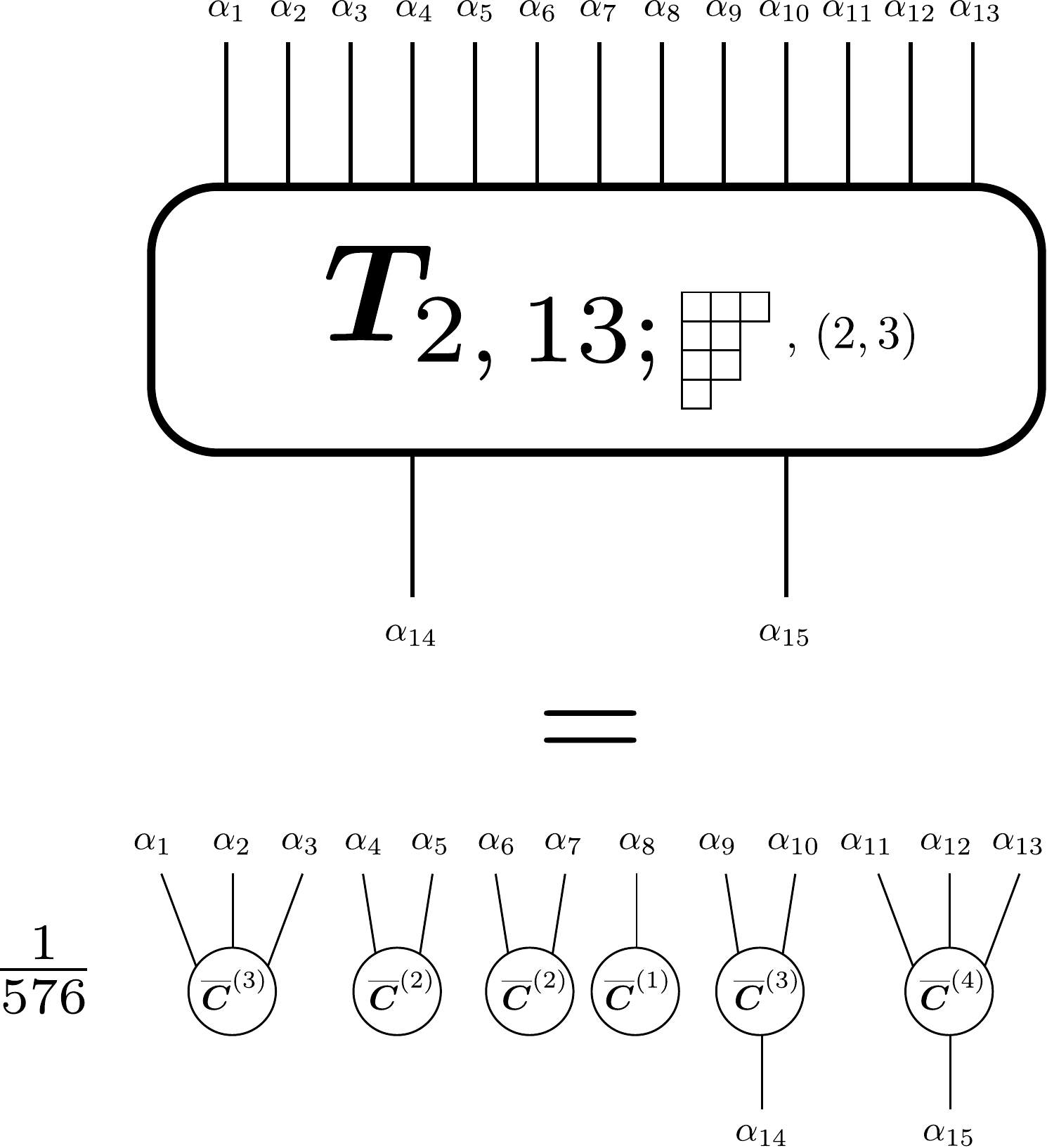}
    \caption{An example contribution to $\bm{T}$ matrix block $\bm{T}_{2, 13}$ ($q = 2, d = 13$). For this contribution, $d' = 8$ with corresponding integer partition $3 + 2 + 2 + 1$, represented on the figure as a Young diagram. As for the tuple, summing to $d - d' = 5$, $\left(l_1, l_2\right) = \left(2, 3\right)$.}
    \label{fig:saddle_point_equation_tensor_interpretation_example_block_contribution}
\end{figure}

In the above expression, $q, d \geq 1$ index the block of $\bm{T}$. Likewise, summation variables $d^{(1)}, \ldots, d^{(m - 1)}$ iterate over block indices. For $m = 1$ ($\bm{T}^m = \bm{T}$), the equation evaluates to $\bm{T}_{q,\,d}$. Now, by definition of $\bm{T}_{q',\,d'}$ (equation \ref{eq:saddle_point_equation_tensor_interpretation_t_block}), each block $\bm{T}_{q',\,d'}$ is of order at least $q' + d' \geq 2$ in $\lambda$. It follows that in the series solution \ref{eq:saddle_point_equation_tensor_interpretation_series_solution} to the saddle-point equation, only a finite number of terms need to be evaluated to obtain an approximation up to a given order in $\lambda$. We now give a closer look to the different terms appearing in the definition of a single block $\bm{T}_{q,\,d}$ in equation \ref{eq:saddle_point_equation_tensor_interpretation_t_block}. We observe that $\bm{T}_{q,\,d}$ is parametrized by a collection of nonnegative integers $\left(\mu_l\right)_{l \geq 1}$ and a $q$-tuple of nonnegative integers $\left(l_1, \ldots, l_q\right)$. These parameters must satisfy constraint:
\begin{align}
    \sum_{l \geq 1}l\mu_l + l_1 + \ldots + l_q & = d.
\end{align}
These can be rephrased as:
\begin{align}
    \left\{\begin{array}{rcl}
         \sum\limits_{l \geq 1}l\mu_l & = & d'\\
         l_1 + \ldots + l_q & = & d - d'\\
         d' & \in & \{0, 1, \ldots, d - 1, d\}
    \end{array}\right..
\end{align}
From this formulation, for any fixed $0 \leq d' \leq d$, $\left(\mu_l\right)_{l \geq 1}$ can be interpreted as a partition of integer $d'$ (where $\mu_l$ counts the multiplicity of $l$ in the partition). As for $\left(l_1, \ldots, l_q\right)$, it is an ordered tuple of integer summing to $d - d'$. This suggests to introduce the following shorthand notation for the terms in the sum of equation \ref{eq:saddle_point_equation_tensor_interpretation_t_block}:
\begin{align}
    \bm{T}_{q,\,d} & = \sum_{\substack{\left(\mu_l\right)_{l \geq 1}\\l_1,\,\ldots,\,l_q\\\sum_{l}l\mu_l + l_1 + \ldots + l_q = d}}\bm{T}_{q,\,d\,;\,\left(\mu_l\right)_{l \geq 1},\,\left(l_1,\,\ldots,\,l_q\right)}.\label{eq:t_block_partitions_decomposition}
\end{align}
That is, we index each term of the sum by a partition and a tuple ($d'$ can be kept implicit as it is the sum of the partition). Each term $\bm{T}_{q, d\,;\,\left(\mu_l\right)_{l \geq 1},\,\left(l_1, \ldots, l_q\right)}$ is a Kronecker product of noninteracting correlation tensors $\overline{\bm{C}}^{(r)}$ ($r \geq 1$). A graphical representation for one such term is given on figure \ref{fig:saddle_point_equation_tensor_interpretation_example_block_contribution}.

\subsubsection{Convergence of saddle point expansion}
\label{sec:convergence_saddle_point_expansion}

In this section, we show correctness and convergence of the heuristic series expansion for the solution to the saddle point equation:
\begin{align}
    \bm{\Theta}^* & = \left(\bm{I} - \bm{T}\right)^{-1}\overline{\bm{\Theta}^*},\label{eq:saddle_point_solution_restated}\\
    \bm{\theta}^* & = \left[\bm{\Theta}^*\right]_1
\end{align}
where operator $\bm{T}$, acting over a normed vector subspace (yet to be specified) of the set of infinite complex number sequences, was defined heuristically by equations \ref{eq:saddle_point_equation_tensor_interpretation_t}, \ref{eq:saddle_point_equation_tensor_interpretation_t_block}. Besides, notation $\left[\bm{\Theta}^*\right]_1$ refers to the first block of ``vector" $\bm{\Theta}^*$. To make this candidate solution rigorous, we will need to specify appropriate normed spaces supporting vectors $\overline{\bm{\Theta}^*}, \bm{\Theta}^*$ and operator $\bm{T}$. This will ultimately allow to prove the existence of the inverse $\left(\bm{I} - \bm{T}\right)^{-1}$ in the heuristic solution ---and this inverse will indeed be given by its Taylor series expansion as postulated in heuristics.

Before reaching this final goal, it will be instructive to prove the existence of the following heuristic formulation of the saddle-point solution. Starting from equation \ref{eq:saddle_point_solution_restated}, assuming the Taylor expansion of the inverse legitimate, and expressing powers of $\bm{T}$ in terms of $\bm{T}$ blocks $\bm{T}_{q,\,d}$ (equation \ref{eq:saddle_point_equation_tensor_interpretation_t_power_from_blocks}), one indeed reaches the following candidate form for $\bm{\theta}^*$:
\begin{align}
    \bm{\theta}^* & = \left[\bm{\Theta}^*\right]_1\nonumber\\
    & = \left[\left(\bm{I} - \bm{T}\right)^{-1}\overline{\bm{\Theta}^*}\right]_1\nonumber\\
    & = \left[\sum_{m \geq 0}\bm{T}^m\overline{\bm{\Theta}^*}\right]_1\nonumber\\
    & = \sum_{m \geq 0}\hspace*{5px}\sum_{d^{(1)},\,d^{(2)},\,\ldots,\,d^{(m - 1)},\,d^{(m)} \geq 1}\bm{T}_{1,\,d^{(1)}}\bm{T}_{d^{(1)},\,d^{(2)}}\ldots\bm{T}_{d^{(m - 2)},\,d^{(m - 1)}}\bm{T}_{d^{(m - 1)},\,d^{(m)}}\left[\overline{\bm{\Theta}^*}\right]_{d^{(m)}}\nonumber\\
    & = \sum_{m \geq 0}\hspace*{5px}\sum_{d^{(1)},\,d^{(2)},\,\ldots,\,d^{(m - 1)},\,d^{(m)} \geq 1}\bm{T}_{1,\,d^{(1)}}\bm{T}_{d^{(1)},\,d^{(2)}}\ldots\bm{T}_{d^{(m - 2)},\,d^{(m - 1)}}\bm{T}_{d^{(m - 1)},\,d^{(m)}}\left(\lambda\overline{\bm{C}}^{(1)}\right)^{\otimes d^{(m)}}.
\end{align}
The convergence of this series will be rather easy to demonstrate and will be the object of proposition \ref{propSaddlePointEquationAnalyticity}. In fact, recalling that each block $\bm{T}_{q,\,d}$ involves a factor $\lambda^{q + d}$, times a matrix independent of $\lambda$, the above equation implicitly gives a power series expansion of $\bm{\theta}^*$ as a function of $\lambda$, and we show the stronger result that $\bm{\theta}^* = \bm{\theta}^*\left(\lambda\right)$ is analytic in $\lambda$ around $0$. This indicates our results can also be applied to imaginary time; however, this also indicates the crudeness of our methods, which do not explicitly exploit the real time assumption relevant to QAOA. To reach this result, we will rely on simple bounds on the entries of blocks $\bm{T}_{q,\,d}$, as stated in proposition \ref{prop:t_block_entrywise_bound}. The derivations will rely on a uniform bound over noninteracting correlations:

\begin{assumption}[Boundedness of time correlations]
\label{assp:gamma_zero_correlations_boundedness}
There exist constants $c_{\mathrm{min}}, c_{\mathrm{max}} > 0$, with $c_{\mathrm{min}} \leq c_{\mathrm{max}}$, such that the following bounds hold:
\begin{align}
    & \left|\overline{\mathcal{Z}^*}\right| > c_{\mathrm{min}},\\
    & \left|\overline{\mathcal{Z}^*}\right|\left|\overline{C}^{(d)}_{\alpha_1,\,\ldots,\,\alpha_d}\right| \leq c_{\mathrm{max}} \qquad \forall d \geq 0,\,\left(\alpha_1, \ldots, \alpha_d\right) \in \mathcal{A}^d.
\end{align}
For $d = 0$, the last constraint should be understood as
\begin{align}
    \left|\overline{\mathcal{Z}^*}\right| & \leq c_{\mathrm{max}}.
\end{align}
\end{assumption}
The non-trivial point is the existence of a constant $c_{\mathrm{max}}$ independent of the tensor degree $d$. Assuming existence of $c_{\mathrm{min}}, c_{\mathrm{max}}$ as in assumption \ref{assp:gamma_zero_correlations_boundedness}, note one may always choose $c_{\mathrm{min}} := \overline{\mathcal{Z}^*}$ at the cost of redefining $c_{\mathrm{max}}$. The reason allowing using a lower bound $c_{\mathrm{min}}$ is that all estimates derived in this section have errors expressible from $c_{\mathrm{min}}, c_{\mathrm{max}}$ only. Therefore, if one can find constants $c_{\mathrm{min}}, c_{\mathrm{max}}$ applying to a \textit{family} of quadratic generalized multinomial sums (rather to a single such sum), we will obtain estimates holding \textit{uniformly} over this family of QGMS.

We now bound the matrix elements of each block $\bm{T}_{q,\,d}$. Note this is clearly well-defined as a finite-dimensional matrix representing a mapping $\left(\mathbf{C}^{\mathcal{A}}\right)^{\otimes d} \longrightarrow \left(\mathbf{C}^{\mathcal{A}}\right)^{\otimes q}$. Using assumption \ref{assp:gamma_zero_correlations_boundedness} and applying the triangular inequality to the sum defining $\bm{T}_{q, d}$ in equation \ref{eq:saddle_point_equation_tensor_interpretation_t_block}, we obtain bound:
\begin{align}
    \left|\left[\bm{T}_{q,\,d}\right]_{\left(\alpha_{d + 1},\,\ldots,\,\alpha_{d + q}\right),\,\left(\alpha_1,\,\ldots,\,\alpha_d\right)}\right| & \leq \left|\lambda\right|^{d + q}\sum_{\substack{\left(\mu_l\right)_{l \geq 1}\\l_1,\,\ldots,\,l_q\\\sum_ll\mu_l + l_1 + \ldots + l_q = d}}\hspace*{-20px}\frac{\left(q - 1 + \sum\limits_{l \geq 1}\mu_{l}\right)!}{\left(q - 1\right)!}\frac{1}{l_1!\ldots l_q!\prod\limits_{l \geq 1}\mu_{l}!l!^{\mu_l}}\left(\frac{c_{\mathrm{max}}}{c_{\mathrm{min}}}\right)^{q + \sum_l\mu_{l}}\label{eq:t_block_bound_step_1},
\end{align}
where we combined the two bounds from assumption \ref{assp:gamma_zero_correlations_boundedness} to obtain 
\begin{align}
    \left\lVert \bm{\overline{C}}^{(d)} \right\rVert_{\infty} & \leq \frac{c_{\mathrm{max}}}{c_{\mathrm{min}}},
\end{align}
uniformly in $d$.

It is convenient to view the sum in equation \ref{eq:t_block_bound_step_1} as the Taylor coefficient of some analytic function of a variable $z$; namely:
\begin{align}
    \left|\left[\bm{T}_{q,\,d}\right]_{\left(\alpha_{d + 1},\,\ldots,\,\alpha_{d + q}\right),\,\left(\alpha_1,\,\ldots,\,\alpha_d\right)}\right| & \leq \left|\lambda\right|^{q + d}\left[f(z)\right]_{z^d},
\end{align}
where $z^d$ refers the coefficient of $z^d$ in the Taylor series expansion of $f(z)$, and
\begin{align}
    f(z) & := \sum_{d \geq 1}\hspace*{5px}z^d\sum_{\substack{\left(\mu_{l}\right)_{l}\\l_1,\,\ldots,\,l_q\\\sum_ll\mu_{l} + l_1 + \ldots + l_q = d}}\frac{\left(q - 1 + \sum\limits_{l \geq 1}\mu_{l}\right)!}{\left(q - 1\right)!}\frac{1}{l_1!\ldots l_q!\prod\limits_{l \geq 1}\mu_{l}!l!^{\mu_{l}}}\left(\frac{c_{\mathrm{max}}}{c_{\mathrm{min}}}\right)^{q + \sum_{l}\mu_{l}}.
\end{align}
We now explicitly compute $f(z)$:
\begin{align}
    f(z) & = \sum_{d \geq 1}\hspace*{5px}z^d\sum_{\substack{\left(\mu_{l}\right)_{l \geq 1}\\l_1,\,\ldots,\,l_q \geq 0\\\sum_ll\mu_{l} + l_1 + \ldots + l_q = d}}\frac{\left(q - 1 + \sum\limits_{l \geq 1}\mu_{l}\right)!}{\left(q - 1\right)!}\frac{1}{l_1!\ldots l_q!\prod\limits_{l \geq 1}\mu_{l}!l!^{\mu_{l}}}\left(\frac{c_{\mathrm{max}}}{c_{\mathrm{min}}}\right)^{q + \sum_l\mu_{l}}\nonumber\\
    & = \sum_{\substack{\left(\mu_{l}\right)_{l \geq 1}\\l_1,\,\ldots,\,l_q \geq 0}}z^{l_1 + \ldots + l_q + \sum_ll\mu_{l}}\left(\frac{c_{\mathrm{max}}}{c_{\mathrm{min}}}\right)^{q + \sum_{l}\mu_{l}}\frac{\left(q - 1 + \sum\limits_{l \geq 1}\mu_{l}\right)!}{\left(q - 1\right)!}\frac{1}{l_1!\ldots l_q!\prod\limits_{l \geq 1}\mu_{l}!l!^{\mu_{l}}}\nonumber\\
    & = e^{qz}\sum_{\substack{\left(\mu_{l}\right)_{l \geq 1}}}z^{\sum_{l}l\mu_{l}}\left(\frac{c_{\mathrm{max}}}{c_{\mathrm{min}}}\right)^{q + \sum_{l}\mu_{l}}\frac{\left(q - 1 + \sum\limits_{l \geq 1}\mu_{l}\right)!}{\left(q - 1\right)!\prod\limits_{l \geq 1}\mu_{l}!l!^{\mu_{l}}}\nonumber\\
    & = e^{qz}\sum_{m \geq 0}\left(\frac{c_{\mathrm{max}}}{c_{\mathrm{min}}}\right)^{q + m}\frac{\left(q - 1 + m\right)!}{\left(q - 1\right)!}\sum_{\substack{\left(\mu_{l}\right)_{l \geq 1}\\\sum_{l}\mu_{l} = m}}z^{\sum_{l}l\mu_{l}}\prod_{l \geq 1}\frac{1}{\mu_{l}!l!^{\mu_{l}}}\nonumber\\
    & = e^{qz}\sum_{m \geq 0}\left(\frac{c_{\mathrm{max}}}{c_{\mathrm{min}}}\right)^{q + m}\frac{\left(q - 1 + m\right)!}{\left(q - 1\right)!}\frac{1}{m!}\left(\sum_{l \geq 1}\frac{z^{l}}{l!}\right)^m\nonumber\\
    & = e^{qz}\sum_{m \geq 0}\left(\frac{c_{\mathrm{max}}}{c_{\mathrm{min}}}\right)^{q + m}\frac{\left(q - 1 + m\right)!}{\left(q - 1\right)!}\frac{1}{m!}\left(e^z - 1\right)^m\nonumber\\
    & = \left(\frac{c_{\mathrm{max}}}{c_{\mathrm{min}}}\right)^qe^{qz}\left(1 - \frac{c_{\mathrm{max}}}{c_{\mathrm{min}}}\left(e^z - 1\right)\right)^{-q}
\end{align}

We can now bound the coefficient of $z^d$ in the Taylor series of $f(z)$ thanks to Cauchy's inequality:
\begin{align}
    \left[f(z)\right]_{z^d} & \leq r^{-d}\sum_{z \in \mathbf{C}\,:\,|z| = r}|f(z)|
\end{align}
holding for arbitrary $r > 0$. In particular, choosing:
\begin{align}
    r & := \log\left(1 + \frac{c_{\mathrm{min}}}{2c_{\mathrm{max}}}\right)
\end{align}
The inequality gives:
\begin{align}
    \left[f(z)\right]_{z^d} & \leq \log\left(1 + \frac{c_{\mathrm{min}}}{2c_{\mathrm{max}}}\right)^{-d}\left(1 + \frac{c_{\mathrm{min}}}{2c_{\mathrm{max}}}\right)^q2^q\nonumber\\
    & \leq c^{q + d},\label{eq:t_block_entry_non_lambda_contribution_bound}
\end{align}
where
\begin{align}
    c & := \max\left\{\log\left(1 + \frac{c_{\mathrm{min}}}{2c_{\mathrm{max}}}\right)^{-1}, 2\left(1 + \frac{c_{\mathrm{min}}}{c_{\mathrm{max}}}\right)\right\}.\label{eq:c_constant_definition}
\end{align}
All in all, we proved the following entrywise bound on matrix blocks $\bm{T}_{q,\,d}$:

\begin{proposition}[Entrywise bound of $\bm{T}_{q,d}$]
\label{prop:t_block_entrywise_bound}
Assume the noninteracting correlations are bounded by $c_{\mathrm{min}}, c_{\mathrm{max}}$ according to assumption \ref{assp:gamma_zero_correlations_boundedness}. Then there exists a constant $c > 0$ such that the following bound holds on all coefficients of $\bm{T}_{q, d}$:
\begin{align}
    \left|\left[\bm{T}_{q, d}\right]_{\left(\alpha_{d + 1},\,\ldots,\,\alpha_{d + q}\right),\,\left(\alpha_1,\,\ldots,\,\alpha_d\right)}\right| & \leq |\lambda|^{q + d}c^{q + d}.
\end{align}
$c$ depends only on $c_{\mathrm{min}}, c_{\mathrm{max}}$ introduced in assumptions \ref{assp:gamma_zero_correlations_boundedness}.
\end{proposition}

From there, we can deduce the convergence of the order-by-order $\lambda$ expansion of the saddle point for sufficiently small $\lambda$.

\begin{repproposition}{propSaddlePointEquationAnalyticity}[Analyticity of candidate solution to saddle-point equation]
Assume correlation tensors to satisfy bound in assumption \ref{assp:gamma_zero_correlations_boundedness}, depending on constants $c_{\mathrm{min}}, c_{\mathrm{max}}$. Consider the non-rigorous saddle-point ``solution" proposed in equation \ref{eq:fixed_point_solution}:
\begin{align}
    \bm{\Theta}^* & = \left(\bm{I} - \bm{T}\right)^{-1}\overline{\bm{\Theta}^*}\nonumber\\
    & = \sum_{m \geq 0}\bm{T}^m\overline{\bm{\Theta}^*},\label{eq:saddle_point_solution_proposition}
\end{align}
where
\begin{align}
    \bm{\Theta^*} & = \begin{pmatrix}
        \bm{\theta}^*\\
        \bm{\theta}^{*\otimes 2}\\
        \bm{\theta}^{*\otimes 2}\\
        \vdots
    \end{pmatrix}
\end{align}
is the block vector of tensor powers of the sought saddle point $\bm{\theta}^*$ and 
\begin{align}
    \overline{\bm{\Theta}^*} & = \begin{pmatrix}
        \lambda\overline{\bm{C}}^{(1)}\\
        \lambda^2\overline{\bm{C}}^{(1)\otimes 2}\\
        \lambda^3\overline{\bm{C}}^{(1) \otimes 3}\\
        \vdots
    \end{pmatrix}
\end{align}
is similarly the block vector of tensor powers of the noninteracting order 1 correlations. Also, recall solution \ref{eq:saddle_point_solution_proposition} is redundant in the sense the left-hand collects the tensor powers of a fixed vector $\bm{\theta}^*$. It is sufficient to consider the unit tensor power, i.e. the first block of this vector to obtain an expression for $\bm{\theta}^*$:
\begin{align}
    \bm{\theta}^* & = \sum_{m \geq 0}\left[\bm{T}^m\overline{\bm{\Theta}^*}\right]_1.\label{eq:saddle_point_solution_proposition_only_first_block}
\end{align}
Note the normed space in which these vectors lived and in which we hoped to solve equation \ref{eq:saddle_point_equation} was not defined.  In this context, solution \ref{eq:saddle_point_solution_proposition} is well-defined in the following sense. First, define the contribution of $\left[\bm{T}^m\overline{\bm{\Theta}^*}\right]_1$ coming from $\bm{T}$ blocks $\bm{T}_{1,\,d^{(2)}}, \bm{T}_{d^{(2)},\,d^{(3)}}, \ldots, \bm{T}_{d^{(m - 1)},\,d^{(m)}}, \bm{T}_{d^{(m)},\,d^{(m + 1)}}$:
\begin{align}
    \bm{\theta}^{*,\,\left(d^{(2)},\,\ldots,\,d^{(m + 1)}\right)} & := 
    \left(\overrightarrow{\prod_{r = 1}^m}\bm{T}_{d^{(r)},\,d^{(r + 1)}}\right)\left(\lambda\overline{\bm{C}}^{(1)}\right)^{\otimes d^{(m + 1)}}\label{eq:saddle_point_t_power_m_and_tuple_contribution}\\
    & = \bm{T}_{1,\,d^{(2)}}\bm{T}_{d^{(2)},\,d^{(3)}}\bm{T}_{d^{(3)},\,d^{(4)}}\ldots \bm{T}_{d^{(m - 1)},\,d^{(m)}}\bm{T}_{d^{(m)},\,d^{(m + 1)}}\left(\lambda\overline{\bm{C}}^{(1)}\right)^{\otimes d^{(m + 1)}},
\end{align}
where we set by convention $d^{(1)} := 1$. For $m = 0$, the equation should then be read (empty product) as:
\begin{align}
    \bm{\theta}^{*,\,\varnothing} := \left(\lambda\overline{\bm{C}}^{(1)}\right)^{\otimes d^{(1)}} = \lambda\overline{\bm{C}}^{(1)},
\end{align}
coinciding with the heuristic expression for $\left[\bm{T}^0\overline{\bm{\Theta}}\right]_1$. Recall this heuristic expression assumed vector of tensor powers $\overline{\bm{\Theta}}$ and operator $\bm{T}$ live and act on well-defined spaces. However, equation \ref{eq:saddle_point_t_power_m_and_tuple_contribution}, for any $m \geq 0$, and choice of tuple $\left(d^{(2)},\,\ldots,\,d^{(m + 1)}\right)$ does not pose such issues as a product of finite-dimensional matrices and vectors. Then, series
\begin{align}
    \sum_{m \geq 1}\sum_{d^{(2)},\,\ldots,\,d^{(m + 1)}}\bm{\theta}^{*,\,\left(d^{(2)},\,\ldots,\,d^{(m + 1)}\right)}
\end{align}
is absolutely convergent in the infinite norm, defining a parametrized vector of $\mathbf{C}^{\mathcal{A}}$ in $\lambda$
\begin{align}
    \bm{\theta}^* := \bm{\theta}^*\left(\lambda\right) = \sum_{m \geq 0}\bm{\theta}^{*,\,m},\label{eq:saddle_point_series}
\end{align}
with 
\begin{align}
    \bm{\theta}^{*,\,m} & := \sum_{d^{(2)},\,\ldots,\,d^{(m + 1)} \geq 1}\bm{\theta}^{*,\,\left(d^{(2)},\,\ldots,\,d^{(m + 1)}\right)}, && m \geq 1,\label{eq:saddle_point_t_power_m_contribution}\\
    \bm{\theta}^{*,\,0} & := \bm{\theta}^{*,\,\varnothing}.
\end{align}
More specifically, absolute convergence and analyticity of this function of $\lambda$ hold in $\lambda$ domain:
\begin{align}
    |\lambda|^2|\mathcal{A}| \leq \max\left(\frac{1}{2c^2}, \frac{c_{\mathrm{min}}}{2c_{\mathrm{max}}}\right)\label{eq:lambda_analyticity_domain}
\end{align}
where $c > 0$ is the constant introduced in proposition \ref{prop:t_block_entrywise_bound}, depending only on $c_{\mathrm{min}}, c_{\mathrm{max}}$ introduced in assumption \ref{assp:gamma_zero_correlations_boundedness}; hence, the upper-bound on $|\lambda|^2|\mathcal{A}|$ also only depends on $c_{\mathrm{min}}, c_{\mathrm{max}}$. Besides, $\bm{\theta}^*\left(\lambda\right)$, seen as a function of $\lambda$, is analytic on this domain and bounded as follows:
\begin{align}
    \left\lVert \bm{\theta}^*\left(\lambda\right) \right\rVert_{\infty} & \leq \frac{2c_{\mathrm{max}}}{c_{\mathrm{min}}}|\lambda|.\label{eq:thetas_star_bound}
\end{align}
\begin{proof}
We are interested in the absolute convergence ---in norm $\left\lVert \cdot \right\rVert_{\infty}$--- of series
\begin{align}
    \sum_{m \geq 1}\sum_{\substack{d^{(2)},\,\ldots,\,d^{(m + 1)} \geq 1}}\bm{\theta}^{*,\,\left(d^{(2)},\,\ldots,\,d^{(m + 1)}\right)}.
\end{align}
Let us bound the norm of a single term of the series. Recalling $\bm{T}_{q,\,d}$ is a matrix mapping from $\left(\mathbf{C}^{\mathcal{A}}\right)^{\otimes d}$ to $\left(\mathbf{C}^{\mathcal{A}}\right)^{\otimes q}$, and using the entrywise bound on this matrix from proposition \ref{prop:t_block_entrywise_bound}:
\begin{align}
    \left\lVert \bm{\theta}^{*,\,\left(d^{(2)},\,\ldots,\,d^{(m + 1)}\right) }\right\rVert_{\infty} & = \left\lVert \bm{T}_{1,\,d^{(2)}}\bm{T}_{d^{(2)},\,d^{(3)}}\bm{T}_{d^{(3)},\,d^{(4)}}\ldots \bm{T}_{d^{(m)},\,d^{(m + 1)}}\left(\lambda \overline{\bm{C}}^{(1)}\right)^{\otimes d^{(m)}} \right\rVert_{\infty}\nonumber\\
    & \leq \left(\left|\lambda\right|c\right)^{1 + d^{(2)}}\left|\mathcal{A}\right|^{d^{(2)}}\left(\left|\lambda\right|c\right)^{d^{(2)} + d^{(3)}}\left|\mathcal{A}\right|^{d^{(3)}}\left(\left|\lambda\right|c\right)^{d^{(3)} + d^{(4)}}\left|\mathcal{A}\right|^{d^{(4)}}\nonumber\\
    & \hspace*{20px} \ldots \left(\left|\lambda\right|c\right)^{d^{(m - 1)} + d^{(m)}}|\mathcal{A}|^{d^{(m)}}\left(\left|\lambda\right|c\right)^{d^{(m)} + d^{(m + 1)}}\left|\mathcal{A}\right|^{d^{(m + 1)}}\left(|\lambda|\frac{c_{\mathrm{max}}}{c_{\mathrm{min}}}\right)^{d^{(m + 1)}}\nonumber\\
    & = |\lambda|c\left(|\lambda|^2|\mathcal{A}|c^2\right)^{d^{(2)} + d^{(3)} + \ldots + d^{(m - 1)} + d^{(m)}}\left(|\lambda|^2|\mathcal{A}|\frac{cc_{\mathrm{max}}}{c_{\mathrm{min}}}\right)^{d^{(m + 1)}}.
\end{align}
Now, the sum of norms of all series terms for fixed $m \geq 1$ is upper-bounded by:
\begin{align}
    \sum_{d^{(2)},\,\ldots,\,d^{(m + 1)} \geq 1}\left\lVert \bm{\theta}^{*,\,\left(d^{(2)},\,\ldots,\,d^{(m + 1)}\right)} \right\rVert_{\infty} & \leq \sum_{d^{(2)},\,\ldots,\,d^{(m + 1)} \geq 1}|\lambda|c\left(|\lambda|^2|\mathcal{A}|c^2\right)^{d^{(2)} + d^{(3)} + \ldots + d^{(m - 1)} + d^{(m)}}\left(|\lambda|^2|\mathcal{A}|\frac{cc_{\mathrm{max}}}{c_{\mathrm{min}}}\right)^{d^{(m + 1)}}\nonumber\\
    & \leq |\lambda|c\left(\frac{|\lambda|^2|\mathcal{A}|c^2}{1 - |\lambda|^2|\mathcal{A}|c^2}\right)^{m - 1}\frac{|\lambda|^2|\mathcal{A}|cc_{\mathrm{max}}/c_{\mathrm{min}}}{1 - |\lambda|^2|\mathcal{A}|cc_{\mathrm{max}}/c_{\mathrm{min}}}\nonumber\\
    & \leq 2^m\left(|\lambda|^2|\mathcal{A}|c^2\right)^m|\lambda|\frac{c_{\mathrm{max}}}{c_{\mathrm{min}}}\label{eq:saddle_point_series_absolute_convergence_bound_inner_sum},
\end{align}
where in the final line, we assumed
\begin{align}
    |\lambda|^2|\mathcal{A}| & \leq \max\left(\frac{1}{2c^2}, \frac{c_{\mathrm{min}}}{2cc_{\mathrm{max}}}\right).
\end{align}
Note that the right-hand-side on the bound is also an upper-bound for $\bm{\theta}^{*,\,0} = \bm{\theta}^{*,\,\varnothing}$ when setting $m = 0$, since
\begin{align}
    \left\lVert\bm{\theta}^{*,\,0}\right\rVert_{\infty} & =  \left\lVert \lambda\overline{\bm{C}}^{(1)}
 \right\rVert_{\infty}\nonumber\\
    & \leq |\lambda|\frac{c_{\mathrm{max}}}{c_{\mathrm{min}}}.
\end{align}
The desired absolute convergence then follows from the fixed $m \geq 1$ bound:
\begin{align}
    \sum_{m \geq 1}\sum_{d^{(2)},\,\ldots,\,d^{(m + 1)} \geq 1}\left\lVert \bm{\theta}^{*,\,\left(d^{(2)},\,\ldots,\,d^{(m + 1)}\right)} \right\rVert_{\infty} & \leq \sum_{m \geq 1}|\lambda|\frac{c_{\mathrm{max}}}{c_{\mathrm{min}}}\left(2|\lambda|^2|\mathcal{A}|c^2\right)^m\nonumber\\
    & \leq |\lambda|\frac{c_{\mathrm{max}}}{c_{\mathrm{min}}}\frac{2c^2|\lambda|^2|\mathcal{A}|}{1 - 2c^2|\lambda|^2|\mathcal{A}|}\nonumber\\
    & \leq |\lambda|\frac{4c^2c_{\mathrm{max}}}{c_{\mathrm{min}}}|\lambda|^2|\mathcal{A}|,
\end{align}
where we used assumption
\begin{align}
    |\lambda|^2|\mathcal{A}| & \leq \frac{1}{4c^2}.
\end{align}
From this geometric absolute convergence, it follows that
\begin{align}
    \bm{\theta}^{*,\,m}\left(\lambda\right) & := \sum_{d^{(2)},\,\ldots,\,d^{(m + 1)} \geq 1}\bm{\theta}^{*,\,\left(d^{(2)},\,\ldots,\,d^{(m + 1)}\right)}\left(\lambda\right), && m \geq 1,
\end{align}
and analytic in $\lambda$. The same statement follows for
\begin{align}
    \bm{\theta}^*\left(\lambda\right) & := \sum_{m \geq 0}\bm{\theta}^{*,\,m}\left(\lambda\right)\\
    & = \bm{\theta}^{*,\,0}\left(\lambda\right) + \sum_{m \geq 1}\bm{\theta}^{*,\,m}\left(\lambda\right)\\
    & = \lambda\overline{\bm{C}}^{(1)} + \sum_{m \geq 1}\sum_{d^{(2)},\,\ldots,\,d^{(m + 1)} \geq 1}\bm{\theta}^{*,\,\left(d^{(2)},\,\ldots,\,d^{(m + 1)}\right)}\left(\lambda\right).
\end{align}
Finally, the following bound can be stated for $\bm{\theta}^*$, recalling the remark that bound \ref{eq:saddle_point_series_absolute_convergence_bound_inner_sum} with $m := 0$ applies to $\bm{\theta}^{*,\,0}$:
\begin{align}
    \left\lVert \bm{\theta}^* \right\rVert_{\infty} & \leq \sum_{m \geq 0}|\lambda|\frac{c_{\mathrm{max}}}{c_{\mathrm{min}}}\left(2|\lambda|^2|\mathcal{A}|c^2\right)^m\nonumber\\
    & = |\lambda|\frac{c_{\mathrm{max}}}{c_{\mathrm{min}}}\frac{1}{1 - 2|\lambda|^2|\mathcal{A}|c^2}\nonumber\\
    & \leq \frac{2c_{\mathrm{max}}}{c_{\mathrm{min}}}|\lambda|,
\end{align}
where we again used
\begin{align}
    |\lambda|^2|\mathcal{A}| & \leq \frac{1}{4c^2}.
\end{align}
\end{proof}
\end{repproposition}

While proposition \ref{propSaddlePointEquationAnalyticity} shows the candidate solution $\bm{\theta}^*$ is well-defined, more work will be required to prove it solves the saddle-point equation. For that purpose, we follow the heuristic intuition of ``linearizing the nonlinear equation" by considering the ``vector of tensor powers". As previously indicated, the bulk of the work is to introduce appropriate normed spaces supporting our infinite-dimensional vectors and objects of interest. The analysis then essentially relies of geometric bounds on the norms of vectors and operators, all following from the earlier more fundamental proposition \ref{prop:t_block_entrywise_bound}.

\subsubsection{Correctness of saddle point expansion}
\label{sec:correctness_saddle_point_expansion}

We now rigorously prove that $\bm{\theta}^*$ defined in equation \ref{eq:saddle_point_series}, an analytic function of $\lambda$, solves the original saddle-point equation \ref{eq:saddle_point_equation}. For this purpose, we need to make the arguments from section \ref{sec:heuristic_pqgms_expansion} rigorous by specifying the space in which the ``vectors of tensor powers"
\begin{align}
    \bm{\Theta}^* = \begin{pmatrix}
        \bm{\theta}^*\\
        \bm{\theta}^{*\otimes 2}\\
        \bm{\theta}^{*\otimes 3}\\
        \vdots
    \end{pmatrix}, && \overline{\bm{\Theta}^*} = \begin{pmatrix}
        \lambda\overline{\bm{C}}^{(1)}\\
        \lambda^2\overline{\bm{C}}^{(1)\otimes 2}\\
        \lambda^3\overline{\bm{C}}^{(1)\otimes 3}\\
        \vdots
    \end{pmatrix}
\end{align}
live. It will be convenient to choose this space as a variant of $\ell^1\left(\mathbf{C}\right)$ ---the space of summable complex number sequences:

\begin{definition}[Space for ``vector of tensor powers"]
\label{def:tensor_powers_vector_space}
Let $c$ be the constant introduced in proposition \ref{prop:t_block_entrywise_bound}, satisfying:
\begin{align}
    \left|\left[\bm{T}_{q,\,d}\right]_{\bm{\alpha'},\,\bm{\alpha''}}\right| & \leq |\lambda|^{q + d}c^{q + d} \qquad \forall q, d \geq 1, \quad \forall \bm{\alpha'}, \bm{\alpha''} \in \mathcal{A}^d.
\end{align}
We then define by $\mathcal{L}$ as the following normed vector space. Vectors $\bm{v}$ of the space are complex number sequences, which may be organized in blocks similar to the informal definition of the ``vectors of tensor powers" used until now:
\begin{align}
    \bm{v} & = \begin{pmatrix}
        \bm{v}^{(1)}\\
        \bm{v}^{(2)}\\
        \bm{v}^{(3)}\\
        \vdots
    \end{pmatrix},\\
    \bm{v}^{(d)} & \in \left(\mathbf{C}^{\mathcal{A}}\right)^{\otimes d} \simeq \mathbf{C}^{|\mathcal{A}|^d}.
\end{align}
The normed vector space $\left(\mathcal{L}, \left\lVert \cdot \right\rVert'\right)$ is defined as the space of sequences $\bm{v}$ satisfying:
\begin{align}
    \left\lVert \bm{v} \right\rVert'_1 & := \sum_{d \geq 1}\left(|\lambda|c\right)^d\left\lVert \bm{v}^{(d)} \right\rVert_1 < \infty.
\end{align}
The $\left\lVert \cdot \right\rVert_1$ symbol on the right-hand side refers to the standard $1$-norm, i.e.
\begin{align}
    \left\lVert \bm{v}^{(d)} \right\rVert_1 & = \sum_{\bm{\alpha} \in \mathcal{A}^d}\left|v^{(d)}_{\bm{\alpha}}\right|.
\end{align}
The completeness of $\left(\mathcal{L}, \left\lVert \cdot \right\rVert'_1\right)$ follows from that of $l^1\left(\mathbf{C}\right)$.
\end{definition}

Space $\mathcal{L}$ introduced in definition \ref{def:tensor_powers_vector_space} is precisely where the ``vectors of tensor powers" $\overline{\bm{\Theta}^*}$ and $\overline{\bm{\Theta}}$, nonrigorously introduced in section \ref{sec:heuristic_pqgms_expansion}, will live. We start with the following lemma establishing a criterion for ``vectors of tensor powers" to effectively belong to $\mathcal{L}$:

\begin{lemma}[Criterion for well-definition of vector of tensor powers]
\label{lemma:vector_tensor_powers_well_defined_criterion}
Let $\bm{\psi} = \left(\psi_{\alpha \in \mathcal{A}}\right) \in \mathbf{C}^{\mathcal{A}}$ a (finite-dimensional) vector indexed by $\mathcal{A}$. Assume
\begin{align}
    \left\lVert \bm{\psi} \right\rVert_1 & \leq |\lambda||\mathcal{A}|c',
\end{align}
where $c'$ is a constant which may only depend on $c_{\mathrm{min}}, c_{\mathrm{max}}$. Let $\delta > 0$ be a free parameter. Then, for
\begin{align}
    |\lambda|^2|\mathcal{A}| \leq \frac{\delta}{\left(1 + \delta\right)cc'}
\end{align}
where the right-hand side only depends on $c', \delta, c_{\mathrm{min}}, c_{\mathrm{max}}$, the ``vector of tensor powers" of $\bm{\psi}$ belongs to $\mathcal{L}$:
\begin{align}
    \bm{\Psi} & := \begin{pmatrix}
        \bm{\psi}\\
        \bm{\psi}^{\otimes 2}\\
        \bm{\psi}^{\otimes 3}\\
        \vdots
    \end{pmatrix} \in \mathcal{L}.
\end{align}
Besides, the norm of this $\mathcal{L}$ vector is related to the $1$-norm of the original vector as follows:
\begin{align}
    \left\lVert \bm{\Psi} \right\rVert_1' & \leq \left(1 + \delta\right)|\lambda|c\left\lVert \bm{\psi} \right\rVert_1.
\end{align}
\begin{proof}
We start by checking the sequence of complex numbers
\begin{align}
    \bm{\Psi} & = \begin{pmatrix}
        \bm{\psi}\\
        \bm{\psi}^{\otimes 2}\\
        \bm{\psi}^{\otimes 3}\\
        \vdots
    \end{pmatrix}
\end{align}
indeed defines a vector from $\mathcal{L}$. We estimate:
\begin{align}
    \left\lVert \bm{\Psi} \right\rVert'_1 & = \sum_{d \geq 1}\left(|\lambda|c\right)^d\left\lVert \bm{\psi}^{\otimes d} \right\rVert_1\nonumber\\
    & = \sum_{d \geq 1}\left(|\lambda|c\right)^d\left\lVert \bm{\psi} \right\rVert_1^d\nonumber\\
    & = \frac{|\lambda|c\left\lVert \bm{\psi}\right\rVert_1}{1 - |\lambda|c\left\lVert \bm{\psi} \right\rVert_1}\nonumber\\
    & = \frac{|\lambda|c\left\lVert \bm{\psi}\right\rVert_1}{1 - |\lambda|^2|\mathcal{A}|cc'}\nonumber\\
    & \leq \left(1 + \delta\right)|\lambda|c\left\lVert \bm{\psi} \right\rVert_1\nonumber\\
    & < \infty.
\end{align}
as long as
\begin{align}
    |\lambda|^2|\mathcal{A}| & \leq \frac{\delta}{\left(1 + \delta\right)cc'},
\end{align}
where the left-hand-side only depends on $\delta, c', c_{\mathrm{min}}, c_{\mathrm{max}}$. These inequalities prove the lemma.
\end{proof}
\end{lemma}

Lemma \ref{lemma:vector_tensor_powers_well_defined_criterion} can be applied to the ``vector of tensor powers" of the lowest-order solution to the saddle-point equation, i.e. the ``vector of tensor powers" of $\overline{\bm{\theta}^*} = \lambda\overline{\bm{C}}^{(1)}$:
\begin{corollary}[Vectors of tensor powers is well-defined for lowest-order solution to saddle-point equation]
\label{corollary:gamma_0_correlations_tensor_powers_vector}
Consider the vector of lowest-order solution to the saddle-point equation:
\begin{align}
    \overline{\bm{\theta}^*} & := \lambda\overline{\bm{C}}^{(1)}.
\end{align}
Then, for $|\lambda|^2|\mathcal{A}|$ bounded by a constant depending only on $c_{\mathrm{min}}, c_{\mathrm{max}}$ introduced in assumption \ref{assp:gamma_zero_correlations_boundedness}, the ``vector of tensor powers" built from this vector belongs to $\mathcal{A}$:
\begin{align}
    \overline{\bm{\Theta}^*} & := \begin{pmatrix}
        \overline{\bm{\theta}^*}\\
        \overline{\bm{\theta}^*}^{\otimes 2}\\
        \overline{\bm{\theta}^*}^{\otimes 3}\\
        \vdots
    \end{pmatrix} \in \mathcal{A}.
\end{align}
\end{corollary}

Now we defined an appropriate space for the ``vector of tensor powers", we have to check that $\bm{T}$, specified by an infinite matrix, is well-defined on this space and maps into it. This is proven in the following lemma:

\begin{lemma}[Boundedness of $\bm{T}$]
\label{lemma:t_boundedness}
Consider $\bm{T}$, defined by an infinite block matrix according to equation \ref{eq:saddle_point_equation_tensor_interpretation_t}:
\begin{align}
    \bm{T} & = \begin{pmatrix}
        \bm{T}_{1,\,1} & \bm{T}_{1,\,2} & \bm{T}_{1,\,3} & \ldots\\
        \bm{T}_{2,\,1} & \bm{T}_{2,\,2} & \bm{T}_{2,\,3} & \ldots\\
        \bm{T}_{3,\,1} & \bm{T}_{3,\,2} & \bm{T}_{3,\,3} & \ldots\\
        \vdots & \vdots & \vdots & \ddots
    \end{pmatrix},
\end{align}
with explicit formula for block $(q, d)$ given by equation \ref{eq:saddle_point_equation_tensor_interpretation_t_block}. Then, for $|\lambda|^2|\mathcal{A}|$ upper-bounded by a constant depending only on $c_{\mathrm{min}}, c_{\mathrm{max}}$ introduced in assumption \ref{assp:gamma_zero_correlations_boundedness}, $\bm{T}$ is a well-defined operator mapping $\mathcal{L}$ into itself; besides, its operator norm is bounded by:
\begin{align}
    \left\lVert \bm{T} \right\rVert_{\mathcal{L} \to \mathcal{L}} & \leq 2|\lambda|^2c^2|\mathcal{A}|,
\end{align}
where constant $c$ depends only on $c_{\mathrm{min}}, c_{\mathrm{max}}$ and relates to the definition of $\mathcal{L}$ (definition \ref{def:tensor_powers_vector_space}).
\begin{proof}
We first check that $\bm{T}$ is well-defined on $\mathcal{L}$. Let then
\begin{align}
    \bm{v} & = \begin{pmatrix}
        \bm{v}^{(1)}\\
        \bm{v}^{(2)}\\
        \bm{v}^{(3)}\\
        \vdots
    \end{pmatrix} \in \mathcal{L}.
\end{align}
We show the series
\begin{align}
    \bm{w}^{(q)} := \sum_{d \geq 1}\bm{T}_{q,\,d}\bm{v}^{(d)}
\end{align}
is absolutely convergent for all $q \geq 1$, and the complex number sequence formed by the $\bm{w}^{(q)}$:
\begin{align}
    \bm{w} & := \begin{pmatrix}
        \bm{w}^{(1)}\\
        \bm{w}^{(2)}\\
        \bm{w}^{(3)}\\
        \vdots
    \end{pmatrix}
\end{align}
is indeed a vector of $\mathcal{L}$. The absolute convergence of the series follows from the entrywise bounds on $\bm{T}_{q, d}$ derived in proposition \ref{prop:t_block_entrywise_bound}:
\begin{align}
    \sum_{d \geq 1}\left\lVert \bm{T}_{q,\,d}\bm{v}^{(d)} \right\rVert_1 & = \sum_{\bm{\alpha} \in \mathcal{A}^q}\left|\sum_{\bm{\beta} \in \mathcal{A}^d}\left[\bm{T}_{q,\,d}\right]_{\bm{\alpha},\,\bm{\beta}}v^{(d)}_{\bm{\beta}}\right|\nonumber\\
    & \leq \sum_{d \geq 1}\sum_{\bm{\alpha} \in \mathcal{A}^q}\sum_{\bm{\beta} \in \mathcal{A}^d}\left|\left[\bm{T}_{q,\,d}\right]_{\bm{\alpha},\,\bm{\beta}}\right|\left|v^{(d)}_{\bm{\beta}}\right|\nonumber\\
    & \leq \sum_{d \geq 1}\sum_{\bm{\alpha} \in \mathcal{A}^q}\sum_{\bm{\beta} \in \mathcal{A}^d}\left(|\lambda|c\right)^{q + d}\left|v^{(d)}_{\bm{\beta}}\right|\nonumber\\
    & \leq \sum_{d \geq 1}\sum_{\bm{\alpha} \in \mathcal{A}^q}\sum_{\bm{\beta} \in \mathcal{A}^d}\left(|\lambda|c\right)^{q + d}\left|v^{(d)}_{\bm{\beta}}\right|\nonumber\\
    & = \sum_{d \geq 1}\left|\mathcal{A}\right|^q\left(|\lambda|c\right)^{q + d}\left\lVert\bm{v}^{(d)}\right\rVert_1\nonumber\\
    & = \left(|\lambda|c|\mathcal{A}|\right)^q\left\lVert \bm{v} \right\rVert_1'\nonumber\\
    & < \infty
\end{align}
as required. Hence, for all $q \geq 1$,
\begin{align}
    \bm{w}^{(q)} & := \sum_{d \geq 1}\bm{T}_{q,\,d}\bm{v}^{(d)} \in \left(\mathbf{C}^{\mathcal{A}}\right)^{\otimes q}
\end{align}
is well-defined. Let us now check the complex sequence $\bm{w} := \left(\bm{w}^{(q)}\right)_{q \geq 1}$ defined by this block lives in $\left(\mathcal{L}, \left\lVert \cdot \right\rVert_1'\right)$. We compute:
\begin{align}
    \left\lVert \bm{w} \right\rVert_1' & = \sum_{q \geq 1}\left(|\lambda|c\right)^q\left\lVert \bm{w}^{(q)} \right\rVert_1\nonumber\\
    & \leq \sum_{q \geq 1}\left(|\lambda|c\right)^q\left(|\lambda|c|\mathcal{A}|\right)^q\left\lVert \bm{v} \right\rVert_1'\\
    & = \sum_{q \geq 1}\left(|\lambda|^2c^2|\mathcal{A}|\right)^q\left\lVert \bm{v} \right\rVert_1'\nonumber\\
    & = \frac{|\lambda|^2c^2|\mathcal{A}|}{1 - |\lambda|^2c^2|\mathcal{A}|}\left\lVert \bm{v} \right\rVert_1'\\
    & < \infty.
\end{align}
This shows that complex sequence $\bm{w}$ defines a vector from $\mathcal{L}$ indeed. Hence, $\bm{T}$ is well-defined and maps $\mathcal{L}$ into itself. Besides, the above inequalities show
\begin{align}
    \left\lVert \bm{T} \right\rVert_{\mathcal{L} \to \mathcal{L}} & \leq \frac{|\lambda|^2c^2|\mathcal{A}|}{1 - |\lambda|^2c^2|\mathcal{A}|}\nonumber\\
    & \leq 2|\lambda|^2c^2|\mathcal{A}|,
\end{align}
assuming for instance
\begin{align}
    |\lambda|^2|\mathcal{A}| & \leq \frac{1}{2c^2},
\end{align}
where the right-hand side is a constant depending only on $c_{\mathrm{min}}, c_{\mathrm{max}}$ from assumption \ref{assp:gamma_zero_correlations_boundedness} as required.
\end{proof}
\end{lemma}

To rephrase the saddle-point equation as as linear equation over vectors in $\mathcal{L}$, we will need the following rigorous generalization of the calculation in equation \ref{eq:tensor_powers_vector}:

\begin{lemma}[Right-hand side of saddle-point equation in terms of $\bm{T}$ operator]
\label{lemma:saddle_point_equation_rhs_generalized_series}
Let $\bm{\psi} = \left(\psi_{\alpha}\right)_{\alpha \in \mathcal{A}} \in \mathbf{C}^{\mathcal{A}}$ a vector bounded as follows in the $1$-norm:
\begin{align}
    \left\lVert \bm{\psi} \right\rVert_1 & \leq |\lambda||\mathcal{A}|c',
\end{align}
where constant $c'$ may only depend on $c_{\mathrm{min}}, c_{\mathrm{max}}$ introduced in assumption \ref{assp:gamma_zero_correlations_boundedness}. In particular, this holds if
\begin{align}
    \left\lVert \bm{\psi} \right\rVert_{\infty} & \leq |\lambda|c'.
\end{align}
Then, for $|\lambda|$ in domain:
\begin{align}
    |\lambda|^2|\mathcal{A}| < \max\left\{\frac{\log(2)}{c'c_{\mathrm{max}}}, \frac{1}{4c'c_{\mathrm{max}}^2}, \frac{1}{c'}\log\left(1 + \frac{c_{\mathrm{min}}}{2c_{\mathrm{max}}}\right), \frac{1}{2cc'}\right\},
\end{align}
the denominator of the right-hand-side of the saddle-point equation:
\begin{align}
    \frac{\sum\limits_{\bm{a} \in \mathcal{S}}Q_{\bm{a}}\exp\left(\lambda\bm{\psi}^T\bm{L}_{:,\,\bm{a}}\right)\lambda\bm{L}_{:,\,\bm{a}}}{\sum\limits_{\bm{a} \in \mathcal{S}}Q_{\bm{a}}\exp\left(\lambda\bm{\psi}^T\bm{L}_{:,\,\bm{a}}\right)} \label{eq:saddle_point_equation_rhs_arbitrary_vector}
\end{align}
is bounded away from zero, implying analyticity in $\lambda$ of this function in equation \ref{eq:saddle_point_equation_rhs_arbitrary_vector}. Besides, for all integer $q \geq 1$, the following equality holds:
\begin{align}
    \left(\frac{\sum\limits_{\bm{a} \in \mathcal{S}}Q_{\bm{a}}\exp\left(\lambda\bm{\psi}^{T}\bm{L}_{:,\,\bm{a}}\right)\lambda\bm{L}_{:,\,\bm{a}}}{\sum\limits_{\bm{a} \in \mathcal{S}}Q_{\bm{a}}\exp\left(\lambda\bm{\psi}^{T}\bm{L}_{:,\,\bm{a}}\right)}\right)^{\otimes q} & = \sum_{\substack{\left(\mu_l\right)_{l \geq 1}\\l_1,\,\ldots,\,l_q}}\lambda^{q + l_1 + \ldots + l_q + \sum_{l}l\mu_l}\frac{\left(q - 1 + \sum\limits_{l \geq 1}\mu_l\right)!}{\left(q - 1\right)!}\frac{(-1)^{\sum\limits_{l \geq 1}\mu_l}}{l_1!\ldots l_q!\prod\limits_{l \geq 1}\mu_l!l!^{\mu_l}}\nonumber\\
    & \hspace*{50px} \times \left\langle \bm{\psi}^{\otimes\left(\sum_ll\mu_l + l_1 + \ldots + l_q\right)}, \bigotimes_{l \geq 1}\overline{\bm{C}}^{(l)\otimes \mu_l} \otimes \overline{\bm{C}}^{\left(l_1 + 1\right)} \otimes \ldots \otimes \overline{\bm{C}}^{\left(l_q + 1\right)}                      \right\rangle\label{eq:saddle_point_equation_rhs_power_q_expansion_repeated},
\end{align}
where the series on the right-hand-side is absolutely convergent. Besides, defining vector:
\begin{align}
    \bm{\chi} & := \frac{\sum\limits_{\bm{a} \in \mathcal{S}}Q_{\bm{a}}\exp\left(\bm{\psi}^T\bm{L}_{:,\,\bm{a}}\right)\lambda\bm{L}_{:,\,\bm{a}}}{\sum\limits_{\bm{a} \in \mathcal{S}}Q_{\bm{a}}\exp\left(\bm{\psi}^T\bm{L}_{:,\,\bm{a}}\right)},
\end{align}
the ``vectors of tensors powers" built from $\bm{\psi}$ and $\bm{\chi}$ are elements of $\mathcal{L}$:
\begin{align}
    \bm{\Psi} := \begin{pmatrix}
        \bm{\psi}\\
        \bm{\psi}^{\otimes 2}\\
        \bm{\psi}^{\otimes 3}\\
        \vdots
    \end{pmatrix} \in \mathcal{L}, && \bm{X} := \begin{pmatrix}
        \bm{\chi}\\
        \bm{\chi}^{\otimes 2}\\
        \bm{\chi}^{\otimes 3}\\
        \vdots
    \end{pmatrix} \in \mathcal{L}.
\end{align}
These vectors are further related by the following linear equation in $\mathcal{L}$:
\begin{align}
    \bm{X} & = \overline{\bm{\Theta}^*} + \bm{T}\bm{\Psi},
\end{align}
where $\overline{\bm{\Theta}^*}$, the ``vector of tensor powers" of noninteracting correlations, was defined in corollary \ref{corollary:gamma_0_correlations_tensor_powers_vector} and shown to belong to $\mathcal{L}$ there.
\begin{proof}
We wish to make rigorous the Taylor series expansion in $\lambda$ carried out in equation \ref{eq:tensor_powers_vector}. For that purpose, it is sufficient to show analyticity of the function to expand:
\begin{align}
    \left(\frac{\sum\limits_{\bm{a} \in \mathcal{S}}Q_{\bm{a}}\exp\left(\lambda\bm{\psi}^{T}\bm{L}_{:,\,\bm{a}}\right)\lambda\bm{L}_{:,\,\bm{a}}}{\sum\limits_{\bm{a} \in \mathcal{S}}Q_{\bm{a}}\exp\left(\lambda\bm{\psi}^{T}\bm{L}_{:,\,\bm{a}}\right)}\right)^{\otimes q}
\end{align}
in a domain of the form:
\begin{align}
    |\lambda|^2|\mathcal{A}| < \textrm{a constant depending only on $c_{\mathrm{min}}$ and $c_{\mathrm{max}}$}
\end{align}
[Note that in this statement, $\bm{\psi}$ is regarded as an arbitrary vector and not as a function of $\lambda$.] To show analyticity, it is in turn sufficient to show nonvanishing of the denominator in such a domain. We then bound:
\begin{align}
    \left|\sum_{\bm{a} \in \mathcal{S}}Q_{\bm{a}}\exp\left(\lambda\bm{\psi}^T\bm{L}_{:,\,\bm{a}}\right)\right| & = \left|\sum_{\bm{a} \in \mathcal{S}}Q_{\bm{a}}\sum_{d \geq 1}\frac{1}{d!}\left(\lambda\bm{\psi}^T\bm{L}_{:,\,\bm{a}}\right)^d\right|\nonumber\\
    & = \left|\sum_{d \geq 0}\frac{\lambda^d}{d!}\left\langle  \bm{\psi}^{\otimes d}, \bm{L}_{:,\,\bm{a}}^{\otimes d}\right\rangle\right|\nonumber\\
    & = \left|1 + \sum_{d \geq 1}\frac{\lambda^d}{d!}\left\langle  \bm{\psi}^{\otimes d}, \bm{L}_{:,\,\bm{a}}^{\otimes d}\right\rangle\right|\nonumber\\
    & = \left|1 + \sum_{d \geq 1}\frac{\lambda^d}{d!}\overline{\mathcal{Z}^*}\left\langle \bm{\psi}^{\otimes d}, \overline{\bm{C}}^{(d)}\right\rangle\right|\nonumber\\
    & \geq 1 - \sum_{d \geq 1}\frac{|\lambda|^d}{d!}\left|\overline{\mathcal{Z}^*}\right|\left\lVert \bm{\psi}^{\otimes d} \right\rVert_1\left\lVert \overline{\bm{C}}^{(d)} \right\rVert_{\infty}\nonumber\\
    & = 1 - \sum_{d \geq 1}\frac{|\lambda|^d}{d!}\left|\overline{\mathcal{Z}^*}\right|\left\lVert \bm{\psi} \right\rVert_1^d\left\lVert \overline{\bm{C}}^{(d)} \right\rVert_{\infty}\nonumber\\
    & \geq 1 - \sum_{d \geq 1}\frac{|\lambda|^d}{d!}c_{\mathrm{max}}^d\left(|\lambda||\mathcal{A}|c'\right)^dc_{\mathrm{max}}\nonumber\\
    & \geq 1 - c_{\mathrm{max}}|\lambda|^2|\mathcal{A}|c'c_{\mathrm{max}}\exp\left(|\lambda|^2|\mathcal{A}|c'c_{\mathrm{max}}\right)\nonumber\\
    & = 1 - 2c_{\mathrm{max}}|\lambda|^2|\mathcal{A}|c'c_{\mathrm{max}}\nonumber\\
    & \geq \frac{1}{2},
\end{align}
where in the final two lines, we use inequalities
\begin{align}
    |\lambda|^2|\mathcal{A}| & \leq \frac{\log(2)}{c'c_{\mathrm{max}}}
\end{align}
and
\begin{align}
    |\lambda|^2|\mathcal{A}| & \leq \frac{1}{4c'c_{\mathrm{max}}^2}.
\end{align}
Hence, for $|\lambda|^2|\mathcal{A}|$ in a neighbourhood of zero depending only $c', c_{\mathrm{min}}, c_{\mathrm{max}}$, the denominator is bounded away from zero. From there, analyticity in $\lambda$ on this domain follows, hence the correctness of the Taylor expansion in equation \ref{eq:saddle_point_equation_rhs_power_q_expansion_repeated}:
\begin{align}
    \left(\frac{\sum\limits_{\bm{a} \in \mathcal{S}}Q_{\bm{a}}\exp\left(\lambda\bm{\psi}^{T}\bm{L}_{:,\,\bm{a}}\right)\lambda\bm{L}_{:,\,\bm{a}}}{\sum\limits_{\bm{a} \in \mathcal{S}}Q_{\bm{a}}\exp\left(\lambda\bm{\psi}^{T}\bm{L}_{:,\,\bm{a}}\right)}\right)^{\otimes q} & = \sum_{\substack{\left(\mu_l\right)_{l \geq 1}\\l_1,\,\ldots,\,l_q}}\lambda^{q + l_1 + \ldots + l_q + \sum_{l}l\mu_l}\frac{\left(q - 1 + \sum\limits_{l \geq 1}\mu_l\right)!}{\left(q - 1\right)!}\frac{(-1)^{\sum\limits_{l \geq 1}\mu_l}}{l_1!\ldots l_q!\prod\limits_{l \geq 1}\mu_l!l!^{\mu_l}}\nonumber\\
    & \hspace*{50px} \times \left\langle \bm{\psi}^{\otimes\left(l_1 + \ldots + l_q + \sum_{l}l\mu_l\right)}, \bigotimes_{l \geq 1}\overline{\bm{C}}^{(l)\otimes \mu_l} \otimes \overline{\bm{C}}^{\left(l_1 + 1\right)} \otimes \ldots \otimes \overline{\bm{C}}^{\left(l_q + 1\right)} \right\rangle.
\end{align}
Absolute convergence of the above series results from bound:
\begin{align}
    & \left|\left\langle \bm{\psi}^{\otimes\left(\sum_{l}l\mu_l + l_1 + \ldots + l_q\right)}, \bigotimes_{l \geq 1}\overline{\bm{C}}^{(l)\otimes \mu_l} \otimes \overline{\bm{C}}^{\left(l_1 + 1\right)} \otimes \ldots \otimes \overline{\bm{C}}^{\left(l_q + 1\right)}\right\rangle\right|\nonumber\\
    & \leq \left|\mathcal{A}\right|^{\sum_ll\mu_l + l_1 + \ldots + l_q}\left\lVert \bm{\psi}^{\otimes \left(\sum_ll\mu_l + l_1 + \ldots + l_q\right)} \right\rVert_{\infty}\left\lVert \bigotimes_{l \geq 1}\overline{\bm{C}}^{(l)\otimes \mu_l} \otimes \overline{\bm{C}}^{\left(l_1 + 1\right)} \otimes \ldots \otimes \overline{\bm{C}}^{\left(l_q + 1\right)} \right\rVert_{\infty}\nonumber\\
    & \leq \left(|\lambda|\left|\mathcal{A}\right|c'\right)^{\sum_ll\mu_l + l_1 + \ldots + l_q}c_{\mathrm{max}}^{q + \sum_l\mu_l}
\end{align}
and a very similar calculation to the one leading to proposition \ref{prop:t_block_entrywise_bound}. Namely
\begin{align}
    & \sum_{\substack{\left(\mu_l\right)_{l \geq 1}\\l_1,\,\ldots,\,l_q \geq 0}}\left|\lambda^{q + l_1 + \ldots + l_q + \sum_{l}l\mu_l}\frac{\left(q - 1 + \sum\limits_{l \geq 1}\mu_l\right)!}{\left(q - 1\right)!}\frac{(-1)^{\sum\limits_{l \geq 1}\mu_l}}{l_1!\ldots l_q!\prod\limits_{l \geq 1}\mu_l!l!^{\mu_l}}\right.\nonumber\\
    & \left. \hspace*{60px} \times \left\langle \bm{\psi}^{\otimes\left(\sum_{l}l\mu_l + l_1 + \ldots + l_q\right)}, \bigotimes_{l \geq 1}\overline{\bm{C}}^{(l)\otimes \mu_l} \otimes \overline{\bm{C}}^{\left(l_1 + 1\right)} \otimes \ldots \otimes \overline{\bm{C}}^{\left(l_q + 1\right)} \right\rangle \right|\nonumber\\
    & \leq \sum_{\substack{\left(\mu_l\right)_{l \geq 1}\\l_1,\,\ldots,\,l_q \geq 0}}|\lambda|^{q + l_1 + \ldots + l_q + \sum_ll\mu_l}\frac{\left(q - 1 + \sum\limits_{l \geq 1}\mu_l\right)!}{(q - 1)!}\frac{1}{l_1!\ldots l_q!\prod\limits_{l \geq 1}\mu_l!l!^{\mu_l}}\nonumber\\
    & \hspace*{60px} \times \left(|\lambda||\mathcal{A}|c'\right)^{\sum_ll\mu_l + l_1 + \ldots + l_q}\left(\frac{c_{\mathrm{max}}}{c_{\mathrm{min}}}\right)^{q + \sum_l\mu_l}\nonumber\\
    & = \left(\frac{c_{\mathrm{min}}}{c_{\mathrm{max}}}|\lambda|\right)^q\exp\left(q|\lambda|^2|\mathcal{A}|c'\right)\left(1 - \frac{c_{\mathrm{max}}}{c_{\mathrm{min}}}\left(\exp\left(|\lambda|^2|A|c'\right) - 1\right)\right)^{-q}\nonumber\\
    & \leq \left(\frac{c_{\mathrm{min}}}{c_{\mathrm{max}}}|\lambda|\right)^q\left(1 + \frac{c_{\mathrm{min}}}{2c_{\mathrm{max}}}\right)^q2^q\nonumber\\
    & < \infty\label{eq:saddle_point_equation_rhs_generalized_absolute_series}
\end{align}
as long as (for instance):
\begin{align}
    |\lambda|^2|\mathcal{A}| \leq \frac{1}{c'}\log\left(1 + \frac{c_{\mathrm{min}}}{2c_{\mathrm{max}}}\right),
\end{align}
a bound depending only on $c', c_{\mathrm{min}}, c_{\mathrm{max}}$ as required. By the absolute convergence result just proven, the ``vector of tensor powers" of $\bm{\chi}$, where
\begin{align}
    \bm{\chi} & := \frac{\sum\limits_{\bm{a} \in \mathcal{S}}Q_{\bm{a}}\exp\left(\lambda\bm{\psi}^T\bm{L}_{:,\,\bm{a}}\right)\lambda\bm{L}_{:,\,\bm{a}}}{\sum\limits_{\bm{a} \in \mathcal{S}}Q_{\bm{a}}\exp\left(\lambda\bm{\psi}^T\bm{L}_{:,\,\bm{a}}\right)},
\end{align}
has its block $q$ given by absolutely convergent series:
\begin{align}
    \bm{\chi}^{\otimes q} & = \sum_{\substack{\left(\mu_l\right)_{l \geq 1}\\l_1,\,\ldots,\,l_q}}\lambda^{q + l_1 + \ldots + l_q + \sum_{l}l\mu_l}\frac{\left(q - 1 + \sum\limits_{l \geq 1}\mu_l\right)!}{\left(q - 1\right)!}\frac{(-1)^{\sum\limits_{l \geq 1}\mu_l}}{l_1!\ldots l_q!\prod\limits_{l \geq 1}\mu_l!l!^{\mu_l}}\nonumber\\
    & \hspace*{50px} \times \left\langle \bm{\psi}^{\otimes\left(\sum_{l}l\mu_l + l_1 + \ldots + l_q\right)}, \bigotimes_{l \geq 1}\overline{\bm{C}}^{(l)\otimes \mu_l} \otimes \overline{\bm{C}}^{\left(l_1 + 1\right)} \otimes \ldots \otimes \overline{\bm{C}}^{\left(l_q + 1\right)} \right\rangle.\label{eq:saddle_point_equation_rhs_generalized_series}
\end{align}
From the bounds obtained while deriving absolute convergence (equation \ref{eq:saddle_point_equation_rhs_generalized_absolute_series}), one could directly show (using definition \ref{def:tensor_powers_vector_space} of $\mathcal{L}$) that the vector of tensor powers of $\bm{\chi}$ belongs to $\mathcal{L}$. However, it will be sufficient to prove that $\bm{\Psi} \in \mathcal{L}$, so that equation \ref{eq:saddle_point_equation_rhs_generalized_series} may be reinterpreted as:
\begin{align}
    \bm{X} & = \overline{\bm{\Theta}^*} + \bm{T}\bm{\Psi},
\end{align}
from which $\bm{X} \in \mathcal{L}$ will follows as $\mathcal{L}$ is stable by $\bm{T}$ (lemma \ref{lemma:t_boundedness}) and $\overline{\bm{\Theta}^*} \in \mathcal{L}$ (corollary \ref{corollary:gamma_0_correlations_tensor_powers_vector}). We then proceed to show:
\begin{align}
    \bm{\Psi} & := \begin{pmatrix}
        \bm{\psi}\\
        \bm{\psi}^{\otimes 2}\\
        \bm{\psi}^{\otimes 3}\\
        \vdots
    \end{pmatrix} \in \mathcal{L}
\end{align}
Given lemma assumption $\left\lVert \bm{\psi} \right\rVert_1 \leq |\lambda||\mathcal{A}|c'$, this follows from earlier lemma \ref{lemma:vector_tensor_powers_well_defined_criterion}, setting $\delta := 1$ there, as long as:
\begin{align}
    |\lambda|^2|\mathcal{A}| < \frac{1}{2cc'}.
\end{align}
Hence, $\bm{\Psi} \in \mathcal{L}$ is well-defined, so is $\bm{T}\bm{\Psi}$ and the right-hand side of the series in equation \ref{eq:saddle_point_equation_rhs_generalized_series} (organized by blocks according to $q$) equals:
\begin{align}
    \overline{\bm{\Theta}^*} + \bm{T}\bm{\Psi}
\end{align}
as claimed.
\end{proof}
\end{lemma}

Informally, a non-trivial consequence of lemma \ref{lemma:saddle_point_equation_rhs_generalized_series} is that for \textit{any} vector $\bm{\psi} \in \mathbf{C}^{\mathcal{A}}$, possibly unrelated to the saddle-point equation solution, the vector from $\mathcal{L}$:
\begin{align}
    \bm{X} & := \overline{\bm{\Theta}^*} + \bm{T}\bm{\Psi}
\end{align}
is a ``vector of tensor powers", i.e. a vector of the form
\begin{align}
    \bm{X} & = \begin{pmatrix}
        \bm{\chi}\\
        \bm{\chi}^{\otimes 2}\\
        \bm{\chi}^{\otimes 3}\\
        \vdots
    \end{pmatrix},\\
    \bm{\chi} & \in \mathbf{C}^{\mathcal{A}}.
\end{align}

We are now ready to rigorously justify correctness of equation \ref{eq:saddle_point_equation_solution_non_rigorous} for the saddle-point equation.

\begin{repproposition}{propSolutionSaddlePointEquation}[Solution of saddle-point equation]
Let $\overline{\bm{\Theta}^*}$ the ``vector of tensor powers" of noninteracting correlations, which is well-defined as an element of $\mathcal{L}$ according to corollary \ref{corollary:gamma_0_correlations_tensor_powers_vector}. Consider:
\begin{align}
    \bm{\Theta}^* & = \left(\bm{I} - \bm{T}\right)^{-1}\overline{\bm{\Theta}^*} \in \mathcal{L}.
\end{align}
By boundedness of $\bm{T}$ (lemma \ref{lemma:t_boundedness}), this is well-defined for $|\lambda|^2|\mathcal{A}|$ upper-bounded by a constant depending only on $c_{\mathrm{min}}, c_{\mathrm{max}}$ from assumption \ref{assp:gamma_zero_correlations_boundedness}. Then, $\bm{\Theta}^*$ is a ``vector of tensor powers", i.e. there exists $\bm{\theta}^* \in \mathbf{C}^{\mathcal{A}}$ such that:
\begin{align}
    \bm{\Theta}^* & = \begin{pmatrix}
        \bm{\theta}^*\\
        \bm{\theta}^{*\otimes 2}\\
        \bm{\theta}^{*\otimes 3}\\
        \vdots
    \end{pmatrix}.
\end{align}
Finally, vector $\bm{\theta}^*$ solves the saddle-point equation, i.e.:
\begin{align}
    \bm{\theta}^* & = \frac{\sum\limits_{\bm{a} \in \mathcal{S}}Q_{\bm{a}}\exp\left(\lambda\bm{\theta}^{*T}\bm{L}_{:,\,\bm{a}}\right)\lambda\bm{L}_{:,\,\bm{a}}}{\sum\limits_{\bm{a} \in \mathcal{S}}Q_{\bm{a}}\exp\left(\lambda\bm{\theta}^{*T}\bm{L}_{:,\,\bm{a}}\right)}.
\end{align}
\begin{proof}
We first ensure that $\left(\bm{I} - \bm{T}\right)^{-1}$ is well-defined as a bounded operator from a Taylor expansion of the inverse series, i.e.:
\begin{align}
    \left(\bm{I} - \bm{T}\right)^{-1} & = \lim_{m \to \infty}\sum_{k \geq 0}^{m}\bm{T}^k
\end{align}
From the ``explicit" bound on the operator norm of $\bm{T}$ in lemma \ref{lemma:t_boundedness},
\begin{align}
    \left\lVert \bm{T} \right\rVert_{\mathcal{L} \to \mathcal{L}} & \leq 2c^2|\lambda|^2|\mathcal{A}|\nonumber\\
    & \leq \frac{1}{4}
\end{align}
as long as
\begin{align}
    |\lambda|^2|\mathcal{A}| & \leq \frac{1}{8c^2},
\end{align}
where the bound only depends on $c$ from lemma \ref{lemma:t_boundedness}, hence only on $c_{\mathrm{min}}, c_{\mathrm{max}}$. As a result, $\left(\bm{I} - \bm{T}\right)^{-1}$ is well-defined. Since furthermore $\overline{\bm{\Theta}^*} \in \mathcal{L}$ (corollary \ref{corollary:gamma_0_correlations_tensor_powers_vector}),
\begin{align}
    \bm{\Theta}^* & = \left(\bm{I} - \bm{T}\right)^{-1}\overline{\bm{\Theta}^*}\nonumber\\
    & = \lim_{m \to \infty}\sum_{k = 0}^m\bm{T}^k\overline{\bm{\Theta}^*} \in \mathcal{L}
\end{align}
is also well-defined. We now show it is a vector of tensor powers as claimed. We first prove by induction on $m$ that for all $m \geq 1$,
\begin{align}
    \sum_{k = 0}^m\bm{T}^k\overline{\bm{\Theta}^*}
\end{align}
is a vector of tensor powers, with first block bounded as follows in $1$-norm:
\begin{align}
    \left\lVert \left[\sum_{k = 0}^m\bm{T}^k\overline{\bm{\Theta}^*}\right]_1 \right\rVert_1 & \leq 2\frac{c_{\mathrm{max}}}{c_{\mathrm{min}}}|\lambda||\mathcal{A}|.
\end{align}

For $m = 1$, observing
\begin{align}
    \left\lVert \overline{\bm{\theta}^*} \right\rVert_1 & \leq \left|\mathcal{A}\right|\left\lVert \overline{\bm{\theta}^*} \right\rVert_{\infty}\nonumber\\
    & = \left|\mathcal{A}\right|\left\lVert 
\lambda\overline{\bm{C}}^{(1)}\right\rVert_{\infty}\nonumber\\
    & \leq \frac{c_{\mathrm{max}}}{c_{\mathrm{min}}}|\lambda||\mathcal{A}|\nonumber\\
    & \leq 2\frac{c_{\mathrm{max}}}{c_{\mathrm{min}}}|\lambda||\mathcal{A}|\nonumber,
\end{align}
one can apply lemma \ref{lemma:saddle_point_equation_rhs_generalized_series} to
\begin{align}
    \bm{\psi} & = \overline{\bm{\Theta}^*}
\end{align}
with $c' = 2c_{\mathrm{max}}/c_{\mathrm{min}}$ to conclude that
\begin{align}
    \sum_{k = 0}^1\bm{T}^k\overline{\bm{\Theta}^*} & = \overline{\bm{\Theta}^*} + \bm{T}\overline{\bm{\Theta}^*}
\end{align}
is a vector of tensor powers. Besides, using lemma \ref{lemma:vector_tensor_powers_well_defined_criterion} with parameters $c' = 2c_{\mathrm{max}}/c_{\mathrm{min}}, \delta = 1$ to relate norms $\left\lVert \cdot \right\rVert_1$ and $\left\lVert \cdot \right\rVert_1'$, as well as bound $\left\lVert \bm{T} \right\rVert_{\mathcal{L} \to \mathcal{L}} \leq 1/4$, it holds:
\begin{align}
    \left\lVert \left[\sum_{k = 0}^1\bm{T}^k\overline{\bm{\Theta}^*}\right]_1 \right\rVert_{\infty} & = \left\lVert \left[\overline{\bm{\Theta}^*} + \bm{T}\overline{\bm{\Theta}^*}\right]_1 \right\rVert_{\infty}\nonumber\\
    & = \left\lVert \overline{\bm{\theta}^*} + \left[\bm{T}\overline{\bm{\Theta}^*}\right]_1 \right\rVert_1\nonumber\\
    & \leq \left\lVert \overline{\bm{\theta}^*}\right\rVert_1 + \left\lVert\left[\bm{T}\overline{\bm{\Theta}^*}\right]_1 \right\rVert_1\nonumber\\
    & \leq \left\lVert \overline{\bm{\theta}^*}\right\rVert_1 + \frac{1}{|\lambda|c}\left\lVert\bm{T}\overline{\bm{\Theta}^*}\right\rVert_{1}'\nonumber\\
    & \leq \left\lVert \overline{\bm{\theta}^*}\right\rVert_1 + \frac{1}{|\lambda|c}\left\lVert\bm{T}\right\rVert_{\mathcal{L} \to \mathcal{L}} \left\lVert\overline{\bm{\Theta}^*}\right\rVert'_{1}\nonumber\\
    & \leq \left\lVert \overline{\bm{\theta}^*}\right\rVert_1 + \frac{1}{|\lambda|c} \times \left\lVert \bm{T} \right\rVert_{\mathcal{L} \to \mathcal{L}} \times 2|\lambda|c\left\lVert \overline{\bm{\theta}^*} \right\rVert_1\nonumber\\
    & = \left\lVert \overline{\bm{\theta}^*}\right\rVert_1 + \frac{1}{|\lambda|c} \times \frac{1}{4} \times 2|\lambda|c\left\lVert \overline{\bm{\theta}^*} \right\rVert_1\nonumber\\
    & \leq \frac{c_{\mathrm{max}}}{c_{\mathrm{min}}}|\lambda|c + \frac{1}{2}\frac{c_{\mathrm{max}}}{c_{\mathrm{min}}}|\lambda|c\nonumber\\
    & \leq 2\frac{c_{\mathrm{max}}}{c_{\mathrm{min}}}|\lambda||\mathcal{A}|,
\end{align}
completing the initial step of the induction. Let us now assume the induction hypothesis up to level $m$ included. We first show that
\begin{align}
    \sum_{k = 0}^{m + 1}\bm{T}^k\overline{\bm{\Theta}^*}
\end{align}
is a vector of tensor powers. Indeed,
\begin{align}
    \sum_{k = 0}^{m + 1}\bm{T}^k\overline{\bm{\Theta}^*} & = \overline{\bm{\Theta}^*} + \bm{T}\sum_{k = 0}^{m}\bm{T}^k\overline{\bm{\Theta}^*}
\end{align}
By the induction hypothesis,
\begin{align}
    \left\lVert \left[\sum_{k = 0}^{m}\bm{T}^k\overline{\bm{\Theta}^*}\right]_1 \right\rVert_1 & \leq 2\frac{c_{\mathrm{max}}}{c_{\mathrm{min}}}|\lambda|\left|\mathcal{A}\right|
\end{align}
Hence, lemma \ref{lemma:saddle_point_equation_rhs_generalized_series} can be invoked, with same constant $c' = 2c_{\mathrm{max}}/c_{\mathrm{min}}$ as in the initialization step, to conclude that
\begin{align}
    \sum_{k = 0}^{m + 1}\bm{T}^k\overline{\bm{\Theta}^*}
\end{align}
is a ``vector of tensor powers". We then bound the norm of the first block of this element of $\mathcal{L}$:
\begin{align}
    \left\lVert \left[\sum_{k = 0}^{m + 1}\bm{T}^k\overline{\bm{\Theta}^*}\right]_1 \right\rVert_1 & = \left\lVert \left[\overline{\bm{\Theta}^*} + \bm{T}\sum_{k = 0}^{m}\bm{T}^k\overline{\bm{\Theta}^*}\right]_1 \right\rVert_1\nonumber\\
    & = \left\lVert \overline{\bm{\theta}^*} + \left[\bm{T}\sum_{k = 0}^m\bm{T}^k\overline{\bm{\Theta}^*}\right]_1 \right\rVert_1\nonumber\\
    & \leq \left\lVert \overline{\bm{\theta}^*} \right\rVert_1 + \left\lVert\left[\bm{T}\sum_{k = 0}^m\bm{T}^k\overline{\bm{\Theta}^*}\right]_1 \right\rVert_1\nonumber\\
    & \leq \left\lVert \overline{\bm{\theta}^*} \right\rVert_1 + \frac{1}{|\lambda|c}\left\lVert \bm{T}\sum_{k = 0}^m\bm{T}^k\overline{\bm{\Theta}^*} \right\rVert_1'\nonumber\\
    & \leq \left\lVert \overline{\bm{\theta}^*} \right\rVert_1 + \frac{1}{|\lambda|c}\left\lVert \bm{T} \right\rVert_{\mathcal{L} \to \mathcal{L}}\left\lVert \sum_{k = 0}^m\bm{T}^k\overline{\bm{\Theta}^*} \right\rVert_1'\nonumber\\
    & \leq \left\lVert \overline{\bm{\theta}^*} \right\rVert_1 + \frac{1}{|\lambda|c}\left\lVert \bm{T} \right\rVert_{\mathcal{L} \to \mathcal{L}} \times 2|\lambda|c\left\lVert \left[\sum_{k = 0}^m\bm{T}^k\overline{\bm{\Theta}^*}\right]_1 \right\rVert_1\nonumber\\
    & \leq \left\lVert \overline{\bm{\theta}^*} \right\rVert_1 + \frac{1}{2}\left\lVert \left[\sum_{k = 0}^m\bm{T}^k\overline{\bm{\Theta}^*}\right]_1 \right\rVert_1\nonumber\\
    & \leq \frac{c_{\mathrm{max}}}{c_{\mathrm{min}}}|\lambda||\mathcal{A}| + \frac{1}{2} \times 2\frac{c_{\mathrm{max}}}{c_{\mathrm{min}}}|\lambda||\mathcal{A}|\nonumber\\
    & \leq 2\frac{c_{\mathrm{max}}}{c_{\mathrm{min}}}|\lambda||\mathcal{A}|,
\end{align}
where, similar to the initialization step, we invoked lemma \ref{lemma:vector_tensor_powers_well_defined_criterion} with parameters $c' = 2c_{\mathrm{max}}/c_{\mathrm{min}}, \delta$ to compare norms $\left\lVert \cdot \right\rVert_1$ and $\left\lVert \cdot \right\rVert_1'$. Note it is crucial that $c', \delta$ stay the same at all steps of the induction, so the induction statement holds for $|\lambda|^2|\mathcal{A}|$ in a neighbourhood of $0$ independent of $m$. We have then proven that for all $m \geq 0$,
\begin{align}
    \sum_{k = 0}^m\bm{T}^k\overline{\bm{\Theta}^*} \in \mathcal{A}
\end{align}
is a ``vector of tensor powers". Now, it is not hard to see that ``vectors of tensor powers" are closed in the normed space topology of $\left(\mathcal{L}, \left\lVert \cdot \right\rVert_1'\right)$; it follows that the $m \to \infty$ limit of this sequence:
\begin{align}
    \bm{\Theta}^* & := \left(\bm{I} - \bm{T}\right)^{-1}\overline{\bm{\Theta}^*}
\end{align}
is a ``vector of tensor powers". More precisely, denoting by $\bm{\theta}^*$ the first block of this vector of $\mathcal{L}$, it holds
\begin{align}
    \bm{\Theta}^* & = \begin{pmatrix}
        \bm{\theta}^*\\
        \bm{\theta}^{*\otimes 2}\\
        \bm{\theta}^{*\otimes 3}\\
        \vdots
    \end{pmatrix}.
\end{align}
We now observe by simple algebra in operator $\bm{T}$ that $\bm{\Theta}^*$ satisfies equation
\begin{align}
    \bm{\Theta}^* & = \overline{\bm{\Theta}^*} + \bm{T}\bm{\Theta}^*.
\end{align}
We finally extract the first block of this linear equation in $\mathcal{L}$ vectors, invoking lemma \ref{lemma:saddle_point_equation_rhs_generalized_series} for the right-hand side thanks to the ``vector of tensor powers" form of $\bm{\Theta}^*$. This gives:
\begin{align}
    \bm{\theta}^* & = \frac{\sum\limits_{\bm{a} \in \mathcal{S}}Q_{\bm{a}}\exp\left(\bm{\theta}^{*T}\bm{L}_{:,\,\bm{a}}\right)\lambda\bm{L}_{:,\,\bm{a}}}{\sum\limits_{\bm{a} \in \mathcal{S}}Q_{\bm{a}}\exp\left(\bm{\theta}^{*T}\bm{L}_{:,\,\bm{a}}\right)},
\end{align}
proving $\bm{\theta}^*$ solves the saddle-point equation.
\end{proof}
\end{repproposition}

\subsubsection{Bounds on pseudo-partition function and correlations}
\label{sec:bounds_pseudo_partition_function_correlations}

The analysis from Sections \ref{sec:convergence_saddle_point_expansion}, \ref{sec:correctness_saddle_point_expansion} showed convergence and correctness of the ``small $\lambda$ expansion" derived heuristically in section \ref{sec:heuristic_pqgms_expansion}, provided $\lambda$ satisfied:
\begin{align}
    |\lambda|^2|\mathcal{A}| < \textrm{a constant depending only on $c_{\mathrm{min}}$, $c_{\mathrm{max}}$},
\end{align}
where $c_{\mathrm{min}}, c_{\mathrm{max}}$ are constants introduced in assumption \ref{assp:gamma_zero_correlations_boundedness}, depending on the parameters of the quadratic generalized multinomial sum. As we will see in concrete examples, in practice they essentially depend on $\bm{Q} = \left(Q_{\bm{a}}\right)_{\bm{a} \in \mathcal{S}}$; besides, the condition in the above equation is synonymous of constant evolution time. In this regime, we proved a bound (equation \ref{eq:thetas_star_bound}) on the solution of the saddle-point equation $\bm{\theta}^*$:
\begin{align}
    \left\lVert \bm{\theta}^* \right\rVert_{\infty} & \leq \frac{2c_{\mathrm{max}}}{c_{\mathrm{min}}}|\lambda|,\label{eq:thetas_star_bound_restated}
\end{align}
provided $|\lambda|^2|\mathcal{A}|$ is upper-bounded by a constant depending only on $c_{\mathrm{min}}, c_{\mathrm{max}}$. In this section, we deduce from this estimate bounds on the pseudo-partition function --as well as its derivatives, recentered about the saddle-point equation solution:
\begin{align}
    \mathcal{Z}\left(\bm{\theta}^* + \frac{1}{\sqrt{n}}\bm{\chi}\right) & = \sum_{\bm{a} \in \mathcal{S}}Q_{\bm{a}}\exp\left(\lambda\left(\bm{\theta}^{*} + \frac{1}{\sqrt{n}}\bm{\chi}\right)^T\bm{L}_{:,\,\bm{a}}\right).
\end{align}
Let us then consider an arbitrary order $m$ derivative of the shifted pseudo-partition function.
\begin{align}
    & \frac{\partial^m\mathcal{Z}\left(\bm{\theta}^* + n^{-1/2}\bm{\chi}\right)}{\partial\chi_{\alpha_1}\ldots\partial\chi_{\alpha_m}}\nonumber\\
    & = n^{-m/2}\sum_{\bm{a} \in \mathcal{S}}Q_{\bm{a}}\exp\left(\lambda\left(\bm{\theta}^{*} + \frac{1}{\sqrt
    n}\bm{\chi}\right)^T\bm{L}_{:,\,\bm{a}}\right)\left(\lambda L_{\alpha_1,\,\bm{a}}\right) \ldots \left(\lambda L_{\alpha_m,\,\bm{a}}\right)\nonumber\\
    & = n^{-m/2}\sum_{\bm{a} \in \mathcal{S}}Q_{\bm{a}}\left(\lambda L_{\alpha_1,\,\bm{a}}\right) \ldots \left(\lambda L_{\alpha_m,\,\bm{a}}\right)\sum_{d \geq 0}\frac{1}{d!}\left(\lambda\left(\bm{\theta}^* + \frac{1}{\sqrt{n}}\bm{\chi}\right)^T\bm{L}_{:,\,\bm{a}}\right)^d\nonumber\\
    & = n^{-m/2}\sum_{d \geq 0}\frac{1}{d!}\left\langle \left(\bm{\theta}^* + \frac{1}{\sqrt{n}}\bm{\chi}\right)^{\otimes d}, \sum_{\bm{a} \in \mathcal{S}}Q_{\bm{a}}\left(\lambda L_{\alpha_1,\,\bm{a}}\right)\ldots \left(\lambda L_{\alpha_m,\,\bm{a}}\right)\left(\lambda \bm{L}_{:,\,\bm{a}}\right)^{\otimes d} \right\rangle\nonumber\\
    & = n^{-m/2}\sum_{d \geq 0}\frac{1}{d!}\left\langle \bm{e}_{\alpha_1} \otimes \ldots \otimes \bm{e}_{\alpha_m} \otimes \left(\bm{\theta}^* + \frac{1}{\sqrt{n}}\bm{\chi}\right)^{\otimes d}, \sum_{\bm{a} \in \mathcal{S}}Q_{\bm{a}}\left(\lambda \bm{L}_{:,\,\bm{a}}\right)^{\otimes \left(m + d\right)} \right\rangle\nonumber\\
    & = n^{-m/2}\sum_{d \geq 0}\frac{\lambda^{m + d}}{d!}\left\langle \bm{e}_{\alpha_1} \otimes \ldots \otimes \bm{e}_{\alpha_m} \otimes \left(\bm{\theta}^* + \frac{1}{\sqrt{n}}\bm{\chi}\right)^{\otimes d}, \overline{\bm{C}}^{\left(m + d\right)} \right\rangle
\end{align}
where we denoted by $\bm{e}_{\alpha}$ the canonical basis vector of $\mathbf{C}^{\mathcal{A}}$ of index $\alpha$. The latter expression can be crudely bounded using the triangle inequality:
\begin{align}
    \left|\frac{\partial^m\mathcal{Z}\left(\bm{\theta}^* + n^{-1/2}\bm{\chi}\right)}{\partial\chi_{\alpha_1}\ldots\partial\chi_{\alpha_m}}\right| & \leq n^{-m/2}\sum_{d \geq 0}\frac{|\lambda|^{m + d}}{d!}\left|\left\langle \bm{e}_{\alpha_1} \otimes \ldots \otimes \bm{e}_{\alpha_m} \otimes \left(\bm{\theta}^* + \frac{1}{\sqrt{n}}\bm{\chi}\right)^{\otimes d}, \overline{\bm{C}}^{\left(m + d\right)} \right\rangle\right|\nonumber\\
    & = n^{-m/2}\sum_{d \geq 0}\frac{|\lambda|^{m + d}}{d!}\left|\left\langle \left(\bm{\theta}^* + \frac{1}{\sqrt{n}}\bm{\chi}\right)^{\otimes n}, \overline{\bm{C}}^{\left(m + d\right)}_{\alpha_1,\,\ldots,\,\alpha_m} \right\rangle\right|\nonumber\\
    & \leq n^{-m/2}\sum_{d \geq 0}\frac{|\lambda|^{m + d}}{d!}\left\lVert \left(\bm{\theta}^* + \frac{1}{\sqrt
    n}\bm{\chi}\right)^{\otimes d} \right\rVert_2 \left\lVert \overline{\bm{C}}^{\left(m + d\right)}_{\alpha_1,\,\ldots,\,\alpha_m} \right\rVert_2\nonumber\\
    & \leq n^{-m/2}\sum_{d \geq 0}\frac{|\lambda|^{m + d}}{d!}\left\lVert \bm{\theta}^* + \frac{1}{\sqrt{n}}\bm{\chi} \right\rVert_2^d\left\lVert \overline{\bm{C}}^{\left(m + d\right)}_{\alpha_1,\,\ldots,\,\alpha_m} \right\rVert_2\nonumber\\
    & \leq n^{-m/2}\sum_{d \geq 0}\frac{|\lambda|^{m + d}}{d!}\left(\left\lVert\bm{\theta}^* \right\rVert_2 + \frac{1}{\sqrt
    n}\left\lVert \bm{\chi} \right\rVert_2\right)^d\left\lVert \overline{\bm{C}}^{\left(m + d\right)}_{\alpha_1,\,\ldots,\,\alpha_m} \right\rVert_2\nonumber\\
    & \leq n^{-m/2}\sum_{d \geq 0}\frac{|\lambda|^{m + d}}{d!}\left(|\mathcal{A}|^{1/2}\left\lVert\bm{\theta}^* \right\rVert_{\infty} + \frac{1}{\sqrt
    n}\left\lVert \bm{\chi} \right\rVert_2\right)^d\left| \mathcal{A} \right|^{d/2}\left\lVert \overline{\bm{C}}^{\left(m + d\right)}_{\alpha_1,\,\ldots,\,\alpha_m} \right\rVert_{\infty}\nonumber\\
    & = n^{-m/2}|\lambda|^m\left\lVert \overline{\bm{C}}^{\left(m + d\right)}_{\alpha_1,\,\ldots,\,\alpha_m} \right\rVert_{\infty}\exp\left(|\lambda||\mathcal{A}|\left\lVert \bm{\theta}^* \right\rVert_{\infty} + \frac{1}{\sqrt{n}}|\lambda||\mathcal{A}|^{1/2}\left\lVert \bm{\chi} \right\rVert_2\right)\nonumber\\
    & \leq n^{-m/2}|\lambda|^mc_{\mathrm{max}}\exp\left(|\lambda||\mathcal{A}|\left\lVert \bm{\theta}^* \right\rVert_{\infty} + \frac{1}{\sqrt{n}}|\lambda||\mathcal{A}|^{1/2}\left\lVert \bm{\chi} \right\rVert_2\right)\nonumber\\
    & \leq n^{-m/2}|\lambda|^mc_{\mathrm{max}}\exp\left(\frac{2c_{\mathrm{max}}}{c_{\mathrm{min}}}|\lambda|^2|\mathcal{A}| + \frac{1}{\sqrt{n}}|\lambda||\mathcal{A}|^{1/2}\left\lVert \bm{\chi} \right\rVert_2\right),\label{eq:pseudo_partition_function_derivative_bound}
\end{align}
where in the final step, we invoked the entrywise bound on $\bm{\theta}^*$ from equation \ref{eq:thetas_star_bound_restated}. Note the argument of the exponential in the final bound only depends on $\lambda$ and $\mathcal{A}$ through product $|\lambda||\mathcal{A}|^2$.

By a similar calculation, one can bound the pseudo-partition function away from $0$ by showing its closeness to $\overline{\mathcal{Z}^*}$ ---which we recall can be understood as the value of the pseudo-partition function in the limit $\lambda = 0$.
\begin{align}
    & \left| \mathcal{Z}\left(\bm{\theta}^* + n^{-1/2}\bm{\chi}\right) - \overline{\mathcal{Z}^*}\right|\nonumber\\
    & = \left|\sum_{d \geq 1}\frac{\lambda^d}{d!}\overline{\mathcal{Z}^*}\left\langle \left(\bm{\theta}^* + \frac{1}{\sqrt{n}}\bm{\chi}\right)^{\otimes d}, \overline{\bm{C}}^{\left(d\right)} \right\rangle\right|\nonumber\\
    & \leq \sum_{d \geq 1}\frac{|\lambda|^d}{d!}\left\lVert \bm{\theta}^* + \frac{1}{\sqrt{n}}\bm{\chi} \right\rVert_2^d\left|\overline{\mathcal{Z}^*}\right|\left\lVert \overline{\bm{C}}^{(d)} \right\rVert_2\nonumber\\
    & \leq \sum_{d \geq 1}\frac{|\lambda|^d}{d!}\left(\left\lVert \bm{\theta}^* \right\rVert_2 + \frac{1}{\sqrt{n}}\left\lVert \bm{\chi} \right\rVert_2\right)^d\left|\overline{\mathcal{Z}^*}\right|\left\lVert \overline{\bm{C}}^{(d)} \right\rVert_2\nonumber\\
    & \leq  \sum_{d \geq 1}\frac{|\lambda|^d}{d!}\left(|\mathcal{A}|^{1/2}\left\lVert \bm{\theta}^* \right\rVert_{\infty} + \frac{1}{\sqrt{n}}\left\lVert \bm{\chi} \right\rVert_2\right)^d\left|\mathcal{A}\right|^{d/2}\left|\overline{\mathcal{Z}^*}\right|\left\lVert \overline{\bm{C}}^{(d)} \right\rVert_{\infty}\nonumber\\
    & \leq c_{\mathrm{max}}\left(|\lambda||\mathcal{A}|\left\lVert \bm{\theta}^* \right\rVert_{\infty} + \frac{1}{\sqrt{n}}|\lambda||\mathcal{A}|^{1/2}\left\lVert \bm{\chi} \right\rVert_2\right)\exp\left(|\lambda||\mathcal{A}|\left\lVert \bm{\theta}^* \right\rVert_{\infty} + \frac{1}{\sqrt{n}}|\lambda||\mathcal{A}|^{1/2}\left\lVert \bm{\chi} \right\rVert_2\right)\nonumber\\
    & \leq c_{\mathrm{max}}\left(\frac{2c_{\mathrm{max}}}{c_{\mathrm{min}}}|\lambda|^2|\mathcal{A}| + \frac{1}{\sqrt{n}}|\lambda||\mathcal{A}|^{1/2}\left\lVert \bm{\chi} \right\rVert_2\right)\exp\left(\frac{2c_{\mathrm{max}}}{c_{\mathrm{min}}}|\lambda|^2|\mathcal{A}| + \frac{1}{\sqrt{n}}|\lambda||\mathcal{A}|^{1/2}\left\lVert \bm{\chi} \right\rVert_2\right)\label{eq:pseudo_partition_function_gamma_0_distance_bound}
\end{align}
Evaluated at $\bm{\chi} = \bm{0}$, the previous bounds gives:
\begin{align}
    \left|\mathcal{Z}^* - \overline{\mathcal{Z}^*}\right| = \left|\mathcal{Z}\left(\bm{\theta}^*\right) - \overline{\mathcal{Z}^*}\right| \leq \frac{2c_{\mathrm{max}}^2}{c_{\mathrm{min}}}|\lambda|^2|\mathcal{A}|\exp\left(\frac{2c_{\mathrm{max}}}{c_{\mathrm{min}}}|\lambda|^2|\mathcal{A}|\right).
\end{align}

From initial bounds \ref{eq:pseudo_partition_function_derivative_bound} and \ref{eq:pseudo_partition_function_gamma_0_distance_bound}, bounds on the correlations (with finite $\lambda$) can be deduced. Indeed:
\begin{align}
    \mathcal{Z}^*C^{\left(m\right)}_{\alpha_1,\,\ldots,\,\alpha_m} & = \sum_{\bm{a} \in \mathcal{S}}Q_{\bm{a}}\exp\left(\lambda\bm{\theta}^{*T}\bm{L}_{:,\,\bm{a}}\right)\left(\lambda L_{\alpha_1,\,\bm{a}}\right) \ldots \left(\lambda L_{\alpha_m,\,\bm{a}}\right)\nonumber\\
    & = n^{m/2}\left(\frac{\partial^m}{\partial\chi_{\alpha_1}\ldots\partial\chi_{\alpha_m}}\sum_{\bm{a} \in \mathcal{S}}Q_{\bm{a}}\exp\left(\lambda\bm{\theta}^{*T}\bm{L}_{:,\,\bm{a}} + \frac{\lambda}{\sqrt{n}}\bm{\chi}^T\bm{L}_{:,\,\bm{a}}\right)\right)\Bigg|_{\bm{\chi} = \bm{0}}\nonumber\\
    & = n^{m/2}\frac{\partial^m\mathcal{Z}\left(\bm{\theta}^* + n^{-1/2}\bm{\chi}\right)}{\partial\chi_{\alpha_1}\ldots\partial\chi_{\alpha_m}}\Bigg|_{\bm{\chi} = \bm{0}}.
\end{align}
Hence, based on the bound in equation \ref{eq:pseudo_partition_function_derivative_bound},
\begin{align}
    \left|\mathcal{Z}^*\right|\cdot\left|C^{\left(m\right)}_{\alpha_1,\,\ldots,\,\alpha_m}\right| & \leq c_{\mathrm{max}}|\lambda|^m\exp\left(\frac{2c_{\mathrm{max}}}{c_{\mathrm{min}}}|\lambda|^2|\mathcal{A}|\right)\label{eq:correlations_tensor_times_partition_function_bound}
\end{align}
Now invoking equation \ref{eq:pseudo_partition_function_gamma_0_distance_bound} to lower-bound $\mathcal{Z}^*$, we obtain the following explicit upper bound on entries of the degree $m$ correlations tensor:
\begin{align}
    \left\lVert \bm{C}^{\left(m\right)} \right\rVert_{\infty} & = \frac{\left\lVert \bm{C}^{(m)} \right\rVert_{\infty} \cdot\left|\mathcal{Z}^*\right|}{\left|\mathcal{Z}^*\right|}\nonumber\\
    & = \frac{\left\lVert \bm{C}^{(m)} \right\rVert_{\infty} \cdot\left|\mathcal{Z}^*\right|}{\left|\overline{\mathcal{Z}^*}\right| - \left|\mathcal{Z}^* - \overline{\mathcal{Z}^*}\right|}\nonumber\\
    & \leq \frac{c_{\mathrm{max}}|\lambda|^m\exp\left(\frac{2c_{\mathrm{max}}}{c_{\mathrm{min}}}|\lambda|^2|\mathcal{A}|\right)}{\left|\overline{\mathcal{Z}^*}\right| - \frac{2c_{\mathrm{max}}^2}{c_{\mathrm{min}}}|\lambda|^2|\mathcal{A}|\exp\left(\frac{2c_{\mathrm{max}}}{c_{\mathrm{min}}}|\lambda|^2|\mathcal{A}|\right)}\nonumber\\
    & \leq \frac{2c_{\mathrm{max}}|\lambda|^m\exp\left(\frac{2c_{\mathrm{max}}}{c_{\mathrm{min}}}|\lambda|^2|\mathcal{A}|\right)}{\left|\overline{\mathcal{Z}^*}\right|}\nonumber\\
    & \leq \frac{2c_{\mathrm{max}}}{c_{\mathrm{min}}}\exp\left(\frac{2c_{\mathrm{max}}}{c_{\mathrm{min}}}|\lambda|^2|\mathcal{A}|\right)|\lambda|^m\label{eq:correlations_tensor_bound}
\end{align}
provided
\begin{align}
    \frac{2c_{\mathrm{max}}^2}{c_{\mathrm{min}}}|\lambda|^2|\mathcal{A}|\exp\left(\frac{2c_{\mathrm{max}}}{c_{\mathrm{min}}}|\lambda|^2|\mathcal{A}|\right) & \leq \frac{1}{2}\left|\overline{\mathcal{Z}^*}\right|\label{eq:correlations_tensor_bound_assumption}
\end{align}
Note the above can be satisfied by choosing $|\lambda|^2|\mathcal{A}|$ smaller than a constant depending only in $c_{\mathrm{min}}, c_{\mathrm{max}}$ from assumption \ref{assp:gamma_zero_correlations_boundedness}, consistent with the usual setting in of the section.

\subsubsection{Simplifying bounds on pseudo-partition function and correlations by rescaling}
\label{sec:simplifying_bounds_pseudo_partition_function_correlations}

In upcoming Section~\ref{sec:qgms_moments_series_expansion}, we will establish a series expansion for the moments of a Parametrized Quadratic Generalized Multinomial Sum (Definition~\ref{def:pqgms-mgf}). Each term of theses series will consist of a contracted tensor network built from correlations tensors $\bm{C}^{(d)}$. To prove convergence of the series, we will need to bound the magnitude of each of these contracted tensor networks. This will in turn require bounds on the norms of correlations tensors, as established in Eq.~\ref{eq:correlations_tensor_times_partition_function_bound}. In the interest of alleviating the notational burden of upcoming Section~\ref{sec:qgms_moments_series_expansion}, the present section will simplify these bounds further. The general idea is to proceed by rescaling: assuming PQGMS parameter $\lambda$ upper-bounded by a sufficiently small constant $\lambda_0$, we reparametrized:
\begin{align}
    \lambda & := \lambda_0\lambda',\\
    \left|\lambda'\right| & \leq 1,
\end{align}
and rephrase and weaken our bounds in terms of $\lambda'$. 

More specifically, we will work on simplifying the bound established in Eq.~\ref{eq:correlations_tensor_times_partition_function_bound}, restated hereafter:
\begin{align}
    \left|\mathcal{Z}^*\right| \left\lVert \bm{C}^{\left(m\right)} \right\rVert_{\infty} & \leq c_{\mathrm{max}}\exp\left(\frac{2c_{\mathrm{max}}}{c_{\mathrm{min}}}|\lambda|^2|\mathcal{A}|\right)|\lambda|^m.\label{eq:correlations_tensor_times_partition_function_bound_restated}
\end{align}
Likewise, we will use the following bound, restated from equation \ref{eq:pseudo_partition_function_gamma_0_distance_bound}, between the pseudo-partition function $\mathcal{Z}^*$ and its $\lambda = 0$ value $\overline{\mathcal{Z}}^*$:
\begin{align}
    \left|\mathcal{Z}^* - \overline{\mathcal{Z}}^*\right| & \leq \frac{2c_{\mathrm{max}}^2}{c_{\mathrm{min}}}|\lambda|^2|\mathcal{A}|\exp\left(\frac{2c_{\mathrm{max}}}{c_{\mathrm{min}}}|\lambda|^2|\mathcal{A}|\right).\label{eq:pseudo_partition_function_gamma_0_distance_bound_restated}
\end{align}
The constant $c$ appearing in these bounds, introduced for the first time in proposition \ref{prop:t_block_entrywise_bound} (more specifically equation \ref{eq:c_constant_definition}) above it, only depends on $c_{\mathrm{min}}, c_{\mathrm{max}}$ introduced in assumption \ref{assp:gamma_zero_correlations_boundedness}, which in turn can be easily estimated from $\bm{L}$. One may now restrict to sufficiently small $\lambda$ to make these bounds easy to handle. This is conveniently done by letting:
\begin{align}
    \lambda & := \lambda_0\lambda'
\end{align}
for $\lambda_0 \in [0, 1]$ a constant to be chosen later, and reasoning in variable $\lambda'$ rather than $\lambda$. Let us restate bound \ref{eq:correlations_tensor_times_partition_function_bound_restated} in terms of this new variable. First, it helps to weaken it to
\begin{align}
    \left|\mathcal{Z}^*\right| \left\lVert \bm{C}^{\left(m\right)} \right\rVert_{\infty} & \leq c_{\mathrm{max}}\exp\left(\frac{2c_{\mathrm{max}}}{c_{\mathrm{min}}}\left(\tau_{\gamma}^*\right)^2\right)|\lambda|^m,
\end{align}
where $\tau_{\gamma}^*$ is an upper bound on $|\lambda||\mathcal{A}|^{1/2}$ (in the case of the SK-QAOA energy QGMS, it corresponds to the maximum total $\gamma$ evolution time). In terms of variable $\lambda'$, the above weakened bound then becomes:
\begin{align}
    \left|\mathcal{Z}^*\right| \left\lVert \bm{C}^{\left(m\right)} \right\rVert_{\infty} & \leq c_{\mathrm{max}}\exp\left(\frac{2c_{\mathrm{max}}}{c_{\mathrm{min}}}\left(\tau^*_{\gamma}\right)^2\right)\lambda_0^m|\lambda'|^m
\end{align}
Choosing
\begin{align}
    \lambda_0 < \mathrm{min}\left\{1, \frac{c_{\mathrm{min}}}{2c_{\mathrm{max}}}\exp\left(-\frac{2c_{\mathrm{max}}}{c_{\mathrm{min}}}\left(\tau^*_{\gamma}\right)^2\right)\right\}
\end{align}
where the upper bounds only depends on $c_{\mathrm{max}}$ and $\mathrm{\tau}_{\mathrm{max}}$ yields weaker bound:
\begin{align}
    \left|\mathcal{Z}^*\right| \left\lVert \bm{C}^{\left(m\right)} \right\rVert_{\infty} & \leq \frac{c_{\mathrm{min}}}{2}\left|\lambda'\right|^m,
\end{align}
uniformly for all $m \geq 1$. Besides, $\tau' := \left|\lambda'\right||\mathcal{A}|^{1/2}$ can be expressed from $\tau := |\lambda||\mathcal{A}|^{1/2}$ as:
\begin{align}
    \tau' & = \left|\lambda'\right||\mathcal{A}|^{1/2}\nonumber\\
    & = \lambda_0^{-1}\left|\lambda\right||\mathcal{A}|^{1/2}\nonumber\\
    & = \lambda_0^{-1}\tau.
\end{align}
That, is $\tau'$ is related to $\tau$ by a constant depending only on $c_{\mathrm{min}}, c_{\mathrm{max}}$ and the original ``maximum total time" $\tau_{\gamma}^*$. Note that when the latter is upper-bounded by a constant depending only on $c_{\mathrm{min}}, c_{\mathrm{max}}$, e.g.
\begin{align}
    \tau_{\gamma}^* & \leq \sqrt{\frac{c_{\mathrm{min}}\log(2)}{2c_{\mathrm{max}}}},
\end{align}
the $\tau_{\gamma}^*$ dependency on proportionality constant $\lambda_0$ vanishes to leading order, e.g.
\begin{align}
    \frac{c_{\mathrm{min}}}{4c_{\mathrm{max}}} \leq \frac{c_{\mathrm{min}}}{2c_{\mathrm{max}}}\exp\left(-\frac{2c_{\mathrm{max}}}{c_{\mathrm{min}}}\left(\tau^*_{\gamma}\right)^2\right) \leq \frac{c_{\mathrm{min}}}{2c_{\mathrm{max}}}.
\end{align}
In particular, assuming $\tau_{\gamma}^*$ smaller than a constant depending only on $c_{\mathrm{min}}, c_{\mathrm{max}}$ is equivalent to a similar statement for $\tau'$. Similarly, after further requiring
\begin{align}
    \lambda_0 < \sqrt{\frac{c_{\mathrm{min}}}{2c_{\mathrm{max}}^2}}\exp\left(-\frac{c_{\mathrm{max}}}{c_{\mathrm{min}}}\left(\tau^*_{\gamma}\right)^2\right),
\end{align}
bound \ref{eq:pseudo_partition_function_gamma_0_distance_bound_restated} can be revised to:
\begin{align}
    \left|\mathcal{Z}^* - \overline{\mathcal{Z}}^*\right| & \leq \tau'.
\end{align}
we have then shown how to rephrase bounds \ref{eq:correlations_tensor_times_partition_function_bound_restated}, \ref{eq:pseudo_partition_function_gamma_0_distance_bound_restated} on correlations and the pseudo-partition functions in terms of new parameter $\lambda'$ (and corresponding ``time parameter" $\tau'$). We now specify how the QGMS:
\begin{align}
    \sum_{\bm{n} \in \mathcal{P}(n)}\binom{n}{\bm{n}}\exp\left(\frac{\lambda^2}{2n}\bm{n}^T\bm{L}^T\bm{L}\bm{n} + \frac{\lambda}{n}\bm{\mu}^T\bm{L}\bm{n}\right)\prod_{\bm{a} \in \mathcal{S}}Q_{\bm{a}}^{n_{\bm{a}}},
\end{align}
parametrized by $\lambda$, can be thought of as parametrized by $\lambda'$. Substituting $\lambda := \lambda_0\lambda'$ in the above expression yields:
\begin{align}
    & \sum_{\bm{n} \in \mathcal{P}(n)}\binom{n}{\bm{n}}\exp\left(\frac{\left(\lambda'\right)^2}{2n}\bm{n}^T\left(\lambda_0\bm{L}\right)^T\left(\lambda_0\bm{L}\right)\bm{n} + \frac{\lambda'}{n}\bm{\mu}^T\left(\lambda_0\bm{L}\right)\bm{n}\right)\prod_{\bm{a} \in \mathcal{S}}Q_{\bm{a}}^{n_{\bm{a}}}\nonumber\\
    & = \sum_{\bm{n} \in \mathcal{P}(n)}\binom{n}{\bm{n}}\exp\left(\frac{\left(\lambda'\right)^2}{2n}\bm{n}^T\bm{L'}^T\bm{L'}\bm{n} + \frac{\lambda'}{n}\bm{\mu}^T\bm{L'}\bm{n}\right)\prod_{\bm{a} \in \mathcal{S}}Q_{\bm{a}}^{n_{\bm{a}}},
\end{align}
where we let $\bm{L'} := \lambda_0\bm{L}$. The latter defines a new (parametrized) QGMS, with $\bm{L'}, \lambda'$ playing the initial roles of $\bm{L}, \lambda$. We now consider constants $c'_{\mathrm{min}}, c'_{\mathrm{max}}$ defined in assumption \ref{assp:gamma_zero_correlations_boundedness}, but relative to the new QGMS. Since $c'_{\mathrm{min}}$ only depends on numbers $Q_{\bm{a}}$, which are identical between the two QGMS, one may take $c'_{\mathrm{min}} := c_{\mathrm{min}}$. As for $c'_{\mathrm{max}}$, it is a bound on
\begin{align}
    \sum_{\bm{a} \in \mathcal{S}}Q_{\bm{a}}L'_{\alpha_1}\ldots L'_{\alpha_m} & = \lambda_0^m\sum_{\bm{a} \in \mathcal{S}}Q_{\bm{a}}L_{\alpha_1}\ldots L_{\alpha_m}.
\end{align}
But $c_{\mathrm{max}}$ is a bound on 
\begin{align}
    \sum_{\bm{a} \in \mathcal{S}}Q_{\bm{a}}L_{\alpha_1}\ldots L_{\alpha_m},
\end{align}
and since $\lambda_0 \in [0, 1]$, one may take $c'_{\mathrm{max}} := c_{\mathrm{max}}$. All in all, we have shown that a QGMS parametrized by $\lambda$, with associated assumption \ref{assp:gamma_zero_correlations_boundedness} constants $c_{\mathrm{min}}, c_{\mathrm{max}}$, can also be viewed as a different QGMS, with same constant $c_{\mathrm{min}}, c_{\mathrm{max}}$ and time parameter $\tau := |\lambda||\mathcal{A}|^{1/2}$ related by a proportionality constant depending only on $c_{\mathrm{min}}, c_{\mathrm{max}}$, but with bounds on correlations and pseudo-partition function now taking simpler form:
\begin{align}
    \left|\mathcal{Z}^*\right|\left\lVert \bm{C}^{\left(m\right)} \right\rVert_{\infty} & \leq \frac{c_{\mathrm{min}}}{2}\left|\lambda'\right|^m\label{eq:correlations_tensor_times_partition_function_bound_rescaled},\\
    \left|\mathcal{Z}^* - \overline{\mathcal{Z}}^*\right| & \leq \left|\lambda'\right||\mathcal{A}|^{1/2}.\label{eq:pseudo_partition_function_gamma_0_distance_bound_rescaled}
\end{align}
From then one, one will implicitly assume this rescaling transformation was performed and the QGMS already satisfies these simpler inequalities. To be explicit, we will then assume
\begin{align}
    \left|\mathcal{Z}^*\right|\left\lVert \bm{C}^{\left(m\right)} \right\rVert_{\infty} & \leq \frac{c_{\mathrm{min}}}{2}\left|\lambda\right|^m,\\
    \left|\mathcal{Z}^* - \overline{\mathcal{Z}}^*\right| & \leq \left|\lambda\right||\mathcal{A}|^{1/2}.
\end{align}
Besides we will assume:
\begin{align}
\label{eq:lambda_bounded_assumption}
    \tau & \leq \frac{c_{\mathrm{min}}}{2},
\end{align}
so that in particular
\begin{align}
    \left\lVert \bm{C}^{\left(m\right)} \right\rVert_{\infty} & = \frac{\left|\mathcal{Z}^*\right|\left\lVert \bm{C}^{\left(m\right)} \right\rVert_{\infty}}{|\mathcal{Z}^*|}\nonumber\\
    & \leq \frac{\left(c_{\mathrm{min}}/2\right)|\lambda|^m}{\left|\overline{\mathcal{Z}}^*\right| - \left|\mathcal{Z}^* - \overline{\mathcal{Z}}^*\right|}\nonumber\\
    & \leq \frac{\left(c_{\mathrm{min}}/2\right)|\lambda|^m}{c_{\mathrm{min}} - \tau'}\nonumber\\
    & \leq \frac{\left(c_{\mathrm{min}}/2\right)|\lambda|^m}{c_{\mathrm{min}} - c_{\mathrm{min}}/2}\nonumber\\
    & \leq \frac{\left(c_{\mathrm{min}}/2\right)|\lambda|^m}{c_{\mathrm{min}}/2} \nonumber\\
    & \leq |\lambda|^m.\label{eq:correlations_bound}
\end{align}

\subsection{Expansion of QGMS moments around the saddle point}
\label{sec:qgms_moments_series_expansion}

In this section, we derive an expansion for the moments of a (parametrized) QGMS (see Definition~\ref{def:qgms-mgf} and motivating discussion) as a series involving correlations tensors ---assuming existence of the saddle point, on which the definition of these tensors depends. Since the SK-QAOA energy can be expressed in terms of QGMS moments (proposition \ref{prop:qgms-mgf_formulation_sk_qaoa_energy}), this will ultimately provide an expansion of this quantity as a series in correlations tensors. For definiteness, we focus on the diagonal QGMS moment of order $2$
\begin{align}
    \frac{\partial^2S_n\left(\lambda, \bm{\mu}\right)}{\partial\mu_{\alpha}^2}\Bigg|_{\bm{\mu} = \bm{0}}.
\end{align}
In Section~\ref{sec:qgms_saddle_point_correlations_tensors}, Eq.~\ref{eq:qgms_second_order_moment_from_qgms_integral_moments} this moment was expressed in terms of the QGMS integral moments tensors $\bm{S}_n^{(k)}$ (Definition~\ref{def:qgms_integral_moments_tensor}); specialized to diagonal second order moments, this identity reads: 
\begin{align}
    \frac{\partial^2S_n\left(\lambda, \bm{\mu}\right)}{\partial\mu_{\alpha}^2}\Bigg|_{\bm{\mu} = \bm{0}} & = \left(\theta^*_{\alpha}\right)^2 + \frac{2}{\sqrt{n}}\theta^*_{\alpha}\left[\bm{S}^{(1)}_n\right]_{\alpha} + \frac{1}{n}\left[\bm{S}^{(2)}_n\right]_{\alpha,\,\alpha},\label{eq:qgms_second_order_diagonal_moment_from_qgms_integral_moments}
\end{align}
where we recall the formula for the QGMS integral moments tensor of order $k$:
\begin{align}
    \bm{S}^{(k)}_n\left(\lambda\right) & := e^{-n\bm{\theta}^*\left(\lambda\right)^T\bm{\theta}^*\left(\lambda\right)/2}\int_{\mathbf{R}^{\mathcal{A}}}\!\mathrm{d}\bm{\chi}\,\bm{\chi}^{\otimes k}\frac{e^{-\bm{\chi}^T\bm{\chi}/2}}{\left(2\pi\right)^{|\mathcal{A}|/2}}\left(\sum_{\bm{a} \in \mathcal{S}}Q_{\bm{a}}\exp\left(\lambda\bm{\theta}^{*}\left(\lambda\right)^T\bm{L}_{:,\,\bm{a}} + \frac{1}{\sqrt{n}}\bm{\chi}^T\left(\bm{L}_{:,\,\bm{a}} - \bm{\theta}^*\left(\lambda\right)\right)\right)\right)^n \in \left(\mathbf{C}^{\mathcal{A}}\right)^{\otimes k}.
\end{align}
The goal of this Section is to introduce a series expansion of these QGMS integral moments tensors. For simplicity, we focus on the integral moments tensor of order $1$, i.e. on term 
\begin{align}
    \nu_{\alpha} & := \frac{2}{\sqrt{n}}\theta^*_{\alpha}\left[\bm{S}_n^{(1)}\right]_{\alpha}
\end{align}
from Eq.~\ref{eq:qgms_second_order_diagonal_moment_from_qgms_integral_moments}.

While we will reason over a parametrized QGMS, with parameter denoted $\lambda$, we will frequently omit the explicit $\lambda$ dependence from the notation, for instance by denoting $\bm{\theta}^*$ for $\bm{\theta}^*\left(\lambda\right)$. We wish to analyze
\begin{align}
    \left[\bm{S}^{(1)}_n\right]_{\alpha} & = e^{-n\bm{\theta}^{*T}\bm{\theta}^*/2}\int_{\mathbf{R}^{\mathcal{A}}}\!\mathrm{d}\bm{\chi}\,\chi_{\alpha}\frac{e^{-\bm{\chi}^T\bm{\chi}/2}}{\left(2\pi\right)^{|\mathcal{A}|/2}}\left(\sum_{\bm{a} \in \mathcal{S}}Q_{\bm{a}}\exp\left(\lambda\bm{\theta}^{*T}\bm{L}_{:,\,\bm{a}} - \frac{1}{\sqrt{n}}\bm{\chi}^T\left(\lambda\bm{L}_{:,\,\bm{a}} - \bm{\theta}^*\right)\right)\right)^n.\label{eq:qgms_analyzed_integral}
\end{align}
We Taylor-expand the exponential inside the $\bm{\chi}$-quantity raised to the power $n$ as follows:
\begin{align}
    \left(\sum_{\bm{a} \in \mathcal{S}}Q_{\bm{a}}\exp\left(\lambda\bm{\theta}^{*T}\bm{L}_{:,\,\bm{a}} + \frac{1}{\sqrt{n}}\bm{\chi}^T\left(\lambda\bm{L}_{:,\,\bm{a}} - \bm{\theta}^*\right)\right)\right)^n  & = \left(\sum_{\bm{a} \in \mathcal{S}}Q_{\bm{a}}\exp\left(\lambda\bm{\theta}^{*T}\bm{L}_{:,\,\bm{a}}\right)\sum_{d \geq 2}\frac{n^{-d/2}}{d!}\left\langle \bm{\chi}, \lambda\bm{L}_{:,\,\bm{a}} - \bm{\theta}^* \right\rangle^d\right)^n\nonumber\\
    & = \left(\sum_{\bm{a} \in \mathcal{S}}Q_{\bm{a}}\exp\left(\bm{\theta}^{*T}\bm{L}_{:,\,\bm{a}}\right)\sum_{d \geq 2}\frac{n^{-d/2}}{d!}\left\langle \bm{\chi}^{\otimes d}, \left(\lambda\bm{L}_{:,\,\bm{a}} - \bm{\theta}^*\right)^{\otimes d} \right\rangle\right)^n\nonumber\\
    & = \left(\sum_{d \geq 2}\frac{n^{-d/2}}{d!}\left\langle \bm{\chi}^{\otimes d}, \sum_{\bm{a} \in \mathcal{S}}Q_{\bm{a}}\exp\left(\lambda\bm{\theta}^{*T}\bm{L}_{:,\,\bm{a}}\right)\left(\lambda\bm{L}_{:\,\bm{a}} - \bm{\theta}^*\right)^{\otimes d} \right\rangle\right)^n\nonumber\\
    & = \left(\sum_{d \geq 2}\frac{n^{-d/2}}{d!}\left\langle \bm{\chi}^{\otimes d}, \mathcal{Z}^*\bm{\delta C}^{\left(d\right)} \right\rangle\right)^n\nonumber\\
    & = \left(\mathcal{Z}^*\right)^n\left(\sum_{d \geq 2}\frac{n^{-d/2}}{d!}\left\langle \bm{\chi}^{\otimes d}, \bm{\delta C}^{\left(d\right)} \right\rangle\right)^n\nonumber\\
    & = \left(\mathcal{Z}^*\right)^n\left(1 + \sum_{d \geq 2}\frac{n^{-d/2}}{d!}\left\langle \bm{\chi}^{\otimes d}, \bm{\delta C}^{\left(d\right)} \right\rangle\right)^n,\label{eq:qgms_integrand_from_centered_correlations_tensor}
\end{align}
From the fifth line, we introduced the ``centered" correlations of degree $d$, specified in Definition~\ref{def:centered_correlations_tensor}. In the final line, we used the special cases $d = 1, 2$ in equations \ref{eq:centered_correlation_degree_1}, \ref{eq:centered_correlation_degree_2} from this definition.

\begin{definition}[Centered correlations tensor]
\label{def:centered_correlations_tensor}
The centered correlations of degree $d$, denoted $\bm{\delta C}^{\left(d\right)}$, is the $d$-dimensional tensor defined by:
\begin{align}
    \bm{\delta C}^{\left(d\right)} & := \frac{\sum\limits_{\bm{a} \in \mathcal{S}}Q_{\bm{a}}\exp\left(\bm{\theta}^{*T}\bm{L}_{:,\,\bm{a}}\right)\left(\bm{L}_{:,\,\bm{a}} - \bm{\theta}^*\right)^{\otimes d}}{\sum\limits_{\bm{a} \in \mathcal{S}}Q_{\bm{a}}\exp\left(\bm{\theta}^{*T}\bm{L}_{:,\,\bm{a}}\right)}\nonumber\\
    & = \left\langle \left(\bm{L}_{:,\,\bm{a}} - \bm{\theta}^*\right)^{\otimes d} \right\rangle_{\bm{a}}.
\end{align}
This can be expressed as a linear combination, with coefficients, $\pm 1$, of tensor products of standard correlation tensors, the degrees of which sum to $d$. Explicitly,
\begin{align}
    \delta C^{(d)}_{\bm{\alpha}_{1:d}} & = \sum_{\substack{S',\,S''\\S' \sqcup S'' = [d]}}(-1)^{|S''|}C^{(|S'|)}_{\bm{\alpha}_{S'}}\prod_{r \in S''}\theta^*_{\alpha_r}\label{eq:centered_correlations_tensor_indices_expression}\\
    & = \sum_{\substack{S',\,S''\\S' \sqcup S'' = [d]}}(-1)^{|S'|}C^{(|S'|)}_{\bm{\alpha}_{S'}}\left[\bm{\theta}^{*\otimes (|S''|)}\right]_{\bm{\alpha}_{S''}}
\end{align}
From this observation and bound \ref{eq:correlations_bound} on the standard correlations tensor, results the following bound on the entries of the centered correlation tensor:
\begin{align}
    \left\lVert \bm{\delta C}^{\left(d\right)} \right\rVert_{\infty} & \leq 2^d|\lambda|^d.\label{eq:centered_correlations_bound}
\end{align}
Besides, note the following important special cases:
\begin{align}
    \bm{\delta C}^{\left(1\right)} & = \bm{0}_{\mathcal{A}},\label{eq:centered_correlation_degree_1}\\
    \bm{\delta C}^{\left(2\right)} & = \bm{C}^{\left(2\right)} - \bm{\theta}^* \otimes \bm{\theta}^*\nonumber\\
    & = \bm{C}^{\left(2\right)} - \bm{C}^{\left(1\right)} \otimes \bm{C}^{\left(1\right)}\nonumber\\
    & = \bm{C}^{\left(2,\,\mathrm{conn}\right)}.\label{eq:centered_correlation_degree_2}
\end{align}
\end{definition}
Note the series raised to the power $n$ in equation \ref{eq:qgms_integrand_from_centered_correlations_tensor} is still manifestly absolutely convergent, since
\begin{align}
    \sum_{d \geq 2}\frac{n^{-d/2}}{d!}\left|\left\langle \bm{\chi}^{\otimes d}, \bm{\delta C}^{\left(d\right)} \right\rangle\right| & \leq \sum_{d \geq 0}\frac{1}{d!}\left\lVert \bm{\chi}^{\otimes d} \right\rVert_2 \left\lVert \bm{\delta C}^{\left(d\right)} \right\rVert_2\nonumber\\
    & \leq \sum_{d \geq 2}\frac{n^{-d/2}}{d!}\left\lVert \bm{\chi} \right\rVert_2^d|\mathcal{A}|^{d/2}\left\lVert \bm{\delta C}^{\left(d\right)} \right\rVert_{\infty}\nonumber\\
    & \leq \sum_{d \geq 2}\frac{n^{-d/2}}{d!}\left\lVert \bm{\chi} \right\rVert_2^d\left|\mathcal{A}\right|^{d/2}2^d|\lambda|^d\nonumber\\
    & = \frac{2}{c_{\mathrm{min}}}\left(\exp\left(\frac{1}{\sqrt{n}}|\lambda||\mathcal{A}|^{1/2}\left\lVert \bm{\chi} \right\rVert_2\right) - 1 - \frac{1}{\sqrt{n}}|\lambda||\mathcal{A}|^{1/2}\left\lVert \bm{\chi} \right\rVert_2\right)\nonumber\\
    & \leq \frac{2}{c_{\mathrm{min}}}\frac{1}{2n}|\lambda|^2|\mathcal{A}|\left\lVert \bm{\chi} \right\rVert_2^2\exp\left(\frac{1}{\sqrt{n}}|\lambda||\mathcal{A}|^{1/2}\left\lVert \bm{\chi} \right\rVert_2\right)\nonumber\\
    & < \infty.
\end{align}
In going from the third to the fourth line, we used the bound in equation \ref{eq:centered_correlations_bound} for the entries of the centered correlations tensor. Given this absolute convergence, we can apply the multinomial theorem for infinite series (theorem \ref{th:multinomial_theorem_infinite_number_terms}) to obtain:
\begin{align}
    \left(\sum_{\bm{a} \in \mathcal{S}}Q_{\bm{a}}\exp\left(\bm{\theta}^{*T}\bm{L}_{:,\,\bm{a}} + \frac{1}{\sqrt{n}}\bm{\chi}^T\left(\bm{L}_{:,\,\bm{a}} - \bm{\theta}^*\right)\right)\right)^n & = \left(\mathcal{Z}^*\right)^n\sum_{\left(n_d\right)_{d \geq 2}}\binom{n}{\left(n_d\right)_{d \geq 2}}\frac{n^{-\sum\limits_{d \geq 2}dn_d/2}}{\prod\limits_{d \geq 2}d!^{n_d}}\prod_{d \geq 2}\left\langle \bm{\chi}^{\otimes d}, \bm{\delta C}^{\left(d\right)} \right\rangle^{n_d}\nonumber\\
    & = \left(\mathcal{Z}^*\right)^n\sum_{\left(n_d\right)_{d \geq 2}}\binom{n}{\left(n_d\right)_{d \geq 2}}\frac{n^{-\sum\limits_{d \geq 2}dn_d/2}}{\prod\limits_{d \geq 2}d!^{n_d}}\left\langle \bm{\chi}^{\otimes \sum\limits_{d \geq 2}dn_d}, \bigotimes\limits_{d \geq 2}\bm{\delta C}^{\left(d\right)\otimes n_d} \right\rangle,\label{eq:integrand_multinomial_expansion}
\end{align}
where the sum is over sequences $\left(n_d\right)_{d \geq 2}$ with only a finite number of nonzero elements (note this constraint is automatically enforced by the multinomial coefficient, as defined in this case by equation \ref{eq:multinomial_coefficient_redefinition}). Coming back to the original goal integral \ref{eq:qgms_analyzed_integral}, we wish to integrate the above over $\bm{\chi}$ against
\begin{align}
    \frac{1}{\left(2\pi\right)^{|\mathcal{A}|/2}}\exp\left(-\frac{1}{2}\bm{\chi}^T\bm{\chi}\right)\chi_{\alpha}.
\end{align}
We wish to do so by inverting integral and summation over $\left(n_d\right)_{d \geq 2}$, since then it will ``suffice" to integrate polynomials in $\bm{\chi}$:
\begin{align}
    \chi_{\alpha}\left\langle \bm{\chi}^{\otimes \sum\limits_{d \geq 2}dn_d}, \bigotimes_{d \geq 2}\bm{\delta C}^{\left(d\right)\otimes n_d} \right\rangle & = \left\langle \bm{\chi}^{\otimes \left(1 + \sum\limits_{d \geq 2}dn_d\right)}, \bm{e}_{\alpha} \otimes \bigotimes_{d \geq 2}\bm{\delta C}^{\left(d\right)\otimes n_d} \right\rangle
\end{align}
against a standard normal distribution. The legitimacy of the sum-integral inversion is proven in the following lemma:

\begin{lemma}[Interchanging Gaussian integral and multinomial expansion]
\label{lemma:gaussian_integral_multinomial_sum_interchange}
The following identity holds:
\begin{align}
    \left[\bm{S}^{(1)}_n\right]_{\alpha} & = \int_{\mathbf{R}^{\mathcal{A}}}\!\mathrm{d}\bm{\chi}\,\frac{1}{\left(2\pi\right)^{|\mathcal{A}|/2}}\exp\left(-\frac{1}{2}\bm{\chi}^T\bm{\chi}\right)\chi_{\alpha}\left(\sum_{\bm{a} \in \mathcal{S}}Q_{\bm{a}}\exp\left(\lambda\bm{\theta}^{*T}\bm{L}_{:,\,\bm{a}} + \frac{1}{\sqrt{n}}\bm{\chi}^T\left(\lambda\bm{L}_{:,\,\bm{a}} - \bm{\theta}^*\right)\right)\right)^n\nonumber\\
    & = \left(\mathcal{Z}^*\right)^n\sum_{\left(n_d\right)_{d \geq 2}}\int_{\mathbf{R}^{\mathcal{A}}}\!\mathrm{d}\bm{\chi}\exp\left(-\frac{1}{2}\bm{\chi}^T\bm{\chi}\right)\chi_{\alpha}\binom{n}{\left(n_d\right)_{d \geq 2}}\frac{n^{-\sum\limits_{d \geq 2}dn_d/2}}{\prod\limits_{d \geq 2}d!^{n_d}}\left\langle \bm{\chi}^{\otimes \sum\limits_{d \geq 2}dn_d}, \bigotimes\limits_{d \geq 2}\bm{\delta C}^{\left(d\right)\otimes n_d} \right\rangle,
\end{align}
and the series on the right-hand side converges absolutely.
\begin{proof}
To justify interchange of summation and integrals, it suffices to prove:
\begin{align}
    \sum_{\left(n_d\right)_{d \geq 2}}\int_{\mathbf{R}^{\mathcal{A}}}\!\mathrm{d}\bm{\chi}\,\frac{1}{\left(2\pi\right)^{|\mathcal{A}|/2}}\left|\exp\left(-\frac{1}{2}\bm{\chi}^T\bm{\chi}\right)\chi_{\alpha}\binom{n}{\left(n_d\right)_{d \geq 2}}\frac{n^{-\sum\limits_{d \geq 2}dn_d/2}}{\prod\limits_{d \geq 2}d!^{n_d}}\left\langle \bm{\chi}^{\otimes \sum\limits_{d \geq 2}dn_d}, \bigotimes\limits_{d \geq 2}\bm{\delta C}^{\left(d\right)\otimes n_d} \right\rangle\right| < \infty.
\end{align}
Note the value of the above series is always defined (possibly equaling $+\infty$) given terms are non-negative. The integrand can be bounded as:
\begin{align}
    & \left|\frac{1}{\left(2\pi\right)^{|\mathcal{A}|/2}}\exp\left(-\frac{1}{2}\bm{\chi}^T\bm{\chi}\right)\chi_{\alpha}\binom{n}{\left(n_d\right)_{d \geq 2}}\frac{n^{-\sum\limits_{d \geq 2}dn_d/2}}{\prod\limits_{d \geq 2}d!^{n_d}}\left\langle \bm{\chi}^{\otimes \sum\limits_{d \geq 2}dn_d}, \bigotimes\limits_{d \geq 2}\bm{\delta C}^{\left(d\right)\otimes n_d} \right\rangle\right|\nonumber\\
    & \leq \frac{1}{\left(2\pi\right)^{|\mathcal{A}|/2}}\exp\left(-\frac{1}{2}\bm{\chi}^T\bm{\chi}\right)|\chi_{\alpha}|\binom{n}{\left(n_d\right)_{d \geq 2}}\frac{n^{-\sum\limits_{d \geq 2}dn_d/2}}{\prod\limits_{d \geq 2}d!^{n_d}}\left|\left\langle \bm{\chi}^{\otimes \sum\limits_{d \geq 2}dn_d/2}, \bigotimes_{d \geq 2}\bm{\delta C}^{\left(d\right)\otimes n_d} \right\rangle\right|
\end{align}
The dot product (which is Euclidean, but can be interpreted as Hermitian given the first argument is real) can be bounded from the Cauchy-Schwartz inequality:
\begin{align}
    \left|\left\langle \bm{\chi}^{\otimes \sum\limits_{d \geq 2}dn_d/2}, \bigotimes_{d \geq 2}\bm{\delta C}^{\left(d\right)\otimes n_d} \right\rangle\right| & \leq \left\lVert \bm{\chi}^{\otimes \sum\limits_{d \geq 2}dn_d/2}\right\rVert_2\left\lVert \bigotimes_{d \geq 2}\bm{\delta C}^{\left(d\right)\otimes n_d} \right\rVert_2\nonumber\\
    & = \left\lVert \bm{\chi}\right\rVert_2^{\sum\limits_{d \geq 2}dn_d/2}\prod_{d \geq 2}\left\lVert\bm{\delta C}^{\left(d\right)} \right\rVert_2^{n_d}\nonumber\\
    & \leq \left\lVert \bm{\chi}\right\rVert_2^{\sum\limits_{d \geq 2}dn_d/2}\prod_{d \geq 2}\left(|\mathcal{A}|^{d/2}\left\lVert\bm{\delta C}^{\left(d\right)} \right\rVert_{\infty}\right)^{n_d}\nonumber\\
    & \leq \left\lVert \bm{\chi}\right\rVert_2^{\sum\limits_{d \geq 2}dn_d/2}\prod_{d \geq 2}\left(2^d|\lambda|^d|\mathcal{A}|^{d/2}\right)^{n_d}\nonumber\\
    & = \left\lVert \bm{\chi}\right\rVert_2^{\sum\limits_{d \geq 2}dn_d/2}\prod_{d \geq 2}\left(2^d\tau^d\right)^{n_d}
\end{align}
Using this bound on the integrand, and interchanging sum and integral for non-negative measurable functions (which is always allowed) yields bound:
\begin{align}
    & \sum_{\left(n_d\right)_{d \geq 2}}\int_{\mathbf{R}^{\mathcal{A}}}\!\mathrm{d}\bm{\chi}\,\left|\frac{1}{\left(2\pi\right)^{|\mathcal{A}|/2}}\exp\left(-\frac{1}{2}\bm{\chi}^T\bm{\chi}\right)\chi_{\alpha}\binom{n}{\left(n_d\right)_{d \geq 2}}\frac{n^{-\sum\limits_{d \geq 2}dn_d/2}}{\prod\limits_{d \geq 2}d!^{n_d}}\left\langle \bm{\chi}^{\otimes \sum\limits_{d \geq 2}dn_d}, \bigotimes\limits_{d \geq 2}\bm{\delta C}^{\left(d\right)\otimes n_d} \right\rangle\right|\nonumber\\
    & \leq \sum_{\left(n_d\right)_{d \geq 2}}\int_{\mathbf{R}^{\mathcal{A}}}\!\mathrm{d}\bm{\chi}\frac{1}{\left(2\pi\right)^{|\mathcal{A}|/2}}\exp\left(-\frac{1}{2}\bm{\chi}^T\bm{\chi}\right)|\chi_{\alpha}|\binom{n}{\left(n_d\right)_{d \geq 2}}\frac{\left(2\tau\left\lVert \bm{\chi} \right\rVert_2n^{-1/2}\right)^{\sum\limits_{d \geq 2}dn_d/2}}{\prod\limits_{d \geq 2}d!^{n_d}}\nonumber\\
    & = \int_{\mathbf{R}^{\mathcal{A}}}\!\mathrm{d}\bm{\chi}\sum_{\left(n_d\right)_{d \geq 2}}\frac{1}{\left(2\pi\right)^{|\mathcal{A}|/2}}\exp\left(-\frac{1}{2}\bm{\chi}^T\bm{\chi}\right)|\chi_{\alpha}|\binom{n}{\left(n_d\right)_{d \geq 2}}\frac{\left(2\tau\left\lVert \bm{\chi} \right\rVert_2n^{-1/2}\right)^{\sum\limits_{d \geq 2}dn_d/2}}{\prod\limits_{d \geq 2}d!^{n_d}}\nonumber\\
    & = \int_{\mathbf{R}^{\mathcal{A}}}\!\mathrm{d}\bm{\chi}\frac{1}{\left(2\pi\right)^{|\mathcal{A}|/2}}\exp\left(-\frac{1}{2}\bm{\chi}^T\bm{\chi}\right)|\chi_{\alpha}|\left(1 + \sum_{d \geq 2}\frac{1}{d!}\left(2\tau\left\lVert \bm{\chi} \right\rVert_2n^{-1/2}\right)^d\right)^n\nonumber\\
    & = \int_{\mathbf{R}^{\mathcal{A}}}\!\mathrm{d}\bm{\chi}\,\frac{1}{\left(2\pi\right)^{|\mathcal{A}|/2}}\exp\left(-\frac{1}{2}\bm{\chi}^T\bm{\chi}\right)|\chi_{\alpha}|\left(\exp\left(\frac{2\tau\left\lVert \bm{\chi} \right\rVert_2}{\sqrt{n}}\right) - \frac{2\tau\left\lVert \bm{\chi} \right\rVert_2}{\sqrt{n}}\right)^n\nonumber\\
    & < \infty
\end{align}
since
\begin{align}
    \left(\exp\left(\frac{2\tau\left\lVert \bm{\chi} \right\rVert_2}{\sqrt{n}}\right) - \frac{2\tau\left\lVert \bm{\chi} \right\rVert_2}{\sqrt{n}}\right)^n
\end{align}
is at most of exponential growth in $\bm{\chi}$. This concludes the proof.
\end{proof}
\end{lemma}

\begin{remark}[Dimension $|\mathcal{A}|$ in absolute value bound for integral-sum inversion]
\label{rem:integral_dimension_dependence_integral_sum_inversion}
Although the proof of lemma \ref{lemma:gaussian_integral_multinomial_sum_interchange} is sufficient as such, using an exponential bound on the absolute value of the integrand to justify sum-integral interchange. However, it will prove instructive to estimate this bound more accurately to understand the role of integral dimension $|\mathcal{A}|$. More precisely, using the bound \ref{eq:exponential_bound_large_r} from lemma \ref{lemma:exponential_bound} with $c = 1$ there,
\begin{align}
    \left(\exp\left(\frac{2\tau\left\lVert \bm{\chi} \right\rVert_2}{\sqrt{n}}\right) - \frac{2\tau\left\lVert \bm{\chi} \right\rVert_2}{\sqrt{n}}\right)^n & \leq \exp\left(2\tau\left\lVert \bm{\chi} \right\rVert_2\sqrt{n}\right),
\end{align}
hence, after switching to spherical coordinates,
\begin{align}
    & \int_{\mathbf{R}^{\mathcal{A}}}\!\mathrm{d}\bm{\chi}\frac{1}{\left(2\pi\right)^{|\mathcal{A}|/2}}\exp\left(-\frac{1}{2}\bm{\chi}^T\bm{\chi}\right)\left|\chi_{\alpha}\right|\left(1 + \sum_{d \geq 2}\frac{1}{d!}\left(2\tau\left\lVert \bm{\chi} \right\rVert_2n^{-1/2}\right)^d\right)^n\nonumber\\
    & \leq \int_{\mathbf{R}^{\mathcal{A}}}\!\mathrm{d}\bm{\chi}\frac{1}{\left(2\pi\right)^{|\mathcal{A}|/2}}\exp\left(-\frac{1}{2}\bm{\chi}^T\bm{\chi}\right)\left|\chi_{\alpha}\right|\exp\left(2\tau\left\lVert \bm{\chi} \right\rVert_2\sqrt{n}\right)\nonumber\\
    & \leq \int_{S^{|\mathcal{A}| - 1}}\!\mathrm{d}\bm{\Omega}_{|\mathcal{A}| - 1}\int_{0}^{+\infty}\!\mathrm{d}r\,r^{|\mathcal{A}| - 1}\frac{1}{\left(2\pi\right)^{|\mathcal{A}|/2}}\exp\left(-\frac{r^2}{2}\right)r\exp\left(2\tau r\sqrt{n}\right)\nonumber\\
    & = \left(\int_{S^{|\mathcal{A}| - 1}}\!\mathrm{d}\bm{\Omega}_{|\mathcal{A}| - 1}\right)\left(\int_{0}^{+\infty}\!\mathrm{d}r\,r^{|\mathcal{A}|}\frac{1}{\left(2\pi\right)^{|\mathcal{A}|/2}}\exp\left(-\frac{r^2}{2}\right)\exp\left(2\tau r\sqrt{n}\right)\right)\nonumber\\
    & = \frac{2\pi^{|\mathcal{A}|/2}}{\Gamma\left(|\mathcal{A}|/2\right)}\int_{0}^{+\infty}\!\mathrm{d}r\,r^{|\mathcal{A}|}\frac{1}{\left(2\pi\right)^{|\mathcal{A}|/2}}\exp\left(-\frac{r^2}{2}\right)\exp\left(2\tau r\sqrt{n}\right)\nonumber\\
    & = \frac{2^{1 - |\mathcal{A}|/2}}{\Gamma\left(|\mathcal{A}|/2\right)}\int_{0}^{+\infty}\!\mathrm{d}r\,r^{|\mathcal{A}|}\exp\left(-\frac{r^2}{2}\right)\exp\left(2\tau r\sqrt{n}\right)\nonumber
\end{align}
where in the last but one line we used the formula for the volume of the $(d - 1)$-dimensional Euclidean sphere (embedded in $d$ dimensional Euclidean space): $2\pi^{d/2}/\Gamma(d/2)$. However, the last bound is diverging in the dimension; indeed, writing
\begin{align}
    \exp\left(2\tau r\sqrt{n}\right) & \geq \frac{1}{\left(2m - 1\right)!}\left(2\tau r\sqrt{n}\right)^{2m - 1}
\end{align}
for arbitrary positive integer $m \geq 1$, we get
\begin{align}
    & \frac{2^{1 - |\mathcal{A}|/2}}{\Gamma\left(|\mathcal{A}|/2\right)}\int_{0}^{+\infty}\!\mathrm{d}r\,r^{|\mathcal{A}|}\exp\left(-\frac{r^2}{2}\right)\exp\left(2\tau r\sqrt{n}\right)\nonumber\\
    & \geq \frac{2^{1 - |\mathcal{A}|/2}}{\Gamma\left(|\mathcal{A}|/2\right)}\int_{0}^{+\infty}\!\mathrm{d}r\,r^{|\mathcal{A}|}\exp\left(-\frac{r^2}{2}\right)\frac{\left(2\tau r\sqrt{n}\right)^{2m - 1}}{\left(2m - 1\right)!}\nonumber\\
    & = \frac{2^{1 - |\mathcal{A}|/2}}{\Gamma\left(|\mathcal{A}|/2\right)}\frac{\left(2\tau\sqrt{n}\right)^{2m - 1}}{\left(2m - 1\right)!}\int_{0}^{+\infty}\!\mathrm{d}r\,r^{|\mathcal{A}| + 2m - 1}\exp\left(-\frac{r^2}{2}\right)\nonumber\\
    & = \frac{2^{1 - |\mathcal{A}|/2}}{\Gamma\left(|\mathcal{A}|/2\right)}\frac{\left(2\tau\sqrt{n}\right)^{2m - 1}}{\left(2m - 1\right)!}2^{|\mathcal{A}|/2 + m - 1}\Gamma\left(|\mathcal{A}|/2 + m\right)\nonumber\\
    & = \frac{2^m\left(2\tau\sqrt{n}\right)^{2m - 1}}{\left(2m - 1\right)!}\frac{\Gamma\left(|\mathcal{A}|/2 + m\right)}{\Gamma\left(|\mathcal{A}|/2\right)}\nonumber\\
    & \geq \frac{2^m\left(2\tau\sqrt{n}\right)^{2m - 1}}{\left(2m - 1\right)!}\left(|\mathcal{A}|/2\right)^m.
\end{align}
This manifestly diverges as $|\mathcal{A}| \to \infty$ (and incidentally, as $n \to \infty$). In fact, optimizing $m$, one could show the divergence is at least of order $\exp\left(\Omega\left(|\mathcal{A}|^{1/2}\right)\right)$. As a result, the bounds on the absolute value of the integrand in the proof of lemma \ref{lemma:gaussian_integral_multinomial_sum_interchange} are insufficient to prove convergence to a limit in the infinite $p$ limit (where $|\mathcal{A}| \to \infty$). This is acceptable, since the only goal was to proven interchange of infinite multinomial sum and Gaussian integral. On the other hand, given the series expansion implied by this interchange, we will be able to bound the terms in the series in a way that does not depend explicitly on the dimension $|\mathcal{A}|$, but only on the total evolution time $\tau$.
\end{remark}

After proving lemma \ref{lemma:gaussian_integral_multinomial_sum_interchange} justifying Gaussian integral and (infinite) multinomial sum interchange, the desired integral can be expressed as

\begin{align}
    \left[\bm{S}^{(1)}_n\right]_{\alpha} & = \int_{\mathbf{R}^{\mathcal{A}}}\!\mathrm{d}\bm{\chi}\,\frac{e^{-\bm{\chi}^T\bm{\chi}/2}}{\left(2\pi\right)^{|\mathcal{A}|/2}}\chi_{\alpha}\left(\sum_{\bm{a} \in \mathcal{S}}Q_{\bm{a}}\exp\left(\lambda\bm{\theta}^{*T}\bm{L}_{:,\,\bm{a}} + \frac{1}{\sqrt{n}}\bm{\chi}^T\left(\lambda\bm{L}_{:,\,\bm{a}} - \bm{\theta}^*\right)\right)\right)^n\nonumber\\
    & = \left(\mathcal{Z}^*\right)^n\sum_{\left(n_d\right)_{d \geq 2}}\binom{n}{\left(n_d\right)_{d \geq 2}}\frac{n^{-\sum\limits_{d \geq 2}dn_d/2}}{\prod\limits_{d \geq 2}d!^{n_d}}\int_{\mathbf{R}^{\mathcal{A}}}\!\mathrm{d}\bm{\chi}\,\frac{e^{-\bm{\chi}^T\bm{\chi}/2}}{\left(2\pi\right)^{|\mathcal{A}|/2}}\chi_{\alpha}\left\langle \bm{\chi}^{\otimes \sum\limits_{d \geq 2}dn_d}, \bigotimes\limits_{d \geq 2}\bm{\delta C}^{\left(d\right)\otimes n_d} \right\rangle,\label{eq:qgms_integral_series_expansion_step_1}
\end{align}

It is now a standard calculation to evaluate the Gaussian integrals of polynomials in $\bm{\chi}$. We state the result in the following proposition for reference.

\begin{proposition}[Moments of standard normal distribution]
\label{prop:standard_normal_distribution_moments}
Consider the standard normal distribution of dimension $d$, given by density function:
\begin{align}
    \begin{array}{ccc}
         \mathbf{R}^d & \longrightarrow & \mathbf{R}\\
         \bm{x} = \left(x_j\right)_{j \in [d]} & \longmapsto & \left(2\pi\right)^{-d/2}\exp\left(-\frac{1}{2}\bm{x}^T\bm{x}\right).
    \end{array}
\end{align}
The moments of odd order $(2m + 1)$ ($m \geq 0$) of this distribution vanish:
\begin{align}
    \int_{\mathbf{R}^d}\!\mathrm{d}\bm{x}\,\left(2\pi\right)^{-d/2}\exp\left(-\frac{1}{2}\bm{x}^T\bm{x}\right)\prod_{1 \leq l \leq 2m + 1}x_{j_l} & = 0
\end{align}
for all $j_1,\,j_2,\,\ldots,\,j_{2m},\,j_{2m + 1} \in [d]$, while the moments of even order $2m$ ($m \geq 0$) can be expressed:
\begin{align}
    \int_{\mathbf{R}^d}\!\mathrm{d}\bm{x}\,\frac{1}{\left(2\pi\right)^{d/2}}\exp\left(-\frac{1}{2}\bm{x}^T\bm{x}\right)\prod_{1 \leq l \leq 2m}x_{j_l} & = \sum_{\substack{\mathcal{M}\mathrm{\,matching\,of\,}[2m]}}\prod_{\{l',\,l''\} \in \mathcal{M}}x_{j_{l'}}x_{j_{l''}},
\end{align}
where the sum is over matchings $\mathcal{M}$ of $[2m]$. A matching of $[2m]$ is a set of pairs of $[2m]$ covering $[2m]$ and such that each element of $[2m]$ occurs in exactly one pair. There are
\begin{align}
    \left(2m - 1\right)!! := (2m - 1)(2m - 3)\ldots 3.1 = \frac{\left(2m\right)!}{2^mm!}
\end{align}
distinct matchings of set $[2m]$. The above equation can be expressed in tensorial form:
\begin{align}
    \int_{\mathbf{R}^d}\!\mathrm{d}\bm{x}\,\frac{1}{\left(2\pi\right)^{d/2}}\exp\left(-\frac{1}{2}\bm{x}^T\bm{x}\right)\bm{x}^{\otimes 2m} & = \bm{\mathcal{I}}^{\left(2m\right)},
\end{align}
where we defined the tensor $\bm{\mathcal{I}}^{\left(2m\right)}$ of degree $2m$ by entries:
\begin{align}
    \mathcal{I}^{\left(2m\right)}_{j_1,\,j_2,\,\ldots,\,j_{2m - 1},\,j_{2m}} & := \sum_{\mathcal{M}\,\mathrm{\,matching\,of\,}[2m]}\prod_{\{l',\,l''\} \in \mathcal{M}}\mathbf{1}\left[j_{l'} = j_{l''}\right].\label{eq:matching_tensor_definition}
\end{align}
In the edge case $m = 0$, the above formula should be understood\footnote{This understanding is natural from the general definition in equation \ref{eq:matching_tensor_definition}, considering the only matching of the empty set is a set with no pair (i.e. the empty set).} as
\begin{align}
    \bm{\mathcal{I}}^{\left(0\right)} & = 1.
\end{align}
We call this tensor the \textnormal{matching tensor of degree $2m$}. One can also define\footnote{Note this choice is consistent with the general definition in equation \ref{eq:matching_tensor_definition} if one understands an odd-size set has no matching, i.e. the set of its matchings is empty.} the matching tensor of odd degree to be the all-zero tensor:
\begin{align}
    \bm{\mathcal{I}}^{\left(2m + 1\right)} & := \bm{0}_{\left(\mathbf{R}^{d}\right)^{\otimes (2m + 1)}} \qquad \forall m \geq 0,
\end{align}
allowing to write the Gaussian moment identities in a fully tensorial form:
\begin{align}
    \int_{\mathbf{R}^d}\!\mathrm{d}\bm{x}\,\exp\left(-\frac{1}{2}\bm{x}^T\bm{x}\right)x^{\otimes m} & = \bm{\mathcal{I}}^{\left(m\right)} \qquad \forall m \geq 0.
\end{align}
\end{proposition}

From proposition \ref{prop:standard_normal_distribution_moments}, the QGMS integral moment tensor of order 1 can be expanded as series:
\begin{align}
    \left[\bm{S}^{(1)}_n\right]_{\alpha} & = \int_{\mathbf{R}^{\mathcal{A}}}\!\mathrm{d}\bm{\chi}\,\frac{e^{-\bm{\chi}^T\bm{\chi}/2}}{\left(2\pi\right)^{|\mathcal{A}|/2}}\exp\left(-\frac{1}{2}\bm{\chi}^T\bm{\chi}\right)\chi_{\alpha}\left(\sum_{\bm{a} \in \mathcal{S}}Q_{\bm{a}}\exp\left(\lambda\bm{\theta}^{*T}\bm{L}_{:,\,\bm{a}} + \frac{1}{\sqrt{n}}\bm{\chi}^T\left(\lambda\bm{L}_{:,\,\bm{a}} - \bm{\theta}^*\right)\right)\right)^n\nonumber\\
    & = \left(\mathcal{Z}^*\right)^n\sum_{\left(n_d\right)_{d \geq 2}}\binom{n}{\left(n_d\right)_{d \geq 2}}\frac{n^{-\sum\limits_{d \geq 2}dn_d/2}}{\prod\limits_{d \geq 2}d!^{n_d}}\left\langle \bm{\mathcal{I}}^{\left(1 + \sum\limits_{d \geq 2}dn_d\right)}_{\alpha}, \bigotimes_{d \geq 2}\bm{\delta C}^{\left(d\right)\otimes n_d}\right\rangle,\label{eq:qgms_integral_series_expansion_step_2}
\end{align}
where $\bm{\mathcal{I}}^{\left(1 + m\right)}_{\alpha}$ refers to tensor $\bm{\mathcal{I}}^{\left(m + 1\right)}$ with first index set to $\alpha$, thereby defining a degree $m$ tensor (more specifically, a slice of the original tensor). Note that since $\bm{\mathcal{I}}^{\left(1 + m\right)}$ is symmetric, the choice of index (first in this case) to project to $\alpha$ is relevant; besides, the resulting tensor $\bm{\mathcal{I}}^{\left(1 + m\right)}_{\alpha}$ is symmetric. We collect this result in the following Proposition ---in fact generalizing to a QGMS moments of any order for completeness:

\begin{proposition}[QGMS second-order moment as series in correlations tensors]
\label{prop:qgms_integral_series_expansion}
Consider the second order moment of the QGMS of index $\left(\alpha, \alpha\right)$, whose representation in terms of QGMS integral moments tensors is given by:
\begin{align}
    \frac{\partial^2S_n\left(\bm{\mu}\right)}{\partial\mu_{\alpha}^2}\Bigg|_{\bm{\mu} = \bm{0}} & = \nu^{(0)}_{\alpha} + \frac{2}{\sqrt{n}}\nu^{(1)}_{\alpha} + \frac{1}{n}\nu^{(2)}_{\alpha},\label{eq:qgms_second_order_moment_integral_additive_contributions}\\
    \nu^{(0)}_{\alpha} & := \left(\theta^*_{\alpha}\right)^2S_n^{(0)},\\
    \nu^{(1)}_{\alpha} & := \theta^*_{\alpha}\left[\bm{S}^{(1)}_n\right]_{\alpha},\\
    \nu^{(2)}_{\alpha} & := \left[\bm{S}^{(2)}_n\right]_{\alpha,\,\alpha} - 1,
\end{align}
where the QGMS integral moments tensors $\bm{S}_n^{(k)}$ admit convergent series expansion:
\begin{align}
    \left[\bm{S}^{(k)}_n\right]_{\bm{\alpha}_{1:k}} & = \left(\mathcal{Z}^*\right)^n\sum_{\substack{\left(n_d\right)_{d \geq 2}}}\binom{n}{\left(n_d\right)_{d \geq 2}}\frac{n^{-\sum\limits_{d \geq 2}dn_d/2}}{\prod\limits_{d \geq 2}d!^{n_d}}\left\langle \bm{\mathcal{I}}_{\bm{\alpha}_{1:k}}^{\left(k + \sum\limits_{d \geq 2}dn_d\right)}, \bigotimes_{d \geq 2}\bm{\delta C}^{(d)\otimes n_d} \right\rangle,\\
    \bm{\alpha}_{1:k} & = \left(\alpha_1, \ldots, \alpha_k\right) \in \mathcal{A}^k.
\end{align}
In particular, the contribution $\nu^{(1)}_{\alpha}$ to the diagonal second order QGMS moment $\left(\alpha, \alpha\right)$ can be expanded as series:
\begin{align}
    \nu^{(1)}_{\alpha} & = \left(\mathcal{Z}^*\right)^n\sum_{\left(n_d\right)_{d \geq 2}}\binom{n}{\left(n_d\right)_{d \geq 2}}\frac{n^{-\sum\limits_{d \geq 2}dn_d/2}}{\prod\limits_{d \geq 2}d!^{n_d}}\theta^*_{\alpha}\left\langle \bm{\mathcal{I}}_{\alpha}, \bigotimes_{d \geq 2}\bm{\delta C}^{(d)\otimes n_d} \right\rangle.\label{eq:qgms_moment_contribution_series_expansion}
\end{align}
\end{proposition}

It is instructive to derive a bound on this expression and compare it with the one from remark \ref{rem:integral_dimension_dependence_integral_sum_inversion}. The bounds described there was diverging superpolynomially with $|\mathcal{A}|$, forbidding to take the infinite number of layers limit (even at constant time). To bound equation \ref{eq:qgms_integral_series_expansion_step_2}, we first consider a single term of the sum $\left(n_d\right)_{d \geq 2}$. For such a term, letting $D := \sum_{d \geq 2}dn_d$ we express the matching tensor $\bm{\mathcal{I}}^{\left(1 + D\right)}$ explicitly as follows:
\begin{align}
    \mathcal{I}^{\left(1 + D\right)}_{\alpha,\,\alpha_1,\,\ldots,\,\alpha_{D}} & = \sum_{l''' \in [D]}\sum_{\substack{\mathcal{M}\textrm{ matching of }[D] - \{l'''\}}}\mathbf{1}\left[\alpha = \alpha_{l'''}\right]\prod_{\{l',\,l''\} \in \mathcal{M}}\mathbf{1}\left[\alpha_{l'} = \alpha_{l''}\right]\\
    & =: \sum_{l''' \in [D]}\sum_{\substack{\mathcal{M}\textrm{ matching of }[D] - \{l'''\}}}\mathcal{I}^{(D + 1),\,l''',\,\mathcal{M}}_{\alpha,\,\alpha_1,\,\ldots,\,\alpha_D}.\label{eq:matching_tensor_slice_decomposition}
\end{align}
Namely, the outer sum over $l''' \in [D]$ refers to the index matched to $\alpha$; the inner sum is over matchings $\mathcal{M}$ of remaining unmatched elements. The number of terms in this decomposition is $D!! = 2^{-(D + 1)/2}(D + 1)!/((D + 1)/2)!$ ---the number of matchings of $(D + 1)$ elements. By linearity, let us consider the contribution of a single term of this decomposition to the dot product from equation \ref{eq:qgms_integral_series_expansion_step_2}:
\begin{align}
    \left\langle \bm{\mathcal{I}}^{(D + 1),\,l''',\,\mathcal{M}}_{\alpha}, \bigotimes_{d \geq 2}\bm{\delta C}^{(d)\otimes n_d} \right\rangle & = \sum_{\bm{\alpha}_{1:D} \in \mathcal{A}^D}\left[\bigotimes_{d \geq 2}\bm{\delta C}^{(d)\otimes n_d}\right]_{\bm{\alpha}_{1:D}}\mathcal{I}^{(D + 1),\,l'',\,\mathcal{M}}_{\alpha,\,\alpha_1,\,\ldots,\,\alpha_D}\nonumber\\
    & = \sum_{\bm{\alpha}_{1:D} \in \mathcal{A}^D}\left[\bigotimes_{d \geq 2}\bm{\delta C}^{(d)\otimes n_d}\right]_{\bm{\alpha}_{1:D}}\mathbf{1}\left[\alpha = \alpha_{l'''}\right]\prod_{\{l',\,l''\} \in \mathcal{M}}\mathbf{1}\left[\alpha_{l'} = \alpha_{l''}\right].
\end{align}
By the uniform bound on centered correlations (equation \ref{eq:centered_correlations_bound}) and the triangular inequality, the above sum can be bounded as:
\begin{align}
    \left|\left\langle \bm{\mathcal{I}}^{(D + 1),\,l''',\,\mathcal{M}}_{\alpha}, \bigotimes_{d \geq 2}\bm{\delta C}^{(d)\otimes n_d} \right\rangle\right| & \leq 2^D|\lambda|^D|\mathcal{A}|^{(D - 1)/2}.
\end{align}
The important point is the exponent of $|\mathcal{A}|$, indicating summation over $(D - 1)/2$ only thanks to the indicator functions. By multiplying this by the number of terms $D!!$ in sum \ref{eq:matching_tensor_slice_decomposition} and incorporating relevant prefactor, we obtain the following bound for a single $(n_d)_{d \geq 2}$ terms of sum \ref{eq:qgms_integral_series_expansion_step_2}:
\begin{align}
    \left|\left(\mathcal{Z}^*\right)^n\binom{n}{\left(n_d\right)_{d \geq 2}}\frac{n^{-\sum\limits_{d \geq 2}dn_d/2}}{\prod\limits_{d \geq 2}d!^{n_d}}\left\langle \bm{\mathcal{I}}^{\left(1 + \sum\limits_{d \geq 2}dn_d\right)}_{\alpha}, \bigotimes_{d \geq 2}\bm{\delta C}^{\left(d\right)\otimes n_d}\right\rangle\right| & \leq |\mathcal{Z}^*|^n|\mathcal{A}|^{-1/2}\frac{\left(|\lambda|^2|\mathcal{A}|n^{-1}\right)^{D/2}}{\prod\limits_{d \geq 2}d!^{n_d}}D!!\nonumber\\
    & = |\mathcal{Z}^*|^n|\mathcal{A}|^{-1/2}\int_{\mathbf{R}}\!\mathrm{d}x\,\frac{e^{-x^2/2}}{\sqrt{2\pi}}\frac{\left(|\lambda|^2|\mathcal{A}|n^{-1}\right)^{D/2}}{\prod\limits_{d \geq 2}d!^{n_d}}x^{D + 1}\nonumber\\
    & = |\mathcal{Z}^*|^n|\mathcal{A}|^{-1/2}\int_{\mathbf{R}}\!\mathrm{d}x\,\frac{e^{-x^2/2}}{\sqrt{2\pi}}\frac{\left(|\lambda|^2|\mathcal{A}|n^{-1}\right)^{D/2}}{\prod\limits_{d \geq 2}d!^{n_d}}|x|^{D + 1},\label{eq:qgms_integral_series_expansion_bound_step_1}
\end{align}
where in the last-but-one line, we introduce a Gaussian integral to express the double factor as an average of exponentials, and in the final line, we used that $(D + 1)$ is even to write $x^{D + 1} = |x|^{D + 1}$ for all $x \in \mathbf{R}$. This concludes the bound for a single term $\left(n_d\right)_{d \geq 2}$ of the sum in equation \ref{eq:qgms_integral_series_expansion_step_2}. We may now  sum the bound over $\left(n_d\right)_{d \geq 2}$ (ignoring the $(D + 1)$ even constraint for simplicity), giving the following bound on equation \ref{eq:qgms_integral_series_expansion_step_2}:
\begin{align}
    & \left|\left(\mathcal{Z}^*\right)^n\sum_{\left(n_d\right)_{d \geq 2}}\binom{n}{\left(n_d\right)_{d \geq 2}}\frac{n^{-D/2}}{\prod\limits_{d \geq 2}d!^{n_d}}\left\langle \bm{\mathcal{I}}^{\left(D + 1\right)}_{\alpha}, \bigotimes_{d \geq 2}\bm{\delta C}^{\left(d\right)\otimes n_d}\right\rangle\right|\nonumber\\
    & \leq |\mathcal{Z}^*|^n|\mathcal{A}|^{-1/2}\int_{\mathbf{R}}\!\mathrm{d}x\,\frac{e^{-x^2/2}}{\sqrt{2\pi}}\sum_{\left(n_d\right)_{d \geq 2}}\binom{n}{\left(n_d\right)_{d \geq 2}}\frac{\left(|\lambda|^2|\mathcal{A}|n^{-1}\right)^{\sum\limits_{d \geq 2}dn_d/2}}{\prod\limits_{d \geq 2}d!^{n_d}}|x|^{1 + D}\nonumber\\
    & = |\mathcal{Z}^*|^n|\mathcal{A}|^{-1/2}\int_{\mathbf{R}}\!\mathrm{d}x\,\frac{e^{-x^2/2}}{\sqrt{2\pi}}|x|\left(1 + \sum_{d \geq 2}\frac{\left(|\lambda||\mathcal{A}|^{1/2}n^{-1/2}x\right)^d}{d!}\right)^n\nonumber\\
    & = |\mathcal{Z}^*|^n|\mathcal{A}|^{-1/2}\int_{\mathbf{R}}\!\mathrm{d}x\,\frac{e^{-x^2/2}}{\sqrt{2\pi}}|x|\left(\exp\left(|\lambda||\mathcal{A}|^{1/2}n^{-1/2}x\right) - |\lambda||\mathcal{A}|^{1/2}n^{-1/2}x\right)^n\nonumber\\
    & = |\mathcal{Z}^*|^n|\mathcal{A}|^{-1/2}\int_{\mathbf{R}}\!\mathrm{d}x\,\frac{e^{-x^2/2}}{\sqrt{2\pi}}|x|\left(\exp\left(\tau_{\gamma}n^{-1/2}x\right) - \tau_{\gamma}n^{-1/2}x\right)^n
\end{align}
We now show that for sufficiently small $\tau_{\gamma}$, the integrand is uniformly bounded by
\begin{align}
    \frac{e^{-x^2/4}}{\sqrt{2\pi}}.
\end{align}
For that purpose, we invoke the bounds from lemma \ref{lemma:exponential_bound}, setting $c = 1$ there. For $|x| \leq \tau_{\gamma}^{-1}n^{1/2}$, using bound \ref{eq:exponential_bound_small_r},
\begin{align}
    \left|\left(\exp\left(\tau_{\gamma}n^{-1/2}x\right) - \tau_{\gamma}n^{-1/2}x\right)^n\right| & \leq \exp\left(\frac{\tau_{\gamma}^2x^2}{2}e^{\tau_{\gamma}n^{-1/2}x}\right)\nonumber\\
    & \leq \exp\left(\frac{e\tau_{\gamma}^2}{2}x^2\right)\nonumber\\
    & \leq \exp\left(\frac{x^2}{4}\right)
\end{align}
assuming
\begin{align}
    \tau_{\gamma} & \leq \frac{1}{\sqrt{2e}}.
\end{align}
This proves the required bound on the integrand in domain $|x| \leq \tau_{\gamma}^{-1}n^{1/2}$. Next, for $|x| \geq \tau_{\gamma}^{-1}n^{1/2}$, using bound \ref{eq:exponential_bound_large_r}
\begin{align}
    \left|\left(\exp\left(\tau_{\gamma}n^{-1/2}x\right) - \tau_{\gamma}n^{-1/2}x\right)^n\right| & \leq \exp\left(\tau_{\gamma}n^{1/2}|x|\right)\nonumber\\
    & \leq \exp\left(\tau_{\gamma}^2x^2\right)\nonumber\\
    & \leq \exp\left(\frac{x^2}{4}\right)
\end{align}
as long as
\begin{align}
    \tau_{\gamma} & \leq \frac{1}{2}.
\end{align}
Plugging these estimates into equation \ref{eq:qgms_integral_series_expansion_bound_step_1} gives the following bound on the integral representation of the QGMS moment:
\begin{align}
    \left|\left(\mathcal{Z}^*\right)^n\sum_{\left(n_d\right)_{d \geq 2}}\binom{n}{\left(n_d\right)_{d \geq 2}}\frac{n^{-D/2}}{\prod\limits_{d \geq 2}d!^{n_d}}\left\langle \bm{\mathcal{I}}^{\left(1 + D\right)}_{\alpha}, \bigotimes_{d \geq 2}\bm{\delta C}^{\left(d\right)\otimes n_d}\right\rangle\right| & \leq \left|\mathcal{Z}^*\right|^n|\mathcal{A}|^{-1/2}\int_{\mathbf{R}}\!\mathrm{d}x\,\frac{e^{-x^2/2}}{\sqrt{2\pi}}|x|e^{x^2/4}\nonumber\\
    & = \left|\mathcal{Z}^*\right|^n|\mathcal{A}|^{-1/2}\int_{\mathbf{R}}\!\mathrm{d}x\,\frac{e^{-x^2/4}}{\sqrt{2\pi}}|x|\nonumber\\
    & = \left|\mathcal{Z}^*\right|^n|\mathcal{A}|^{-1/2}\frac{4}{\sqrt{2\pi}}.
\end{align}
The bound no longer has exponential dependence on the QMGS integral's dimension $|\mathcal{A}|$. The difference with the bound from remark \ref{rem:integral_dimension_dependence_integral_sum_inversion} is that in the latter, we used the Cauchy-Schwartz inequality to bound dot product
\begin{align}
    \left\langle \bm{\mathcal{I}}^{\left(D + 1\right)}_{\alpha}, \bigotimes_{d \geq 2}\bm{\delta C}^{(d)\otimes n_d} \right\rangle.
\end{align}
Using this bound was convenient and sufficient to prove the sum-integral inversion result in lemma \ref{lemma:gaussian_integral_multinomial_sum_interchange}. However, in doing so, we did not use the sparsity of the matching tensor, leading to overestimate the dependency in the dimension. With that said, note the remaining polynomial dependence $|\mathcal{A}|^{-1/2}$. When specializing the analysis to the QGMS evaluation the SK-QAOA energy (section \ref{sec:main_theorem_derivation}), we will see this dependency ultimately cancels out of the energy.

\section{The continuum limit of the SK-QAOA energy}
\label{sec:sk_qaoa_energy_continuum_limit}

In this section, we specialize to the SK-QAOA the series expansion of Parametrized Quadratic Generalized Multinomial Sums (PQGMS) obtained in Appendix~\ref{sec:qgms_moments_series_expansion} in terms of correlations tensors $\bm{C}^{(d)}$. We apply this expansion to the QGMS producing the SK-QAOA energy, as derived in Section~\ref{sec:sk_qaoa_qgms}. By establishing a continuum limit term by term for this expansion, we then reach the main technical result of the manuscript Theorem~\ref{th:approximation_continuous_time_annealing_qaoa}. The exposition is organized as follows. First, Section~\ref{sec:qgms_moments_series_expansion_sk_qaoa} specializes to the SK-QAOA energy QGMS the series expansion of PQGMS around the noninteracting limit developed for a generic PQGMS in earlier Appendix Section~\ref{sec:qgms_moments_series_expansion}. Next, Section~\ref{sec:main_theorem_derivation} outlines the main argument in the proof of main Theorem~\ref{th:approximation_continuous_time_annealing_qaoa}. For clarity, we omit the tedious the tedious part of establishing continuum limits for all intermediate objects defining the QGMS. Instead, we rely on the subsequently proven result that QGMS moments $\left(\nu_{\alpha}\right)_{\alpha \in \mathcal{A}}$ admit continuum limits in the case of the SK-QAOA energy QGMS. The more technical part of establishing continuum limits is deferred to Section~\ref{sec:sk_qaoa_continuum_limit}. The content of the present Appendix is specific to the QGMS representing the SK-QAOA instance-averaged energy, although we consider it plausibly extensible to a wider variety of QGMS.

\subsection{Expansion of QMGS moments around the saddle point: specialization to SK-QAOA}
\label{sec:qgms_moments_series_expansion_sk_qaoa}

In section \ref{sec:qgms_moments_series_expansion}, we established a representation for (particular contribution to) a QGMS moment as a series involving dot products of correlation tensors (proposition \ref{prop:qgms_integral_series_expansion}). In this section, we specialize this representation to the QGMS evaluating the SK-QAOA energy. Note that while the representation was derived for a special contribution to the order 1 moment, calculations extend straightfowardly to other moments; we will therefore continue with the same special example in the current section.

We begin by defining the parameters of the (parametrized) QGMS under consideration. We assume an angles schedule $\bm\gamma, \bm\beta$ derived from continuum schedules $\gamma^{\mathrm{opt}}, \widetilde{\beta}^{\mathrm{opt}}$ as per definition \ref{def:angles_from_continuum}:
\begin{align}
    \gamma_t & := \frac{1}{p + 1}\gamma^{\mathrm{cont}}\left(\frac{t - 1}{p + 1/2}\right) & \forall 1 \leq t \leq p,\\
    \beta_t & := -\int_{(t - 1)/(p + 1/2)}^{t/(p + 1/2)}\!\mathrm{d}x\,\widetilde{\beta}^{\mathrm{cont}}\left(x\right) & \forall 1 \leq t \leq p,
\end{align}
where $\gamma^{\mathrm{cont}}: [0, 1] \longrightarrow \mathbf{R}$ is a $M_{\gamma}$-Lipschitz (hence continuous) function and $\widetilde{\beta}^{\mathrm{cont}}: [0, 2] \longrightarrow \mathbf{R}$ is a function odd about $1$:
\begin{align}
    \widetilde{\beta}^{\mathrm{cont}}\left(1 - x\right) = -\widetilde{\beta}^{\mathrm{cont}}\left(x\right),
\end{align}
and continuous on $[0, 1)$ and $(1, 2]$ separately; thanks to this separate continuity assumption, one needs not assume $\widetilde{\beta}^{\mathrm{cont}}\left(1^-\right) = 0$. The dissymmetry between definitions of $\bm\gamma$ and $\bm\beta$ angles may be perplexing, but will actually simplify the analysis. This proper specification of the schedule now makes the statement of theorem \ref{th:approximation_continuous_time_annealing_qaoa} from the main text fully rigorous. Assuming a QAOA angles schedule defined by the above prescription, it holds
\begin{align}
    \Gamma_j & = \frac{1}{p + 1}\Gamma^{\mathrm{cont}}\left(\frac{j}{p + 1/2}\right) & \forall 0 \leq j \leq 2p + 1.
\end{align}
for some function $\Gamma^{\mathrm{cont}}$ defined in \ref{eq:Gamma_continuum} (see also proposition \ref{prop:angle_functions_discrete_from_continuum}). This function satisfies the same symmetries and continuity properties as $\widetilde{\beta}^{\mathrm{cont}}$ ---in particular it $M_{\gamma}$-Lipschitz separately on $[0, 1)$ and $(1, 2]$. From there, we introduce the maximum continuum $\gamma$:
\begin{align}
    \mathrm{\gamma}_{\mathrm{max}} & := \max_{x \in [0, 1]}\left|\gamma^{\mathrm{cont}}\left(x\right)\right|.\label{eq:gamma_max}
\end{align}

Throughout the analysis, it proves convenient to recast the SK-QAOA QGMS as a parametrized QGMS:
\begin{align}
    \sum_{\bm{n}}\binom{n}{\bm{n}}\exp\left(\frac{\lambda^2}{2n}\bm{n}^T\bm{L}^T\bm{L}\bm{n} + \frac{\lambda}{n}\bm{\mu}^T\bm{L}\bm{n}\right)\prod_{\bm{a} \in \mathcal{S}}Q_{\bm{a}}^{n_{\bm{a}}},
\end{align}
with QGMS vector $\bm{Q}$, QGMS matrix $\bm{L}$, and parametrized QGMS variable $\lambda$ defined as follows:
\begin{align}
    \mathcal{I} & := \{0, 1, \ldots, 2p, 2p + 1\},\label{eq:sk_qaoa_qgms_redefinition_rescaled_I_main_text}\\
    \mathcal{S} & := \{1, -1\}^{\mathcal{I}},\label{eq:sk_qaoa_qgms_redefinition_rescaled_S_main_text}\\
    \mathcal{A} & := \mathcal{I}^2,\label{eq:sk_qaoa_qgms_redefinition_rescaled_A_main_text}\\
    Q_{\bm{a}} & := \frac{1}{2}\mathbf{1}\left[a_p = a_{p + 1}\right]\prod_{1 \leq t \leq p}\braket{a_{2p + 2 - t}|e^{i\beta_tX}|a_{2p + 1 - t}}\braket{a_t|e^{-i\beta_tX}|a_{t - 1}},\label{eq:sk_qaoa_qgms_redefinition_rescaled_Q_main_text}\\
    L_{\left(r,\,s\right),\,\bm{a}} & = \sqrt{-\frac{\left(p + 1\right)^2\Gamma_r\Gamma_s}{\gamma_{\mathrm{max}}^2} + i\varepsilon}\,a_ra_s,\label{eq:sk_qaoa_qgms_redefinition_rescaled_L_main_text},\\
    \lambda & = \frac{2^{-1/2}\gamma_{\mathrm{max}}}{p + 1}.\label{eq:sk_qaoa_qgms_redefinition_rescaled_lambda_main_text}.
\end{align}
Given the special form of the $\mathcal{A}$ set: $\mathcal{A} = \mathcal{I}^2 = \{0, 1, \ldots, 2p, 2p + 1\}^2$, degree $d$ tensors indexed by $\mathcal{A}$ (for instance, the order $d$ correlations) can equivalently be regarded as degree $2d$ tensors indexed by $\mathcal{I}$. Hence, we may for instance write
\begin{align}
    C^{(d)}_{\alpha_1,\,\ldots,\,\alpha_d} & = C^{(d)}_{j_1,\,j_2,\,\ldots,\,j_{2d - 1},\,j_{2d}},
\end{align}
or (in vectorized notation)
\begin{align}
    C^{(d)}_{\bm{\alpha}_{1:d}} & = C^{(d)}_{\bm{j}_{1:2d}},
\end{align}
where
\begin{align}
    \bm{\alpha}_{1:d} & = \left(\alpha_1,\,\ldots,\,\alpha_d\right) \in \mathcal{A}^d,\\
    \bm{j}_{1:2d} & = \left(j_1,\,\ldots,\,j_{2d}\right) \in \mathcal{I}^{2d},\\
    \alpha_r & = \left(j_{2r - 1}, j_{2r}\right) \in \mathcal{A} = \mathcal{I}^2
\end{align}
To avoid confusion, we use Greek letters for $\mathcal{A}$ indices, and Roman ones for $\mathcal{I}$ indices.

A first convenient adaptation for the SK-QAOA energy QGMS is to work with $\bm{G}$ correlation tensors rather than correlation tensors. As we will see, the $\bm{G}$ correlation tensors have greater symmetry than standard correlation tensors. Besides, we will see $\bm{G}$ correlation tensors converge to continuous multivariate function in the continuum limit developed in section \ref{sec:sk_qaoa_continuum_limit}, unlike the correlations tensors which only converge to piecewise continuous functions. Specifically, the $\bm{G}$ correlations tensor of order $2d$ is related to the correlations tensor $\bm{C}^{(d)}$ by elementwise multiplication with a tensor $\bm{\mathcal{G}}^{(2d)}$. Restating appendix definition \ref{def:g_correlations_tensor_rescaled}, we let
\begin{align}
    \mathcal{G}^{(2d)}_{\bm{j}_{1:2d}} & := \prod_{1 \leq r \leq d}\sqrt{-\frac{\Gamma_{j_{2r - 1}}\Gamma_{j_{2r}}}{\gamma_{\mathrm{max}}^2} + i\varepsilon},\label{eq:curvy_g_tensor_definition_rescaled_main_text}
\end{align}
and also define a closely related tensor
\begin{align}
    \mathcal{G}_j & := \frac{i\Gamma_j}{\gamma_{\mathrm{max}}}.\label{eq:curvy_g_vector_definition_rescaled_main_text}
\end{align}
Namely,
\begin{align}
    \left(\mathcal{G}^{(2d)}_{\bm{j}_{1:2d}}\right)^2 & = \prod_{1 \leq r \leq d}\mathcal{G}_{j_{2r - 1}}\mathcal{G}_{j_{2r}}.
\end{align}
From there, the $\bm{G}$ correlations tensor of order $2d$ is defined by
\begin{align}
    C^{(d)}_{\bm{\alpha}_{1:d}} & =: \lambda^d\mathcal{G}^{(2d)}_{\bm{\alpha}_{1:d}}G^{(2d)}_{\bm{\alpha}_{1:d}} \qquad \forall \bm{\alpha}_{1:d} \in \mathcal{A}^d\label{eq:g_correlations_tensor_definition_rescaled_alpha_indices_main_text}
\end{align}
using $\mathcal{A}$ indices, or equivalently
\begin{align}
    C^{(2d)}_{\bm{j}_{1:2d}} & =: \lambda^d\mathcal{G}^{(2d)}_{\bm{j}_{1:2d}}G^{(2d)}_{\bm{j}_{1:2d}} \qquad \forall \bm{j}_{1:2d} \in \mathcal{I}^{2d}\label{eq:g_correlations_tensor_definition_rescaled_main_text}
\end{align}
using $\mathcal{I}$ indices. Expanding $\bm{\mathcal{G}}$ from its definition in equation \ref{eq:curvy_g_tensor_definition_rescaled_main_text}, the last equation also reads
\begin{align}
    C^{\left(d\right)}_{\bm{j}_{1:2d}} & =: G^{(2d)}_{\bm{j}_{1:2d}}\prod_{1 \leq r \leq d}\sqrt{-\frac{\Gamma_{j_{2r - 1}}\Gamma_{j_{2r}}}{2} + i\varepsilon}.
\end{align}
It will be instructive to rephrase the saddle-point equation:
\begin{align}
    \bm{\theta}^* & = \frac{\sum\limits_{\bm{a} \in \mathcal{S}}Q_{\bm{a}}\exp\left(\lambda\bm{\theta}^{*T}\bm{L}_{:,\,\bm{a}}\right)\lambda\bm{L}_{:,\,\bm{a}}}{\sum\limits_{\bm{a} \in \mathcal{S}}Q_{\bm{a}}\exp\left(\lambda\bm{\theta}^{*T}\bm{L}_{:,\,\bm{a}}\right)}.
\end{align}
in terms of the $\bm{G}$ correlations. First, the saddle-point equation can be read as:
\begin{align}
    \bm{\theta}^* & = \bm{C}^{(1)},
\end{align}
where the $\bm{C}^{(1)}$ correlations tensor is understood as a function of $\bm{\theta}^*$. Letting then
\begin{align}
    \theta^*_{j_1,\,j_2} & := \lambda\mathcal{G}^{(2)}_{j_1,\,j_2}G^{(2)}_{j_1,\,j_2}\nonumber\\
    & = \sqrt{-\frac{\Gamma_{j_1}\Gamma_{j_2}}{2} + i\varepsilon}\,G^{(2)}_{j_1,\,j_2}
\end{align}
in the saddle-point equation, and well as plugging the explicit expressions of $\lambda$ (equation \ref{eq:sk_qaoa_qgms_redefinition_rescaled_lambda_main_text}) and $\bm{L}$, it becomes
\begin{align}
    \sqrt{-\frac{\Gamma_{j_1}\Gamma_{j_2}}{2} + i\varepsilon}\,G^{(2)}_{j_1,\,j_2} & = \frac{\sum\limits_{\bm{a} \in \mathcal{S}}Q_{\bm{a}}\exp\left(-\frac{1}{2}\sum\limits_{j_3,\,j_4 \in \mathcal{I}}G^{(2)}_{j_3,\,j_4}\Gamma_{j_3}\Gamma_{j_4}a_{j_3}a_{j_4}\right)\sqrt{-\frac{\Gamma_{j_1}\Gamma_{j_2}}{2} + i\varepsilon}\,a_{j_1}a_{j_2}}{\sum\limits_{\bm{a} \in \mathcal{S}}Q_{\bm{a}}\exp\left(-\frac{1}{2}\sum\limits_{j_3,\,j_4 \in \mathcal{I}}G^{(2)}_{j_3,\,j_4}\Gamma_{j_3}\Gamma_{j_4}a_{j_3}a_{j_4}\right)}.
\end{align}
Simplifying by $\sqrt{-\Gamma_{j_1}\Gamma_{j_2}/2 + i\varepsilon}$ and renaming indices, this is equivalent to:
\begin{align}
    G^{(2)}_{j,\,k} & = \frac{\sum\limits_{\bm{a} \in \mathcal{S}}Q_{\bm{a}}\exp\left(-\frac{1}{2}\sum\limits_{r,\,s \in \mathcal{I}}G^{(2)}_{r,\,s}\Gamma_r\Gamma_sa_ra_s\right)a_ja_k}{\sum\limits_{\bm{a} \in \mathcal{S}}Q_{\bm{a}}\exp\left(-\frac{1}{2}\sum\limits_{r,\,s \in \mathcal{I}}G^{(2)}_{r,\,s}\Gamma_r\Gamma_sa_ra_s\right)}.\label{eq:g_matrix_fixed_point_equation_current_work_step_1}
\end{align}
Hence, the saddle-point equation can be regarded as an equation in variable $\bm{G}^{(2)}$ rather than $\bm{\theta}^*$. Given a solution $\bm{G}^{(2)}$ to this equation, a solution $\bm{\theta}^*$ to the original saddle-point equation is given by setting
\begin{align}
    \theta^*_{j_1,\,j_2} & := \sqrt{-\frac{\Gamma_{j_1}\Gamma_{j_2}}{2} + i\varepsilon}\,G^{(2)}_{j_1,\,j_2}.
\end{align}

Up to a relabelling of indices, this is very similar to the fixed-point equation for the $\bm{G}$ matrix characterizing the energy of constant-$p$ SK-QAOA in the infinite size limit \cite[adapted from eq. (3.8)]{qaoa_maxcut_high_girth}:
\begin{align}
    G_{j,\,k} & = \sum_{\bm{a} \in \{1, -1\}^{2p + 1}}a_ja_kf\left(\bm{a}\right)\exp\left(-\frac{1}{2}\sum_{-p \leq r, s \leq p}G_{r, s}\Gamma_r\Gamma_sa_ra_s\right).\label{eq:g_matrix_fixed_point_equation_earlier_work}
\end{align}
The quantity $f\left(\bm{a}\right)$ defined in \cite{qaoa_maxcut_high_girth} is the same as the quantity $Q_{\bm{a}}$ in this work. Besides, \cite{qaoa_maxcut_high_girth} consistently indexes bitstring bits (and related tensors) with index set $\{-p, \ldots, -1, 0, 1, \ldots, p\}$, unlike the present section of this work where index set $\{0, \ldots, 2p + 1\}$ makes it more convenient to take the continuum limit. To see that the $\bm{G}$ matrix computed in \cite{qaoa_maxcut_high_girth} satisfies equation \ref{eq:g_matrix_fixed_point_equation_current_work_step_1} in the present work, we use \cite[preprint version, lemma 5]{qaoa_maxcut_high_girth} establishing:
\begin{align}
    \sum_{\bm{a} \in \{1, -1\}^{2p + 1}}f\left(\bm{a}\right)\exp\left(-\frac{1}{2}\sum_{-p \leq r, s\leq p}G_{r,\,s}\Gamma_r\Gamma_sa_ra_s\right) & = 1.
\end{align}
Taking the $\bm{G}$ matrix defined in \cite{qaoa_maxcut_high_girth} as candidate for $\bm{G}^{(2)}$ in the current work (with appropriate reindexing) then gives corresponding identity 
\begin{align}
    \sum_{\bm{a} \in \mathcal{S}}Q_{\bm{a}}\exp\left(-\frac{1}{2}\sum_{j_3, j_4 \in \mathcal{I}}G^{(2)}_{j_3,\,j_4}\Gamma_{j_3}\Gamma_{j_4}a_{j_3}a_{j_4}\right) & = 1.
\end{align}
Note this statement is equivalent to
\begin{align}
    \mathcal{Z}^* & = \sum_{\bm{a} \in \mathcal{S}}Q_{\bm{a}}\exp\left(\lambda\bm{\theta}^{*T}\bm{L}_{:,\,\bm{a}}\right) = 1\label{eq:sk_qaoa_zstar_1}.
\end{align}
This equality turns saddle-point equation \ref{eq:g_matrix_fixed_point_equation_current_work_step_1} (rephrased in variable $\bm{G}^{(2)}$) into:
\begin{align}
    G^{(2)}_{j,\,k} & = \sum_{\bm{a} \in \mathcal{S}}a_ja_kQ_{\bm{a}}\exp\left(-\frac{1}{2}\sum_{r, s \in \mathcal{I}}G^{(2)}_{r,\,s}\Gamma_r\Gamma_sa_ra_s\right),\label{eq:g_matrix_fixed_point_equation_current_work}
\end{align}
which is indeed equivalent to equation \ref{eq:g_matrix_fixed_point_equation_earlier_work} adapted from \cite{qaoa_maxcut_high_girth} up to notation changes. All in all, the $\bm{G}$ matrix defined in \cite{qaoa_maxcut_high_girth} solves the saddle-point equation \ref{eq:g_matrix_fixed_point_equation_current_work_step_1} (rephrased in variable $\bm{G}^{(2)}$) for the SK-QAOA energy QGMS. We now derive a formula for the $\bm{G}$ correlations of order $2d$ (equation \ref{eq:g_correlations_tensor_definition_rescaled_main_text}) in terms of $\bm{G}^{(2)}$. Starting with the general definition of the correlation tensor of order $d$:
\begin{align}
    C^{(d)}_{\bm{\alpha}_{1:d}} & = \sum_{\bm{a} \in \mathcal{S}}Q_{\bm{a}}\exp\left(\bm{\theta}^{*T}\bm{L}_{:,\,\bm{a}}\right)\prod_{1 \leq r \leq d}\lambda\bm{L}_{\alpha_r,\,\bm{a}},
\end{align}
letting
\begin{align}
    \alpha_r & := \left(j_{2r - 1}, j_{2r}\right) \in \mathcal{I}^2
\end{align}
and plugging the formula of correlations in terms of $\bm{G}$ correlations (equation \ref{eq:g_correlations_tensor_definition_rescaled_main_text}), the last equation becomes:
\begin{align}
    G^{(2d)}_{\bm{j}_{1:2d}} & = \sum_{\bm{a} \in \mathcal{S}}Q_{\bm{a}}\exp\left(-\frac{1}{2}\sum_{k_1, k_2 \in \mathcal{I}}G^{(2)}_{k_1,\,k_2}\Gamma_{k_1}\Gamma_{k_2}a_{k_1}a_{k_2}\right)\prod_{1 \leq r \leq d}a_{j_{2r - 1}}a_{j_{2r}}\nonumber\\
    & = \sum_{\bm{a} \in \mathcal{S}}Q_{\bm{a}}\exp\left(-\frac{1}{2}\sum_{k_1, k_2 \in \mathcal{I}}G^{(2)}_{k_1,\,k_2}\Gamma_{k_1}\Gamma_{k_2}a_{k_1}a_{k_2}\right)\prod_{1 \leq r \leq 2d}a_{j_r}.
\end{align}
From the last expression, it appears $\bm{G}^{(2d)}$ is entirely symmetric in its $2d$ indices. On the other hand $C^{(2d)}_{\bm{j}_{1:2d}}$ is only symmetric under permutation of tuples $\left(j_1, j_2\right),\,\ldots,\,\left(j_{2d - 1}, j_{2d}\right)$ between themselves, and permutation of elements within a given tuple. Also, note the equality also makes sense for $d = 1$, where it corresponds to the saddle-point equation in variable (equation \ref{eq:g_matrix_fixed_point_equation_current_work}). Having defined $\bm{G}$ correlations tensors, one may now introduce the centered $\bm{G}$ correlations tensor of order $2d$, related to the centered correlations tensor of order $d$ (definition \ref{def:centered_correlations_tensor}) by the same elementwise multiplication by $\mathcal{\bm{G}}$:

\begin{definition}[Centered $\bm{G}$ correlations tensor for SK-QAOA]
\label{def:centered_g_correlations_tensor}
In the context of the SK-QAOA QGMS, the centered $\bm{G}$ correlations tensor of order $2d$, denoted $\bm{\delta C}^{(2d)}$, is related to the centered correlations tensor of order $d$ (definition \ref{def:centered_correlations_tensor}) as follows:
\begin{align}
    \delta C^{(d)}_{\bm{j}_{1:2d}} & =: \lambda^d\mathcal{G}^{(2d)}_{\bm{j}_{1:2d}}\delta G^{(d)}_{\bm{j}_{1:2d}}.
\end{align}
From the fundamental definition of centered correlations (definition \ref{def:centered_correlations_tensor}), and the expression of correlation themselves (definition \ref{def:c_tensor}), a generic entry of the centered $\bm{G}$ correlations tensor can be expressed:
\begin{align}
\delta G^{(2d)}_{\bm{j}_{1:2d}} & := \sum_{\bm{a} \in \mathcal{S}}Q_{\bm{a}}\exp\left(\lambda\bm{\theta}^{*T}\bm{L}_{:,\,\bm{a}}\right)\prod_{1 \leq r \leq d}\left(a_{j_{2r - 1}}a_{j_{2r}} - G^{(2)}_{j_{2r - 1},\,j_{2r}}\right).
\end{align}
From the explicit combinatorial expression of the centered correlations tensor's entries (equation \ref{eq:centered_correlations_tensor_indices_expression}), the following combinatorial expression of the centered $\bm{G}$ correlations tensor can be deduced:
\begin{align}
    \delta G^{(2d)}_{\bm{\alpha}_{1:d}} & = \sum_{\substack{S',\,S''\\S' \sqcup S'' = [d]}}(-1)^{\left|S'\right|}\left[\bm{G}^{\left(2\left|S'\right|\right)}\right]_{\bm{\alpha}_{S'}}\left[\bm{G}^{(2)\otimes \left|S''\right|}\right]_{\bm{\alpha}_{S''}}, \qquad \bm{\alpha}_{1:d} \in \mathcal{A}^d.\label{eq:centered_g_correlations_tensor_indices_expression}
\end{align}
In the above equation, for any set $S' = \{s_1, \ldots, s_{d'}\} \subset [d]$, $\bm{\alpha}_{S'} := \left(\alpha_{s_1},\,\ldots,\,\alpha_{s_{d'}}\right)$ and similarly for $\bm{\alpha}_{S''}$. The implicit ordering of the set defining the tuple does not matter by symmetry of $\bm{G}$ correlations.
\end{definition}

\begin{remark}[Symmetry of centered $\bm{G}$ correlations]
While the $\bm{G}$ correlations are symmetric under all permutation of their indices, this is not true of centered $\bm{G}$ correlations tensors. These only have the symmetries of $\bm{C}$ correlation tensors, i.e. invariance under permutation of pairs of consecutive $\mathcal{I}$ indices, or equivalently invariance under permutation of $\mathcal{A}$ indices.
\end{remark}

We now rewrite in terms of $\bm{\delta G}$ the series expansion of QGMS (integral) moments stated in terms of $\bm{C}$ in Proposition~\ref{prop:qgms_integral_series_expansion}. In this proposition, we observed the diagonal QGMS moment of order 2, in terms of which the instance-averaged SK-QAOA energy is expressed (Proposition~\ref{prop:qgms-mgf_formulation_sk_qaoa_energy}), decomposes as:
\begin{align}
    \frac{\partial^2S_n\left(\bm{\mu}\right)}{\partial\mu_{\alpha}^2}\Bigg|_{\bm{\mu} = \bm{0}} & = \nu^{(0)}_{\alpha} + \frac{2}{\sqrt{n}}\nu^{(1)}_{\alpha} + \frac{1}{n}\nu^{(2)}_{\alpha},
\end{align}
with $\nu^{(1)}_{\alpha}$ (for instance) admitting series expansion
\begin{align}
    \nu^{(1)}_{\alpha} & = \sum_{\left(n_d\right)_{d \geq 2}}\binom{n}{\left(n_d\right)_{d \geq 2}}\frac{n^{-\sum\limits_{d \geq 2}dn_d/2}}{\prod\limits_{d \geq 2}d!^{n_d}}\theta^*_{\alpha}\left\langle \bm{\mathcal{I}}_{\alpha}^{\left(1 + \sum\limits_{d \geq 2}dn_d\right)}, \bigotimes_{d \geq 2}\bm{\delta C}^{(d)\otimes n_d} \right\rangle.
\end{align}
In the above equation, we plugged in simplification $\mathcal{Z}^* = 1$ (Eq.~\ref{eq:sk_qaoa_zstar_1}) holding for the SK-QAOA energy QGMS. The rest of the Section and current Appendix will be dedicated to the study of this object, which we will abbreviate $\nu_{\alpha}$. It will be convenient to break down the above series for $\nu_{\alpha}$ according to multinomial numbers $\left(n_d\right)_{d \geq 2}$, and for all fixed such numbers, by matchings $\left(l''',\,\mathcal{M}\right)$ (labelling the decomposition of the matching tensor slice in equation \ref{eq:matching_tensor_slice_decomposition}):
\begin{align}
    \nu_{\alpha} & := \sum_{\left(n_d\right)_{d \geq 2}}\nu^{\left(n_d\right)_{d \geq 2}}_{\alpha},\label{eq:qgms_integral_series_expansion_as_sum_multinomial_contributions}\\
    \nu^{\left(n_d\right)_{d \geq 2}}_{\alpha} & := \binom{n}{\left(n_d\right)_{d \geq 2}}\frac{n^{-D/2}}{\prod\limits_{d \geq 2}d!^{n_d}}\sum_{l''' \in [D]}\hspace*{5px}\sum_{\mathcal{M}\textrm{ matching of } [D] - \{l'''\}}\nu^{\left(n_d\right)_{d \geq 2},\,l''',\,\mathcal{M}}_{\alpha},\label{eq:qgms_integral_series_expansion_multinomial_contribution}\\
    \nu^{\left(n_d\right)_{d \geq 2},\,l''',\,\mathcal{M}}_{\alpha} & := \theta^*_{\alpha}\left\langle \bm{\mathcal{I}}^{\left(1 + D\right),\,l''',\,\mathcal{M}}_{\alpha}, \bigotimes_{d \geq 2}\bm{\delta C}^{(d)\otimes n_d} \right\rangle,\label{eq:qgms_integral_series_expansion_multinomial_matching_contribution}
\end{align}
where we introduced simplification $\mathcal{Z}^* = 1$ (equation \ref{eq:sk_qaoa_zstar_1}) in the special case of the SK-QAOA energy QGMS, and let $D := \sum_{d \geq 2}dn_d$ for brevity. Let us fix a specific collection of multinomial numbers $\left(n_d\right)_{d \geq 2}$ and matching $\left(l''', \mathcal{M}\right)$, and let $D := \sum_{d \geq 2}dn_d$. The dot product in equation \ref{eq:qgms_integral_series_expansion_multinomial_matching_contribution} can then be expanded as:
\begin{align}
    \left\langle \bm{\mathcal{I}}_{\alpha}^{\left(D + 1\right),\,l''',\,\mathcal{M}}, \bigotimes_{d \geq 2}\bm{\delta C}^{\left(d\right)\otimes n_d} \right\rangle & = \sum_{\bm{\alpha}_{1:D} \in \mathcal{A}^D}\left[\bigotimes_{d \geq 2}\bm{\delta C}^{(d)\otimes n_d}\right]_{\bm{\alpha}_{1:D}}\mathcal{I}^{\left(D + 1\right),\,l''',\,\mathcal{M}}_{\alpha,\,\bm{\alpha}_{1:D}}\nonumber\\
    & = \sum_{\bm{\alpha}_{1:D} \in \mathcal{A}^D}\left[\bigotimes_{d \geq 2}\bm{\delta C}^{(d)\otimes n_d}\right]_{\bm{\alpha}_{1:D}}\mathbf{1}\left[\alpha = \alpha_{l'''}\right]\prod_{\{l',\,l''\} \in \mathcal{M}}\mathbf{1}\left[\alpha_{l'} = \alpha_{l''}\right]\nonumber\\
    & = \sum_{\bm{\alpha}_{1:D} \in \mathcal{A}^D}\left[\bigotimes_{d \geq 2}\left(\lambda^d\bm{\mathcal{G}}^{(2d)}\right)^{\otimes n_d}\right]_{\bm{\alpha}_{1:D}}\left[\bigotimes_{d \geq 2}\bm{\delta G}^{(2d)\otimes n_d}\right]_{\bm{\alpha}_{1:D}}\mathbf{1}\left[\alpha = \alpha_{l'''}\right]\prod_{\{l',\,l''\} \in \mathcal{M}}\mathbf{1}\left[\alpha_{l'} = \alpha_{l''}\right]\nonumber\\
    & = \sum_{\bm{\alpha}_{1:D} \in \mathcal{A}^D}\left[\bigotimes_{d \geq 2}\left(\lambda\bm{\mathcal{G}}^{(2)}\right)^{\otimes dn_d}\right]_{\bm{\alpha}_{1:D}}\left[\bigotimes_{d \geq 2}\bm{\delta G}^{(2d)\otimes n_d}\right]_{\bm{\alpha}_{1:D}}\mathbf{1}\left[\alpha = \alpha_{l'''}\right]\prod_{\{l',\,l''\} \in \mathcal{M}}\mathbf{1}\left[\alpha_{l'} = \alpha_{l''}\right]\nonumber\\
    & = \lambda^D\sum_{\bm{\alpha}_{1:D} \in \mathcal{A}^D}\left(\prod_{1 \leq r \leq D}\mathcal{G}^{(2)}_{\alpha_r}\right)\left[\bigotimes_{d \geq 2}\bm{\delta G}^{(2d)\otimes n_d}\right]_{\bm{\alpha}_{1:D}}\mathbf{1}\left[\alpha = \alpha_{l'''}\right]\prod_{\{l',\,l''\} \in \mathcal{M}}\mathbf{1}\left[\alpha_{l'} = \alpha_{l''}\right]\label{eq:qgms_integral_series_expansion_multinomial_matching_contribution_dot_product_expansion_step_1}
\end{align}
We observe that the indicator function combine factors $\mathcal{G}^{(2)}_{\alpha_r}$ in pairs, yielding contributions of the form
$\left(\mathcal{G}^{(2)}_{\alpha_r}\right)^2$. To express it precisely, it will help to see a matching as a one-to-one mapping between sets of identical sizes. To explain the correspondence, assume without loss of generality that $\mathcal{M}$ is a matching of $[D - 1]$ (where $(D - 1)$ is even by assumption). One may write
\begin{align}
    \mathcal{M} & := \left\{\left\{x_1,\,y_1\right\},\,\left(x_2,\,y_2\right),\,\ldots,\,\left(x_{(D - 1)/2 - 1},\,y_{(D - 1)/2 - 1}\right),\,\left(x_{(D - 1)/2},\,y_{(D - 1)/2}\right)\right\},
\end{align}
with $x_j < y_j$ for all $j \in [(D - 1)/2]$ and $x_1 < x_2 < \ldots < x_{(D - 1)/2 - 1} < x_{(D - 1)/2}$. Then, matching $\mathcal{M}$ can be identified to one-to-one mapping:
\begin{align}
\label{eq:matching_one_one_mapping_interpretation}
    \mathcal{M}: \left\{\begin{array}{ccc}
         \mathcal{D}\left(\mathcal{M}\right) & \longrightarrow & \mathcal{R}\left(\mathcal{M}\right)\\
         x_j & \longrightarrow & y_j
    \end{array}\right.,
\end{align}
with domain $\mathcal{D}\left(\mathcal{M}\right) = \left\{x_1,\,x_2,\,\ldots,\,x_{(D - 1)/2 - 1},\,x_{(D - 1)/2}\right\}$ and range $\mathcal{R}\left(\mathcal{M}\right) = \left\{y_1,\,y_2,\,\ldots,\,y_{(D - 1)/2 - 1},\,y_{(D - 1)/2}\right\}$. From these notations, and now considering a general matching of $[D] - \{l'''\}$ rather than $[D - 1]$, the product of $\bm{\mathcal{G}}^{(2)}$ tensor entries in equation \ref{eq:qgms_integral_series_expansion_multinomial_matching_contribution_dot_product_expansion_step_1} can be expressed as follows:
\begin{align}
    \prod_{1 \leq r \leq D}\mathcal{G}^{(2)}_{\alpha_r} & = \mathcal{G}^{(2)}_{\alpha_{l'''}}\prod_{l' \in \mathcal{D}\left(\mathcal{M}\right)}\mathcal{G}^{(2)}_{\alpha_{l'}}\mathcal{G}^{(2)}_{\alpha_{\mathcal{M}\left(l'\right)}}\nonumber\\
    & = \mathcal{G}^{(2)}_{\alpha}\prod_{l' \in \mathcal{D}\left(\mathcal{M}\right)}\left(\mathcal{G}^{(2)}_{\alpha_{l'}}\right)^2\nonumber\\
    & = \mathcal{G}^{(2)}_{\alpha}\prod_{l' \in \mathcal{D}\left(\mathcal{M}\right)}\mathcal{G}_{j_{2l' - 1}}\mathcal{G}_{j_{2l'}}\nonumber\\
    & = \mathcal{G}^{(2)}_{\alpha}\left[\bm{\mathcal{G}}^{\otimes (D - 1)}\right]_{\bm{\alpha}_{\mathcal{D}(\mathcal{M})}}
\end{align}
where in the final lines, we expanded the $\mathcal{A}$ indices into pairs of $\mathcal{I}$ indices: $\alpha := \left(k_1, k_2\right)$, $\alpha_{l'} := \left(j_{2l' - 1}, j_{2l'}\right)$ for $l' \in \mathcal{D}(\mathcal{M})$. We can also reexpress the tensor product of $\bm{G}$ correlation tensors using the same mapping interpretation of a matching. Namely, from decomposition of $[D]$ into disjoint sets:
\begin{align}
    \left[D\right] & = \left\{l'''\right\} \sqcup \mathcal{D}\left(\mathcal{M}\right) \sqcup \mathcal{R}\left(\mathcal{M}\right),
\end{align}
one can identify
\begin{align}
    \bm{\alpha}_{1:D} & \simeq \left(\alpha_{l'''},\,\bm{\alpha}_{\mathcal{D}(M)},\,\bm{\alpha}_{\mathcal{R}(\mathcal{M})}\right),
\end{align}
where $\bm{\alpha}_{\mathcal{D}(\mathcal{M})}$ is the tuple of elements $\alpha_{l'}$ for $l' \in \mathcal{D}\left(\mathcal{M}\right)$ listed in increasing order, and similarly for the other vector notations. From this identification, and for an appropriate permutation $\pi_{l''',\,\mathcal{M}}$ permuting factors of the tensor product space $\left(\mathbf{C}^{\mathcal{A}}\right)^{\otimes D}$, one can then write:
\begin{align}
    \left[\bigotimes_{d \geq 2}\bm{\delta C}^{(d)\otimes n_d}\right]_{\bm{\alpha}_{1:D}} & = \left[\pi_{l''',\,\mathcal{M}} \cdot \bigotimes_{d \geq 2}\bm{\delta C}^{(d)\otimes n_d}\right]_{\left(\alpha_{l'''},\,\bm{\alpha}_{\mathcal{D}(\mathcal{M})},\,\bm{\alpha}_{\mathcal{R}(\mathcal{M})}\right)},\label{eq:matching_permutation_defining_equation}
\end{align}
which, thanks for the indicator functions in equation \ref{eq:qgms_integral_series_expansion_multinomial_matching_contribution_dot_product_expansion_step_1}, becomes
\begin{align}
    \left[\bigotimes_{d \geq 2}\bm{\delta C}^{(d)\otimes n_d}\right]_{\bm{\alpha}_{1:D}} & = \left[\pi_{l''',\,\mathcal{M}} \cdot \bigotimes_{d \geq 2}\bm{\delta C}^{(d)\otimes n_d}\right]_{\left(\alpha,\,\bm{\alpha}_{\mathcal{D}(\mathcal{M})},\,\bm{\alpha}_{\mathcal{D}(\mathcal{M})}\right)}
\end{align}
in this context. All in all, the sum in equation \ref{eq:qgms_integral_series_expansion_multinomial_matching_contribution_dot_product_expansion_step_1}, representing the dot product in equation \ref{eq:qgms_integral_series_expansion_multinomial_matching_contribution}, can be rewritten
\begin{align}
    \left\langle \bm{\mathcal{I}}_{\alpha}^{\left(D + 1\right),\,l''',\,\mathcal{M}},\, \bigotimes_{d \geq 2}\bm{\delta C}^{(d)\otimes n_d}\right\rangle & = \lambda^D\mathcal{G}^{(2)}_{\alpha}\sum_{\substack{\bm{\alpha}_{\mathcal{D}(\mathcal{M})} \in \mathcal{A}^{\mathcal{D}(\mathcal{M})}}}\left[\pi_{l''',\,\mathcal{M}} \cdot \bigotimes_{d \geq 2}\bm{\delta G}^{(2d)\otimes n_d}\right]_{\left(\alpha,\,\bm{\alpha}_{\mathcal{D}(\mathcal{M})},\,\bm{\alpha}_{\mathcal{D}(\mathcal{M})}\right)}\left[\bm{\mathcal{G}}^{\otimes (D - 1)}\right]_{\bm{\alpha}_{\mathcal{D}(\mathcal{M})}}.\label{eq:qgms_integral_series_expansion_multinomial_matching_contribution_dot_product_expansion_step_2}
\end{align}
From this representation of the dot product results the following representation for contribution $\nu^{\left(n_d\right)_{d \geq 2},\,l''',\,\mathcal{M}}_{\alpha}$ to the QGMS moment:
\begin{align}
    \nu^{\left(n_d\right)_{d \geq 2},\,l''',\,\mathcal{M}}_{\alpha} & = \theta^*_{\alpha}\left\langle \bm{\mathcal{I}}_{\alpha}^{\left(D + 1\right),\,l''',\,\mathcal{M}},\, \bigotimes_{d \geq 2}\bm{\delta C}^{(d)\otimes n_d}\right\rangle\nonumber\\
    & = \lambda^{1 + D}G^{(2)}_{\alpha}\mathcal{G}^{(2)}_{\alpha}\left\langle \bm{\mathcal{I}}_{\alpha}^{\left(D + 1\right),\,l''',\,\mathcal{M}},\, \bigotimes_{d \geq 2}\bm{\delta C}^{(d)\otimes n_d}\right\rangle\nonumber\\
    & = \lambda^{1 + D}G^{(2)}_{\alpha}\left(\mathcal{G}^{(2)}_{\alpha}\right)^2\sum_{\substack{\bm{\alpha}_{\mathcal{D}(\mathcal{M})} \in \mathcal{A}^{\mathcal{D}(\mathcal{M})}}}\left[\pi_{l''',\,\mathcal{M}} \cdot \bigotimes_{d \geq 2}\bm{\delta G}^{(2d)\otimes n_d}\right]_{\left(\alpha,\,\bm{\alpha}_{\mathcal{D}(\mathcal{M})},\,\bm{\alpha}_{\mathcal{D}(\mathcal{M})}\right)}\left[\bm{\mathcal{G}}^{\otimes (D - 1)}\right]_{\bm{\alpha}_{\mathcal{D}(\mathcal{M})}}\nonumber\\
    & = \lambda^{1 + D}G^{(2)}_{k_1,\,k_2}\mathcal{G}_{k_1}\mathcal{G}_{k_2}\sum_{\substack{\bm{\alpha}_{\mathcal{D}(\mathcal{M})} \in \mathcal{A}^{\mathcal{D}(\mathcal{M})}}}\left[\pi_{l''',\,\mathcal{M}} \cdot \bigotimes_{d \geq 2}\bm{\delta G}^{(2d)\otimes n_d}\right]_{\left(\alpha,\,\bm{\alpha}_{\mathcal{D}(\mathcal{M})},\,\bm{\alpha}_{\mathcal{D}(\mathcal{M})}\right)}\left[\bm{\mathcal{G}}^{\otimes (D - 1)}\right]_{\bm{\alpha}_{\mathcal{D}(\mathcal{M})}}\label{eq:qgms_integral_series_expansion_multinomial_matching_contribution_expansion}
\end{align}
Collecting some prefactors separately, the last equation can be rewritten:
\begin{align}
    \nu_{\alpha}^{\left(n_d\right)_{d \geq 2},\,l''',\,\mathcal{M}} & = \lambda G^{(2)}_{k_1,\,k_2}\mathcal{G}_{k_1}\mathcal{G}_{k_2}\widetilde{\nu}_{\alpha}^{\left(n_d\right)_{d \geq 2},\,l''',\,\mathcal{M}},
\end{align}
with
\begin{align}
    \widetilde{\nu}_{\alpha}^{\left(n_d\right)_{d \geq 2},\,l''',\,\mathcal{M}} & := \lambda^D\sum_{\bm{\alpha}_{\mathcal{D}(\mathcal{M})} \in \mathcal{A}^{\mathcal{D}(\mathcal{M})}}\left[\pi_{l''',\,\mathcal{M}} \cdot \bigotimes_{d \geq 2}\bm{\delta G}^{\left(2d\right)\otimes n_d}\right]_{\left(\alpha,\,\bm{\alpha}_{\mathcal{D}(\mathcal{M})},\,\bm{\alpha}_{\mathcal{D}(\mathcal{M})}\right)}\left[\bm{\mathcal{G}}^{\otimes (D - 1)}\right]_{\bm{\alpha}_{\mathcal{D}(\mathcal{M})}}.\label{eq:qgms_integral_series_expansion_multinomial_matching_contribution_expansion_tilde_first_statement}
\end{align}
The interest of definition \ref{eq:qgms_integral_series_expansion_multinomial_matching_contribution_expansion_tilde_first_statement} is, $(p + 1)\widetilde{\nu}_{\alpha}^{\left(n_d\right)_{d \geq 1},\,l''',\,\mathcal{M}}$ converges to an integral in the $p \to \infty$ limit. To infer that, we invoke equation \ref{eq:curvy_g_vector_discrete_from_continuum_rescaled}, stating that $\bm{\mathcal{G}}$ is a perfect discretization of $\mathcal{G}^{\mathrm{cont}}$:
\begin{align}
    \mathcal{G}_j & = \mathcal{G}^{\mathrm{cont}}\left(\frac{j}{p + 1/2}\right).
\end{align}
We also invoke proposition \ref{prop:g_higher_order_correlations_continuum_approximation}, showing that $\bm{G}$ correlations are approximate discretizations of continuum functions, without magnitude rescaling factor:
\begin{align}
    G^{(2d)}_{\bm{\alpha}_{1:d}} & \approx G^{(2d),\,\mathrm{cont}}\left(\frac{\bm{\alpha}_{1:d}}{p + 1/2}\right).
\end{align}
Finally, we consider the scaling in $p$ of prefactor $\lambda^D$, combined with rescaling $(p + 1)$:
\begin{align}
    \left(p + 1\right)\lambda^D & = (p + 1)\left(\frac{2^{-1/2}\gamma_{\mathrm{max}}}{p + 1}\right)^D\nonumber\\
    & = \frac{\left(2^{-1/2}\gamma_{\mathrm{max}}\right)^D}{\left(p + 1\right)^{D - 1}},
\end{align}
and compare this to the number of terms $|\mathcal{A}|^{\left|\mathcal{D}(\mathcal{M})\right|} = \left((2p + 2)^2\right)^{(D - 1)/2} = 2^{D - 1}\left(p + 1\right)^{D - 1}$ in the discrete sum Eq.~\ref{eq:qgms_integral_series_expansion_multinomial_matching_contribution_expansion_tilde_first_statement}. This suggests the following limit for the quantity defined in equation \ref{eq:qgms_integral_series_expansion_multinomial_matching_contribution_expansion_tilde_first_statement}:
\begin{align}
    (p + 1)\widetilde{\nu}_{\alpha}^{\left(n_d\right)_{d \geq 2},\,l''',\,\mathcal{M}} & \underset{p \to \infty}{\sim} \widetilde{\nu}^{\left(n_d\right)_{d \geq 2},\,l''',\,\mathcal{M},\,\mathrm{cont}}\left(\frac{\alpha}{p + 1/2}\right),   
\end{align}
where
\begin{align}
    \widetilde{\nu}^{\left(n_d\right)_{d \geq 2},\,l''',\,\mathcal{M},\,\mathrm{cont}}\left(\xi\right) & := \left(2^{-1/2}\gamma_{\mathrm{max}}\right)^D\int\limits_{\bm{\xi} \in \left([0, 2]^2\right)^{\mathcal{D}(\mathcal{M})}}\!\mathrm{d}\bm{\xi}\,\left(\pi_{l''',\,\mathcal{M}} \cdot \bigotimes_{d \geq 2}\left(\delta G^{(d),\,\mathrm{cont}}\right)^{\otimes n_d}\right)\left(\xi, \bm{\xi}, \bm{\xi}\right)\left(\mathcal{G}^{\mathrm{cont}}\right)^{\otimes (D - 1)}\left(\bm{\xi}\right).\label{eq:qgms_integral_series_expansion_multinomial_matching_contribution_tilde_continuum_first_statement}
\end{align}
Note the integration interval $[0, 2]$ in each of the $(D - 1)$ scalar coordinates accounts for $2^{D - 1}$ factor in the expansion of $\left|\mathcal{A}\right|^{\left|\mathcal{D}(\mathcal{M})\right|}$. The continuum analogue $\delta G^{\left(2d\right),\,\mathrm{cont}}$ of the centered discrete correlations tensor $\bm{\delta G}^{\left(2d\right)}$ is defined from the continuum $\bm{G}$ correlations tensors $G^{(2m),\,\mathrm{cont}}$ according to the same combinatorial relation linking centered and non-centered discrete $\bm{G}$ correlations tensors:

\begin{definition}[Continuum centered $\bm{G}$ correlations tensor for SK-QAOA]
\label{def:centered_g_correlations_tensor_continuum}
The continuum analogue of the centered $\bm{G}$ correlations tensor of order $2d$, introduced in definition \ref{def:centered_g_correlations_tensor}, is a continuous function: $[0, 2]^{2d} \longmapsto \mathbf{C}$, defined as follows from continuum $\bm{G}$ correlations (in analogy to equation \ref{eq:centered_g_correlations_tensor_indices_expression}):
\begin{align}
    \delta G^{(2d),\,\mathrm{cont}}\left(\bm{\xi}\right) & := \sum_{\substack{S',\,S''\\S' \sqcup S'' = [d]}}(-1)^{\left|S'\right|}\left(\bm{G}^{\left(2\left|S'\right|\right),\,\mathrm{cont}}\right)\left(\bm{\xi}_{S'}\right)\left(\bm{G}^{(2),\,\mathrm{cont}}\right)^{\otimes \left|S''\right|}\left(\bm{\xi}_{S''}\right),\label{eq:centered_g_correlations_tensor_indices_expression_continuum}\\
    \bm{\xi}_{1:d} & \in \left(\left[0, 2\right]^2\right)^d.
\end{align}
From these, we define the continuum analogue of contribution $\nu^{\left(n_d\right)_{d \geq 2},\,l''',\,\mathcal{M}}_{\alpha}$ (equation \ref{eq:qgms_integral_series_expansion_multinomial_matching_contribution}) to the QGMS moment:
\begin{align}
    \nu^{\left(n_d\right)_{d \geq 2},\,l''',\,\mathcal{M},\,\mathrm{cont}}\left(\xi\right) & := \left(2^{-1/2}\gamma_{\mathrm{max}}\right)G^{(2),\,\mathrm{cont}}\left(x_1, x_2\right)\mathcal{G}\left(x_1\right)\mathcal{G}\left(x_2\right)\widetilde{\nu}^{\left(n_d\right)_{d \geq 2},\,l''',\,\mathcal{M}}\left(\xi\right),\label{eq:qgms_integral_series_expansion_multinomial_matching_contribution_continuum}\\
    \widetilde{\nu}^{\left(n_d\right)_{d \geq 2},\,l''',\,\mathcal{M},\,\mathrm{cont}}\left(\xi\right) & := \left(2^{-1/2}\gamma_{\mathrm{max}}\right)^D\int\limits_{\substack{\bm{\xi} \in \left([0, 2]^2\right)^{\mathcal{D}(\mathcal{M})}}}\!\mathrm{d}\bm{\xi}\,\left(\pi_{l''',\,\mathcal{M}} \cdot \bigotimes_{d \geq 2}\left(\delta G^{\left(2d\right),\,\mathrm{cont}}\right)^{\otimes n_d}\right)\left(\xi,\,\bm{\xi},\,\bm{\xi}\right)\left(\mathcal{G}^{\mathrm{cont}}\right)^{\otimes (D - 1)}\left(\bm{\xi}\right)\label{eq:qgms_integral_series_expansion_multinomial_matching_contribution_tilde_continuum},\\
    \xi & := \left(x_1, x_2\right) \in [0, 2]^2.
\end{align}
This mirrors the representation of (discrete) $\nu^{\left(n_d\right)_{d \geq 2},\,l''',\,\mathcal{M}}_{\alpha}$ previously derived, namely
\begin{align}
    \nu^{\left(n_d\right)_{d \geq 2},\,l''',\,\mathcal{M}}_{\alpha} & = \lambda G^{(2)}_{k_1,\,k_2}\mathcal{G}_{k_1}\mathcal{G}_{k_2}\widetilde{\nu}^{\left(n_d\right)_{d \geq 2},\,l''',\,\mathcal{M}}_{\alpha},\\
    \widetilde{\nu}^{\left(n_d\right)_{d \geq 2},\,l''',\,\mathcal{M}}_{\alpha} & := \lambda^D\sum_{\substack{\bm{\alpha}_{\mathcal{D}(\mathcal{M})} \in \mathcal{A}^{\mathcal{D}(\mathcal{M})}}}\left[\pi_{l''',\,\mathcal{M}} \cdot \bigotimes_{d \geq 2}\bm{\delta G}^{(2d)\otimes n_d}\right]_{\alpha,\,\bm{\alpha}_{\mathcal{D}(\mathcal{M})},\,\bm{\alpha}_{\mathcal{D}(\mathcal{M})}}\left[\bm{\mathcal{G}}^{\otimes (D - 1)}\right]_{\bm{\alpha}_{\mathcal{D}(\mathcal{M})}}\label{eq:qgms_integral_series_expansion_multinomial_matching_contribution_expansion_tilde},\\
    \alpha & := \left(k_1,\,k_2\right) \in \mathcal{I}^2.
\end{align}
In the above formulae, we let $D := \sum_{d \geq 2}dn_d$ and defined $\pi_{l''',\,\mathcal{M}}$, parametrized by $l''' \in [D]$ and a matching $\mathcal{M}$ of $[D] - \{l'''\}$, as the unique permutation (acting over tensor product factors) such that:
\begin{align}
    \left[\pi_{l''',\,\mathcal{M}} \cdot \bm{T}\right]_{\bm{\alpha}_{1:D}} & := \left[\bm{T}\right]_{l''',\,\bm{\alpha}_{\mathcal{D}(\mathcal{M})},\,\bm{\alpha}_{\mathcal{R}(\mathcal{M})}}
\end{align}
for all tensor $\bm{T} \in \left(\mathbf{C}^{\mathcal{A}}\right)^{\otimes D}$ and $D$-dimensional $\mathcal{A}$ index $\bm{\alpha}_{1:D} \in \mathcal{A}^{D}$.
\end{definition}

Note that unlike in the definition of continuum $\bm{G}^{(2)}$ correlations (definition \ref{def:g_correlations_continuum}) or the continuum higher-order $\bm{G}$ correlations (definition \ref{def:g_higher_order_correlations}), the defining sum in equation \ref{eq:centered_g_correlations_tensor_indices_expression_continuum} is finite, hence clearly defines a continuous function.

\subsection{Proof outline of main theorem}
\label{sec:main_theorem_derivation}

In this section, we outline the proof of the main theoretical result of the manuscript (Theorem~\ref{th:approximation_continuous_time_annealing_qaoa}). For space reason, we only provide the final step of the argument, deferring the details to upcoming Section~\ref{sec:sk_qaoa_continuum_limit}. We nonetheless sketch the general program of proof outlined in the introduction to this more technical section. Namely, the analysis relies on establishing convergence of different tensors to continuous multivariate functions in the limit $p \to \infty$. In order, these tensors are the noninteracting correlations tensors, the saddle point vector, the higher-order correlations tensors, and the QGMS moment vector $\bm{\nu} = \left(\nu_{\alpha}\right)_{\alpha \in \mathcal{A}}$. For instance, the existence of a continuum limit for the $\bm{G}$ correlations tensor of order $2d$ means there exists a continuous function $G^{(2d),\,\mathrm{cont}}$, independent of $p$, such that in the limit $p \to \infty$,
\begin{align}
    \left|G^{(2d)}_{\bm{j}_{1:2d}} - G^{(2d),\,\mathrm{cont}}\left(\frac{\bm{j}_{1:2d}}{p + 1/2}\right)\right| & \leq \frac{\mathcal{O}(1)}{p} \qquad \forall \bm{j}_{1:2d} \in \mathcal{I}^{2d},
\end{align}
with implicit constant $\mathcal{O}(1)$ uniform in tensor entry $\bm{j}_{1:2d} \in \mathcal{I}^{2d}$. Besides these uniform approximation bounds between discrete tensors and continuous functions, the analysis requires uniform bounds on their absolute magnitudes, i.e. bounds on quantities
\begin{align}
    \left\lVert \bm{G}^{\left(2d\right)} \right\rVert_{\infty}, \quad \left\lVert G^{(2d),\,\mathrm{cont}} \right\rVert_{\infty}
\end{align}
for the example previously considered. The current section limits itself to exploiting these results to prove convergence of the SK-QAOA to a well-defined value in the $p \to \infty$ limit. 

We start by recalling the formula for the energy of $p$-layers QAOA at arbitrary size $n$ and depth $p$, as implicitly established in \cite{qaoa_sk} and explicitly rederived in proposition \ref{prop:qgms-mgf_formulation_sk_qaoa_energy}:
\begin{align}
    V_{p, n} & = -\frac{i}{\Gamma_{p + 1}}\sum_{0 \leq r \leq 2p + 1}\frac{\partial^2S_n\left(\bm\mu\right)}{\partial\mu_{(r,\,p + 1)}^2}\Bigg|_{\bm{\mu} = \bm{0}},
\end{align}
where we adapted indexation from $\{-1, -2, \ldots, -(p - 1), -p, 0, p, p - 1, \ldots, 2, 1\}$ in proposition \ref{prop:qgms-mgf_formulation_sk_qaoa_energy} to $\{0, 1, \ldots, p - 2, p - 1, p, p + 1, p + 2, \ldots, 2p, 2p + 1\}$ in this appendix section. For conciseness, we considered a single additive contribution of this quantity (see Eq.~\ref{eq:qgms_second_order_moment_integral_additive_contributions} of Proposition \ref{prop:qgms_integral_series_expansion}), namely
\begin{align}
    \frac{\partial^2S_n\left(\bm{\mu}\right)}{\partial\mu_{(r,\,p + 1)}^2}\Bigg|_{\bm{\mu} = \bm{0}} & \supset \frac{2}{\Gamma_{p + 1}\sqrt{n}}\sum_{0 \leq r \leq 2p + 1}\nu^{(1)}_{\left(r,\,p + 1\right)}.
\end{align}
While other terms Equation~\ref{eq:qgms_second_order_moment_integral_additive_contributions} can be similarly treated, this requires some further introductory discussion, deferred to Appendix~\ref{sec:other_moments_concentration}. For conciseness, we will abbreviate $\nu^{(1)}_{\alpha}$ as $\nu_{\alpha}$ in the rest of the proof. We reexpress this sum in terms of $\bm{\widetilde{\nu}}$ (definition \ref{def:qgms_integral_series_expansion_tilde_quantities}):
\begin{align}
    -\frac{1}{\Gamma_{p + 1}}\sum_{0 \leq r \leq 2p + 1}\nu_{\left(r,\,p + 1\right)} & = -\frac{1}{\Gamma_{p + 1}}\sum_{0 \leq r \leq 2p + 1}\theta^*_{(r,\,p + 1)}\widetilde{\nu}_{\left(r,\,p + 1\right)}\nonumber\\
    & = -\frac{1}{\Gamma_{p + 1}}\sum_{0 \leq r \leq 2p + 1}\lambda G^{(2)}_{r,\,p + 1}\mathcal{G}_{r}\mathcal{G}_{p + 1}\widetilde{\nu}_{\left(r,\,p + 1\right)}\nonumber\\
    & = \frac{1}{\Gamma_{p + 1}}\sum_{0 \leq r \leq 2p + 1}\lambda G^{(2)}_{r,\,p + 1}\frac{(p + 1)^2\Gamma_r\Gamma_{p + 1}}{\gamma_{\mathrm{max}}^2}\widetilde{\nu}_{\left(r,\,p + 1\right)}\nonumber\\
    & = \sum_{0 \leq r \leq 2p + 1}\lambda G^{(2)}_{\left(r,\,p + 1\right)}\frac{(p + 1)^2\Gamma_r}{\gamma_{\mathrm{max}}^2}\widetilde{\nu}_{\left(r,\,p + 1\right)}\nonumber\\
    & = -\frac{1}{\sqrt{2}}\sum_{0 \leq r \leq 2p + 1}G^{(2)}_{\left(r,\,p + 1\right)}\frac{\left(p + 1\right)\Gamma_r}{\gamma_{\mathrm{max}}}\widetilde{\nu}_{\left(r,\,p + 1\right)}\nonumber\\
    & = -\frac{1}{\sqrt{2}\gamma_{\mathrm{max}}}\sum_{0 \leq r \leq 2p + 1}G^{(2)}_{(r,\,p + 1)}\Gamma^{\mathrm{cont}}\left(\frac{r}{p + 1/2}\right)\widetilde{\nu}_{\left(r,\,p + 1\right)}.
\end{align}
We now claim the latter is approximated by
\begin{align}
    -\frac{1}{\sqrt{2}\gamma_{\mathrm{max}}}\int_{[0, 2]}\!\mathrm{d}x\,G^{(2),\,\mathrm{cont}}\left(x, 1\right)\Gamma^{\mathrm{cont}}\left(x\right)\widetilde{\nu}^{\mathrm{cont}}\left(x, 1\right).
\end{align}
For that purpose, we separate between discretization and Riemann sum approximation errors:
\begin{align}
    \frac{1}{\sqrt{2}\gamma_{\mathrm{max}}}\sum_{0 \leq r \leq 2p + 1}G^{(2)}_{(r,\,p + 1)}\Gamma^{\mathrm{cont}}\left(\frac{r}{p + 1/2}\right)\widetilde{\nu}_{\left(r,\,p + 1\right)} - \frac{1}{\sqrt{2}\gamma_{\mathrm{max}}}\int_{[0, 2]}\!\mathrm{d}x\,G^{(2),\,\mathrm{cont}}\left(x, 1\right)\Gamma^{\mathrm{cont}}\left(x\right)\widetilde{\nu}^{\mathrm{cont}}\left(x, 1\right) & = A + B,
\end{align}
where the discretization error is
\begin{align}
    A & := \frac{1}{p + 1}\frac{1}{\sqrt{2}\gamma_{\mathrm{max}}}\sum_{0 \leq r \leq 2p + 1}\left\{G^{(2)}_{(r,\,p + 1)}\Gamma^{\mathrm{cont}}\left(\frac{r}{p + 1/2}\right)(p + 1)\widetilde{\nu}_{\left(r,\,p + 1\right)}\right.\nonumber\\
    & \left. \hspace*{130px} - G^{(2),\,\mathrm{cont}}\left(\frac{r}{p + 1}, 1\right)\Gamma^{\mathrm{cont}}\left(\frac{r}{p + 1/2}\right)\widetilde{\nu}^{\mathrm{cont}}\left(\frac{r}{p + 1/2}, 1\right)\right\},
\end{align}
and the Riemann sum approximation error is
\begin{align}
    B & := \frac{1}{p + 1}\frac{1}{\sqrt{2}\gamma_{\mathrm{max}}}\sum_{0 \leq r \leq 2p + 1}G^{(2),\,\mathrm{cont}}\left(\frac{r}{p + 1/2}, 1\right)\Gamma^{\mathrm{cont}}\left(\frac{r}{p + 1/2}\right)\widetilde{\nu}^{\mathrm{cont}}\left(\frac{r}{p + 1/2},\,1\right)\nonumber\\
    & \hspace*{20px} - \frac{1}{\sqrt{2}\gamma_{\mathrm{max}}}\int_{[0, 2]}\!\mathrm{d}x\,G^{(2),\,\mathrm{cont}}\left(x, 1\right)\Gamma^{\mathrm{cont}}\left(x, 1\right)\widetilde{\nu}^{\mathrm{cont}}\left(x\right)
\end{align}

We first bound $A$, starting with a single term of the sum. For that purpose, we invoke lemma \ref{lemma:bound_variation_product} to bound the variation of a product, applying it to factors
\begin{gather}
    G^{(2)}_{\left(r,\,p + 1\right)}, \qquad \Gamma^{\mathrm{cont}}\left(\frac{r}{p + 1/2}\right), \qquad (p + 1)\widetilde{\nu}_{\left(r,\,p + 1\right)}\label{eq:main_theorem_discrete_triplet}
\end{gather}
and 
\begin{align}
    G^{(2),\,\mathrm{cont}}\left(\frac{r}{p + 1/2}, 1\right), \qquad \Gamma^{\mathrm{cont}}\left(\frac{r}{p + 1/2}\right), \qquad \widetilde{\nu}^{\mathrm{cont}}\left(\frac{r}{p + 1/2}, 1\right)\label{eq:main_theorem_contunuous_triplet}
\end{align}
These triplets are both respectively bounded by
\begin{gather}
    2, \qquad \gamma_{\mathrm{max}}, \qquad \mathcal{O}(1)
\end{gather}
(propositions \ref{prop:g_correlations_continuum_well_definiteness} and corollary \ref{cor:qgms_integral_series_expansion_contributions_continuum_approximation} respectively). 
Using the same propositions, the differences between the elements of Eq.~\ref{eq:main_theorem_discrete_triplet} and the corresponding ones in Eq.~\ref{eq:main_theorem_contunuous_triplet}
are bounded by
\begin{gather}
    \frac{\mathcal{O}(1)}{p + 1}\gamma_{\mathrm{max}}^2\max\left(1, 2\beta_{\mathrm{max}}, \frac{M_{\gamma}}{\gamma_{\mathrm{max}}}\right) + \frac{2\beta_{\mathrm{max}}}{p}, \qquad 0, \qquad \frac{\mathcal{O}(1)}{p + 1}\gamma_{\mathrm{max}}^2\max\left(4\beta_{\mathrm{max}}, \frac{2M_{\gamma}}{\gamma_{\mathrm{max}}}, \gamma'_{\mathrm{max}}\right) + \frac{\mathcal{O}(1)\beta_{\mathrm{max}}}{p}.
\end{gather}
Specifically, in establishing the first bound, we used that $G^{(2),\,\mathrm{cont}}$ is $4\beta_{\mathrm{max}}$-Lipschitz in each of its variables (proposition \ref{prop:g_correlations_continuum_well_definiteness}) and the continuum approximation bound for $\bm{G}^{(2)}$ (proposition \ref{prop:g_correlations_continuum_approximation}). In establishing the third bound, we used corollary \ref{cor:qgms_integral_series_expansion_contributions_continuum_approximation}, stating that $\widetilde{\nu}^{\mathrm{cont}}$ is $\mathcal{O}(1)\beta_{\mathrm{max}}$-Lipschitz in each variable, and providing a discretization bound between $\bm{\widetilde{\nu}}$ and $\widetilde{\nu}^{\mathrm{cont}}$. It results in the following bound for a single term of $A$:
\begin{align}
    & \left|G^{(2)}_{(r,\,p + 1)}\Gamma^{\mathrm{cont}}\left(\frac{r}{p + 1/2}\right)(p + 1)\widetilde{\nu}_{\left(r,\,p + 1\right)} - G^{(2),\,\mathrm{cont}}\left(\frac{r}{p + 1}, 1\right)\Gamma^{\mathrm{cont}}\left(\frac{r}{p + 1/2}\right)\widetilde{\nu}^{\mathrm{cont}}\left(\frac{r}{p + 1/2}, 1\right)\right|\nonumber\\
    & \leq \frac{\mathcal{O}(1)\gamma_{\mathrm{max}}}{p + 1}\max\left(\gamma_{\mathrm{max}}^2, \beta_{\mathrm{max}}, \beta_{\mathrm{max}}\gamma_{\mathrm{max}}^2, M_{\gamma}\gamma_{\mathrm{max}}, \gamma_{\mathrm{max}}^2\gamma'_{\mathrm{max}}\right)\nonumber\\
    & \leq \frac{\mathcal{O}\left(1\right)\gamma_{\mathrm{max}}}{p + 1}\max\left(\gamma_{\mathrm{max}}^2, \beta_{\mathrm{max}}, M_{\gamma}\gamma_{\mathrm{max}}\right),
\end{align}
where in the final line, we assumed $\gamma_{\mathrm{max}} \leq 1$ as we only need to prove the result for a small constant $\gamma$. By the triangular inequality, the last bound implies the following bound on $A$:
\begin{align}
    |A| & \leq \frac{\mathcal{O}(1)}{p + 1}\max\left(\gamma_{\mathrm{max}}^2, \beta_{\mathrm{max}}, M_{\gamma}\gamma_{\mathrm{max}}\right).
\end{align}
We now turn to the Riemann sum approximation error $B$. For that purpose, we use lemma \ref{lemma:riemann_sum_approximation} to bound the Riemann sum-integral error. We apply it to Lipschitz functions
\begin{gather}
    G^{(2),\,\mathrm{cont}}\left(\,\cdot,\,1\right), \qquad \Gamma^{\mathrm{cont}}, \qquad \widetilde{\nu}^{\mathrm{cont}}\left(\,\cdot,\,1\right),
\end{gather}
which are uniformly bounded (as stated in the analysis of $A$) by
\begin{gather}
    2, \qquad \gamma_{\mathrm{max}}, \qquad \mathcal{O}\left(1\right).
\end{gather}
Besides (as also observed in the analysis of $A$), they have Lipschitz constants in each variable
\begin{align}
    4\beta_{\mathrm{max}}, \qquad M_{\gamma}, \qquad \mathcal{O}(1)\beta_{\mathrm{max}}.
\end{align}
Lemma \ref{lemma:riemann_sum_approximation} then gives bound
\begin{align}
    |B| & \leq \frac{\mathcal{O}(1)}{p + 1}\max\left(\beta_{\mathrm{max}}, \frac{M_{\gamma}}{\gamma_{\mathrm{max}}}\right)
\end{align}
on the sum-integral error. All in all, still assuming $\gamma_{\mathrm{max}} \leq 1$ for simplicity,
\begin{align}
    \frac{1}{\Gamma_{p + 1}}\sum_{0 \leq r \leq 2p + 1}\nu_{\left(r,\,p + 1\right)}
\end{align}
is bounded away from its $p \to \infty$ limit:
\begin{align}
    \frac{1}{\sqrt{2}\gamma_{\mathrm{max}}}\int_{[0, 2]}\!\mathrm{d}x\,G^{(2),\,\mathrm{cont}}\left(x, 1\right)\Gamma^{\mathrm{cont}}\left(x\right)\widetilde{\nu}^{\mathrm{cont}}\left(x, 1\right)
\end{align}
by at most:
\begin{align}
    & \left|\frac{1}{\Gamma_{p + 1}}\sum_{0 \leq r \leq 2p + 1}\nu_{\left(r,\,p + 1\right)} - \frac{1}{\sqrt{2}\gamma_{\mathrm{max}}}\int_{[0, 2]}\!\mathrm{d}x\,G^{(2),\,\mathrm{cont}}\left(x, 1\right)\Gamma^{\mathrm{cont}}\left(x\right)\widetilde{\nu}^{\mathrm{cont}}\left(x, 1\right)\right|\nonumber\\
    & \leq |A| + |B|\nonumber\\
    & \leq \frac{\mathcal{O}(1)}{p + 1}\max\left(\beta_{\mathrm{max}}, \gamma_{\mathrm{max}}^2, \frac{M_{\gamma}}{\gamma_{\mathrm{max}}}\right).
\end{align}
Recalling the relationship between this quantity and the relevant additive contribution to the energy density:
\begin{align}
    \nu_{p, n} & \supset -\frac{2i}{\Gamma_{p + 1}\sqrt{n}}\sum_{0 \leq r \leq 2p + 1}\nu_{(r,\,p + 1)},
\end{align}
the discrepancy between the finite and infinite $p$ energy densities due to this additive contribution to the energy density is at most:
\begin{align}
    n^{-1/2}\frac{\mathcal{O}(1)}{p + 1}\max\left(\beta_{\mathrm{max}}, \gamma_{\mathrm{max}}^2, \frac{M_{\gamma}}{\gamma_{\mathrm{max}}}\right).
\end{align}
Similarly analyzing other additive contributions (see proposition \ref{prop:qgms_integral_series_expansion}) to the QGMS second order moments, hence to the energy, yields main theorem \ref{th:approximation_continuous_time_annealing_qaoa}. The analysis of these remaining additive contributions is sketched in Appendix~\ref{sec:other_moments_concentration}.

\subsection{The continuum limit of the SK-QAOA QGMS objects}
\label{sec:sk_qaoa_continuum_limit}

In this section, we detail the construction of the continuum limit of correlations tensors, on which the derivation of the main result (theorem \ref{th:approximation_continuous_time_annealing_qaoa}) in Section~\ref{sec:main_theorem_derivation} is based. As established in proposition \ref{prop:qgms-mgf_formulation_sk_qaoa_energy}, the SK-QAOA energy can be expressed from the order 2 moments (definition \ref{def:pqgms-mgf}) of a specific quadratic generalized multinomial sum, i.e. from complex numbers
\begin{align}
    & \frac{\partial^2S_n\left(\bm{\mu}\right)}{\partial\mu_{\alpha}^2}\Bigg|_{\bm{\mu} = \bm{0}}, \quad \alpha \in \mathcal{A}.
\end{align}
Section \ref{sec:qgms_moments_series_expansion} in turn provides a series expansion of such a moment; the explicit series is given in proposition \ref{prop:qgms_integral_series_expansion} for a generic order 2 diagonal moment. For instance, it states $\frac{\partial^2S_n\left(\bm{\mu}\right)}{\partial\mu_{\alpha}^2}\Bigg|_{\bm{\mu} = \bm{0}}$ has an additive contribution:
\begin{align}
    \frac{\partial^2S_n\left(\bm{\mu}\right)}{\partial\mu_{\alpha}^2}\Bigg|_{\bm{\mu} = \bm{0}} & \supset \frac{2}{\sqrt{n}}\nu_{\alpha},
\end{align}
where
\begin{align}
    \nu_{\alpha} & = \sum_{\left(n_d\right)_{d \geq 2}}\binom{n}{\left(n_d\right)_{d \geq 2}}\frac{n^{-\sum\limits_{d \geq 2}dn_d/2}}{\prod\limits_{d \geq 2}d!^{n_d}}\theta^*_{\alpha}\left\langle \bm{\mathcal{I}}_{\alpha}^{\left(1 + \sum\limits_{d \geq 2}dn_d\right)}, \bigotimes_{d \geq 2}\bm{\delta C}^{(d)\otimes n_d} \right\rangle.\label{eq:qgms_integral_series_expansion_restated}
\end{align}
Each term in the above series presents as a combinatorial constant, times a dot product between a matching tensor $\bm{\mathcal{I}}^{(D)}$ (defined in proposition \ref{prop:standard_normal_distribution_moments}) and tensor products of correlation tensors $\bm{C}^{(d)}$. Fixing a specific set of multinomial numbers $\left(n_d\right)_{d \geq 2}$ and letting $D := \sum_{d \geq 2}dn_d$, the dot product is merely a shorthand notation for sum:
\begin{align}
    \left\langle \bm{\mathcal{I}}^{\left(1 + D\right)}, \bigotimes_{d \geq 2}\bm{\delta C}^{(d)\otimes n_d} \right\rangle & = \sum_{\bm{\alpha}_{1:D} \in \mathcal{A}^D}\mathcal{I}^{(1 + D)}_{\alpha,\,\bm{\alpha}_{1:D}}\left[\bigotimes_{d \geq 2}\bm{\delta C}^{(d)\otimes n_d}\right]_{\bm{\alpha}_{1:D}}.\label{eq:qgms_series_expansion_dot_product_example}
\end{align}
The fundamental idea behind the derivation of the continuum limit is to show convergence of this discrete sum to an integral. This will in turn be obtained by establishing convergence of correlations tensors\footnote{More accurately, the $\bm{G}$ correlations tensors, related to the $\bm{C}^{(d)}$ ones through multiplication by  a simple function $\bm{\gamma}$ angles, converge to continuous functions.} $\bm{C}^{(d)}$ to corresponding continuous multivariate functions $C^{(d),\,\mathrm{cont}}$. More accurately, we will establish that in the large $p$ limit and for a family of schedules $\bm{\gamma}^{(p)}, \bm{\beta}^{(p)}$ parametrized according to definition \ref{def:continuous_schedule_informal}, $\bm{C}^{(d)} \in \left(\mathbf{C}^{\mathcal{A}}\right)^{\otimes d}$ is an approximate discretization of a function $C^{(d),\,\mathrm{cont}}: \left([0, 2]^2\right)^d \longrightarrow \mathbf{C}$, depending on $d$ and annealing schedules $\gamma^{\mathrm{cont}}$, $\beta^{\mathrm{cont}}$, but not on $p$:
\begin{align}
    C^{(d)}_{\alpha_1,\,\ldots,\,\alpha_d} & \approx C^{(d),\,\mathrm{cont}}\left(\frac{\alpha_1}{p + 1/2}, \ldots, \frac{\alpha_d}{p + 1/2}\right)\label{eq:correlations_tensor_discretization_continuum_correlations_tensor_informal}
\end{align}
To establish approximation \ref{eq:correlations_tensor_discretization_continuum_correlations_tensor_informal}, we recall the representation of $\bm{C}^{(d)}$ as a series involving the saddle point $\bm{\theta}^*$ and noninteracting correlations tensors $\bm{\overline{C}}^{(m)}$, derived in Proposition~\ref{prop:c_tensors_noninteracting_c_tensors_series_expansion}:
\begin{align}
    \bm{C}^{(d)} & = \sum_{m \geq 0}\frac{1}{m!}\left\langle \bm{\theta}^{* \otimes m}, \bm{\overline{C}}^{\left(d + m\right)} \right\rangle\label{eq:c_tensors_noninteracting_c_tensors_series_expansion_restatement}\\
    & =: \sum_{m \geq 0}\bm{C}^{(d),\,m},\\
    \bm{C}^{\left(d\right),\,m} & \in \left(\mathbf{C}^{\mathcal{A}}\right)^{\otimes d},
\end{align}
where we adapted the result to the SK-QAOA energy case ($\mathcal{Z}^* = \overline{Z} = 1$) and omitted the $\lambda$ parameter for simplicity. To establish the existence of continuum limit Eq.~\ref{eq:correlations_tensor_discretization_continuum_correlations_tensor_informal}, we show convergence to a continuum limit of each term $C^{(d),\,m}$ of the series. This is in turn achieved by showing convergence to a continuum limit of the noninteracting correlations tensors $\bm{\overline{C}}^{(m)}$, the saddle point $\bm{\theta}^*$, and interpreting discrete sum:
\begin{align}
    \left[\left\langle \bm{\theta}^{* \otimes m}, \bm{\overline{C}}^{\left(d + m\right)} \right\rangle\right]_{\alpha_1,\,\ldots,\,\alpha_d} & = \sum_{\alpha_{d + 1},\,\ldots,\,\alpha_{d + m} \in \mathcal{A}}\theta^*_{\alpha_{d + 1}}\ldots\theta^*_{\alpha_{d + m}}\overline{C}^{\left(d + m\right)}_{\alpha_1,\,\ldots,\,\alpha_d,\,\alpha_{d + 1},\,\ldots,\,\alpha_{d + m}}.
\end{align}
as approximating a partial integral of continuous functions in the $p \to \infty$ limit; the partial integral is then itself a continuous function

We formalize these ideas according to the following steps. In section \ref{sec:noninteracting_correlations_tensors_continuum_limit}, we show convergence of noninteracting correlations to a continuum limit. This step is rather straightforward, relying on elementary calculations, and shows noninteracting correlations are in fact \textit{exact discretizations} of their continuum couterpart. In section \ref{sec:error_bounds_discrete_continuum_iterations}, we deduce from the previous results convergence of the saddle point $\bm{\theta}^*$ to a continuum limit. This step relies on analyzing the series expansion of $\bm{\theta}^*$ established in Appendix Section~\ref{sec:pqgms_saddle_point_expansion}. In this discussion, $\bm{\theta}^*$ was expressed as the first block of an infinite-dimensional vector $\left(\bm{I} - \bm{T}\right)^{-1}\bm{\overline{\Theta}}$, with $\bm{\overline{\Theta}}$ an infinite-dimensional vector and $\bm{T}$ an infinite-dimensional operator. Since both $\bm{\overline{\Theta}}$ and $\bm{T}$ are expressible in terms of noninteracting correlation tensors, one can invoke the results on the continuum limit of noninteracting correlation tensors established in section \ref{sec:noninteracting_correlations_tensors_continuum_limit}. However, unlike the latter section where the discretization was exact, section \ref{sec:error_bounds_discrete_continuum_iterations} only establishes $\bm{\theta}^*$ to be an approximate discretization of a limiting function $\theta^{*,\,\mathrm{cont}}$. With a continuum limit for $\bm{\theta}^*$ established, section \ref{sec:continuum_limit_higher_order_correlations} then invokes series expansion Eq.~\ref{eq:c_tensors_noninteracting_c_tensors_series_expansion_restatement} of correlations tensors to deduce their convergence to a continuum limit, following the sum-integral comparison argument previously discussed. Similar to the saddle point, the discretization of higher-order correlations tensor is only approximate. With the continuum limit of all correlations tensors in hand, section \ref{sec:continuum_limit_qgms_moments} establishes the continuum limit of a generic QGMS moment contribution $\nu_{\alpha}$, as recalled in equation \ref{eq:qgms_integral_series_expansion_restated}.

\subsubsection{The continuum limit of noninteracting correlation tensors}
\label{sec:noninteracting_correlations_tensors_continuum_limit}

In this section, we specialize to the QGMS describing the SK-QAOA energy. For convenience, we will use an indexing of bitstrings distinct from the usual 
\begin{align}
    \{1, 2, \ldots, p - 1, p, p + 1, -p - 1, -p, \ldots, -2, -1\}
\end{align}
(introduced in \cite{qaoa_maxcut_high_girth} and used in Appendix \ref{sec:sk_qaoa_qgms} of the present text). We replace these indices respectively by:
\begin{align}
    \{2p + 1, 2p, \ldots, p + 2, p + 1, p, p - 1, \ldots, 1, 0\}
\end{align}
The reason for this zero-based indexing is, in the continuum limit, bitstring index $0 \leq t \leq 2p + 1$ will be mapped to a real variable $x := j/(2p + 1) \in [0, 1]$. Under this new indexing, the SK-QAOA energy QGMS parameters can be rephrased as

\begin{align}
    \mathcal{I} & = \left\{0,\,1,\,\ldots,\,2p,\,2p + 1\right\},\label{eq:sk_qaoa_qgms_redefinition_I}\\
    \mathcal{S} & := \{1, -1\}^{\mathcal{I}}\nonumber\\
    & = \left\{\left(a_{2p + 1}, a_{2p}, \ldots, a_1, a_0\right)\,:\,a_{2p + 1},\,a_{2p},\,\ldots,\,a_1,\,a_0 \in \{1, -1\}\right\},\label{eq:sk_qaoa_qgms_redefinition_S}\\
    \mathcal{A} & := \mathcal{I}^2\nonumber\\
    & = \left\{\left(j, k\right)\,:\,j, k \in \mathcal{I}\right\},\label{eq:sk_qaoa_qgms_redefinition_A}\\
    Q_{\bm{a}} & := \frac{1}{2}\mathbf{1}\left[a_p = a_{p + 1}\right]\prod_{1 \leq t \leq p}\braket{a_{2p + 2 - t}|e^{i\beta_tX}|a_{2p + 1 - t}}\braket{a_t|e^{-i\beta_tX}|a_{t - 1}},\label{eq:sk_qaoa_qgms_redefinition_Q}\\
    L_{\left(r,\,s\right),\,\bm{a}} & := \sqrt{-\frac{\Gamma_r\Gamma_s}{2} + i\varepsilon}\,a_ra_s \quad \forall r, s \in \mathcal{I}\label{eq:sk_qaoa_qgms_redefinition_L}
\end{align}
As a side effect of this bitstring reindexing, $\bm{\Gamma}$ in the last line is now indexed and defined as:
\begin{align}
    \bm{\Gamma} & := \left(\Gamma_0, \Gamma_1, \ldots, \Gamma_{p - 2}, \Gamma_{p - 1}, \Gamma_p, \Gamma_{p + 1}, \Gamma_{p + 2}, \ldots, \Gamma_{2p}, \Gamma_{2p + 1}\right)\\
    & = \left(-\gamma_1, -\gamma_2, \ldots, -\gamma_{p - 1}, -\gamma_p, -\gamma_{p + 1}, \gamma_{p + 1}, \gamma_p, \gamma_{p - 1}, \ldots, \gamma_2, \gamma_1\right).\label{eq:sk_qaoa_Gamma_redefinition}
\end{align}

Observe that for this specific QGMS, multiplying $\bm{L}$ by a non-negative number $\lambda$:
\begin{align}
    \bm{L} & \longrightarrow \lambda\bm{L}
\end{align}
can be achieved by rescaling all $\bm{\gamma}$ angles by the same value. In other words, the series expansion around $\lambda = 0$ can be interpreted as an expansion where the shape of the $\bm{\gamma}$ schedule is fixed, but the absolute magnitude of the $\bm{\gamma}$ angles varies. In particular, the $\lambda \to 0$ limit corresponds to the limit where all $\bm{\gamma}$ angles tend to zero, in which case the QAOA ansatz degenerates to a product of independent $X$ rotations. In this section, we will use a notational variant for correlations. Namely, consdering correlations of degree $d$ for definiteness, if 
\begin{align}
    \bm{\alpha} & = \left(\alpha_1, \ldots, \alpha_d\right) \in \mathcal{A}^d,\\
    \alpha_r & = \left(j_{2r - 1}, j_{2r}\right), \qquad j_{2r - 1}, j_{2r - 1}\in \mathcal{I},\\
    \bm{j}_{1:2d} & := \left(j_1, \ldots, j_{2d}\right) \in \mathcal{I}^{2d},
\end{align}
we may use equivalent notations
\begin{align}
    C^{\left(d\right)}_{\bm{\alpha}} = C^{\left(d\right)}_{\left(\alpha_1, \ldots, \alpha_d\right)} = C^{\left(d\right)}_{j_1,\,j_2,\,\ldots,\,j_{2d - 1},\,j_{2d}} = C^{\left(d\right)}_{\bm{j}_{1:2d}}.
\end{align}
Of course, we also extend these notations to the noninteracting correlations. In the upcoming derivations, it will be convenient to separate correlation tensors $\bm{C}^{(d)}$ ---as well as their non-interacting counterpart $\overline{\bm{C}}^{(d)}$--- into an elementwise product of some matrix tensor $\bm{G}^{(2d)}$ ---or $\overline{\bm{G}}^{(2d)}$ for the noninteracting case--- and a simple tensor depending on the $\bm\gamma$ angles only.

\begin{definition}[$\bm{G}$ correlations tensor]
\label{def:g_correlations_tensor}
Let us define, for all $d \geq 1$, a tensor $\bm{\mathcal{G}}^{(2d)}$ of degree $2d$ index by $\mathcal{I}$, with entries defined by:
\begin{align}
    \mathcal{G}^{\left(2d\right)}_{\bm{j}_{1:2d}} & := \prod_{1 \leq r \leq d}\sqrt{-\frac{1}{2}\Gamma_{j_{2r - 1}}\Gamma_{j_{2r}} + i\varepsilon}.\label{eq:curvy_g_tensor_definition}
\end{align}
We also define a related tensor of degree $1$, $\bm{\mathcal{G}}$, with entries given by:
\begin{align}
    \mathcal{G}_{j} & := \frac{i\Gamma_j}{\sqrt{2}}, \qquad \forall j \in \mathcal{I}.\label{eq:curvy_g_vector_definition}
\end{align}
We note the following elementary identitities:
\begin{align}
    \bm{\mathcal{G}}^{\left(2d\right)} & = \left(\bm{\mathcal{G}}^{\left(2\right)}\right)^{\otimes d},\\
    \left(\mathcal{G}^{(2d)}_{\bm{j}_{1:2d}}\right)^2 & = \prod_{1 \leq r \leq 2d}\mathcal{G}_{j_{r}}.
\end{align}
We now define the $\bm{G}$ correlations tensor of order $2d$, and denote by $\bm{G}^{(2d)}$ the tensor of order $2d$ indexed by $\mathcal{I}$, such that $\bm{C}^{(d)}$ is the element-wise product of $\bm{G}^{\left(2d\right)}$ and $\bm{\mathcal{G}}^{(2d)}$:
\begin{align}
    C^{(d)}_{\bm{j}_{1:2d}} & =: G^{(2d)}_{\bm{j}_{1:2d}}\mathcal{G}^{(2d)}_{\bm{j}_{1:2d}}.\label{eq:g_higher_order_correlations_tensor}
\end{align}
We naturally extend this to non-interacting correlations, defining the noninteracting $\bm{G}$ correlations tensor as:
\begin{align}
    \overline{C}^{(d)}_{\bm{j}_{1:2d}} & =: \overline{G}^{(2d)}_{\bm{j}_{1:2d}}\mathcal{G}^{(2d)}_{\bm{j}_{1:2d}}.
\end{align}
\end{definition}

Definition \ref{def:g_correlations_tensor} of the (noninteracting) $\bm{G}$ correlations tensor imply the following explicit formula, following from the definition of (noninteracting) correlations tensors (definition \ref{def:c_tensor} and equation \ref{eq:noninteracting_correlations_tensor}), the specification of $\bm{L}$ for the present QGMS (equation \ref{eq:sk_qaoa_qgms_redefinition_L}), the fact $\mathcal{Z}^* = \overline{\mathcal{Z}^*} = 1$ for the SK-QAOA QGMS, and the definition of tensors $\bm{G}^{(2d)}$ above:
\begin{align}
    G^{(2d)}_{\bm{j}_{1:2d}} & = \sum_{\bm{a} \in \{1, -1\}^{\mathcal{I}}}Q_{\bm{a}}\exp\left(-\frac{1}{2}\sum_{r, s \in \mathcal{I}}G^{(2)}_{r,\,s}\Gamma_r\Gamma_sa_ra_s\right)\prod_{1 \leq r \leq d}a_{j_{2r - 1}}a_{j_{2r}},\\
    \overline{G}^{(2d)}_{\bm{j}_{1:2d}} & = \sum_{\bm{a} \in \{1, -1\}^{\mathcal{I}}}Q_{\bm{a}}\prod_{1 \leq r \leq d}a_{j_{2r - 1}}a_{j_{2r}}.
\end{align}

We now proceed to computing the noninteracting correlations $\overline{\bm{G}}^{(2d)}$ tensor in the following proposition:

\begin{proposition}[Computing noninteracting correlations]
\label{prop:gamma_0_correlations}
For all $d \geq 0$ and multi-index
\begin{align}
    \bm{j}_{1:2d} = \left(j_1, j_2, \ldots, j_{2d - 1}, j_{2d}\right) \in \mathcal{I}^{2d},
\end{align}
the noninteracting $\bm{G}$ correlations tensor of order $2d$ has entry $\bm{j}_{1:2d}$ given by:
\begin{align}
    \overline{G}^{\left(2d\right)}_{\bm{j}_{1:2d}} & = \exp\left(-2i\sum_{1 \leq r \leq d}\left(B_{j^{\left(2r\right)}} - B_{j^{\left(2r - 1\right)}}\right)\right).\label{eq:noninteracting_g_higher_order_correlations_tensor}
\end{align}
In the last equation,
\begin{align}
    j^{\left(1\right)}, j^{\left(2\right)}, \ldots, j^{\left(2d - 1\right)}, j^{\left(2d\right)}
\end{align}
is the ordering of numbers $j_1, j_2, \ldots, j_{2d - 1}, j_{2d}$ in increasing order. The notation is from order statistics and will be frequently used in the following. Besides,
\begin{align}
    \bm{B} = \left(B_0, B_1, \ldots, B_{2p}, B_{2p + 1}\right)
\end{align}
is the sequence of partial sums of the $\bm{\beta}$ angles, accounting both the QAOA unitaries (minus signs) and their inverse (plus signs):
\begin{align}
    B_r & := \sum_{0 \leq s \leq r}\widetilde{\beta}_s \qquad \forall 0 \leq r \leq 2p + 1,\label{eq:B_definition}
\end{align}
with the signed $\beta$ angles given by:
\begin{align}
    \bm{\widetilde\beta} & := \left(\widetilde{\beta}_0, \widetilde{\beta}_1, \widetilde{\beta}_2, \ldots, \widetilde{\beta}_{p - 1}, \widetilde{\beta}_p, \widetilde{\beta}_{p + 1}, \widetilde{\beta}_{p + 2}, \beta_{p + 3}, \ldots, \widetilde{\beta}_{2p}, \widetilde{\beta}_{2p + 1}\right)\nonumber\\
    & = \left(0, -\beta_1, -\beta_2, \ldots, -\beta_{p - 1}, -\beta_p, 0, \beta_p, \beta_{p - 1}, \ldots, \beta_2, \beta_1\right)\label{eq:beta_tilde_definition}.
\end{align}
In the following, we may denote $\overline{\bm{G}}^{\left(2d\right)}$ omitting the superscript when the implied dimension is clear from the number of indices.
\begin{proof}
Let us consider case $d = 2$ for notational simplicity; from the proof, the generalization to higher $d$ will be straightforward. We then aim to compute:
\begin{align}
    G^{(4)}_{j_1,\,j_2,\,j_3,\,j_4} & := \sum_{\bm{a} \in \{1, -1\}^{\mathcal{I}}}Q_{\bm{a}}a_{j_1}a_{j_2}a_{j_3}a_{j_4}.
\end{align}
This is manifestly symmetric in permutations of indices $j_1, j_2, j_3, j_4$, so one may assume without loss of generality $j_1 \leq j_2 \leq j_3 \leq j_4$, hence $\left(j^{(1)}, j^{(2)}, j^{(3)}, j^{(4)}\right) = \left(j_1, j_2, j_3, j_4\right)$. Recalling the definition of $\bm{Q}$ for the QGMS under consideration (equation \ref{eq:sk_qaoa_qgms_redefinition_Q}):
\begin{align}
    Q_{\bm{a}} & = \frac{1}{2}\braket{a_{2p + 1}|e^{i\beta_1X}|a_{2p}}\ldots\braket{a_{p + 2}|e^{i\beta_pX}|a_{p + 1}}\mathbf{1}\left[a_p = a_{p + 1}\right]\braket{a_p|e^{-i\beta_pX}|a_{p - 1}}\ldots \braket{a_1|e^{-i\beta_1X}|a_0}\\
    & = \frac{1}{2}\prod_{1 \leq t \leq 2p + 1}\braket{a_t|e^{i\widetilde{\beta}_tX}|a_{t - 1}}\\
    & = \braket{+|a_{2p + 1}}\left(\prod_{1 \leq t \leq 2p + 1}\braket{a_t|e^{i\widetilde{\beta}_tX}|a_{t - 1}}\right)\braket{a_0|+},
\end{align}
$Q_{\bm{a}}$ can be interpreted as the computational basis path integral weight of single-qubit matrix element:
\begin{align}
    \sum_{\bm{a} \in \{1, -1\}^{\mathcal{I}}}Q_{\bm{a}} & = \braket{+|e^{i\beta_1X}\ldots e^{i\beta_pX}e^{-i\beta_pX} \ldots e^{-i\beta_1X}|+}\\
    & = \bra{+}\overleftarrow{\prod_{t = 1}^{2p + 1}}e^{i\widetilde{\beta}_tX}\ket{+}
\end{align}
Likewise, $Q_{\bm{a}}$ can be interpreted as the path integral weight of the same matrix element but with $Z$ inserted at positions $j_1, j_2, j_3, j_4$:
\begin{align}
    \sum_{\bm{a} \in \{1, -1\}^{\mathcal{I}}}Q_{\bm{a}}a_{j_1}a_{j_2}a_{j_3}a_{j_4} & = \bra{+}\left(\overleftarrow{\prod_{t = j_4 + 1}^{2p + 1}}e^{i\widetilde{\beta}_tX}\right)Z\left(\overleftarrow{\prod_{t = j_3 + 1}^{j_4}}e^{i\widetilde{\beta}_tX}\right)Z\left(\overleftarrow{\prod_{t = j_2 + 1}^{j_3}}e^{i\widetilde{\beta}_tX}\right)Z\left(\overleftarrow{\prod_{t = j_1 + 1}^{j_2}}e^{i\widetilde{\beta}_tX}\right)Z\left(\overleftarrow{\prod_{t = 1}^{j_1}}e^{i\widetilde{\beta}_tX}\right)\ket{+}
\end{align}
This single qubit matrix element can then be evaluated using elementary relations:
\begin{align}
    e^{i\widetilde{\beta}_tX}\ket{\pm} & = e^{\pm i\widetilde{\beta}}\ket{\pm},\\
    Z\ket{\pm} & = \ket{\mp}.
\end{align}
This gives:
\begin{align}
    \sum_{\bm{a} \in \{1, -1\}^{\mathcal{I}}}Q_{\bm{a}}a_{j_1}a_{j_2}a_{j_3}a_{j_4} & = \left(\prod_{t = j_4 + 1}^{2p + 1}e^{i\widetilde{\beta}_t}\right)\left(\prod_{t = j_3 + 1}^{j_4}e^{-i\widetilde{\beta}_t}\right)\left(\prod_{t = j_2 + 1}^{j_3}e^{  i\widetilde{\beta}_t}\right)\left(\prod_{t = j_1 + 1}^{j_2}e^{-i\widetilde{\beta}_t}\right)\left(\prod_{t = 1}^{j_1}e^{i\widetilde{\beta}_t}\right)\nonumber\\
    & = \exp\left(\underbrace{i\sum_{t = 0}^{2p + 1}\widetilde{\beta}_t}_{= 0} - 2i\sum_{t = j_3 + 1}^{j_4}\widetilde{\beta}_t - 2i\sum_{t = j_1 + 1}^{j_2}\widetilde{\beta}_t\right)\nonumber\\
    & = \exp\left(-2i\left(B_{j_4} - B_{j_3}\right) - 2i\left(B_{j_2} - B_{j_1}\right)\right)\\
    & = \exp\left(-2i\left(B_{j^{(4)}} - B_{j^{(3)}}\right) - 2i\left(B_{j^{(2)}} - B_{j^{(1)}}\right)\right),
\end{align}
which is the claimed formula for $d = 2$. The generalization to arbitrary $d$ is straightforward from the proof.
\end{proof}
\end{proposition}

We now introduce a ``continuum limit" for QAOA objects ---starting with angles, which we will show finite $p$ QAOA objects converge to in the constant time, $p \to \infty$, for an appropriate scaling of angles.

\begin{definition}[Continuum version of QAOA angles and related angle functions]
\label{def:angles_from_continuum}
A ``continuum limit" for QAOA angles is defined by a continuous function, $M_{\gamma}$-Lipschitz function:
\begin{align}
    \gamma^{\mathrm{cont}}: \begin{array}{ccc}
         [0, 1] & \longrightarrow & \mathbf{R}
    \end{array}\label{eq:gamma_continuum}
\end{align}
and a function odd about $1$:
\begin{align}
    \widetilde{\beta}^{\mathrm{cont}}: & \begin{array}{ccc}
         [0, 2] & \longrightarrow & \mathbf{R}
    \end{array}\label{eq:beta_tilde_continuum},\\
    \widetilde{\beta}^{\mathrm{cont}}\left(1 - x\right) & = -\widetilde{\beta}^{\mathrm{cont}}\left(x\right),\\
    \widetilde{\beta}^{\mathrm{cont}} & \textrm{ continuous on $[0, 1)$ and $(1, 2]$}
\end{align}
Informally, $\widetilde{\beta}^{\mathrm{cont}}$ should be regarded as a continuum analogue of sequence $\bm{\beta}$ introduced in equation \ref{eq:beta_tilde_definition}. From these functions, discrete $\bm{\gamma}$ and $\bm{\beta}$ angles are defined as follows at depth $p$:
\begin{align}
    \gamma_t & := \frac{1}{p + 1}\gamma^{\mathrm{cont}}\left(\frac{t - 1}{p + 1/2}\right) & \forall 1 \leq t \leq p + 1,\label{eq:gamma_from_continuum}\\
    \beta_t & := -\int_{(t - 1)/(p + 1/2)}^{t/(p + 1/2)}\mathrm{d}x\,\widetilde{\beta}^{\mathrm{cont}}\left(x\right) & \forall 1 \leq t \leq p.\label{eq:beta_from_continuum}
\end{align}
Given a function $\gamma^{\mathrm{opt}}$ in equation \ref{eq:gamma_continuum}, we define a related function 
\begin{align}
    \Gamma^{\left(\mathrm{cont}\right)}: \left\{\begin{array}{ccc}
        [0, 2] & \longrightarrow & \mathbf{R}\\
        x & \longmapsto & \left\{\begin{array}{cc}
             -\gamma^{\mathrm{cont}}\left(x\right) & \textrm{if } x \in [0, 1)\\
             \gamma^{\mathrm{cont}}\left(2 - x\right) & \textrm{if } x \in [1, 2] 
        \end{array}\right.
    \end{array}\right.,\label{eq:Gamma_continuum}
\end{align}
We also define a continuum analogue of sequence $\bm{B}$ (equation \ref{eq:B_definition}) as the integral of $\widetilde{\beta}^{\mathrm{cont}}$:
\begin{align}
    B^{\mathrm{cont}}: \left\{\begin{array}{ccc}
         [0, 2] & \longrightarrow & \mathbf{R}\\
         x & \longmapsto & \int_0^x\!\mathrm{d}y\,\widetilde{\beta}^{\mathrm{cont}}\left(y\right)
    \end{array}\right..\label{eq:B_continuum}
\end{align}
From the assumption that $\widetilde{\beta}^{\mathrm{cont}}$ is continuous everywhere except perhaps at $1$, $B^{\mathrm{cont}}$ is continuous, and in fact differentiable everywhere with continuous derivative except perhaps at $1$.
\end{definition}

Definition \ref{def:angles_from_continuum} then introduces a prescription to define $\bm{\gamma}, \bm{\beta}$ angles at any $p$ by discretization of continuous functions $\gamma^{\mathrm{opt}}, \beta^{\mathrm{opt}}$. The following proposition shows that for $\bm{\gamma}, \bm{\beta}$ angles defined by continuous functions $\gamma^{\mathrm{cont}}, \beta^{\mathrm{cont}}$ according to equation \ref{eq:gamma_from_continuum}, \ref{eq:beta_from_continuum}, $\bm{\Gamma}$ (as indexed in equation \ref{eq:sk_qaoa_Gamma_redefinition}) and sequence $\bm{B}$ (equation \ref{eq:B_definition}) are discretizations of functions $\Gamma^{\mathrm{cont}}, B^{\mathrm{cont}}$ introduced in equation \ref{eq:Gamma_continuum}, \ref{eq:B_continuum}.

\begin{proposition}[Finite $p$ angle functions $\bm{\Gamma}$ and $\bm{B}$ are discretizations of $\Gamma^{\mathrm{cont}}$, $B^{\mathrm{cont}}$]
\label{prop:angle_functions_discrete_from_continuum}
Let $\bm{\gamma} = \left(\gamma_t\right)_{t \in [p]}$ be a QAOA $\bm\gamma$ angles schedule defined by discretization of a continuum schedule $\gamma^{\mathrm{cont}}$ as per equation \ref{eq:gamma_from_continuum}. Then, $\bm{\Gamma} = \left(\Gamma_0, \ldots, \Gamma_{2p}\right)$, as indexed and specified from $\bm{\gamma}$ in equation \ref{eq:sk_qaoa_Gamma_redefinition}, is the discretization of function $\Gamma^{\mathrm{cont}}$ defined by equation \ref{eq:Gamma_continuum}:
\begin{align}
    \Gamma_r & = \frac{1}{p + 1}\Gamma^{\mathrm{cont}}\left(\frac{r}{p + 1/2}\right) \qquad \forall 0 \leq r \leq 2p + 1.\label{eq:Gamma_from_continuum}
\end{align}
Likewise, given a $\bm\beta$ schedule at finite $p$ arising from a continuum $\widetilde{\beta}^{\mathrm{cont}}$ schedule (according to equation \ref{eq:beta_from_continuum}), the sequence $\bm{B}$ defined from these $\bm\beta$ angles (equation \ref{eq:B_definition}) is a discretization of function $B^{\mathrm{cont}}$ introduced in definition \ref{def:angles_from_continuum} (equation \ref{eq:B_continuum}):
\begin{align}
    B_r & = B^{\mathrm{cont}}\left(\frac{r}{p + 1/2}\right), \qquad \forall 0 \leq r \leq 2p + 1.\label{eq:B_from_continuum}
\end{align}
\begin{proof}
We start by proving equality \ref{eq:Gamma_from_continuum}, relating to the $\bm\gamma$ angles. We check the equality for $0 \leq r \leq p$ and $p + 1 \leq r \leq 2p + 1$ respectively, starting with the definition of $\bm{\Gamma}$ in equation \ref{eq:sk_qaoa_Gamma_redefinition}.
For $0 \leq r \leq p$,
\begin{align}
    \Gamma_r & = -\gamma_{r + 1}\nonumber\\
    & = -\frac{1}{p + 1}\gamma^{\mathrm{opt}}\left(\frac{r}{p + 1/2}\right)\nonumber\\
    & = \frac{1}{p + 1}\Gamma^{\mathrm{cont}}\left(\frac{r}{p + 1/2}\right) && \textrm{since } \frac{r}{p + 1/2} \in [0, 1).
\end{align}

For $p + 1 \leq r \leq 2p + 1$,
\begin{align}
    \Gamma_r & = \gamma_{2p + 2 - r}\nonumber\\
    & = \frac{1}{p + 1}\gamma^{\mathrm{cont}}\left(\frac{2p + 1 - r}{p + 1/2}\right)\nonumber\\
    & = \frac{1}{p + 1}\gamma^{\mathrm{cont}}\left(2 - \frac{r}{p + 1/2}\right)\nonumber\\
    & = \frac{1}{p + 1}\Gamma^{\mathrm{cont}}\left(\frac{r}{p + 1/2}\right) && \textrm{since } \frac{r}{p + 1/2} \in (1, 2].
\end{align}

We now prove equality \ref{eq:B_from_continuum}, relating to the $\bm\beta$ angles. We first observe that for all $1 \leq r \leq 2p + 1$,
\begin{align}
    \widetilde{\beta}_r & = \int_{(r - 1)/(p + 1/2)}^{r/(p + 1/2)}\!\mathrm{d}y\,\widetilde{\beta}^{\mathrm{cont}}\left(y\right),
\end{align}
where $\widetilde{\bm{\beta}} = \left(\widetilde{\beta}_r\right)_{0 \leq r \leq 2p}$ and $\widetilde{\beta}^{\mathrm{cont}}$ are defined in equations \ref{eq:beta_tilde_definition}, \ref{eq:beta_tilde_continuum} respectively. First, for $1 \leq r \leq p$, the integration interval $\left[(r - 1)/(p + 1/2), r/(p + 1/2)\right] \subset [0, 1)$, hence
\begin{align}
    \int_{(r - 1)/(p + 1/2)}^{r/(p + 1/2)}\!\mathrm{d}y\,\widetilde{\beta}^{\mathrm{cont}}\left(y\right) & = -\int_{(r - 1)/(p +  1/2)}^{r/(p + 1/2)}\!\mathrm{d}y\,\beta^{\mathrm{cont}}\left(y\right)\nonumber\\
    & = -\beta_r\nonumber\\
    & = \widetilde{\beta}_r.
\end{align}
Next, for $r = p + 1$,
\begin{align}
    \int_{(r - 1)/(p + 1/2)}^{r/(p + 1/2)}\!\mathrm{d}y\,\widetilde{\beta}^{\mathrm{cont}}\left(y\right) & = \int_{p/(p + 1/2)}^{(p + 1)/(p + 1/2)}\!\mathrm{d}y\,\widetilde{\beta}^{\mathrm{cont}}\left(y\right)\nonumber\\
    & = 0\nonumber\\
    & = \widetilde{\beta}_{p + 1},
\end{align}
where the vanishing in the second line results from the symmetry of the interval about $1$, relative to which $\widetilde{\beta}^{\mathrm{cont}}$ is odd. Finally, for $p + 2 \leq r \leq 2p + 1$, the integration interval $[(r - 1)/(p + 1/2), r/(p + 1/2)] \subset (1, 2]$, hence
\begin{align}
    \int_{(r - 1)/(p + 1/2)}^{r/(p + 1/2)}\!\mathrm{d}y\,\widetilde{\beta}^{\mathrm{cont}}\left(y\right) & = \int_{(r - 1)/(p + 1/2)}^{r/(p + 1/2)}\!\mathrm{d}y\,\beta^{\mathrm{cont}}\left(2 - y\right)\nonumber\\
    & = \int_{2 - r/(p + 1/2)}^{2 - (r - 1)/(p + 1/2)}\!\mathrm{d}y\,\beta^{\mathrm{cont}}\left(y\right)\nonumber\\
    & = \int_{(2p + 1 - r)/(p + 1/2)}^{(2p + 2 - r)/(p + 1/2)}\!\mathrm{d}y\,\beta^{\mathrm{cont}}\left(y\right)\nonumber\\
    & = \beta_{2p + 2 - r}\nonumber\\
    & = \widetilde{\beta}_r.
\end{align}
We can now write, for all $0 \leq r \leq 2p + 1$,
\begin{align}
    B_r & = \sum_{1 \leq s \leq r}\widetilde{\beta}_s\nonumber\\
    & = \sum_{1 \leq s \leq r}\int_{(s - 1)/(p + 1/2)}^{s/(p + 
 1/2)}\!\mathrm{d}y\,\widetilde{\beta}^{\mathrm{cont}}\left(y\right)\nonumber\\
    & = \int_0^{r/(p + 1/2)}\!\mathrm{d}y\,\widetilde{\beta}^{\mathrm{cont}}\left(y\right)\nonumber\\
    & = B^{\mathrm{cont}}\left(\frac{r}{p + 1/2}\right),
\end{align}
which is equation \ref{eq:B_from_continuum}.
\end{proof}
\end{proposition}

Inspired by the explicit formulae equations \ref{eq:noninteracting_g_higher_order_correlations_tensor} for the noninteracting correlations in the SK-QAOA energy QGMS, one can introduce continuum analogues of these objects:

\begin{definition}[Continuum version of the noninteracting correlations]
\label{def:continuum_gamma_0_correlations}
The continuum version of the noninteracting $\bm{G}$ matrix of order $2d$ is a function of $2d$ variables: $[0, 2]^{2d} \longrightarrow \mathbf{R}$, defined by:
\begin{align}
    \overline{G}^{(2d),\,\mathrm{cont}}\left(\bm{x}_{1:2d}\right) & := \exp\left(-2i\sum_{1 \leq r \leq d}\left(B^{\mathrm{cont}}\left(x^{\left(2r\right)}\right) - B^{\mathrm{cont}}\left(x^{\left(2r - 1\right)}\right)\right)\right)\label{eq:continuum_gamma_0_g_correlations}
\end{align}
In the above definition, similar to proposition \ref{prop:gamma_0_correlations}, we used notation (similar to order statistics)
\begin{align}
    x^{\left(1\right)}, x^{\left(2\right)}, \ldots, x^{\left(2d - 1\right)}, x^{\left(2d\right)}
\end{align}
for the ordering of sequence
\begin{align}
    \bm{x}_{1:2d} = \left(x_1, x_2, \ldots, x_{2d - 1}, x_{2d}\right)
\end{align}
in increasing order. The continuum version of the $\bm{\mathcal{G}}$ angles tensor (equation \ref{eq:curvy_g_tensor_definition}) is defined as a $2d$-variables function:
\begin{align}
    \mathcal{G}^{\left(2d\right),\,\mathrm{cont}}\left(\bm{x}_{1:2d}\right) & := \prod_{1 \leq r \leq d}\sqrt{-\frac{1}{2}\Gamma^{\mathrm{cont}}\left(x_{2r - 1}\right)\Gamma^{\mathrm{cont}}\left(x_{2r}\right) + i\varepsilon}.
\end{align}
Similar to the discrete case, note identity
\begin{align}
    \mathcal{G}^{\left(2d\right),\,\mathrm{cont}} & = \left(\mathcal{G}^{\left(2\right),\,\mathrm{cont}}\right)^{\otimes d},
\end{align}
where the tensor product is meant in the usual sense for functions, that is $\left(f \otimes g\right)\left(x, y\right) := f\left(x\right)g\left(y\right)$ for functions $f, g$ of variables $x, y$ respectively. The continuum version of the noninteracting correlations of order $d$ is a $2d$-variables function defined as follows:
\begin{align}
    \overline{C}^{\left(d\right),\,\mathrm{cont}}\left(\bm{x}_{1:2d}\right) & := \mathcal{G}^{\left(2d\right),\,\mathrm{cont}}\left(\bm{x}_{1:2d}\right)\overline{G}^{\left(2d\right),\,\mathrm{cont}}\left(\bm{x}_{1:2d}\right).\label{eq:gamma_0_correlations_continuum}
\end{align}
\end{definition}

The interest of definition \ref{def:continuum_gamma_0_correlations} is, noninteracting  correlations at finite $p$ (computed in proposition \ref{prop:gamma_0_correlations}) can be shown to be discretizations of the noninteracting correlations in the continuum limit:

\begin{proposition}[Noninteracting correlations at finite $p$ are discretized continuum noninteracting correlations]
\label{prop:gamma_0_correlations_discrete_from_continuum}
Let $\bm{\gamma}, \bm{\beta}$ a QAOA angles schedule at finite $p$ derived from continuum schedules $\gamma^{\mathrm{cont}}, \widetilde{\beta}^{\mathrm{cont}}$ according to equations \ref{eq:gamma_from_continuum}, \ref{eq:beta_from_continuum}. Then, the noninteracting correlations computed for this schedule are discretizations of the continuum noninteracting correlations defined for the continuum schedule in definition \ref{def:continuum_gamma_0_correlations}:
\begin{align}
    \overline{C}^{\left(d\right)}_{\bm{j}_{1:2d}} & = \frac{1}{\left(p + 1\right)^d}\overline{C}^{\left(d\right),\,\mathrm{cont}}\left(\frac{\bm{j}_{1:2d}}{p + 1/2}\right).
\end{align}
More specifically, the noninteracting $\bm{G}$ correlations of order $2d$: $\bm{G}^{(2d)}$ are discretized from their continuum counterpart $G^{(2d),\,\mathrm{cont}}$ as follows:
\begin{align}
    \overline{G}^{(2d)}_{\bm{j}_{1:2d}} & = \overline{G}^{(2d),\,\mathrm{cont}}\left(\frac{\bm{j}_{1:2d}}{p + 1/2}\right),
\end{align}
and the $\bm{\mathcal{G}}^{(2d)}$ tensor is discretized from its continuum counterpart $\mathcal{G}^{\left(2d\right),\,\mathrm{cont}}$ as follows:
\begin{align}
    \mathcal{G}^{(2d)}_{\bm{j}_{1:2d}} & = \frac{1}{\left(p + 1\right)^d}\mathcal{G}^{(2d),\,\mathrm{cont}}\left(\frac{\bm{j}_{1:2d}}{p + 1/2}\right).\label{eq:noninteracting_g_correlations_discrete_from_continuum}
\end{align}
\begin{proof}
We first show 
\begin{align}
    \overline{G}^{\left(2d\right)}_{\bm{j}_{1:2d}} & = \overline{G}^{\left(2d\right),\,\mathrm{cont}}\left(\frac{\bm{j}_{1:2d}}{p + 1/2}\right).\label{eq:curvy_g_tensor_discrete_from_continuum}
\end{align}
Indeed,
\begin{align}
    \overline{G}^{\left(2d\right)}_{\bm{j}_{1:2d}} & = \exp\left(-2i\sum_{1 \leq r \leq d}\left(B_{j^{(2r)}} - B_{j^{\left(2r - 1\right)}}\right)\right) \qquad \textrm{(proposition \ref{prop:gamma_0_correlations})}\nonumber\\
    & = \exp\left(-2i\sum_{1 \leq r \leq d}\left(B^{\mathrm{cont}}\left(\frac{j^{\left(2r\right)}}{p + 1/2}\right) - B^{\mathrm{cont}}\left(\frac{j^{\left(2r - 1\right)}}{p + 1/2}\right)\right)\right) \qquad \textrm{(equation \ref{eq:B_from_continuum})}\nonumber\\
    & = \overline{G}^{\left(2d\right),\,\mathrm{cont}}\left(\frac{\bm{j}_{1:2d}}{p + 1/2}\right).
\end{align}
Next,
\begin{align}
    \mathcal{G}^{\left(2d\right)}_{\bm{j}_{1:2d}} & = \prod_{1 \leq r \leq d}\sqrt{-\frac{1}{2}\Gamma_{j_{2r - 1}}\Gamma_{j_{2r}} + i\varepsilon}\nonumber\\
    & = \prod_{1 \leq r \leq d}\sqrt{-\frac{1}{2\left(p + 1\right)^2}\Gamma^{\mathrm{cont}}\left(\frac{j_{2r - 1}}{p + 1/2}\right)\Gamma^{\mathrm{cont}}\left(\frac{j_{2r}}{p + 1/2}\right) + i\varepsilon}\nonumber\\
    & = \frac{1}{\left(p + 1\right)^d}\mathcal{G}^{\left(2d\right),\,\mathrm{cont}}\left(\frac{\bm{j}_{1:2d}}{p + 1/2}\right).
\end{align}
Hence,
\begin{align}
    C^{\left(d\right)}_{\bm{j}_{1:2d}} & = \mathcal{G}^{\left(2d\right)}_{\bm{j}_{1:2d}}\overline{\bm{G}}^{\left(2d\right)}_{\bm{j}_{1:2d}}\nonumber\\
    & = \frac{1}{\left(p + 1\right)^d}\mathcal{G}^{\left(2d\right),\,\mathrm{cont}}\left(\frac{\bm{j}_{1:2d}}{p + 1/2}\right)\overline{G}^{\left(2d\right),\,\mathrm{cont}}\left(\frac{\bm{j}_{1:2d}}{p + 1/2}\right)\nonumber\\
    & = \overline{C}^{\left(d\right),\,\mathrm{cont}}\left(\frac{\bm{j}_{1:2d}}{p + 1/2}\right).
\end{align}
\end{proof}
\end{proposition}

Now we introduced continuum versions of noninteracting correlations, we are able to introduce continuum versions of $\bm{T}$ operator blocks ---which are defined from noninteracting correlations.

\begin{definition}[Continuum version of $\bm{T}$ operator block]
\label{def:t_block_continuum}
For any $q, d \geq 1$, the continuum version $T^{\mathrm{cont}}_{q,\,d}$ of $\bm{T}$ matrix block  $\bm{T}_{q,\,d}$ is a $2(q + d)$-variables function defined as follows. First, define the continuum contribution of partition
\begin{align}
    \left(\mu_l\right)_{l \geq 1},\,\left(l_1, \ldots, l_q\right)
\end{align}
as follows:
\begin{align}
    T^{\mathrm{cont}}_{q,\,d\,;\,\left(\mu_l\right)_{l \geq 1},\,\left(l_1, \ldots, l_q\right)}\left(\bm{x}_{1:2q},\,\cdot\,\right) & := (-1)^{\sum_l\mu_l}\frac{\left(q - 1 + 
    \sum\limits_{l \geq 1}\mu_l\right)!}{\left(q - 1\right)!}\frac{1}{l_1!\ldots l_q!\prod\limits_{l \geq 1}l!^{\mu_l}\mu_l!}\nonumber\\
    & \hspace*{20px} \times \bigotimes_{l \geq 1}\left(\overline{C}^{\left(l\right),\,\mathrm{cont}}\right)^{\otimes \mu_l} \otimes \bigotimes_{r = 1}^q\overline{C}^{\left(l_r + 1\right),\,\mathrm{cont}}\left(x_{2r - 1}, x_{2r},\,\cdot\,\right).\label{eq:t_block_partition_contribution_continuum}
\end{align}
In the above equation, the interpretation of tensor products of multivariate functions is the usual one, where each tensor power represents the space of functions over a single variable. For instance, for two single-variable functions $f$ and $g$, $\left(f \otimes g\right)\left(x, y\right) = f(x)g(y)$. Besides, $\overline{C}^{\left(l_1 + 1\right)}(x_1, x_2,\,\cdot)$ (for instance) is the functions of $l_1$ variables defined by partial evaluation of $\overline{C}^{\left(l_1 + 1\right)}$, setting the first two variables to $x_1, x_2$. Next, define the continuum version of block $(q, d)$ as the sum of these over all relevant partitions:
\begin{align}
    T^{\mathrm{cont}}_{q,\,d} & = \sum_{\substack{\left(\mu_l\right)_{l \geq 1}\\l_1, \ldots, l_q\\\sum_ll\mu_l + l_1 + \ldots + l_q = d}}T^{\mathrm{cont}}_{q,\,d\,;\,\left(\mu_l\right)_{l \geq 1},\,\left(l_1, \ldots, l_q\right)}.\label{eq:t_block_continuum}
\end{align}
\end{definition}

We now observe the $\bm{T}$ block at finite $p$ can be interpreted as discretizations of the continuum ones:

\begin{proposition}[$\bm{T}$ blocks at finite $p$ are discretized continuum $\bm{T}$ blocks]
\label{prop:t_block_discrete_from_continuum}
Let $\bm{\gamma}^{(p)}$ and $\bm{\beta}^{(p)}$ be defined from a continuum schedule as prescribed in definition \ref{def:angles_from_continuum}. Then, the $\bm{T}_{q,\,d}$ block corresponding to this finite $p$ QAOA is a discretization of the continuum $\bm{T}$ block of index $\left(q, d\right)$ introduced in definition \ref{def:t_block_continuum}:
\begin{align}
    \left[T_{q,\,d}\right]_{\bm{j}_{1:2q},\,\bm{k}_{1:2d}} & = \frac{1}{\left(p + 1\right)^{q + d}}T^{\mathrm{cont}}_{q,\,d}\left(\frac{\bm{j}_{1:2q}}{p + 1/2}, \frac{\bm{k}_{1:2d}}{p + 1/2}\right).
\end{align}
In the above equation, $\left[T_{q,\,d}\right]_{\bm{j}_{1:2q},\,\bm{k}_{1:2d}}$ on the left-hand-side is a simplified notation for $\left[T_{q,\,d}\right]_{\bm{\alpha}_{d + 1:d + q}\,\bm{\alpha}_{1:d}}$, where
\begin{align}
    \bm{\alpha}_{d + 1:d + q} = \left(\alpha_{d + 1}, \ldots, \alpha_{d + q}\right) = \left(\left(j_1, j_2\right), \ldots, \left(j_{2q - 1}, j_{2q}\right)\right) \simeq \left(j_1, j_2, \ldots, j_{2q - 1}, j_{2q}\right) = \bm{j}_{1:2q}
\end{align}
and
\begin{align}
    \bm{\alpha}_{1:d} = \left(\alpha_{1}, \ldots, \alpha_{d}\right) = \left(\left(k_1, k_2\right), \ldots, \left(k_{2q - 1}, k_{2q}\right)\right) \simeq \left(k_1, k_2, \ldots, k_{2d - 1}, k_{2d}\right) = \bm{k}_{1:2d}.
\end{align}
\begin{proof}
\begin{align}
    & \left[\bm{T}_{q,\,d\,;\,\left(\mu_l\right)_{l \geq 1},\,\left(l_1, \ldots, l_q\right)}\right]_{\bm{j}_{1:2q},\,\bm{k}_{1:2d}}\nonumber\\
    & = (-1)^{\sum_l\mu_l}\frac{\left(q - 1 + \sum\limits_{l \geq 1}\mu_l\right)!}{\left(q - 1\right)!}\frac{1}{l_1!\ldots l_q!\prod\limits_{l \geq 1}l!^{\mu_l}\mu_l!} \left[\bigotimes_{l \geq 1}\overline{\bm{C}}^{\left(l\right)\otimes \mu_l}\right]_{\bm{k}_{1:2\sum_ll\mu_l}}\left[\bigotimes_{r = 1}^q\overline{\bm{C}}^{\left(l_r + 1\right)}_{j_{2r - 1},\,j_{2r}}\right]_{\bm{k}_{2\sum_ll\mu_l + 1:2d}}\nonumber\\
    & = (-1)^{\sum_l\mu_l}\frac{\left(q - 1 + \sum\limits_{l \geq 1}\mu_l\right)!}{\left(q - 1\right)!}\frac{1}{l_1!\ldots l_q!\prod\limits_{l \geq 1}l!^{\mu_l}\mu_l!}\frac{1}{\left(p + 1\right)^{\sum_ll\mu_l}}\left(\bigotimes_{l \geq 1}\left(\overline{\bm{C}}^{\left(l\right),\,\mathrm{cont}}\right)^{\otimes \mu_l}\right)\left(\frac{\bm{k}_{1:2\sum_ll\mu_l}}{p + 1/2}\right)\nonumber\\
    & \hspace*{20px} \times \frac{1}{\left(p + 1\right)^{\sum_{r = 1}^q\left(l_r + 1\right)}}\left(\bigotimes_{r = 1}^q\overline{\bm{C}}^{\left(l_r + 1\right),\,\mathrm{cont}}\left(\frac{j_{2r - 1}}{p + 1/2}, \frac{j_{2r}}{p + 1/2},\,\cdot\,\right)\right)\left(\frac{\bm{k}_{2\sum_ll\mu_l + 1:2d}}{p + 1/2}\right)\nonumber\\
    & = (-1)^{\sum_l\mu_l}\frac{\left(q - 1 + \sum\limits_{l \geq 1}\mu_l\right)!}{\left(q - 1\right)!}\frac{1}{l_1!\ldots l_q!\prod\limits_{l \geq 1}l!^{\mu_l}\mu_l!}\nonumber\\
    & \hspace*{20px} \times \frac{1}{\left(p + 1\right)^{d + q}}\left(\bigotimes_{l \geq 1}\left(\overline{\bm{C}}^{\left(l\right),\,\mathrm{cont}}\right)^{\otimes \mu_l} \otimes \bigotimes_{r = 1}^q\overline{\bm{C}}^{\left(l_r + 1\right)}\left(\frac{j_{2r - 1}}{p + 1/2}, \frac{j_{2r}}{p + 1/2},\,\cdot\,\right)\right)\left(\frac{\bm{k}_{1:2d}}{p + 1/2}\right)\nonumber\\
    & = \frac{1}{\left(p + 1\right)^{q + d}}T^{\mathrm{cont}}_{q,\,d\,;\,\left(\mu_l\right)_{l \geq 1},\,\left(l_1, \ldots, l_q\right)}\left(\frac{\bm{j}_{1:2q}}{p + 1/2}, \frac{\bm{k}_{1:2d}}{p + 1/2}\right).
\end{align}
For the first equality, we merely conveniently rewrote the definition of (discrete) block $(q, d)$, evaluated at row indexed
\begin{align}
    \bm{\alpha}_{d + 1:d + q} = \left(\alpha_{d + 1}, \ldots, \alpha_{d + q}\right) & = \left(\left(j_1, j_2\right), \ldots, \left(j_{2q - 1}, j_{2q}\right)\right) = \bm{j}_{1:2q}
\end{align}
and column index
\begin{align}
    \bm{\alpha}_{1:d} = \left(\alpha_1, \ldots, \alpha_d\right) = \left(\left(k_1, k_2\right), \ldots, \left(k_{2d - 1}, k_{2d}\right)\right) = \bm{k}_{1:2d}.
\end{align}
For the second equality, we used the relation between discrete and continuum noninteracting correlations (proposition \ref{prop:gamma_0_correlations_discrete_from_continuum}). For the third equality, we gathered tensor products together and made some algebraic simplications. For the final equality, we recalled the continuum definition of block $(q, d)$ (definition \ref{def:t_block_continuum}).

Summing the previous equality over $\left(\mu_l\right)_{l \geq 1}$ and $\left(l_1, \ldots, l_q\right)$ then gives the desired equality.
\end{proof}
\end{proposition}

\subsubsection{The continuum limit of the saddle point}
\label{sec:error_bounds_discrete_continuum_iterations}

In the light of results from section \ref{sec:noninteracting_correlations_tensors_continuum_limit}, we start by redefining the QGMS as a parametrized QGMS ---using a simple rescaling--- to conveniently apply the small parameter saddle point expansion results from Appendix Section~\ref{sec:pqgms_saddle_point_expansion}. We assume a finite $p$ angles schedule $\bm\gamma, \bm\beta$ derived from continuum schedules $\gamma^{\mathrm{opt}}, \widetilde{\beta}^{\mathrm{opt}}$ as per definition \ref{def:angles_from_continuum}:
\begin{align}
    \gamma_t & := \frac{1}{p + 1}\gamma^{\mathrm{cont}}\left(\frac{t - 1}{p + 1/2}\right) & \forall 1 \leq t \leq p,\\
    \beta_t & := -\int_{(t - 1)/(p + 1/2)}^{t/(p + 1/2)}\!\mathrm{d}x\,\widetilde{\beta}^{\mathrm{cont}}\left(x\right) & \forall 1 \leq t \leq p,
\end{align}
implying 
\begin{align}
    \Gamma_j & = \frac{1}{p + 1}\Gamma^{\mathrm{cont}}\left(\frac{j}{p + 1/2}\right) & \forall 0 \leq j \leq 2p + 1.
\end{align}
(see equation \ref{eq:Gamma_continuum} defining $\Gamma^{\mathrm{cont}}$, as well as proposition \ref{prop:angle_functions_discrete_from_continuum}). From there, we introduce the maximum continuum $\gamma$:
\begin{align}
    \mathrm{\gamma}_{\mathrm{max}} & := \max_{x \in [0, 1]}\left|\gamma^{\mathrm{cont}}\left(x\right)\right|.\label{eq:gamma_max_repeated}
\end{align}
and redefine the QGMS as a parametrized QGMS:
\begin{align}
    \sum_{\bm{n}}\binom{n}{\bm{n}}\exp\left(\frac{\lambda^2}{2n}\bm{n}^T\bm{L}^T\bm{L}\bm{n} + \frac{\lambda}{n}\bm{\mu}^T\bm{L}\bm{n}\right)\prod_{\bm{a} \in \mathcal{S}}Q_{\bm{a}}^{n_{\bm{a}}},
\end{align}
with parameters now given by 
\begin{align}
    \mathcal{I} & := \{0, 1, \ldots, 2p, 2p + 1\},\label{eq:sk_qaoa_qgms_redefinition_rescaled_I}\\
    \mathcal{S} & := \{1, -1\}^{\mathcal{I}},\label{eq:sk_qaoa_qgms_redefinition_rescaled_S}\\
    \mathcal{A} & := \mathcal{I}^2,\label{eq:sk_qaoa_qgms_redefinition_rescaled_A}\\
    Q_{\bm{a}} & := \frac{1}{2}\mathbf{1}\left[a_p = a_{p + 1}\right]\prod_{1 \leq t \leq p}\braket{a_{2p + 2 - t}|e^{i\beta_tX}|a_{2p + 1 - t}}\braket{a_t|e^{-i\beta_tX}|a_{t - 1}},\label{eq:sk_qaoa_qgms_redefinition_rescaled_Q}\\
    L_{\left(r,\,s\right),\,\bm{a}} & = \sqrt{-\frac{\left(p + 1\right)^2\Gamma_r\Gamma_s}{\gamma_{\mathrm{max}}^2} + i\varepsilon}\,a_ra_s,\label{eq:sk_qaoa_qgms_redefinition_rescaled_L}
\end{align}
and parameter $\lambda$ defined as:
\begin{align}
    \lambda & = \frac{2^{-1/2}\gamma_{\mathrm{max}}}{p + 1}.\label{eq:sk_qaoa_qgms_redefinition_rescaled_lambda}
\end{align}
Comparing this QGMS parameters definition with those of section \ref{sec:noninteracting_correlations_tensors_continuum_limit} (equations \ref{eq:sk_qaoa_qgms_redefinition_I} to \ref{eq:sk_qaoa_qgms_redefinition_L}), we merely pulled out a constant factor $\lambda$ from the definition of $\bm{L}$. The definition of the $\bm{Q} = \left(Q_{\bm{a}}\right)_{\bm{a} \in \mathcal{S}}$ remains unchanged. However, given this reparametrization, it will prove convenient to amend the formulae of tensors $\bm{\mathcal{G}}^{(2d)}$, initially defined in definition \ref{def:g_correlations_tensor} and relating the (noninteracting) correlations tensors and $\bm{G}$ (noninteracting) correlations tensors. This redefinition will be such that the formula for noninteracting $\bm{G}$ correlation tensors remains unchanged.

\begin{definition}[$\bm{\mathcal{G}}^{(2d)}$ tensors and $\bm{G}$ correlations tensors, adapted for parametrized QGMS]
\label{def:g_correlations_tensor_rescaled}
Let us define, for all $d \geq 1$, a tensor $\bm{\mathcal{G}}^{(2d)}$ of degree $2d$ index by $\mathcal{I}$, with entries defined by:
\begin{align}
    \mathcal{G}^{\left(2d\right)}_{\bm{j}_{1:2d}} & := \prod_{1 \leq r \leq d}\sqrt{-\frac{(p + 1)^2}{\gamma_{\mathrm{max}}^2}\Gamma_{j_{2r - 1}}\Gamma_{j_{2r}} + i\varepsilon}.\label{eq:curvy_g_tensor_definition_rescaled}
\end{align}
Let us also define a related vector $\bm{\mathcal{G}}$ by entries
\begin{align}
    \mathcal{G}_j & := \frac{i(p + 1)\Gamma_j}{\gamma_{\mathrm{max}}}, \qquad \forall j \in \mathcal{I}.\label{eq:curvy_g_vector_definition_rescaled}
\end{align}
Observe the following elementary identities:
\begin{align}
    \bm{\mathcal{G}}^{\left(2d\right)} & = \left(\bm{\mathcal{G}}^{\left(2\right)}\right)^{\otimes d},\\
    \left(\mathcal{G}^{(2d)}_{\bm{j}_{1:2d}}\right)^2 & = \prod_{1 \leq r \leq 2d}\mathcal{G}_{j_r}.
\end{align}
This differs from definition \ref{def:g_correlations_tensor} by constant factors; this definition ensures $\bm{\mathcal{G}}^{(2d)}$ have entries bounded by $1$. We now define the $\bm{G}$ correlations tensor of order $2d$, and denote by $\bm{G}^{(2d)}$ the tensor of order $2d$ indexed by $\mathcal{I}$, such that $\bm{C}^{(d)}$ is the element-wise product of $\bm{G}^{\left(2d\right)}$ and $\bm{\mathcal{G}}^{(2d)}$, with an additional rescaling by the $\lambda$ parameter of the parametrized QGMS:
\begin{align}
    C^{(d)}_{\bm{j}_{1:2d}} & =: \lambda^dG^{(2d)}_{\bm{j}_{1:2d}}\mathcal{G}^{(2d)}_{\bm{j}_{1:2d}}.\label{eq:g_correlations_tensor_definition_rescaled}
\end{align}
We naturally extend this to non-interacting correlations, defining the noninteracting $\bm{G}$ correlations tensor as:
\begin{align}
    \overline{C}^{(d)}_{\bm{j}_{1:2d}} & =: \overline{G}^{(2d)}_{\bm{j}_{1:2d}}\mathcal{G}^{(2d)}_{\bm{j}_{1:2d}}.\label{eq:g_noninteracting_correlations_tensor_definition_rescaled}
\end{align}
Note the extra factor $\lambda^d$ in the definition of interacting correlations as compared to definition \ref{def:g_correlations_tensor}.
\end{definition}

Under this new definition of the $\bm{\mathcal{G}}^{(2d)}$ tensors (equation \ref{eq:curvy_g_tensor_definition_rescaled}), proposition \ref{prop:gamma_0_correlations} for the explicit expression of noninteracting correlation tensors still holds:
\begin{align}
    \overline{G}^{\left(2d\right)}_{\bm{j}_{1:2d}} & = \exp\left(-2i\sum_{1 \leq r \leq d}\left(B_{j^{(2r)}} - B_{j^{(2r - 1)}}\right)\right).
\end{align}

Due to the updated definition of tensor $\bm{\mathcal{G}}^{(2d)}$, with entries now bounded by $1$, noninteracting correlation tensors are then bounded by $1$:
\begin{align}
    \left|\overline{C}^{\left(d\right)}_{\bm{j}_{1:2d}}\right| & \leq 1,
\end{align}
so that one may choose
\begin{align}
    c_{\mathrm{max}} & := 1.\label{eq:sk_qaoa_qgms_cmax}
\end{align}
Besides,
\begin{align}
    \sum_{\bm{a} \in \mathcal{S}}Q_{\bm{a}} = 1,
\end{align}
allowing choice
\begin{align}
    c_{\mathrm{min}} & := 1.\label{eq:sk_qaoa_qgms_cmin}
\end{align}
Finally, the sufficiently small time condition translates to the following being smaller than a constant depending only on $c_{\mathrm{min}}, c_{\mathrm{max}}$ (both of which absolute constants, in particular independent\footnote{The claim of independence of $p$ may be misleading. It is true only because for arbitrary $p$, we define the angles from a continuum schedule, with constant evolution time in the limit $p \to \infty$. If the time is allowed to grow, i.e. $\gamma_{\mathrm{max}}$ was allowed to grow with $p$, the rescaling of the QGMS parameters would still be valid, indeed allowing $c_{\mathrm{min}} = c_{\mathrm{max}} = 1$, but one will ultimately fail to satisfy the sufficiently small time assumption.} of $p$):
\begin{align}
    |\lambda||\mathcal{A}|^{1/2} & = \frac{2^{-1/2}\gamma_{\mathrm{max}}}{\left(p + 1\right)}\sqrt{\left(2p + 2\right)^2}\nonumber\\
    & = 2^{1/2}\gamma_{\mathrm{max}}.
\end{align}
This is in turn equivalent to assuming $\gamma_{\mathrm{max}}$ smaller than an absolute constant, implying bounded total evolution time under the phase separator unitary. All in all, for this reformulation as a $\lambda$-parametrized QGMS, the small $\lambda$ expansion results will hold for $\gamma_{\mathrm{max}}$ smaller than an absolute constant (independent of the QAOA angles and $p$).

We now provide appropriate redefinition of the continuum correlations and related objects for this reformulation of the original QGMS as a parametrized one.

\begin{definition}[Redefinition of continuum correlations and related objects]
\label{def:objects_redefinition_rescaled}
Given continuum angle schedules $\gamma^{\mathrm{opt}}, \widetilde{\beta}^{\mathrm{cont}}$, with $\gamma^{\mathrm{cont}}$ achieving maximum value $\gamma_{\mathrm{max}}$, the continuum noninteracting correlations of order $d$ are defined as:
\begin{align}
    \overline{C}^{\left(d\right),\,\mathrm{cont}}\left(\bm{x}_{1:2d}\right) & := \mathcal{G}^{\left(2d\right),\,\mathrm{cont}}\left(\bm{x}_{1:2d}\right)\overline{G}^{\left(2d\right),\,\mathrm{cont}}\left(\bm{x}_{1:2d}\right),\label{eq:gamma_0_correlations_continuum_rescaled}
\end{align}
where the continuum noninteracting $\bm{G}$ correlations of order $2d$ are defined as:
\begin{align}
    \overline{G}^{\left(2d\right),\,\mathrm{cont}}\left(\bm{x}_{1:2d}\right) & := \exp\left(-2i\sum_{1 \leq r \leq d}\left(B\left(x^{\left(2r\right)}\right) - B\left(x^{\left(2r - 1\right)}\right)\right)\right),\label{eq:sk_qaoa_gamma_0_correlations_continuum_rescaled}
\end{align}
and the continuum analogue of the $\bm{\mathcal{G}}^{(2d)}$ tensor (equation \ref{eq:curvy_g_tensor_definition_rescaled}) is defined as:
\begin{align}
    \mathcal{G}^{\left(2d\right),\,\mathrm{cont}}\left(\bm{x}_{1:2d}\right) & := \prod_{1 \leq r \leq d}\sqrt{-\frac{\Gamma^{\mathrm{cont}}\left(x_{2r - 1}\right)\Gamma^{\mathrm{cont}}\left(x_{2r}\right)}{\gamma_{\mathrm{max}}^2} + i\varepsilon}.\label{eq:curvy_g_tensor_continuum_definition_rescaled}
\end{align}
The discrete definition of $\bm{\mathcal{G}}$ (equation \ref{eq:curvy_g_vector_definition}) is likewise given a continuum analogue:
\begin{align}
    \mathcal{G}^{\mathrm{cont}}\left(x\right) & := \frac{i\Gamma^{\mathrm{cont}}(x)}{\gamma_{\mathrm{max}}}, \qquad \forall x \in [0, 2].\label{eq:curvy_g_vector_continuum_definition_rescaled}
\end{align}
Note elementary relations
\begin{align}
    \mathcal{G}^{\left(2d\right),\,\mathrm{cont}} & = \left(\mathcal{G}^{\left(2\right),\,\mathrm{cont}}\right)^{\otimes d},\\
    \mathcal{G}^{(2d),\,\mathrm{cont}}\left(\bm{x}_{1:2d}\right)^2 & = \prod_{1 \leq r \leq 2d}\mathcal{G}^{\mathrm{cont}}\left(x_r\right).
\end{align}
The definition of continuum $\bm{T}$ block $(q, d)$ as a function of noninteracting correlations is slightly adapted compared to definition \ref{def:t_block_continuum}, namely:
\begin{align}
    T^{\mathrm{cont}}_{q,\,d} & := \sum_{\substack{\left(\mu_l\right)_{l \geq 1}\\l_1, \ldots, l_q\\\sum_ll\mu_l + l_1 + \ldots + l_q = d}}T^{\mathrm{cont}}_{q,\,d\,;\,\left(\mu_l\right)_{l \geq 1},\,\left(l_1, \ldots, l_q\right)},\label{eq:t_block_continuum_rescaled}
\end{align}
with
\begin{align}
    T^{\mathrm{cont}}_{q,\,d\,;\,\left(\mu_l\right)_{l \geq 1},\,\left(l_1, \ldots, l_q\right)}\left(\bm{x}_{1:2q},\,\cdot\,\right) & := (-1)^{\sum_l\mu_l}\frac{\left(q - 1 + 
    \sum\limits_{l \geq 1}\mu_l\right)!}{\left(q - 1\right)!}\frac{\left(2^{-1/2}\gamma_{\mathrm{max}}\right)^{q + d}}{l_1!\ldots l_q!\prod\limits_{l \geq 1}l!^{\mu_l}\mu_l!}\nonumber\\
    & \hspace*{20px} \times \bigotimes_{l \geq 1}\left(\overline{C}^{\left(l\right),\,\mathrm{cont}}\right)^{\otimes \mu_l} \otimes \bigotimes_{r = 1}^q\overline{C}^{\left(l_r + 1\right),\,\mathrm{cont}}\left(x_{2r - 1}, x_{2r},\,\cdot\,\right),\label{eq:t_block_partition_contribution_continuum_rescaled}
\end{align}
where the continuum noninteracting correlations should now be defined by equation \ref{eq:gamma_0_correlations_continuum_rescaled}. The other difference compared to equation \ref{eq:t_block_partition_contribution_continuum} is the inclusion of geometric sequence $\left(2^{-1/2}\gamma_{\mathrm{max}}\right)^{q + d}$ in the definition.
\end{definition}

The correlations and $\bm{T}$ operator blocks associated to the new parametrized QGMS are now related by the following identities to their continuum counterparts:

\begin{proposition}[Relation between discrete and continuum noninteracting correlations and $\bm{T}$ blocks]
\label{prop:t_block_discrete_from_continuous_rescaled}
Consider the parametrized QGMS defined in this section by equations \ref{eq:sk_qaoa_qgms_redefinition_rescaled_I}-\ref{eq:sk_qaoa_qgms_redefinition_rescaled_L}, and the $\lambda$ parameter from equation \ref{eq:sk_qaoa_qgms_redefinition_rescaled_lambda}. Then, the noninteracting correlations (computed in equation \ref{eq:sk_qaoa_gamma_0_correlations_continuum_rescaled}) relate to their continuum counterparts (defined in equation \ref{eq:gamma_0_correlations_continuum_rescaled}) by:
\begin{align}
    \overline{C}^{\left(d\right)}_{\bm{j}_{1:2d}} & = \overline{C}^{\left(d\right),\,\mathrm{cont}}\left(\frac{\bm{j}_{1:2d}}{p + 1/2}\right).
\end{align}
More specifically, the continuum noninteracting $\bm{G}$ correlations and continuum $\bm{\mathcal{G}}^{(2d)}$ in the continuum setting are related as follows to their discrete counterparts:
\begin{align}
    G^{(2d)}_{\bm{j}_{1:2d}} & = \overline{G}^{(2d),\,\mathrm{cont}}\left(\frac{\bm{j}_{1:2d}}{p + 1/2}\right),\label{eq:noninteracting_g_correlations_discrete_from_continuum_rescaled}\\
    \mathcal{G}^{(2d)}_{\bm{j}_{1:2d}} & = \mathcal{G}^{(2d),\,\mathrm{cont}}\left(\frac{\bm{j}_{1:2d}}{p + 1/2}\right),\label{eq:curvy_g_tensor_discrete_from_continuum_rescaled}\\
    \mathcal{G}_{j} & = \mathcal{G}^{\mathrm{cont}}\left(\frac{j}{p + 1/2}\right),\label{eq:curvy_g_vector_discrete_from_continuum_rescaled}
\end{align}
Besides, $\bm{T}$ operator block $(q, d)$ associated to the parametrized QGMS relates to is continuum counterpart by:
\begin{align}
    \left[\bm{T}_{q,\,d}\right]_{\bm{j}_{1:2q},\,\bm{k}_{1:2d}} & = \frac{1}{\left(p + 1\right)^{q + d}}T^{\mathrm{cont}}_{q,\,d}\left(\frac{\bm{j}_{1:2q}}{p + 1/2}, \frac{\bm{k}_{1:2d}}{p + 1/2}\right).
\end{align}
\end{proposition}
Note the relation between discrete and continuum $\bm{T}$ blocks is the same as in proposition \ref{prop:t_block_discrete_from_continuum}, relying on a different parametrization of the QGMS (related by a simple rescaling). The proportionality factor $(p + 1)^{-q - d}$ nonetheless comes from a different place, and the continuum version of $\bm{T}_{q,\,d}$ further contains a geometric factor $\left(2^{-1/2}\gamma_{\mathrm{max}}\right)^{q + d}$, similar to the discrete $\lambda^{q + d}$, which will be crucial to prove convergence. It will be convenient to decompose matrix block $\bm{T}_{q,\,d}$ into the product of a closely related matrix $\bm{U}_{q,\,d}$ (depending only on the $\bm\beta$ angles), and diagonal matrices depending only on the $\bm{\gamma}$ angles. We also introduce the continuum analogue of $\bm{U}_{q,\,d}$.

\begin{definition}[$\bm{U}_{q,\,d}$ matrix block, discrete and continuum]
\label{def:u_block}
Given integers $q, d \geq 1$, the discrete $\bm{U}_{q,\,d}$ matrix block an operator with same domain and image space as $\bm{T}_{q,\,d}$ ($\mathcal{A}^q \longrightarrow \mathcal{A}^d$), defined in the same way as $\bm{T}_{q,\,d}$, but replacing $\overline{C}^{\left(d\right)}$ by $\overline{G}^{\left(2d\right)}$:
\begin{align}
    \bm{U}_{q,\,d} & := \sum_{\substack{\left(\mu_l\right)_{l \geq 1}\\l_1, \ldots, l_q\\\sum_{l}l\mu_l + l_1 + \ldots + l_q = d}}\bm{U}_{q,\,d\,;\,\left(\mu_l\right)_{l \geq 1},\,\left(l_1, \ldots, l_q\right)},\label{eq:u_block_definition}
\end{align}
with row $\bm{j}_{1:2q} \in \mathcal{I}^{2q} \simeq \mathcal{A}^q$ of $\bm{U}_{q,\,d\,;\,\left(\mu_l\right)_{l \geq 1},\,\left(l_1, \ldots, l_q\right)}$ given by:
\begin{align}
    \left[\bm{U}_{q,\,d\,;\,\left(\mu_l\right)_{l \geq 1},\,\left(l_1, \ldots, l_q\right)}\right]_{\bm{j}_{1:2q},\,:} & := \left(-1\right)^{\sum_l\mu_l}\frac{\left(q - 1 + \sum\limits_{l \geq 1}\mu_l\right)!}{\left(q - 1\right)!}\frac{\lambda^{q + d}}{l_1!\ldots l_q!\prod\limits_{l \geq 1}l!^{\mu_l}\mu_l!}\bigotimes_{l \geq 1}\overline{\bm{G}}^{\left(2l\right)\otimes \mu_l} \otimes \bigotimes_{r = 1}^q\overline{\bm{G}}^{\left(2l_r + 2\right)}_{j_{2r - 1},\,j_{2r}},\label{eq:u_block_partition_contribution_definition}
\end{align}
where we recall $\lambda = 2^{-1/2}\gamma_{\mathrm{max}}/(p + 1)$ (equation \ref{eq:sk_qaoa_qgms_redefinition_rescaled_lambda}). We also introduced a continuum analogue of $\bm{U}_{q,\,d}$ as a function $[0, 2]^{2q} \longrightarrow [0, 2]^{2d}$ defined by formula:
\begin{align}
    U^{\mathrm{cont}}_{q,\,d} & := \sum_{\substack{\left(\mu_l\right)_{l \geq 1}\\l_1, \ldots, l_q\\\sum_{l}l\mu_l + l_1 + \ldots + l_q = d}}U^{\mathrm{cont}}_{q,\,d\,;\,\left(\mu_l\right)_{l \geq 1},\,\left(l_1,\,\ldots,\,l_q\right)}.\label{eq:u_block_continuum}
\end{align}
Each term of this sum is defined as follows for fixed first variable $\bm{x}_{1:2q} \in [0, 2]^{2q}$:
\begin{align}
    U^{\mathrm{cont}}_{q,\,d\,;\,\left(\mu_l\right)_{l \geq 1},\,\left(l_1,\,\ldots,\,l_q\right)}\left(\bm{x}_{1:2d},\,\cdot\,\right) & := \left(-1\right)^{\sum_l\mu_l}\frac{\left(q - 1 + \sum\limits_{l \geq 1}\mu_l\right)!}{\left(q - 1\right)!}\frac{\left(2^{-1/2}\gamma_{\mathrm{max}}\right)^{q + d}}{l_1!\ldots l_q!\prod\limits_{l \geq 1}l!^{\mu_l}\mu_l!}\nonumber\\
    & \hspace*{20px} \times \bigotimes_{l \geq 1}\left(\overline{\bm{G}}^{\left(2l\right),\,\mathrm{cont}}\right)^{\otimes \mu_l} \otimes \bigotimes_{r = 1}^q\overline{\bm{G}}^{\left(2l_r + 2\right),\,\mathrm{cont}}\left(x_{2r - 1}, x_{2r},\,\cdot\,\right)\label{eq:u_block_partition_contribution_continuum}
\end{align}
From the relation between $\overline{\bm{C}}^{(d)}$ and $\overline{\bm{G}}^{\left(2d\right)}$ (Eq.~\ref{eq:noninteracting_g_higher_order_correlations_tensor}), the following identity between $\bm{T}_{q,\,d}$ and $\bm{U}_{q,\,d}$ holds:
\begin{align}
    \left[\bm{T}_{q,\,d}\right]_{\bm{j}_{1:2q},\,\bm{k}_{1:2d}} & = \mathcal{G}^{\left(2q\right)}_{\bm{j}_{1:2q}}\left[\bm{U}_{q,\,d}\right]_{\bm{j}_{1:2q},\,\bm{k}_{1:2d}}\mathcal{G}^{\left(2d\right)}_{\bm{k}_{1:2d}}.\label{eq:t_block_u_block_relation}
\end{align}
Likewise, from the relation between $\overline{C}^{\left(d\right),\,\mathrm{cont}}$ and $\overline{G}^{(2d),\,\mathrm{cont}}$ (equation \ref{eq:gamma_0_correlations_continuum}), the following identity between $T^{\mathrm{cont}}_{q,\,d}$ and $U^{\mathrm{cont}}_{q,\,d}$ holds:
\begin{align}
    T^{\mathrm{cont}}_{q,\,d}\left(\bm{x}_{1:2q}, \bm{y}_{1:2d}\right) & = \mathcal{G}^{\left(2q\right)}\left(\bm{x}_{1:2q}\right)U^{\mathrm{cont}}_{q,\,d}\left(\bm{x}_{1:2q}, \bm{y}_{1:2d}\right)\mathcal{G}^{\left(2d\right)}\left(\bm{y}_{1:2d}\right).\label{eq:t_block_continuum_u_block_continuum_relation}
\end{align}
Finally, the following discretization identity holds between $\bm{U}_{q,\,d}$ and its continuum counterpart:
\begin{align}
    \left[\bm{U}_{q,\,d}\right]_{\bm{j}_{1:2q},\,\bm{k}_{1:2d}} & = \frac{1}{\left(p + 1\right)^{q + d}}U^{\mathrm{cont}}_{q,\,d}\left(\frac{\bm{j}_{1:2q}}{p + 1/2}, \frac{\bm{k}_{1:2d}}{p + 1/2}\right).\label{eq:u_block_from_continuum}
\end{align}
\end{definition}
Identity \ref{eq:t_block_u_block_relation} means that $\bm{T}_{q,\,d}$ is related to $\bm{U}_{q,\,d}$ by left- and right-multiplication by matrices involving only the $\bm{\gamma}$ angles. Based on a simple adaptation of the proof of proposition \ref{prop:t_block_entrywise_bound}, one can state bounds on the entries of matrix $\bm{U}_{q,\,d}$ and the values of its continuum counterpart $U_{q,\,d}^{\mathrm{cont}}$:

\begin{proposition}[Uniform bounds on $\bm{U}_{q,\,d}$ and $U^{\mathrm{cont}}_{q,\,d}$]
\label{prop:u_block_entrywise_bound}
The following bound holds uniformly over entries of $\bm{U}_{q,\,d}$:
\begin{align}
    \left|\left[\bm{U}_{q,\,d}\right]_{\bm{j}_{1:2q},\,\bm{k}_{1:2d}}\right| & \leq \left(|\lambda|c\right)^{q + d} \qquad \forall \bm{j}_{1:2q} \in \mathcal{I}^{2q},\,\bm{k}_{1:2d} \in \mathcal{I}^{2d}.
\end{align}
for some universal constant $c$, which in this case can be taken $c = \log(3/2)^{-1}$. Similarly, the following uniform bound holds over the continuum counterpart of this matrix, continuous function $\bm{U}_{q,\,d}^{\mathrm{cont}}$:
\begin{align}
    \left|U^{\mathrm{cont}}_{q,\,d}\left(\bm{x}_{1:2q}, \bm{y}_{1:2d}\right)\right| \leq \left(2^{-1/2}\gamma_{\mathrm{max}}c\right)^{q + d} \qquad \forall \bm{x}_{1:2q} \in [0, 2]^{2q},\,\bm{y}_{1:2d} \in [0, 2]^{2d}.
\end{align}
\end{proposition}

The goal of the discretization identities in proposition \ref{prop:t_block_discrete_from_continuous_rescaled} and definition \ref{def:u_block} (equation \ref{eq:u_block_from_continuum}) will be to replace discrete sums by integrals. For that purpose, we will also need to show certain functions are Lipschitz. The relevant results are collected in the following proposition:

\begin{proposition}[Continuum noninteracting $\bm{G}$ correlations and related functions are Lipschitz in each variable]
\label{prop:gamma_0_g_correlations_continuum_lipschitz_bound}
Consider, for any $d \geq 1$, the continuum $\overline{\bm{G}}$ correlations of order $\overline{G}^{\left(2d\right),\,\mathrm{cont}}$ introduced in definition \ref{def:continuum_gamma_0_correlations}, equation \ref{eq:continuum_gamma_0_g_correlations}. Then, this function is $2\beta_{\mathrm{max}}$-Lipschitz in each of its variables, where
\begin{align}
    \beta_{\mathrm{max}} & := \max_{x \in [0, 1]}\left|\widetilde{\beta}^{\mathrm{cont}}(x)\right|.
\end{align}
Besides, $U^{\mathrm{cont}}_{q,\,d}$, the continuum analogue of matrix block $\bm{U}_{q,\,d}$ introduced in definition \ref{def:u_block}, is $2\beta_{\mathrm{max}}\left(2^{-1/2}c\gamma_{\mathrm{max}}\right)^{q + d}$-Lipschitz in each of its variables, with $c$ an absolute constant.
\begin{proof}
We start proving that $\overline{G}^{(2d),\,\mathrm{cont}}$ is Lipschitz in each of its variables. Recall
\begin{align}
    \overline{G}^{(2d),\,\mathrm{cont}}\left(\bm{x}_{1:2d}\right) & = \exp\left(-2i\sum_{1 \leq r \leq d}\left(B^{\mathrm{cont}}\left(x^{\left(2r\right)}\right) - B^{\mathrm{cont}}\left(x^{\left(2r - 1\right)}\right)\right)\right),
\end{align}
where we remind $x^{(1)}, x^{(2)}, \ldots, x^{(2d - 1)}, x^{(2d)}$ is the sorting of sequence $x_1, x_2, \ldots, x_{2d - 1}, x_{2d}$ in increasing order. This function is continuous everywhere and differentiable everywhere with continuous derivative except on the hyperplanes where two coordinates coincide, e.g. $x_1 = x_2$. Hence, a Lipschitz constant with respect to any variable $x_r$ ($1 \leq r \leq 2d$) is given by an upper bound on the derivative with respect to $x_r$, maximized over all the differentiability domain. For definiteness, consider the derivative with respect to $x_{2s}$ in domain $x_1 < x_2 < \ldots < x_{2d - 1} < x_{2d}$:
\begin{align}
    \frac{\partial}{\partial x_{2s}}\overline{G}^{(2d),\,\mathrm{cont}}\left(\bm{x}_{1:2d}\right) & = \frac{\partial}{\partial x_{2s}}\exp\left(-2i\sum_{1 \leq r \leq d}\left(B^{\mathrm{cont}}\left(x_{2r}\right) - B^{\mathrm{cont}}\left(x_{2r - 1}\right)\right)\right)\nonumber\\
    & = -2i\frac{\partial B^{\mathrm{cont}}}{\partial x_{2s}}\exp\left(-2i\sum_{1 \leq r \leq d}\left(B^{\mathrm{cont}}\left(x_{2r}\right) - B^{\mathrm{cont}}\left(x_{2r - 1}\right)\right)\right)\nonumber\\
    & = -2i\widetilde{\beta}^{\mathrm{cont}}\left(x_{2s}\right)\exp\left(-2i\sum_{1 \leq r \leq d}\left(B^{\mathrm{cont}}\left(x_{2r}\right) - B^{\mathrm{cont}}\left(x_{2r - 1}\right)\right)\right),
\end{align}
which is bounded by $2\beta_{\mathrm{max}}$. This proves the claim concerning $\overline{G}^{\left(2d\right),\,\mathrm{cont}}$.

Let us now consider $U^{\mathrm{cont}}_{q,\,d}$. Recalling equation \ref{eq:u_block_continuum}, we decompose it according to partitions:
\begin{align}
    U^{\mathrm{cont}}_{q,\,d} & := \sum_{\substack{\left(\mu_l\right)_{l \geq 1}\\l_1, \ldots, l_q\\\sum_{l}l\mu_l + l_1 + \ldots + l_q = d}}U^{\mathrm{cont}}_{q,\,d\,;\,\left(\mu_l\right)_{l \geq 1},\,\left(l_1,\,\ldots,\,l_q\right)}
\end{align}
and first consider a specific partitions $\left(\mu_l\right)_{l \geq 1}, \left(l_1, \ldots, l_q\right)$. Recalling equation \ref{eq:u_block_partition_contribution_continuum}, $U^{\mathrm{cont}}_{q,\,d\,;\,\left(\mu_l\right)_{l \geq 1},\,\left(l_1,\,\ldots,\,l_q\right)}\left(\bm{x}_{1:2d}\right)$ is the product of constant
\begin{align}
    \left(-1\right)^{\sum_l\mu_l}\frac{\left(q - 1 + \sum\limits_{l \geq 1}\mu_l\right)!}{\left(q - 1\right)!}\frac{\left(2^{-1/2}\gamma_{\mathrm{max}}\right)^{q + d}}{l_1!\ldots l_q!\prod\limits_{l \geq 1}l!^{\mu_l}\mu_l!},
\end{align}
times a product of $\overline{G}^{\left(2d'\right),\,\mathrm{cont}}$ functions evaluated at disjoint variables. Since each of these functions is $2\beta_{\mathrm{max}}$-Lipschitz in each variable and also bounded by $1$, it follows $U^{\mathrm{cont}}_{q,\,d\,;\,\left(\mu_l\right)_{l \geq 1},\,\left(l_1,\,\ldots,\,l_q\right)}$ is Lipschitz in each variable, with Lipschitz constant upper-bounded as:
\begin{align}
    2\beta_{\mathrm{max}}\frac{\left(q - 1 + \sum\limits_{l \geq 1}\mu_l\right)!}{\left(q - 1\right)!}\frac{\left(2^{-1/2}\gamma_{\mathrm{max}}\right)^{q + d}}{l_1!\ldots l_q!\prod\limits_{l \geq 1}l!^{\mu_l}\mu_l!}.
\end{align}
Summing this over partitions, $U^{\mathrm{cont}}_{q,\,d}$ is Lipschitz in each variable, with Lipschitz constant upper-bounded as
\begin{align}
    & \leq \sum_{\substack{\left(\mu_l\right)_{l \geq 1}\\l_1,\,\ldots,\,l_q\\\sum_ll\mu_l + l_1 + \ldots + l_q = d}}2\beta_{\mathrm{max}}\frac{\left(q - 1 + \sum\limits_{l \geq 1}\mu_l\right)!}{\left(q - 1\right)!}\frac{\left(2^{-1/2}\gamma_{\mathrm{max}}\right)^{q + d}}{l_1!\ldots l_q!\prod\limits_{l \geq 1}l!^{\mu_l}\mu_l!}\nonumber\\
    & \leq 2\beta_{\mathrm{max}}\left(2^{-1/2}\gamma_{\mathrm{max}}c\right)^{q + d},
\end{align}
where the simpler bound in the last line results from equation \ref{eq:t_block_entry_non_lambda_contribution_bound}; in this simpler bound $c$ is an absolute constant, which is in this case may be taken to $c := \log(3/2)^{-1}$.
\end{proof}
\end{proposition}

We now wish to give a continuum analogue of the saddle-point $\bm{\theta}^*$. Referring to proposition \ref{propSaddlePointEquationAnalyticity}, this is given by absolutely convergent series:
\begin{align}
    \bm{\theta}^* & = \sum_{m \geq 0}\bm{\theta}^{*,\,m}\label{eq:thetas_star_t_power_m_decomposition}\\
    & = \sum_{m \geq 0}\sum_{d^{(2)},\,\ldots,\,d^{(m + 1)} \geq 1}\bm{\theta}^{*,\,\left(d^{(2)},\,\ldots,\,d{(m + 1)}\right)}\label{eq:thetas_star_t_power_m_contribution_tuples_decomposition}
\end{align}
where
\begin{align}
    \bm{\theta}^{*,\,m} & := \left[\bm{T}^m\overline{\bm{\Theta}^*}\right]_1,\label{eq:saddle_point_t_power_m_contribution_repeated}\\
    \bm{\theta}^{*,\,\left(d^{(2)},\,\ldots,\,d^{(m + 1)}\right)} & := \left(\overrightarrow{\prod_{r = 1}^m}\bm{T}_{d^{\left(r\right)},\,d^{\left(r + 1\right)}}\right)\left(\lambda \overline{\bm{C}}^{\left(1\right)}\right)^{\otimes d^{\left(m + 1\right)}}.\label{eq:saddle_point_t_power_m_and_tuple_contribution_repeated}
\end{align}
In equation \ref{eq:saddle_point_t_power_m_and_tuple_contribution_repeated}, we let $d^{(1)} := 1$ (reflecting that we extract the first block of the vector obtained by applying $\bm{T}^m$ to $\overline{\bm{\Theta}^*}$). Let us write matrix products as index summations to infer the continuum limit of this object. For fixed $m \geq 0$ and $d^{\left(2\right)}, \ldots, d^{\left(m + 1\right)}$, and letting $\alpha^{(1)} = \bm{\alpha}^{\left(1\right)} \in \mathcal{A}$
\begin{align}
    \theta^{*,\,\left(d^{(2)},\,\ldots,\,d^{(m + 1)}\right)}_{\alpha^{(1)}} & = \left[\left(\overrightarrow{\prod_{r = 1}^m}\bm{T}_{d^{\left(r\right)},\,d^{\left(r + 1\right)}}\right)\left(\lambda \overline{\bm{C}}^{\left(1\right)}\right)^{\otimes d^{\left(m + 1\right)}}\right]_{\alpha^{\left(1\right)}}\nonumber\\
    & = \sum_{\substack{\forall 2 \leq r \leq m + 1,\,\bm{\alpha}^{\left(r\right)} \in \mathcal{A}^{d^{\left(r\right)}}}}\left(\prod_{1 \leq r \leq m}\left[\bm{T}_{d^{\left(r\right)},\,d^{\left(r + 1\right)}}\right]_{\bm{\alpha}^{\left(r\right)},\,\bm{\alpha}^{\left(r + 1\right)}}\right)\left[\left(\lambda\overline{\bm{C}}^{\left(1\right)}\right)^{\otimes d^{\left(m + 1\right)}}\right]_{\bm{\alpha}^{\left(m + 1\right)}}\nonumber\\
    & = \sum_{\substack{\forall 2 \leq r \leq m + 1,\,\bm{j}^{\left(r\right)} \in \mathcal{I}^{2d^{\left(r\right)}}}}\left(\prod_{1 \leq r \leq m}\left[\bm{T}_{d^{\left(r\right)},\,d^{\left(r + 1\right)}}\right]_{\bm{j}^{\left(r\right)},\,\bm{j}^{\left(r + 1\right)}}\right)\left[\left(\lambda\overline{\bm{C}}^{\left(1\right)}\right)^{\otimes d^{\left(m + 1\right)}}\right]_{\bm{j}^{\left(m + 1\right)}}.\label{eq:theta_star_matrix_vector_product_contribution_step_1}
\end{align}
Plugging in the relation between $\bm{T}_{q,\,d}$ and $\bm{U}_{q,\,d}$ from definition \ref{def:u_block}, the previous equation can be expressed in terms of $\bm{U}_{q,\,d}$ rather than $\bm{T}_{q,\,d}$. We also expand $\alpha^{(1)} = \left(j^{(1)}_1, j^{(1)}_2\right)$ into a pair of $\mathcal{I}$ indices.
\begin{align}
    & \theta^{*,\,\left(d^{(2)},\,\ldots,\,d^{(m + 1)}\right)}_{\alpha^{(1)}} \nonumber\\
    & = \left[\left(\overrightarrow{\prod_{r = 1}^m}\bm{T}_{d^{\left(r\right)},\,d^{\left(r + 1\right)}}\right)\left(\lambda \overline{\bm{C}}^{\left(1\right)}\right)^{\otimes d^{\left(m + 1\right)}}\right]_{\bm{\alpha}^{\left(1\right)}}\nonumber\\
    & = \sum_{\substack{\forall 2 \leq r \leq m + 1,\,\bm{j}^{\left(r\right)} \in \mathcal{I}^{2d^{\left(r\right)}}}}\left(\prod_{1 \leq r \leq m}\mathcal{G}^{\left(2d^{\left(r\right)}\right)}_{\bm{j}^{\left(r\right)}}\right)\left(\prod_{1 \leq r \leq m}\mathcal{G}^{\left(2d^{\left(r + 1\right)}\right)}_{\bm{j}^{\left(r + 1\right)}}\right)\left(\prod_{1 \leq r \leq m}\left[\bm{U}_{d^{\left(r\right)},\,d^{\left(r + 1\right)}}\right]_{\bm{j}^{\left(r\right)},\,\bm{j}^{\left(r + 1\right)}}\right)\left[\left(\lambda\overline{\bm{C}}^{\left(1\right)}\right)^{\otimes d^{\left(m + 1\right)}}\right]_{\bm{j}^{\left(m + 1\right)}}\nonumber\\
    & = \sum_{\substack{\forall 2 \leq r \leq m + 1,\,\bm{j}^{\left(r\right)} \in \mathcal{I}^{2d^{\left(r\right)}}}}\left(\prod_{1 \leq r \leq m + 1}\mathcal{G}^{\left(2d^{\left(r\right)}\right)}_{\bm{j}^{\left(r\right)}}\right)\left(\prod_{1 \leq r \leq m}\mathcal{G}^{\left(2d^{\left(r + 1\right)}\right)}_{\bm{j}^{\left(r + 1\right)}}\right)\left(\prod_{1 \leq r \leq m}\left[\bm{U}_{d^{\left(r\right)},\,d^{\left(r + 1\right)}}\right]_{\bm{j}^{\left(r\right)},\,\bm{j}^{\left(r + 1\right)}}\right)\left[\left(\lambda\overline{\bm{G}}^{\left(2\right)}\right)^{\otimes d^{\left(m + 1\right)}}\right]_{\bm{j}^{\left(m + 1\right)}}\nonumber\\
    & = \mathcal{G}^{\left(2\right)}_{\bm{j}^{\left(1\right)}}\sum_{\substack{\forall 2 \leq r \leq m + 1,\,\bm{j}^{\left(r\right)} \in \mathcal{I}^{2d^{\left(r\right)}}}}\left(\prod_{2 \leq r \leq m + 1}\left(\mathcal{G}^{\left(2d^{\left(r\right)}\right)}_{\bm{j}^{\left(r\right)}}\right)^2\right)\left(\prod_{1 \leq r \leq m}\left[\bm{U}_{d^{\left(r\right)},\,d^{\left(r + 1\right)}}\right]_{\bm{j}^{\left(r\right)},\,\bm{j}^{\left(r + 1\right)}}\right) \left[\left(\lambda\overline{\bm{G}}^{\left(2\right)}\right)^{\otimes d^{\left(m + 1\right)}}\right]_{\bm{j}^{\left(m + 1\right)}}\nonumber\\
    & = \mathcal{G}^{\left(2\right)}_{\bm{j}^{(1)}}\sum_{\substack{\forall 2 \leq r \leq m + 1,\,\bm{j}^{\left(r\right)} \in \mathcal{I}^{2d^{\left(r\right)}}}}\left(\prod_{2 \leq r \leq m + 1}\left[\bm{\mathcal{G}}^{\otimes 2d^{(r)}}\right]_{\bm{j}^{(r)}}\right)\left(\prod_{1 \leq r \leq m}\left[\bm{U}_{d^{\left(r\right)},\,d^{\left(r + 1\right)}}\right]_{\bm{j}^{\left(r\right)},\,\bm{j}^{\left(r + 1\right)}}\right) \left[\left(\lambda\overline{\bm{G}}^{\left(2\right)}\right)^{\otimes d^{\left(m + 1\right)}}\right]_{\bm{j}^{\left(m + 1\right)}}\nonumber\\
    & = \mathcal{G}^{\left(2\right)}_{\bm{j}^{(1)}}\sum_{\substack{\forall 2 \leq r \leq m + 1,\,\bm{j}^{\left(r\right)} \in \mathcal{I}^{2d^{\left(r\right)}}}}\left(\prod_{\substack{2 \leq r \leq m + 1\\1 \leq s \leq 2d^{(r)}}}\mathcal{G}_{j^{(r)}_s}\right)\left(\prod_{1 \leq r \leq m}\left[\bm{U}_{d^{\left(r\right)},\,d^{\left(r + 1\right)}}\right]_{\bm{j}^{\left(r\right)},\,\bm{j}^{\left(r + 1\right)}}\right)\left[\left(\lambda\overline{\bm{G}}^{\left(2\right)}\right)^{\otimes d^{\left(m + 1\right)}}\right]_{\bm{j}^{\left(m + 1\right)}}\nonumber\\
    & = \frac{\left(2^{-1/2}\gamma_{\mathrm{max}}\right)^{d^{(m + 1)}}}{\left(p + 1\right)^{d^{(1)} + 2d^{(2)} + \ldots + 2d^{(m + 1)}}}\mathcal{G}^{\left(2\right),\,\mathrm{cont}}\left(\frac{\bm{j}^{\left(1\right)}}{p + 1/2}\right)\sum_{\forall 2 \leq r \leq m + 1,\,\bm{j}^{\left(r\right)} \in \mathcal{I}^{2d^{(r)}}}\left(\prod_{\substack{2 \leq r \leq m + 1\\1 \leq s \leq 2d^{(r)}}}\mathcal{G}^{\mathrm{cont}}\left(\frac{j^{(r)}_s}{p + 1/2}\right)\right)\nonumber\\
    & \hspace*{60px} \times \left(\prod_{1 \leq r \leq m}U^{\mathrm{cont}}_{d^{\left(r\right)},\,d^{\left(r + 1\right)}}\left(\frac{\bm{j}^{(r)}}{p + 1/2}, \frac{\bm{j}^{(r + 1)}}{p + 1/2}\right)\right)\left(\overline{G}^{\left(2\right),\,\mathrm{cont}}\right)^{\otimes d^{\left(m + 1\right)}}\left(\frac{\bm{j}^{\left(m + 1\right)}}{p + 1/2}\right)\label{eq:theta_star_matrix_vector_product_contribution_step_2}
\end{align}
From then on, it will be more convenient ---for instance, to state continuity properties--- to work with the $\bm{G}^{(2)}$ correlations tensor rather than with correlation tensors, including $\bm{\theta}^*$. Recalling the relation between the two (definition \ref{def:g_correlations_tensor_rescaled}):
\begin{align}
    \theta^*_{j_1,\,j_2} & = \lambda\mathcal{G}^{(2)}_{j_1,\,j_2}G^{(2)}_{j_1,\,j_2},
\end{align}
this is because in the continuum limit $\bm{G}^{(2)}$ discretizes a continuous function, whereas $\bm{\mathcal{G}}^{(2)}$ discretizes a piecewise continuous function $\mathcal{G}^{(2),\,\mathrm{cont}}$ with a 4 jumps (due to the jump of $\Gamma^{\mathrm{cont}}$ around $1$).
In terms of $\bm{G}^{(2)}$, equation \ref{eq:theta_star_matrix_vector_product_contribution_step_2} can be rewritten as:
\begin{align}
    G^{(2),\,\left(d^{(2)},\,\ldots,\,d^{(m + 1)}\right)}_{j_1,\,j_2} & = \frac{\left(2^{-1/2}\gamma_{\mathrm{max}}\right)^{d^{(m + 1)} - 1}}{\left(p + 1\right)^{2d^{(2)} + \ldots + 2d^{(m + 1)}}}\sum_{\forall 2 \leq r \leq m + 1,\,\bm{j}^{\left(r\right)} \in \mathcal{I}^{2d^{(r)}}}\left(\prod_{\substack{2 \leq r \leq m + 1\\1 \leq s \leq 2d^{(r)}}}\mathcal{G}^{\mathrm{cont}}\left(\frac{j^{(r)}_s}{p + 1/2}\right)\right)\nonumber\\
    & \hspace*{20px} \times \left(\prod_{1 \leq r \leq m}U^{\mathrm{cont}}_{d^{\left(r\right)},\,d^{\left(r + 1\right)}}\left(\frac{\bm{j}^{(r)}}{p + 1/2}, \frac{\bm{j}^{(r + 1)}}{p + 1/2}\right)\right)\left(\overline{G}^{\left(2\right),\,\mathrm{cont}}\right)^{\otimes d^{\left(m + 1\right)}}\left(\frac{\bm{j}^{\left(m + 1\right)}}{p + 1/2}\right).\label{eq:g_correlations_matrix_vector_product_contribution}
\end{align}

$G^{(2)}$ can then be expressed as follows given this newly defined quantity:
\begin{align}
    \bm{G}^{(2)} & = \sum_{m \geq 0}\bm{G}^{(2),\,m},\label{eq:g_correlations_as_sum_t_power_m_contributions}\\
    \bm{G}^{\left(2\right),\,m} & := \sum_{d^{(2)},\,\ldots,\,d^{(m + 1)} \geq 1}\bm{G}^{(2),\,\left(d^{(2)},\,\ldots,\,d^{(m + 1)}\right)}\label{eq:g_correlations_t_power_m_contribution}
\end{align}
These are simply reformulations of equation \ref{eq:thetas_star_t_power_m_decomposition}, \ref{eq:thetas_star_t_power_m_contribution_tuples_decomposition} --dividing these by $\lambda\bm{\mathcal{G}}^{(2)}$. Finally, it will be convenient to use the following uniform bound on $\bm{G}^{(2)}$, which is a simple revision of bound \ref{eq:thetas_star_bound} on $\bm{\theta}^*$, established in proposition \ref{propSaddlePointEquationAnalyticity}:

\begin{align}
\label{eq:g_correlations_bounds}
    \left\lVert \bm{G} \right\rVert_{\infty} & \leq 2.
\end{align}

We now interpret equation \ref{eq:g_correlations_matrix_vector_product_contribution} for a single contribution of $\bm{G}^{(2)}$ as a discretized integral. Indeed, the summation variables are $\bm{j}^{(2)}, \ldots, \bm{j}^{(m + 1)}$, of respective dimensions $2d^{(2)}, \ldots, 2d^{(m + 1)}$, and iterating over $\mathcal{I}$. On the other hand, $\alpha^{(1)} = \bm{j}^{(1)} = \left(j^{(1)}_1, j^{(1)}_2\right)$ is fixed and of dimension $2d^{(1)} = 2$. The summed-over function:
\begin{align}
   & \left(\prod_{\substack{2 \leq r \leq m + 1\\1 \leq s \leq 2d^{(r)}}}\mathcal{G}^{\mathrm{cont}}\left(\frac{j^{(r)}_s}{p + 1/2}\right)\right)\left(\prod_{1 \leq r \leq m}U^{\mathrm{cont}}_{d^{\left(r\right)},\,d^{\left(r + 1\right)}}\left(\frac{\bm{j}^{(r)}}{p + 1/2}, \frac{\bm{j}^{(r + 1)}}{p + 1/2}\right)\right)\left(\overline{G}^{\left(2\right),\,\mathrm{cont}}\right)^{\otimes d^{\left(m + 1\right)}}\left(\frac{\bm{j}^{\left(m + 1\right)}}{p + 1/2}\right)
\end{align}
depends on summed-over variables $\bm{j}^{(2)}, \ldots, \bm{j}^{(m + 1)}$ only through ratios
\begin{align}
    \frac{\bm{j}^{(2)}}{p + 1/2}, \ldots, \frac{\bm{j}^{(m + 1)}}{p + 1/2},
\end{align}
whose coordinates lie in $[0, 1]$. Besides, the denominator in prefactor
\begin{align}
    \frac{1}{\left(p + 1\right)^{\sum\limits_{2 \leq r \leq m + 1}2d^{(r)}}}
\end{align}
is exactly the number of values taken by $\bm{j}^{(2)}, \ldots, \bm{j}^{(m + 1)}$. This suggests to approximate this discrete sum as an integral. We then introduce the following continuum analogue of $\bm{G}^{(2),\,\left(d^{(2)},\,\ldots,\,d^{(m + 1)}\right)}$:

\begin{definition}[Continuum analogue of $\bm{G}^{(2)}$ contribution]
\label{def:g_correlations_continuum}
We define the continuum analogue of the tuple $\left(d^{(2)},\,\ldots,\,d^{(m + 1)}\right)$ contribution to $\bm{G}^{(2)}$ (equation \ref{eq:g_correlations_matrix_vector_product_contribution} for the discrete case) as the following function of two variables $\bm{x}^{\left(1\right)} = \left(x^{(1)}_1,\,x^{\left(1\right)}_2\right) \in [0, 2]^2$: 
\begin{align}
    & G^{(2),\,\left(d^{(2)},\,\ldots,\,d^{(m + 1)}\right),\,\mathrm{cont}}\left(\bm{x}^{(1)}\right)\nonumber\\
    & := G^{(2),\,\left(d^{(2)},\,\ldots,\,d^{(m + 1)}\right),\,\mathrm{cont}}\left(x^{(1)}_1, x^{(1)}_2\right)\nonumber\\
    & = \left(2^{-1/2}\gamma_{\mathrm{max}}\right)^{d^{\left(m + 1\right)} - 1}\int\limits_{[0, 2]^{2d^{(2)}} \times \ldots \times [0, 2]^{2d^{(m + 1)}}}\!\prod_{2 \leq r \leq m + 1}\mathrm{d}\bm{x}^{(r)}\,\left(\prod_{1 \leq r \leq m}U^{\mathrm{cont}}_{d^{\left(r\right)},\,d^{\left(r + 1\right)}}\left(\bm{x}^{(r)}, \bm{x}^{(r + 1)}\right)\right)\nonumber\\
    & \hspace*{200px} \times \left(\overline{G}^{\left(2\right),\,\mathrm{cont}}\right)^{\otimes d^{\left(m + 1\right)}}\left(\bm{x}^{\left(m + 1\right)}\right)\prod_{\substack{2 \leq r \leq m + 1}}\left(\mathcal{G}^{\mathrm{cont}}\right)^{\otimes 2d^{(r)}}\left(\bm{x}^{(r)}\right)\label{eq:g_correlations_t_power_m_and_tuple_contribution_continuum}
\end{align}
Note this function is well-defined as the integral of a continuous bounded function over a bounded domain in a finite number of variables. Besides, by continuity of $U^{\mathrm{cont}}_{1,\,d^{\left(2\right)}}\left(\bm{x}^{(1)}, \bm{x}^{(2)}\right)$ in variable $\bm{x}^{(1)}$, the function is continuous in $\bm{x}^{(1)}$.

Likewise, we define the continuum analogue of contribution $\bm{G}^{(2),\,m}$ to the $\bm{G}^{(2)}$ correlations tensor (equation \ref{eq:g_correlations_t_power_m_contribution} in the discrete case):
\begin{align}
    G^{(2),\,m,\,\mathrm{cont}}\left(\bm{x}^{(1)}\right) & := \sum_{d^{(2)},\,\ldots,\,d^{(m + 1)} \geq 1}G^{(2),\,\left(d^{(2)},\,\ldots,\,d^{(m + 1)}\right),\,\mathrm{cont}}\left(\bm{x}^{(1)}\right),\qquad \forall m \geq 1.\label{eq:g_correlations_t_power_m_contribution_continuum}
\end{align}
We also define edge case $G^{(2),\,0,\,\mathrm{cont}}$ by the following formula, consistent with discrete formula $\bm{\theta}^{*,\,0} = \left[\overline{\bm{\Theta}^*}\right]_1 = \lambda \overline{\bm{C}}^{\left(1\right)}/\overline{C}^{(0)} = \lambda \overline{\bm{C}}^{(1)}$, equivalent to $\bm{G}^{(2),\,0} = \lambda \overline{\bm{G}}^{(2)}$: 
\begin{align}
    G^{(2),\,0,\,\mathrm{cont}}\left(\bm{x}^{(1)}\right) & := \overline{G}^{(2),\,\mathrm{cont}}\left(\bm{x}^{(1)}\right)\label{eq:saddle_point_t_power_0_contribution_continuum}.
\end{align}
Finally, we define the continuum analogue of $\bm{G}^{(2)}$ correlation tensor as the following sum over $m$ (see equation \ref{eq:g_correlations_as_sum_t_power_m_contributions} in the discrete case):
\begin{align}
    G^{(2),\,\mathrm{cont}}\left(\bm{x}^{(1)}\right) & := \sum_{m \geq 0}G^{(2),\,m,\,\mathrm{cont}}\left(\bm{x}^{(1)}\right).\label{eq:g_correlations_continuum}
\end{align}
Since equations \ref{eq:g_correlations_t_power_m_contribution_continuum} and \ref{eq:g_correlations_continuum} involve infinite sums, it is less obvious they are well-defined. However, proposition \ref{prop:g_correlations_continuum_well_definiteness} shows they indeed define continuous functions.
\end{definition}

The following proposition shows the continuum analogue $G^{(2),\,\mathrm{cont}}$ of the $\bm{G}^{(2)}$ correlations introduced in definition \ref{def:g_correlations_continuum} indeed define continuous functions. Since the proof uses uniform convergence, it also gives a uniform bound on the function as a byproduct:

\begin{proposition}[Well-definiteness and continuity of continuum $\bm{G}^{(2)}$ correlations]
\label{prop:g_correlations_continuum_well_definiteness}
The following uniform bound holds on the continuum analogue of the $\left(d^{(2)},\,\ldots,\,d^{(m + 1)}\right)$ tuple contribution to $\bm{G}^{(2)}$ correlations (equation \ref{eq:g_correlations_t_power_m_and_tuple_contribution_continuum}):
\begin{align}
    \left|G^{(2),\,\left(d^{(2)},\,\ldots,\,d^{(m + 1)}\right),\,\mathrm{cont}}\left(\bm{x}^{(1)}\right)\right| & \leq c^{1 -d^{(m + 1)}}\left(2^{1/2}\gamma_{\mathrm{max}}c\right)^{2d^{(2)} + \ldots + 2d^{(m + 1)}} \qquad \forall \bm{x}^{\left(1\right)} \in [0, 2]^2.\label{eq:g_correlations_continuum_t_power_m_and_tuple_contribution_bound}
\end{align}
As a result, the infinite sum specifying the continuum analogue of the $\bm{G}^{(2),\,m}$ contribution to $\bm{G}^{(2)}$ (equation \ref{eq:g_correlations_t_power_m_contribution_continuum}) is well-defined and defines a continuous function, uniformly bounded as:
\begin{align}
    \left|G^{(2),\,m,\,\mathrm{cont}}\left(\bm{x}^{(1)}\right)\right| & \leq \left(2\gamma_{\mathrm{max}}c\right)^{2m} \qquad \forall \bm{x}^{\left(1\right)} \in [0, 2]^2.\label{eq:g_correlations_continuum_t_power_m_contribution_bound}
\end{align}
as long as
\begin{align}
    \gamma_{\mathrm{max}} \leq \max\left(\frac{1}{2^{3/2}c}, \frac{1}{2\sqrt{c}}\right).
\end{align}
As a result, the series defining the continuum analogue $G^{(2),\,\mathrm{cont}}$ of the $\bm{G}^{(2)}$ correlations (equation \ref{eq:g_correlations_continuum}) is absolutely convergent, defining a continuous function bounded by:
\begin{align}
    \left|G^{(2),\,\mathrm{cont}}\left(\bm{x}^{(1)}\right)\right| & \leq 2 \qquad \forall \bm{x}^{(1)} \in [0, 2]^2.\label{eq:g_correlations_continuum_bound}  
\end{align}
Additionally, $G^{(2),\,\mathrm{cont}}$ is $4\beta_{\mathrm{max}}$-Lipschitz in each of its variables. Finally, the discrete versions of $\bm{G}^{\left(2\right),\,\left(d^{(2)},\,\ldots,\,d^{(m + 1)}\right)}$, $\bm{G}^{(2),\,m}$, $\bm{G}^{(2)}$ (equations \ref{eq:g_correlations_matrix_vector_product_contribution}, \ref{eq:g_correlations_t_power_m_contribution}, \ref{eq:g_correlations_as_sum_t_power_m_contributions}) satisfy the same bounds:
\begin{align}
    \left\lVert \bm{G}^{(2),\,\left(d^{(2)},\,\ldots,\,d^{(m + 1)}\right)} \right\rVert_{\infty} & \leq c^{1 - d^{(m + 1)}}\left(2^{1/2}\gamma_{\mathrm{max}}c\right)^{2d^{(2)} + \ldots + 2d^{(m + 1)}},\label{eq:g_correlations_t_power_m_and_tuple_contribution_bound}\\
    \left\lVert \bm{G}^{(2),\,m} \right\rVert_{\infty} & \leq \left(2\gamma_{\mathrm{max}}c\right)^{2m},\label{eq:g_correlations_t_power_m_contribution_bound}\\
    \left\lVert \bm{G}^{(2)} \right\rVert_{\infty} & \leq 2,\label{eq:g_correlations_bound}
\end{align}
where the infinite norms now refer to standard vector norms rather than the uniform norm of continuous functions.
\begin{proof}
We start by proving inequality \ref{eq:g_correlations_continuum_t_power_m_and_tuple_contribution_bound}. For that purpose, we bound the integrand in the integral defining $G^{(2),\,m,\,\left(d^{(m)},\,\ldots,\,d^{(m + 1)}\right),\,\mathrm{cont}}$ (definition \ref{def:g_correlations_continuum}):
\begin{align}
    \left|\left(\prod_{1 \leq r \leq m}U^{\mathrm{cont}}_{d^{(r)},\,d^{(r + 1)}}\left(\bm{x}^{(r)}, \bm{x}^{(r + 1)}\right)\right)\left(G^{(2),\,\mathrm{cont}}\right)^{\otimes d^{(m + 1)}}\left(\bm{x}^{(m + 1)}\right)\prod_{\substack{2 \leq r \leq m + 1\\1 \leq s \leq d^{(r)}}}\mathcal{G}^{\mathrm{cont}}\left(\bm{x}^{\left(r\right)}\right)\right|  & \leq \prod_{1 \leq r \leq m}\left(2^{-1/2}\gamma_{\mathrm{max}}c\right)^{d^{(r)} + d^{(r + 1)}}.
\end{align}
Here, we bounded $\overline{G}^{\left(2\right),\,\mathrm{cont}}$ and $\mathcal{G}^{\mathrm{cont}}$ uniformly by $1$, which follows from their explicit expressions; besides, we used the uniform bound on $U^{\mathrm{cont}}_{q,\,d}$ from proposition \ref{prop:u_block_entrywise_bound}. Multiplying this by the volume of the  integration domain: $2^{2d^{(2)} + \ldots + 2d^{(m + 1)}}$ and incorporating integral's prefactor by $\left(2^{-1/2}\gamma_{\mathrm{max}}\right)^{d^{(m + 1)} - 1}$ gives bound
\begin{align}
    \left|G^{(2),\,m,\,\left(d^{(2)},\,\ldots,\,d^{(m + 1)}\right)}\left(\bm{x}^{\left(1\right)}\right)\right| & \leq 2^{2d^{(2)} + \ldots + 2d^{(m + 1)}}\left(\prod_{1 \leq r \leq m}\left(2^{-1/2}\gamma_{\mathrm{max}}c\right)^{d^{(r)} + d^{(r + 1)}}\right)\left(2^{-1/2}\gamma_{\mathrm{max}}\right)^{d^{(m + 1)} - 1}\nonumber\\
    & = 2^{2d^{(2)} + \ldots + 2d^{(m + 1)}}c\left(2^{-1/2}\gamma_{\mathrm{max}}c\right)^{2d^{(2)} + \ldots + 2d^{(m)}}\left(2^{-1/2}\gamma_{\mathrm{max}}c^{1/2}\right)^{2d^{(m + 1)}}\nonumber\\
    & = c^{1 - d^{(m + 1)}}\left(2^{1/2}\gamma_{\mathrm{max}}c\right)^{2d^{(2)} + \ldots + 2d^{(m + 1)}}.
\end{align}
We now turn to inequality \ref{eq:g_correlations_continuum_t_power_m_contribution_bound}. We first focus case $m \geq 1$, where $G^{(2),\,m,\,\mathrm{cont}}$ is defined by equation \ref{eq:g_correlations_t_power_m_contribution_continuum}). The inequality then results from summing the previous inequality over $d^{\left(2\right)}, \ldots, d^{\left(m + 1\right)}$, namely
\begin{align}
    \sum_{d^{(2)},\,\ldots,\,d^{(m + 1)} \geq 1}\left\lVert G^{(2),\,\left(d^{(2)},\,\ldots,\,d^{(m + 1)}\right),\,\mathrm{cont}}\right\rVert_{\infty} & \leq c\sum_{d^{(2)},\,\ldots,\,d^{(m + 1)} \geq 1}\left(2^{1/2}\gamma_{\mathrm{max}}c\right)^{2d^{(2)} + \ldots + 2d^{(m + 1)}}c^{-d^{(m + 1)}}\nonumber\\
    & = c\frac{\left(2^{1/2}\gamma_{\mathrm{max}}c\right)^{2m - 2}}{\left(1 - 2\gamma_{\mathrm{max}}^2c^2\right)^{m - 1}}\frac{\left(2^{1/2}\gamma_{\mathrm{max}}c^{1/2}\right)^2}{1 - 2\gamma_{\mathrm{max}}^2c}\nonumber\\
    & \leq \left(2^{1/2}\gamma_{\mathrm{max}}c\right)^{2m}2^m\nonumber\\
    & \leq \left(2\gamma_{\mathrm{max}}c\right)^{2m},
\end{align}
where in the third and fourth lines, we assumed:
\begin{align}
    \gamma_{\mathrm{max}} & \leq \max\left(\frac{1}{2c}, \frac{1}{2\sqrt{c}}\right).
\end{align}
Finally, note that each function $G^{(2),\,\left(d^{(2)},\,\ldots,\,d^{(m + 1)}\right),\,\mathrm{cont}}$ is continuous by continuity of parametrized integrals, given the dependence in variable $\bm{x}^{(1)}$ is restricted to continuous factor $U_{1,\,d^{(2)}}^{\mathrm{cont}}\left(\bm{x}^{(1)}, \bm{x}^{(2)}\right)$ of the integrand\footnote{In particular, no dependence in $\bm{x}^{(1)}$ occurs in the $\mathcal{G}^{\mathrm{cont}}$, which have a jump around $1$.}. The previous bound shows that the series of continuous functions
\begin{align}
    \sum_{d^{(2)},\,\ldots,\,d^{(m + 1)} \geq 1}G^{(2),\,\left(d^{(2)},\,\ldots,\,d^{(m + 1)}\right),\,\mathrm{cont}}
\end{align}
is uniformly absolutely convergent, hence defines a continuous function
\begin{align}
    G^{(2),\,m,\,\mathrm{cont}} & := \sum_{d^{(2)},\,\ldots,\,d^{(m + 1)} \geq 1}G^{(2),\,\left(d^{(2)},\,\ldots,\,d^{(m + 1)}\right),\,\mathrm{cont}},
\end{align}
uniformly bounded by the sum of uniform norms:
\begin{align}
    \left\lVert G^{(2),\,m,\,\mathrm{cont}}\right\rVert_{\infty} & \leq \left(2\gamma_{\mathrm{max}}c\right)^{2m}.
\end{align}
This bound also trivially holds for $G^{(2),\,0,\,\mathrm{cont}} = \overline{G}^{(2),\,\mathrm{cont}}$ from the explicit expression in proposition \ref{def:continuum_gamma_0_correlations}. This concludes the proof of inequality \ref{eq:g_correlations_continuum_t_power_m_contribution_bound}. We now consider inequality \ref{eq:g_correlations_continuum_bound}. From inequality \ref{eq:g_correlations_continuum_t_power_m_contribution_bound} just proven,
\begin{align}
    \sum_{m \geq 0}\left\lVert G^{(2),\,m,\,\mathrm{cont}}\right\rVert_{\infty} & \leq \sum_{m \geq 0}\left(2\gamma_{\mathrm{max}}c\right)^{2m}\nonumber\\
    & = \frac{1}{1 - \left(2\gamma_{\mathrm{max}}c\right)^2}\nonumber\\
    & \leq 2.
\end{align}
This proves uniform absolute convergence of the series (equation \ref{eq:g_correlations_continuum}) defining $G^{(2),\,\mathrm{cont}}$, so that this function is continuous and bounded as stated.

To estimate a Lipschitz constant for $G^{(2),\,\mathrm{cont}}$, we essentially repeat the proof leading to the uniform bound. However, rather than using the uniform bound on $U^{\mathrm{cont}}_{1,\,d^{(2)}}\left(\bm{x}^{(1)},\,\bm{x}^{(2)}\right)$ as a starting point:
\begin{align}
    \left|U^{\mathrm{cont}}\left(\bm{x}^{(1)},\,\bm{x}^{(2)}\right)\right| & \leq \left(2^{-1/2}\gamma_{\mathrm{max}}\right)^{1 + d^{(2)}},
\end{align}
we resort to the very similar bound on its Lipschitz constant (with respect to each variable):
\begin{align}
    2\beta_{\mathrm{max}}\left(2^{-1/2}\gamma_{\mathrm{max}}c\right)^{1 + d^{(2)}},
\end{align}
established in proposition \ref{prop:gamma_0_g_correlations_continuum_lipschitz_bound}.

The bounds on the discrete version of $\bm{G}^{(2),\,\left(d^{(2)},\,\ldots,\,d^{(m + 1)}\right)}, \bm{G}^{(2),\,m}, \bm{G}^{(2)}$ are adaptations of the proof of proposition \ref{propSaddlePointEquationAnalyticity}, which reasoned over $\bm{\theta}^*$ rather than $\bm{G}^{(2)}$. Alternatively, they can be obtained by replaying the current proof, with the role of integration volume $2^{2d^{(2)} + \ldots + 2d^{(m + 1)}}$ in bound \ref{eq:g_correlations_continuum_t_power_m_and_tuple_contribution_bound} now being played by the number of discrete sum terms $|\mathcal{A}|^{d^{(2)} + \ldots + d^{(m + 1)}} = \left(2p + 2\right)^{2d^{(2)} + \ldots + 2d^{(m + 1)}} $.
\end{proof}
\end{proposition}

Proposition \ref{prop:g_correlations_continuum_well_definiteness} proves the ``continuum $\bm{G}^{(2)}$ correlations" $G^{(2),\,\mathrm{cont}}$ are well-defined and continuous as a sum of series
\begin{align}
    G^{(2),\,\mathrm{cont}} & := \sum_{m \geq 0}G^{(2),\,m},\\
    G^{(2),\,m} & := \sum_{d^{(2)},\,\ldots,\,d^{(m + 1)} \geq 1}G^{(2),\,\left(d^{(2)},\,\ldots,\,d^{(m + 1)}\right),\,\mathrm{cont}},
\end{align}
provides a uniform bound on $G^{(2),\,\mathrm{cont}}$ and an estimate of its Lipschitz constant.

Having introduced the continuum analogues of $\bm{G}^{(2),\,\left(d^{(2)},\,\ldots,\,d^{(m + 1)}\right)}$, $\bm{G}^{(2),\,m}$ in definition \ref{def:g_correlations_continuum}, and established their well-definiteness as continuous functions in proposition \ref{prop:g_correlations_continuum_well_definiteness}, we are now ready to relate the discrete versions of these objects to continuum ones. We start by approximating $\bm{G}^{(2),\,\left(d^{(2)},\,\ldots,\,d^{(m + 1)}\right)}$ in terms of $G^{(2),\,\left(d^{(2)},\,\ldots,\,d^{\left(m + 1\right)}\right),\,\mathrm{cont}}$. For that purpose, we recall the expression of $G_{\alpha^{(1)}}^{(2),\,\left(d^{(2)},\,\ldots,\,d^{(m + 1)}\right)}$ as a discrete sum involving (continuum) functions $U^{\mathrm{cont}}_{d^{(r)},\,d^{(r + 1)}}$, $\mathcal{G}^{\mathrm{cont}}$ and $G^{(2),\,\mathrm{cont}}$ in equation \ref{eq:g_correlations_matrix_vector_product_contribution}. As suggested in the discussion following that equation, we regard the discrete sum as the approximation of an integral ---namely, the integral defining $G^{(2),\,\left(d^{(2)},\,\ldots,\,d^{(m + 1)}\right),\,\mathrm{cont}}$ in equation \ref{eq:g_correlations_t_power_m_and_tuple_contribution_continuum}. This Riemann sum approximation is made quantitative by lemma \ref{lemma:riemann_sum_approximation}. The roles of discrete variables $\bm{n}$ in the lemma are played by discrete variables
\begin{align}
    \bm{j}^{(2)} \in \mathcal{I}^{2d^{(2)}},\,\ldots,\,\bm{j}^{(m + 1)} \in \mathcal{I}^{2d^{(m + 1)}},
\end{align}
of total dimension
\begin{align}
    D & := 2d^{(2)} + \ldots + 2d^{(m + 1)},
\end{align}
and where we recall
\begin{align}
    \mathcal{I} & := \left\{0, 1, \ldots, 2p, 2p + 1\right\},
\end{align}
consistent with the lemma's conventions. These discrete variables are associated to continuum variables
\begin{align}
    \bm{x}^{(2)} \in [0, 2]^{2d^{(2)}},\,\ldots,\,\bm{x}^{(m + 1)} \in [0, 2]^{2d^{(m + 1)}}.
\end{align}
The functions $f_1, \ldots, f_m$ to which we apply the lemma are
\begin{gather}
    U^{\mathrm{cont}}_{d^{(1)},\,d^{(2)}}\left(\bm{x}^{(1)}, \bm{x}^{(2)}\right),\,U^{\mathrm{cont}}_{d^{(2)},\,d^{(3)}}\left(\bm{x}^{(2)},\,\bm{x}^{(3)}\right)\,\ldots,\,U^{\mathrm{cont}}_{d^{(m)},\,d^{(m + 1)}}\left(\bm{x}^{(m)}, \bm{x}^{(m + 1)}\right),\\
    \mathcal{G}^{\mathrm{cont}}\left(x^{(2)}_1\right), \ldots, \mathcal{G}^{\mathrm{cont}}\left(x^{(2)}_{2d^{(2)}}\right),\,\ldots,\,\mathcal{G}^{\mathrm{cont}}\left(x^{(m + 1)}_1\right), \ldots, \mathcal{G}^{\mathrm{cont}}\left(x^{(m + 1)}_{2d^{(m + 1)}}\right),\\
    \overline{G}^{\left(2\right),\,\mathrm{cont}}\left(\bm{x}^{(m + 1)}\right),
\end{gather}
where for convenience, we have categorized the functions $f_l$ into 3 ``kinds" and listed the functions of the same ``kind" on the same line. Note that in this context, $\bm{x}^{(1)}$ is regarded as fixed rather than an integration variable. The above functions then have respective numbers of variables:
\begin{gather}
    d^{(2)},\,d^{(2)} + d^{(3)},\,\ldots,\,d^{(m)} + d^{(m + 1)},\\
    1,\,\ldots,\,1,\,\ldots,\,1,\,\ldots,\,1,\\
    d^{(m + 1)}.
\end{gather}
We can therefore compute the relevant parameter in the lemma's error bound (equation \ref{eq:riemann_sum_approximation_error_bound}):
\begin{align}
    \sum_{1 \leq l \leq m}d_l & = 2d^{(2)} + \left(2d^{(2)} + 2d^{(3)}\right) + \ldots + \left(2d^{(m)} + 2d^{(m + 1)}\right)\nonumber\\
    & \hspace*{10px} + 2d^{(2)} + \ldots + 2d^{(m + 1)}\nonumber\\
    & \hspace*{10px} + 2d^{(m + 1)}\nonumber\\
    & = 6d^{(2)} + \ldots + 6d^{(m + 1)}.
\end{align}
Respective bounds on these functions (constants $K_l$ in the lemma's statement) are
\begin{gather}
    \left(2^{-1/2}\gamma_{\mathrm{max}}c\right)^{d^{(1)} + d^{(2)}},\,\left(2^{-1/2}\gamma_{\mathrm{max}}c\right)^{d^{(2)} + d^{(3)}},\,\ldots,\,\left(2^{-1/2}\gamma_{\mathrm{max}}c\right)^{d^{(m)} + d^{(m + 1)}},\\
    1,\,\ldots,\,1,\\
    1.
\end{gather}
The bounds on the first line follows from proposition \ref{prop:u_block_entrywise_bound}, the ones on the last two lines from the definitions of $\mathcal{G}^{\mathrm{cont}}$ and $\overline{G}^{(2),\,\mathrm{cont}}$. One may further use the following Lipschitz constants (constants $M_l$ in the lemma's statement) for these functions (recalling definition \ref{def:angles_from_continuum} proposition \ref{prop:gamma_0_g_correlations_continuum_lipschitz_bound}):
\begin{gather}
    2\beta_{\mathrm{max}}\left(2^{-1/2}\gamma_{\mathrm{max}}c\right)^{d^{(1)} + d^{(2)}},\,2\beta_{\mathrm{max}}\left(2^{-1/2}\gamma_{\mathrm{max}}c\right)^{d^{(2)} + d^{(3)}},\,\ldots,\,2\beta_{\mathrm{max}}\left(2^{-1/2}\gamma_{\mathrm{max}}c\right)^{d^{(m)} + d^{(m + 1)}},\\
    \frac{M_{\gamma}}{\gamma_{\mathrm{max}}},\,\ldots,\,\frac{M_{\gamma}}{\gamma_{\mathrm{max}}},\\
    2\beta_{\mathrm{max}}.
\end{gather}
From these uniform bounds and Lipschitz constants, one may bound the relevant parameter in the lemma's inequality, namely
\begin{align}
    \max_{l \in [m]}\left(M_l\prod_{l' \in [m] - \{l\}}K_{l'}\right) & \leq \max\left(1, 2\beta_{\mathrm{max}}, \frac{M_{\gamma}}{\gamma_{\max}}\right)\left(2^{-1/2}\gamma_{\mathrm{max}}c\right)^{d^{(1)} + d^{(m + 1)} + 2d^{(2)} + \ldots + 2d^{(m)}}.
\end{align}
From these estimates, it results the following approximation between $\bm{G}^{(2),\,\left(d^{(2)},\,\ldots,\,d^{(m + 1)}\right)}$ and $G^{(2),\,\left(d^{(2)},\,\ldots,\,d^{(m + 1)}\right),\,\mathrm{cont}}$:

\begin{proposition}[Approximation of $\bm{G}^{(2)}$ correlations by continuum counterpart]
\label{prop:g_correlations_continuum_approximation}
Assume the same bound on $\gamma_{\mathrm{max}}$ as in proposition \ref{prop:g_correlations_continuum_well_definiteness}, namely
\begin{align}
    \gamma_{\mathrm{max}} & \leq \max\left(\frac{1}{2^{3/2}c}, \frac{1}{2\sqrt{c}}\right).
\end{align}
Then, the following bound holds between contribution $\bm{G}^{(2),\,\left(d^{(2)},\,\ldots,\,d^{(m + 1)}\right)}$ to the discrete $\bm{G}^{(2)}$ correlations, and its continuum counterpart $G^{(2),\,\left(d^{(2)},\,\ldots,\,d^{(m + 1)}\right),\,\mathrm{cont}}$:
\begin{align}
    & \left|G^{(2),\,\left(d^{(2)},\,\ldots,\,d^{(m + 1)}\right)}_{\bm{j}^{(1)}} - \frac{1}{p + 1}G^{(2),\,\left(d^{(2)},\,\ldots,\,d^{(m + 1)}\right),\,\mathrm{cont}}\left(\frac{\bm{j}^{(1)}}{p + 1/2}\right)\right|\nonumber\\
    & \leq \frac{12}{p + 1}\max\left(1, 2\beta_{\mathrm{max}}, \frac{M_{\gamma}}{\gamma_{\mathrm{max}}}\right)c^{1 - d^{(m + 1)}}\left(2^{1/2}\gamma_{\mathrm{max}}c\right)^{\sum\limits_{2 \leq r \leq m + 1}2d^{(r)}}\sum_{2 \leq r \leq m + 1}d^{(r)},\label{eq:saddle_point_t_power_m_and_tuple_contribution_continuum_approximation}
\end{align}
for all $\bm{j}^{(1)} = \left(j^{(1)}_1, j^{(1)}_2\right) \in \mathcal{I}^{2}$. From this inequality, it follows 
\begin{align}
    \left|G^{(2),\,m}_{\bm{j}^{(1)}} - \frac{1}{p + 1}G^{(2),\,m,\,\mathrm{cont}}\left(\frac{\bm{j}^{(1)}}{p + 1/2}\right)\right| & \leq \frac{24}{p + 1}\max\left(1, 2\beta_{\mathrm{max}}, \frac{M_{\gamma}}{\gamma_{\mathrm{max}}}\right)\left(2\gamma_{\mathrm{max}}c\right)^{2m}m,\label{eq:g_correlations_t_power_m_contribution_continuum_approximation}
\end{align}
and
\begin{align}
    \left|G^{(2)}_{\bm{j}^{(1)}} - \frac{1}{p + 1}G^{(2),\,\mathrm{cont}}\left(\frac{\bm{j}^{(1)}}{p + 1/2}\right)\right| & \leq \frac{384}{p + 1}\max\left(1, 2\beta_{\mathrm{max}}, \frac{M_{\gamma}}{\gamma_{\mathrm{max}}}\right)\gamma_{\mathrm{max}}^2c^2.\label{eq:g_correlations_continuum_approximation}
\end{align}
\begin{proof}
We start by proving inequality \ref{eq:g_correlations_t_power_m_contribution_continuum_approximation}. It follows from the triangular inequality and summing inequality \ref{eq:g_correlations_continuum_t_power_m_and_tuple_contribution_bound}:
\begin{align}
    & \left|G^{(2),\,m}_{\bm{j}^{(1)}} - \frac{1}{p + 1}G^{(2),\,m,\,\mathrm{cont}}\left(\frac{\bm{j}^{(1)}}{p + 1/2}\right)\right|\nonumber\\
    & = \left|\sum_{d^{(2)},\,\ldots,\,d^{(m + 1)} \geq 1}\left(G^{(2),\,\left(d^{(2)},\,\ldots,\,d^{(m + 1)}\right)}_{\bm{j}^{(1)}} - \frac{1}{p + 1}G^{(2),\,\left(d^{(2)},\,\ldots,\,d^{(m + 1)}\right),\,\mathrm{cont}}\left(\frac{\bm{j}^{(1)}}{p + 1/2}\right)\right)\right|\nonumber\\
    & \leq \sum_{d^{(2)},\,\ldots,\,d^{(m + 1)} \geq 1}\left|G^{(2),\,\left(d^{(2)},\,\ldots,\,d^{(m + 1)}\right)}_{\bm{j}^{(1)}} - \frac{1}{p + 1}G^{(2),\,\left(d^{(2)},\,\ldots,\,d^{(m + 1)}\right),\,\mathrm{cont}}\left(\frac{\bm{j}^{(1)}}{p + 1/2}\right)\right|\nonumber\\
    & \leq \sum_{d^{(2)},\,\ldots,\,d^{(m + 1)} \geq 1}\frac{12}{p + 1}\max\left(1, 2\beta_{\mathrm{max}}, \frac{M_{\gamma}}{\gamma_{\mathrm{max}}}\right)c^{1 - d^{(m + 1)}}\left(2^{1/2}\gamma_{\mathrm{max}}c\right)^{\sum\limits_{2 \leq r \leq m + 1}2d^{(r)}}\sum_{2 \leq r \leq m + 1}d^{(r)}\nonumber\\
    & \leq \frac{24}{p + 1}\max\left(1, 2\beta_{\mathrm{max}}, \frac{M_{\gamma}}{\gamma_{\mathrm{max}}}\right)\left(2\gamma_{\mathrm{max}}c\right)^{2m}m,
\end{align}
where to go from the fourth to the fifth line, we used geometric sums bounds
\begin{align}
    \sum_{d^{(r)} \geq 1}\left(2^{1/2}\gamma_{\mathrm{max}}c\right)^{2d^{(r)}} & = \frac{\left(2^{1/2}\gamma_{\mathrm{max}}c\right)^2}{1 - \left(2^{1/2}\gamma_{\mathrm{max}}c\right)^2}  \leq \left(2\gamma_{\mathrm{max}}c\right)^2,\\
    \sum_{d^{(m + 1)} \geq 1}\left(2^{1/2}\gamma_{\mathrm{max}}c^{1/2}\right)^{2d^{(m + 1)}} & = \frac{\left(2^{1/2}\gamma_{\mathrm{max}}c^{1/2}\right)^2}{1 - \left(2^{1/2}\gamma_{\mathrm{max}}c^{1/2}\right)^2} \leq \left(2\gamma_{\mathrm{max}}c^{1/2}\right)^2,\\
    \sum_{d^{(r)} \geq 1}\left(2^{1/2}\gamma_{\mathrm{max}}c\right)^{2d^{(r)}}d^{(r)} & = \frac{\left(2^{1/2}\gamma_{\mathrm{max}}c\right)^{2}}{\left(1 - \left(2^{1/2}\gamma_{\mathrm{max}}c\right)^{2}\right)^2} \leq \left(2^{3/2}\gamma_{\mathrm{max}}c\right)^2,\\
    \sum_{d^{(m + 1)} \geq 1}\left(2^{1/2}\gamma_{\mathrm{max}}c^{1/2}\right)^{2d^{(m + 1)}}d^{(m + 1)} & = \frac{\left(2^{1/2}\gamma_{\mathrm{max}}c^{1/2}\right)^2}{1 - \left(2^{1/2}\gamma_{\mathrm{max}}c^{1/2}\right)^2} \leq \left(2^{3/2}\gamma_{\mathrm{max}}c^{1/2}\right)^2.
\end{align}
This establishes inequality \ref{eq:g_correlations_t_power_m_contribution_continuum_approximation}. Inequality \ref{eq:g_correlations_continuum_approximation} is proven in a similar way:
\begin{align}
    \left|G^{(2)}_{\bm{j}^{(1)}} - \frac{1}{p + 1}G^{(2),\,\mathrm{cont}}\left(\frac{\bm{j}^{(1)}}{p + 1/2}\right)\right| & \leq \left|\sum_{m \geq 0}\left(G^{(2),\,m}_{\bm{j}^{(1)}} - \frac{1}{p + 1}G^{(2),\,m,\,\mathrm{cont}}\left(\frac{\bm{j}^{(1)}}{p + 1/2}\right)\right)\right|\nonumber\\
    & \leq \sum_{m \geq 0}\left|G^{(2),\,m}_{\bm{j}^{(1)}} - \frac{1}{p + 1}G^{(2),\,m,\,\mathrm{cont}}\left(\frac{\bm{j}^{(1)}}{p + 1/2}\right)\right|\nonumber\\
    & \leq \sum_{m \geq 0}\frac{24}{p + 1}\max\left(1, 2\beta_{\mathrm{max}}, \frac{M_{\gamma}}{\gamma_{\mathrm{max}}}\right)\left(2\gamma_{\mathrm{max}}c\right)^{2m}m\nonumber\\
    & \leq \frac{384}{p + 1}\max\left(1, 2\beta_{\mathrm{max}}, \frac{M_{\gamma}}{\gamma_{\mathrm{max}}}\right)\gamma_{\mathrm{max}}^2c^2.
\end{align}
\end{proof}
\end{proposition}
All in all, proposition \ref{prop:g_correlations_continuum_approximation} established a continuum approximation for saddle point $\bm{\theta}^*$.

\subsubsection{The continuum limit of higher-order correlation tensors}
\label{sec:continuum_limit_higher_order_correlations}

We now define a continuum analogue of higher-order correlation tensors. For that purpose, we use the series expansion of higher-order correlation tensors in terms of non-interacting correlation tensors derived in Proposition~\ref{prop:t_block_discrete_from_continuous_rescaled}, i.e.
\begin{align}
    \bm{C}^{\left(d\right)} & = \frac{\overline{\mathcal{Z}}^*}{\mathcal{Z}}\sum_{m \geq 0}\bm{C}^{(d),\,m},
\end{align}
where we defined
\begin{align}
    \bm{C}^{(d),\,m} & = \frac{\lambda^{m + d}}{m!}\left\langle \bm{\theta}^{*\otimes m}, \overline{\bm{C}}^{(m + d)} \right\rangle.
\end{align}
Specializing to SK-QAOA, $\mathcal{Z}^* = \overline{\mathcal{Z}}^* = 1$, and the $m$ contribution $\bm{C}^{(d),\,m}$, evaluated at index
\begin{align}
    \bm{j}^{(1)} = \bm{j}^{(1)}_{1:2d} = \left(j^{(1)}_1, j^{(1)}_2, \ldots, j^{(1)}_{2d - 1}, j^{(1)}_{2d}\right) \in \mathcal{I}^{2d},
\end{align}
expands as
\begin{align}
    C^{(d),\,m}_{\bm{j}^{(1)}} & = \frac{\lambda^{m + d}}{m!}\left\langle \bm{\theta}^{*\otimes m}, \overline{\bm{C}}^{(m + d)} \right\rangle_{\bm{\alpha}^{(1)}}\nonumber\\
    & = \frac{\lambda^{m + d}}{m!}\sum_{\bm{j}^{(2)} = \bm{j}^{(2)}_{1:2m} \in \mathcal{A}^{2m}}\overline{C}^{\left(m + d\right)}_{\bm{j}^{(1)},\,\bm{j}^{(2)}}\left[\bm{\theta}^{*\otimes m}\right]_{\bm{j}^{(2)}}\nonumber\\
    & = \frac{\lambda^{m + d}}{m!}\sum_{\bm{j}^{(2)} \in \mathcal{I}^{2m}}\overline{C}^{\left(m + d\right)}_{\bm{j}^{(1)},\,\bm{j}^{(2)}}\prod_{1 \leq r \leq m}\theta^*_{j^{(2)}_{2r - 1},\,j^{(2)}_{2r}}\nonumber\\
    & = \frac{\lambda^{m + d}}{m!}\sum_{\bm{j}^{(2)} \in \mathcal{I}^{2m}}\mathcal{G}^{(2m + 2d)}_{\bm{j}^{(1)},\,\bm{j}^{(2)}}\overline{G}^{\left(2m + 2d\right)}_{\bm{j}^{(1)},\,\bm{j}^{(2)}}\prod_{1 \leq r \leq m}\lambda\mathcal{G}^{(2)}_{j^{(2)}_{2r - 1},\,j^{(2)}_{2r}}G^{(2)}_{j^{(2)}_{2r - 1},\,j^{(2)}_{2r}}\nonumber\\
    & = \frac{\lambda^{m + d}}{m!}\sum_{\bm{j}^{(2)} \in \mathcal{I}^{2m}}\mathcal{G}^{(2d)}_{\bm{j}^{(1)}}\left(\prod_{1 \leq r \leq m}\mathcal{G}^{(2)}_{j^{(2)}_{2r - 1},\,j^{(2)}_{2r}}\right)\overline{G}^{(2m + 2d)}_{\bm{j}^{(1)},\,\bm{j}^{(2)}}\lambda^m\prod_{1 \leq r \leq m}\mathcal{G}^{(2)}_{j^{(2)}_{2r - 1},\,j^{(2)}_{2r}}G^{(2)}_{j^{(2)}_{2r - 1},\,j^{(2)}_{2r}}\nonumber\\
    & = \frac{\lambda^{2m + d}}{m!}\mathcal{G}^{(2d)}_{\bm{j}^{(1)}}\sum_{\bm{j}^{(2)} \in \mathcal{I}^{2m}}\left(\prod_{1 \leq r \leq 2m}\mathcal{G}_{j^{(2)}_r}\right)\overline{G}^{(2m + 2d)}_{\bm{j}^{(1)},\,\bm{j}^{(2)}}\prod_{1 \leq r \leq m}G^{(2)}_{j^{(2)}_{2r - 1},\,j^{(2)}_{2r}}.\label{eq:higher_order_correlations_expansion}
\end{align}
By analogy with the relation between correlations and $\bm{G}$ correlations (definition \ref{def:g_correlations_tensor}):
\begin{align}
    C^{(d)}_{\bm{j}} & = \lambda^d\mathcal{G}^{(2d)}_{\bm{j}}G^{(2d)}_{\bm{j}}, \qquad \bm{j} \in \mathcal{I}^{2d},
\end{align}
one may then define $\bm{G}^{(2d),\,m}$, the order $m$ contribution to the $\bm{G}$ correlations of order $2d$, by:
\begin{align}
    C^{(d),\,m}_{\bm{j}} & =: \lambda^d\mathcal{G}^{(2d)}_{\bm{j}}G^{(2d),\,m}_{\bm{j}}, \qquad \bm{j} \in \mathcal{I}^{2d}.
\end{align}
In terms of $\bm{G}^{(2d),\,m}$, equation \ref{eq:higher_order_correlations_expansion} can be rewritten:
\begin{align}
    G^{(2d),\,m}_{\bm{j}^{(1)}} & = \frac{\lambda^{2m}}{m!}\sum_{\bm{j}^{(2)} \in \mathcal{I}^{2m}}\overline{G}^{(2d + 2m)}_{\bm{j}^{(1)},\,\bm{j}^{(2)}}\left(\prod_{1 \leq r \leq m}G^{(2)}_{j^{(2)}_{2r - 1},\,j^{(2)}_{2r}}\right)\prod_{1 \leq r \leq 2m}\mathcal{G}_{j^{(2)}_r}\nonumber\\
    & = \frac{\lambda^{2m}}{m!}\sum_{\bm{j}^{(2)} \in \mathcal{I}^{2m}}\overline{G}^{(2d + 2m)}_{\bm{j}^{(1)},\,\bm{j}^{(2)}}\left[\left(\bm{G}^{(2)}\right)^{\otimes m}\right]_{\bm{j}^{(2)}}\left[\bm{\mathcal{G}}^{\otimes 2m}\right]_{\bm{j}^{(2)}}.\label{eq:g_correlations_m_contribution}
\end{align}
The $\bm{G}$ correlation of order $2d$ (definition \ref{def:g_correlations_tensor}, equation \ref{eq:g_higher_order_correlations_tensor}) can then be expressed as the sum of order $m$ contributions:
\begin{align}
    \bm{G}^{(2d)} & = \sum_{m \geq 0}\bm{G}^{(2d),\,m}.\label{eq:g_correlations_from_m_contributions}
\end{align}
Note these formulae hold for $d = 1$, since by virtue of the saddle-point equation, $\bm{\theta}^* = \bm{C}^{(1)}$. In that case, equations \ref{eq:g_correlations_from_m_contributions} and \ref{eq:g_correlations_m_contribution} read:
\begin{align}
    \bm{G}^{(2)} & = \sum_{m \geq 0}\bm{G}^{(2),\,m},\\
    \bm{G}^{(2),\,m} & := \frac{\lambda^{2m}}{m!}\sum_{\bm{j}^{(2)} \in \mathcal{I}^{2m}}\overline{G}^{(2 + 2m)}_{\bm{j}^{(1)},\,\bm{j}^{(2)}}\left[\bm{G}^{(2)\otimes m}\right]_{\bm{j}^{(2)}}\left[\bm{\mathcal{G}}^{\otimes 2m}\right]_{\bm{j}^{(2)}}.
\end{align}

Interpreting the sum in equation \ref{eq:g_correlations_m_contribution} as a discretized integral, this suggests the following continuum definition of the order $d$ correlations tensor:

\begin{definition}[Continuum higher-order $\bm{G}$ correlations]
\label{def:g_higher_order_correlations}
For all $d \geq 2$, the continuum $\bm{G}$ correlations tensor of order $2d$ is a continuous function $G^{(2d),\,\mathrm{cont}}: [0, 2]^{2d} \longrightarrow \mathbf{R}$ defined by series:
\begin{align}
    G^{(2d),\,\mathrm{cont}} & := \sum_{m \geq 0}G^{(2d),\,m,\,\mathrm{cont}},\label{eq:g_higher_order_correlations_continuum_definition}
\end{align}
where $G^{(2d),\,m}$ is a continuous function defined as
\begin{align}
    G^{(2d),\,m,\,\mathrm{cont}}\left(\bm{x}_{1:2d}\right)  := \frac{\left(2^{-1/2}\gamma_{\mathrm{max}}\right)^{2m}}{m!} \int\limits_{[0, 2]^{2m}}\!\mathrm{d}\bm{y}\,\overline{G}^{(2d + 2m),\,\mathrm{cont}}\left(\bm{x}_{1:2d}, \bm{y}_{1:2m}\right)\left(G^{(2),\,\mathrm{cont}}\right)^{\otimes m}\left(\bm{y}_{1:2m}\right)\left(\mathcal{G}^{\mathrm{cont}}\right)^{\otimes 2m}\left(\bm{y}_{1:2m}\right).\label{eq:g_higher_order_correlations_m_contribution_definition}
\end{align}
For all $d \geq 2, m \geq 0$, $G^{(2d),\,m,\,\mathrm{cont}}$ is well-defined and continuous by integration of (piecewise) continuous bounded functions. It is less obvious that the series defining $G^{(2d),\,\mathrm{cont}}$ (equation \ref{eq:g_higher_order_correlations_continuum_definition}) is well-defined, but this will be established in proposition \ref{prop:g_higher_order_correlations_well_definiteness}.
\end{definition}

\begin{proposition}[Continuum higher-order $\bm{G}$ correlations are well-defined]
\label{prop:g_higher_order_correlations_well_definiteness}
For all $d \geq 1$, the continuum analogue of the higher-order $\bm{G}^{(2d),\,\mathrm{cont}}$ is well-defined and continuous as the the sum of a uniformly absolutely convergent series of continuous functions. Besides, the following uniform bound holds:
\begin{align}
    \left\lVert G^{(2d),\,\mathrm{cont}} \right\rVert_{\infty} & \leq e^{4\gamma_{\mathrm{max}}^2},\label{eq:g_higher_order_correlations_continuum_bound},
\end{align}
and the function is $2\beta_{\mathrm{max}}e^{4\gamma_{\mathrm{max}}^2}$-Lipschitz in each of its variables. Finally, the discrete higher-order correlations satisfy the same higher uniform bound:
\begin{align}
    \left\lVert \bm{G}^{(2d)} \right\rVert_{\infty} & \leq e^{4\gamma_{\mathrm{max}}^2},\label{eq:g_higher_order_correlations_bound}
\end{align}
where the infinite norm now refers to the standard vector infinite norm rather than the uniform norm over continuous functions.
\begin{proof}
Recalling the bound on $G^{(2),\,\mathrm{cont}}$ (proposition \ref{prop:g_correlations_continuum_well_definiteness}), as well as bounding $\overline{G}^{(2d + 2m)}$ and $\mathcal{G}^{\mathrm{cont}}$ by $1$ (following from their definition), yields the following uniform bound on series term $G^{(2d),\,m,\,\mathrm{cont}}$ defined in equation \ref{eq:g_higher_order_correlations_m_contribution_definition}:
\begin{align}
    \left\lVert G^{(2d),\,m,\,\mathrm{cont}} \right\rVert_{\infty} & \leq \frac{\left(2^{-1/2}\gamma_{\mathrm{max}}\right)^{2m}}{m!}\int\limits_{[0, 2]^{2m}}\!\mathrm{d}\bm{y}\,1 \times 2^m \times 1\nonumber\\
    & = \frac{\left(2\gamma_{\mathrm{max}}\right)^{2m}}{m!}.
\end{align}
These establishes the uniform absolute convergence of the series of continuous functions defining $G^{(2d),\,\mathrm{cont}}$ (equation \ref{eq:g_higher_order_correlations_continuum_definition}), and provides the following uniform bound on these functions:
\begin{align}
    \left\lVert G^{(2d),\,\mathrm{cont}} \right\rVert_{\infty} & \leq \sum_{m \geq 0}\frac{\left(2\gamma_{\mathrm{max}}\right)^{2m}}{m!}\nonumber\\
    & \leq \exp\left(4\gamma_{\mathrm{max}}^2\right).
\end{align}
The Lipschitz constant estimate is a variation of this bound, using that $\overline{G}^{(2m + 2d)}$ is $2\beta_{\mathrm{max}}$-Lipschitz in each of its variables.

Similar to the proof of proposition \ref{prop:g_correlations_continuum_well_definiteness}, the proof of the discrete bound equation \ref{eq:g_higher_order_correlations_bound} can either be seen as a replay of general bound \ref{eq:correlations_tensor_bound} on the correlations tensor ---now reasoning over $\bm{G}$ correlations instead of correlations, or a rewriting of the current proof, with the role of the integration volume being played the number of terms in the discrete sum.
\end{proof}
\end{proposition}

We are now ready to relate the discrete and continuum higher order $\bm{G}$ correlations. This result from the bounds between discrete and continuum $\bm{G}^{(2)}$ correlations, as well as higher-order non-interacting correlations $\bm{\overline{G}}^{(2d + 2m)}$, $m \geq 0$.

\begin{proposition}[Approximation of discrete higher-order $\bm{G}$ correlations by continuum ones]
\label{prop:g_higher_order_correlations_continuum_approximation}
The following approximation holds between discrete higher-order $\bm{G}$ correlations (equation \ref{eq:g_higher_order_correlations_tensor}) and their continuum analogue (equation \ref{eq:g_higher_order_correlations_continuum_definition}):
\begin{align}
    \left|G^{(2d),\,m}_{\bm{j}_{1:2d}} - G^{(2d),\,m,\,\mathrm{cont}}\left(\frac{\bm{j}_{1:2d}}{p + 1/2}\right)\right| & \leq \frac{\left(2\gamma_{\mathrm{max}}\right)^{2m}}{m!}\frac{34m}{p + 1}\max\left(1, 2\beta_{\mathrm{max}}, \frac{M_{\gamma}}{\gamma_{\mathrm{max}}}\right).\label{eq:g_higher_order_correlations_m_contribution_continuum_approximation}
\end{align}
\begin{align}
    \left|G^{(2d)}_{\bm{j}_{1:2d}} - G^{(2d),\,\mathrm{cont}}\left(\frac{\bm{j}_{1:2d}}{p + 1/2}\right)\right| & \leq \frac{136\gamma_{\mathrm{max}}^2e^{4\gamma_{\mathrm{max}}^2}}{p + 1}\max\left(1, 2\beta_{\mathrm{max}}, \frac{M_{\gamma}}{\gamma_{\mathrm{max}}}\right).\label{eq:g_higher_order_correlations_continuum_approximation}
\end{align}
\begin{proof}
We start by proving equation \ref{eq:g_higher_order_correlations_m_contribution_continuum_approximation}. For that purpose, we start with explicit expression of the order $m$ contribution $\bm{G}^{(2d),\,m}$ in equation \ref{eq:g_higher_order_correlations_m_contribution_definition}. We divide the error into two contributions:
\begin{align}
    G^{(2d),\,m}_{\bm{j}_{1:2d}} - G^{(2d),\,m,\,\mathrm{cont}}\left(\frac{\bm{j}_{1:2d}}{p + 1/2}\right) & = A + B,
\end{align}
where
\begin{align}
    A & := \frac{\lambda^{2m}}{m!}\sum_{\bm{j}^{(2)} \in \mathcal{I}^{2m}}\overline{G}^{(2d + 2m)}_{\bm{j}_{1:2m}}\left[\left(\bm{G}^{(2)}\right)^{\otimes m}\right]_{\bm{j}^{(2)}}\left[\bm{\mathcal{G}}^{\otimes 2m}\right]_{\bm{j}^{(2)}}\nonumber\\
    & \hspace*{30px} - \frac{\lambda^{2m}}{m!}\sum_{\bm{j}^{(2)} \in \mathcal{I}^{2m}}\overline{G}^{(2d + 2m)}_{\bm{j}_{1:2m}}\left(G^{(2),\,\mathrm{cont}}\right)^{\otimes m}\left(\frac{\bm{j}^{(2)}}{p + 1/2}\right)\left(\mathcal{G}^{\mathrm{cont}}\right)^{\otimes 2m}\left(\frac{\bm{j}^{(2)}}{p + 1/2}\right)
\end{align}
and
\begin{align}
    B & := \frac{\lambda^{2m}}{m!}\sum_{\bm{j}^{(2)} \in \mathcal{I}^{2m}}\overline{G}^{(2d + 2m)}_{\bm{j}_{1:2m}}\left(G^{(2),\,\mathrm{cont}}\right)^{\otimes m}\left(\frac{\bm{j}^{(2)}}{p + 1/2}\right)\left(\mathcal{G}^{\mathrm{cont}}\right)^{\otimes 2m}\left(\frac{\bm{j}^{(2)}}{p + 1/2}\right)\nonumber\\
    & \hspace*{20px} - \frac{\left(2^{-1/2}\gamma_{\mathrm{max}}\right)^{2m}}{m!} \int\limits_{[0, 2]^{2m}}\!\mathrm{d}\bm{y}\,\overline{G}^{(2d + 2m),\,\mathrm{cont}}\left(\frac{\bm{j}_{1:2d}}{p + 1/2}, \bm{y}_{1:2m}\right)\left(G^{(2),\,\mathrm{cont}}\right)^{\otimes m}\left(\bm{y}_{1:2m}\right)\left(\mathcal{G}^{\mathrm{cont}}\right)^{\otimes 2m}\left(\bm{y}_{1:2m}\right)
\end{align}
We start by estimating $A$. For that purpose, we estimate the error commited in each term $\bm{j}^{(2)} \in \mathcal{I}^{2m}$ of the sum, and multiply by the number of terms $\left(2p + 2\right)^{2m}$. The error in a single term results from replacing 
\begin{gather}
    G^{\left(2\right)}_{j^{(2)}_1,\,j^{(2)}_2},\,\ldots,\,G^{\left(2\right)}_{j^{(2)}_{2m - 1},\,j^{(2)}_{2m}}\\
    \mathcal{G}_{j^{(2)}_1},\,\mathcal{G}_{j^{(2)}_2},\,\ldots,\,\mathcal{G}_{j^{(2)}_{2m - 1}},\,\mathcal{G}_{j^{(2)}_{2m}}
\end{gather}
by
\begin{gather}
    G^{\left(2\right),\,\mathrm{cont}}\left(\frac{j^{(2)}_{1}}{p + 1/2}, \frac{j^{(2)}_{2}}{p + 1/2}\right),\,\ldots,\,G^{\left(2\right),\,\mathrm{cont}}\left(\frac{j^{(2)}_{2m - 1}}{p + 1/2}, \frac{j^{(2)}_{2m}}{p + 1/2}\right)\\
    \mathcal{G}^{\mathrm{cont}}\left(\frac{j^{(2)}}{p + 1/2}\right),\,\mathcal{G}^{\mathrm{cont}}\left(\frac{j^{(2)}_2}{p + 1/2}\right),\,\ldots,\,\mathcal{G}^{\mathrm{cont}}\left(\frac{j^{(2)}_{2m - 1}}{p + 1/2}\right),\,\mathcal{G}^{\mathrm{cont}}\left(\frac{j^{(2)}_{2m}}{p + 1/2}\right)
\end{gather}
We estimate the variation of the product resulting from this replacement using lemma \ref{lemma:bound_variation_product}. The terms in the list are bounded by
\begin{gather}
    2,\,\ldots,\,2,\\
    1,\,1,\,\ldots,1,\,1,
\end{gather}
where the bounds on the first line result from equation \ref{eq:g_correlations_bounds} (discrete case) and proposition \ref{eq:g_correlations_continuum_bound} (continuum case), and the bounds in the second line are by definition of $\bm{\mathcal{G}}$ and $\mathcal{G}^{\mathrm{cont}}$ (equation \ref{eq:curvy_g_vector_definition_rescaled} in the discrete case, \ref{eq:curvy_g_tensor_continuum_definition_rescaled} in the continuum case). Also, invoking proposition \ref{prop:g_correlations_continuum_approximation} for the continuum approximation of $\bm{G}^{(2)}$ and using that discretization of $\mathcal{G}^{\mathrm{cont}}$ is exact:
\begin{align}
    \mathcal{G}_{j_r} & := \mathcal{G}^{\mathrm{cont}}\left(\frac{j_r}{p + 1/2}\right),
\end{align}
the difference between the relevant quantities are upper-bounded by
\begin{gather}
    \frac{384}{p + 1}\max\left(1, 2\beta_{\mathrm{max}}, \frac{M_{\gamma}}{\gamma_{\mathrm{max}}}\right)\gamma_{\mathrm{max}}^2c^2,\,\ldots,\,\frac{384}{p + 1}\max\left(1, 2\beta_{\mathrm{max}}, \frac{M_{\gamma}}{\gamma_{\mathrm{max}}}\right)\gamma_{\mathrm{max}}^2c^2\\
    0,\,0,\,\ldots,\,0,\,0.
\end{gather}
Applying lemma \ref{lemma:bound_variation_product} then gives a bound
\begin{align}
    2^{m - 1}m\frac{384}{p + 1}\max\left(1, 2\beta_{\mathrm{max}}, \frac{M_{\gamma}}{\gamma_{\mathrm{max}}}\right)\gamma_{\mathrm{max}}^2c^2
\end{align}
on a single term of the sum defining $A$, hence
\begin{align}
    \left|A\right| & \leq \left(2p + 2\right)^{2m}\frac{\lambda^{2m}}{m!}2^{m - 1}m\frac{384}{p + 1}\max\left(1, 2\beta_{\mathrm{max}}, \frac{M_{\gamma}}{\gamma_{\mathrm{max}}}\right)\gamma_{\mathrm{max}}^2c^2\nonumber\\
    & = \frac{\left(2\gamma_{\mathrm{max}}\right)^{2m}}{m!}\frac{192m}{p + 1}\max\left(1, 2\beta_{\mathrm{max}}, \frac{M_{\gamma}}{\gamma_{\mathrm{max}}}\right)\gamma_{\mathrm{max}}^2c^2\nonumber\\
    & \leq \frac{\left(2\gamma_{\mathrm{max}}\right)^{2m}}{m!}\frac{24m}{p + 1}\max\left(1, 2\beta_{\mathrm{max}}, \frac{M_{\gamma}}{\gamma_{\mathrm{max}}}\right),
\end{align}
where in the final line, we plugged assumption
\begin{align}
    \gamma_{\mathrm{max}} & \leq \frac{1}{2^{3/2}c}.
\end{align}
We now turn to bounding $B$, which is a difference between a sum and an integral. Recalling the definition of $\lambda$ in equation \ref{eq:sk_qaoa_qgms_redefinition_rescaled_lambda}, the prefactor of the sum expands as
\begin{align}
    \frac{\lambda^{2m}}{m!} & = \frac{\left(2^{-1/2}\gamma_{\mathrm{max}}\right)^{2m}}{m!}\frac{1}{\left(p + 1\right)^{2m}}.
\end{align}
Given the summation vector $\bm{j}^{(2)}$ is of dimension $2m$, with coordinate iterating in $\{0, 1, \ldots, 2p, 2p + 1\}$, this provides the correct scaling for applying sum-integral comparison lemma \ref{lemma:riemann_sum_approximation}. The lemma is applied to discrete summation variable $\bm{j}^{(2)}$, with $\bm{j}^{(1)}$ being regarded as a constant. These respectively correspond to a real variable $\bm{x}^{(2)} \in [0, 2]^{2m}$, and a real constant $\bm{x} := \bm{j}^{(1)}/(p + 1/2)$. The functions we apply the lemma to are
\begin{gather}
    \overline{G}^{(2d + 2m),\,\mathrm{cont}}\left(\bm{x}, \bm{x}^{(2)}\right),\\
    G^{\left(2\right),\,\mathrm{cont}}\left(x^{(2)}_1,\,x^{(2)}_2\right),\,\ldots,\,G^{\left(2\right),\,\mathrm{cont}}\left(x^{(2)}_{2m - 1},\,x^{(2)}_{2m}\right),\\
    \mathcal{G}^{\mathrm{cont}}\left(x^{(2)}_1\right),\,\mathcal{G}^{\mathrm{cont}}\left(x^{(2)}_2\right),\,\ldots,\,\mathcal{G}^{\mathrm{cont}}\left(x^{(2)}_{2m - 1}\right),\,\mathcal{G}^{\mathrm{cont}}\left(x^{(2)}_{2m}\right).
\end{gather}
The total dimension of the vaiiables occurring in these functions is:
\begin{align}
    D & := m\nonumber\\
    & \hspace*{20px} + 2 + \ldots + 2\\
    & \hspace*{20px} + 1 + 1 + \ldots + 1 + 1\nonumber\\
    & = 5m.
\end{align}
The functions are respectively bounded (constants $K_l$ from the lemma) by
\begin{gather}
    1,\\
    2,\,\ldots,\,2,\\
    1,\,1,\,\ldots,\,1,\,1.
\end{gather}
Besides, they have respective Lipschitz constants in each variable (constants $M_l$ from the lemma):
\begin{gather}
    2\beta_{\mathrm{max}},\\
    4\beta_{\mathrm{max}},\,\ldots,\,4\beta_{\mathrm{max}},\\
    \frac{M_{\gamma}}{\gamma_{\mathrm{max}}},\,\frac{M_{\gamma}}{\gamma_{\mathrm{max}}},\,\ldots,\,\frac{M_{\gamma}}{\gamma_{\mathrm{max}}},\,\frac{M_{\gamma}}{\gamma_{\mathrm{max}}}.
\end{gather}
(for $G^{(2),\,\mathrm{cont}}$, the Lipschitz constant was estimated in proposition \ref{prop:g_correlations_continuum_well_definiteness}). From these estimates, lemma \ref{lemma:riemann_sum_approximation} provides the following estimate on the sum-integral error $B$:
\begin{align}
    |B| & \leq \frac{\left(2\gamma_{\mathrm{max}}\right)^{2m}}{m!}\frac{10m}{p + 1}\max\left(2\beta_{\mathrm{max}},\,\frac{M_{\gamma}}{\gamma_{\mathrm{max}}}\right)
\end{align}
Combining bounds on $A$ and $B$ then gives the desired bound (equation \ref{eq:g_higher_order_correlations_m_contribution_continuum_approximation}) on the continuum approximation of the order $m$ contribution to the order $2d$ correlation:
\begin{align}
    \left|G^{(2d),\,m}_{\bm{j}_{1:2d}} - G^{(2d),\,m,\,\mathrm{cont}}\left(\frac{\bm{j}_{1:2d}}{p + 1/2}\right)\right| & \leq |A| + |B|\nonumber\\
    & \leq \frac{\left(2\gamma_{\mathrm{max}}\right)^{2m}}{m!}\frac{34m}{p + 1}\max\left(1, 2\beta_{\mathrm{max}}, \frac{M_{\gamma}}{\gamma_{\mathrm{max}}}\right).
\end{align}
Summing this bound over $m \geq 0$ in turn gives bound \ref{eq:g_higher_order_correlations_continuum_approximation} on the continuum approximation of higher-order correlations:
\begin{align}
    \left|G^{(2d)}_{\bm{j}_{1:2d}} - G^{(2d),\,\mathrm{cont}}\left(\frac{\bm{j}_{1:2d}}{p + 1/2}\right)\right| & \leq \sum_{m \geq 0}\frac{\left(2\gamma_{\mathrm{max}}\right)^{2m}}{m!}\frac{34m}{p + 1}\max\left(1, 2\beta_{\mathrm{max}}, \frac{M_{\gamma}}{\gamma_{\mathrm{max}}}\right)\nonumber\\
    & \leq \frac{136\gamma_{\mathrm{max}}^2e^{4\gamma_{\mathrm{max}}^2}}{p + 1}\max\left(1, 2\beta_{\mathrm{max}}, \frac{M_{\gamma}}{\gamma_{\mathrm{max}}}\right).
\end{align}
\end{proof}
\end{proposition}

We will now compute bounds on the discrete and continuum centered $\bm{G}$ correlations, as well as derive discretization bounds. To achieve that, we will use uniform bounds (proposition \ref{prop:g_higher_order_correlations_well_definiteness}) and approximation bounds (proposition \ref{prop:g_higher_order_correlations_continuum_approximation}) on discrete and continuum (noncentered) $\bm{G}$ correlations. To make formulae less cumbersome, it will pay to simplify these bounds by redefining $\gamma_{\mathrm{max}}$ up to a rescaling. This is done in the following proposition:

\begin{proposition}[Uniform bounds and discretization bounds on discrete and continuum correlations, simplified]
\label{prop:g_higher_order_correlations_bounds_simplified}
Let 
\begin{align}
    \gamma_{\mathrm{max}} & := \gamma_0\gamma'_{\mathrm{max}},
\end{align}
where $\gamma_0 > 0$ is a universal constant to be determined. We claim that for $\gamma_0$ sufficiently small and $\gamma'_{\mathrm{max}} \leq 1$, the following bounds hold. First, the discrete and continuum higher-order $\bm{G}$ correlations are uniformly bounded as follows:
\begin{align}
    \left\lVert \bm{G}^{(2d)} \right\rVert_{\infty} & \leq 2\label{eq:g_higher_order_correlations_continuum_bound_simplified},\\
    \left\lVert G^{(2d),\,\mathrm{cont}} \right\rVert_{\infty} & \leq 2.\label{eq:g_higher_order_correlations_bound_simplified}
\end{align}
Besides, the continuum higher-order $\bm{G}$ correlations of order $2d$ are $4\beta_{\mathrm{max}}$-Lipschitz in each of their variables. Finally, the following discretization bounds hold between continuum and discrete higher-order $\bm{G}$ correlations:
\begin{align}
    \left|G^{(2d),\,\mathrm{cont}}_{\bm{j}_{1:2d}} - G^{(2d),\,\mathrm{cont}}\left(\frac{\bm{j}_{1:2d}}{p + 1/2}\right)\right| & \leq \frac{\gamma'_{\mathrm{max}}}{p + 1}.\label{eq:g_higher_order_correlations_continuum_approximation_simplified}
\end{align}
The parameter $\gamma_0$ for these bounds to hold depends on $M_{\gamma}$ ---the Lipschitz constant of $\gamma^{\mathrm{cont}}$--- as well as $\beta_{\mathrm{max}}$; however, it can be bounded by a universal constant if $M_{\gamma}$ and $\beta_{\mathrm{max}}$ are bounded by universal constants. Given $M_{\gamma}$ and $\beta_{\mathrm{max}}$, $\gamma_0$ is nonetheless independent of $d \geq 1$ and $\bm{j}_{1:2d} \in \mathcal{I}^d$ in the above bounds.
\begin{proof}
Bounds \ref{eq:g_higher_order_correlations_continuum_bound_simplified}, \ref{eq:g_higher_order_correlations_bound_simplified} result from bounds \ref{eq:g_higher_order_correlations_continuum_bound}, \ref{eq:g_higher_order_correlations_bound} in proposition \ref{prop:g_higher_order_correlations_well_definiteness}: $\left\lVert \bm{G}^{(2d),\,\mathrm{cont}} \right\rVert_{\infty}, \left\lVert G^{(2),\,\mathrm{cont}} \right\rVert_{\infty} \leq e^{4\gamma_{\mathrm{max}}^2}$, assuming $\gamma_0 \leq \sqrt{\log(2)/4}$.

As for the discretization bound in equation \ref{eq:g_higher_order_correlations_continuum_approximation_simplified}, it follows from original bound in proposition \ref{prop:g_higher_order_correlations_continuum_approximation} (equation \ref{eq:g_higher_order_correlations_continuum_approximation}):
\begin{align}
    \left|G^{(2d)}_{\bm{j}_{1:2d}} - G^{(2d),\,\mathrm{cont}}\left(\frac{\bm{j}_{1:2d}}{p + 1/2}\right)\right| & \leq \frac{136e^{4\gamma_{\mathrm{max}}^2}\gamma_{\mathrm{max}}}{p + 1}\max\left(1, 2\beta_{\mathrm{max}}, \frac{M_{\gamma}}{\gamma_{\mathrm{max}}}\right)\nonumber\\
    & = \frac{136e^{4\gamma_0^2\left(\gamma'_{\mathrm{max}}\right)^2}\gamma_0^2\left(\gamma'_{\mathrm{max}}\right)^2}{p + 1}\max\left(1, 2\beta_{\mathrm{max}}, \frac{M_{\gamma}}{\gamma_0\gamma'_{\mathrm{max}}}\right)\nonumber\\
    & = \frac{136e^{4\gamma_0^2\left(\gamma'_{\mathrm{max}}\right)^2}\gamma_0\gamma'_{\mathrm{max}}}{p + 1}\max\left(\gamma_0\gamma'_{\mathrm{max}}, 2\gamma_0\gamma'_{\mathrm{max}}\beta_{\mathrm{max}}, M_{\gamma}\right)\nonumber\\
    & \leq \frac{136e^{4\gamma_0^2}\gamma_0\gamma'_{\mathrm{max}}}{p + 1}\max\left(\gamma_0, 2\gamma_0\beta_{\mathrm{max}}, M_{\gamma}\right).
\end{align}
Choosing $\gamma_0$ smaller than an a constant depending only on $\beta_{\mathrm{max}}$ and $M_{\gamma}$ guarantees the above is bounded by $\gamma'_{\mathrm{max}}/(p + 1)$ as required.
\end{proof}
\end{proposition}

From the simplified bounds on $\bm{G}$ correlations introduced in proposition \ref{prop:g_higher_order_correlations_bounds_simplified}, it will be reasonably convenient to produce uniform bounds on the centered $\bm{G}$ correlations and quantity the continuum approximation error. This is done in the following proposition:

\begin{proposition}[Uniform and discretization bounds on centered $\bm{G}$ correlations]
\label{prop:centered_g_correlations_bounds_and_continuum_approximation}
Similar to the setting of proposition \ref{prop:g_higher_order_correlations_bounds_simplified}, let $\gamma_{\mathrm{max}} := \gamma_0\gamma'_{\mathrm{max}}$, with $\gamma_0$ smaller than an absolute constant and $\gamma'_{\mathrm{max}} \leq 1$. Then, the following uniform bounds hold on the centered $\bm{G}$ correlations tensor of order $d$ and its continuum analogue:
\begin{align}
    \left\lVert \bm{\delta G}^{(2d)} \right\rVert_{\infty} & \leq 2.3^{d},\label{eq:centered_g_correlations_tensor_bound}\\
    \left\lVert \delta G^{(2d),\,\mathrm{cont}} \right\rVert_{\infty} & \leq 2.3^{d}\label{eq:centered_g_correlations_tensor_bound_continuum}.
\end{align}
Besides, the continuum $\bm{G}$ correlations of order $d$ are $4.3^d\beta_{\mathrm{max}}$-Lipschitz in each of their variables; in this context, individual variables are seen to be real numbers $\in [0, 2]$, i.e. $\delta C^{(2d),\,\mathrm{cont}}$ is regarded as a function of $2d$ real variables. Finally, the following discretization bound holds between the discrete tensor and the continuous function:
\begin{align}
    \left| \delta G^{(2d)}_{\bm{\alpha}_{1:d}} - \delta G^{(2d),\,\mathrm{cont}}\left(\frac{\bm{\alpha}_{1:d}}{p + 1/2}\right) \right| & \leq \frac{3^{d + 1}d\gamma'_{\mathrm{max}}}{p + 1}, \qquad \forall \bm{\alpha} \in \mathcal{A}^{d}.\label{eq:centered_g_correlations_tensor_continuum_approximation}
\end{align}
\begin{proof}
We start with the uniform bounds, for instance with discrete bound \ref{eq:centered_g_correlations_tensor_bound}. This follows from the explicit expression of the tensor entries in equation \ref{eq:centered_g_correlations_tensor_indices_expression}, the triangular inequality, and using the uniform bounds on (non-centered) correlations from proposition \ref{prop:g_higher_order_correlations_bounds_simplified}:
\begin{align}
    \left|\delta G^{(2d)}_{\bm{\alpha}_{1:d}}\right| & = \left|\sum_{\substack{S',\,S''\\S' \sqcup S'' = [d]}}(-1)^{\left|S'\right|}\left[\bm{G}^{\left(2\left|S'\right|\right)}\right]_{\bm{\alpha}_{S'}}\left[\bm{G}^{(2)\otimes \left|S''\right|}\right]_{\bm{\alpha}_{S''}}\right|\nonumber\\
    & \leq \sum_{\substack{S',\,S''\\S' \sqcup S'' = [d]}}\left\lVert \bm{G}^{\left(2\left|S'\right|\right)} \right\rVert_{\infty}\left\lVert \bm{G}^{\left(2\right)\otimes \left|S''\right|} \right\rVert_{\infty}\nonumber\\
    & = \sum_{\substack{S',\,S''\\S' \sqcup S'' = [d]}}\left\lVert \bm{G}^{\left(2\left|S'\right|\right)} \right\rVert_{\infty}\left\lVert \bm{G}^{\left(2\right)} \right\rVert_{\infty}^{\left|S''\right|}\nonumber\\
    & \leq \sum_{\substack{S',\,S''\\S' \sqcup S'' = [d]}}2.2^{\left|S''\right|}\nonumber\\
    & = \sum_{\substack{s',\,s''\\s' + s'' = d}}\binom{d}{s',\,s''}2.2^{s''}\nonumber\\
    & = 2.3^d,
\end{align}
which is equation \ref{eq:centered_g_correlations_tensor_bound}. The proof of the continuum analogue equation \ref{eq:centered_g_correlations_tensor_bound_continuum} is identical up to the interpretation of the infinite norm. The Lipschitz bound follows from the one of the (noncentered) $\bm{G}$ correlation stated in proposition \ref{prop:g_higher_order_correlations_bounds_simplified}.

Let us now turn to the approximation bound in equation \ref{eq:centered_g_correlations_tensor_continuum_approximation}. We wish to show closeness between
\begin{align}
    \sum_{\substack{S',\,S''\\S' \sqcup S'' = [d]}}(-1)^{\left|S'\right|}\left[\bm{G}^{\left(2\left|S'\right|\right)}\right]_{\bm{\alpha}_{S'}}\left[\bm{G}^{(2)\otimes \left|S''\right|}\right]_{\bm{\alpha}_{S''}}
\end{align}
and
\begin{align}
    \sum_{\substack{S',\,S''\\S' \sqcup S'' = [d]}}(-1)^{\left|S'\right|}G^{\left(2\left|S'\right|\right),\,\mathrm{cont}}\left(\frac{\bm{\alpha}_{S'}}{p + 1/2}\right)\left(G^{(2),\,\mathrm{cont}}\right)^{\otimes \left|S''\right|}\left(\frac{\bm{\alpha}_{S''}}{p + 1/2}\right).
\end{align}
We start by showing closeness between a single term $\left(S',\,S''\right)$ of the two sums. For this purpose, we invoke lemma \ref{lemma:bound_variation_product} bounding the variation of a product; in this context, we view each tensor factor of tensor products $\bm{G}^{(2)\otimes \left|S''\right|}, \left(G^{(2),\,\mathrm{cont}}\right)^{\otimes \left|S''\right|}$ as a separate factor. Recalling the simplified uniform and discretization bounds on $\bm{G}$ correlations from proposition \ref{prop:g_higher_order_correlations_bounds_simplified}, lemma \ref{lemma:bound_variation_product} gives:
\begin{align}
    \left|\left[\bm{G}^{\left(2\left|S'\right|\right)}\right]_{\bm{\alpha}_{S'}}\left[\bm{G}^{(2)\otimes \left|S''\right|}\right]_{\bm{\alpha}_{S''}} - G^{\left(2\left|S'\right|\right),\,\mathrm{cont}}\left(\frac{\bm{\alpha}_{S'}}{p + 1/2}\right)\left(G^{(2),\,\mathrm{cont}}\right)^{\otimes \left|S''\right|}\left(\frac{\bm{\alpha}_{S''}}{p + 1/2}\right)\right| & \leq \left(\left|S''\right| + 1\right)2^{\left|S''\right|}\frac{\gamma'_{\mathrm{max}}}{p + 1}.
\end{align}
Summing this bound over $S', S''$ gives a discretization bound between $\bm{\delta G}^{(d)}$ and its continuum analogue:
\begin{align}
    \left|\delta G^{(2d)}_{\bm{\alpha}_{1:d}} - \delta G^{(2d),\,\mathrm{cont}}\left(\frac{\bm{\alpha}_{1:d}}{p + 1/2}\right)\right| & \leq \sum_{\substack{S',\,S''\\S' \sqcup S'' = [d]}}\left(\left|S''\right| + 1\right)2^{\left|S''\right|}\frac{\gamma'_{\mathrm{max}}}{p + 1}\nonumber\\
    & = \sum_{\substack{s',\,s''\\s' + s'' = d}}\binom{d}{s', s''}\left(s'' + 1\right)2^{s''}\frac{\gamma'_{\mathrm{max}}}{p + 1}\nonumber\\
    & = \frac{3^d\left(2d + 1\right)\gamma'_{\mathrm{max}}}{p + 1}\nonumber\\
    & \leq \frac{3^{d + 1}d\gamma'_{\mathrm{max}}}{p + 1}.
\end{align}
\end{proof}
\end{proposition}

\subsubsection{The continuum limit of QGMS moments}
\label{sec:continuum_limit_qgms_moments}

Thanks to the uniform and continuum approximation bounds established for the centered correlations in proposition \ref{prop:centered_g_correlations_bounds_and_continuum_approximation}, we are now in position to bound the discrete and continuum versions of contribution $\nu^{\left(n_d\right)_{d \geq 2},\,l''',\,\mathcal{M}}$ (definition \ref{def:centered_g_correlations_tensor_continuum}) and the discretization error between them:

\begin{proposition}[Uniform and discretization bounds on multinomial numbers and matching contribution to QGMS moment]
\label{prop:qgms_integral_series_expansion_multinomial_matching_contribution_continuum_approximation}
Let $\gamma_{\mathrm{max}} := \gamma_0\gamma'_{\mathrm{max}}$ with $\gamma_0$ chosen smaller than an absolute constant (so proposition \ref{prop:g_higher_order_correlations_bounds_simplified} holds) and $\gamma'_{\mathrm{max}} \leq 1$. Recall the discrete: $\bm{\nu}^{\left(n_d\right)_{d \geq 2},\,l''',\,\mathcal{M}} = \left(\nu^{\left(n_d\right)_{d \geq 2},\,l''',\,\mathcal{M}}_{\alpha}\right)_{\alpha \in \mathcal{A}}$ and continuum: $\nu^{\left(n_d\right)_{d \geq 2},\,l''',\,\mathcal{M},\,\mathrm{cont}}: [0, 2]^2 \longrightarrow \mathbf{C}$ versions of the contribution of multinomial numbers $\left(n_d\right)_{d \geq 2}$ and matching $\left(l''', \mathcal{M}\right)$ to the QGMS moment, recapitulated or introduced in definition \ref{def:centered_g_correlations_tensor_continuum}. Finally, recall the related notations $\bm{\widetilde{\nu}}^{\left(n_d\right)_{d \geq 2},\,l''',\,\mathcal{M}}$, $\widetilde{\nu}^{\left(n_d\right)_{d \geq 2},\,l''',\,\mathcal{M},\,\mathrm{cont}}$ (equations \ref{eq:qgms_integral_series_expansion_multinomial_matching_contribution_expansion_tilde} and \ref{eq:qgms_integral_series_expansion_multinomial_matching_contribution_tilde_continuum}) introduced there, factoring out elementwise multiplications by $\bm{G}^{(2)}$ and $\bm{\mathcal{G}}$:
\begin{align}
    \nu^{\left(n_d\right)_{d \geq 2},\,l''',\,\mathcal{M},\,\mathrm{cont}}\left(\xi\right) & = G^{(2),\,\mathrm{cont}}\left(x_1, x_2\right)\mathcal{G}^{\mathrm{cont}}\left(x_1\right)\mathcal{G}^{\mathrm{cont}}\left(x_2\right)\widetilde{\nu}^{\left(n_d\right)_{d \geq 2},\,l''',\,\mathcal{M}}\left(\xi\right), && \xi = \left(x_1, x_2\right),\\
    \nu^{\left(n_d\right)_{d \geq 2},\,l''',\,\mathcal{M}}_{\alpha} & = G^{(2)}_{k_1,\,k_2}\mathcal{G}_{k_1}\mathcal{G}_{k_2}\widetilde{\nu}^{\left(n_d\right)_{d \geq 2},\,l''',\,\mathcal{M}}_{\alpha}, && \alpha = \left(k_1,\,k_2\right).
\end{align}
Then, the following uniform bounds hold on vector $\bm{\widetilde{\nu}}^{\left(n_d\right)_{d \geq 2},\,l''',\,\mathcal{M}}$ and function $\widetilde{\nu}^{\left(n_d\right)_{d \geq 2},\,l''',\,\mathcal{M},\,\mathrm{cont}}$:
\begin{align}
    \left\lVert \bm{\widetilde{\nu}}^{\left(n_d\right)_{d \geq 2},\,l''',\,\mathcal{M}} \right\rVert_{\infty} & \leq  \frac{\left(3\sqrt{2}\gamma_{\mathrm{max}}\right)^D}{2p + 2}2^{\sum\limits_{d \geq 2}n_d},\label{eq:qgms_integral_series_expansion_multinomial_matching_contribution_tilde_bound}\\
    \left\lVert \widetilde{\nu}^{\left(n_d\right)_{d \geq 2},\,l''',\,\mathcal{M},\,\mathrm{cont}} \right\rVert_{\infty} & \leq \frac{\left(3\sqrt{2}\gamma_{\mathrm{max}}\right)^D}{2}2^{\sum\limits_{d \geq 2}n_d},\label{eq:qgms_integral_series_expansion_multinomial_matching_contribution_tilde_continuum_bound}
\end{align}
where we let $D := \sum_{d \geq 2}dn_d$ for conciseness. Besides, $\widetilde{\nu}^{\left(n_d\right)_{d \geq 2},\,l''',\,\mathcal{M},\,\mathrm{cont}}$ is $2^{\sum_{d \geq 2}n_d}\left(3\sqrt{2}\gamma_{\mathrm{max}}\right)^D\beta_{\mathrm{max}}$-Lipschitz in each of its (two) variables. Finally, the following discretization bound holds between $\bm{\widetilde{\nu}}^{\left(n_d\right)_{d \geq 2},\,l''',\,\mathcal{M}}$ and  $\widetilde{\nu}^{\left(n_d\right)_{d \geq 2},\,l''',\,\mathcal{M},\,\mathrm{cont}}$:
\begin{align}
    \left|\widetilde{\nu}^{\left(n_d\right)_{d \geq 2},\,l''',\,\mathcal{M}}_{k_1,\,k_2} - \frac{1}{p + 1}\widetilde{\nu}^{\left(n_d\right)_{d \geq 2},\,l''',\,\mathcal{M},\,\mathrm{cont}}\left(\frac{k_1}{p + 1/2}, \frac{k_2}{p + 1/2}\right)\right| & \leq \frac{\left(12\gamma_{\mathrm{max}}\right)^DD}{\left(p + 1\right)^2}\max\left(4\beta_{\mathrm{max}}, \frac{2M_{\gamma}}{\gamma_{\mathrm{max}}}, \gamma'_{\mathrm{max}}\right).\label{eq:qgms_integral_series_expansion_multinomial_matching_contribution_tilde_approximation_error}
\end{align}
\begin{proof}
We start by proving the uniform bounds in equations \ref{eq:qgms_integral_series_expansion_multinomial_matching_contribution_tilde_bound}, \ref{eq:qgms_integral_series_expansion_multinomial_matching_contribution_tilde_continuum_bound}, for instance the discrete one. We start with the explicit expression of $\widetilde{\nu}^{\left(n_d\right)_{d \geq 2},\,l''',\,\mathcal{M}}_{\alpha}$ in equation \ref{eq:qgms_integral_series_expansion_multinomial_matching_contribution_expansion_tilde}. We then apply the triangle inequality and uniform bounds on $\bm{\delta C}^{(d)}$ from proposition \ref{prop:centered_g_correlations_bounds_and_continuum_approximation}:
\begin{align}
    \left|\widetilde{\nu}_{\alpha}^{\left(n_d\right)_{d \geq 2},\,l''',\,\mathcal{M}}\right| & = \lambda^D\left|\sum_{\substack{\bm{\alpha}_{\mathcal{D}(\mathcal{M})} \in \mathcal{A}^{\mathcal{D}(\mathcal{M})}}}\left[\pi_{l''',\,\mathcal{M}} \cdot \bigotimes_{d \geq 2}\bm{\delta G}^{(2d)\otimes n_d}\right]_{\alpha_{l'''},\,\bm{\alpha}_{\mathcal{D}(\mathcal{M})},\,\bm{\alpha}_{\mathcal{D}(\mathcal{M})}}\left[\bm{\mathcal{G}}^{\otimes (D - 1)}\right]_{\bm{\alpha}_{\mathcal{D}(\mathcal{M})}}\right|\nonumber\\
    & \leq \lambda^D\sum_{\bm{\alpha}_{\mathcal{D}(\mathcal{M})} \in \mathcal{A}^{\mathcal{D}(\mathcal{M})}}\left\lVert \bigotimes_{d \geq 2}\bm{\delta G}^{(2d)\otimes n_d} \right\rVert_{\infty}\left\lVert \bm{\mathcal{G}}^{\otimes (D - 1)} \right\rVert_{\infty}\nonumber\\
    & \leq \lambda^D\sum_{\bm{\alpha}_{\mathcal{D}(\mathcal{M})} \in \mathcal{A}^{\mathcal{D}(\mathcal{M})}}\left\lVert \bm{\mathcal{G}} \right\rVert_{\infty}^{D - 1}\prod_{d \geq 2}\left\lVert \bm{\delta G}^{(2d)} \right\rVert_{\infty}^{n_d}\nonumber\\
    & \leq \lambda^D\sum_{\bm{\alpha}_{\mathcal{D}(\mathcal{M})} \in \mathcal{A}^{\mathcal{D}(\mathcal{M})}}\prod_{d \geq 2}\left(2.3^{d}\right)^{n_d}\nonumber\\
    & = \lambda^D\left|\mathcal{A}\right|^{\left|\mathcal{D}(\mathcal{M})\right|}2^{\sum\limits_{d \geq 2}n_d}3^D\nonumber\\
    & = \lambda^D|\mathcal{A}|^{(D - 1)/2}2^{\sum\limits_{d \geq 2}n_d}3^D\nonumber\\
    & = \left(\frac{2^{-1/2}\gamma_{\mathrm{max}}}{p + 1}\right)^D\left(\left(2p + 2\right)^2\right)^{(D - 1)/2}2^{\sum\limits_{d \geq 2}n_d}3^{D}\nonumber\\
    & = \frac{\left(3\sqrt{2}\gamma_{\mathrm{max}}\right)^D}{2p + 2}2^{\sum\limits_{d \geq 2}n_d}.
\end{align}
This establishes the discrete uniform bound equation \ref{eq:qgms_integral_series_expansion_multinomial_matching_contribution_tilde_bound}; the derivation of the continuum one (equation \ref{eq:qgms_integral_series_expansion_multinomial_matching_contribution_tilde_continuum_bound}) is very similar. For the Lipschitzness (in each variable), we combined the uniform bound and Lipschitz constant bounds on $\bm{\delta G}^{(d)}$ from proposition \ref{prop:centered_g_correlations_bounds_and_continuum_approximation} to conclude that the integrand in the definition of $\widetilde{\nu}^{\left(n_d\right)_{d \geq 2},\,\mathcal{M},\,l''',\,\mathrm{cont}}$ is $2^{1 + \sum_{d \geq 2}n_d}3^D\beta_{\mathrm{max}}$-Lipschitz in $x_1, x_2$. Multiplying by the integration volume and the constant prefactor gives the claimed Lipschitz constant.

We now turn to the discretization bound equation \ref{eq:qgms_integral_series_expansion_multinomial_matching_contribution_tilde_approximation_error}. We decompose the discretization error into one term $A$ coming from the discretization of centered correlations, and one terms $B$ reflecting the Riemann sum approximation error to the integral defining $\widetilde{\nu}^{\left(n_d\right)_{d \geq 2},\,l''',\,\mathcal{M}}$.
\begin{align}
    \nu_{\alpha}^{\left(n_d\right)_{d \geq 2},\,l''',\,\mathcal{M}} - \frac{1}{p + 1}\widetilde{\nu}^{\left(n_d\right)_{d \geq 2},\,l''',\,\mathcal{M}}\left(\frac{\alpha}{p + 1/2}\right) & = A + B,
\end{align}
with
\begin{align}
    A & := \lambda^D\sum_{\bm{\alpha}_{\mathcal{D}(\mathcal{M})} \in \mathcal{A}^{\mathcal{D}(\mathcal{M})}}\left\{\left[\pi_{l''',\,\mathcal{M}} \cdot \bigotimes_{d \geq 2}\bm{\delta G}^{(2d)\otimes n_d}\right]_{\alpha,\,\bm{\alpha}_{\mathcal{D}(\mathcal{M})},\,\bm{\alpha}_{\mathcal{D}(\mathcal{M})}}\left[\bm{\mathcal{G}}^{\otimes (D - 1)}\right]_{\bm{\alpha}_{\mathcal{D}(\mathcal{M})}}\right.\nonumber\\
    & \hspace*{20px} - \left(\pi_{l''',\,\mathcal{M}} \cdot \bigotimes_{d \geq 2}\left(\delta G^{(2d),\,\mathrm{cont}}\right)^{\otimes n_d}\right)\left(\frac{\alpha}{p + 1/2},\,\frac{\bm{\alpha}_{\mathcal{D}(\mathcal{M})}}{p + 1/2},\,\frac{\bm{\alpha}_{\mathcal{D}(\mathcal{M})}}{p + 1/2}\right)\left(\mathcal{G}^{\mathrm{cont}}\right)^{\otimes (D - 1)}\left(\frac{\bm{\alpha}_{\mathcal{D}(\mathcal{M})}}{p + 1/2}\right)\vast\}
\end{align}
and 
\begin{align}
    B & := \lambda^D\sum_{\bm{\alpha}_{\mathcal{D}(\mathcal{M})} \in \mathcal{A}^{\mathcal{D}(\mathcal{M})}}\left(\pi_{l''',\,\mathcal{M}} \cdot \bigotimes_{d \geq 2}\left(\delta G^{(2d),\,\mathrm{cont}}\right)^{\otimes n_d}\right)\left(\frac{\alpha}{p + 1/2},\,\frac{\bm{\alpha}_{\mathcal{D}(\mathcal{M})}}{p + 1/2},\,\frac{\bm{\alpha}_{\mathcal{D}(\mathcal{M})}}{p + 1/2}\right)\left(\mathcal{G}^{\mathrm{cont}}\right)^{\otimes (D - 1)}\left(\frac{\bm{\alpha}_{\mathcal{D}(\mathcal{M})}}{p + 1/2}\right)\nonumber\\
    & \hspace*{20px} - \frac{\left(2^{-1/2}\gamma_{\mathrm{max}}\right)^D}{p + 1}\int\limits_{\bm{\xi} \in \left([0, 2]^2\right)^{\mathcal{D}(\mathcal{M})}}\!\mathrm{d}\bm{\xi}\,\left(\pi_{l''',\,\mathcal{M}} \cdot \bigotimes_{d \geq 2}\left(\delta G^{(2d),\,\mathrm{cont}}\right)^{\otimes n_d}\right)\left(\frac{\alpha}{p + 1/2},\,\bm{\xi},\,\bm{\xi}\right)\left(\mathcal{G}^{\mathrm{cont}}\right)^{\otimes (D - 1)}\left(\bm{\xi}\right)
\end{align}
We start by bounding term $A$. We first focus on a single term $\bm{\alpha}_{\mathcal{D}(\mathcal{M})}$ of the sum defining $A$. For such a term, the uniform bounds and discretization bounds on $\bm{\delta G}$ (proposition \ref{prop:centered_g_correlations_bounds_and_continuum_approximation}), combined with lemma \ref{lemma:bound_variation_product} bounding the variation of a product, give:
\begin{align}
    & \left|\left[\pi_{l''',\,\mathcal{M}} \cdot \bigotimes_{d \geq 2}\bm{\delta G}^{(2d)\otimes n_d}\right]_{\alpha,\,\bm{\alpha}_{\mathcal{D}(\mathcal{M})},\,\bm{\alpha}_{\mathcal{D}(\mathcal{M})}}\left[\bm{\mathcal{G}}^{\otimes (D - 1)}\right]_{\bm{\alpha}_{\mathcal{D}(\mathcal{M})}}\right.\nonumber\\
    & \hspace*{20px} - \left(\pi_{l''',\,\mathcal{M}} \cdot \bigotimes_{d \geq 2}\left(\delta G^{(2d),\,\mathrm{cont}}\right)^{\otimes n_d}\right)\left(\frac{\alpha}{p + 1/2},\,\frac{\bm{\alpha}_{\mathcal{D}(\mathcal{M})}}{p + 1/2},\,\frac{\bm{\alpha}_{\mathcal{D}(\mathcal{M})}}{p + 1/2}\right)\left(\mathcal{G}^{\mathrm{cont}}\right)^{\otimes (D - 1)}\left(\frac{\bm{\alpha}_{\mathcal{D}(\mathcal{M})}}{p + 1/2}\right)\vast|\nonumber\\
    & \leq \sum_{d \geq 2}n_d\frac{3}{2}2^{\sum\limits_{d \geq 2}n_d}3^{D}\frac{d\gamma'_{\mathrm{max}}}{p + 1}\nonumber\\
    & \leq \sum_{d \geq 2}n_d\frac{3}{2}2^{\sum\limits_{d \geq 2}dn_d/2}3^{D}\frac{d\gamma'_{\mathrm{max}}}{p + 1}\nonumber\\
    & = \frac{3}{2}\left(3\sqrt{2}\right)^D\sum_{d \geq 2}n_d\frac{d\gamma'_{\mathrm{max}}}{p + 1}\nonumber\\
    & = \frac{3}{2}\left(3\sqrt{2}\right)^DD\frac{\gamma'_{\mathrm{max}}}{p + 1}.
\end{align}
Multiplying this by the number terms in the sum defining $A$, as well as prefactor $\lambda^D$, gives the following bound on $A$:
\begin{align}
    \left|A\right| & \leq \frac{3}{2}\left(3\sqrt{2}\right)^DD\frac{\gamma'_{\mathrm{max}}}{p + 1}\lambda^D|\mathcal{A}|^{|\mathcal{D}(\mathcal{M})|}\nonumber\\
    & = \frac{3}{2}\left(3\sqrt{2}\right)^DD\frac{\gamma'_{\mathrm{max}}}{p + 1}\frac{\left(2^{-1/2}\gamma_{\mathrm{max}}\right)^D}{\left(p + 1\right)^D}\left(\left(2p + 2\right)^2\right)^{(D - 1)/2}\nonumber\\
    & = \frac{3}{4}\frac{\gamma'_{\mathrm{max}}}{\left(p + 1\right)^2}\left(6\gamma_{\mathrm{max}}\right)^DD
\end{align}

We then bound $B$. For that purpose, we observe the $\lambda^D$ scaling before the sum:
\begin{align}
    \lambda^D & = \frac{\left(2^{-1/2}\gamma_{\mathrm{max}}\right)^D}{\left(p + 1\right)^D} = \frac{\left(2^{-1/2}\gamma_{\mathrm{max}}\right)^D}{p + 1}\frac{1}{\left(p + 1\right)^{(D - 1)/2}}
\end{align}
has the correct dimensional scaling in $(p + 1)$ to apply sum-integral approximation lemma \ref{lemma:riemann_sum_approximation}. The lemma is applied to list of functions
\begin{gather}
    \delta G^{(2d),\,\mathrm{cont}} \textrm{ ($n_d$ times)}, \quad \mathcal{G}^{\mathrm{cont}} \textrm{ ($(D - 1)$ times)}
\end{gather}
The sum and integral are over
\begin{align}
    2(D - 1) & = 2D - 2
\end{align}
scalar variables, where the $2$ prefactor accounts for the fact each $\xi_{l'} \in [0, 2]^2, l' \in \mathcal{D}(\mathcal{M})$, consists of two scalar variables. The sum of variable dimensions in these functions is
\begin{align}
    2\left(\sum_{d \geq 2}dn_d + (D - 1) - 1\right) & = 4D - 4,
\end{align}
where the final $-1$ inside the parenthesis accounts for fixed $\alpha$. Next (see proposition \ref{prop:centered_g_correlations_bounds_and_continuum_approximation}), uniform bounds on these functions are given by
\begin{gather}
    2.3^d \textrm{ ($n_d$ times)}, \qquad 1 \textrm{ ($(D - 1)$ times)},
\end{gather}
and Lipschitz constants (in each variable) are given by
\begin{gather}
    4.3^d\beta_{\mathrm{max}} \textrm{ ($n_d$ times)}, \quad \frac{M_{\gamma}}{\gamma_{\mathrm{max}}} \textrm{ ($(D - 1)$ times)}.
\end{gather}
Lemma \ref{lemma:riemann_sum_approximation} then produces bound
\begin{align}
    \left|B\right| & \leq \frac{\left(2^{-1/2}\gamma_{\mathrm{max}}\right)^D}{p + 1}\frac{2^{2D - 1}}{p + 1}(4D - 4)2^{\sum\limits_{d \geq 2}n_d}3^D\max\left(2\beta_{\mathrm{max}}, \frac{M_{\gamma}}{\gamma_{\mathrm{max}}}\right)
\end{align}
This can in turn be weakened to
\begin{align}
    \left|B\right| & \leq \frac{1}{\left(p + 1\right)^2}2^{1 + \sum\limits_{d \geq 2}n_d}\left(6\sqrt{2}\gamma_{\mathrm{max}}\right)^D(D - 1)\max\left(2\beta_{\mathrm{max}}, \frac{M_{\gamma}}{\gamma_{\mathrm{max}}}\right)\nonumber\\
    & \leq \frac{2}{\left(p + 1\right)^2}2^{D/2}\left(6\sqrt{2}\gamma_{\mathrm{max}}\right)^D(D - 1)\max\left(2\beta_{\mathrm{max}}, \frac{M_{\gamma}}{\gamma_{\mathrm{max}}}\right)\nonumber\\
    & \leq \frac{2}{\left(p + 1\right)^2}\left(12\gamma_{\mathrm{max}}\right)^DD\max\left(2\beta_{\mathrm{max}}, \frac{M_{\gamma}}{\gamma_{\mathrm{max}}}\right).
\end{align}
Combining and weakening bounds on $A, B$ we arrive at
\begin{align}
    \left|\nu_{\alpha}^{\left(n_d\right)_{d \geq 2},\,l''',\,\mathcal{M}} - \frac{1}{p + 1}\widetilde{\nu}^{\left(n_d\right)_{d \geq 2},\,l''',\,\mathcal{M},\,\mathrm{cont}}\left(\frac{\alpha}{p + 1/2}\right)\right| & = \left|A + B\right|\nonumber\\
    & \leq |A| + |B|\nonumber\\
    & \leq \frac{\left(12\gamma_{\mathrm{max}}\right)^DD}{(p + 1)^2}\max\left(4\beta_{\mathrm{max}}, \frac{2M_{\gamma}}{\gamma_{\mathrm{max}}}, \gamma'_{\mathrm{max}}\right),
\end{align}
which is equation \ref{eq:qgms_integral_series_expansion_multinomial_matching_contribution_tilde_approximation_error}.
\end{proof}
\end{proposition}

Proposition \ref{prop:qgms_integral_series_expansion_multinomial_matching_contribution_continuum_approximation} then established uniform bounds and a continuum approximation for the contribution $\nu^{\left(n_d\right)_{d \geq 2},\,l''',\,\mathcal{M}}_{\alpha}$ to the QGMS moment, parametrized by multinomial numbers $\left(n_d\right)_{d \geq 2}$, an integer $l''' \in [D]$ and a matching $\mathcal{M}$ of $[D] - \{l'''\}$, where $D := \sum_{d \geq 2}dn_d$. Recalling equations \ref{eq:qgms_integral_series_expansion_as_sum_multinomial_contributions}, \ref{eq:qgms_integral_series_expansion_multinomial_contribution}, \ref{eq:qgms_integral_series_expansion_multinomial_matching_contribution}, the desired moment $\nu_{\alpha}$ is given by summation of these contributions over matchings $\left(l''', \mathcal{M}\right)$, then over multinomial numbers $\left(n_d\right)_{d \geq 2}$. From the bounds in proposition \ref{prop:qgms_integral_series_expansion_multinomial_matching_contribution_continuum_approximation}, we then deduce bounds on these partial sums. For that purpose, it will help to introduce additional preliminary notation for the partial sums

\begin{definition}[Contributions to QGMS moment, with $\bm{G}^{(2)}$ and $\bm{\mathcal{G}}$ factored out]
\label{def:qgms_integral_series_expansion_tilde_quantities}
In the discrete case, define vectors $\bm{\widetilde{\nu}}^{\left(n_d\right)_{d \geq 2}}$ and $\bm{\widetilde{\nu}}$ by the following equations:
\begin{align}
    \bm{\widetilde{\nu}}^{\left(n_d\right)_{d \geq 2}} & := \left(\widetilde{\nu}^{\left(n_d\right)_{d \geq 2}}_{\alpha}\right)_{\alpha \in \mathcal{A}},\\
    \bm{\widetilde{\nu}} & := \left(\widetilde{\nu}_{\alpha}\right)_{\alpha \in \mathcal{A}},\\
    \widetilde{\nu}_{\alpha} & := \sum_{\left(n_d\right)_{d \geq 2}}\widetilde{\nu}^{\left(n_d\right)_{d \geq 2}}_{\alpha},\label{eq:qgms_integral_series_expansion_as_sum_multinomial_contributions_tilde}\\
    \widetilde{\nu}^{\left(n_d\right)_{d \geq 2}}_{\alpha} & := \binom{n}{\left(n_d\right)_{d \geq 2}}\frac{n^{-D/2}}{\prod\limits_{d \geq 2}d!^{n_d}}\sum_{l''' \in [D]}\hspace*{5px}\sum_{\mathcal{M}\mathrm{\,matching\,of\,} [D] - \{l'''\}}\widetilde{\nu}^{\left(n_d\right)_{d \geq 2},\,l''',\,\mathcal{M}}_{\alpha}.\label{eq:qgms_integral_series_expansion_multinomial_contribution_tilde}
\end{align}
Comparing these definitions with equations \ref{eq:qgms_integral_series_expansion_as_sum_multinomial_contributions}, \ref{eq:qgms_integral_series_expansion_multinomial_contribution}, it holds
\begin{align}
    \nu_{k_1,\,k_2} = G^{\left(2\right)}_{k_1,\,k_2}\mathcal{G}_{k_1}\mathcal{G}_{k_2}\widetilde{\nu}_{k_1,\,k_2}, && \nu^{\left(n_d\right)_{d \geq 2}}_{k_1,\,k_2} = G^{\left(2\right)}_{k_1,\,k_2}\mathcal{G}_{k_1}\mathcal{G}_{k_2}\widetilde{\nu}^{\left(n_d\right)_{d \geq 2}}_{k_1,\,k_2} .
\end{align}
Similarly, from the continuum analogue $\widetilde{\nu}^{\left(n_d\right)_{d \geq 2},\,l''',\,\mathcal{M},\,\mathrm{cont}}$ of $\bm{\widetilde{\nu}}^{\left(n_d\right)_{d \geq 2},\,l''',\,\mathcal{M}}$ introduced in definition \ref{def:centered_g_correlations_tensor_continuum}, we define functions $\widetilde{\nu}^{\left(n_d\right)_{d \geq 2},\,\mathrm{cont}}$ and $\widetilde{\nu}^{\mathrm{cont}}$ by:
\begin{align}
    \widetilde{\nu}^{\mathrm{cont}}: & \quad [0, 2]^2 \longrightarrow \mathbf{C},\\
    \widetilde{\nu}^{\left(n_d\right)_{d \geq 2},\,\mathrm{cont}}: & \quad [0, 2]^2 \longrightarrow \mathbf{C},\\
    \widetilde{\nu}^{\mathrm{cont}}\left(x_1, x_2\right) & := \sum_{\left(n_d\right)_{d \geq 2}}\widetilde{\nu}^{\left(n_d\right)_{d \geq 2},\,\mathrm{cont}}\left(x_1, x_2\right),\label{eq:qgms_integral_series_expansion_as_sum_multinomial_contributions_tilde_continuum}\\
    \widetilde{\nu}^{\left(n_d\right)_{d \geq 2},\,\mathrm{cont}}\left(x_1, x_2\right) & := \binom{n}{\left(n_d\right)_{d \geq 2}}\frac{n^{-D/2}}{\prod\limits_{d \geq 2}d!^{n_d}}\sum_{l''' \in [D]}\hspace*{5px}\sum_{\mathcal{M}\mathrm{\,matching\,of\,} [D] - \{l'''\}}\widetilde{\nu}^{\left(n_d\right)_{d \geq 2},\,l''',\,\mathcal{M},\,\mathrm{cont}}\left(x_1, x_2\right).\label{eq:qgms_integral_series_expansion_multinomial_contribution_tilde_continuum}
\end{align}
In all these definitions, we used notation
\begin{align}
    D & := \sum_{d \geq 2}dn_d
\end{align}
as an implicit function of $\left(n_d\right)_{d \geq 2}$ for brevity.
\end{definition}

\begin{corollary}[Uniform and discretization bounds on QGMS moments]
\label{cor:qgms_integral_series_expansion_contributions_continuum_approximation}
For all multinomial numbers $\left(n_d\right)$, the following uniform bounds hold on the discrete and continuum versions of the moment contribution indexed by $\left(n_d\right)_{d \geq 2}$:
\begin{align}
    \left\lVert \bm{\widetilde{\nu}}^{\left(n_d\right)_{d \geq 2}} \right\rVert_{\infty} & \leq D!!\frac{n^{-D/2}}{\prod\limits_{d \geq 2}d!^{n_d}}\binom{n}{\left(n_d\right)_{d \geq 2}}\frac{\left(3\sqrt{2}\gamma_{\mathrm{max}}\right)^D}{2p + 2}2^{\sum\limits_{d \geq 2}n_d},\label{eq:qgms_integral_series_expansion_multinomial_contribution_tilde_bound}\\
    \left\lVert \widetilde{\nu}^{\left(n_d\right)_{d \geq 2},\,\mathrm{cont}} \right\rVert_{\infty} & \leq D!!\frac{n^{-D/2}}{\prod\limits_{d \geq 2}d!^{n_d}}\binom{n}{\left(n_d\right)_{d \geq 2}}\frac{\left(3\sqrt{2}\gamma_{\mathrm{max}}\right)^D}{2}2^{\sum\limits_{d \geq 2}n_d},\label{eq:qgms_integral_series_expansion_multinomial_contribution_tilde_continuum_bound}
\end{align}
and $\widetilde{\nu}^{\left(n_d\right)_{d \geq 2},\,\mathrm{cont}}$ is Lipschitz in each of its (two) variables, with Lipschitz constant bounded by
\begin{align}
    D!!\frac{n^{-D/2}}{\prod\limits_{d \geq 2}\left(d!/2\right)^{n_d}}\binom{n}{\left(n_d\right)_{d \geq 2}}\left(3\sqrt{2}\gamma_{\mathrm{max}}\right)^D\beta_{\mathrm{max}}.\label{eq:qgms_integral_series_expansion_multinomial_contribution_tilde_continuum_lipschitz_constant}
\end{align}
Besides, the following discretization bound holds between this discrete vector and continuous function:
\begin{align}
    \left|\widetilde{\nu}^{\left(n_d\right)_{d \geq 2}}_{\alpha} - \frac{1}{p + 1}\widetilde{\nu}^{\left(n_d\right)_{d \geq 2},\,\mathrm{cont}}\left(\frac{\alpha}{p + 1/2}\right)\right| & \leq D!!\frac{n^{-D/2}}{\prod\limits_{d \geq 2}d!^{n_d}}\binom{n}{\left(n_d\right)_{d \geq 2}}\frac{\left(12\gamma_{\mathrm{max}}\right)^DD}{\left(p + 1\right)^2}\max\left(4\beta_{\mathrm{max}}, \frac{2M_{\gamma}}{\gamma_{\mathrm{max}}}, \gamma'_{\mathrm{max}}\right).
\end{align}
for all $\alpha \in \mathcal{A}$. Next, the following uniform bounds hold on the discrete and continuum moments
\begin{align}
    \left\lVert \bm{\widetilde{\nu}} \right\rVert_{\infty} & \leq \frac{\mathcal{O}(1)}{p + 1},\label{eq:qgms_integral_series_expansion_as_sum_multinomial_contributions_tilde_bound}\\
    \left\lVert \widetilde{\nu}^{\mathrm{cont}} \right\rVert_{\infty} & \leq \mathcal{O}(1),\label{eq:qgms_integral_series_expansion_as_sum_multinomial_contributions_tilde_continuum_bound}
\end{align}
and $\widetilde{\nu}^{\mathrm{cont}}$ is $\mathcal{O}(1)\gamma_{\mathrm{max}}^2\beta_{\mathrm{max}}$-Lipschitz in each of its (two) variables. Finally, the following discretization bound holds between the discrete vector and continuous function:
\begin{align}
    \left|\widetilde{\nu}_{\alpha} - \frac{1}{p + 1}\widetilde{\nu}^{\mathrm{cont}}\left(\frac{\alpha}{p + 1/2}\right)\right| & \leq \frac{\mathcal{O}(1)}{\left(p + 1\right)^2}\gamma_{\mathrm{max}}^2\max\left(4\beta_{\mathrm{max}}, \frac{2M_{\gamma}}{\gamma_{\mathrm{max}}}, \gamma'_{\mathrm{max}}\right)
\end{align}
These bounds hold for $\gamma_{\mathrm{max}}$ smaller than an absolute constant. Likewise, $\mathcal{O}(1)$ in these bounds refers to absolute constants for brevity. 
\begin{proof}
Bounds \ref{eq:qgms_integral_series_expansion_multinomial_contribution_tilde_bound}, \ref{eq:qgms_integral_series_expansion_multinomial_contribution_tilde_continuum_bound} follow from counting pairs $\left(l''', \mathcal{M}\right)$ as $D!! = 2^{-D - 1}(2D + 2)!/(D + 1)!^2$, as well as uniform bounds \ref{eq:qgms_integral_series_expansion_multinomial_matching_contribution_tilde_bound}, \ref{eq:qgms_integral_series_expansion_multinomial_matching_contribution_tilde_continuum_bound} on $\bm{\widetilde{\nu}}^{\left(n_d\right)_{d \geq 2},\,l''',\,\mathcal{M}}$, $\widetilde{\nu}^{\left(n_d\right)_{d \geq 2},\,l''',\,\mathcal{M},\,\mathrm{cont}}$. The Lipschitz constant claimed in equation \ref{eq:qgms_integral_series_expansion_multinomial_contribution_tilde_continuum_lipschitz_constant} for $\widetilde{\nu}^{\left(n_d\right)_{d \geq 2}\,\mathrm{cont}}$ follows from that on $\widetilde{\nu}^{\left(n_d\right)_{d \geq 2}\,l''',\,\mathcal{M},\,\mathrm{cont}}$ (proposition \ref{prop:qgms_integral_series_expansion_multinomial_matching_contribution_continuum_approximation}) and the same combinatorics. Likewise, the discretization error for $\widetilde{\nu}^{\left(n_d\right)_{d \geq 2}}$ claimed in equation \ref{eq:qgms_integral_series_expansion_multinomial_matching_contribution_tilde_approximation_error} follows from discretization bound \ref{eq:qgms_integral_series_expansion_multinomial_matching_contribution_tilde_approximation_error} for $\widetilde{\nu}^{\left(n_d\right)_{d \geq 2}\,l''',\,\mathcal{M},\,\mathrm{cont}}$.

The uniform, Lipschitz, and discretization bounds on $\bm{\widetilde{\nu}}$ and $\widetilde{\nu}^{\mathrm{cont}}$ then follow from summing the previous bounds over $\left(n_d\right)_{d \geq 2}$, i.e. applying the triangular inequality to sums over multinomial numbers $\left(n_d\right)_{d \geq 2}$. The summation of over multinomial numbers is then bounded using Lemma~\ref{lemma:qgms_integral_moment_series_multinomial_sum_bound}. Let us, for instance, consider the case of the discretization bound.
\begin{align}
    & \left|\widetilde{\nu}_{\alpha} - \frac{1}{p + 1}\widetilde{\nu}^{\mathrm{cont}}\left(\frac{\alpha}{p + 1/2}\right)\right|\nonumber\\
    & \leq \sum_{\left(n_d\right)_{d \geq 2}}\mathbf{1}\left[2\,|\,D + 1\right]\left|\widetilde{\nu}^{\left(n_d\right)_{d \geq 2}}_{\alpha} - \widetilde{\nu}^{\left(n_d\right)_{d \geq 2},\,\mathrm{cont}}\left(\frac{\alpha}{p + 1/2}\right)\right|\nonumber\\
    & \leq \sum_{\left(n_d\right)_{d \geq 2}}\mathbf{1}\left[2\,|\,D + 1\right]D!!\frac{n^{-D/2}}{\prod\limits_{d \geq 2}\left(d!/2\right)^{n_d}}\binom{n}{\left(n_d\right)_{d \geq 2}}\frac{\left(12\gamma_{\mathrm{max}}\right)^DD}{\left(p + 1\right)^2}\max\left(4\beta_{\mathrm{max}}, \frac{2M_{\gamma}}{\gamma_{\mathrm{max}}}, \gamma'_{\mathrm{max}}\right)\nonumber\\
    & \leq \sum_{\left(n_d\right)_{d \geq 2}}\mathbf{1}\left[2\,|\,D + 1\right]D!!\frac{2^{D/2}n^{-D/2}}{\prod\limits_{d \geq 2}d!^{n_d}}\binom{n}{\left(n_d\right)_{d \geq 2}}\frac{\left(12\gamma_{\mathrm{max}}\right)^DD}{\left(p + 1\right)^2}\max\left(4\beta_{\mathrm{max}}, \frac{2M_{\gamma}}{\gamma_{\mathrm{max}}}, \gamma'_{\mathrm{max}}\right)\nonumber\\
    & = \sum_{\left(n_d\right)_{d \geq 2}}D!!\frac{2^{D/2}n^{-D/2}}{\prod\limits_{d \geq 2}d!^{n_d}}\binom{n}{\left(n_d\right)_{d \geq 2}}\frac{\left(12\gamma_{\mathrm{max}}\right)^DD}{\left(p + 1\right)^2}\max\left(4\beta_{\mathrm{max}}, \frac{2M_{\gamma}}{\gamma_{\mathrm{max}}}, \gamma'_{\mathrm{max}}\right).
\end{align}
We then invoke Lemma~\ref{lemma:qgms_integral_moment_series_multinomial_sum_bound} to conclude the above is upper-bounded by
\begin{align}
    \frac{\mathcal{O}(1)}{(p + 1)^2}\gamma_{\mathrm{max}}^2\max\left(4\beta_{\mathrm{max}}, \frac{2M_{\gamma}}{\gamma_{\mathrm{max}}}, \gamma'_{\mathrm{max}}\right).
\end{align}
\end{proof}
\end{corollary}

\section{Analysis of additional QGMS moments contributions and concentration of energy}
\label{sec:other_moments_concentration}

In this Appendix, we complement the proof of our main result Theorem~\ref{th:approximation_continuous_time_annealing_qaoa}, as partially carried out in Appendix~\ref{sec:sk_qaoa_energy_continuum_limit}. More specifically, Appendix~\ref{sec:sk_qaoa_energy_continuum_limit} analyzed a single additive contribution to the SK-QAOA disorder-averaged energy by introducing various techniques. The present Appendix sketches how the same techniques can be generalized to the other additive contributions to the disorder-averaged QAOA expected cost $\mathbb{E}\bra{\Psi_{p,\,n}}C_n/n\ket{\Psi_{p,\,n}}$. Still using the same techniques, the Appendix also considers the average of the square cost: $\mathbb{E}\bra{\Psi_{p,\,n}}\left(C_n/n\right)^2\ket{\Psi_{p,\,n}}$, with the goal of establishing concentration of the expected QAOA cost according to the Gaussian disorder. Concentration is non-trivial in this setting: while it was proven unconditionally at fixed $p$ and in the $n \to \infty$ limit in Ref.~\cite{qaoa_sk}, in the present work $p$ is sent to infinity while $n$ is fixed, and we require uniform bounds in this limit.

\subsection{Reminders and additional notation}

In Section~\ref{sec:sk_qaoa_first_order_moment_qgms}, a representation of the disorder-averaged SK-QAOA expected cost function in terms of QGMS moments was established:
\begin{align}
    \mathbb{E}\bra{\Psi_{p,\,n}}C_n/n\ket{\Psi_{p,\,n}} & = -\frac{i}{\Gamma_{p + 1}}\sum_{0 \leq r \leq 2p + 1}\frac{\partial^2S_n\left(\bm{\mu}\right)}{\partial\mu_{(p + 1, r)}^2}\Bigg|_{\bm{\mu} = \bm{0}}.
\end{align}
Reexpressing second order moments in terms of QGMS integral moments tensors (Eq.~\ref{eq:qgms_second_order_moment_from_qgms_integral_moments}) yielded:
\begin{align}
    \frac{\partial^2S_n\left(\bm{\mu}\right)}{\partial\mu_{\alpha}^2}\Bigg|_{\bm{\mu} = \bm{0}} & = \nu^{(0)}_{\alpha} + \frac{2}{\sqrt{n}}\nu^{(1)}_{\alpha} + \frac{1}{n}\nu^{(2)}_{\alpha},
\end{align}
where
\begin{align}
    \nu^{(0)}_{\alpha} & := \left(\theta^*_{\alpha}\right)^2,\\
    \nu^{(1)}_{\alpha} & := \theta^*_{\alpha}\left[\bm{S}_n^{(1)}\right]_{\alpha},\\
    \nu^{(2)}_{\alpha} & := \left[\bm{S}_n^{(2)}\right]_{\alpha,\,\alpha} - 1.
\end{align}
In Appendix Section~\ref{sec:sk_qaoa_energy_continuum_limit}, dedicated to the proof of Theorem~\ref{th:approximation_continuous_time_annealing_qaoa}, we analyzed in detail a single additive contribution to the QGMS second order moment, hence to the disorder-averaged expected SK-QAOA energy:
\begin{align}
    \frac{\partial^2S_n\left(\bm{\mu}\right)}{\partial\mu_{\alpha}^2}\Bigg|_{\bm{\mu} = \bm{0}} & \supset \frac{2}{\sqrt{n}}\nu^{(1)}_{\alpha}.
\end{align}
In that context, we denoted $\nu_{\alpha}$ for $\nu^{(1)}_{\alpha}$ for brevity. Bounds and a continuum approximation for vector $\bm{\widetilde{\nu}}^{(1)} = \left(\widetilde{\nu}^{(1)}_{\alpha}\right)_{\alpha \in \mathcal{A}}$ were ultimately established in Proposition~\ref{prop:centered_g_correlations_bounds_and_continuum_approximation}, where $\nu^{(1)}$ is related to $\widetilde{\nu}^{(1)}$ by elementwise multiplication:
\begin{align}
    \nu^{(1)}_{\alpha} & := \mathcal{G}_{\alpha}\widetilde{\nu}^{(1)}_{\alpha},
\end{align}
where $\bm{\mathcal{G}} = \left(\mathcal{G}_{\alpha}\right)_{\alpha \in \mathcal{A}}$ was introduced in Eq.~\ref{eq:curvy_g_vector_definition_rescaled}. In this Section, we sketch how a similar analysis would work for other contributions to the second order QGMS moment:
\begin{align}
    \frac{\partial^2S_n\left(\bm{\mu}\right)}{\partial\mu_{\alpha}^2}\Bigg|_{\bm{\mu} = \bm{0}} \supset \nu^{(0)}_{\alpha}, && \frac{\partial^2S_n\left(\bm{\mu}\right)}{\partial\mu_{\alpha}^2}\Bigg|_{\bm{\mu} = \bm{0}} \supset \frac{1}{n}\nu^{(2)}_{\alpha},
\end{align}
We also sketch the analysis for the contributions to the partially diagonal quartic QGMS moment, from which the disorder-average of the expected square cost $\mathbb{E}\bra{\Psi_{p,\,n}}\left(C/n\right)^2\ket{\Psi_{p,\,n}}$ can be expressed according to Section~\ref{sec:sk_qaoa_second_order_moment_qgms}. The contributions to the quartic moments read:
\begin{align}
    \frac{\partial^4S_n\left(\bm{\mu}\right)}{\partial\mu_{\alpha}^2\partial\mu_{\beta}^2}\Bigg|_{\bm{\mu} = \bm{0}} & = \frac{1}{n^2}\nu^{(7)}_{\alpha,\,\beta} + \frac{1}{n^{3/2}}\nu^{(6)}_{\alpha,\,\beta} + \frac{1}{n}\nu^{(5)}_{\alpha,\,\beta} + \frac{1}{\sqrt{n}}\nu^{(4)}_{\alpha,\,\beta} + \nu^{(3)}_{\alpha,\,\beta},
\end{align}
where
\begin{align}
    \nu^{(7)}_{\alpha,\,\beta} & := 1 + 2\delta_{\alpha\beta} - 4\delta_{\alpha\beta}\left[\bm{S}^{(2)}_n\right]_{\alpha,\,\alpha} - \left[\bm{S}^{(2)}_n\right]_{\alpha,\,\alpha} - \left[\bm{S}^{(2)}_n\right]_{\beta,\,\beta} + \left[\bm{S}^{(4)}_n\right]_{\alpha,\,\alpha,\,\beta,\,\beta},\\
    \nu^{(6)}_{\alpha,\,\beta} & := -8\delta_{\alpha\beta}\theta^*_{\alpha}\left[\bm{S}^{(1)}_n\right]_{\alpha} - 2\theta^*_{\alpha}\left[\bm{S}^{(1)}_n\right]_{\alpha} - 2\theta^*_{\beta}\left[\bm{S}^{(1)}_n\right]_{\beta} + 2\theta^*_{\alpha}\left[\bm{S}^{(3)}_n\right]_{\alpha,\,\beta,\,\beta} + 2\theta^*_{\beta}\left[\bm{S}^{(3)}_n\right]_{\beta,\,\alpha,\,\alpha},\\
    \nu^{(5)}_{\alpha,\,\beta} & := -4\delta_{\alpha,\,\beta}\left(\theta^*_{\alpha}\right)^2 - \left(\theta^*_{\alpha}\right)^2 - \left(\theta^*_{\beta}\right)^2 + 4\theta^*_{\alpha}\theta^*_{\beta}\left[\bm{S}^{(2)}_n\right]_{\alpha,\,\beta} + \left(\theta^*_{\alpha}\right)^2\left[\bm{S}^{(2)}\right]_{\beta,\,\beta} + \left(\theta^*_{\beta}\right)^2\left[\bm{S}^{(2)}_n\right]_{\alpha,\,\alpha},\\
    \nu^{(4)}_{\alpha,\,\beta} & := 2\left(\theta^*_{\beta}\right)^2\theta^*_{\alpha}\left[\bm{S}^{(1)}_n\right]_{\alpha} + 2\left(\theta^*_{\alpha}\right)^2\theta^*_{\beta}\left[\bm{S}^{(1)}_n\right]_{\beta},\\
    \nu^{(3)}_{\alpha,\,\beta} & = \left(\theta^*_{\alpha}\right)^2\left(\theta^*_{\beta}\right)^2.
\end{align}
At a high level, the important point is to show that when summed over $\alpha$ (resp. $\alpha, \beta$) to give an additive contribution to the disorder-averaged expected cost $\mathbb{E}\bra{\Psi_{p,\,n}}C_n/n\ket{\Psi_{p,\,n}}$ (resp. to the disorder-averaged expected cost squared $\mathbb{E}\bra{\Psi_{p,\,n}}\left(C_n/n\right)^2\ket{\Psi_{p,\,n}}$), each of these contributions remains bounded as $p \to \infty$. This in turns requires to identify the correct scaling of each $\nu^{(k)}_{\alpha}$, $0 \leq k \leq 2$ (resp. each $\nu^{(k)}_{\alpha,\,\beta}$, $3 \leq k \leq 7$) in $p$.

The analysis proceeds along similar line to that of $\bm{\nu}^{(1)}$. In particular, the starting point is a series expansion of this quantity in terms of centered correlations tensors. In the case of $\bm{\nu}^{(1)}$, the expansion read:
\begin{align}
    \nu^{(1)}_{\alpha} & = \left(\mathcal{Z}^*\right)^n\sum_{\left(n_d\right)_{d \geq 2}}\binom{n}{\left(n_d\right)_{d \geq 2}}\frac{n^{-\sum\limits_{d \geq 2}dn_d/2}}{\prod\limits_{d \geq 2}d!^{n_d}}\theta^*_{\alpha}\left\langle \bm{\mathcal{I}}^{\left(1 + \sum\limits_{d \geq 2}dn_d\right)}, \bigotimes_{d \geq 2}\bm{\delta C}^{(d)\otimes n_d} \right\rangle,\label{eq:nu_1_series_expansion}
\end{align}
and followed from the general series expansion QGMS integral moments tensors stated in Proposition~\ref{prop:qgms_integral_series_expansion}:
\begin{align}
    \left[\bm{S}^{(k)}_n\right]_{\bm{\alpha}_{1:k}} & = \left(\mathcal{Z}^*\right)^n\sum_{\left(n_d\right)_{d \geq 2}}\binom{n}{\left(n_d\right)_{d \geq 2}}\frac{n^{-\sum\limits_{d \geq 2}dn_d/2}}{\prod\limits_{d \geq 2}d!^{n_d}}\left\langle \bm{\mathcal{I}}^{\left(k + \sum\limits_{d \geq 2}dn_d\right)}_{\bm{\alpha}_{1:k}}, \bigotimes_{d \geq 2}\bm{\delta C}^{(d)\otimes n_d} \right\rangle,\\
    \bm{\alpha}_{1:k} & = \left(\alpha_1, \ldots, \alpha_k\right) \in \mathcal{A}^k.
\end{align}
Applying this formula yields a series expansion of all $\nu^{(k)}_{\alpha}$ ($0 \leq k \leq 2$) and $\nu^{(k)}_{\alpha,\,\beta}$ ($3 \leq k \leq 7$) similar to Eq.~\ref{eq:nu_1_series_expansion}, where
\begin{align}
    \bm{\mathcal{I}}^{\left(1 + \sum\limits_{d \geq 2}dn_d\right)}_{\alpha}
\end{align}
on the left-hand side of the dot product is replaced by a linear combination of matching tensor slices. For instance,
\begin{align}
    \nu^{(2)}_{\alpha} & = \left(\mathcal{Z}^*\right)^n\sum_{\left(n_d\right)_{d \geq 2}}\binom{n}{\left(n_d\right)_{d \geq 2}}\frac{n^{-\sum\limits_{d \geq 2}dn_d/2}}{\prod\limits_{d \geq 2}d!^{n_d}}\left\langle \bm{\mathcal{I}}^{\left(2 + \sum\limits_{d \geq 2}dn_d\right)}_{\alpha,\,\alpha} - \bm{\mathcal{I}}^{\left(\sum\limits_{d \geq 2}dn_d\right)}, \bigotimes_{d \geq 2}\bm{\delta C}^{(d)\otimes n_d} \right\rangle.
\end{align}
It turns out these linear combinations of matching tensors slices will cancel in a convenient way. To express this, we introduce the following variant of the matching tensor, where external edges are allowed besides internally matched edges:

\begin{definition}[Matching tensor with external edges]
\label{def:matching_tensor_with_edges}
Let $D, k$ be non-negative integers and $\alpha_{D + 1}, \ldots, \alpha_{D + k} \in \mathcal{A}$ a list of $\mathcal{A}$ indices. Then, the matching tensor of degree $D$ with external edges $\alpha_{D + 1}, \ldots, \alpha_{D + k}$ is defined by:
\begin{align}
    \bm{\mathcal{I}}^{\left(D,\,\alpha_{D + 1},\,\ldots,\,\alpha_{D + k}\right)} & = \left(\mathcal{I}_{\alpha_1,\,\ldots,\,\alpha_D}^{\left(D,\,\alpha_{D + 1},\,\ldots,\,\alpha_{D + k}\right)}\right)_{\alpha_1,\,\ldots,\,\alpha_D \mathcal{A}} \in \left(\mathbf{C}^{\mathcal{A}}\right)^{\otimes d},\\
    \mathcal{I}^{\left(D,\,\alpha_{D + 1},\,\ldots,\,\alpha_{D + k}\right)}_{\alpha_1,\,\ldots,\,\alpha_D} & := \sum_{\substack{j_1,\,\ldots,\,j_k \in [D]\\j_1,\,\ldots,\,j_k\,\mathrm{pairwise\,distinct}}}\mathcal{I}^{\left(D - k\right)}_{\bm{\alpha}_{[D] - \{j_1,\,\ldots,\,j_k\}}}\prod_{1 \leq r \leq k}\mathbf{1}\left[\alpha_{D + r} = \alpha_{j_r}\right].
\end{align}
\end{definition}

Intuitively, the matching tensor with external edges introduced in Definition~\ref{def:matching_tensor_with_edges} enumerates the distinct ways of matching its $D$ indices with $k$ external indices $\alpha_{D + 1},\,\ldots,\,\alpha_{D + k}$ and within themselves for the remaining $D - k$ indices. Furthermore, by standard conventions on empty sums and products, $\bm{\mathcal{I}}^{(D,\,\alpha_{D + 1},\,\ldots,\,\alpha_{D + k})}$ is understood to vanish whenever $k > D$ or $(D - k)$ is odd. Finally, note that for $k = 0$, the definition coincides with that of the ordinary matching tensor $\bm{\mathcal{I}}^{(D)}$.

The following lemma allows to express a slice of a matching tensor in terms of matching tensors with external edges:

\begin{lemma}[Matching tensor slice from matching tensors with external edges]
\label{lemma:matching_tensor_slice_from_matching_tensor_with_edges}
Let $D, k \geq 0$ be non-negative integers and $\bm{\eta} := \bm{\eta}_{1:k} = \left(\eta_1, \ldots, \eta_k\right) \in \mathcal{A}^{k}$ a list of $k$ indices. Then, the following relation holds between slice $\bm{\alpha}$ of the matching tensor of degree $D$ (introduced in Eq.~\ref{eq:matching_tensor_definition}) and matching tensors with external edges (introduced in Definition~\ref{def:matching_tensor_with_edges}):
\begin{align}
    \bm{\mathcal{I}}^{(D + k)}_{\bm{\eta}_{1:k}} & = \sum_{S \subset [k]}\hspace*{5px}\sum_{\mathcal{M}\mathrm{\,matching\,of\,}S}\bm{\mathcal{I}}^{\left(D,\,\bm{\eta}_{[k]\backslash S}\right)}\prod_{\{a, b\} \in \mathcal{M}}\mathbf{1}\left[\eta_a = \eta_b\right].\label{eq:matching_tensor_slice_from_matching_tensor_with_edges}
\end{align}
\begin{proof}
We start with the definition of the matching tensor (Eq.~\ref{eq:matching_tensor_definition}). For all $\bm{\alpha} = \bm{\alpha}_{1:D + k} = \left(\alpha_1, \ldots, \alpha_{D + k}\right)$,
\begin{align}
    \mathcal{I}^{(D + k)}_{\bm{\alpha}_{1:D + k}} & = \sum_{\mathcal{M}\mathrm{\,matching\,of\,}[D + k]}\hspace*{5px}\prod_{\{a, b\} \in \mathcal{M}}\mathbf{1}\left[\alpha_a = \alpha_b\right].\label{eq:matching_tensor_definition_reproduced}
\end{align}
The above: entry $\bm{\alpha}_{1:D + k}$ of $\bm{\mathcal{I}}^{(D + k)}$ can also be seen as entry $\bm{\alpha}_{k + 1:D + k}$ of slice $\bm{\alpha}_{1:k}$ of the same tensor:
\begin{align}
    \mathcal{I}^{(D + k)}_{\bm{\alpha}_{1:D + k}} & = \left[\bm{\mathcal{I}}^{(D + k)}_{\bm{\alpha}_{1:k}}\right]_{\bm{\alpha}_{k + 1:D + k}},
\end{align}
which is up to index relabelling the quantity on the left-hand-side of Eq.~\ref{eq:matching_tensor_slice_from_matching_tensor_with_edges}. We now decompose each matching $\mathcal{M}$ of $[D + k]$ in the sum according to set partition $[D + k] = \{1,\,\ldots,\,k\} \sqcup \{k + 1,\,\ldots,\,D + k\}$; we distinguish between matching edges internal to each of these sets, and edges connecting these two sets. Explicitly:
\begin{align}
    \mathcal{M} & = \mathcal{M}' \sqcup \mathcal{M}'' \sqcup \mathcal{M}''',\\
    \mathcal{M}' & := \left\{\{a, b\}\,:\,\{a, b\} \in \mathcal{M},\,a \in \{1, \ldots, k\},\,b \in \{1,\,\ldots,\,k\}\right\},\label{eq:matching_set1_intra_edges}\\
    \mathcal{M}'' & := \left\{\{a, b\}\,:\,\{a, b\} \in \mathcal{M},\,a \in \{k + 1, \ldots, D + k\},\,b \in \{k + 1,\,\ldots,\,D + k\}\right\},\label{eq:matching_set2_intra_edges}\\
    \mathcal{M}''' & := \left\{\{a, b\}\,:\,\{a, b\} \in \mathcal{M},\,a \in \{1,\,\ldots,\,k\},\,b \in \{k + 1,\,\ldots,\,D + k\}\right\}.\label{eq:matching_inter_edges}
\end{align}
This decomposition suggests an alternative combinatorial procedure of enumerating matchings $\mathcal{M}$:
\begin{itemize}
    \item Pick a subset $S' \subset \{1,\,\ldots,\,k\}$.
    \item Pick a subset $S'' \subset \{k + 1,\,\ldots,\,D + k\}$ of same size as $S'$.
    \item Given subsets $S', S''$, independently choose:
    \begin{itemize}
        \item A matching $\mathcal{M}'$ of set $\{1,\,\ldots,\,k\} \backslash S'$.
        \item A matching $\mathcal{M}''$ of set $\{k + 1,\,\ldots,\,D + k\} \backslash S''$.
        \item A one-to-one mapping $\sigma$ from set $S'$ to set $S''$.
    \end{itemize}
\end{itemize}
If it easily seen that valid choices $\left(S', S'', \mathcal{M}', \mathcal{M}'', \sigma\right)$ in this construction are in one-to-one correspondence with matchings $\mathcal{M}$. Namely, $\mathcal{M}', \mathcal{M}''$ directly give the sub-matchings of same name defined by Eqs.~\ref{eq:matching_set1_intra_edges}, \ref{eq:matching_set2_intra_edges}, while sub-matching $\mathcal{M}'''$ defined in Eq.~\ref{eq:matching_inter_edges} is related to $\sigma$ by:
\begin{align}
    \mathcal{M}''' & \longleftarrow \left\{\{a, \sigma(a)\}\,:\,a \in S'\right\}.
\end{align}
Informally, we will write
\begin{align}
    \mathcal{M} & \longleftarrow \mathcal{M}' \sqcup \mathcal{M}'' \sqcup \sigma.
\end{align}
Rewriting Eq.~\ref{eq:matching_tensor_definition_reproduced} in terms of this parametrization of matchings:
\begin{align}
    \mathcal{I}^{\left(D + k\right)}_{\bm{\alpha}_{1:D + k}} & = \sum_{\mathcal{M}}\prod_{\{a, b\} \in \mathcal{M}}\mathbf{1}\left[\alpha_a = \alpha_b\right]\nonumber\\
    & = \sum_{S',\,S''\,\mathcal{M}',\,\mathcal{M}'',\,\sigma}\hspace*{5px}\prod_{\{a, b\} \in \left(\mathcal{M}' \sqcup \mathcal{M}'' \sqcup \sigma\right)}\mathbf{1}\left[\alpha_a = \alpha_b\right]\nonumber\\
    & = \sum_{S',\,S''\,\mathcal{M}',\,\mathcal{M}'',\,\sigma}\hspace*{5px}\prod_{\{a, b\} \in \mathcal{M}'}\mathbf{1}\left[\alpha_a = \alpha_b\right]\prod_{\{a, b\} \in \mathcal{M}''}\mathbf{1}\left[\alpha_a = \alpha_b\right]\prod_{a \in S'}\mathbf{1}\left[\alpha_a = \alpha_{\sigma(a)}\right]\nonumber\\
    & = \sum_{S',\,\mathcal{M}'}\hspace*{5px}\left(\prod_{\{a, b\} \in \mathcal{M}'}\mathbf{1}\left[\alpha_a = \alpha_b\right]\right)\sum_{S''}\sum_{\sigma}\left(\prod_{a \in S'}\mathbf{1}\left[\alpha_a = \alpha_{\sigma(a)}\right]\right)\sum_{\mathcal{M}''}\hspace*{5px}\prod_{\{a, b\} \in \mathcal{M}''}\mathbf{1}\left[\alpha_a = \alpha_b\right]\nonumber\\
    & = \sum_{S',\,\mathcal{M}'}\hspace*{5px}\left(\prod_{\{a, b\} \in \mathcal{M}'}\mathbf{1}\left[\alpha_a = \alpha_b\right]\right)\sum_{S''}\hspace*{5px}\mathcal{I}^{\left(D - \left|S''\right|\right)}_{\bm{\alpha}_{\{k + 1,\,\ldots,\,D + k\} \backslash S''}}\sum_{\sigma}\prod_{a \in S'}\mathbf{1}\left[\alpha_a = \alpha_{\sigma(a)}\right]\nonumber\\
    & = \sum_{S',\,\mathcal{M}'}\hspace*{5px}\left(\prod_{\{a, b\} \in \mathcal{M}'}\mathbf{1}\left[\alpha_a = \alpha_b\right]\right)\mathcal{I}^{\left(D,\,\bm{\alpha}_{S'}\right)}_{\bm{\alpha}_{\{k + 1,\,\ldots,\,D + k\}}},
\end{align}
which, up to change of summation index $S' \longrightarrow \{1,\,\ldots,\,k\} \backslash S'$ and relabelling, is the desired identity.
\end{proof}
\end{lemma}
For concreteness, we explicitly write the identity proven in Lemma~\ref{lemma:matching_tensor_slice_from_matching_tensor_with_edges} (Eq.~\ref{eq:matching_tensor_slice_from_matching_tensor_with_edges}) for small $k$. For $k = 1$,
\begin{align}
    \bm{\mathcal{I}}^{(1 + D)}_{\alpha_1} & = \bm{\mathcal{I}}^{\left(D,\,\alpha_1\right)}.\label{eq:matching_tensor_slice_from_matching_tensor_with_edges_1_index}
\end{align}
For $k = 2$,
\begin{align}
    \bm{\mathcal{I}}^{(2 + D)}_{\alpha_1,\,\alpha_2} & = \delta_{\alpha_1,\,\alpha_2}\bm{\mathcal{I}}^{\left(D\right)} + \bm{\mathcal{I}}^{\left(D,\,\alpha_1,\,\alpha_2\right)}\label{eq:matching_tensor_slice_from_matching_tensor_with_edges_2_indices}
\end{align}
For $k = 3$,
\begin{align}
    \bm{\mathcal{I}}^{(3 + D)}_{\alpha_1,\,\alpha_2,\,\alpha_3} & = \delta_{\alpha_1,\,\alpha_2}\bm{\mathcal{I}}^{\left(D,\,\alpha_3\right)} + \delta_{\alpha_1,\,\alpha_3}\bm{\mathcal{I}}^{\left(D,\,\alpha_2\right)} + \delta_{\alpha_2,\,\alpha_3}\bm{\mathcal{I}}^{\left(D,\,\alpha_1\right)} + \bm{\mathcal{I}}^{\left(D,\,\alpha_1,\,\alpha_2,\,\alpha_3\right)}.\label{eq:matching_tensor_slice_from_matching_tensor_with_edges_3_indices}
\end{align}
For $k = 4$,
\begin{align}
    \bm{\mathcal{I}}^{(4 + D)}_{\alpha_1,\,\alpha_2,\,\alpha_3,\,\alpha_4} & = \delta_{\alpha_1,\,\alpha_2}\delta_{\alpha_3,\,\alpha_4}\bm{\mathcal{I}}^{\left(D\right)} + \delta_{\alpha_1,\,\alpha_3}\delta_{\alpha_2,\,\alpha_4}\bm{\mathcal{I}}^{\left(D\right)} + \delta_{\alpha_1,\,\alpha_4}\delta_{\alpha_2,\,\alpha_3}\bm{\mathcal{I}}^{\left(D\right)}\nonumber\\
    & \hspace*{20px} + \delta_{\alpha_1,\,\alpha_2}\bm{\mathcal{I}}^{\left(D,\,\alpha_3,\,\alpha_4\right)} + \delta_{\alpha_1,\,\alpha_3}\bm{\mathcal{I}}^{\left(D,\,\alpha_2,\,\alpha_4\right)} + \delta_{\alpha_1,\,\alpha_4}\bm{\mathcal{I}}^{\left(D,\,\alpha_2,\,\alpha_3\right)}\nonumber\\
    & \hspace*{20px} + \delta_{\alpha_2,\,\alpha_3}\bm{\mathcal{I}}^{\left(D,\,\alpha_1,\,\alpha_4\right)} + \delta_{\alpha_2,\,\alpha_4}\bm{\mathcal{I}}^{\left(D,\,\alpha_1,\,\alpha_3\right)} + \delta_{\alpha_3,\,\alpha_4}\bm{\mathcal{I}}^{\left(D,\,\alpha_1,\,\alpha_2\right)}\nonumber\\
    & \hspace*{20px} + \bm{\mathcal{I}}^{\left(D,\,\alpha_1,\,\alpha_2,\,\alpha_3,\,\alpha_4\right)}.\label{eq:matching_tensor_slice_from_matching_tensor_with_edges_4_indices}
\end{align}

From these identities, one can deduce series expansions for each the contributions $\nu^{(k)}_{\alpha}$ ($0 \leq k \leq 2$) and $\nu^{(k)}_{\alpha,\,\beta}$ ($3 \leq k \leq 7$) to the QGMS moments; they are collected in the following two Propositions for convenience. We note they are not specific to the SK-QAOA QGMS but apply to any QGMS, similar to the expansion in Proposition~\ref{prop:qgms_integral_series_expansion}

\begin{proposition}[Series expansions of contributions to diagonal quadratic QGMS moments]
\label{prop:qgms_diagonal_quadratic_moment_contributions_series_expansions}
The diagonal quadratic QGMS moment can be decomposed as:
\begin{align}
    \frac{\partial^2S_n\left(\bm{\mu}\right)}{\partial\mu_{\alpha}^2}\Bigg|_{\bm{\mu} = \bm{0}} & = \nu^{(0)}_{\alpha} + \frac{2}{\sqrt{n}}\nu^{(1)}_{\alpha} + \frac{1}{n}\nu^{(2)}_{\alpha}.
\end{align}
The following series expansion holds for each of the $\nu^{(k)}_{\alpha}$ ($0 \leq k \leq 2$):
\begin{align}
    \nu^{(k)}_{\alpha} & = \left(\mathcal{Z}^*\right)^n\sum_{\substack{\left(n_d\right)_{d \geq 2}}}\binom{n}{\left(n_d\right)_{d \geq 2}}\frac{n^{-\sum\limits_{d \geq 2}dn_d/2}}{\prod\limits_{d \geq 2}d!^{n_d}}\left\langle \bm{\mathcal{T}}^{(k),\,\left(\sum\limits_{d \geq 2}dn_d,\,\alpha\right)}, \bigotimes_{d \geq 2}\bm{\delta C}^{(d)\otimes n_d} \right\rangle,
\end{align}
with tensors $\bm{\mathcal{T}}^{(k),\,\left(D,\,\alpha\right)}$ given by:
\begin{align}
    \bm{\mathcal{T}}^{(0),\,\left(D,\,\alpha\right)} & := \left(\theta^*_{\alpha}\right)^2\bm{\mathcal{I}}^{(D)},\\
    \bm{\mathcal{T}}^{(1),\,\left(D,\,\alpha\right)} & := \theta^*_{\alpha}\bm{\mathcal{I}}^{\left(D,\,\alpha\right)},\\
    \bm{\mathcal{T}}^{(2),\,\left(D,\,\alpha\right)} & := \bm{\mathcal{I}}^{\left(D,\,\alpha,\,\alpha\right)}.
\end{align}
\end{proposition}

\begin{proposition}[Series expansions of contributions to diagonal quartic moment]
\label{prop:qgms_diagonal_quartic_moment_contributions_series_expansions}
The diagonal quartic QGMS moment can be decomposed as:
\begin{align}
    \frac{\partial^4S_n\left(\bm{\mu}\right)}{\partial\mu_{\alpha}^2\partial\mu_{\beta}^2}\Bigg|_{\bm{\mu} = \bm{0}} & = \frac{1}{n^2}\nu^{(7)}_{\alpha,\,\beta} + \frac{1}{n^{3/2}}\nu^{(6)}_{\alpha,\,\beta} + \frac{1}{n}\nu^{(5)}_{\alpha,\,\beta} + \frac{1}{\sqrt{n}}\nu^{(4)}_{\alpha,\,\beta} + \nu^{(3)}_{\alpha,\,\beta}.
\end{align}
The following series expansion holds for each of the $\nu^{(k)}_{\alpha,\,\beta}$ ($3 \leq k \leq 7$):
\begin{align}
    \nu^{(k)}_{\alpha,\,\beta} & = \left(\mathcal{Z}^*\right)^n\sum_{\left(n_d\right)_{d \geq 2}}\binom{n}{\left(n_d\right)_{d \geq 2}}\frac{n^{-\sum\limits_{d \geq 2}dn_d/2}}{\prod\limits_{d \geq 2}d!^{n_d}}\left\langle \bm{\mathcal{T}}^{(k),\,\left(\sum\limits_{d \geq 2}dn_d,\,\alpha,\,\beta\right)}, \bigotimes_{d \geq 2}\bm{\delta C}^{(d)\otimes n_d} \right\rangle, 
\end{align}
with tensors $\bm{\mathcal{T}}^{(k),\,(D,\,\alpha,\,\beta)}$ given by:
\begin{align}
    \bm{\mathcal{T}}^{(3),\,\left(D,\,\alpha,\,\beta\right)} & := \left(\theta^*_{\alpha}\right)^2\left(\theta^*_{\beta}\right)^2,\\
    \bm{\mathcal{T}}^{(4),\,\left(D,\,\alpha,\,\beta\right)} & := 2\left(\theta^*_{\beta}\right)^2\theta^*_{\alpha}\bm{\mathcal{I}}^{\left(D,\,\alpha\right)} + 2\left(\theta^*_{\alpha}\right)^2\theta^*_{\beta}\bm{\mathcal{I}}^{\left(D,\,\beta\right)},\\
    \bm{\mathcal{T}}^{(5),\,\left(D,\,\alpha,\,\beta\right)} & := 4\theta^*_{\alpha}\theta^*_{\beta}\bm{\mathcal{I}}^{\left(D,\,\alpha,\,\beta\right)} + \left(\theta^*_{\alpha}\right)^2\bm{\mathcal{I}}^{\left(D,\,\beta,\,\beta\right)} + \left(\theta^*_{\beta}\right)^2\bm{\mathcal{I}}^{\left(D,\,\alpha,\,\alpha\right)},\\
    \bm{\mathcal{T}}^{(6),\,\left(D,\,\alpha,\,\beta\right)} & := 2\theta^*_{\alpha}\bm{\mathcal{I}}^{\left(D,\,\alpha,\,\beta,\,\beta\right)} + 2\theta^*_{\beta}\bm{\mathcal{I}}^{\left(D,\,\beta,\,\alpha,\,\alpha\right)},\\
    \bm{\mathcal{T}}^{(7),\,\left(D,\,\alpha,\,\beta\right)} & := \bm{\mathcal{I}}^{(D,\,\alpha,\,\alpha,\,\beta,\,\beta)}.
\end{align}
\begin{proof}
This is proven by direct calculation. For concreteness, we illustrate the calculation in the case of $\nu^{(5)}_{\alpha,\,\beta}$. We start with the expression of this quantity in terms of QGMS integral moments tensors $\bm{S}^{(k)}_n$, as recalled at the beginning of the Section:
\begin{align}
    \nu^{(5)}_{\alpha,\,\beta} & = -8\delta_{\alpha\beta}\theta^*_{\alpha}\left[\bm{S}^{(1)}_n\right]_{\alpha} - 2\theta^*_{\alpha}\left[\bm{S}^{(1)}_n\right]_{\alpha} - 2\theta^*_{\beta}\left[\bm{S}^{(1)}_n\right]_{\beta} + 2\theta^*_{\alpha}\left[\bm{S}^{(3)}_n\right]_{\alpha,\,\beta,\,\beta} + 2\theta^*_{\beta}\left[\bm{S}^{(3)}_n\right]_{\beta,\,\alpha,\,\alpha}.
\end{align}
From Proposition~\ref{prop:qgms_integral_series_expansion} stating the series expansion of QGMS integral moments tensors, and by linearity, the above expands as a similar series, with general term $\left(n_d\right)_{d \geq 2}$ given by:
\begin{align}
    \nu^{(5)}_{\alpha,\,\beta} \supset \left(\mathcal{Z}^*\right)^n\binom{n}{\left(n_d\right)_{d \geq 2}}\frac{n^{-D/2}}{\prod\limits_{d \geq 2}d!^{n_d}}\left\langle -8\delta_{\alpha\beta}\theta^*_{\alpha}\bm{\mathcal{I}}^{(1 + D)}_{\alpha} - 2\theta^*_{\alpha}\bm{\mathcal{I}}^{(1 + D)}_{\alpha} - 2\theta^*_{\beta}\bm{\mathcal{I}}^{(1 + D)}_{\beta} + 2\theta^*_{\alpha}\bm{\mathcal{I}}^{(3 + D)}_{\alpha,\,\beta,\,\beta} + 2\theta^*_{\beta}\bm{\mathcal{I}}^{(3 + D)}_{\beta,\,\alpha,\,\alpha}, \bigotimes_{d \geq 2}\bm{\delta C}^{(d)\otimes n_d} \right\rangle,
\end{align}
where for brevity we let $D := \sum_{d \geq 2}dn_d$. We then simplify the tensor on the left-hand side of the dot product bracket by rewriting matching tensors slices in terms of matching tensors with external edges following Eqs.~\ref{eq:matching_tensor_slice_from_matching_tensor_with_edges_1_index}-\ref{eq:matching_tensor_slice_from_matching_tensor_with_edges_3_indices}:
\begin{align}
    & -8\delta_{\alpha\beta}\theta^*_{\alpha}\bm{\mathcal{I}}^{\left(1 + D\right)}_{\alpha} - 2\theta^*_{\alpha}\bm{\mathcal{I}}^{\left(1 + D\right)}_{\alpha} - 2\theta^*_{\beta}\bm{\mathcal{I}}^{\left(1 + D\right)}_{\beta} + 2\theta^*_{\alpha}\bm{\mathcal{I}}^{\left(D,\,\alpha,\,\beta,\,\beta\right)} + 2\theta^*_{\beta}\bm{\mathcal{I}}^{\left(3 + D\right)}_{\beta,\,\alpha,\,\alpha}\nonumber\\
    & = -8\delta_{\alpha\beta}\theta^*_{\alpha}\bm{\mathcal{I}}^{\left(D,\,\alpha\right)} - 2\theta^*_{\alpha}\bm{\mathcal{I}}^{\left(D,\,\alpha\right)} - 2\theta^*_{\beta}\bm{\mathcal{I}}^{\left(D,\,\beta\right)} + 2\theta^*_{\alpha}\left(2\delta_{\alpha\beta}\bm{\mathcal{I}}^{\left(D,\,\beta\right)} + \delta_{\beta\beta}\bm{\mathcal{I}}^{\left(D,\,\alpha\right)} + \bm{\mathcal{I}}^{\left(D,\,\alpha,\,\beta,\,\beta\right)}\right)\nonumber\\
    & \hspace*{20px} + 2\theta^*_{\beta}\left(2\delta_{\alpha\beta}\bm{\mathcal{I}}^{\left(D,\,\alpha\right)} + \delta_{\alpha\alpha}\bm{\mathcal{I}}^{\left(D,\,\beta\right)} + \bm{\mathcal{I}}^{\left(D,\,\beta,\,\alpha,\,\alpha\right)}\right)\nonumber\\
    & = 2\theta^*_{\alpha}\bm{\mathcal{I}}^{\left(D,\,\alpha,\,\beta,\,\beta\right)} + 2\theta^*_{\beta}\bm{\mathcal{I}}^{\left(D,\,\beta,\,\alpha,\,\alpha\right)},
\end{align}
proving the claimed expression for $\bm{\mathcal{T}}^{(5),\,\left(D,\,\alpha,\,\beta\right)}$.
\end{proof}
\end{proposition}

The continuum limit of tensors $\bm{\nu}^{(k)} = \left(\nu^{(k)}_{\alpha}\right)_{\alpha \in \mathcal{A}} \in \mathbf{C}^{\mathcal{A}}$ ($0 \leq k \leq 2$) and $\bm{\nu}^{(k)} = \left(\nu^{(k)}_{\alpha,\,\beta}\right)_{\alpha, \beta \in \mathcal{A}} \in \left(\mathbf{C}^{\mathcal{A}}\right)^{\otimes 2}$ can be derived similar to that of $\bm{\nu} = \bm{\nu}^{(1)}$ discussed in Section~\ref{sec:continuum_limit_qgms_moments}. Recall that in the case of $\bm{\nu} = \bm{\nu}^{(1)}$, we analyzed
\begin{align}
    \nu_{\alpha} & = \sum_{\left(n_d\right)_{d \geq 2}}\nu^{\left(n_d\right)_{d \geq 2}}_{\alpha},\\
    \nu^{(n_d)_{d \geq 2}}_{\alpha} & := \binom{n}{\left(n_d\right)_{d \geq 2}}\frac{n^{-\sum\limits_{d \geq 2}dn_d/2}}{\prod\limits_{d \geq 2}d!^{n_d}}\left\langle \bm{\mathcal{I}}_{\alpha}^{\left(1 + \sum\limits_{d \geq 2}dn_d\right)}, \bigotimes_{d \geq 2}\bm{\delta C}^{(d)\otimes n_d} \right\rangle
\end{align}
by decomposing the matching tensor slice $\bm{\mathcal{I}}^{\left(1 + D\right)}_{\alpha} \in \left(\mathbf{C}^{\mathcal{A}}\right)^{\otimes (D - 1)}$ according to the index $l'''$ matched to $\alpha$, and the matching $\mathcal{M}$ between remaining indices:
\begin{align}
    \left[\bm{\mathcal{I}}^{\left(1 + D\right)}_{\alpha}\right]_{\bm{\alpha}_{1:D}} & = \sum_{l''' \in [D]}\mathbf{1}\left[\alpha = \alpha_{l'''}\right]\sum_{\mathcal{M}\textrm{ matching of }[D] - \{l'''\}}\prod_{\{a, b\} \in \mathcal{M}}\mathbf{1}\left[\alpha_a = \alpha_b\right],
\end{align}
corresponding to special case $k = 1$ of Lemma~\ref{lemma:matching_tensor_slice_from_matching_tensor_with_edges}. From this linear decomposition of the matching tensor slice, we introduce the corresponding linear decomposition of $\nu^{\left(n_d\right)}_{\alpha}$:
\begin{align}
    \nu^{\left(n_d\right)_{d \geq 2}}_{\alpha} & := \binom{n}{\left(n_d\right)_{d \geq 2}}\frac{n^{-D/2}}{\prod\limits_{d \geq 2}d!^{n_d}}\sum_{l''' \in [D]}\sum_{\mathcal{M}\textrm{ matching of }[D] - \{l'''\}}\nu^{\left(n_d\right)_{d \geq 2},\,l''',\,\mathcal{M}}_{\alpha},\\
    \nu^{\left(n_d\right)_{d \geq 2},\,l''',\,\mathcal{M}}_{\alpha} & := \theta^*_{\alpha}\sum_{\bm{\alpha}_{1:D} \in \mathcal{A}^{D}}\left[\bigotimes_{d \geq 2}\bm{\delta C}^{(d)\otimes n_d}\right]_{\bm{\alpha}_{1:D}}\mathbf{1}\left[\alpha = \alpha_{l'''}\right]\prod_{\{a, b \in \mathcal{M}\}}\mathbf{1}\left[\alpha_a = \alpha_b\right],
\end{align}
where we let $D := \sum_{d \geq 2}dn_d$ for brevity. We then rewrote the definition of $\nu^{\left(n_d\right)_{d \geq 2},\,l''',\,\mathcal{M}}_{\alpha}$ above in a more tensorial form:
\begin{align}
    \nu^{\left(n_d\right)_{d \geq 2},\,l''',\,\mathcal{M}}_{\alpha} & = \theta^*_{\alpha}\sum_{\bm{\alpha}_{\mathcal{D}(\mathcal{M})} \in \mathcal{A}^{\mathcal{D}(M)}}\left[\pi_{l''',\,\mathcal{M}} \cdot \bigotimes_{d \geq 2}\bm{\delta C}^{(d)\otimes n_d} \right]_{\alpha,\,\bm{\alpha}_{\mathcal{D}(\mathcal{M})},\,\bm{\alpha}_{\mathcal{D}(\mathcal{M})}},
\end{align}
where we interpreted a matching as a one-to-one mapping $\mathcal{D}(\mathcal{M}) \longrightarrow \mathcal{R}(\mathcal{M})$, such for all $\{a, b\} \in \mathcal{M}$, $a < b$, $a$ is mapped to $b$. The summation in the right-hand side of the above equation is over multi-$\mathcal{A}$-indices indexed by the ``domain" $\mathcal{D}(\mathcal{M})$ of $\mathcal{M}$. One can further disjointly decompose $[D] = \{\alpha\} \sqcup \mathcal{D}(\mathcal{M}) \sqcup \mathcal{R}(\mathcal{M})$. From these considerations, we introduced $\pi_{l''',\,\mathcal{M}}$ as the unique permutation of $[D]$ such that, when acting over tensor product factors of $\left(\mathbf{C}^{\mathcal{A}}\right)^{\otimes D}$,
\begin{align}
    \left[\pi_{l''',\,\mathcal{M}} \cdot \bm{T}\right]_{\bm{\alpha}_{1:D}} & = \bm{T}_{\alpha,\,\bm{\alpha}_{\mathcal{D}(\mathcal{M})},\,\bm{\alpha}_{\mathcal{R}(\mathcal{M})}},\\
    \bm{\alpha}_{1:D} & = \left(\alpha_1, \ldots, \alpha_D\right) \in \mathcal{A}^D,
\end{align}
for all tensor $\bm{T} \in \left(\mathbf{C}^{\mathcal{A}}\right)^{\otimes D}$. The advantage of the tensorial notation is to factor in equality constraints $\alpha_a = \alpha_b$, $\alpha_{l'''} = \alpha$ enforced by indicator functions in the matching tensor; in this rewriting, $\bm{\alpha}_{\mathcal{D}(\mathcal{M})}$ are proper independent variables.

To formulate similar expansions for $\nu^{(k)}_{\alpha}$ ($0 \leq k \leq 2$) and $\nu^{(k)}_{\alpha,\,\beta}$ ($3 \leq k \leq 7$), we extend the definition of this permutation as follows:

\begin{definition}[Permutation associated to matching]
Let $k$ and $D$ nonnegative integers. Let $l_1, \ldots, l_k \in [D]$ pairwise distinct and $\mathcal{M}$ a perfect matching of $[D] - \{l_1, \ldots, l_k\}$. We define $\pi_{l_1,\,\ldots,\,l_k,\,\mathcal{M}}$ as the unique permutation of $[D]$ such that
\begin{align}
    \left[\pi_{l_1,\,\ldots,\,l_k,\,\mathcal{M}} \cdot \bm{T}\right]_{\bm{\alpha}_{1:D}} & = \bm{T}_{\alpha_{l_1},\,\ldots,\,\alpha_{l_2},\,\bm{\alpha}_{\mathcal{D}(\mathcal{M})},\bm{\alpha}_{\mathcal{R}(\mathcal{M})}}
\end{align}
for all tensor $\bm{T} \in \left(\mathbf{C}^{\mathcal{A}}\right)^{\otimes D}$.
\end{definition}

\subsection{The continuum limits of additional moments contributions}
\label{sec:other_moments_contributions_continuum_limits}

We are now ready to state continuum limits for tensors $\bm{\nu}^{(k)} = \left(\nu^{(k)}_{\alpha}\right)_{\alpha \in \mathcal{A}}$ ($0 \leq k \leq 2$) and $\bm{\nu}^{(k)} = \left(\nu^{(k)}_{\alpha,\,\beta}\right)_{\alpha, \beta \in \mathcal{A}}$ ($3 \leq k \leq 7$) along the lines of Section~\ref{sec:continuum_limit_qgms_moments}, which was limited to $\bm{\nu}^{(1)}$. The case of $\bm{\nu}^{(0)}$ is special, and covered in the following Proposition:

\begin{proposition}[Simplification of quadratic QGMS moment contribution $\bm{\nu}^{(0)}$]
\label{prop:nu0_continuum_limit}
Recall the expansion of the QGMS diagonal quadratic moments in terms of QGMS integral moments tensors, giving the following for contribution $\bm{\nu}^{(0)}$:
\begin{align}
    \nu^{(0)}_{\alpha} & = \left(\theta^*_{\alpha}\right)^2S_n^{(0)}.
\end{align}
Since $S_n^{(0)} = 1$ for the SK-QAOA energy QGMS, this further simplifies to
\begin{align}
    \nu^{(0)}_{\alpha} = \left(\theta^*_{\alpha}\right)^2
\end{align}
for all $\alpha \in \mathcal{A} = \mathcal{I}^2$, which can also be written
\begin{align}
    \nu^{(0)}_{j_1,\,j_2} & = \lambda^2\mathcal{G}_{j_1}\mathcal{G}_{j_2}\left(G^{(2)}_{j_1,\,j_2}\right)^2
\end{align}
for all $j_1, j_2 \in \mathcal{I}$.
\begin{proof}
First, observe that $S_n^{(0)} = S_n\left(\bm{0}\right)$ according to the integral representation of QGMS-MGF established in Proposition~\ref{prop:qgms_mgf_saddle_point_centered_integral_representation} and the definition of QGMS integral moments tensor Definition~\ref{def:qgms_integral_moments_tensor}. Then, by a reasoning similar to the evaluation of the order 1 QAOA cost function moment 
\begin{align}
    \mathbb{E}\bra{\Psi_{p,\,n}}C_n/n\ket{\Psi_{p,\,n}} & = -\frac{1}{\Gamma_{p + 1}}\sum_{0 \leq r \leq 2p + 1}\frac{\partial^2S_n\left(\bm{\mu}\right)}{\partial\mu_{(r,\,p + 1)}^2}\Bigg|_{\bm{\mu} = \bm{0}}
\end{align}
(Section~\ref{sec:sk_qaoa_first_order_moment_qgms}), or the order 2 QAOA cost function moment
\begin{align}
    \mathbb{E}\bra{\Psi_{p,\,n}}\left(C_n/n\right)^2\ket{\Psi_{p,\,n}} & = -\frac{1}{\Gamma_{p + 1}^2}\sum_{0 \leq r, s \leq 2p + 1}\frac{\partial^4S_n\left(\bm{\mu}\right)}{\partial\mu_{(r,\,p + 1)}^2\partial\mu_{(s,\,p + 1)}^2}\Bigg|_{\bm{\mu} = \bm{0}}.
\end{align}
(Section~\ref{sec:sk_qaoa_second_order_moment_qgms}), one finds the ``order zero moment" of the QAOA cost function is
\begin{align}
    \mathbb{E}\bra{\Psi_{p,\,n}}1\ket{\Psi_{p,\,n}} & = S_n\left(\bm{0}\right),
\end{align}
and the left-hand side is trivially 1 by normalization of quantum state. This proves $S_n\left(\bm{0}\right) = 1$ and the statement.
\end{proof}
\end{proposition}

From the previous Proposition, the continuum limit of $\bm{\nu}^{(0)}$ from that of $\bm{G}^{(2),\,\mathrm{cont}}$, without the need for an additional tensor and its continuum limit like $\bm{\widetilde{\nu}}^{(1)}$. The following Proposition establishes the continuum limit of the remaining contribution $\bm{\nu}^{(2)}$ to the diagonal order 2 quadratic QGMS moment $\frac{\partial^2S_n\left(\bm{\mu}\right)}{\partial\mu_{\alpha}^2}\bigg|_{\bm{\mu} = \bm{0}}$.

\begin{proposition}[Continuum limit of quadratic QGMS moment contribution $\bm{\nu}^{(2)}$]
\label{prop:nu2_continuum_limit}
Starting from the series expansion of $\bm{\nu}^{(2)}$ established in Proposition~\ref{prop:qgms_diagonal_quadratic_moment_contributions_series_expansions}, let us write:
\begin{align}
    \nu^{(2)}_{\alpha} & := \sum_{\left(n_d\right)_{d \geq 2}}\binom{n}{\left(n_d\right)_{d \geq 2}}\frac{n^{-D/2}}{\prod\limits_{d \geq 2}d!^{n_d}}\left\langle \bm{\mathcal{T}}^{(2),\,\left(D,\,\alpha\right)}, \bigotimes_{d \geq 2}\bm{\delta C}^{(d)\otimes n_d} \right\rangle,\\
    \bm{\mathcal{T}}^{(2),\,\left(D,\,\alpha\right)} & := \bm{\mathcal{I}}^{\left(D,\,\alpha,\,\alpha\right)},
\end{align}
where we let $D := \sum_{d \geq 2}dn_d$ as an implicit function of $\left(n_d\right)_{d \geq 2}$. Define auxiliary tensor $\bm{\widetilde\nu}^{(2)} = \left(\widetilde{\nu}^{(2)}_{\alpha}\right)_{\alpha \in \mathcal{A}}$ by:
\begin{align}
    \widetilde{\nu}^{(2)}_{\alpha} & := \sum_{\left(n_d\right)_{d \geq 2}}\widetilde{\nu}_{\alpha}^{(2),\,\left(n_d\right)_{d \geq 2}},\\
    \widetilde{\nu}_{\alpha}^{(2),\,\left(n_d\right)_{d \geq 2}} & := \binom{n}{\left(n_d\right)_{d \geq 2}}\frac{n^{-D/2}}{\prod\limits_{d \geq 2}d!^{n_d}}\sum_{\{l_1, l_2\} \subset [D]}\hspace*{5px}\sum_{\mathcal{M}\textrm{ matching of }[D] - \{l_1, l_2\}}\widetilde{\nu}_{\alpha}^{(2),\,\left(n_d\right)_{d \geq 2},\,l_1,\,l_2,\,\mathcal{M}},\\
    \widetilde{\nu}_{\alpha}^{(2),\,\left(n_d\right)_{d \geq 2},\,l_1,\,l_2,\,\mathcal{M}} & := \lambda^D\sum_{\bm{\alpha}_{\mathcal{D}(\mathcal{M})} \in \mathcal{A}^{\mathcal{D}(\mathcal{M})}}\left[\pi_{l_1,\,l_2,\,\mathcal{M}} \cdot \bigotimes_{d \geq 2}\bm{\delta G}^{(2d)\otimes n_d}\right]_{\alpha,\,\alpha,\,\bm{\alpha}_{\mathcal{D}(\mathcal{M})},\,\bm{\alpha}_{\mathcal{D}(\mathcal{M})}}\left[\bm{\mathcal{G}}^{\otimes 2\left|\mathcal{D}(\mathcal{M})\right|}\right]_{\bm{\alpha}_{\mathcal{D}(\mathcal{M})}}
\end{align}
for all $\alpha \in \mathcal{I}$. Original tensor $\bm{\nu}^{(2)}$ can be written as follows in terms of $\bm{\widetilde{\nu}}^{(2)}$:
\begin{align}
    \nu^{(2)}_{j_1,\,j_2} & = \mathcal{G}_{j_1}\mathcal{G}_{j_2}\widetilde{\nu}^{(2)}_{j_1,\,j_2}
\end{align}
for all $j_1, j_2 \in \mathcal{I}$. Then, tensor $\bm{\widetilde{\nu}}^{(2)}$ admits continuum limit $\widetilde{\nu}^{(2),\,\mathrm{cont}}$, defined by series:
\begin{align}
    \widetilde{\nu}^{(2),\,\mathrm{cont}}\left(\xi\right) & := \sum_{\left(n_d\right)_{d \geq 2}}\widetilde{\nu}^{(2),\,\left(n_d\right)_{d \geq 2},\,\mathrm{cont}}\left(\xi\right),\\
    \widetilde{\nu}^{(2),\,\left(n_d\right)_{d \geq 2},\,\mathrm{cont}}\left(\xi\right) & := \binom{n}{\left(n_d\right)_{d \geq 2}}\frac{n^{-D/2}}{\prod\limits_{d \geq 2}d!^{n_d}}\sum_{\{l_1, l_2\} \subset [D]}\hspace*{5px}\sum_{\mathcal{M}\textrm{ matching of }[D] - \{l_1, l_2\}}\widetilde{\nu}^{(2),\,\left(n_d\right)_{d \geq 2},\,l_1,\,l_2,\,\mathcal{M},\,\mathrm{cont}}\left(\xi_1, \xi_2\right),\\
    \widetilde{\nu}^{(2),\,\left(n_d\right)_{d \geq 2},\,l_1,\,l_2,\,\mathcal{M},\,\mathrm{cont}}\left(\xi\right) & := \left(\frac{\gamma_{\mathrm{max}}}{\sqrt{2}}\right)^D\int\limits_{\left([0, 2]^2\right)^{\mathcal{D}(\mathcal{M})}}\!\mathrm{d}\bm{\xi}\,\left(\pi_{l_1,\,l_2,\,\mathcal{M}} \cdot \bigotimes_{d \geq 2}\left(\delta G^{(2d),\,\mathrm{cont}}\right)^{\otimes n_d}\right)\left(\xi, \xi, \bm{\xi}, \bm{\xi}\right)\left(\mathcal{G}^{\mathrm{cont}}\right)^{\otimes 2\left|\mathcal{D}(\mathcal{M})\right|}\left(\bm{\xi}\right)
\end{align}
for all $\xi \in [0, 2]^2$, where we let $D := \sum_{d \geq 2}dn_d$ as an implicit function of $\left(n_d\right)_{d \geq 2}$. The following uniform and discretization bounds hold over $\bm{\widetilde{\nu}}^{(2),\,\mathrm{cont}}$:
\begin{align}
    \left\lVert \bm{\widetilde{\nu}}^{(2)} \right\rVert_{\infty} & \leq \frac{\mathcal{O}(1)}{\left(p + 1\right)^2}\gamma_{\mathrm{max}}^2\max\left(4\beta_{\mathrm{max}}, \frac{2M_{\gamma}}{\gamma_{\mathrm{max}}}, \gamma'_{\mathrm{max}}\right),\\
    \left\lVert \widetilde{\nu}^{(2),\,\mathrm{cont}} \right\rVert_{\infty} & \leq \mathcal{O}(1)\gamma_{\mathrm{max}}^2\max\left(4\beta_{\mathrm{max}}, \frac{2M_{\gamma}}{\gamma_{\mathrm{max}}}, \gamma'_{\mathrm{max}}\right),\\
    \left|\widetilde{\nu}^{(2)}_{\alpha} - \frac{1}{(p + 1)^2}\widetilde{\nu}^{(2),\,\mathrm{cont}}\left(\frac{\alpha}{p + 1/2}\right)\right| & \leq \frac{\mathcal{O}(1)}{\left(p + 1\right)^3}\gamma_{\mathrm{max}}^2\max\left(4\beta_{\mathrm{max}}, \frac{2M_{\gamma}}{\gamma_{\mathrm{max}}}, \gamma'_{\mathrm{max}}\right)
\end{align}
for all $\alpha \in \mathcal{A} = \mathcal{I}^2$.
\end{proposition}

The now turn to stating the continuum limit of additive contributions $\bm{\nu}^{(k)}$, $3 \leq k \leq 7$ to the diagonal quartic QGMS moment $\frac{\partial^4S_n\left(\bm{\mu}\right)}{\partial\mu_{\alpha}^2\partial\mu_{\beta}^2}\Bigg|_{\bm{\mu} = \bm{0}}$. Contribution $\bm{\nu}^{(3)}$, similar to $\bm{\nu}^{(0)}$ for the quadratic QGMS moment, is special and covered in the following Proposition:

\begin{proposition}[Simplification of quartic QGMS moment contribution $\bm{\nu}^{(3)}$]
\label{prop:nu3_continuum_limit}
Recall the expansion of the contributions to the diagonal quartic QGMS moments in terms of QGMS integral moments tensors, giving the following for $\bm{\nu}^{(3)}$:
\begin{align}
    \nu^{(3)}_{\alpha,\,\beta} & = \left(\theta^*_{\alpha}\right)^2\left(\theta^*_{\beta}\right)^2S_n^{(0)}.
\end{align}
Since $S_n^{(0)} = 1$ for the SK-QAOA QGMS, this further simplifies to:
\begin{align}
    \nu^{(3)}_{\alpha,\,\beta} & = \left(\theta^*_{\alpha}\right)^2\left(\theta^*_{\beta}\right)^2
\end{align}
for $\alpha, \beta \in \mathcal{A} = \mathcal{I}^2$, which can also be written as:
\begin{align}
    \nu^{(3)}_{j_1,\,j_2,\,j_3,\,j_4} & = \lambda^4\mathcal{G}_{j_1}\mathcal{G}_{j_2}\mathcal{G}_{j_3}\mathcal{G}_{j_4}\left(G^{(2)}_{j_1,\,j_2}\right)^2\left(G^{(2)}_{j_3,\,j_4}\right)^2
\end{align}
for all $j_1, j_2, j_3, j_4 \in \mathcal{I}$.
\end{proposition}

The following 3 Propositions state the continuum limits of the remaining contributions $\bm{\nu}^{(k)}$, $4 \leq k \leq 7$ to the diagonal quartic QGMS moments tensor.

\begin{proposition}[Continuum limit of quartic QGMS moment contribution $\bm{\nu}^{(4)}$]
\label{prop:nu4_continuum_limit}
Starting from the series expansion of $\bm{\nu}^{(4)}$ established in Proposition~\ref{prop:qgms_diagonal_quartic_moment_contributions_series_expansions}, let us write:
\begin{align}
    \nu^{(4)}_{\alpha,\,\beta} & = \nu^{(4,\,1)}_{\alpha,\,\beta} + \nu^{(4,\,2)}_{\alpha,\,\beta},\\
    \nu^{(4,\,r)}_{\alpha,\,\beta} & := \sum_{\left(n_d\right)_{d \geq 2}}\binom{n}{\left(n_d\right)_{d \geq 2}}\frac{n^{-D/2}}{\prod\limits_{d \geq 2}d!^{n_d}}\left\langle \bm{\mathcal{T}}^{(4,\,r),\,\left(D,\,\alpha,\,\beta\right)}, \bigotimes_{d \geq 2}\bm{\delta C}^{(d)\otimes n_d} \right\rangle,\\
    \bm{\mathcal{T}}^{(4,\,1),\,\left(D,\,\alpha,\,\beta\right)} & := 2\left(\theta^*_{\beta}\right)^2\theta^*_{\alpha}\bm{\mathcal{I}}^{(D,\,\alpha)},\\
    \bm{\mathcal{T}}^{(4,\,2),\,\left(D,\,\alpha,\,\beta\right)} & := 2\left(\theta^*_{\alpha}\right)^2\theta^*_{\beta}\bm{\mathcal{T}}^{(D,\,\beta)},
\end{align}
where we let $D := \sum_{d \geq 2}dn_d$ as an implicit function of $\left(n_d\right)_{d \geq 2}$. Define auxiliary tensors $\bm{\widetilde\nu}^{(4,\,1)} = \left(\widetilde{\nu}^{(4,\,1)}_{\alpha,\,\beta}\right)_{\alpha,\,\beta \in \mathcal{A}}$, $\bm{\widetilde\nu}^{(4,\,2)} = \left(\widetilde{\nu}^{(4,\,2)}_{\alpha,\,\beta}\right)_{\alpha,\,\beta \in \mathcal{A}}$ by:
\begin{align}
    \widetilde{\nu}^{(4,\,r)}_{\alpha,\,\beta} & := \sum_{\left(n_d\right)_{d \geq 2}}\widetilde{\nu}_{\alpha,\,\beta}^{(4,\,r),\,\left(n_d\right)_{d \geq 2}},\\
    \widetilde{\nu}_{\alpha,\,\beta}^{(4,\,r),\,\left(n_d\right)_{d \geq 2}} & := \binom{n}{\left(n_d\right)_{d \geq 2}}\frac{n^{-D/2}}{\prod\limits_{d \geq 2}d!^{n_d}}\sum_{l \in [D]}\hspace*{5px}\sum_{\mathcal{M}\textrm{ matching of }[D] - \{l\}}\widetilde{\nu}_{\alpha,\,\beta}^{(4,\,r),\,\left(n_d\right)_{d \geq 2},\,l,\,\mathcal{M}},\\
    \widetilde{\nu}_{\alpha,\,\beta}^{(4,\,1),\,\left(n_d\right)_{d \geq 2},\,l,\,\mathcal{M}} & := \lambda^D\sum_{\bm{\alpha}_{\mathcal{D}(\mathcal{M})} \in \mathcal{A}^{\mathcal{D}(\mathcal{M})}}\left[\pi_{l,\,\mathcal{M}} \cdot \bigotimes_{d \geq 2}\bm{\delta G}^{(2d)\otimes n_d}\right]_{\alpha,\,\bm{\alpha}_{\mathcal{D}(\mathcal{M})},\,\bm{\alpha}_{\mathcal{D}(\mathcal{M})}}\left[\bm{\mathcal{G}}^{\otimes 2\left|\mathcal{D}(\mathcal{M})\right|}\right]_{\bm{\alpha}_{\mathcal{D}(\mathcal{M})}},\\
    \widetilde{\nu}_{\alpha,\,\beta}^{(4,\,2),\,\left(n_d\right)_{d \geq 2},\,l,\,\mathcal{M}} & := \lambda^D\sum_{\bm{\alpha}_{\mathcal{D}(\mathcal{M})} \in \mathcal{A}^{\mathcal{D}(\mathcal{M})}}\left[\pi_{l,\,\mathcal{M}} \cdot \bigotimes_{d \geq 2}\bm{\delta G}^{(2d)\otimes n_d}\right]_{\beta,\,\bm{\alpha}_{\mathcal{D}(\mathcal{M})},\,\bm{\alpha}_{\mathcal{D}(\mathcal{M})}}\left[\bm{\mathcal{G}}^{\otimes 2\left|\mathcal{D}(\mathcal{M})\right|}\right]_{\bm{\alpha}_{\mathcal{D}(\mathcal{M})}}
\end{align}
for all $\alpha, \beta \in \mathcal{A} = \mathcal{I}^2$. Original tensors $\bm{\nu}^{(4,\,r)}$ can be written as follows in terms of $\bm{\widetilde{\nu}}^{(4,\,r)}$:
\begin{align}
    \nu^{(4,\,1)}_{j_1,\,j_2,\,j_3,\,j_4} & = 2\lambda^3\mathcal{G}_{j_1}\mathcal{G}_{j_2}\mathcal{G}_{j_3}\mathcal{G}_{j_4}G^{(2)}_{j_1,\,j_2}\left(G^{(2)}_{j_3,\,j_4}\right)^2\widetilde{\nu}^{(4,\,1)}_{j_1,\,j_2,\,j_3,\,j_4},\\
    \nu^{(4,\,2)}_{j_1,\,j_2,\,j_3,\,j_4} & = 2\lambda^3\mathcal{G}_{j_1}\mathcal{G}_{j_2}\mathcal{G}_{j_3}\mathcal{G}_{j_4}\left(G^{(2)}_{j_1,\,j_2}\right)^2G^{(2)}_{j_3,\,j_4}\widetilde{\nu}^{(4,\,2)}_{j_1,\,j_2,\,j_3,\,j_4}.
\end{align}
Then, tensors $\bm{\widetilde{\nu}}^{(4,\,r)}$ admit continuum limits $\widetilde{\nu}^{(4,\,r),\,\mathrm{cont}}$, defined by series:
\begin{align}
    \widetilde{\nu}^{(4,\,r),\,\mathrm{cont}}\left(\xi_1, \xi_2\right) & := \sum_{\left(n_d\right)_{d \geq 2}}\widetilde{\nu}^{(4,\,r),\,\left(n_d\right)_{d \geq 2},\,\mathrm{cont}}\left(\xi_1, \xi_2\right),\\
    \widetilde{\nu}^{(4,\,r),\,\left(n_d\right)_{d \geq 2},\,\mathrm{cont}}\left(\xi_1, \xi_2\right) & := \binom{n}{\left(n_d\right)_{d \geq 2}}\frac{n^{-D/2}}{\prod\limits_{d \geq 2}d!^{n_d}}\sum_{l \in [D]}\hspace*{5px}\sum_{\mathcal{M}\textrm{ matching of }[D] - \{l\}}\widetilde{\nu}^{(4,\,r),\,\left(n_d\right)_{d \geq 2},\,l,\,\mathcal{M},\,\mathrm{cont}}\left(\xi_1, \xi_2\right),\\
    \widetilde{\nu}^{(4,\,1),\,\left(n_d\right)_{d \geq 2},\,l,\,\mathcal{M},\,\mathrm{cont}}\left(\xi_1, \xi_2\right) & := \left(\frac{\gamma_{\mathrm{max}}}{\sqrt{2}}\right)^D\int\limits_{\left([0, 2]^2\right)^{\mathcal{D}(\mathcal{M})}}\!\mathrm{d}\bm{\xi}\,\left(\pi_{l,\,\mathcal{M}} \cdot \bigotimes_{d \geq 2}\left(\delta G^{(2d),\,\mathrm{cont}}\right)^{\otimes n_d}\right)\left(\xi_1, \bm{\xi}, \bm{\xi}\right)\left(\mathcal{G}^{\mathrm{cont}}\right)^{\otimes 2\left|\mathcal{D}(\mathcal{M})\right|}\left(\bm{\xi}\right),\\
    \widetilde{\nu}^{(4,\,2),\,\left(n_d\right)_{d \geq 2},\,l,\,\mathcal{M},\,\mathrm{cont}}\left(\xi_1, \xi_2\right) & := \left(\frac{\gamma_{\mathrm{max}}}{\sqrt{2}}\right)^D\int\limits_{\left([0, 2]^2\right)^{\mathcal{D}(\mathcal{M})}}\!\mathrm{d}\bm{\xi}\,\left(\pi_{l,\,\mathcal{M}} \cdot \bigotimes_{d \geq 2}\left(\delta G^{(2d),\,\mathrm{cont}}\right)^{\otimes n_d}\right)\left(\xi_2, \bm{\xi}, \bm{\xi}\right)\left(\mathcal{G}^{\mathrm{cont}}\right)^{\otimes 2\left|\mathcal{D}(\mathcal{M})\right|}\left(\bm{\xi}\right)
\end{align}
for all $\xi_1, \xi_2 \in [0, 2]^2$, where we let $D := \sum_{d \geq 2}dn_d$ as an implicit function of $\left(n_d\right)_{d \geq 2}$. The following uniform and discretization bounds hold over $\bm{\widetilde{\nu}}^{(4,\,r)}$ and their continuum counterparts $\widetilde{\nu}^{(4,\,r),\,\mathrm{cont}}$:
\begin{align}
    \left\lVert \bm{\widetilde{\nu}}^{(4,\,r)} \right\rVert_{\infty} & \leq \frac{\mathcal{O}(1)}{\left(p + 1\right)}\gamma_{\mathrm{max}}^2\max\left(4\beta_{\mathrm{max}}, \frac{2M_{\gamma}}{\gamma_{\mathrm{max}}}, \gamma'_{\mathrm{max}}\right),\\
    \left\lVert \widetilde{\nu}^{(4,\,r),\,\mathrm{cont}} \right\rVert_{\infty} & \leq \mathcal{O}(1)\gamma_{\mathrm{max}}^2\max\left(4\beta_{\mathrm{max}}, \frac{2M_{\gamma}}{\gamma_{\mathrm{max}}}, \gamma'_{\mathrm{max}}\right),\\
    \left|\widetilde{\nu}^{(4,\,r)}_{\alpha,\,\beta} - \frac{1}{p + 1}\widetilde{\nu}^{(4,\,r),\,\mathrm{cont}}\left(\frac{\alpha}{p + 1/2}, \frac{\beta}{p + 1/2}\right)\right| & \leq \frac{\mathcal{O}(1)}{\left(p + 1\right)^2}\gamma_{\mathrm{max}}^2\max\left(4\beta_{\mathrm{max}}, \frac{2M_{\gamma}}{\gamma_{\mathrm{max}}}, \gamma'_{\mathrm{max}}\right)
\end{align}
for all $\alpha, \beta \in \mathcal{A} = \mathcal{I}^2$.
\end{proposition}

\begin{proposition}[Continuum limit of quartic QGMS moment contribution $\bm{\nu}^{(5)}$]
\label{prop:nu5_continuum_limit}
Starting from the series expansion of $\bm{\nu}^{(5)}$ established in Proposition~\ref{prop:qgms_diagonal_quartic_moment_contributions_series_expansions}, let us write:
\begin{align}
    \nu^{(5)}_{\alpha,\,\beta} & = \nu^{(5,\,1)}_{\alpha,\,\beta} + \nu^{(5,\,2)}_{\alpha,\,\beta} + \nu^{(5,\,3)}_{\alpha,\,\beta},\\
    \nu^{(5,\,r)}_{\alpha,\,\beta} & := \sum_{\left(n_d\right)_{d \geq 2}}\binom{n}{\left(n_d\right)_{d \geq 2}}\frac{n^{-D/2}}{\prod\limits_{d \geq 2}d!^{n_d}}\left\langle \bm{\mathcal{T}}^{(5,\,r),\,\left(D,\,\alpha,\,\beta\right)}, \bigotimes_{d \geq 2}\bm{\delta C}^{(d)\otimes n_d} \right\rangle,\\
    \bm{\mathcal{T}}^{(5,\,1),\,\left(D,\,\alpha,\,\beta\right)} & := 4\theta^*_{\alpha}\theta^*_{\beta}\bm{\mathcal{I}}^{(D,\,\alpha,\,\beta)},\\
    \bm{\mathcal{T}}^{(5,\,2),\,\left(D,\,\alpha,\,\beta\right)} & := \left(\theta^*_{\alpha}\right)^2\bm{\mathcal{I}}^{(D,\,\beta,\,\beta)},\\
    \bm{\mathcal{T}}^{(5,\,3),\,\left(D,\,\alpha,\,\beta\right)} & := \left(\theta^*_{\beta}\right)^2\bm{\mathcal{I}}^{(D,\,\alpha,\,\alpha)},
\end{align}
where we let $D := \sum_{d \geq 2}dn_d$ as an implicit function of $\left(n_d\right)_{d \geq 2}$. Define auxiliary tensors $\bm{\widetilde\nu}^{(5,\,1)} = \left(\widetilde{\nu}^{(5,\,1)}_{\alpha,\,\beta}\right)_{\alpha,\,\beta \in \mathcal{A}}$, $\bm{\widetilde\nu}^{(5,\,2)} = \left(\widetilde{\nu}^{(5,\,3)}_{\alpha,\,\beta}\right)_{\alpha,\,\beta \in \mathcal{A}}$, $\bm{\widetilde\nu}^{(5,\,3)} = \left(\widetilde{\nu}^{(5,\,3)}_{\alpha,\,\beta}\right)_{\alpha,\,\beta \in \mathcal{A}}$ by:
\begin{align}
    \widetilde{\nu}^{(5,\,r)}_{\alpha,\,\beta} & := \sum_{\left(n_d\right)_{d \geq 2}}\widetilde{\nu}_{\alpha,\,\beta}^{(5,\,r),\,\left(n_d\right)_{d \geq 2}},\\
    \widetilde{\nu}_{\alpha,\,\beta}^{(5,\,r),\,\left(n_d\right)_{d \geq 2}} & := \binom{n}{\left(n_d\right)_{d \geq 2}}\frac{n^{-D/2}}{\prod\limits_{d \geq 2}d!^{n_d}}\sum_{\{l_1, l_2\} \subset [D]}\hspace*{5px}\sum_{\mathcal{M}\textrm{ matching of }[D] - \{l_1, l_2\}}\widetilde{\nu}_{\alpha,\,\beta}^{(5,\,r),\,\left(n_d\right)_{d \geq 2},\,l_1,\,l_2,\,\mathcal{M}},\\
    \widetilde{\nu}_{\alpha,\,\beta}^{(5,\,1),\,\left(n_d\right)_{d \geq 2},\,l_1,\,l_2,\,\mathcal{M}} & := \lambda^D\sum_{\bm{\alpha}_{\mathcal{D}(\mathcal{M})} \in \mathcal{A}^{\mathcal{D}(\mathcal{M})}}\left[\pi_{l_1,\,l_2,\,\mathcal{M}} \cdot \bigotimes_{d \geq 2}\bm{\delta G}^{(2d)\otimes n_d}\right]_{\alpha,\,\beta,\,\bm{\alpha}_{\mathcal{D}(\mathcal{M})},\,\bm{\alpha}_{\mathcal{D}(\mathcal{M})}}\left[\bm{\mathcal{G}}^{\otimes 2\left|\mathcal{D}(\mathcal{M})\right|}\right]_{\bm{\alpha}_{\mathcal{D}(\mathcal{M})}},\\
    \widetilde{\nu}_{\alpha,\,\beta}^{(5,\,2),\,\left(n_d\right)_{d \geq 2},\,l_1,\,l_2,\,\mathcal{M}} & := \lambda^D\sum_{\bm{\alpha}_{\mathcal{D}(\mathcal{M})} \in \mathcal{A}^{\mathcal{D}(\mathcal{M})}}\left[\pi_{l_1,\,l_2,\,\mathcal{M}} \cdot \bigotimes_{d \geq 2}\bm{\delta G}^{(2d)\otimes n_d}\right]_{\beta,\,\beta,\,\bm{\alpha}_{\mathcal{D}(\mathcal{M})},\,\bm{\alpha}_{\mathcal{D}(\mathcal{M})}}\left[\bm{\mathcal{G}}^{\otimes 2\left|\mathcal{D}(\mathcal{M})\right|}\right]_{\bm{\alpha}_{\mathcal{D}(\mathcal{M})}},\\
    \widetilde{\nu}_{\alpha,\,\beta}^{(5,\,3),\,\left(n_d\right)_{d \geq 2},\,l_1,\,l_2,\,\mathcal{M}} & := \lambda^D\sum_{\bm{\alpha}_{\mathcal{D}(\mathcal{M})} \in \mathcal{A}^{\mathcal{D}(\mathcal{M})}}\left[\pi_{l_1,\,l_2,\,\mathcal{M}} \cdot \bigotimes_{d \geq 2}\bm{\delta G}^{(2d)\otimes n_d}\right]_{\alpha,\,\alpha,\,\bm{\alpha}_{\mathcal{D}(\mathcal{M})},\,\bm{\alpha}_{\mathcal{D}(\mathcal{M})}}\left[\bm{\mathcal{G}}^{\otimes 2\left|\mathcal{D}(\mathcal{M})\right|}\right]_{\bm{\alpha}_{\mathcal{D}(\mathcal{M})}}
\end{align}
for all $\alpha, \beta \in \mathcal{A} = \mathcal{I}^2$. Original tensors $\bm{\nu}^{(5,\,r)}$ can be written as follows in terms of $\bm{\widetilde{\nu}}^{(5,\,r)}$
\begin{align}
    \nu^{(5,\,1)}_{j_1,\,j_2,\,j_3,\,j_4} & = 4\lambda^2\mathcal{G}_{j_1}\mathcal{G}_{j_2}\mathcal{G}_{j_3}\mathcal{G}_{j_4}G^{(2)}_{j_1,\,j_2}G^{(2)}_{j_3,\,j_4}\widetilde{\nu}^{(5,\,1)}_{j_1,\,j_2,\,j_3,\,j_4},\\
    \nu^{(5,\,2)}_{j_1,\,j_2,\,j_3,\,j_4} & = \lambda^2\mathcal{G}_{j_1}\mathcal{G}_{j_2}\mathcal{G}_{j_3}\mathcal{G}_{j_4}\left(G^{(2)}_{j_1,\,j_2}\right)^2\widetilde{\nu}^{(5,\,2)}_{j_1,\,j_2,\,j_3,\,j_4},\\
    \nu^{(5,\,3)}_{j_1,\,j_2,\,j_3,\,j_4} & = \lambda^2\mathcal{G}_{j_1}\mathcal{G}_{j_2}\mathcal{G}_{j_3}\mathcal{G}_{j_4}\left(G^{(2)}_{j_3,\,j_4}\right)^2\widetilde{\nu}^{(5,\,3)}_{j_1,\,j_2,\,j_3,\,j_4}
\end{align}
for all $j_1, j_2, j_3, j_4 \in \mathcal{I}$. Then, tensors $\bm{\widetilde{\nu}}^{(5,\,r)}$ admit continuum limits $\widetilde{\nu}^{(5,\,r),\,\mathrm{cont}}$, defined by series:
\begin{align}
    \widetilde{\nu}^{(5,\,r),\,\mathrm{cont}}\left(\xi_1, \xi_2\right) & := \sum_{\left(n_d\right)_{d \geq 2}}\widetilde{\nu}^{(5,\,r),\,\left(n_d\right)_{d \geq 2},\,\mathrm{cont}}\left(\xi_1, \xi_2\right),\\
    \widetilde{\nu}^{(5,\,r),\,\left(n_d\right)_{d \geq 2},\,\mathrm{cont}}\left(\xi_1, \xi_2\right) & := \binom{n}{\left(n_d\right)_{d \geq 2}}\frac{n^{-D/2}}{\prod\limits_{d \geq 2}d!^{n_d}}\sum_{\{l_1, l_2\} \subset [D]}\hspace*{5px}\sum_{\mathcal{M}\textrm{ matching of }[D] - \{l_1, l_2\}}\widetilde{\nu}^{(5,\,r),\,\left(n_d\right)_{d \geq 2},\,l_1,\,l_2,\,\mathcal{M},\,\mathrm{cont}}\left(\xi_1, \xi_2\right),\\
    \widetilde{\nu}^{(5,\,1),\,\left(n_d\right)_{d \geq 2},\,l_1,\,l_2,\,\mathcal{M},\,\mathrm{cont}}\left(\xi_1, \xi_2\right) & := \left(\frac{\gamma_{\mathrm{max}}}{\sqrt{2}}\right)^D\int\limits_{\left([0, 2]^2\right)^{\mathcal{D}(\mathcal{M})}}\!\mathrm{d}\bm{\xi}\,\left(\pi_{l_1,\,l_2,\,\mathcal{M}} \cdot \bigotimes_{d \geq 2}\left(\delta G^{(2d),\,\mathrm{cont}}\right)^{\otimes n_d}\right)\left(\xi_1, \xi_2, \bm{\xi}, \bm{\xi}\right)\left(\mathcal{G}^{\mathrm{cont}}\right)^{\otimes 2\left|\mathcal{D}(\mathcal{M})\right|}\left(\bm{\xi}\right),\\
    \widetilde{\nu}^{(5,\,2),\,\left(n_d\right)_{d \geq 2},\,l_1,\,l_2,\,\mathcal{M},\,\mathrm{cont}}\left(\xi_1, \xi_2\right) & := \left(\frac{\gamma_{\mathrm{max}}}{\sqrt{2}}\right)^D\int\limits_{\left([0, 2]^2\right)^{\mathcal{D}(\mathcal{M})}}\!\mathrm{d}\bm{\xi}\,\left(\pi_{l_1,\,l_2,\,\mathcal{M}} \cdot \bigotimes_{d \geq 2}\left(\delta G^{(2d),\,\mathrm{cont}}\right)^{\otimes n_d}\right)\left(\xi_2, \xi_2, \bm{\xi}, \bm{\xi}\right)\left(\mathcal{G}^{\mathrm{cont}}\right)^{\otimes 2\left|\mathcal{D}(\mathcal{M})\right|}\left(\bm{\xi}\right),\\
    \widetilde{\nu}^{(5,\,3),\,\left(n_d\right)_{d \geq 2},\,l_1,\,l_2,\,\mathcal{M},\,\mathrm{cont}}\left(\xi_1, \xi_2\right) & := \left(\frac{\gamma_{\mathrm{max}}}{\sqrt{2}}\right)^D\int\limits_{\left([0, 2]^2\right)^{\mathcal{D}(\mathcal{M})}}\!\mathrm{d}\bm{\xi}\,\left(\pi_{l_1,\,l_2,\,\mathcal{M}} \cdot \bigotimes_{d \geq 2}\left(\delta G^{(2d),\,\mathrm{cont}}\right)^{\otimes n_d}\right)\left(\xi_1, \xi_1, \bm{\xi}, \bm{\xi}\right)\left(\mathcal{G}^{\mathrm{cont}}\right)^{\otimes 2\left|\mathcal{D}(\mathcal{M})\right|}\left(\bm{\xi}\right)
\end{align}
for all $\xi_1, \xi_2 \in [0, 2]^2$, where we let $D := \sum_{d \geq 2}dn_d$ as an implicit function of $\left(n_d\right)_{d \geq 2}$. The following uniform and discretization bounds hold over $\bm{\widetilde{\nu}}^{(5,\,r)}$ and their continuum counterparts $\widetilde{\nu}^{(5,\,r),\,\mathrm{cont}}$:
\begin{align}
    \left\lVert \bm{\widetilde{\nu}}^{(5,\,r)} \right\rVert_{\infty} & \leq \frac{\mathcal{O}(1)}{\left(p + 1\right)^2}\gamma_{\mathrm{max}}^2\max\left(4\beta_{\mathrm{max}}, \frac{2M_{\gamma}}{\gamma_{\mathrm{max}}}, \gamma'_{\mathrm{max}}\right),\\
    \left\lVert \widetilde{\nu}^{(5,\,r),\,\mathrm{cont}} \right\rVert_{\infty} & \leq \mathcal{O}(1)\gamma_{\mathrm{max}}^2\max\left(4\beta_{\mathrm{max}}, \frac{2M_{\gamma}}{\gamma_{\mathrm{max}}}, \gamma'_{\mathrm{max}}\right),\\
    \left|\widetilde{\nu}^{(5,\,r)}_{\alpha,\,\beta} - \frac{1}{\left(p + 1\right)^2}\widetilde{\nu}^{(5,\,r),\,\mathrm{cont}}\left(\frac{\alpha}{p + 1/2}, \frac{\beta}{p + 1/2}\right)\right| & \leq \frac{\mathcal{O}(1)}{\left(p + 1\right)^3}\gamma_{\mathrm{max}}^2\max\left(4\beta_{\mathrm{max}}, \frac{2M_{\gamma}}{\gamma_{\mathrm{max}}}, \gamma'_{\mathrm{max}}\right)
\end{align}
for all $\alpha, \beta \in \mathcal{A} = \mathcal{I}^2$.
\end{proposition}

\begin{proposition}[Continuum limit of quartic QGMS moment contribution $\bm{\nu}^{(6)}$]
\label{prop:nu6_continuum_limit}
Starting from the series expansion of $\bm{\nu}^{(6)}$ established in Proposition~\ref{prop:qgms_diagonal_quartic_moment_contributions_series_expansions}, let us write:
\begin{align}
    \nu^{(6)}_{\alpha,\,\beta} & = \nu^{(6,\,1)}_{\alpha,\,\beta} + \nu^{(6,\,2)}_{\alpha,\,\beta},\\
    \nu^{(6,\,r)}_{\alpha,\,\beta} & := \sum_{\left(n_d\right)_{d \geq 2}}\binom{n}{\left(n_d\right)_{d \geq 2}}\frac{n^{-D/2}}{\prod\limits_{d \geq 2}d!^{n_d}}\left\langle \bm{\mathcal{T}}^{(6,\,r),\,\left(D,\,\alpha,\,\beta\right)}, \bigotimes_{d \geq 2}\bm{\delta C}^{(d)\otimes n_d} \right\rangle,\\
    \bm{\mathcal{T}}^{(6,\,1),\,\left(D,\,\alpha,\,\beta\right)} & := 2\theta^*_{\alpha}\bm{\mathcal{I}}^{(D,\,\alpha,\,\beta,\,\beta)},\\
    \bm{\mathcal{T}}^{(6,\,2),\,\left(D,\,\alpha,\,\beta\right)} & := 2\theta^*_{\beta}\bm{\mathcal{I}}^{(D,\,\beta,\,\alpha,\,\alpha)},
\end{align}
where we let $D := \sum_{d \geq 2}dn_d$ as an implicit function of $\left(n_d\right)_{d \geq 2}$. Define auxiliary tensors $\bm{\widetilde\nu}^{(6,\,1)} = \left(\widetilde{\nu}^{(6,\,1)}_{\alpha,\,\beta}\right)_{\alpha,\,\beta \in \mathcal{A}}$, $\bm{\widetilde\nu}^{(6,\,2)} = \left(\widetilde{\nu}^{(6,\,2)}_{\alpha,\,\beta}\right)_{\alpha,\,\beta \in \mathcal{A}}$ by:
\begin{align}
    \widetilde{\nu}^{(6,\,r)}_{\alpha,\,\beta} & := \sum_{\left(n_d\right)_{d \geq 2}}\widetilde{\nu}_{\alpha,\,\beta}^{(6,\,r),\,\left(n_d\right)_{d \geq 2}},\\
    \widetilde{\nu}_{\alpha,\,\beta}^{(6,\,r),\,\left(n_d\right)_{d \geq 2}} & := \binom{n}{\left(n_d\right)_{d \geq 2}}\frac{n^{-D/2}}{\prod\limits_{d \geq 2}d!^{n_d}}\sum_{\{l_1, l_2, l_3\} \subset [D]}\hspace*{5px}\sum_{\mathcal{M}\textrm{ matching of }[D] - \{l_1, l_2, l_3\}}\widetilde{\nu}_{\alpha,\,\beta}^{(6,\,r),\,\left(n_d\right)_{d \geq 2},\,l_1,\,l_2,\,l_3,\,\mathcal{M}},\\
    \widetilde{\nu}_{\alpha,\,\beta}^{(6,\,1),\,\left(n_d\right)_{d \geq 2},\,l_1,\,l_2,\,l_3,\,\mathcal{M}} & := \lambda^D\sum_{\bm{\alpha}_{\mathcal{D}(\mathcal{M})} \in \mathcal{A}^{\mathcal{D}(\mathcal{M})}}\left[\pi_{l_1,\,l_2,\,l_3,\,\mathcal{M}} \cdot \bigotimes_{d \geq 2}\bm{\delta G}^{(2d)\otimes n_d}\right]_{\alpha,\,\beta,\,\beta,\,\bm{\alpha}_{\mathcal{D}(\mathcal{M})},\,\bm{\alpha}_{\mathcal{D}(\mathcal{M})}}\left[\bm{\mathcal{G}}^{\otimes 2\left|\mathcal{D}(\mathcal{M})\right|}\right]_{\bm{\alpha}_{\mathcal{D}(\mathcal{M})}},\\
    \widetilde{\nu}_{\alpha,\,\beta}^{(6,\,2),\,\left(n_d\right)_{d \geq 2},\,l_1,\,l_2,\,l_3,\,\mathcal{M}} & := \lambda^D\sum_{\bm{\alpha}_{\mathcal{D}(\mathcal{M})} \in \mathcal{A}^{\mathcal{D}(\mathcal{M})}}\left[\pi_{l_1,\,l_2,\,l_3,\,\mathcal{M}} \cdot \bigotimes_{d \geq 2}\bm{\delta G}^{(2d)\otimes n_d}\right]_{\beta,\,\alpha,\,\alpha,\,\bm{\alpha}_{\mathcal{D}(\mathcal{M})},\,\bm{\alpha}_{\mathcal{D}(\mathcal{M})}}\left[\bm{\mathcal{G}}^{\otimes 2\left|\mathcal{D}(\mathcal{M})\right|}\right]_{\bm{\alpha}_{\mathcal{D}(\mathcal{M})}}
\end{align}
for all $\alpha, \beta \in \mathcal{A} = \mathcal{I}^2$. Original tensors $\bm{\nu}^{(6,\,r)}$ can be written as follows in terms of $\bm{\widetilde{\nu}}^{(6,\,r)}$
\begin{align}
    \nu^{(6,\,1)}_{j_1,\,j_2,\,j_3,\,j_4} & = 2\lambda\mathcal{G}_{j_1}\mathcal{G}_{j_2}\mathcal{G}_{j_3}\mathcal{G}_{j_4}G^{(2)}_{j_1,\,j_2}\widetilde{\nu}^{(6,\,1)}_{j_1,\,j_2,\,j_3,\,j_4},\\
    \nu^{(6,\,2)}_{j_1,\,j_2,\,j_3,\,j_4} & = 2\lambda\mathcal{G}_{j_1}\mathcal{G}_{j_2}\mathcal{G}_{j_3}\mathcal{G}_{j_4}G^{(2)}_{j_3,\,j_4}\widetilde{\nu}^{(6,\,2)}_{j_1,\,j_2,\,j_3,\,j_4}
\end{align}
for all $j_1, j_2, j_3, j_4 \in \mathcal{I}$. Then, tensors $\bm{\widetilde{\nu}}^{(6,\,r)}$ admit continuum limits $\widetilde{\nu}^{(6,\,r),\,\mathrm{cont}}$, defined by series:
\begin{align}
    \widetilde{\nu}^{(6,\,r),\,\mathrm{cont}}\left(\xi_1, \xi_2\right) & := \sum_{\left(n_d\right)_{d \geq 2}}\widetilde{\nu}^{(6,\,r),\,\left(n_d\right)_{d \geq 2},\,\mathrm{cont}}\left(\xi_1, \xi_2\right),\\
    \widetilde{\nu}^{(6,\,r),\,\left(n_d\right)_{d \geq 2},\,\mathrm{cont}}\left(\xi_1, \xi_2\right) & := \binom{n}{\left(n_d\right)_{d \geq 2}}\frac{n^{-D/2}}{\prod\limits_{d \geq 2}d!^{n_d}}\sum_{l \in [D]}\hspace*{5px}\sum_{\mathcal{M}\textrm{ matching of }[D] - \{l\}}\widetilde{\nu}^{(6,\,r),\,\left(n_d\right)_{d \geq 2},\,l,\,\mathcal{M},\,\mathrm{cont}}\left(\xi_1, \xi_2\right),\\
    \widetilde{\nu}^{(6,\,1),\,\left(n_d\right)_{d \geq 2},\,l_1,\,l_2,\,l_3,\,\mathcal{M},\,\mathrm{cont}}\left(\xi_1, \xi_2\right) & := \left(\frac{\gamma_{\mathrm{max}}}{\sqrt{2}}\right)^D\int\limits_{\left([0, 2]^2\right)^{\mathcal{D}(\mathcal{M})}}\!\mathrm{d}\bm{\xi}\,\left(\pi_{l_1,\,l_2,\,l_3,\,\mathcal{M}} \cdot \bigotimes_{d \geq 2}\left(\delta G^{(2d),\,\mathrm{cont}}\right)^{\otimes n_d}\right)\left(\xi_1, \xi_2, \xi_2, \bm{\xi}, \bm{\xi}\right)\nonumber\\
    & \hspace*{120px} \times \left(\mathcal{G}^{\mathrm{cont}}\right)^{\otimes 2\left|\mathcal{D}(\mathcal{M})\right|}\left(\bm{\xi}\right),\\
    \widetilde{\nu}^{(6,\,2),\,\left(n_d\right)_{d \geq 2},\,l_1,\,l_2,\,l_3,\,\mathcal{M},\,\mathrm{cont}}\left(\xi_1, \xi_2\right) & := \left(\frac{\gamma_{\mathrm{max}}}{\sqrt{2}}\right)^D\int\limits_{\left([0, 2]^2\right)^{\mathcal{D}(\mathcal{M})}}\!\mathrm{d}\bm{\xi}\,\left(\pi_{l_1,\,l_2,\,l_3,\,\mathcal{M}} \cdot \bigotimes_{d \geq 2}\left(\delta G^{(2d),\,\mathrm{cont}}\right)^{\otimes n_d}\right)\left(\xi_2, \xi_1, \xi_1, \bm{\xi}, \bm{\xi}\right)\nonumber\\
    & \hspace*{120px} \times \left(\mathcal{G}^{\mathrm{cont}}\right)^{\otimes 2\left|\mathcal{D}(\mathcal{M})\right|}\left(\bm{\xi}\right)
\end{align}
for all $\xi_1, \xi_2 \in [0, 2]^2$, where we let $D := \sum_{d \geq 2}dn_d$ as an implicit function of $\left(n_d\right)_{d \geq 2}$. The following uniform and discretization bounds hold over $\bm{\widetilde{\nu}}^{(6,\,r)}$ and their continuum counterparts $\widetilde{\nu}^{(6,\,r),\,\mathrm{cont}}$:
\begin{align}
    \left\lVert \bm{\widetilde{\nu}}^{(6,\,r)} \right\rVert_{\infty} & \leq \frac{\mathcal{O}(1)}{\left(p + 1\right)^3}\gamma_{\mathrm{max}}^2\max\left(4\beta_{\mathrm{max}}, \frac{2M_{\gamma}}{\gamma_{\mathrm{max}}}, \gamma'_{\mathrm{max}}\right),\\
    \left\lVert \widetilde{\nu}^{(6,\,r),\,\mathrm{cont}} \right\rVert_{\infty} & \leq \mathcal{O}(1)\gamma_{\mathrm{max}}^2\max\left(4\beta_{\mathrm{max}}, \frac{2M_{\gamma}}{\gamma_{\mathrm{max}}}, \gamma'_{\mathrm{max}}\right),\\
    \left|\widetilde{\nu}^{(6,\,r)}_{\alpha,\,\beta} - \frac{1}{(p + 1)^3}\widetilde{\nu}^{(6,\,r),\,\mathrm{cont}}\left(\frac{\alpha}{p + 1/2}, \frac{\beta}{p + 1/2}\right)\right| & \leq \frac{\mathcal{O}(1)}{\left(p + 1\right)^4}\gamma_{\mathrm{max}}^2\max\left(4\beta_{\mathrm{max}}, \frac{2M_{\gamma}}{\gamma_{\mathrm{max}}}, \gamma'_{\mathrm{max}}\right)
\end{align}
for all $\alpha, \beta \in \mathcal{A} = \mathcal{I}^2$.
\end{proposition}

\begin{proposition}[Continuum limit of quartic QGMS moment contribution $\bm{\nu}^{(7)}$]
\label{prop:nu7_continuum_limit}
Starting from the series expansion of $\bm{\nu}^{(6)}$ established in Proposition~\ref{prop:qgms_diagonal_quartic_moment_contributions_series_expansions}, let us write:
\begin{align}
    \nu^{(7)}_{\alpha,\,\beta} & := \sum_{\left(n_d\right)_{d \geq 2}}\binom{n}{\left(n_d\right)_{d \geq 2}}\frac{n^{-D/2}}{\prod\limits_{d \geq 2}d!^{n_d}}\left\langle \bm{\mathcal{T}}^{(7),\,\left(D,\,\alpha,\,\beta\right)}, \bigotimes_{d \geq 2}\bm{\delta C}^{(d)\otimes n_d} \right\rangle,\\
    \bm{\mathcal{T}}^{(7),\,\left(D,\,\alpha,\,\beta\right)} & := \bm{\mathcal{I}}^{\left(D,\,\alpha,\,\alpha,\,\beta,\,\beta\right)},
\end{align}
where we let $D := \sum_{d \geq 2}dn_d$ as an implicit function of $\left(n_d\right)_{d \geq 2}$. Define auxiliary tensor $\bm{\widetilde\nu}^{(7)} = \left(\widetilde{\nu}^{(7)}_{\alpha,\,\beta}\right)_{\alpha,\,\beta \in \mathcal{A}}$ by:
\begin{align}
    \widetilde{\nu}^{(7)}_{\alpha,\,\beta} & := \sum_{\left(n_d\right)_{d \geq 2}}\widetilde{\nu}_{\alpha,\,\beta}^{(7),\,\left(n_d\right)_{d \geq 2}},\\
    \widetilde{\nu}_{\alpha,\,\beta}^{(7),\,\left(n_d\right)_{d \geq 2}} & := \binom{n}{\left(n_d\right)_{d \geq 2}}\frac{n^{-D/2}}{\prod\limits_{d \geq 2}d!^{n_d}}\sum_{\{l_1, l_2, l_3, l_4\} \subset [D]}\hspace*{5px}\sum_{\mathcal{M}\textrm{ matching of }[D] - \{l_1, l_2, l_3, l_4\}}\widetilde{\nu}_{\alpha,\,\beta}^{(7),\,\left(n_d\right)_{d \geq 2},\,l_1,\,l_2,\,l_3,\,l_4,\,\mathcal{M}},\\
    \widetilde{\nu}_{\alpha,\,\beta}^{(7),\,\left(n_d\right)_{d \geq 2},\,l_1,\,l_2,\,l_3,\,l_4,\,\mathcal{M}} & := \lambda^D\sum_{\bm{\alpha}_{\mathcal{D}(\mathcal{M})} \in \mathcal{A}^{\mathcal{D}(\mathcal{M})}}\left[\pi_{l_1,\,l_2,\,l_3,\,l_4,\,\mathcal{M}} \cdot \bigotimes_{d \geq 2}\bm{\delta G}^{(2d)\otimes n_d}\right]_{\alpha,\,\alpha,\,\beta,\,\beta,\,\bm{\alpha}_{\mathcal{D}(\mathcal{M})},\,\bm{\alpha}_{\mathcal{D}(\mathcal{M})}}\left[\bm{\mathcal{G}}^{\otimes 2\left|\mathcal{D}(\mathcal{M})\right|}\right]_{\bm{\alpha}_{\mathcal{D}(\mathcal{M})}}
\end{align}
for all $\alpha, \beta \in \mathcal{A} = \mathcal{I}^2$. Original tensor $\bm{\nu}^{(7)}$ can be written as follows in terms of $\bm{\widetilde{\nu}}^{(7)}$
\begin{align}
    \nu^{(7)}_{j_1,\,j_2,\,j_3,\,j_4} & = \mathcal{G}_{j_1}\mathcal{G}_{j_2}\mathcal{G}_{j_3}\mathcal{G}_{j_4}G^{(2)}_{j_1,\,j_2}\widetilde{\nu}^{(7)}_{j_1,\,j_2,\,j_3,\,j_4}
\end{align}
for all $j_1, j_2, j_3, j_4 \in \mathcal{I}$. Then, tensor $\bm{\widetilde{\nu}}^{(7)}$ admits continuum limit $\widetilde{\nu}^{(7),\,\mathrm{cont}}$, defined by series:
\begin{align}
    \widetilde{\nu}^{(7),\,\mathrm{cont}}\left(\xi_1, \xi_2\right) & := \sum_{\left(n_d\right)_{d \geq 2}}\widetilde{\nu}^{(7),\,\left(n_d\right)_{d \geq 2},\,\mathrm{cont}}\left(\xi_1, \xi_2\right),\\
    \widetilde{\nu}^{(7),\,\left(n_d\right)_{d \geq 2},\,\mathrm{cont}}\left(\xi_1, \xi_2\right) & := \binom{n}{\left(n_d\right)_{d \geq 2}}\frac{n^{-D/2}}{\prod\limits_{d \geq 2}d!^{n_d}}\sum_{\{l_1, l_2, l_3, l_4\} \subset [D]}\hspace*{5px}\sum_{\substack{\mathcal{M}\textrm{ matching}\\\textrm{of }[D] - \{l_1, l_2, l_3, l_4\}}}\hspace*{-20px}\widetilde{\nu}^{(7),\,\left(n_d\right)_{d \geq 2},\,l_1,\,l_2,\,l_3,\,l_4,\,\mathcal{M},\,\mathrm{cont}}\left(\xi_1, \xi_2\right),\\
    \widetilde{\nu}^{(7),\,\left(n_d\right)_{d \geq 2},\,l_1,\,l_2,\,l_3,\,l_4,\,\mathcal{M},\,\mathrm{cont}}\left(\xi_1, \xi_2\right) & := \left(\frac{\gamma_{\mathrm{max}}}{\sqrt{2}}\right)^D\int\limits_{\left([0, 2]^2\right)^{\mathcal{D}(\mathcal{M})}}\!\mathrm{d}\bm{\xi}\,\left(\pi_{l_1,\,l_2,\,l_3,\,l_4,\,\mathcal{M}} \cdot \bigotimes_{d \geq 2}\left(\delta G^{(2d),\,\mathrm{cont}}\right)^{\otimes n_d}\right)\left(\xi_1, \xi_1, \xi_2, \xi_2, \bm{\xi}, \bm{\xi}\right)\nonumber\\
    & \hspace*{120px} \times \left(\mathcal{G}^{\mathrm{cont}}\right)^{\otimes 2\left|\mathcal{D}(\mathcal{M})\right|}\left(\bm{\xi}\right)
\end{align}
for all $\xi_1, \xi_2 \in [0, 2]^2$, where we let $D := \sum_{d \geq 2}dn_d$ as an implicit function of $\left(n_d\right)_{d \geq 2}$. The following uniform and discretization bounds hold over $\bm{\widetilde{\nu}}^{(7),\,\mathrm{cont}}$:
\begin{align}
    \left\lVert \bm{\widetilde{\nu}}^{(7)} \right\rVert_{\infty} & \leq \frac{\mathcal{O}(1)}{\left(p + 1\right)^4}\gamma_{\mathrm{max}}^2\max\left(4\beta_{\mathrm{max}}, \frac{2M_{\gamma}}{\gamma_{\mathrm{max}}}, \gamma'_{\mathrm{max}}\right),\\
    \left\lVert \widetilde{\nu}^{(7),\,\mathrm{cont}} \right\rVert_{\infty} & \leq \mathcal{O}(1)\gamma_{\mathrm{max}}^2\max\left(4\beta_{\mathrm{max}}, \frac{2M_{\gamma}}{\gamma_{\mathrm{max}}}, \gamma'_{\mathrm{max}}\right),\\
    \left|\widetilde{\nu}^{(7)}_{\alpha,\,\beta} - \frac{1}{(p + 1)^4}\widetilde{\nu}^{(7),\,\mathrm{cont}}\left(\frac{\alpha}{p + 1/2}, \frac{\beta}{p + 1/2}\right)\right| & \leq \frac{\mathcal{O}(1)}{\left(p + 1\right)^5}\gamma_{\mathrm{max}}^2\max\left(4\beta_{\mathrm{max}}, \frac{2M_{\gamma}}{\gamma_{\mathrm{max}}}, \gamma'_{\mathrm{max}}\right)
\end{align}
for all $\alpha, \beta \in \mathcal{A} = \mathcal{I}^2$.
\end{proposition}

We illustrate how the derivation of these continuum limits and uniform bounds proceeds for the example of tensors $\bm{\nu}^{(6)}$ and $\bm{\nu}^{(7)}$. This follows the same lines as for the analysis of $\bm{\nu}^{(1)}$ in Section~\ref{sec:continuum_limit_qgms_moments}.
\\
\paragraph{Analysis of $\bm{\nu}^{(6)}$}\mbox{}

We start with the analysis of $\bm{\nu}^{(6)}$. The start by rewriting its series expansion from Proposition~\ref{prop:qgms_diagonal_quartic_moment_contributions_series_expansions} as:
\begin{align}
    \nu^{(6)}_{\alpha,\,\beta} & = \nu^{(6,\,1)}_{\alpha,\,\beta} + \nu^{(6,\,2)}_{\alpha,\,\beta},\\
    \nu^{(6,\,r)}_{\alpha,\,\beta} & = \sum_{\left(n_d\right)_{d \geq 2}}\nu_{\alpha,\,\beta}^{(6,\,r),\,\left(n_d\right)_{d \geq 2}},\\
    \nu^{(6,\,r),\,\left(n_d\right)_{d \geq 2}}_{\alpha,\,\beta} & := \binom{n}{\left(n_d\right)_{d \geq 2}}\frac{n^{-\sum\limits_{d \geq 2}dn_d/2}}{\prod\limits_{d \geq 2}d!^{n_d}}\theta^*_{\alpha}\left\langle \bm{\mathcal{T}}^{(6,\,r),\,\left(\sum\limits_{d \geq 2}dn_d,\,\alpha,\,\beta\right)}, \bigotimes_{d \geq 2}\bm{\delta C}^{(d)\otimes n_d} \right\rangle,\\
    \bm{\mathcal{T}}^{(6,\,1),\,\left(D,\,\alpha,\,\beta\right)} & := 2\theta^*_{\alpha}\bm{\mathcal{I}}^{\left(D,\,\alpha,\,\beta,\,\beta\right)},\\
    \bm{\mathcal{T}}^{(6,\,2),\,\left(D,\,\alpha,\,\beta\right)} & := 2\theta^*_{\beta}\bm{\mathcal{I}}^{\left(D,\,\beta,\,\alpha,\,\alpha\right)}.
\end{align}
Decomposing matching tensors with external edges according to their Definition~\ref{def:matching_tensor_with_edges}, a multinomial number contribution to $\bm{\nu}^{(6)}$ can be decomposed (letting $D := \sum_{d \geq 2}dn_d$):
\begin{align}
    \nu_{\alpha,\,\beta}^{(6,\,r),\,\left(n_d\right)_{d \geq 2}} & = \binom{n}{\left(n_d\right)_{d \geq 2}}\frac{n^{D/2}}{\prod\limits_{d \geq 2}d!^{n_d}}\sum_{\substack{\{l_1, l_2, l_3\} \subset [D]}}\hspace*{5px}\sum_{\mathcal{M}\textrm{ matching of }[D] - \{l_1, l_2, l_3\}}\nu_{\alpha,\,\beta}^{(6, r),\,\left(n_d\right)_{d \geq 2},\,l_1,\,l_2,\,l_3,\,\mathcal{M}},\\
    \nu_{\alpha,\,\beta}^{(6,\,1),\,\left(n_d\right)_{d \geq 2},\,l_1,\,l_2,\,l_3} & := \theta^*_{\alpha}\sum_{\bm{\alpha}_{\mathcal{D}(\mathcal{M})} \in \mathcal{A}^{\mathcal{D}(\mathcal{M})}}\left[\pi_{l_1,\,l_2,\,l_3,\,\mathcal{M}} \cdot \bigotimes_{d \geq 2}\bm{\delta C}^{(d)\otimes n_d}\right]_{\alpha,\,\beta,\,\beta,\,\bm{\alpha}_{\mathcal{D}(\mathcal{M})},\,\bm{\alpha}_{\mathcal{D}(\mathcal{M})}},\\
    \nu_{\alpha,\,\beta}^{(6,\,2),\,\left(n_d\right)_{d \geq 2},\,l_1,\,l_2,\,l_3} & := \theta^*_{\beta}\sum_{\bm{\alpha}_{\mathcal{D}(\mathcal{M})} \in \mathcal{A}^{\mathcal{D}(\mathcal{M})}}\left[\pi_{l_1,\,l_2,\,l_3,\,\mathcal{M}} \cdot \bigotimes_{d \geq 2}\bm{\delta C}^{(d)\otimes n_d}\right]_{\beta,\,\alpha,\,\alpha,\,\bm{\alpha}_{\mathcal{D}(\mathcal{M})},\,\bm{\alpha}_{\mathcal{D}(\mathcal{M})}}
\end{align}
Reexpressing (centered) correlations tensors in terms of (centered) $\bm{G}$ correlations tensors, the two additive contributions to $\bm{\nu}^{(6)}$ can be expressed:
\begin{align}
    \nu^{(6,\,1)}_{j_1,\,j_2,\,j_3,\,j_4} & = 2\lambda\mathcal{G}_{j_1}\mathcal{G}_{j_2}\mathcal{G}_{j_3}\mathcal{G}_{j_4}G^{(2)}_{j_1,\,j_2}\widetilde{\nu}^{(6,\,1)}_{j_1,\,j_2,\,j_3,\,j_4},\\
    \nu^{(6,\,2)}_{\alpha,\,\beta} & = 2\lambda\mathcal{G}_{j_1}\mathcal{G}_{j_2}\mathcal{G}_{j_3}\mathcal{G}_{j_4}G^{(2)}_{j_3,\,j_4}\widetilde{\nu}^{(6,\,2)}_{j_1,\,j_2,\,j_3,\,j_4},
\end{align}
for all $j_1, j_2, j_3, j_4 \in \mathcal{I}$, where the tilded $\bm{\nu}^{(6)}$ tensors are defined by:
\begin{align}
    \widetilde{\nu}^{(6,\,r)}_{\alpha,\,\beta} & = \sum_{\left(n_d\right)_{d \geq 2}}\widetilde{\nu}_{\alpha,\,\beta}^{(6,\,r),\,\left(n_d\right)_{d \geq 2}},\\
    \widetilde{\nu}_{\alpha,\,\beta}^{(6,\,r),\,\left(n_d\right)_{d \geq 2}} & = \binom{n}{\left(n_d\right)_{d \geq 2}}\frac{n^{-D/2}}{\prod\limits_{d \geq 2}d!^{n_d}}\sum_{\substack{\{l_1, l_2, l_3\} \subset [D]}}\hspace*{5px}\sum_{\mathcal{M}\textrm{ matching of }[D] - \{l_1, l_2, l_3\}}\widetilde{\nu}_{\alpha,\,\beta}^{(6,\,r),\,\left(n_d\right)_{d \geq 2},\,l_1,\,l_2,\,l_3,\,\mathcal{M}},\\
    \widetilde{\nu}_{j_1,\,j_2,\,j_3,\,j_4}^{(6,\,1),\,\left(n_d\right)_{d \geq 2},\,l_1,\,l_2,\,l_3,\,\mathcal{M}} & := \lambda^D\sum_{\bm{\alpha}_{\mathcal{D}(\mathcal{M})} \in \mathcal{A}^{\mathcal{D}(\mathcal{M})}}\left[\pi_{l_1,\,l_2,\,l_3,\,\mathcal{M}} \cdot \bigotimes_{d \geq 2}\bm{\delta G}^{(2d)\otimes n_d}\right]_{\alpha,\,\beta,\,\beta,\,\bm{\alpha}_{\mathcal{D}(\mathcal{M})},\,\bm{\alpha}_{\mathcal{D}(\mathcal{M})}}\left[\bm{\mathcal{G}}^{\otimes 2\left|\mathcal{D}(\mathcal{M})\right|}\right]_{\bm{\alpha}_{\mathcal{D}(\mathcal{M})}},\\
    \widetilde{\nu}_{\alpha,\,\beta}^{(6,\,2),\,\left(n_d\right)_{d \geq 2},\,l_1,\,l_2,\,l_3,\,\mathcal{M}} & := \lambda^D\sum_{\bm{\alpha}_{\mathcal{D}(\mathcal{M})} \in \mathcal{A}^{\mathcal{D}(\mathcal{M})}}\left[\pi_{l_1,\,l_2,\,l_3,\,\mathcal{M}} \cdot \bigotimes_{d \geq 2}\bm{\delta G}^{(2d)\otimes n_d}\right]_{\beta,\,\alpha,\,\alpha,\,\bm{\alpha}_{\mathcal{D}(\mathcal{M})},\,\bm{\alpha}_{\mathcal{D}(\mathcal{M})}}\left[\bm{\mathcal{G}}^{\otimes 2\left|\mathcal{D}(\mathcal{M})\right|}\right]_{\bm{\alpha}_{\mathcal{D}(\mathcal{M})}},
\end{align}
for all $\alpha, \beta \in \mathcal{A}$. With this decomposition at hand, the uniform bounds and continuum limit result established in Proposition~\ref{prop:qgms_integral_series_expansion_multinomial_matching_contribution_continuum_approximation} for a single matching-indexed contribution of $\bm{\nu}^{(1)}$ generalize straightforwardly to $\bm{\nu}^{(6)}$. The only difference is the relation of $\left|\mathcal{D}(\mathcal{M})\right|$ to $D$, which is now $\left|\mathcal{D}(\mathcal{M})\right| = (D - 3)/2$ instead of $\left|\mathcal{D}(\mathcal{M})\right| := (D - 1)/2$. This gives uniform and discretization bounds:
\begin{align}
    \left\lVert \bm{\widetilde{\nu}}^{(6,\,r),\,\left(n_d\right)_{d \geq 2},\,l_1,\,l_2,\,l_3,\,\mathcal{M}} \right\rVert_{\infty} & \leq \frac{\left(3\sqrt{2}\gamma_{\mathrm{max}}\right)^D}{\left(2p + 2\right)^3}2^{\sum\limits_{d \geq 2}n_d},\\
    \left\lVert \widetilde{\nu}^{(6,\,r),\,\left(n_d\right)_{d \geq 2},\,l_1,\,l_2,\,l_3,\,\mathcal{M},\,\mathrm{cont}} \right\rVert_{\infty} & \leq \frac{\left(3\sqrt{2}\gamma_{\mathrm{max}}\right)^D}{8}2^{\sum\limits_{d \geq 2}n_d},\\
    \left|\widetilde{\nu}^{(6,\,r),\,\left(n_d\right)_{d \geq 2},\,l_1,\,l_2,\,l_3,\,\mathcal{M}}_{k_1,\,k_2,\,k_3,\,k_4} - \frac{1}{(p + 1)^3}\widetilde{\nu}^{(6,\,r),\,\left(n_d\right)_{d \geq 2},\,l_1,\,l_2,\,l_3,\,\mathcal{M},\,\mathrm{cont}}\left(\frac{\bm{k}_{1:4}}{p + 1/2}\right)\right| & \leq \frac{\left(12\gamma_{\mathrm{max}}\right)^DD}{\left(p + 1\right)^4}\max\left(4\beta_{\mathrm{max}}, \frac{2M_{\gamma}}{\gamma_{\mathrm{max}}}, \gamma'_{\mathrm{max}}\right)
\end{align}
for all $\bm{k}_{1:4} \in \mathcal{I}^4 = \mathcal{A}^2$. In the above equation, the continuum limits of the matching-indexed contributions are explicitly defined by:
\begin{align}
    & \widetilde{\nu}^{(6,\,1),\,\left(n_d\right)_{d \geq 2},\,l_1,\,l_2,\,l_3,\,\mathcal{M},\,\mathrm{cont}}\left(x_1, x_2, x_3, x_4\right)\nonumber\\
    & := \left(2^{-1/2}\gamma_{\mathrm{max}}\right)^D\int\limits_{\left([0, 2]^2\right)^{\mathcal{D}\left(\mathcal{M}\right)}}\!\mathrm{d}\bm{\xi}\,\left(\pi_{l_1,\,l_2,\,l_3,\,\mathcal{M}} \cdot \bigotimes_{d \geq 2}\bm{\delta G}^{(2d)\otimes n_d}\right)\left(x_1, x_2,x_3, x_4, x_3, x_4, \bm{\xi}, \bm{\xi}\right)\left(\mathcal{G}^{\mathrm{cont}}\right)^{\otimes 2\left|\mathcal{D}(\mathcal{M})\right|}\left(\bm{\xi}\right),
\end{align}
and
\begin{align}
    & \widetilde{\nu}^{(6,\,2),\,\left(n_d\right)_{d \geq 2},\,l_1,\,l_2,\,l_3,\,\mathcal{M},\,\mathrm{cont}}\left(x_1, x_2, x_3, x_4\right)\nonumber\\
    & := \left(2^{-1/2}\gamma_{\mathrm{max}}\right)^D\int\limits_{\left([0, 2]^2\right)^{\mathcal{D}\left(\mathcal{M}\right)}}\!\mathrm{d}\bm{\xi}\,\left(\pi_{l_1,\,l_2,\,l_3,\,\mathcal{M}} \cdot \bigotimes_{d \geq 2}\bm{\delta G}^{(2d)\otimes n_d}\right)\left(x_3, x_4, x_1, x_2, x_1, x_2, \bm{\xi}, \bm{\xi}\right)\left(\mathcal{G}^{\mathrm{cont}}\right)^{\otimes 2\left|\mathcal{D}(\mathcal{M})\right|}\left(\bm{\xi}\right),
\end{align}
where $x_1, x_2, x_3, x_4 \in [0, 2]$. Adding these bounds over $l_1, l_2, l_3$ and $\mathcal{M}$ the produces similar bounds similar to those of Corollary~\ref{cor:qgms_integral_series_expansion_contributions_continuum_approximation} for each multinomial number-indexed contribution:
\begin{align}
    \left\lVert \bm{\widetilde{\nu}}^{(6,\,r),\,\left(n_d\right)_{d \geq 2}} \right\rVert_{\infty} & \leq \binom{D}{3}(D - 4)!!\frac{\left(3\sqrt{2}\gamma_{\mathrm{max}}\right)^D}{\left(2p + 2\right)^3}2^{\sum\limits_{d \geq 2}n_d},\\
    \left\lVert \widetilde{\nu}^{(6,\,r),\,\left(n_d\right)_{d \geq 2},\,\mathrm{cont}} \right\rVert_{\infty} & \leq \binom{D}{3}(D - 4)!!\frac{\left(3\sqrt{2}\gamma_{\mathrm{max}}\right)^D}{8}2^{\sum\limits_{d \geq 2}n_d},
\end{align}
and
\begin{align}
    & \left|\widetilde{\nu}^{(6,\,r),\,\left(n_d\right)_{d \geq 2}}_{\alpha,\,\beta} - \frac{1}{\left(p + 1\right)^3}\widetilde{\nu}^{(6,\,r),\,\left(n_d\right)_{d \geq 2},\,\mathrm{cont}}\left(\frac{\alpha}{p + 1/2}, \frac{\beta}{p + 1/2}\right)\right|\nonumber\\
    & \leq \binom{D}{3}(D - 4)!!\frac{n^{-D/2}}{\prod\limits_{d \geq 2}d!^{n_d}}\frac{\left(12\gamma_{\mathrm{max}}\right)^DD}{\left(p + 1\right)^4}\max\left(4\beta_{\mathrm{max}}, \frac{2M_{\gamma}}{\gamma_{\mathrm{max}}}, \gamma'_{\mathrm{max}}\right)\nonumber\\
    & = \frac{D - 1}{6}D!!\frac{n^{-D/2}}{\prod\limits_{d \geq 2}d!^{n_d}}\frac{\left(12\gamma_{\mathrm{max}}\right)^DD}{\left(p + 1\right)^4}\max\left(4\beta_{\mathrm{max}}, \frac{2M_{\gamma}}{\gamma_{\mathrm{max}}}, \gamma'_{\mathrm{max}}\right)\nonumber\\
    & \leq \frac{1}{6}D!!\frac{n^{-D/2}}{\prod\limits_{d \geq 2}d!^{n_d}}\frac{\left(12\gamma_{\mathrm{max}}\right)^DD^2}{\left(p + 1\right)^4}\max\left(4\beta_{\mathrm{max}}, \frac{2M_{\gamma}}{\gamma_{\mathrm{max}}}, \gamma'_{\mathrm{max}}\right)
\end{align}
for all $\alpha, \beta \in \mathcal{I}$. Similar to the proof of Corollary~\ref{cor:qgms_integral_series_expansion_contributions_continuum_approximation}, we can then invoke Lemma~\ref{lemma:qgms_integral_moment_series_multinomial_sum_bound} to sum these bounds over multinomial numbers $\left(n_d\right)_{d \geq 2}$. Note this summation indeed excludes $\left(n_d\right)_{d \geq 2} = \bm{0}$ as required by the Lemma, since then $D = \sum_{d \geq 2}dn_d = 0$, which gives a vanishing series term. Application of the Lemma then provides bounds:
\begin{align}
    \left\lVert \bm{\widetilde{\nu}}^{(6,\,r)} \right\rVert_{\infty} & \leq \frac{\mathcal{O}(1)}{\left(p + 1\right)^3}\gamma_{\mathrm{max}}^2\max\left(4\beta_{\mathrm{max}}, \frac{2M_{\gamma}}{\gamma_{\mathrm{max}}}, \gamma'_{\mathrm{max}}\right),\\
    \left\lVert \widetilde{\nu}^{(6,\,r),\,\mathrm{cont}} \right\rVert_{\infty} & \leq \mathcal{O}(1)\gamma_{\mathrm{max}}^2\max\left(4\beta_{\mathrm{max}}, \frac{2M_{\gamma}}{\gamma_{\mathrm{max}}}, \gamma'_{\mathrm{max}}\right),\\
    \left|\widetilde{\nu}^{(6,\,r)}_{\alpha,\,\beta} - \frac{1}{\left(p + 1\right)^3}\widetilde{\nu}^{(6,\,r),\,\mathrm{cont}}\left(\frac{\alpha}{p + 1/2}, \frac{\beta}{p + 1/2}\right)\right| & \leq \frac{\mathcal{O}(1)}{\left(p + 1\right)^4}\gamma_{\mathrm{max}}^2\max\left(4\beta_{\mathrm{max}}, \frac{2M_{\gamma}}{\gamma_{\mathrm{max}}}, \gamma'_{\mathrm{max}}\right),
\end{align}
Letting $\widetilde{\nu}^{(6),\,\mathrm{cont}} := \widetilde{\nu}^{(6,\,0),\,\mathrm{cont}} + \widetilde{\nu}^{(6,\,1),\,\mathrm{cont}}$, the following uniform and discretization bounds for $\bm{\widetilde{\nu}}^{(6)}$ immediately follows:
\begin{align}
    \left\lVert \bm{\widetilde{\nu}}^{(6)} \right\rVert_{\infty} & \leq \frac{\mathcal{O}(1)}{\left(p + 1\right)^3}\gamma_{\mathrm{max}}^2\max\left(4\beta_{\mathrm{max}}, \frac{2M_{\gamma}}{\gamma_{\mathrm{max}}}, \gamma'_{\mathrm{max}}\right),\label{eq:nu6_uniform_bound}\\
    \left\lVert \widetilde{\nu}^{(6),\,\mathrm{cont}} \right\rVert_{\infty} & \leq \mathcal{O}(1)\gamma_{\mathrm{max}}^2\max\left(4\beta_{\mathrm{max}}, \frac{2M_{\gamma}}{\gamma_{\mathrm{max}}}, \gamma'_{\mathrm{max}}\right),\label{eq:nu6_continuum_uniform_bound}\\
    \left| \widetilde{\nu}^{(6)}_{\alpha,\,\beta} - \frac{1}{\left(p + 1\right)^3}\widetilde{\nu}^{(6),\,\mathrm{cont}}\left(\frac{\alpha}{p + 1/2}, \frac{\beta}{p + 1/2}\right) \right| & \leq \frac{\mathcal{O}(1)}{\left(p + 1\right)^4}\gamma_{\mathrm{max}}^2\max\left(4\beta_{\mathrm{max}}, \frac{2M_{\gamma}}{\gamma_{\mathrm{max}}}, \gamma'_{\mathrm{max}}\right).\label{eq:nu6_discretization_bound}
\end{align}
Using the relation between $\bm{\nu}^{(6,\,r)}$, $\bm{\widetilde{\nu}}^{(6,\,r)}$ as well as:
\begin{align}
    \left\lVert \bm{\mathcal{G}} \right\rVert_{\infty} \leq 1, &&  \left\lVert \bm{G}^{(2)} \right\rVert_{\infty} \leq 2, &&\lambda = \frac{2^{-1/2}\gamma_{\mathrm{max}}}{p + 1},
\end{align}
the uniform bounds on $\bm{\widetilde{\nu}}^{(6,\,r)}$ imply, for all $j_1, j_2, j_3, j_4 \in \mathcal{I}$:
\begin{align}
    \left|\nu^{(6)}_{j_1,\,j_2,\,j_3,\,j_4}\right| & = \left|\nu^{(6,\,1)}_{j_1,\,j_2,\,j_3,\,j_4} + \nu^{(6,\,2)}_{j_1,\,j_2,\,j_3,\,j_4}\right|\nonumber\\
    & = \left|\lambda\mathcal{G}_{j_1}\mathcal{G}_{j_2}\mathcal{G}_{j_3}\mathcal{G}_{j_4}G^{(2)}_{j_1,\,j_2}\widetilde{\nu}^{(6,\,1)}_{j_1,\,j_2,\,j_3,\,j_4} + \lambda\mathcal{G}_{j_1}\mathcal{G}_{j_2}\mathcal{G}_{j_3}\mathcal{G}_{j_4}G^{(2)}_{j_3,\,j_4}\widetilde{\nu}^{(6,\,2)}_{j_1,\,j_2,\,j_3,\,j_4}\right|\nonumber\\
    & \leq \frac{\mathcal{O}(1)}{\left(p + 1\right)^4}\gamma_{\mathrm{max}}^2\max\left(4\beta_{\mathrm{max}}, \frac{2M_{\gamma}}{\gamma_{\mathrm{max}}}, \gamma'_{\mathrm{max}}\right).
\end{align}
\\
\paragraph{Analysis of $\bm{\nu}^{(7)}$}\mbox{}

After sketching the analysis of $\bm{\nu}^{(6)}$ following the same approach as Section~\ref{sec:continuum_limit_qgms_moments} focusing on $\bm{\nu}^{(1)}$, we now similarly discuss the continuum limit of $\bm{\nu}^{(7)}$. Both $\bm{\nu}^{(6)}$ and $\bm{\nu}^{(7)}$ are additive contributions to the diagonal QGMS quartic moments tensor $\frac{\partial^4S_n\left(\bm{\mu}\right)}{\partial\mu_{\alpha}^2\partial\mu_{\beta}^2}\bigg|_{\bm{\mu} = \bm{0}}$. The focus will be on showing that this contribution has a similar $p$ scaling to $\bm{\nu}^{(6)}$, i.e.
\begin{align}
    \left\lVert \bm{\nu}^{(6)} \right\rVert_{\infty} & = \mathcal{O}\left(p^{-4}\right).
\end{align}
We start from the series decomposition of $\nu^{(7)}$ stated in Proposition~\ref{prop:qgms_diagonal_quartic_moment_contributions_series_expansions}, rephrased as:
\begin{align}
    \nu^{(7)}_{j_1,\,j_2,\,j_3,\,j_4} & = \mathcal{G}_{j_1}\mathcal{G}_{j_2}\mathcal{G}_{j_3}\mathcal{G}_{j_4}\widetilde{\nu}^{(7)}_{j_1,\,j_2,\,j_3,\,j_4},\\
    \widetilde{\nu}^{(7)}_{\alpha,\,\beta} & := \sum_{\left(n_d\right)_{d \geq 2}}\widetilde{\nu}_{\alpha,\,\beta}^{(7),\,\left(n_d\right)_{d \geq 2}},\\
    \widetilde{\nu}_{\alpha,\,\alpha,\,\beta,\,\beta}^{(7),\,\left(n_d\right)_{d \geq 2}} & := \binom{n}{\left(n_d\right)_{d \geq 2}}\frac{n^{-D/2}}{\prod\limits_{d \geq 2}d!^{n_d}}\sum_{\{l_1,\,l_2,\,l_3,\,l_4\} \subset [D]}\hspace*{5px}\sum_{\mathcal{M}\textrm{ matching of }[D] - \{l_1,\,l_2,\,l_3,\,l_4\}}\widetilde{\nu}_{\alpha,\,\beta}^{(7),\,\left(n_d\right)_{d \geq 2},\,l_1,\,l_2,\,l_3,\,l_4,\,\mathcal{M}},\\
    \widetilde{\nu}_{\alpha,\,\beta}^{(7),\,\left(n_d\right)_{d \geq 2},\,l_1,\,l_2,\,l_3,\,l_4,\,\mathcal{M}} & := \lambda^D\sum_{\bm{\alpha}_{\mathcal{D}(\mathcal{M})} \in \mathcal{A}^{\mathcal{D}(\mathcal{M})}}\left[\pi_{l_1,\,l_2,\,l_3,\,l_4,\,\mathcal{M}} \cdot \bigotimes_{d \geq 2}\bm{\delta G}^{(2d)\otimes n_d}\right]_{\alpha,\,\beta,\,\bm{\alpha}_{\mathcal{D}(\mathcal{M}),\,\bm{\alpha}_{\mathcal{D}(\mathcal{M})}}}\left[\mathcal{G}^{\otimes 2\left|\mathcal{D}(\mathcal{M})\right|}\right]_{\bm{\alpha}_{\mathcal{D}(\mathcal{M})}},
\end{align}
for all $j_1, j_2, j_3, j_4 \in \mathcal{I}$ and $\alpha, \beta \in \mathcal{A} = \mathcal{I}^2$. Comparing this to $\widetilde{\nu}^{(6,\,r),\,\left(n_d\right),\,l_1,\,l_2,\,l_3,\,\mathcal{M}}$, $\widetilde{\nu}_{\alpha,\,\beta}^{(7),\,\left(n_d\right)_{d \geq 2},\,l_1,\,l_2,\,l_3,\,l_4,\,\mathcal{M}}$ does not have a $\bm{\theta}^*$ prefactor but has one additional external index $l_4$; as we will see, these differences compensate each other, leading to a scaling $\left\lVert \bm{\nu}^{(7)} \right\rVert_{\infty} \leq \mathcal{O}\left(p^{-4}\right)$ identical to $\left\lVert \bm{\nu}^{(6)} \right\rVert_{\infty}$. Proposition~\ref{prop:qgms_integral_series_expansion_multinomial_matching_contribution_continuum_approximation} is now adapted by noting the new relation $\left|\mathcal{D}(\mathcal{M})\right| = (D - 4)/2$ between the matched set size and $D := \sum_{d \geq 2}dn_d$. This gives the following uniform and discretization bounds on multinomial numbers and matching-indexed contributions to $\bm{\widetilde{\nu}}^{(7)}$ as well as their continuum counterparts:
\begin{align}
    \left\lVert \bm{\widetilde{\nu}}^{(7),\,\left(n_d\right)_{d \geq 2},\,l_1,\,l_2,\,l_3,\,l_4,\,\mathcal{M}} \right\rVert_{\infty} & \leq \frac{\left(3\sqrt{2}\gamma_{\mathrm{max}}\right)^D}{\left(2p + 2\right)^4}2^{\sum\limits_{d \geq 2}n_d},\\
    \left\lVert \widetilde{\nu}^{(7),\,\left(n_d\right)_{d \geq 2},\,l_1,\,l_2,\,l_3,\,l_4,\,\mathcal{M},\,\mathrm{cont}} \right\rVert_{\infty} & \leq \frac{\left(3\sqrt{2}\gamma_{\mathrm{max}}\right)^D}{16}2^{\sum\limits_{d \geq 2}n_d},\\
    \left|\widetilde{\nu}^{(7),\,\left(n_d\right)_{d \geq 2},\,l_1,\,l_2,\,l_3,\,l_4,\,\mathcal{M}}_{k_1,\,k_2,\,k_3,\,k_4} - \frac{1}{(p + 1)^4}\widetilde{\nu}^{(7),\,\left(n_d\right)_{d \geq 2},\,l_1,\,l_2,\,l_3,\,l_4,\,\mathcal{M},\,\mathrm{cont}}\left(\frac{\bm{k}_{1:4}}{p + 1/2}\right)\right| & \leq \frac{\left(12\gamma_{\mathrm{max}}\right)^DD}{\left(p + 1\right)^5}\max\left(4\beta_{\mathrm{max}}, \frac{2M_{\gamma}}{\gamma_{\mathrm{max}}}, \gamma'_{\max}\right),
\end{align}
for all $\bm{k}_{1:4} \in \mathcal{I}^4 = \mathcal{A}^2$. These only differ from the analogous bounds for $\bm{\nu}^{(6)}$ by the power of $(p + 1)$. The continuum limit of a matching-indexed contribution is defined by:
\begin{align}
    & \widetilde{\nu}^{(7),\,\left(n_d\right)_{d \geq 2},\,l_1,\,l_2,\,l_3,\,l_4,\,\mathcal{M},\,\mathrm{cont}}\left(x_1, x_2, x_3, x_4\right)\nonumber\\
    & := \left(2^{-1/2}\gamma_{\mathrm{max}}\right)^D\int\limits_{\left([0, 2]^2\right)^{\mathcal{D}\left(\mathcal{M}\right)}}\!\mathrm{d}\bm{\xi}\,\left(\pi_{l_1,\,l_2,\,l_3,\,l_4,\,\mathcal{M}} \cdot \bigotimes_{d \geq 2}\bm{\delta G}^{(2d)\otimes n_d}\right)\left(x_1, x_2, x_1, x_2, x_3, x_4, x_3, x_4, \bm{\xi}, \bm{\xi}\right)\left(\mathcal{G}^{\mathrm{cont}}\right)^{\otimes 2\left|\mathcal{D}(\mathcal{M})\right|}\left(\bm{\xi}\right),
\end{align}
Summing these bounds over $l_1, l_2, l_3, l_4$ and matching $\mathcal{M}$ gives the following uniform and discretization bounds for the multinomial numbers-indexed contribution to $\bm{\widetilde{\nu}}^{(7)}$:
\begin{align}
    \left\lVert \bm{\widetilde{\nu}}^{(7),\,\left(n_d\right)_{d \geq 2}} \right\rVert_{\infty} & \leq \binom{D}{4}(D - 5)!!\frac{\left(3\sqrt{2}\gamma_{\mathrm{max}}\right)^D}{\left(2p + 2\right)^4}2^{\sum\limits_{d \geq 2}n_d}\\
    & \leq \frac{1}{24}(D - 1)!!D^2\frac{\left(3\sqrt{2}\gamma_{\mathrm{max}}\right)^D}{\left(2p + 2\right)^4}2^{\sum\limits_{d \geq 2}n_d},\\
    \left\lVert \widetilde{\nu}^{(7),\,\left(n_d\right)_{d \geq 2},\,\mathrm{cont}} \right\rVert_{\infty} & \leq \binom{D}{4}(D - 5)!!\frac{\left(3\sqrt{2}\gamma_{\mathrm{max}}\right)^D}{16}2^{\sum\limits_{d \geq 2}n_d}\\
    & \leq \frac{1}{24}(D - 1)!!D^2\frac{\left(3\sqrt{2}\gamma_{\mathrm{max}}\right)^D}{16}2^{\sum\limits_{d \geq 2}n_d},
\end{align}
and
\begin{align}
    & \left|\widetilde{\nu}_{\alpha,\,\beta}^{(7),\,\left(n_d\right)_{d \geq 2}} - \frac{1}{(p + 1)^4}\widetilde{\nu}^{(7),\,\left(n_d\right)_{d \geq 2},\,\mathrm{cont}}\left(\frac{\alpha}{p + 1/2}, \frac{\beta}{p + 1/2}\right)\right|\nonumber\\
    & \leq \binom{D}{4}\left(D - 5\right)!!\frac{\left(12\gamma_{\mathrm{max}}\right)^DD}{\left(p + 1\right)^5}\max\left(4\beta_{\mathrm{max}}, \frac{2M_{\gamma}}{\gamma_{\mathrm{max}}}, \gamma'_{\max}\right)\\
    & \leq \frac{1}{24}(D - 1)!!D^3\frac{\left(12\gamma_{\mathrm{max}}\right)^D}{\left(p + 1\right)^5}\max\left(4\beta_{\mathrm{max}}, \frac{2M_{\gamma}}{\gamma_{\mathrm{max}}}, \gamma'_{\max}\right).
\end{align}
Applying Lemma~\ref{lemma:qgms_integral_moment_series_multinomial_sum_bound} allows to sum these bounds over multinomial numbers $\left(n_d\right)_{d \geq 2}$ (with the all-zero tuple being excluded due to bounds involving a multiple of $D$), yielding the following bounds $\bm{\widetilde{\nu}}^{(7)}$ and its continuum counterpart:
\begin{align}
    \left\lVert \bm{\widetilde{\nu}}^{(7)} \right\rVert_{\infty} & \leq \frac{\mathcal{O}(1)}{(p + 1)^4}\gamma_{\mathrm{max}}^2\max\left(4\beta_{\mathrm{max}}, \frac{2M_{\gamma}}{\gamma_{\mathrm{max}}}, \gamma'_{\max}\right),\\
    \left\lVert \nu^{(7),\,\mathrm{cont}} \right\rVert_{\infty} & \leq \mathcal{O}(1)\gamma_{\mathrm{max}}^2\max\left(4\beta_{\mathrm{max}}, \frac{2M_{\gamma}}{\gamma_{\mathrm{max}}}, \gamma'_{\max}\right),\\
    \left|\widetilde{\nu}^{(7)}_{\alpha,\,\beta} - \frac{1}{(p + 1)^4}\widetilde{\nu}^{(7),\,\mathrm{cont}}\left(\frac{\alpha}{p + 1/2}, \frac{\beta}{p + 1/2}\right)\right| & \leq \frac{\mathcal{O}(1)}{(p + 1)^5}\gamma_{\mathrm{max}}^2\max\left(4\beta_{\mathrm{max}}, \frac{2M_{\gamma}}{\gamma_{\mathrm{max}}}, \gamma'_{\max}\right),
\end{align}
for all $\alpha, \beta \in \mathcal{A} = \mathcal{I}^2$. Recalling the relation between tensors $\bm{\nu}^{(7)}$ and $\bm{\widetilde{\nu}}^{(7)}$, the following uniform bound results on $\bm{\nu}^{(7)}$; letting $j_1, j_2, j_3, j_4 \in \mathcal{I}$,
\begin{align}
    \left|\nu^{(7)}_{j_1,\,j_2,\,j_3,\,j_4}\right| & = \left|\mathcal{G}_{j_1}\mathcal{G}_{j_2}\mathcal{G}_{j_3}\mathcal{G}_{j_4}\widetilde{\nu}_{j_1,\,j_2,\,j_3,\,j_4}^{(7)}\right|\nonumber\\
    & \leq \left|\widetilde{\nu}_{j_1,\,j_2,\,j_3,\,j_4}^{(7)}\right|\nonumber\\
    & = \mathcal{O}\left(p^{-4}\right).
\end{align}
All in all the scaling in $p$ is identical to $\left\lVert \bm{\nu}^{(7)} \right\rVert_{\infty}$. This results from the following 2 compensating changes. On the one hand, comparing the relations between tilded and untilded tensors:
\begin{align}
    \nu^{(6,\,1)}_{\alpha,\,\beta} = \lambda G^{(2)}_{\alpha}\mathcal{G}_{\alpha}\mathcal{G}_{\beta}\widetilde{\nu}^{(6,\,1)}_{\alpha,\,\beta}, && \nu^{(6,\,2)}_{\alpha,\,\beta} = \lambda G^{(2)}_{\alpha}\mathcal{G}_{\alpha}\mathcal{G}_{\beta}\widetilde{\nu}^{(6,\,2)}_{\alpha,\,\beta}, && \nu^{(7)}_{\alpha,\,\beta} = \mathcal{G}_{\alpha}\mathcal{G}_{\beta}\widetilde{\nu}^{(7)}_{\alpha,\,\beta},
\end{align}
the latter expression misses the $\lambda := 2^{-1/2}\gamma_{\mathrm{max}}/(p + 1)$ factor, causing it to be $p$ times greater. On the other hand, in the summation defining each multinomial numbers and matching-indexed contribution:
\begin{align}
     \widetilde{\nu}_{\alpha,\,\beta}^{(6,\,2),\,\left(n_d\right)_{d \geq 2},\,l_1,\,l_2,\,l_3,\,\mathcal{M}} & := \lambda^D\sum_{\bm{\alpha}_{\mathcal{D}(\mathcal{M})} \in \mathcal{A}^{\mathcal{D}(\mathcal{M})}}\left[\pi_{l_1,\,l_2,\,l_3,\,\mathcal{M}} \cdot \bigotimes_{d \geq 2}\bm{\delta G}^{(2d)\otimes n_d}\right]_{\beta,\,\alpha,\,\alpha,\,\bm{\alpha}_{\mathcal{D}(\mathcal{M})},\,\bm{\alpha}_{\mathcal{D}(\mathcal{M})}}\left[\bm{\mathcal{G}}^{\otimes 2\left|\mathcal{D}(\mathcal{M})\right|}\right]_{\bm{\alpha}_{\mathcal{D}(\mathcal{M})}},\\
    \widetilde{\nu}_{\alpha,\,\beta}^{(7),\,\left(n_d\right)_{d \geq 2},\,l_1,\,l_2,\,l_3,\,l_4,\,\mathcal{M}} & := \lambda^D\sum_{\bm{\alpha}_{\mathcal{D}(\mathcal{M})} \in \mathcal{A}^{\mathcal{D}(\mathcal{M})}}\left[\pi_{l_1,\,l_2,\,l_3,\,l_4,\,\mathcal{M}} \cdot \bigotimes_{d \geq 2}\bm{\delta G}^{(2d)\otimes n_d}\right]_{\alpha,\,\beta,\,\bm{\alpha}_{\mathcal{D}(\mathcal{M}),\,\bm{\alpha}_{\mathcal{D}(\mathcal{M})}}}\left[\mathcal{G}^{\otimes 2\left|\mathcal{D}(\mathcal{M})\right|}\right]_{\bm{\alpha}_{\mathcal{D}(\mathcal{M})}},
\end{align}
the summation has $p$ times less terms in the case of $\bm{\nu}^{(7)}$ due to the difference relation between $\left|\mathcal{D}(M)\right|$ and $D$. All in all, the additional and missing factors of $p$ compensation, leading to a bound $\mathcal{O}\left(p^{-4}\right)$ on both $\left\lVert \bm{\widetilde{\nu}}^{(6)} \right\rVert_{\infty}$ and $\left\lVert \bm{\widetilde{\nu}}^{(7)} \right\rVert_{\infty}$.

\subsection{The continuum limit of the QAOA cost function's moments and concentration}

In Section~\ref{sec:main_theorem_derivation}, we sketched the proof of our main theoretical result Theorem~\ref{th:approximation_continuous_time_annealing_qaoa}, by focusing on the analysis of a single additive contribution to the disorder-averaged QAOA energy:
\begin{align}
    \nu_{p, n} & = -\frac{i}{\Gamma_{p + 1}}\sum_{0 \leq r \leq 2p + 1}\frac{\partial^2S_n\left(\bm{\mu}\right)}{\partial\mu_{(r, p + 1)}^2}\Bigg|_{\bm{\mu} = \bm{0}}\nonumber\\
    & = -\frac{i}{\Gamma_{p + 1}}\sum_{0 \leq r \leq 2p + 1}\left(\nu^{(0)}_{(r,\,p + 1)} + \frac{2}{\sqrt{n}}\nu^{(1)}_{(r,\,p + 1)} + \frac{1}{n}\nu^{(2)}_{(r,\,p + 1)}\right)\nonumber\\
    & \supset -\frac{i}{\Gamma_{p + 1}}\sum_{0 \leq r \leq 2p + 1}\frac{2}{\sqrt{n}}\nu^{(1)}_{(r,\,p + 1)}.
\end{align}
By writing
\begin{align}
    \nu^{(1)}_{j_1,\,j_2} & = \lambda\mathcal{G}_{j_1}\mathcal{G}_{j_2}G^{(2)}_{j_1,\,j_2}\widetilde{\nu}^{(1)}_{j_1,\,j_2},\nonumber\\
    & = \frac{2^{-1/2}\gamma_{\mathrm{max}}}{p + 1}\mathcal{G}_{j_1}\mathcal{G}_{j_2}G^{(2)}_{j_1,\,j_2}\widetilde{\nu}^{(1)}_{j_1,\,j_2},
\end{align}
and invoking Corollary~\ref{cor:qgms_integral_series_expansion_contributions_continuum_approximation} for the continuum limit of $\bm{\nu}^{(1)}$, Proposition~\ref{prop:g_correlations_continuum_approximation} for the continuum limit of $\bm{G}^{(2)}$, and Definition~\ref{eq:curvy_g_vector_definition_rescaled} of $\bm{\mathcal{G}}$ from discretization of the continuous $\bm{\gamma}$ schedule, we established
\begin{align}
    -\frac{i}{\Gamma_{p + 1}}\sum_{0 \leq r \leq 2p + 1}\frac{2}{\sqrt{n}}\nu^{(1)}_{(r,\,p + 1)} & \xlongrightarrow[p \to \infty]{} \frac{i\sqrt{2}}{\gamma_{\mathrm{max}}\sqrt{n}}\int_{[0, 2]}\mathrm{d}x\,G^{(2),\,\mathrm{cont}}\left(x, 1\right)\Gamma^{\mathrm{cont}}\left(x\right)\widetilde{\nu}^{(1),\,\mathrm{cont}}\left(x, 1\right),
\end{align}
up to an error $\mathcal{O}(1)/(p\sqrt{n})$, where the implicit constant in the $\mathcal{O}(1)$ depends on the continuous schedules parameters ($\beta_{\mathrm{max}}$, $\gamma_{\mathrm{max}}$, $M_{\gamma}$), but can be taken uniform in both $n$ and $p$. By invoking Propositions~\ref{prop:nu0_continuum_limit}, \ref{prop:nu2_continuum_limit} for the continuum limits of $\bm{\nu}^{(0)}, \bm{\nu}^{(1)}$, and reproducing the reasoning from Section~\ref{sec:main_theorem_derivation}, similar statements can be made for the other two additive contributions of the disorder-average QAOA energy. The convergence results for all three contributions are recapitulated in the following Proposition:

\begin{proposition}[Continuum limit of disorder-averaged expected QAOA cost function, contribution by contribution]
\label{prop:first_order_moment_contributions_continuum_limit}
Consider the representation of the disorder-averaged expected QAOA cost function as the following 3 additive contributions:
\begin{align}
    \nu_{p, n} & := \mathbb{E}\bra{\Psi_{p,\,n}}C_n/n\ket{\Psi_{p,\,n}}\nonumber\\
    & = -\frac{i}{\Gamma_{p + 1}}\sum_{0 \leq r \leq 2p + 1}\left(\nu^{(0)}_{(r,\,p + 1)} + \frac{2}{\sqrt{n}}\nu^{(1)}_{(r,\,p + 1)} + \frac{1}{n}\nu^{(2)}_{(r,\,p + 1)}\right).
\end{align}
Then, the following $p \to \infty$ limits hold for these individual 3 contributions. For the contribution involving $\bm{\nu}^{(0)}$,
\begin{align}
    -\frac{i}{\Gamma_{p + 1}}\sum_{0 \leq r \leq 2p + 1}\nu^{(0)}_{(r,\,p + 1)} & \xlongrightarrow[p \to \infty]{} \frac{i}{2}\int_{[0, 2]}\!\mathrm{d}x\,\Gamma^{\mathrm{cont}}(x)G^{(2),\,\mathrm{cont}}(x)^2,\label{eq:first_order_moment_nu0_contribution_continuum_limit}
\end{align}
with error
\begin{align}
    \varepsilon^{(0)} & := c^{(0)}\left(\beta_{\mathrm{max}}, \gamma_{\mathrm{max}}, \frac{M_{\gamma}}{\gamma_{\mathrm{max}}}\right)\frac{\gamma_{\mathrm{max}}}{p}.
\end{align}
For the contribution involving $\bm{\nu}^{(1)}$,
\begin{align}
    -\frac{i}{\Gamma_{p + 1}}\sum_{0 \leq r \leq 2p + 1}\frac{2}{\sqrt{n}}\nu^{(1)}_{(r,\,p + 1)} & \xlongrightarrow[p \to \infty]{} \frac{i\sqrt{2}}{\gamma_{\mathrm{max}}\sqrt{n}}\int_{[0, 2]}\!\mathrm{d}x\,\Gamma^{\mathrm{cont}}\left(x\right)G^{(2),\,\mathrm{cont}}\left(x, 1\right)\widetilde{\nu}^{(1),\,\mathrm{cont}}\left(x, 1\right),\label{eq:first_order_moment_nu1_contribution_continuum_limit}
\end{align}
with error:
\begin{align}
    \varepsilon^{(1)} & := c^{(1)}\left(\beta_{\mathrm{max}}, \gamma_{\mathrm{max}}, \frac{M_{\gamma}}{\gamma_{\mathrm{max}}}\right)\frac{\gamma_{\mathrm{max}}^2}{p\sqrt{n}}.
\end{align}
For the contribution involving $\bm{\nu}^{(2)}$,
\begin{align}
    -\frac{i}{\Gamma_{p + 1}}\sum_{0 \leq r \leq 2p + 1}\frac{1}{n}\nu^{(2)}_{(r,\,p + 1)} & \xlongrightarrow[p \to \infty]{} \frac{i}{\gamma_{\mathrm{max}}^2n}\int_{[0, 2]}\!\mathrm{d}x\,\Gamma^{\mathrm{cont}}(x)\widetilde{\nu}^{(2),\,\mathrm{cont}}\left(x, 1\right),\label{eq:first_order_moment_nu2_contribution_continuum_limit}
\end{align}
with error:
\begin{align}
    \varepsilon^{(2)} & := c^{(2)}\left(\beta_{\mathrm{max}}, \gamma_{\mathrm{max}}, \frac{M_{\gamma}}{\gamma_{\mathrm{max}}}\right)\frac{\gamma_{\mathrm{max}}}{pn}.
\end{align}
In the above estimates, $c^{(k)}\left(\beta_{\mathrm{max}}, \gamma_{\mathrm{max}}, M_{\gamma}/\gamma_{\mathrm{max}}\right)$, $0 \leq k \leq 2$, are constants depending only on the maximum continuum $\beta$ angle $\beta_{\mathrm{max}}$, the maximum continuum $\gamma$ angle $\gamma_{\mathrm{max}}$ and the normalized Lipschitz constant $M_{\gamma}/\gamma_{\mathrm{max}}$ of the continuous $\gamma$ schedules; these constants are bounded when all these parameters are.  
\end{proposition}
A presumably surprising aspect of the additive contributions Eqs.~\ref{eq:first_order_moment_nu1_contribution_continuum_limit}, \ref{eq:first_order_moment_nu2_contribution_continuum_limit} is the negative power of $\gamma_{\mathrm{max}}$, suggesting a potential ``unphysical" singularity in the non-interacting limit $\gamma_{\mathrm{max}} \to 0$. This is however not the case due to $\Gamma^{\mathrm{cont}}$ and $\widetilde{\nu}^{(k),\,\mathrm{cont}}$, $k \in \{1, 2\}$, having absolute magnitudes and Lipschitz constants bounded as $\mathcal{O}(1)\gamma_{\mathrm{max}}^2$ provided parameters $\beta_{\mathrm{max}}, \gamma_{\mathrm{max}}, M_{\gamma}/\gamma_{\mathrm{max}}$ are maintained bounded (Corollary~\ref{cor:qgms_integral_series_expansion_contributions_continuum_approximation} and Propositions~\ref{prop:nu2_continuum_limit}). These bounds allow application of the main sum-integral comparison Lemma~\ref{lemma:riemann_sum_approximation}, yielding error bounds $\varepsilon^{(1)}, \varepsilon^{(2)}$ in fact vanishing with $\gamma_{\mathrm{max}}$. We also note that the contributions involving $\bm{\nu}^{(1)}$, $\bm{\nu}^{(2)}$ and their continuum counterparts are suppressed in $n$, unlike the contribution involving $\bm{\nu}^{(0)}$. In the finite $p$ case, this contribution gives the familiar formula (Ref.~\cite{qaoa_maxcut_high_girth}) for the energy density of SK-QAOA in the infinite-size limit:
\begin{align}
    -\frac{1}{\Gamma_{p + 1}}\sum_{0 \leq r \leq 2p + 1}\nu^{(0)}_{(r,\,p + 1)} & = \frac{i}{2}\sum_{0 \leq r \leq 2p + 1}\Gamma_r\left(G^{(2)}_{r,\,p + 1}\right)^2,
\end{align}
converging in the continuum limit $p \to \infty$ to
\begin{align}
    \frac{i}{2}\int_{[0, 2]}\!\mathrm{d}x\,\Gamma^{\mathrm{cont}}\left(x\right)G^{(2),\,\mathrm{cont}}\left(x, 1\right)^2.
\end{align}

Proposition~\ref{prop:first_order_moment_contributions_continuum_limit} stated the limits of all additive contribution to the first-order moment of the cost function, with average taken over the QAOA distribution and disorder: $\mathbb{E}\bra{\Psi_{p,\,n}}C_n/n\ket{\Psi_{p,\,n}}$. A similar result can be produced for the second-order moment $\mathbb{E}\bra{\Psi_{p,\,n}}(C_n/n)^2\ket{\Psi_{p,\,n}}$ of the cost function. For definiteness, we denote by $\omega_{p, n}$ the cost function squared averaged over QAOA distribution and Gaussian disorder at depth $p$ and size $n$. A representation of this quantity in terms of QGMS moments was produced in Section~\ref{sec:sk_qaoa_second_order_moment_qgms} (Eq.~\ref{eq:second_order_cost_function_moment}):
\begin{align}
    \mathbb{E}\bra{\Psi_{p,\,n}}\left(C_n/n\right)^2\ket{\Psi_{p,\,n}} & = \frac{1}{n^3}\binom{n}{2} - \frac{1}{\Gamma_{p + 1}^2}\sum_{0 \leq r, s \leq 2p + 1}\frac{\partial^4S_n\left(\bm{\mu}\right)}{\partial\mu_{(p + 1,\,r)}^2\partial\mu_{(p + 1,\,s)}^2}\Bigg|_{\bm{\mu} = \bm{0}}
\end{align}
Plugging in the additive decomposition of the quartic QGMS moments stated in Proposition~\ref{prop:qgms_diagonal_quartic_moment_contributions_series_expansions} yields an expression of the expectation in terms of tensors $\bm{\nu}^{(k)}$, $3 \leq k \leq 7$, which the following Proposition analyzes:

\begin{proposition}[Continuum limit of disorder-averaged QAOA cost function squared, contribution by contribution]
\label{prop:second_order_moment_contributions_continuum_limit}
Consider the following additive decomposition of the cost function squared averaged over QAOA distribution and Gaussian disorder, given by Proposition~\ref{prop:qgms_diagonal_quartic_moment_contributions_series_expansions}:
\begin{align}
    \omega_{p,\,n}& := \mathbb{E}\bra{\Psi_{p,\,n}}\left(C_n/n\right)^2\ket{\Psi_{p,\,n}}\nonumber\\
    & = \frac{1}{n^3}\binom{n}{2} - \frac{1}{\Gamma_{p + 1}^2}\sum_{0 \leq r, s \leq 2p + 1}\nu^{(3)}_{(p + 1,\,r),\,(p + 1,\,s)} - \frac{1}{\Gamma_{p + 1}^2}\sum_{0 \leq r, s \leq 2p + 1}\frac{1}{\sqrt{n}}\nu^{(4)}_{(p + 1,\,r),\,(p + 1,\,s)}\nonumber\\
    & \hspace*{20px} - \frac{1}{\Gamma_{p + 1}^2}\sum_{0 \leq r, s \leq 2p + 1}\frac{1}{n}\nu^{(5)}_{(p + 1,\,r),\,(p + 1,\,s)} - \frac{1}{\Gamma_{p + 1}^2}\sum_{0 \leq r, s \leq 2p + 1}\frac{1}{n^{3/2}}\nu^{(6)}_{(p + 1,\,r),\,(p + 1,\,s)}\nonumber\\
    & \hspace*{20px} - \frac{1}{\Gamma_{p + 1}^2}\sum_{0 \leq r, s \leq 2p + 1}\frac{1}{n^2}\nu^{(7)}_{(p + 1,\,r),\,(p + 1,\,s)}.
\end{align}
Consider the finer additive decomposition given by plugging in Propositions~\ref{prop:nu4_continuum_limit}, \ref{prop:nu5_continuum_limit}, \ref{prop:nu6_continuum_limit}, breaking down each of $\bm{\nu}^{(4)}, \bm{(5)}, \bm{\nu}^{(6)}$ into several terms:
\begin{align}
    \mathbb{E}\bra{\Psi_{p,\,n}}\left(C_n/n\right)^2\ket{\Psi_{p,\,n}} & = \frac{1}{n^3}\binom{n}{2} - \frac{1}{\Gamma_{p + 1}^2}\sum_{0 \leq r, s \leq 2p + 1}\nu^{(3)}_{(p + 1,\,r),\,(p + 1,\,s)}\nonumber\\
    & \hspace*{20px} - \frac{1}{\Gamma_{p + 1}^2}\sum_{0 \leq r, s \leq 2p + 1}\nu^{(4,\,1)}_{(p + 1,\,r),\,(p + 1,\,s)} - \frac{1}{\Gamma_{p + 1}^2}\sum_{0 \leq r, s \leq 2p + 1}\nu^{(4,\,2)}_{(p + 1,\,r),\,(p + 1,\,s)}\nonumber\\
    & \hspace*{20px} - \frac{1}{\Gamma_{p + 1}^2}\sum_{0 \leq r, s \leq 2p + 1}\nu^{(5,\,1)}_{(p + 1,\,r),\,(p + 1,\,s)} - \frac{1}{\Gamma_{p + 1}^2}\sum_{0 \leq r, s \leq 2p + 1}\nu^{(5,\,2)}_{(p + 1,\,r),\,(p + 1,\,s)}\nonumber\\
    & \hspace*{50px} - \frac{1}{\Gamma_{p + 1}^2}\sum_{0 \leq r, s \leq 2p + 1}\nu^{(5,\,3)}_{(p + 1,\,r),\,(p + 1,\,s)}\nonumber\\
    & \hspace*{20px} - \frac{1}{\Gamma_{p + 1}^2}\sum_{0 \leq r, s \leq 2p + 1}\nu^{(6,\,1)}_{(p + 1,\,r),\,(p + 1,\,s)} - \frac{1}{\Gamma_{p + 1}^2}\sum_{0 \leq r, s \leq 2p + 1}\nu^{(6,\,2)}_{(p + 1,\,r),\,(p + 1,\,s)}\nonumber\\
    & \hspace*{20px} - \frac{1}{\Gamma_{p + 1}^2}\sum_{0 \leq r, s \leq 2p + 1}\nu^{(7)}_{(p + 1,\,r),\,(p + 1,\,s)}.
\end{align}
The following limits hold for reach term of this additive expansion. For the term involving $\bm{\nu}^{(3)}$,
\begin{align}
    -\frac{1}{\Gamma_{p + 1}^2}\sum_{0 \leq r, s \leq 2p + 1}\nu^{(3)}_{(p + 1,\,r),\,(p + 1,\,s)} \xlongrightarrow[p \to \infty]{} & -\frac{1}{4}\int_{[0, 2]^2}\mathrm{d}x_1\mathrm{d}x_2\,\Gamma^{\mathrm{cont}}\left(x_1\right)\Gamma^{\mathrm{cont}}\left(x_2\right)G^{(2,\,\mathrm{cont})}\left(x_1, 1\right)^2G^{(2),\,\mathrm{cont}}\left(x_2, 1\right)^2\nonumber\\
    & = \left(\frac{i}{2}\int_{[0, 2]}\!\mathrm{d}x\,\Gamma^{\mathrm{cont}}\left(x\right)G^{(2),\,\mathrm{cont}}\left(x, 1\right)^2\right)^2,
\end{align}
with error:
\begin{align}
    \varepsilon^{(3)} & = c^{(3)}\left(\beta_{\mathrm{max}}, \gamma_{\mathrm{max}}, \frac{M_{\gamma}}{\gamma_{\mathrm{max}}}\right)\frac{\gamma_{\mathrm{max}}}{p}.
\end{align}
For the term involving $\bm{\nu}^{(4,\,1)}$,
\begin{align}
    -\frac{1}{\Gamma_{p + 1}^2}\sum_{0 \leq r, s \leq 2p + 1}\frac{1}{\sqrt{n}}\nu^{(4,\,1)}_{(p + 1,\,r),\,(p + 1,\,s)} \xlongrightarrow[p \to \infty]{} & -\frac{1}{\sqrt{2}\gamma_{\mathrm{max}}\sqrt{n}}\int_{[0, 2]^2}\!\mathrm{d}x_1\mathrm{d}x_2\,\Gamma^{\mathrm{cont}}\left(x_1\right)\Gamma^{\mathrm{cont}}\left(x_2\right)G^{(2),\,\mathrm{cont}}\left(x_1, 1\right)G^{(2),\,\mathrm{cont}}\left(x_2, 1\right)^2\nonumber\\
    & \hspace*{140px} \times \widetilde{\nu}^{(4,\,1),\,\mathrm{cont}}\left(x_1, 1, x_2, 1\right),
\end{align}
with error:
\begin{align}
    \varepsilon^{(4,\,1)} & = c^{(4,\,1)}\left(\beta_{\mathrm{max}}, \gamma_{\mathrm{max}}, \frac{M_{\gamma}}{\gamma_{\mathrm{max}}}\right)\frac{\gamma_{\mathrm{max}}^3}{p\sqrt{n}}.
\end{align}
For the term involving $\bm{\nu}^{(4,\,2)}$,
\begin{align}
    -\frac{1}{\Gamma_{p + 1}^2}\sum_{0 \leq r, s \leq 2p + 1}\frac{1}{\sqrt{n}}\nu^{(4,\,2)}_{(p + 1,\,r),\,(p + 1,\,s)} \xlongrightarrow[p \to \infty]{} & -\frac{1}{\sqrt{2}\gamma_{\mathrm{max}}\sqrt{n}}\int_{[0, 2]^2}\!\mathrm{d}x_1\mathrm{d}x_2\,\Gamma^{\mathrm{cont}}\left(x_1\right)\Gamma^{\mathrm{cont}}\left(x_2\right)G^{(2),\,\mathrm{cont}}\left(x_1, 1\right)^2G^{(2),\,\mathrm{cont}}\left(x_2, 1\right)\nonumber\\
    & \hspace*{120px} \times \widetilde{\nu}^{(4,\,2),\,\mathrm{cont}}\left(x_1, 1, x_2, 1\right),
\end{align}
with error:
\begin{align}
    \varepsilon^{(4,\,2)} & = c^{(4,\,2)}\left(\beta_{\mathrm{max}}, \gamma_{\mathrm{max}}, \frac{M_{\gamma}}{\gamma_{\mathrm{max}}}\right)\frac{\gamma_{\mathrm{max}}^3}{p\sqrt{n}}.
\end{align}
For the term involving $\bm{\nu}^{(5,\,1)}$,
\begin{align}
    -\frac{1}{\Gamma_{p + 1}^2}\sum_{0 \leq r, s \leq 2p + 1}\frac{1}{n}\nu^{(5,\,1)}_{(p + 1,\,r),\,(p + 1,\,s)} \xlongrightarrow[p \to \infty]{} & -\frac{2}{\gamma_{\mathrm{max}}^2n}\int_{[0, 2]^2}\!\mathrm{d}x_1\mathrm{d}x_2\,\Gamma^{\mathrm{cont}}\left(x_1\right)\Gamma^{\mathrm{cont}}\left(x_2\right)G^{(2),\,\mathrm{cont}}\left(x_1, 1\right)G^{(2),\,\mathrm{cont}}\left(x_2, 1\right)\nonumber\\
    & \hspace*{140px} \times \widetilde{\nu}^{(5,\,1)\,\mathrm{cont}}\left(x_1, 1, x_2, 1\right),
\end{align}
with error:
\begin{align}
    \varepsilon^{(5,\,1)} & = c^{(5,\,1)}\left(\beta_{\mathrm{max}}, \gamma_{\mathrm{max}}, \frac{M_{\gamma}}{\gamma_{\mathrm{max}}}\right)\frac{\gamma_{\mathrm{max}}^2}{pn}.
\end{align}
For the term involving $\bm{\nu}^{(5,\,2)}$,
\begin{align}
    -\frac{1}{\Gamma_{p + 1}^2}\sum_{0 \leq r, s \leq 2p + 1}\frac{1}{n}\nu^{(5,\,2)}_{(p + 1,\,r),\,(p + 1,\,s)} \xlongrightarrow[p \to \infty]{} & -\frac{2}{\gamma_{\mathrm{max}}^2n}\int_{[0, 2]^2}\!\mathrm{d}x_1\mathrm{d}x_2\,\Gamma^{\mathrm{cont}}\left(x_1\right)\Gamma^{\mathrm{cont}}\left(x_2\right)G^{(2),\,\mathrm{cont}}\left(x_1, 1\right)^2\widetilde{\nu}^{(5,\,2)\,\mathrm{cont}}\left(x_1, 1, x_2, 1\right),
\end{align}
with error:
\begin{align}
    \varepsilon^{(5,\,2)} & = c^{(5,\,2)}\left(\beta_{\mathrm{max}}, \gamma_{\mathrm{max}}, \frac{M_{\gamma}}{\gamma_{\mathrm{max}}}\right)\frac{\gamma_{\mathrm{max}}^2}{pn}.
\end{align}
For the term involving $\bm{\nu}^{(5,\,3)}$,
\begin{align}
    -\frac{1}{\Gamma_{p + 1}^2}\sum_{0 \leq r, s \leq 2p + 1}\frac{1}{n}\nu^{(5,\,3)}_{(p + 1,\,r),\,(p + 1,\,s)} \xlongrightarrow[p \to \infty]{} & -\frac{2}{\gamma_{\mathrm{max}}^2n}\int_{[0, 2]^2}\!\mathrm{d}x_1\mathrm{d}x_2\,\Gamma^{\mathrm{cont}}\left(x_1\right)\Gamma^{\mathrm{cont}}\left(x_2\right)G^{(2),\,\mathrm{cont}}\left(x_2, 1\right)^2\widetilde{\nu}^{(5,\,3)\,\mathrm{cont}}\left(x_1, 1, x_2, 1\right),
\end{align}
with error:
\begin{align}
    \varepsilon^{(5,\,3)} & = c^{(5,\,3)}\left(\beta_{\mathrm{max}}, \gamma_{\mathrm{max}}, \frac{M_{\gamma}}{\gamma_{\mathrm{max}}}\right)\frac{\gamma_{\mathrm{max}}^2}{pn}.
\end{align}
For the term involving $\bm{\nu}^{(6,\,1)}$,
\begin{align}
    -\frac{1}{\Gamma_{p + 1}^2}\sum_{1 \leq r, s \leq 2p + 1}\frac{1}{n^{3/2}}\nu^{(6,\,1)}_{(p + 1,\,r),\,(p + 1,\,s)} \xlongrightarrow[p \to \infty]{} & -\frac{1}{\sqrt{2}\gamma_{\mathrm{max}}^3n^{3/2}}\int_{[0, 2]^2}\!\mathrm{d}x_1\mathrm{d}x_2\,\Gamma^{\mathrm{cont}}\left(x_1\right)\Gamma^{\mathrm{cont}}\left(x_2\right)G^{(2),\,\mathrm{cont}}\left(x_1, 1\right)\nonumber\\
    & \hspace*{130px} \times \widetilde{\nu}^{(6,\,1)\,\mathrm{cont}}\left(x_1, 1, x_2, 1\right),
\end{align}
with error:
\begin{align}
    \varepsilon^{(6,\,1)} & = c^{(6,\,1)}\left(\beta_{\mathrm{max}}, \gamma_{\mathrm{max}}, \frac{M_{\gamma}}{\gamma_{\mathrm{max}}}\right)\frac{\gamma_{\mathrm{max}}}{pn^{3/2}}.
\end{align}
For the term involving $\bm{\nu}^{(6,\,2)}$,
\begin{align}
    -\frac{1}{\Gamma_{p + 1}^2}\sum_{1 \leq r, s \leq 2p + 1}\frac{1}{n^{3/2}}\nu^{(6,\,2)}_{(p + 1,\,r),\,(p + 1,\,s)} \xlongrightarrow[p \to \infty]{} & -\frac{1}{\sqrt{2}\gamma_{\mathrm{max}}^3n^{3/2}}\int_{[0, 2]^2}\!\mathrm{d}x_1\mathrm{d}x_2\,\Gamma^{\mathrm{cont}}\left(x_1\right)\Gamma^{\mathrm{cont}}\left(x_2\right)G^{(2),\,\mathrm{cont}}\left(x_2, 1\right)\nonumber\\
    & \hspace*{130px} \times \widetilde{\nu}^{(6,\,2)\,\mathrm{cont}}\left(x_1, 1, x_2, 1\right),
\end{align}
with error:
\begin{align}
    \varepsilon^{(6,\,2)} & = c^{(6,\,2)}\left(\beta_{\mathrm{max}}, \gamma_{\mathrm{max}}, \frac{M_{\gamma}}{\gamma_{\mathrm{max}}}\right)\frac{\gamma_{\mathrm{max}}}{pn^{3/2}}.
\end{align}
For the term involving $\bm{\nu}^{(7)}$,
\begin{align}
    -\frac{1}{\Gamma_{p + 1}^2}\sum_{0 \leq r, s \leq 2p + 1}\frac{1}{n^2}\nu^{(7)}_{(p + 1,\,r),\,(p + 1,\,s)} \xlongrightarrow[p \to \infty]{} & -\frac{1}{\gamma_{\mathrm{max}}^2n}\int_{[0, 2]^2}\!\mathrm{d}x_1\mathrm{d}x_2\,\Gamma^{\mathrm{cont}}\left(x_1\right)\Gamma^{\mathrm{cont}}\left(x_2\right)\widetilde{\nu}^{(7),\,\mathrm{cont}}\left(x_1, 1, x_2, 1\right),
\end{align}
with error:
\begin{align}
    \varepsilon^{(7)} & = c^{(7)}\left(\beta_{\mathrm{max}}, \gamma_{\mathrm{max}}, \frac{M_{\gamma}}{\gamma_{\mathrm{max}}}\right)\frac{1}{pn}.
\end{align}
\end{proposition}

With the expressions of the QAOA cost function first and second moments at hand, one may now establish concentration of the QAOA expectation across Gaussian disorder, uniformly in $p$. This is expressed in the following Proposition:

\begin{proposition}[Concentration of QAOA cost across disorder, uniformly in $p$]
\label{prop:uniform_concentration_qaoa}
Let continuum schedule parameters $\beta_{\mathrm{max}}, \gamma_{\mathrm{max}}, M_{\gamma}/\gamma_{\mathrm{max}}$ be sufficiently small. Then, for all $p \geq 2$, the following uniform in $p$ concentration bound holds for the expected QAOA cost with respect to Gaussian disorder $\left(J_{j, k}\right)_{1 \leq j < k \leq n}$:
\begin{align}
    \mathbb{P}\left[\left|\bra{\Psi_{p,\,n}}C_n/n\ket{\Psi_{p,\,n}} - \nu_{p,\,n}\right| \geq \delta\right] & \leq \frac{c\left(\beta_{\mathrm{max}}, \gamma_{\mathrm{max}}, M_{\gamma}/\gamma_{\mathrm{max}}\right)}{n^{1/2}\delta^2},
\end{align}
for some constant $c\left(\beta_{\mathrm{max}}, \gamma_{\mathrm{max}}, M_{\gamma}/\gamma_{\mathrm{max}}\right)$ depending only on the continuum schedules magnitudes and the normalized Lipschitz constant of the $\gamma$ continuum schedule, and remaining bounded when these 3 parameters are.
\begin{proof}
The main idea is to apply Chebyshev's inequality, using estimates for the second-order moment of the QAOA cost function established in Proposition~\ref{prop:second_order_moment_contributions_continuum_limit}. Explicitly,
\begin{align}
    \mathbb{P}\left[\left|\bra{\Psi_{p,\,n}}C_n/n\ket{\Psi_{p,\,n}} - \nu_{p,\,n}\right| \geq \delta\right] & \leq \frac{1}{\delta^2}\mathbb{E}\left[\left(\bra{\Psi_{p,\,n}}C_n/n\ket{\Psi_{p,\,n}} - \nu_{p,\,n}\right)^2\right]
\end{align}
Following Ref.~\cite{qaoa_sk}, we then estimate the expectation as:
\begin{align}
    \mathbb{E}\left[\left(\bra{\Psi_{p,\,n}}C_n/n\ket{\Psi_{p,\,n}} - \nu_{p,\,n}\right)^2\right]  & = \mathbb{E}\left[\left(\bra{\Psi_{p,\,n}}C_n/n\ket{\Psi_{p,\,n}} - \mathbb{E}\bra{\Psi_{p,\,n}}C_n/n\ket{\Psi_{p,\,n}}\right)^2\right]\nonumber\\
    & = \mathbb{E}\left[\bra{\Psi_{p,\,n}}C_n/n\ket{\Psi_{p,\,n}}^2\right] - \left(\mathbb{E}\bra{\Psi_{p,\,n}}C_n/n\ket{\Psi_{p,\,n}}\right)^2\nonumber\\
    & \leq \mathbb{E}\bra{\Psi_{p,\,n}}\left(C_n/n\right)^2\ket{\Psi_{p,\,n}} - \left(\mathbb{E}\bra{\Psi_{p,\,n}}C_n/n\ket{\Psi_{p,\,n}}\right)^2,
\end{align}
where in the last line, we applied Jensen's inequality with respect to the QAOA output cost probability distribution for each realization of the Gaussian disorder $\left(J_{j, k}\right)_{1 \leq j < k \leq n}$. The difference between the disorder-averaged expected QAOA square cost and the square of the disorder-averaged expected QAOA cost can then be expressed from their decompositions stated in Propositions~\ref{prop:first_order_moment_contributions_continuum_limit}, \ref{prop:second_order_moment_contributions_continuum_limit}. This gives:
\begin{align}
    & \mathbb{E}\bra{\Psi_{p,\,n}}\left(C_n/n\right)^2\ket{\Psi_{p,\,n}} - \left(\mathbb{E}\bra{\Psi_{p,\,n}}C_n/n\ket{\Psi_{p,\,n}}\right)^2\nonumber\\
    & = -\frac{1}{\Gamma_{p + 1}^2}\sum_{0 \leq r, s \leq 2p + 1}\left(\nu^{(3)}_{(r,\,p + 1)} + \frac{1}{\sqrt{n}}\nu^{(4,\,1)}_{(r,\,p + 1)} + \frac{1}{\sqrt{n}}\nu^{(4,\,2)}_{(r,\,p + 1)} + \frac{1}{n}\nu^{(5,\,1)}_{(r,\,p + 1)} + \frac{1}{n}\nu^{(5,\,2)}_{(r,\,p + 1)} + \frac{1}{n}\nu^{(5,\,3)}_{(r,\,p + 1)}\right.\nonumber\\
    & \left. \hspace*{110px} + \frac{1}{n^{3/2}}\nu^{(6,\,1)}_{(r,\,p + 1)} + \frac{1}{n^{3/2}}\nu^{(6,\,2)}_{(r,\,p + 1)} + \frac{1}{n^2}\nu^{(7)}_{(r,\,p + 1)}\right)\nonumber\\
    & \hspace*{20px} - \left(-\frac{i}{\Gamma_{p + 1}}\sum_{0 \leq r \leq 2p + 1}\left(\nu^{(0)}_{(r,\,p + 1)} + \frac{2}{\sqrt{n}}\nu^{(1)}_{(r,\,p + 1)} + \frac{1}{n}\nu^{(2)}_{(r,\,p + 1)}\right)\right)^2\nonumber\\
    & = -\frac{1}{\Gamma_{p + 1}^2}\sum_{0 \leq r, s \leq 2p + 1}\left(\frac{1}{\sqrt{n}}\nu^{(4,\,1)}_{(r,\,p + 1)} + \frac{1}{\sqrt{n}}\nu^{(4,\,2)}_{(r,\,p + 1)} + \frac{1}{n}\nu^{(5,\,1)}_{(r,\,p + 1)} + \frac{1}{n}\nu^{(5,\,2)}_{(r,\,p + 1)} + \frac{1}{n}\nu^{(5,\,3)}_{(r,\,p + 1)}\right.\nonumber\\
    & \left. \hspace*{110px} + \frac{1}{n^{3/2}}\nu^{(6,\,1)}_{(r,\,p + 1)} + \frac{1}{n^{3/2}}\nu^{(6,\,2)}_{(r,\,p + 1)} + \frac{1}{n^2}\nu^{(7)}_{(r,\,p + 1)} - \frac{4}{n^2}\nu^{(4)}_{(r,\,p + 1)}\nu^{(4)}_{(s,\,p + 1)} - \frac{1}{n^2}\nu^{(2)}_{(r,\,p + 1)}\nu^{(2)}_{(s,\,p + 1)}\right.\nonumber\\
    & \left. \hspace*{110px} - \frac{4}{\sqrt{n}}\nu^{(0)}_{(r,\,p + 1)}\nu^{(1)}_{(s,\,p + 1)} - \frac{2}{n}\nu^{(0)}_{(r,\,p + 1)}\nu^{(2)}_{(s,\,p + 1)} - \frac{4}{n^{3/2}}\nu^{(1)}_{(r,\,p + 1)}\nu^{(2)}_{(s,\,p + 1)}\right).
\end{align}
The important point is, the $\nu^{(3)}_{(r,\,p + 1),\,(s,\,p + 1)}$ contribution from the disorder-averaged QAOA cost squared cancels with the $\nu^{(0)}_{(r,\,p + 1)}\nu^{(0)}_{(s,\,p + 1)}$ from the squared of the disorder-averaged QAOA cost, leaving only contributions of magnitude at most $n^{-1/2}$ in instance size. From the estimates in Propositions~\ref{prop:first_order_moment_contributions_continuum_limit}, \ref{prop:second_order_moment_contributions_continuum_limit}, the above quantity is bounded by
\begin{align}
    \frac{c\left(\beta_{\mathrm{max}}, \gamma_{\mathrm{max}}, M_{\gamma}/\gamma_{\mathrm{max}}\right)}{\sqrt{n}}.
\end{align}
for some constant $c$ depending only on the above 3 parameters, remaining bounded as these parameters are. This completes the proof.
\end{proof}
\end{proposition}

Proposition~\ref{prop:uniform_concentration_qaoa} establishes concentration of the QAOA cost function across random Gaussian disorder for all finite $p$, uniformly in $p$, when depth $p$ QAOA schedules $\left(\bm{\gamma}^{(p)}, \bm{\beta}^{(p)}\right)$ are discretizations of continuum schedules according to the prescription of Definition~\ref{def:continuous_schedule_informal}. This improves on original concentration result Ref.~\cite{qaoa_sk} by providing a rate in $n$ for the concentration, as well as pinning down the dependence in $p$ for this schedule regime. On the other hand, a major weakness of our results is the requirement of an absolute bound on the continuum $\gamma$ schedule magnitude $\gamma_{\mathrm{max}}$. Concentration shows that at large size $n$, the cost function of any SK-QAOA instance is close to its disorder-averaged with high probability. We would now like to prove a similar result for the state produced by constant-time quantum annealing
\begin{align}
    \ket{\Psi_{\infty,\,n}} & = \lim_{p \to \infty}\ket{\Psi_{p,\,n}}.
\end{align}
Since $p$ is no longer finite, it is not even immediate that the disorder-averaged QAOA cost (squared) exists. However, one intuitively expects that the disorder-averaged QAOA cost and cost squared to be respectively given by:
\begin{align}
    \nu_{\infty,\,n} := \lim_{p \to \infty}\nu_{p,\,\infty}, && \omega_{\infty,\,n} := \lim_{p \to \infty}\omega_{p,\,n}.
\end{align}
Taking the example of the disorder-averaged QAOA cost, this is equivalent to assuming correctness of limit-expectation inversion:
\begin{align}
    \lim_{p \to \infty}\nu_{p,\,n} & = \lim_{p \to \infty}\mathbb{E}\left[\bra{\Psi_{p,\,n}}C_n/n\ket{\Psi_{p,\,n}}\right]\nonumber\\
    & = \mathbb{E}\left[\lim_{p \to \infty}\bra{\Psi_{p,\,n}}C_n/n\ket{\Psi_{p,\,n}}\right]\nonumber\\
    & = \mathbb{E}\left[\bra{\Psi_{\infty,\,n}}C_n/n\ket{\Psi_{\infty,\,n}}\right].
\end{align}
The following Lemma establishes the validity of this calculation via a dominated convergence argument:

\begin{lemma}[Existence of disorder-averaged quantum annealing cost and cost squared]
\label{lemma:qa_disorder_averages}
Consider the state produced by quantum annealing as prescribed in the main Theorem~\ref{th:approximation_continuous_time_annealing_qaoa}, i.e. the final state $\ket{\Psi(1)}$ of the following Hamiltonian evolution:
\begin{align}
    \ket{\Psi(0)} := \ket{+}^{\otimes n}, && i\frac{\mathrm{d}\ket{\Psi(u)}}{\mathrm{d}u} = H(u)\ket{\Psi(u)}, && H(u) := \gamma^{\mathrm{cont}}\left(u\right)C + \beta^{\mathrm{cont}}\left(u\right)B,
\end{align}
By Trotterization, for all finite size $n$, this corresponds to the $p \to \infty$ limit of the QAOA state with schedules derived from $\gamma^{\mathrm{cont}}, \beta^{\mathrm{cont}}$:
\begin{align}
    \ket{\Psi(1)} & = \ket{\Psi_{\infty,\,n}}\nonumber\\
    & = \lim_{p \to \infty}\ket{\Psi_{p,\,n}}.
\end{align}
Then, the disorder-averaged cost and cost squared under $\ket{\Psi_{\infty,\,n}}$ exist and are given by the $p \to \infty$ limits of their finite $p$ counterparts:
\begin{align}
    \mathbb{E}\bra{\Psi_{\infty,\,n}}C_n/n\ket{\Psi_{\infty,\,n}} & = \lim_{p \to \infty}\mathbb{E}\bra{\Psi_{p,\,n}}C_n/n\ket{\Psi_{p,\,n}},\\
    \mathbb{E}\bra{\Psi_{\infty,\,n}}\left(C_n/n\right)^2\ket{\Psi_{\infty,\,n}} & = \lim_{p \to \infty}\mathbb{E}\bra{\Psi_{p,\,n}}\left(C_n/n\right)^2\ket{\Psi_{p,\,n}}.
\end{align}
\begin{proof}
Let us focus on the disorder average of the QAOA cost function ---the treatment of the cost function squared being very similar. We wish to apply dominated convergence to show that
\begin{align}
    \bra{\Psi_{\infty,\,n}}C_n/n\ket{\Psi_{\infty,\,n}} & = \lim_{p \to \infty}\bra{\Psi_{p,\,n}}C_n/n\ket{\Psi_{p,\,n}}
\end{align}
is integrable (``the expectation exists") and that the expectation of the limit is the limit of the expectation. For that purpose, it only remains to verify the domination assumption. We use simple bound:
\begin{align}
    \left|\bra{\Psi_{p,\,n}}C_n/n\ket{\Psi_{p,\,n}}\right| & \leq \left\lVert C_n/n \right\rVert_{\infty}\nonumber\\
    & = \max_{\bm{\sigma} \in \{1, -1\}^n}\left|C_n\left(\bm{\sigma}\right)/n\right|.
\end{align}
We then need to show the maximum, seen as a function of random Gaussian disorder $\left(J_{j, k}\right)_{1 \leq j < k \leq n}$, is integrable with respect to this randomness. That is, given the function is non-negative (so the expectation is always defined, possibly infinite), one wants to show:
\begin{align}
    \mathbb{E}\left[\max_{\bm{\sigma} \in \{1, -1\}^n}\left|C_n\left(\bm{\sigma}\right)/n\right|\right] < \infty.
\end{align}
This follows from standard techniques in Gaussian calculus. First, expressing the expectation from the tail sum formula:
\begin{align}
    \mathbb{E}\left[\max_{\bm{\sigma} \in \{1, -1\}^n}\left|C_n\left(\bm{\sigma}\right)/n\right|\right] & = \int_0^{\infty}\!\mathrm{d}t\,\mathbb{P}\left[\max_{\bm{\sigma} \in \{1, -1\}^n}\left|C_n\left(\bm{\sigma}\right)/n\right| > t\right].
\end{align}
The integrability then follows from controlling the cumulative distribution function by union bound:
\begin{align}
    \mathbb{P}\left[\max_{\bm{\sigma} \in \{1, -1\}^n}\left|C_n\left(\bm{\sigma}\right)/n\right| > t\right] & = \mathbb{P}\left[\bigcup_{\bm{\sigma} \in \{1, -1\}^n}\left\{\left|C_n\left(\bm{\sigma}\right)/n\right| > t\right\}\right]\nonumber\\
    & \leq \sum_{\bm{\sigma} \in \{1, -1\}^n}\mathbb{P}\left[\left|C_n\left(\bm{\sigma}\right)/n\right| > t\right].
\end{align}
Now, for each sum term $\bm{\sigma} \in \{1, -1\}^n$,
\begin{align}
    C_n\left(\bm{\sigma}\right)/n & \overset{\mathrm{dist}}{=} \mathcal{N}\left(0, \frac{1}{n^3}\binom{n}{2}\right),
\end{align}
so that
\begin{align}
    \mathbb{P}\left[\left|C_n\left(\bm{\sigma}\right)/n\right| > t\right] & \leq 2\exp\left(-\frac{t^2}{2\frac{1}{n^3}\binom{n}{2}}\right)\nonumber\\
    & \leq 2\exp\left(-nt^2\right).
\end{align}
Hence,
\begin{align}
    \mathbb{P}\left[\max_{\bm{\sigma} \in \{1, -1\}^n}\left|C_n\left(\bm{\sigma}\right)/n\right| > t\right] & \leq 2^{n + 1}\exp\left(-nt^2\right),
\end{align}
whence
\begin{align}
    \mathbb{E}\left[\max_{\bm{\sigma} \in \{1, -1\}^n}\left|C_n\left(\bm{\sigma}\right)/n\right|\right] & \leq \int_0^{\infty}\!\mathrm{d}t\,2^{n + 1}\exp\left(-nt^2\right)\nonumber\\
    & < \infty.
\end{align}
Therefore, the domination hypothesis holds= in dominated convergence, and the claims follow.
\end{proof}
\end{lemma}

Using Lemma~\ref{lemma:qa_disorder_averages} to express the quantum annealing disorder averages as the limits of QAOA ones, the following cost concentration result can be obtained for linear-time quantum annealing by essentially repeating the proof of Proposition~\ref{prop:uniform_concentration_qaoa}:

\begin{proposition}[Concentration of linear-time quantum annealing cost across disorder]
\label{prop:concentration_linear_time_qa}
Let continuum schedule parameters $\beta_{\mathrm{max}}, \gamma_{\mathrm{max}}, M_{\gamma}/\gamma_{\mathrm{max}}$ be sufficiently small. Then, for all $p \geq 2$, the following concentration bound holds for the expected quantum annealing cost with respect to Gaussian disorder $\left(J_{j, k}\right)_{1 \leq j < k \leq n}$:
\begin{align}
    \mathbb{P}\left[\left|\bra{\Psi_{p,\,n}}C_n/n\ket{\Psi_{p,\,n}} - \nu_{p,\,n}\right| \geq \delta\right] & \leq \frac{c\left(\beta_{\mathrm{max}}, \gamma_{\mathrm{max}}, M_{\gamma}/\gamma_{\mathrm{max}}\right)}{n^{1/2}\delta^2},
\end{align}
for some constant $c\left(\beta_{\mathrm{max}}, \gamma_{\mathrm{max}}, M_{\gamma}/\gamma_{\mathrm{max}}\right)$ depending only on the continuum schedules magnitudes and the normalized Lipschitz constant of the $\gamma$ continuum schedule, and remaining bounded when these 3 parameters are.
\end{proposition}

\section{Technical results}

We conclude the appendix with a paragraph collecting technical results frequently used in the proofs from Appendix~\ref{sec:sk_qaoa_energy_continuum_limit}.

\begin{theorem}[Multinomial theorem with infinite number of terms]
\label{th:multinomial_theorem_infinite_number_terms}
Let $\left(x_k\right)_{k \geq 0}$ be an absolutely convergent series of complex numbers:
\begin{align}
    \sum_{k \geq 0}|x_k| & < \infty.
\end{align}
Then, for all integer $n \geq 0$, the following multinomial identity holds:
\begin{align}
    \left(\sum_{k \geq 0}x_k\right)^n & = \sum_{\substack{\left(n_k\right)_{k \geq 0}\\\sum\limits_{k \geq 0}n_k = n}}\binom{n}{\left(n_k\right)_{k \geq 0}}\prod_{k \geq 0}x_k^{n_k},
\end{align}
where the sum of the right-hand side is over sequence of integers $\left(n_k\right)_{k \geq 0}$, and the series on the right-hand side is absolutely convergent. It follows that
\begin{align}
    \left(1 + \sum_{k \geq 0}x_k\right)^n & = \sum_{\left(n_k\right)_{k \geq 0}}\binom{n}{\left(n_k\right)_{k \geq 0}}\prod\limits_{k \geq 0}x_k^{n_k},
\end{align}
where the sum on the right-hand side is over sequences of integers $\left(n_k\right)_{k \geq 0}$ with a finite number of nonzero integers ---without the constraint of summing to $n$--- and we extended the definition of the multinomial coefficient to the case where the bottom numbers do not sum to the top one:
\begin{align}
    \binom{n}{\left(n_k\right)_{k \geq 0}} & := \frac{n!}{\left(n - \sum\limits_{k \geq 0}n_k\right)!\prod\limits_{k \geq 0}n_k!}.\label{eq:multinomial_coefficient_redefinition}
\end{align}
[The definition does coincide with the usual one if the bottom numbers sum to the top one.]
\begin{proof}
By multiplication of absolutely convergent series, we compute,
\begin{align}
    \left(\sum_{k \geq 0}x_k\right)^n & = \prod_{1 \leq r \leq n}\sum_{k_r \geq 0}x_{k_r}\nonumber\\
    & = \sum_{k_1,\,\ldots,\,k_n \geq 0}\prod_{1 \leq r \leq n}x_{k_r},
\end{align}
Since the series on the right-hand side is absolutely convergent:
\begin{align}
    \sum_{k_1,\,\ldots,\,k_n \geq 0}\prod_{1 \leq r \leq n}|x_{k_r}| & = \left(\sum_{k \geq 0}|x_k|\right)^n < \infty,
\end{align}
we may reorder its terms arbitrarily. One then groups terms $\prod_{1 \leq r \leq n}x_{k_r}$ according to the number of occurrences $N\left(k,\,\left(k_r\right)_{1 \leq r \leq n}\right)$ of each integer $k \geq 0$ in $n$-tuple $\left(k_r\right)_{1 \leq r \leq n}$, where $k \geq 0$. The value of the term is only determined by the $N\left(k, \left(k_r\right)_{1 \leq r \leq n}\right)$ for all $k \geq 0$, namely:
\begin{align}
    \prod_{1 \leq r \leq n}x_{k_r} & = \prod_{k \geq 0}x_k^{N\left(k,\,\left(k_r\right)_{1 \leq r \leq n}\right)},
\end{align}
where the product is finite. Besides, given non-negative integers $\left(n_k\right)_{k \geq 0}$ summing to $n$: $\sum_{k \geq 0}n_k = n$, there are exactly
\begin{align}
    \binom{n}{\left(n_k\right)_{k \geq 0}}
\end{align}
$n$-tuples $\left(k_r\right)_{1 \leq r \leq n}$ satisfying $N\left(k, \left(k_r\right)_{1 \leq r \leq n}\right) = n_k$ for all $k \geq 0$. Hence,
\begin{align}
    \sum_{k_1,\,\ldots,\,k_n \geq 0}\prod_{1 \leq r \leq n}x_{k_r} & = \sum_{k_1,\,\ldots,\,k_n \geq 0}\prod_{k \geq 0}x_k^{N\left(k,\,\left(k_r\right)_{1 \leq r \leq n}\right)}\\
    & = \sum_{\substack{\left(n_k\right)_{k \geq 0}\\\sum\limits_{k \geq 0}n_k = n}}\binom{n}{\left(n_k\right)_{k \geq 0}}\prod_{k \geq 0}x_k^{n_k}.
\end{align}
\end{proof}
\end{theorem}

\begin{lemma}[Bound on a function related to the exponential]
\label{lemma:exponential_bound}
Let $c \geq 0$ an arbitrary non-negative real number, $n \geq 0$ an integer, and consider function:
\begin{align}
    \begin{array}{ccc}
         \mathbf{R}_+ & \longrightarrow & \mathbf{R}_+\\
         r & \longmapsto & \left(1 + c\left(e^r - 1 - r\right)\right)^n 
    \end{array}.
\end{align}
The following bounds hold on this function:
\begin{align}
    \left(1 + c\left(e^r - 1 - r\right)\right)^n & \leq \exp\left(\frac{cr^2n}{2}e^r\right)\label{eq:exponential_bound_small_r},\\
    \left(1 + c\left(e^r - 1 - r\right)\right)^n & \leq \left\{\begin{array}{cc}
         e^{rn} & \textrm{for } 0 \leq c \leq 1\\
         c^ne^{rn} &  \textrm{for } c > 1
    \end{array}\right.\label{eq:exponential_bound_large_r}.
\end{align}
\begin{proof}
The bound in equation \ref{eq:exponential_bound_small_r} follows from standard bound
\begin{align}
    1 + x \leq \exp\left(x\right) \qquad \forall x \geq 0.
\end{align}
From this bound,
\begin{align}
    \left(1 + c\left(e^r - 1 - r\right)\right)^n & \leq \exp\left(c\left(e^r - 1 - r\right)\right)^n\nonumber\\
    & = \exp\left(c\left(e^r - 1 - r\right)n\right).
\end{align}
We next use bound
\begin{align}
    0 \leq e^r - 1 - r \leq \frac{r^2}{2}e^r.
\end{align}
following from Taylor's integral (in)equality, yielding equation \ref{eq:exponential_bound_small_r}.

For equation \ref{eq:exponential_bound_large_r}, we distinguish cases $0 \leq c \leq 1$ and $c > 1$. In the former, we write
\begin{align}
    1 + c\left(e^r - 1 - r\right) & = (1 - c) + c\left(e^r - r\right)
\end{align}
Regarding $\left(1 - c\right), c \in [0, 1]$ as average weights, and using $e^r - r \geq 1$, we obtain,
\begin{align}
    \left(1 - c\right) + c\left(e^r - r\right) & \leq e^r - r\nonumber\\
    & \leq e^r,
\end{align}
hence
\begin{align}
    \left(1 + c\left(e^r - 1 - r\right)\right)^n & = \left( \left(1 - c\right) + c\left(e^r - r\right)\right)^n\nonumber\\
    & \leq e^{rn}.
\end{align}

We now look at case $c > 1$. The statement to prove amounts to
\begin{align}
    1 & \geq \frac{\left(1 + c\left(e^r - 1 - r\right)\right)^n}{c^ne^{rn}}\nonumber\\
    & = \left(\frac{1}{c}e^{-r} + 1 - e^{-r} - re^{-r}\right)^n,
\end{align}
i.e.
\begin{align}
    1 \geq \frac{1}{c}e^{-r} + 1 - e^{-r} - re^{-r}
\end{align}
or equivalently
\begin{align}
    0 \geq \frac{1}{c}e^{-r} - e^{-r} - re^{-r}.
\end{align}
The statement then results from elementary function analysis. Namely, the $r$ derivative of the above function is:
\begin{align}
    \left(r - \frac{1}{c}\right)e^{-r},
\end{align}
which is non-positive on $\left[0, 1/c\right]$ and non-negative on $\left[1/c, +\infty\right)$; also, the function assumes special non-positive values
\begin{align}
    1/c - 1, -e^{-1/c}, 0
\end{align}
at $r = 0, 1/c, +\infty$ respectively.
\end{proof}
\end{lemma}

\begin{lemma}[Bound on variation of product]
\label{lemma:bound_variation_product}
Let
\begin{align}
    x_1, x_2, \ldots, x_{n - 1}, x_n \in \mathbf{C}
\end{align}
and
\begin{align}
    y_1, y_2, \ldots, y_{n - 1}, y_n \in \mathbf{C}
\end{align}
two length-$n$ sequences of complex numbers. Then, the following bound holds on the difference of their product:
\begin{align}
    \left|\prod_{j \in [n]}x_j - \prod_{j \in [n]}y_j\right| & \leq \sum_{j \in [n]}|x_j - y_j|\prod_{k \in [n] - \{j\}}\max\left(|x_k|, |y_k|\right)\label{eq:bound_variation_product_stronger}\\
    & \leq \max_{j \in [n]}\left(\prod_{k \in [n] - \{j\}}\max\left(|x_k|, |y_k|\right)\right)\sum_{j \in [n]}|x_j - y_j|.\label{eq:bound_variation_product_weaker}
\end{align}
\begin{proof}
The result follows from elementary identity:
\begin{align}
    \prod_{j \in [n]}x_j - \prod_{j \in [n]}y_j & = \sum_{1 \leq j \leq n}\left(\prod_{1 \leq k < j}x_k\right)\left(x_j - y_j\right)\left(\prod_{j < k \leq n}y_k\right).
\end{align}
Applying the triangular inequality,
\begin{align}
    \left|\prod_{j \in [n]}x_j - \prod_{j \in [n]}y_j\right| & \leq \sum_{1 \leq j \leq n}\left(\prod_{1 \leq k < j}\left|x_k\right|\right)|x_j - y_j|\left(\prod_{j < k \leq n}\left|y_k\right|\right)\nonumber\\
    & \leq \sum_{1 \leq j \leq n}|x_j - y_j|\prod_{k \in [n] - \{j\}}\max\left(|x_k|, |y_k|\right)\nonumber\\
    & \leq \max_{j \in [n]}\left(\prod_{k \in [n] - \{j\}}\max\left(|x_k|, |y_k|\right)\right)\sum_{j \in [n]}|x_j - y_j|.
\end{align}
\end{proof}
\end{lemma}

The following lemma, which is a Riemann sum approximation result, is the central statement we invoke to justify approximation of discrete sums by integrals.

\begin{lemma}[Riemann sum approximation of integrals]
\label{lemma:riemann_sum_approximation}
Consider $m$ complex-valued functions $f_1, \ldots, f_m$ defined over a Cartesian power of $[0, 2]$:
\begin{align}
    f_l: & \, [0, 2]^{d_l} \longrightarrow \mathbf{C}, \qquad \forall 1 \leq l \leq m,
\end{align}
Besides, assume each $f_l$ is bounded by some constant $K_l$
\begin{align}
    \left|f_l\left(\bm{x}\right)\right| & \leq K_l \qquad \forall \bm{x} \in [0, 2]^{d_l},
\end{align}
and furthermore $M_l$-Lipschitz in each of its variables on intervals $[0, 1), [1, 2)$, separately, i.e.
\begin{align}
    & \left|f_l\left(\bm{x}\right) - f_l\left(\bm{y}\right)\right| \leq M_l\sum_{1 \leq j \leq d_l}\left|x_j - y_j\right| = M_l\left\lVert \bm{x} - \bm{y} \right\rVert_1,\\
    & \hspace*{20px} \forall \bm{x} = \left(x_j\right)_{j \in [d_l]}, \bm{y}= \left(y_j\right)_{j \in [d_l]}\,:\,\forall j \in [d_l],\,x_j, y_j \in [0, 1)\textrm{ or } x_j, y_j \in [1, 2).
\end{align}
Let now a function $f$ be defined by:
\begin{align}
f: \left\{\begin{array}{ccc}
     [0, 2]^D & \longrightarrow & \mathbf{C}\\
     \bm{x} & \longmapsto & \prod\limits_{1 \leq l \leq m}f_l\left(\bm{x}_{I_l}\right)
\end{array}\right.,
\end{align}
where for each $1 \leq l \leq m$, $I_l \subset [D]^{d_l}$ is a tuple of indices of $\bm{x}$, and for $I_l = \left(j_1, \ldots, j_{d_l}\right)$, $\bm{x}_{I_l} := \left(x_{j_1}, \ldots, x_{j_{d_l}}\right)$. Then, for all integer $p \geq 1$ the following Riemann sum estimate holds:
\begin{align}
    \left|\frac{1}{\left(p + 1\right)^D}\sum_{\bm{n} \in \{0,\,\ldots,\,2p + 1\}^D}f\left(\frac{\bm{n}}{p + 1/2}\right) - \int_{[0, 2]^D}\!\mathrm{d}\bm{x}\,f\left(\bm{x}\right)\right| & \leq \frac{2^{D + 1}}{p + 1}\max_{l \in [m]}\left(M_l\prod_{l' \in [m] - \{l\}}K_{l'}\right)\sum_{1 \leq l \leq m}d_l.\label{eq:riemann_sum_approximation_error_bound}
\end{align}
\begin{proof}
We start by expressing the sum as the integral of a piecewise constant function. Namely,
\begin{align}
    \frac{1}{\left(p + 1\right)^D}\sum_{\bm{n} \in \{0,\,\ldots,\,2p + 1\}^D}f\left(\frac{\bm{n}}{p + 1/2}\right) & = \int_{[0, 2]^D}\!\mathrm{d}\bm{x}\,\widetilde{f}\left(\bm{x}\right),
\end{align}
where
\begin{align}
    \widetilde{f}\left(\bm{x}\right) & := f\left(\frac{\bm{n}\left(\bm{x}\right)}{p + 1/2}\right),\\
    \bm{n}\left(\bm{x}\right) & := \left(n_j\left(x_j\right)\right)_{j \in [D]},\\
    n_j\left(x_j\right) & := k \qquad \forall x_j \in \left[\frac{k}{p + 1}, \frac{k + 1}{p + 1}\right), \quad \forall 0 \leq k \leq 2p + 1.
\end{align}
The sum-integral approximation error can then be bounded using the triangular inequality:
\begin{align}
    \left|\frac{1}{\left(p + 1\right)^D}\sum_{\bm{n} \in \{0,\,\ldots,\,2p + 1\}^D}f\left(\frac{\bm{n}}{p + 1/2}\right) - \int_{[0, 2]^D}\!\mathrm{d}\bm{x}\,f\left(\bm{x}\right)\right| & = \left|\int_{[0, 2]^D}\!\mathrm{d}\bm{x}\,\left(\widetilde{f}\left(\bm{x}\right) - f\left(\bm{x}\right)\right)\right|\nonumber\\
    & \leq \int_{[0, 2]^D}\!\mathrm{d}\bm{x}\,\left|\widetilde{f}\left(\bm{x}\right) - f\left(\bm{x}\right)\right|.
\end{align}
It therefore remains to find a uniform bound on $\widetilde{f} - f$. For that purpose, we start by using lemma \ref{lemma:bound_variation_product} (equation \ref{eq:bound_variation_product_stronger}) to bound the variation of a product, giving:
\begin{align}
    \left|\widetilde{f}\left(\bm{x}\right) - f\left(\bm{x}\right)\right| & = \left|\prod_{1 \leq l \leq m}\widetilde{f}_l\left(\bm{x}_{I_l}\right) - \prod_{1 \leq l \leq m}f_l\left(\bm{x}_{I_l}\right)\right|\nonumber\\
    & \leq \sum_{1 \leq l \leq m}\left|\widetilde{f}_l\left(\bm{x}_{I_l}\right) - f_l\left(\bm{x}_{I_l}\right)\right|\prod_{l' \in [m] - \{l\}}\max\left(\left|\widetilde{f}_{l'}\left(\bm{x}_{I_{l'}}\right)\right|, \left|f_{l'}\left(\bm{x}_{I_{l'}}\right)\right|\right)\nonumber\\
    & \leq \sum_{1 \leq l \leq m}\left|\widetilde{f}_l\left(\bm{x}_{I_l}\right) - f_l\left(\bm{x}_{I_l}\right)\right|\prod_{l' \in [m] - \{l\}}K_{l'}\nonumber\\
    & \leq \sum_{1 \leq l \leq m}\left|f_l\left(\frac{\bm{n}\left(\bm{x}\right)_{I_l}}{p}\right) - f_l\left(\bm{x}_{I_l}\right)\right|\prod_{l' \in [m] - \{l\}}K_{l'}\nonumber\\
    & \leq \sum_{1 \leq l \leq m}M_l\left\lVert \frac{\bm{n}\left(\bm{x}\right)_{I_l}}{p} - \bm{x}_{I_l} \right\rVert_1\prod_{l' \in [m] - \{l\}}K_{l'}\nonumber\\
    & \leq \max_{l \in [m]}\left(M_l\prod_{l' \in [m] - \{l\}}K_{l'}\right)\sum_{1 \leq l \leq m}\left\lVert \frac{\bm{n}\left(\bm{x}\right)_{I_l}}{p} - \bm{x}_{I_l} \right\rVert_1.
\end{align}
Consistent with notation in the statement, $\bm{n}\left(\bm{x}\right)_{I_l}$ denotes the projection of $\bm{n}\left(\bm{x}\right)$ onto coordinates tuple $I_l$ (some coordinates may be repeated). In the last but one line, we used that $f_l$ (hence $\widetilde{f}_l$) is bounded by $K_l$, and that $f_l$ is $M_l$-Lipschitz in each of its variables, separately on $[0, 1)$ and $[1, 2)$ [Separate Lipschitz monotonicity is sufficient since for all $j \in [D]$, $x_j$ and $n_j\left(x_j\right)$ lie both in $[0, 1)$ or $[1, 2)$ by construction.] Now, letting $I_l := \left(j_1,\,\ldots,\,j_{d_l}\right)$,
\begin{align}
    \left\lVert \frac{\bm{n}\left(\bm{x}\right)_{I_l}}{p + 1/2} - \bm{x}_{I_l} \right\rVert_1 & = \left|\frac{n_{j_1}\left(x_{j_1}\right)}{p + 1/2} - x_{j_1}\right| + \ldots + \left|\frac{n_{j_{d_l}}\left(x_{j_{d_l}}\right)}{p + 1/2} - x_{j_{d_l}}\right|.
\end{align}
We can now bound each term in the last equation (taking the first for illustration) as follows:
\begin{align}
    \left|\frac{n_{j_1}\left(x_{j_1}\right)}{p + 1/2} - x_{j_1}\right| & = \left|\frac{n_{j_1}\left(x_{j_1}\right)}{p + 1} - x_{j_1} + n_{j_1}\left(x_{j_1}\right)\left(\frac{1}{p + 1/2} - \frac{1}{p + 1}\right)\right|\nonumber\\
    & = \left|\frac{n_{j_1}\left(x_{j_1}\right)}{p + 1} - x_{j_1}\right| + \frac{1}{2}\left|\frac{n_{j_1}\left(x_{j_1}\right)}{\left(p + 1/2\right)(p + 1)}\right|\nonumber\\
    & \leq \frac{1}{p + 1} + \frac{1}{2}\frac{2p + 1}{\left(p + 1/2\right)(p + 1)}\nonumber\\
    & \leq \frac{2}{p + 1}.
\end{align}
Hence,
\begin{align}
    \left\lVert \frac{\bm{n}\left(\bm{x}\right)_{I_l}}{p} - \bm{x}_{I_l} \right\rVert_1 & \leq \frac{2d_l}{p + 1},
\end{align}
from where
\begin{align}
    \left|\widetilde{f}\left(\bm{x}\right) - f\left(\bm{x}\right)\right| & \leq \frac{2}{p + 1}\max_{l \in [m]}\left(M_l\prod_{l' \in [m] - \{l\}}K_{l'}\right)\sum_{1 \leq l \leq m}d_l,
\end{align}
and
\begin{align}
    & \left|\frac{1}{\left(p + 1\right)^D}\sum_{\bm{n} \in \{0,\,\ldots,\,2p + 1\}^D}f\left(\frac{\bm{n}}{p + 1/2}\right) - \int_{[0, 2]^D}\!\mathrm{d}\bm{x}\,f\left(\bm{x}\right)\right| \nonumber\\
    & \leq \frac{2^{D + 1}}{p + 1}\max_{l \in [m]}\left(M_l\prod_{l' \in [m] - \{l\}}K_{l'}\right)\sum_{1 \leq l \leq m}d_l,
\end{align}
which is the claim.
\end{proof}
\end{lemma}

To control the magnitude of series terms in the series expansion of QGMS moments introduced in Section~\ref{sec:qgms_moments_series_expansion} (specifically, Proposition~\ref{prop:qgms_integral_series_expansion}), we will be interested in the following combinatorial sum:
\begin{align}
    \sum_{\left(n_d\right)_{d \geq 2} \neq \bm{0}}\binom{n}{\left(n_d\right)_{d \geq 2}}\frac{n^{-D/2}}{\prod\limits_{d \geq 2}d!^{n_d}}D^kD!!x^D,
\end{align}
where for conciseness, we defined inside the sum $D$ as the following implicit function of multinomial numbers $\left(n_d\right)_{d \geq 2}$:
\begin{align}
    D & := \sum_{d \geq 2}dn_d.
\end{align}
We are interested in the regime where $k$ is considered a small integer, and $x$ is a sufficiently small non-negative constant. The following Lemma controls this quantity:

\begin{lemma}[Bound on multinomial sum in QGMS integral moments series expansion]
\label{lemma:qgms_integral_moment_series_multinomial_sum_bound}
There exist absolute constants $r_* > 0$ and $c > 0$ such for all complex $z$ bounded by $r_*$ and integer $k \geq 0$, the following inequality holds:
\begin{align}
    \left|\sum_{\left(n_d\right)_{d \geq 2} \neq \bm{0}}\binom{n}{\left(n_d\right)_{d \geq 2}}\frac{n^{-D/2}}{\prod\limits_{d \geq 2}d!^{n_d}}D!!D^kz^D\right| & \leq c\frac{k!}{\left(\log 2\right)^k}|z|^2.
\end{align}
Similarly,
\begin{align}
    \left|\sum_{\left(n_d\right)_{d \geq 2} \neq \bm{0}}\binom{n}{\left(n_d\right)_{d \geq 2}}\frac{n^{-D/2}}{\prod\limits_{d \geq 2}d!^{n_d}}(D - 1)!!D^kz^D\right| & \leq c\frac{k!}{\left(\log 2\right)^k}|z|^2,\\
    \left|\sum_{\left(n_d\right)_{d \geq 2}}\binom{n}{\left(n_d\right)_{d \geq 2}}\frac{n^{-D/2}}{\prod\limits_{d \geq 2}d!^{n_d}}D!!D^kz^D\right| & \leq c\frac{k!}{\left(\log 2\right)^k}.
\end{align}
In all the above inequalities, we introduced the following shorthand notation as an implicit function of the multinomial numbers $\left(n_d\right)_{d \geq 2}$:
\begin{align}
    D & := \sum_{d \geq 2}dn_d
\end{align}
Besides,
\begin{align}
    D!! & := \prod_{0 \leq k \leq \left\lfloor D/2 \right\rfloor}(D - 2k)
\end{align}
is the double factorial, which vanishes for even $D$.
\begin{proof}
We focus on proving the first inequality, with the other 2 being easy variants. We first show that:
\begin{align}
    f\left(z\right) & := \sum_{\left(n_d\right)_{d \geq 2} \neq \bm{0}}\binom{n}{\left(n_d\right)_{d \geq 2}}\frac{n^{-D/2}}{\prod\limits_{d \geq 2}d!^{n_d}}D!!z^D
\end{align}
is analytic over $B(0, r)$ for sufficiently small $r > 0$. For that purpose, it suffices to show:
\begin{align}
    \sum_{\left(n_d\right)_{d \geq 2} \neq \bm{0}}\binom{n}{\left(n_d\right)_{d \geq 2}}\frac{n^{-D/2}}{\prod\limits_{d \geq 2}d!^{n_d}}D!!r^D < \infty.
\end{align}
We compute:
\begin{align}
    \sum_{\left(n_d\right)_{d \geq 2} \neq \bm{0}}\binom{n}{\left(n_d\right)_{d \geq 2}}\frac{n^{-D/2}}{\prod\limits_{d \geq 2}d!^{n_d}}D!!r^D & = \sum_{\substack{\left(n_d\right)_{d \geq 2} \neq \bm{0}}}\binom{n}{\left(n_d\right)_{d \geq 2}}\frac{n^{-D/2}}{\prod\limits_{d \geq 2}d!^{n_d}}D!!r^D\nonumber\\
    & = 2\sum_{\substack{\left(n_d\right)_{d \geq 2} \neq \bm{0}\\D\textrm{ odd}}}\binom{n}{\left(n_d\right)_{d \geq 2}}\frac{n^{-D/2}}{\prod\limits_{d \geq 2}d!^{n_d}}r^D\int_0^{\infty}\!\mathrm{d}y\,\frac{e^{-y^2/2}}{\sqrt{2\pi}}\,y^{D + 1}\nonumber\\
    & \leq 2\sum_{\substack{\left(n_d\right)_{d \geq 2} \neq \bm{0}}}\binom{n}{\left(n_d\right)_{d \geq 2}}\frac{n^{-D/2}}{\prod\limits_{d \geq 2}d!^{n_d}}r^D\int_0^{\infty}\!\mathrm{d}y\,\frac{e^{-y^2/2}}{\sqrt{2\pi}}\,y^{D + 1}\nonumber\\
    & = 2\int_0^{\infty}\!\mathrm{d}y\,\frac{e^{-y^2/2}}{\sqrt{2\pi}}y\sum_{\left(n_d\right)_{d \geq 2} \neq \bm{0}}\binom{n}{\left(n_d\right)_{d \geq 2}}\frac{n^{-D/2}}{\prod\limits_{d \geq 2}d!^{n_d}}r^D\nonumber\\
    & = 2\int_0^{\infty}\!\mathrm{d}y\,\frac{e^{-y^2/2}}{\sqrt{2\pi}}y\sum_{\left(n_d\right)_{d \geq 2} \neq \bm{0}}\binom{n}{\left(n_d\right)_{d \geq 2}}\prod_{d \geq 2}\left(\frac{n^{-d/2}r^dy^d}{d!}\right)^{n_d}\nonumber\\
    & = 2\int_0^{\infty}\!\mathrm{d}y\,\frac{e^{-y^2/2}}{\sqrt{2\pi}}y\left(\left(1 + \sum_{d \geq 2}\frac{n^{-d/2}r^dy^d}{d!}\right)^n - 1\right)\nonumber\\
    & = 2\int_0^{\infty}\!\mathrm{d}y\,\frac{e^{-y^2/2}}{\sqrt{2\pi}}y\left(\left(\exp\left(n^{-1/2}ry\right) - n^{-1/2}ry\right)^n - 1\right).
\end{align}
We now bound the quantity raised to the power $n$ in the integrand. First, on $y \in I_1 := \left[0, n^{1/2}r^{-1}\right]$, using general bound
\begin{align}
    \left|e^z - z\right| & = \left|1 + \left(e^z - 1 - z\right)\right|\nonumber\\
    & \leq 1 + \left|e^z - 1 - z\right|\nonumber\\
    & \leq 1 + \frac{|z|^2}{2}e^{|z|}\nonumber\\
    & \leq \exp\left(\frac{|z|^2}{2}e^{|z|}\right),
\end{align}
valid for all complex number $z \in \mathbf{C}$, we obtain:
\begin{align}
    \left(\exp\left(n^{-1/2}ry\right) - n^{-1/2}ry\right)^n & = \left|\left(\exp\left(n^{-1/2}ry\right) - n^{-1/2}ry\right)^n\right|\nonumber\\
    & \leq \exp\left(\frac{r^2e}{2}y^2\right),
\end{align}
hence
\begin{align}
    \left|\left(\exp\left(n^{-1/2}ry\right) - n^{-1/2}ry\right)^n - 1\right| & = \left(\exp\left(n^{-1/2}ry\right) - n^{-1/2}ry\right)^n - 1\nonumber\\
    & = \exp\left(\frac{r^2r}{2}y^2\right) - 1\nonumber\\
    & \leq \frac{r^2e}{2}y^2\exp\left(\frac{r^2e}{2}y^2\right).\label{eq:qgms_integral_moment_series_multinomial_sum_bound_proof_interval_1}
\end{align}
This proves a bound of the integrand on interval $I_1 = \left[0, n^{1/2}r^{-1}\right]$. Now, for $y \in I_2 := \left[n^{1/2}r^{-1}, +\infty\right)$, using bound
\begin{align}
    \left|e^z - z\right| & \leq e^{|z|}
\end{align}
holding for all $z \in \mathbf{C}$, we bound
\begin{align}
    \left(\exp\left(n^{-1/2}ry\right) - n^{-1/2}ry\right)^n & = \left|\left(\exp\left(n^{-1/2}ry\right) - n^{-1/2}ry\right)^n\right|\nonumber\\
    & \leq \exp\left(n^{1/2}ry\right)\nonumber\\
    & \leq \exp\left(r^2y^2\right).
\end{align}
From there, we deduce bound:
\begin{align}
    \left|\left(\exp\left(n^{-1/2}ry\right) - n^{-1/2}ry\right)^n - 1\right| & = \left(\exp\left(n^{-1/2}ry\right) - n^{-1/2}ry\right)^n - 1\nonumber\\
    & \leq \exp\left(r^2y^2\right) - 1\nonumber\\
    & \leq r^2y^2\exp\left(r^2y^2\right)\label{eq:qgms_integral_moment_series_multinomial_sum_bound_proof_interval_2}
\end{align}
For $r$ smaller than an absolute constant $r_*$, bounds in Eqs.~\ref{eq:qgms_integral_moment_series_multinomial_sum_bound_proof_interval_1}, \ref{eq:qgms_integral_moment_series_multinomial_sum_bound_proof_interval_2} can be combined to ensure:
\begin{align}
    \left|\left(\exp\left(n^{-1/2}ry\right) - n^{-1/2}ry\right)^n - 1\right| & \leq r^2y^2e^{y^2/4} \qquad \forall y \in \mathbf{R}_+.
\end{align}
Hence,
\begin{align}
    \sum_{\left(n_d\right)_{d \geq 2}}\binom{n}{\left(n_d\right)_{d \geq 2}}\frac{n^{-D/2}}{\prod\limits_{d \geq 2}d!^{n_d}}D!!r^D & \leq 2r^2\int_0^{+\infty}\!\mathrm{d}y\,\frac{e^{-y^2/4}}{\sqrt{2\pi}}y^3\nonumber\\
    & =: cr^2\nonumber\\
    & < \infty,
\end{align}
which is the desired statement. Hence, we proved analyticity of $z$ on $B(0, r_*/2)$ for a sufficiently small absolute constant $r_* > 0$. By replaying the previous calculation with $r \to |z|$, we have also implicitly proven bound:
\begin{align}
    \left|f\left(z\right)\right| & \leq c|z|^2.
\end{align}
For all $z \in B\left(0, r_*\right)$, one can write:
\begin{align}
    \sum_{\left(n_d\right)_{d \geq 2}}\binom{n}{\left(n_d\right)_{d \geq 2}}\frac{n^{-D/2}}{\prod\limits_{d \geq 2}}D!!D^kz^D & = \frac{\mathrm{d}^kf\left(ze^u\right)}{\mathrm{d}u^k}\Bigg|_{u = 0},
\end{align}
and
\begin{align}
    u & \longmapsto f\left(ze^u\right)
\end{align}
is analytic on an open ball containing $\overline{B}\left(0, \log(2)\right)$. Applying the Cauchy inequality to bound the derivative then gives:
\begin{align}
    \left|\frac{\mathrm{d}^kf\left(ze^u\right)}{\mathrm{d}u^k}\Bigg|_{u = 0}\right| & \leq \frac{k!}{\left(\log 2\right)^k}\sup_{\substack{u \in \mathbf{C}\\|u| = \log(2)}}\left|f\left(ze^u\right)\right|\nonumber\\
    & \leq \frac{k!}{\left(\log 2\right)^k}c\left(2|z|\right)^2,
\end{align}
proving the result.
\end{proof}
\end{lemma}

When studying QAOA on random optimization problems depending on Gaussian disorder (as is the case of the quadratic Sherrington-Kirkpatrick model), Gaussian integration by parts proves very useful. The following lemma states this fundamental result:

\begin{proposition}[Gaussian integration by parts]
\label{prop:gaussian_integration_by_parts}
Consider a Gaussian-distributed vector
\begin{align}
    \bm{g} & = \left(g_1, \ldots, g_d\right),
\end{align}
with mean $\bm{0}$ and arbitrary covariance matrix
\begin{align}
    \mathbb{E}\bm{g} = 0, && \mathbb{E}\bm{g}\bm{g}^T =: \bm{\Sigma}.
\end{align}
Then, for all sufficiently nice function $f$ of the Gaussian vector and all index $1 \leq j \leq d$, the following integration by parts formula holds:
\begin{align}
    \mathbb{E}\left[g_jf\left(\bm{g}\right)\right] & = \sum_{1 \leq k \leq d}\mathbf{E}\left[g_jg_k\right]\partial_kf\left(\bm{g}\right).
\end{align}
\end{proposition}

Schematically, Proposition~\ref{prop:gaussian_integration_by_parts} turns a linear Gaussian factor $g_j$ into a derivative inside the expectation. From applying this Proposition twice, the following corollary, applying to a quadratic Gaussian monomial, can be deduced:

\begin{corollary}[Gaussian integration by part, quadratic version]
\label{cor:gaussian_integration_by_parts_quadratic}
Consider a Gaussianly distributed vector
\begin{align}
    \bm{g} & = \left(g_1, \ldots, g_d\right),
\end{align}
with mean $\bm{0}$ and arbitrary covariance matrix:
\begin{align}
    \mathbb{E}\bm{g} = 0, && \mathbb{E}\bm{g}\bm{g}^T =: \bm{\Sigma}.
\end{align}
Then, for all sufficiently nice function $f$ of the Gaussian vector and all pair of indices $1 \leq j, k \leq d$, the following integration by parts formula holds:
\begin{align}
    \mathbb{E}\left[g_jg_k\varphi\left(\bm{g}\right)\right] & = \mathbb{E}\left[\mathbb{E}\left[g_jg_k\right]\varphi\left(\bm{g}\right)\right] + \sum_{1 \leq l, m \leq d}\mathbb{E}\left[\mathbb{E}\left[g_jg_l\right]\mathbb{E}\left[g_kg_m\right]\partial_{lm}\varphi\left(\bm{g}\right)\right]
\end{align}
\end{corollary}

\section{Expansion of parametrized QGMS saddle point around the noninteracting limit: lowest order}

After establishing general formulae for the expansion of the saddle-point $\bm{\theta}^*$ order-by-order in $\lambda$, let us specialize these to the lowest-order contributions in $\lambda$. As observed earlier, the series expansion \ref{eq:saddle_point_equation_tensor_interpretation_series_solution} of the saddle-point can be truncated to a finite number of terms for a given expansion order in $\lambda$. Evaluating each such term involves computing the power of some matrix, where the matrix is defined by blocks according to equation \ref{eq:saddle_point_equation_tensor_interpretation_t_block} and each block is in turn expressed as a sum over partitions and tuples. Despite the apparent combinatorial complexity of this procedure, it can be carried out manually for smallest orders, leading to a simple low order expansion of $\bm{\theta}^*$ in terms of the noninteracting correlation tensors $\overline{\bm{C}}^{(r)}$.

This section concretely implements this approach. The general method is to start with the series expansion of $\bm{\theta}^*$ in equation \ref{eq:saddle_point_equation_tensor_interpretation_series_solution}. After truncating this to the appropriate order, remaining powers of $\bm{T}$ are expressed in terms of $\bm{T}$ blocks $\bm{T}_{q,\,d}$. Finally, each block is expressed a the sum of contributions indexed by a partition and tuple, as explicitly defined in equation \ref{eq:saddle_point_equation_tensor_interpretation_t_block}.

\subsection{Order $1$ expansion}

This order can easily be derived by assuming existence of the series expansion and expanding the right-hand side of the saddle point equation:
\begin{align}
    \bm{\theta}^*\left(\lambda\right) & = \frac{\sum\limits_{\bm{a} \in \mathcal{S}}Q_{\bm{a}}\exp\left(\lambda\bm{\theta}^{*}\left(\lambda\right)^T\bm{L}_{:,\,\bm{a}}\right)\lambda\bm{L}_{:,\,\bm{a}}}{\sum\limits_{\bm{a} \in \mathcal{S}}Q_{\bm{a}}\exp\left(\lambda\bm{\theta}^{*}\left(\lambda\right)^T\bm{L}_{:,\bm{a}}\right)}\nonumber\\
    & = \frac{\sum\limits_{\bm{a} \in \mathcal{S}}Q_{\bm{a}}\lambda\bm{L}_{:,\,\bm{a}} + \mathcal{O}\left(\lambda^2\right)}{\sum\limits_{\bm{a} \in \mathcal{S}}Q_{\bm{a}} + \mathcal{O}\left(\lambda\right)}\nonumber\\
    & = \lambda\frac{\sum_{\bm{a} \in \mathcal{S}}Q_{\bm{a}}\bm{L}_{:,\,\bm{a}}}{\sum\limits_{\bm{a} \in \mathcal{S}}Q_{\bm{a}}} + \mathcal{O}\left(\lambda^2\right).
\end{align}
Let us verify consistency of this direct calculation with equation \ref{eq:saddle_point_equation_tensor_interpretation_series_solution}.

The desired quantity $\bm{\theta}^*$ is the first block row of the left-hand-side $\bm{\Theta}^*$ of equation \ref{eq:saddle_point_equation_tensor_interpretation_series_solution}. Therefore, using indices to index blocks rather than scalar entries,
\begin{align}
    \bm{\theta}^* & = \left[\bm{\Theta}^*\right]_{1}\nonumber\\
    & = \left[\overline{\bm{\Theta}^*} + \bm{T}\overline{\bm{\Theta}^*} + \bm{T}^2\overline{\bm{\Theta}^*} + \ldots\right]_1\nonumber\\
    & = \left[\overline{\bm{\Theta}^*} + \mathcal{O}\left(\lambda^3\right)\right]_1\nonumber\\
    & = \overline{\bm{\Theta}^*}_1 + \mathcal{O}\left(\lambda^3\right)\nonumber\\
    & = \overline{\bm{\theta}^*} + \mathcal{O}\left(\lambda^3\right)\nonumber\\
    & = \lambda\overline{\bm{C}}^{(1)} + \mathcal{O}\left(\lambda^3\right),
\end{align}
where in going from the second to the third line we used that all blocks of $\bm{T}$ are of order at most $\lambda^2$.

\clearpage

\subsection{Order $3$ expansion}

\begin{figure}[!htbp]
    \centering
    \includegraphics[width=0.6\textwidth]{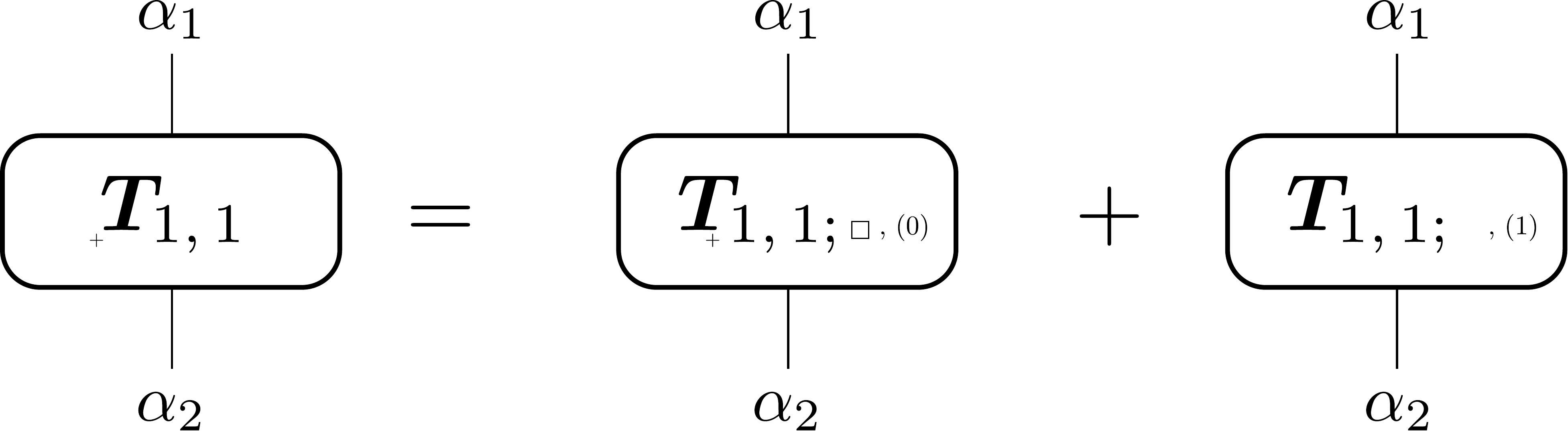}
    \caption{Contributions to $\bm{T}$ matrix block $\bm{T}_{1,\,1}$, listed according to partitions and tuples (see equation \ref{eq:t_block_partitions_decomposition})}
    \label{fig:t_1_1_contributions}
\end{figure}

\begin{figure}[!htbp]
    \centering
    \begin{subfigure}{0.4\textwidth}
        \includegraphics[width=\textwidth]{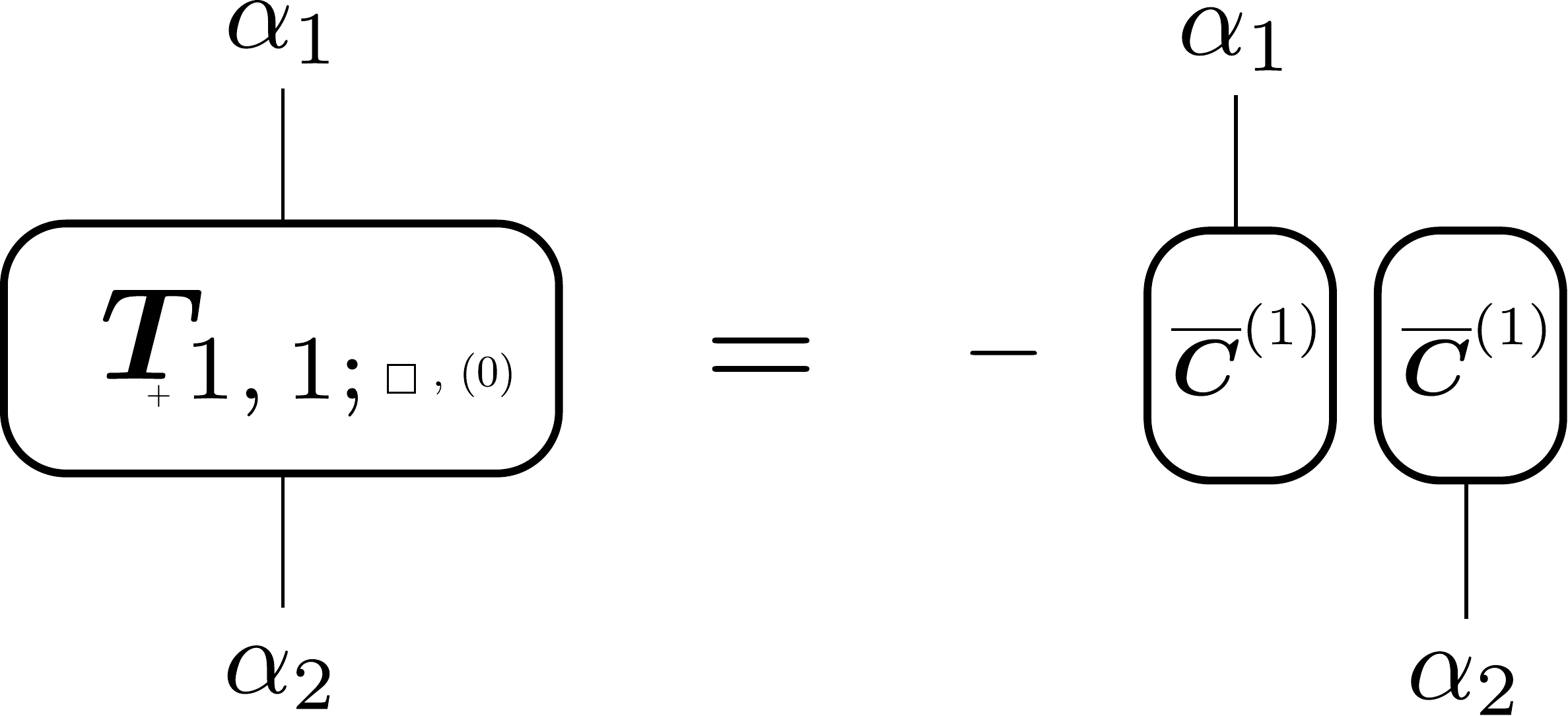}
        \caption{Contribution 1}
    \end{subfigure}
    \hspace*{0.08\textwidth}
    \begin{subfigure}{0.28\textwidth}
        \includegraphics[width=\textwidth]{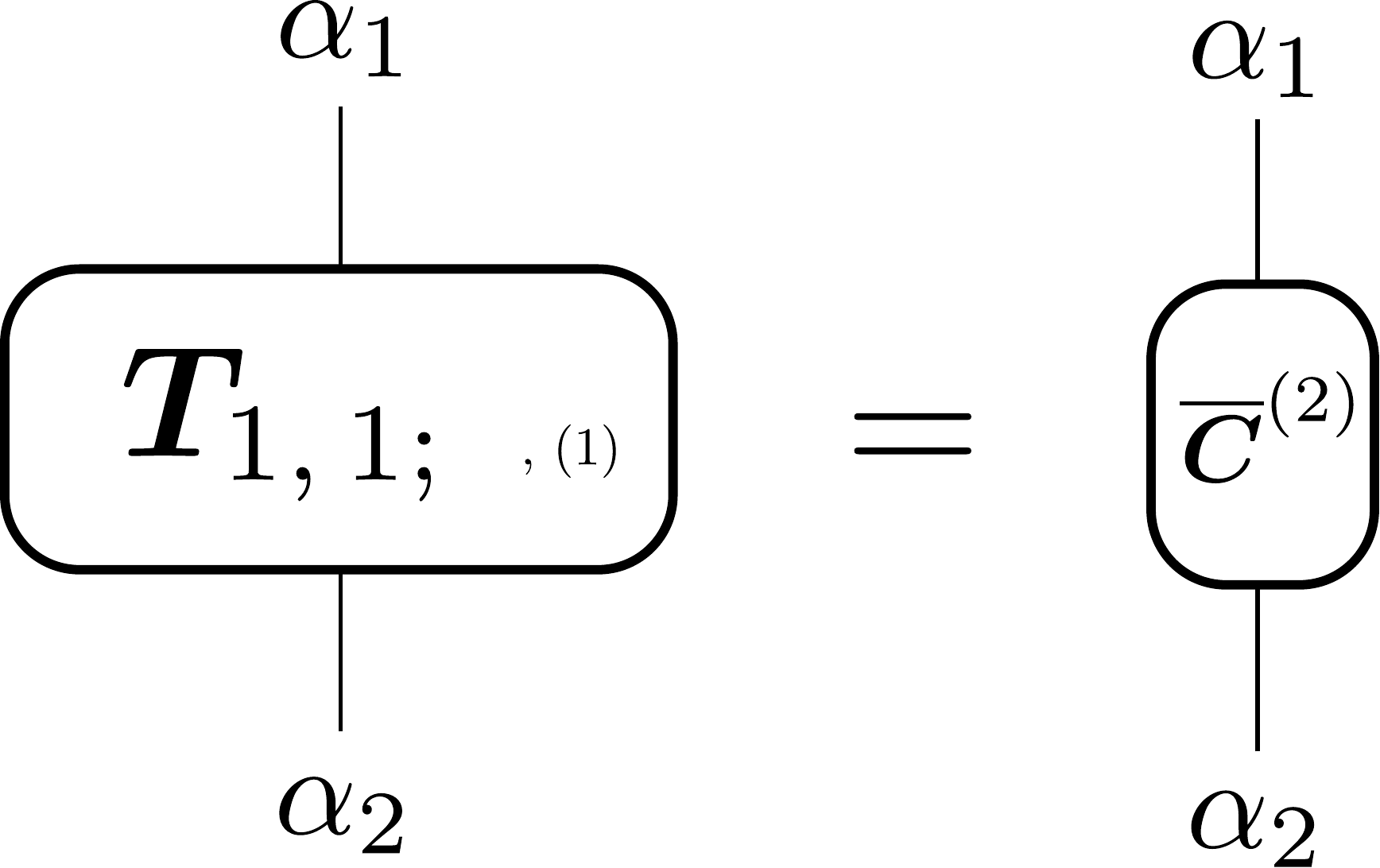}
        \caption{Contribution 2}
    \end{subfigure}
    \caption{Expression of each contribution of $\bm{T}_{1,\,1}$ in terms of noninteracting correlation tensors $\overline{\bm{C}}^{(d)}$.}
    \label{fig:t_1_1_contribution_by_contribution}
\end{figure}

Similar to the previous calculation, we now have

\begin{align}
    \bm{\theta}^* & = \left[\overline{\bm{\Theta}^*} + \bm{T}\overline{\bm{\Theta}^*} + \mathcal{O}\left(\lambda^5\right)\right]_{1}
\end{align}
  Now, recalling block  $\bm{T}_{q,\,d}$ of $\bm{T}$ has order $q + d$ (see equation \ref{eq:saddle_point_equation_tensor_interpretation_t_block}), only block $\bm{T}_{1,\,1}$ contributes to the desired order, i.e.:
\begin{align}
    \left[\bm{T}\overline{\bm{\Theta}^*}\right]_1 & = \bm{T}_{1, 1}\left[\overline{\bm{\Theta}^*}\right]_1 + \mathcal{O}\left(\lambda^5\right)\nonumber\\
    & = \bm{T}_{1, 1}\overline{\bm{\theta}^*} + \mathcal{O}\left(\lambda^5\right)
\end{align}
We then require the expansion of $\bm{T}_{1, 1}$ as a sum over partitions and tuples following equation \ref{eq:saddle_point_equation_tensor_interpretation_t_block}. This decomposition, shown as a tensor network on figures \ref{fig:t_1_1_contributions} and \ref{fig:t_1_1_contribution_by_contribution}, is rather simple due to the smallness of block indices.

\begin{figure}[!htbp]
    \centering
    \begin{subfigure}{0.49\textwidth}
        \centering
        \includegraphics[width=0.4\textwidth]{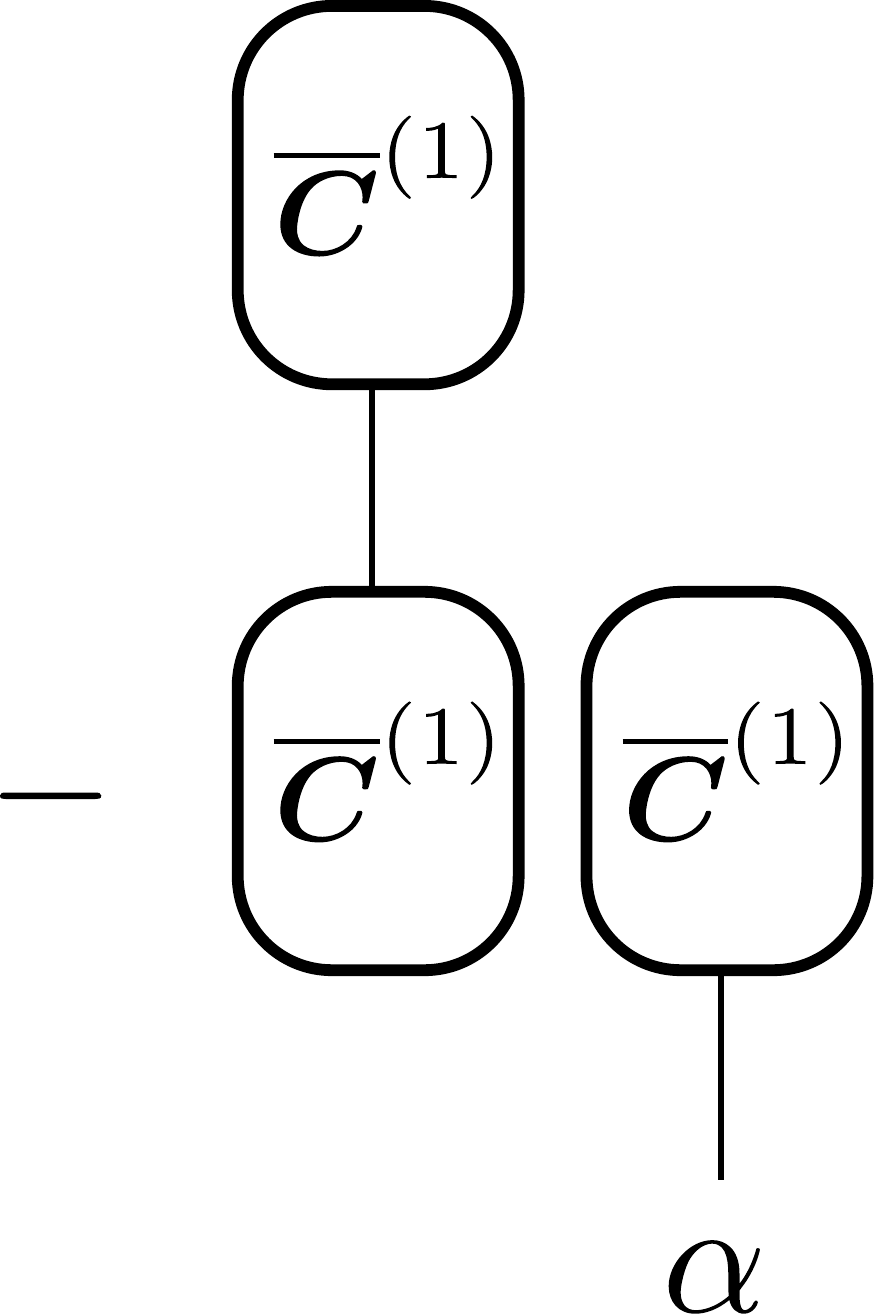}
        \caption{Contribution 1}
    \end{subfigure}
    \begin{subfigure}{0.49\textwidth}
        \centering
        \includegraphics[width=0.125\textwidth]{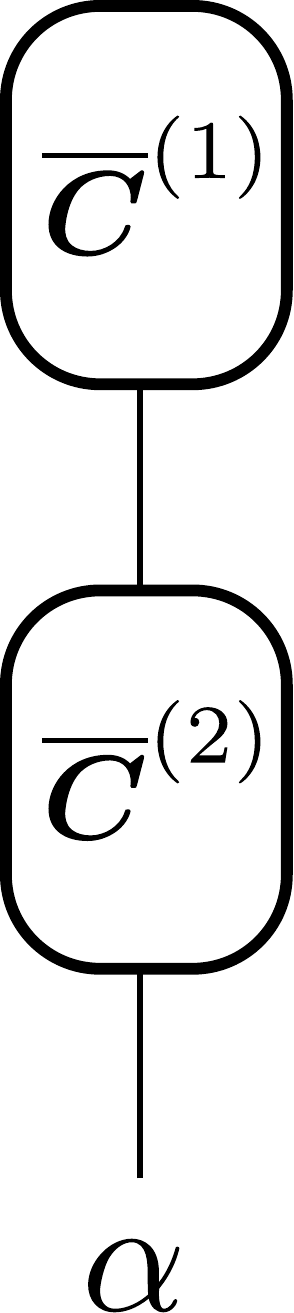}
        \caption{Contribution 2}
    \end{subfigure}
    \caption{The order 3 contributions to the saddle point $\bm{\theta}^* = \left(\theta_{\alpha}\right)_{\alpha \in \mathcal{A}}$. The $\lambda^3$ factor is omitted in the diagrams for brevity.}
    \label{fig:order_3_contributions}
\end{figure}

Multiplying these matrix blocks together, we obtain all the order 3 contributions to the saddle point $\bm{\theta}^*$, as represented on figure \ref{fig:order_3_contributions}.

\clearpage

\subsection{Order $5$ expansion}

\begin{align}
    \bm{\theta}^* & = \left[\overline{\bm{\Theta}^*} + \bm{T}\overline{\bm{\Theta}^*} + \bm{T}^2\overline{\bm{\Theta}^*} + \mathcal{O}\left(\lambda^7\right)\right]_1.
\end{align}

\begin{figure}[!htbp]
    \centering
    \includegraphics[width=0.6\textwidth]{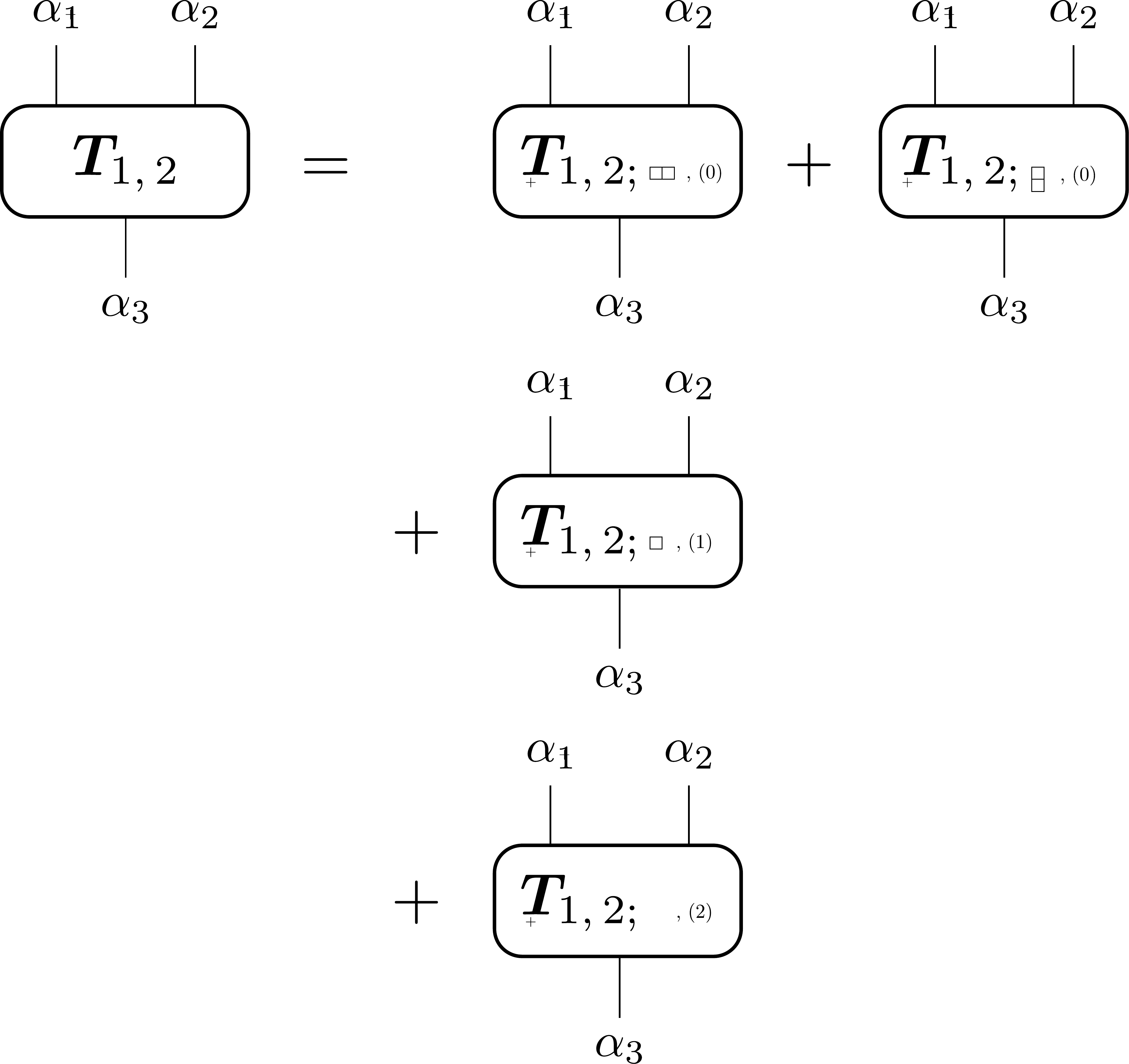}
    \caption{Contributions to $\bm{T}$ matrix block $\bm{T}_{1,\,2}$.}
    \label{fig:t_1_2_contributions}
\end{figure}

\begin{figure}[!htbp]
    \centering
    \begin{subfigure}{0.42\textwidth}
        \includegraphics[width=\textwidth]{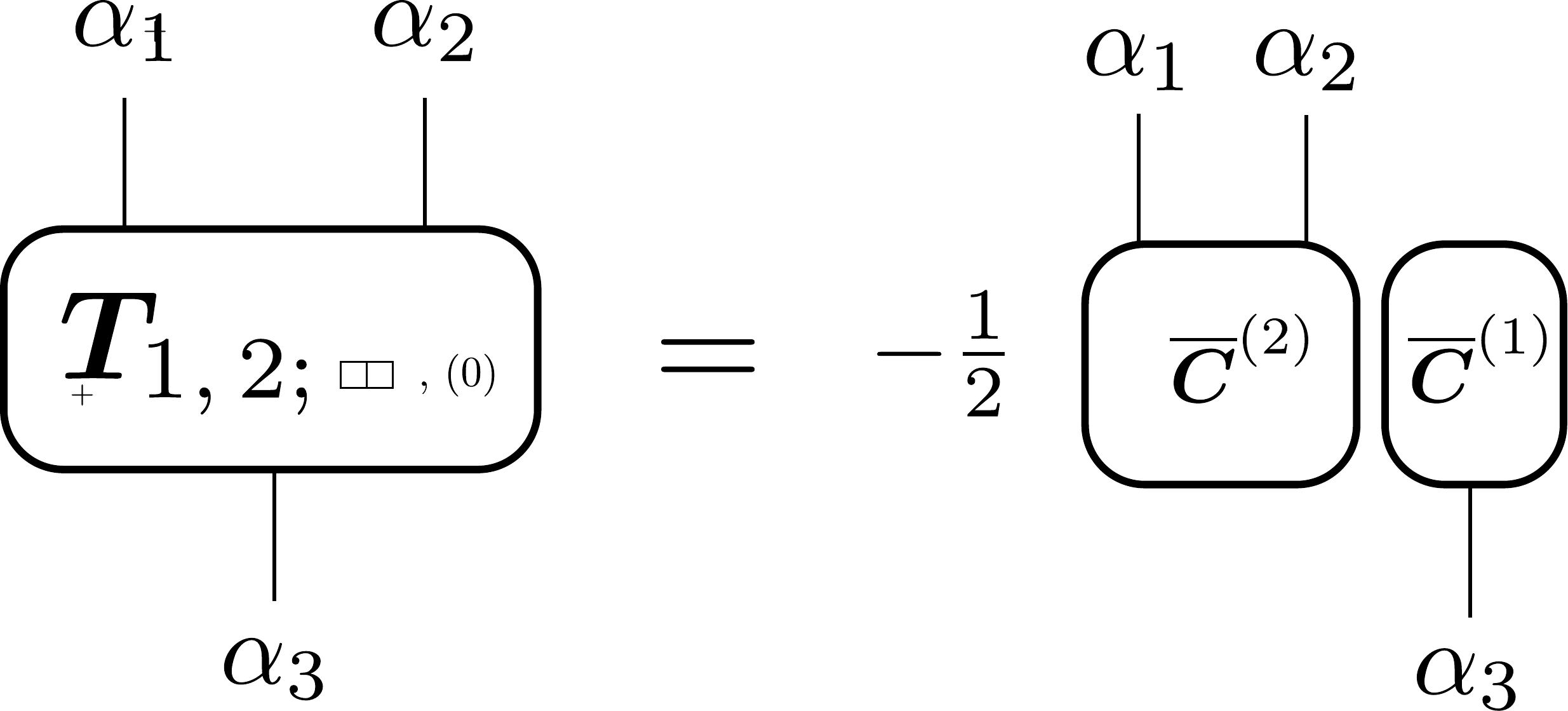}
        \caption{Contribution 1}
    \end{subfigure}
    \hspace*{0.08\textwidth}
    \begin{subfigure}{0.42\textwidth}
        \includegraphics[width=\textwidth]{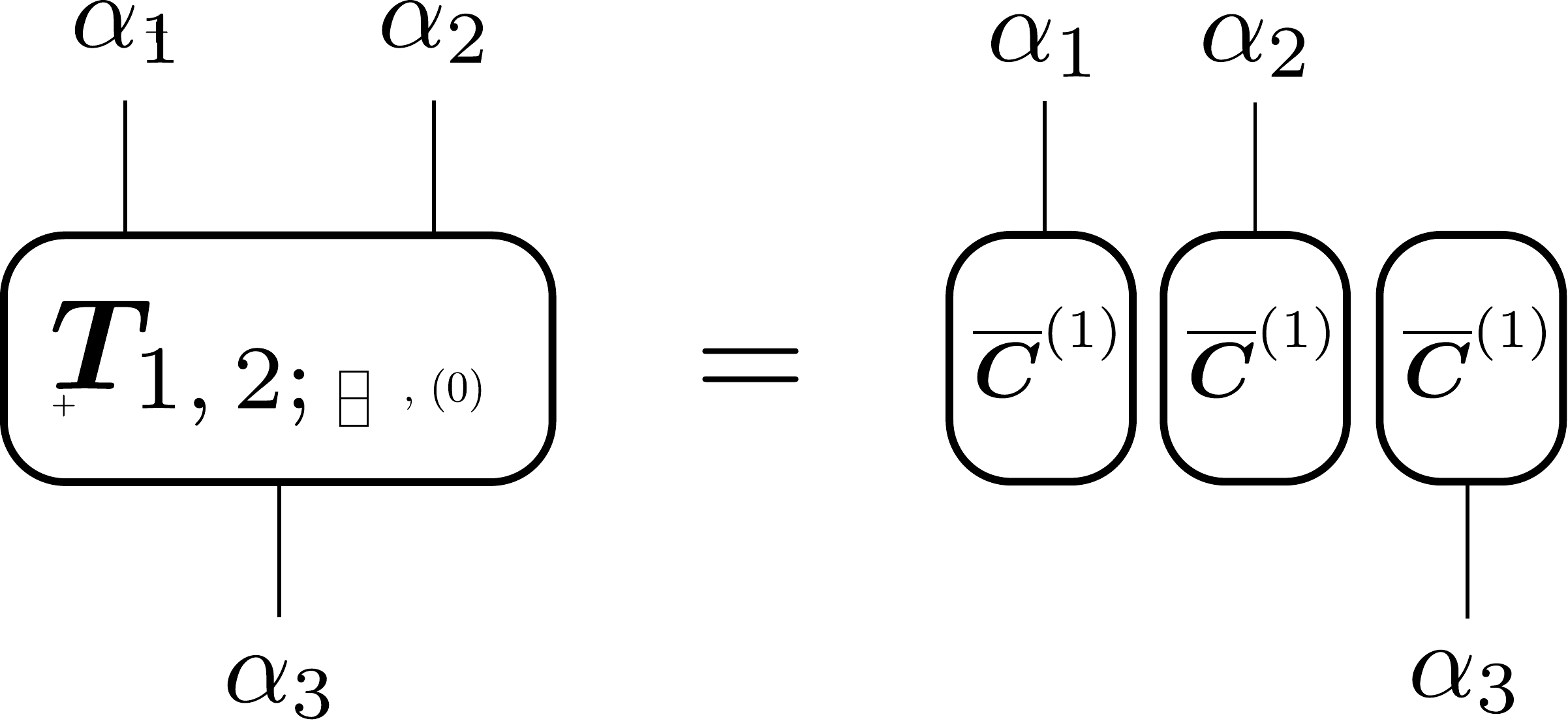}
        \caption{Contribution 2}
    \end{subfigure}\\
    \vspace*{10px}
    \begin{subfigure}{0.4\textwidth}
        \includegraphics[width=\textwidth]{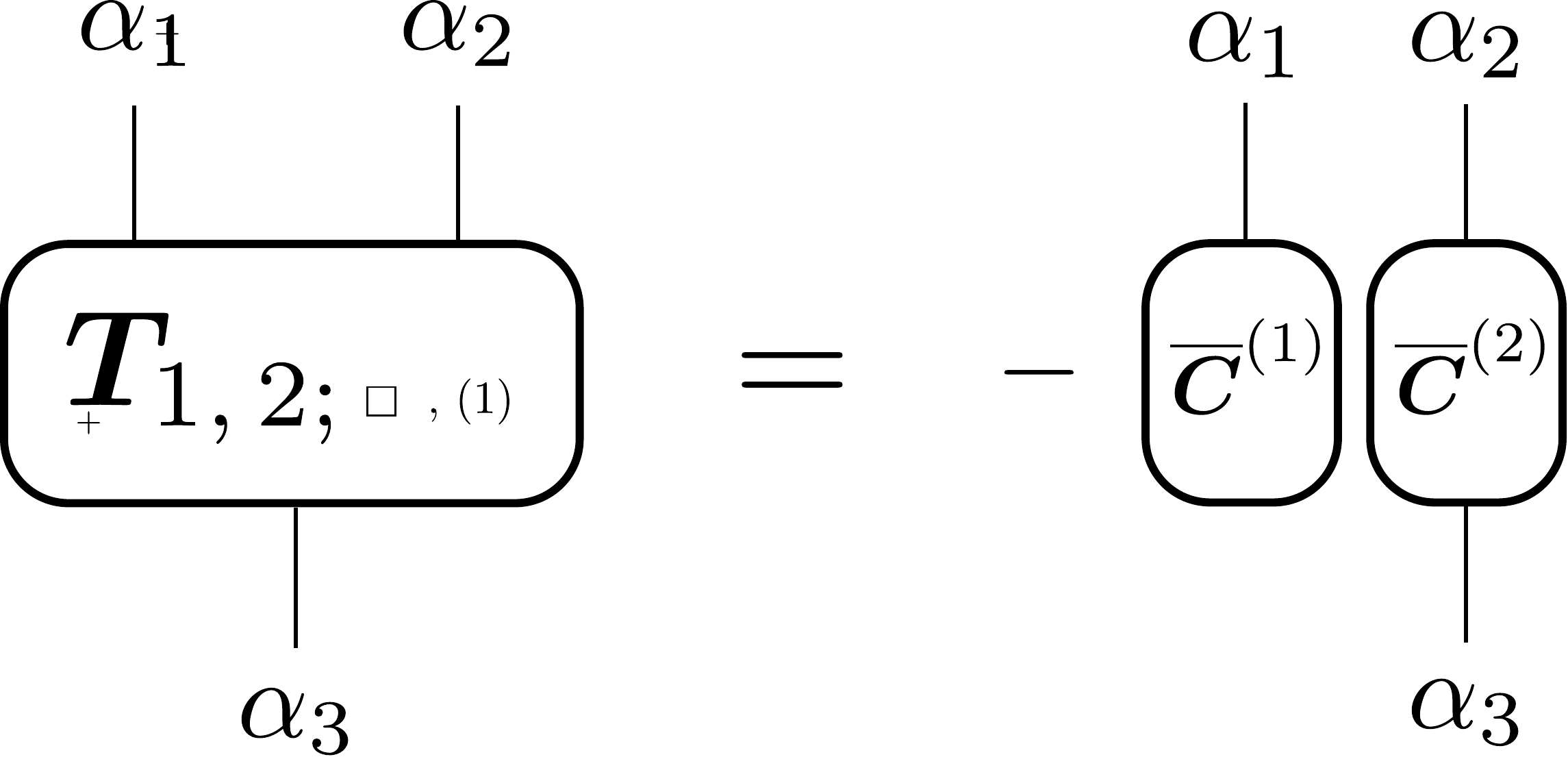}
        \caption{Contribution 3}
    \end{subfigure}
    \hspace*{0.08\textwidth}
    \begin{subfigure}{0.39\textwidth}
        \includegraphics[width=0.85\textwidth]{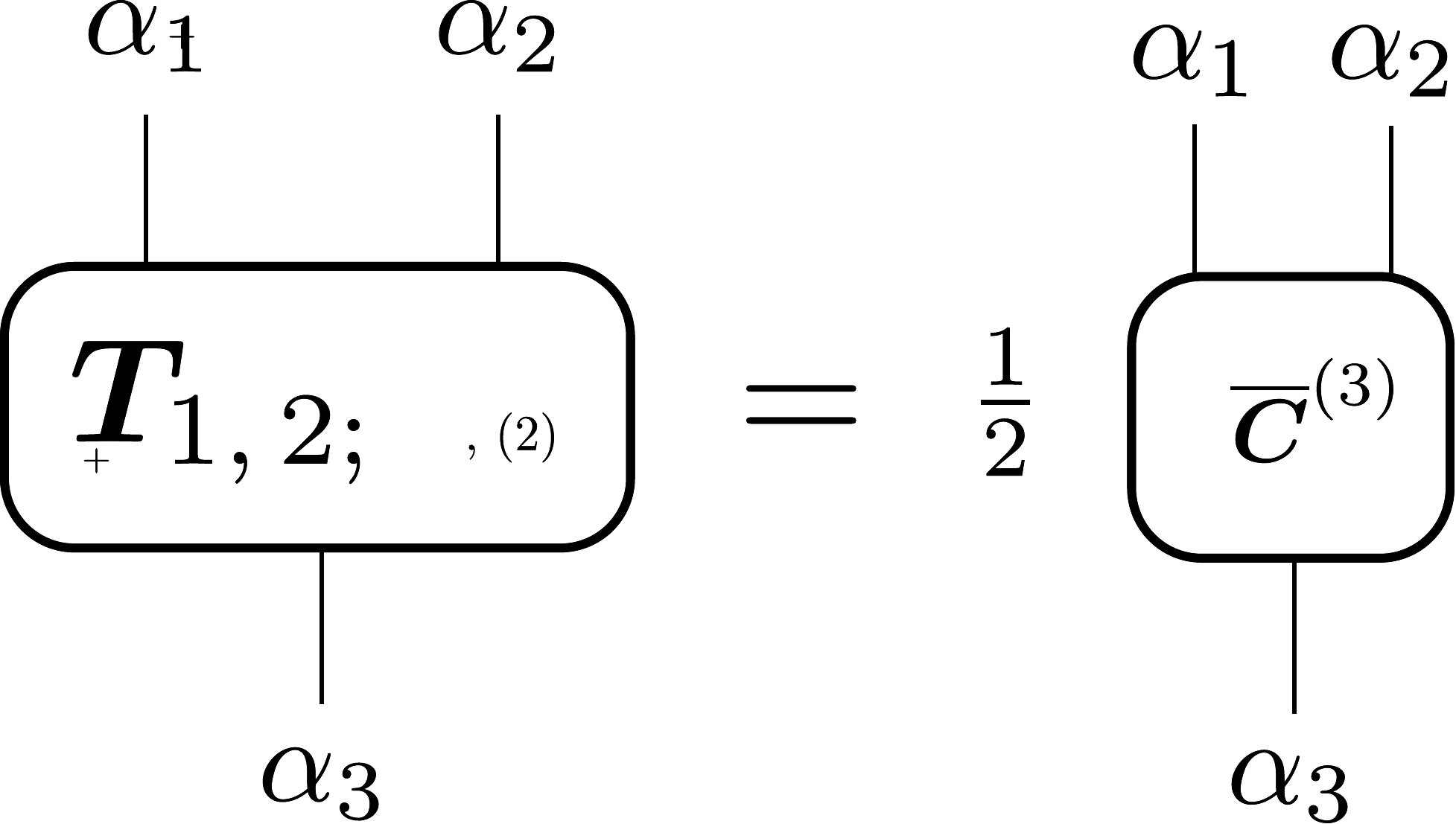}
        \caption{Contribution 4}
    \end{subfigure}
    \caption{Expression of each contribution to $\bm{T}_{1,\,2}$ in terms of noninteracting correlation tensors}
    \label{fig:t_1_2_contribution_by_contribution}
\end{figure}

The single $\bm{T}$ power has now more contributions:
\begin{align}
    \left[\bm{T}\overline{\bm{\Theta}^*}\right]_1 & = \bm{T}_{1, 1}\left[\overline{\bm{\Theta}^*}\right]_1 + \bm{T}_{1, 2}\left[\overline{\bm{\Theta}^*}\right]_2 + \mathcal{O}\left(\lambda^7\right)\nonumber\\
    & = \bm{T}_{1,\,1}\overline{\bm{\theta}^*} + \bm{T}_{1,\,2}\overline{\bm{\theta}^*}^{\otimes 2},
\end{align}
where we used that block coordinate $q$ of $\bm{\Theta}^*$, which is $\left(\lambda\overline{\bm{\theta^*}}\right)^{\otimes q}$ is of order at most $\lambda^{q}$. The only contribution of order exactly $5$ in the above equation (given contributions of order exactly $3$ were already accounted for in the previous paragraph) is
\begin{align}
    \bm{T}_{1,\,2}\overline{\bm{\theta}^*}^{\otimes 2}
\end{align}
We then need the expansion of $\bm{T}_{1, 2}$ as a sum over partitions and tuples, which is provided on figures \ref{fig:t_1_2_contributions} and \ref{fig:t_1_2_contribution_by_contribution}.

The contributions of the square of $\bm{T}$ are rather trivial:
\begin{align}
    \left[\bm{T}^2\overline{\bm{\Theta}^*}\right]_1 & = \bm{T}_{1, 1}\bm{T}_{1, 1}\left[\overline{\bm{\Theta}^*}\right]_1 + \mathcal{O}\left(\lambda^7\right)\\
    & = \bm{T}_{1,\,1}\bm{T}_{1,\,1}\overline{\bm{\theta}^*}
\end{align}

\begin{figure}[!htbp]
    \centering
    \begin{subfigure}{0.24\textwidth}
        \centering
        \includegraphics[width=0.85\textwidth]{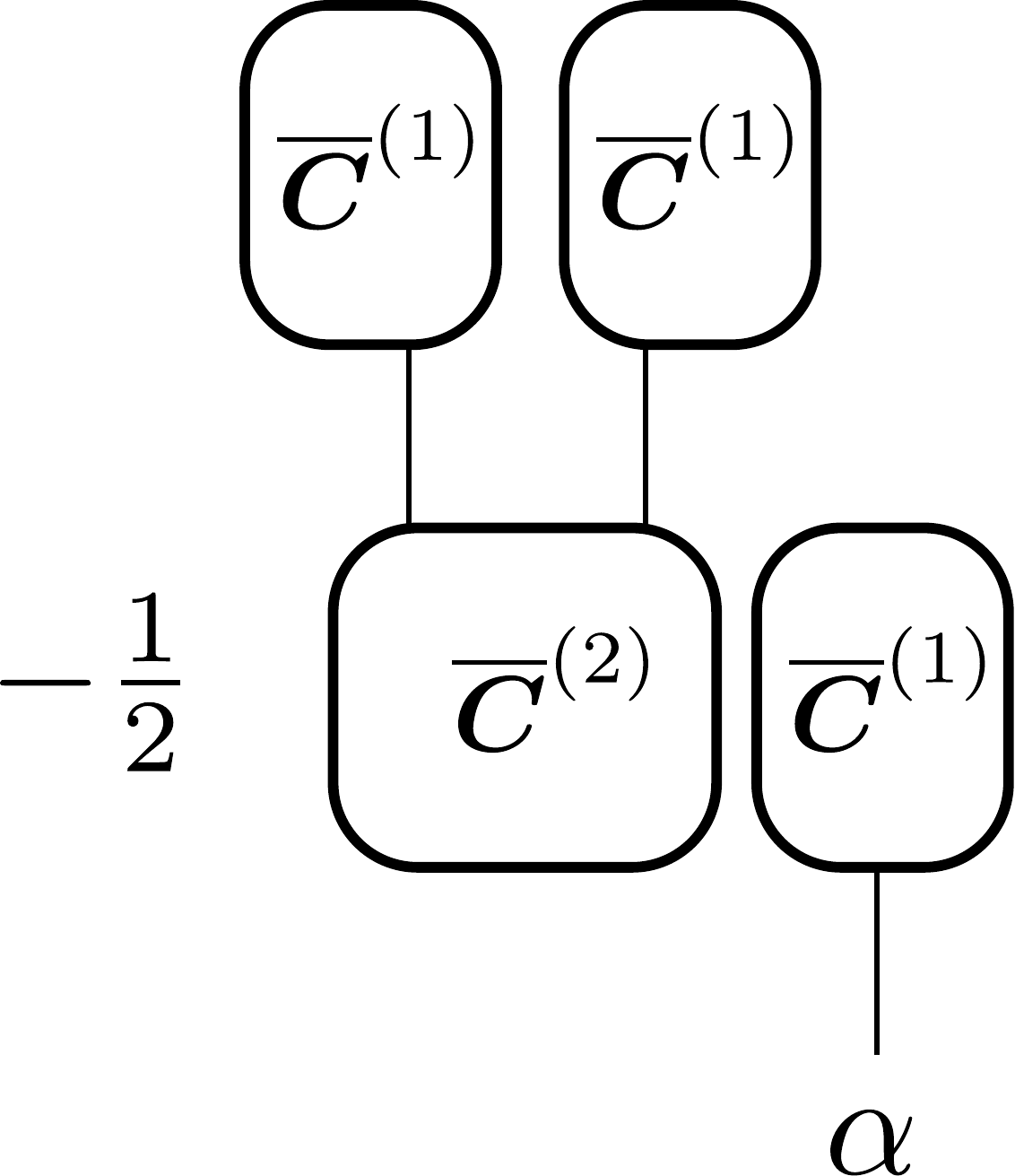}
        \caption{Contribution 1}
    \end{subfigure}
    \begin{subfigure}{0.24\textwidth}
        \centering
        \includegraphics[width=0.75\textwidth]{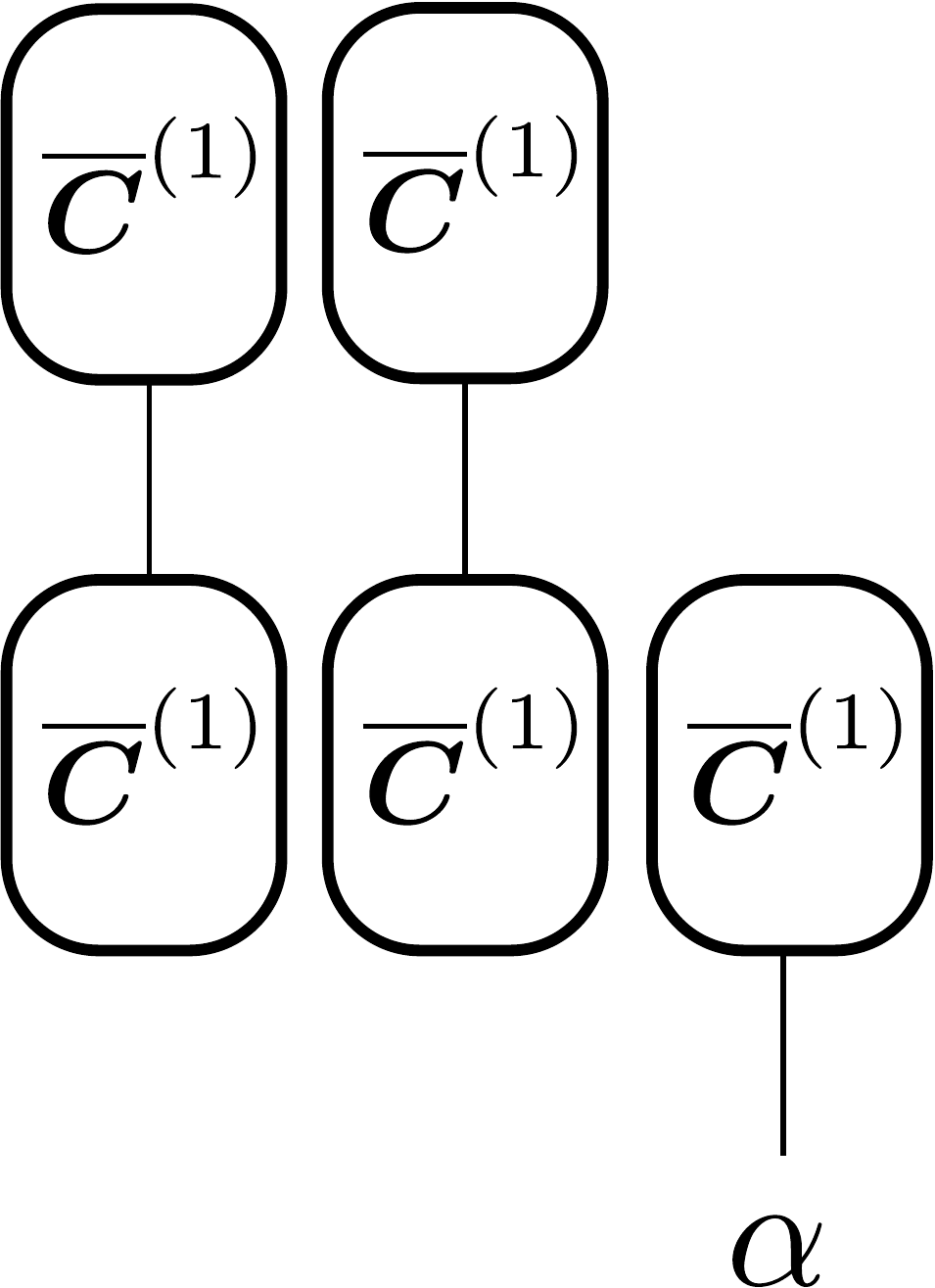}
        \caption{Contribution 2}
    \end{subfigure}
    \begin{subfigure}{0.24\textwidth}
        \centering
        \includegraphics[width=0.65\textwidth]{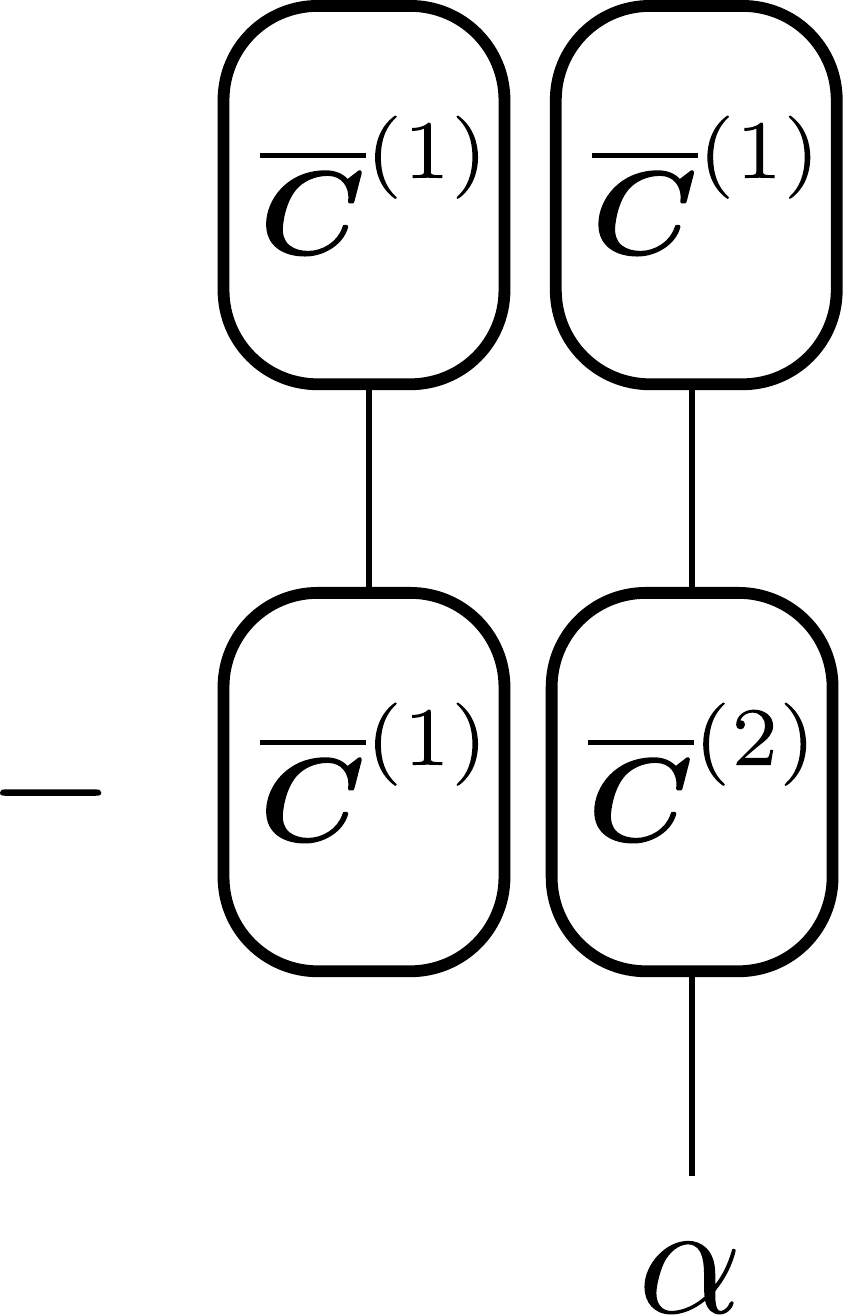}
        \caption{Contribution 3}
    \end{subfigure}
    \begin{subfigure}{0.24\textwidth}
        \centering
        \includegraphics[width=0.6\textwidth]{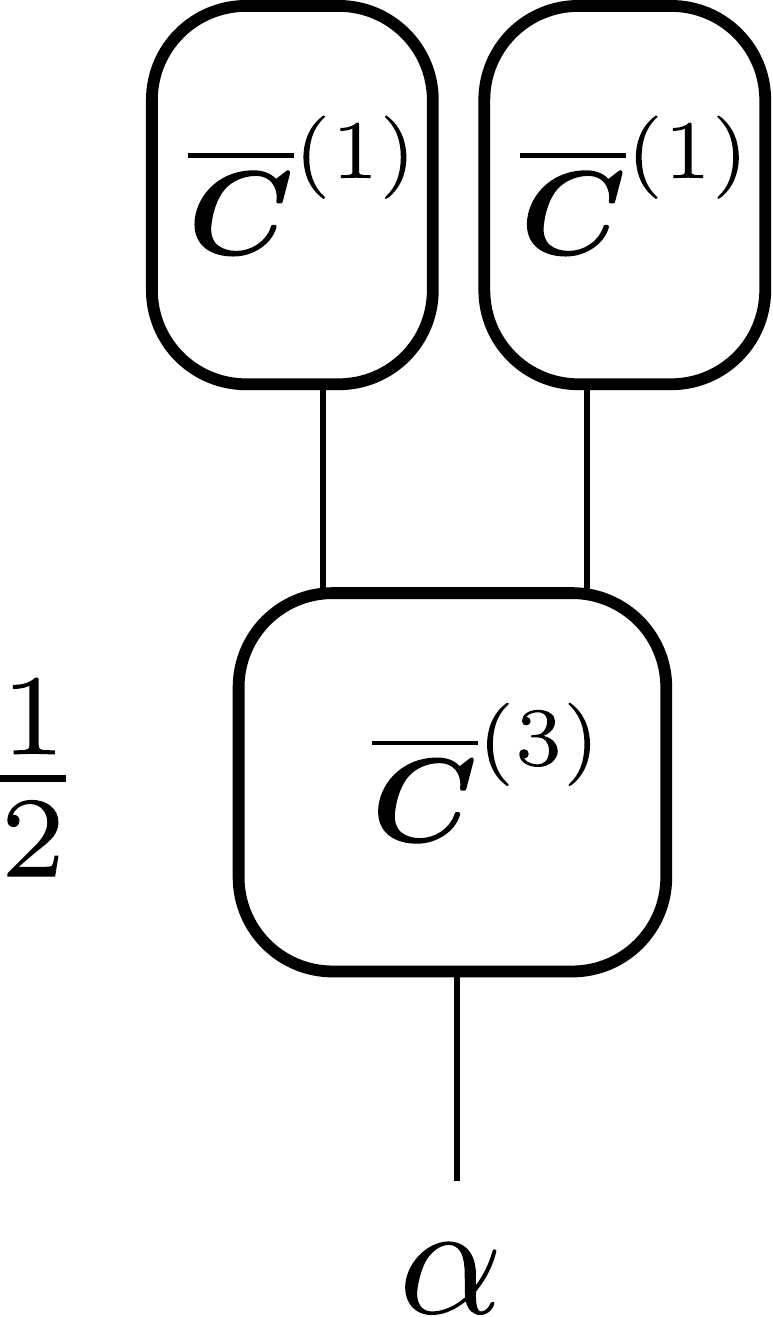}
        \caption{Contribution 4}
    \end{subfigure}
    \caption{Order 5 contributions to saddle point $\bm{\theta}^* = \left(\theta^*_{\alpha}\right)_{\alpha \in \mathcal{A}}$ from linear $\bm{T}$ term $\bm{T}\overline{\bm{\Theta}^*}$}
    \label{fig:order_5_t_contributions}
\end{figure}

\begin{figure}[!htbp]
    \centering
    \begin{subfigure}{0.24\textwidth}
        \centering
        \includegraphics[width=0.75\textwidth]{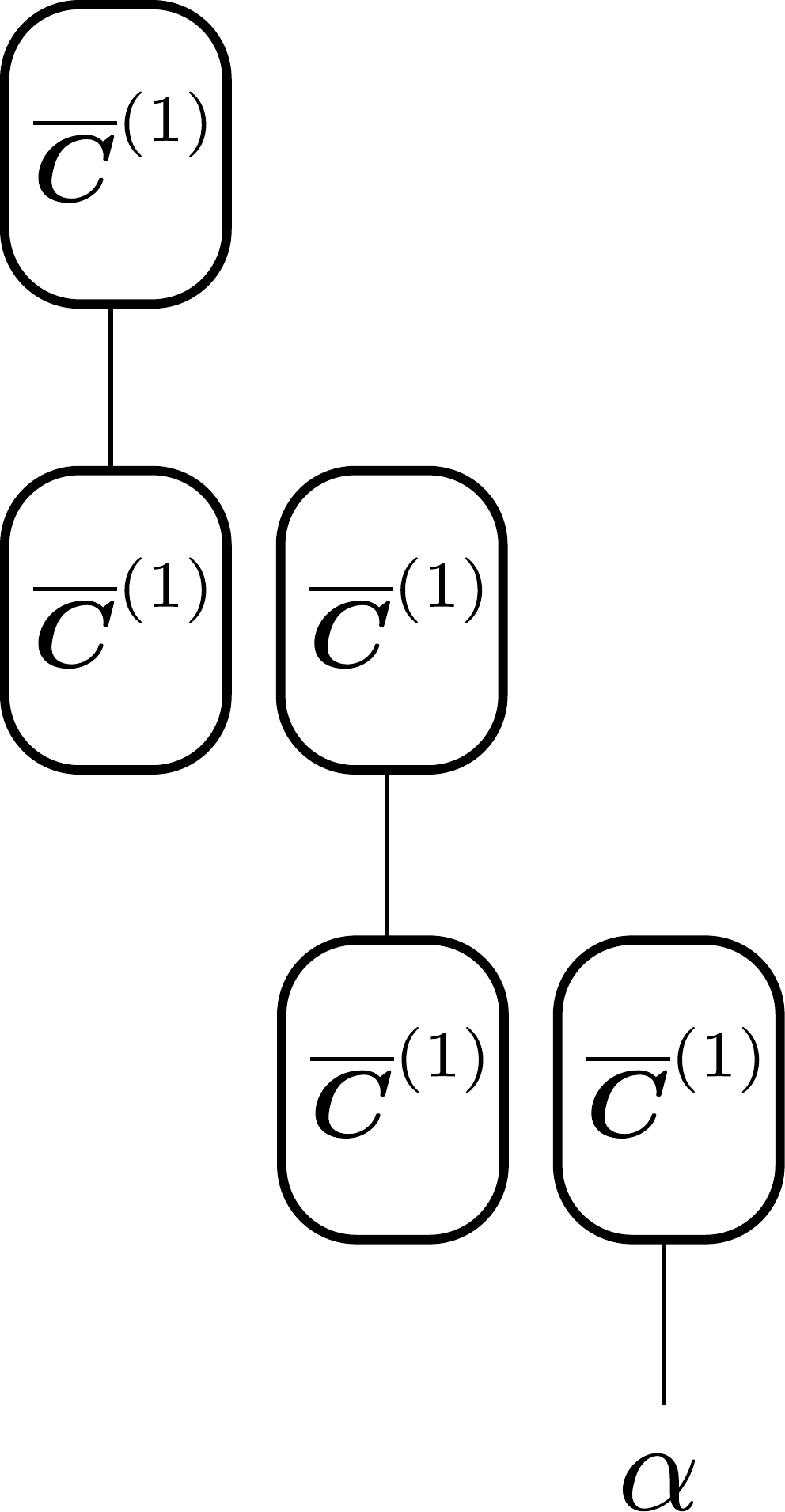}
        \caption{Contribution 1}
    \end{subfigure}
    \begin{subfigure}{0.24\textwidth}
        \centering
        \includegraphics[width=0.65\textwidth]{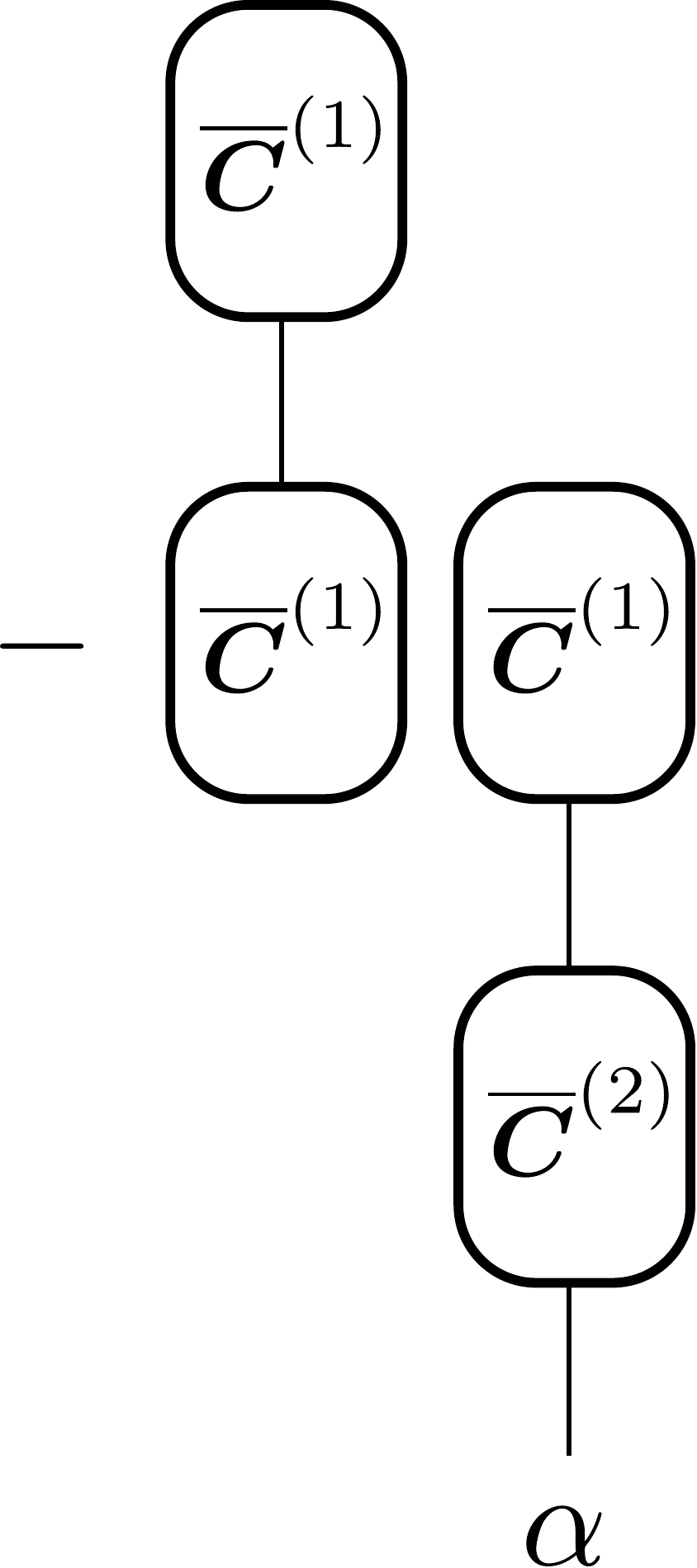}
        \caption{Contribution 2}
    \end{subfigure}
    \begin{subfigure}{0.24\textwidth}
        \centering
        \includegraphics[width=0.7\textwidth]{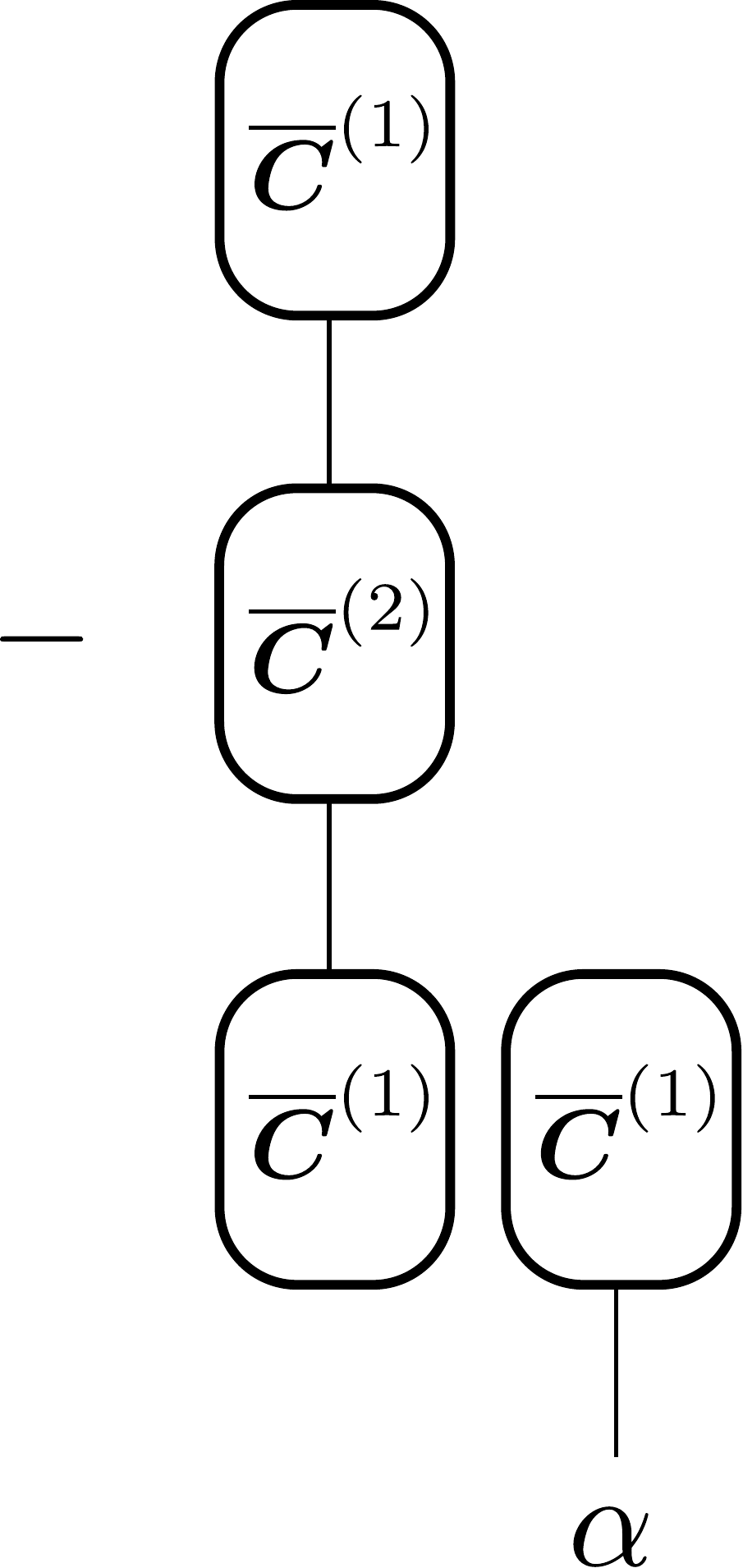}
        \caption{Contribution 3}
    \end{subfigure}
    \begin{subfigure}{0.24\textwidth}
        \centering
        \includegraphics[width=0.23\textwidth]{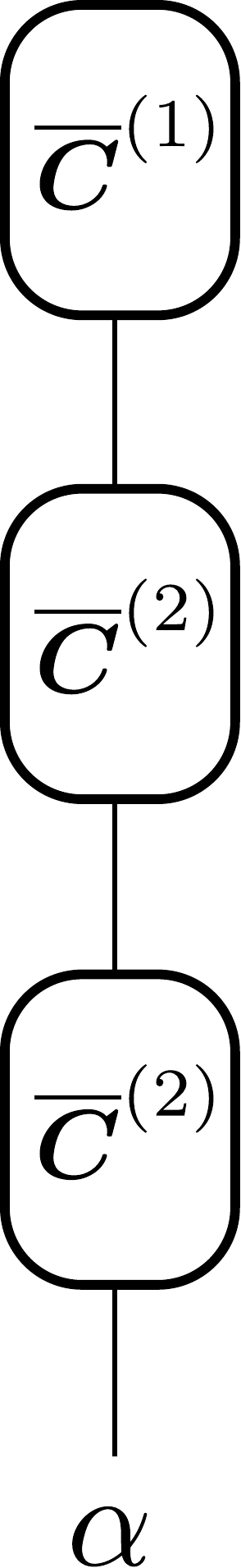}
        \caption{Contribution 4}
    \end{subfigure}
    \caption{Order 5 contributions to saddle point $\bm{\theta}^* = \left(\theta^*_{\alpha}\right)_{\alpha \in \mathcal{A}}$ from quadratic $\bm{T}$ term $\bm{T}^2\overline{\bm{\Theta}^*}$}
    \label{fig:order_5_t_squared_contributions}
\end{figure}

Multiplying matrix block together, one can again obtain explicit graphical representations of the order 5 contributions to the saddle point. Since these are numerous than at order 3, it will be helpful to separate them in two groups: on the one hand, contributions arising from the linear $\bm{T}$ temr $\bm{T}\overline{\bm{\Theta}^*}$; on the other hand, contributions from the quadratic term $\bm{T}^2\overline{\bm{\Theta}^*}$. These are respectively presented on figures \ref{fig:order_5_t_contributions}, \ref{fig:order_5_t_squared_contributions}.

\clearpage

\subsection{Order $7$ expansion}

\begin{align}
    \bm{\theta}^* & = \left[\overline{\bm{\Theta}^*} + \bm{T}\overline{\bm{\Theta}^*} + \bm{T}^2\overline{\bm{\Theta}^*} + \bm{T}^3\overline{\bm{\Theta}^*} + \mathcal{O}\left(\lambda^7\right)\right]_1.
\end{align}

\begin{figure}[!htbp]
    \centering
    \includegraphics[width=0.6\textwidth]{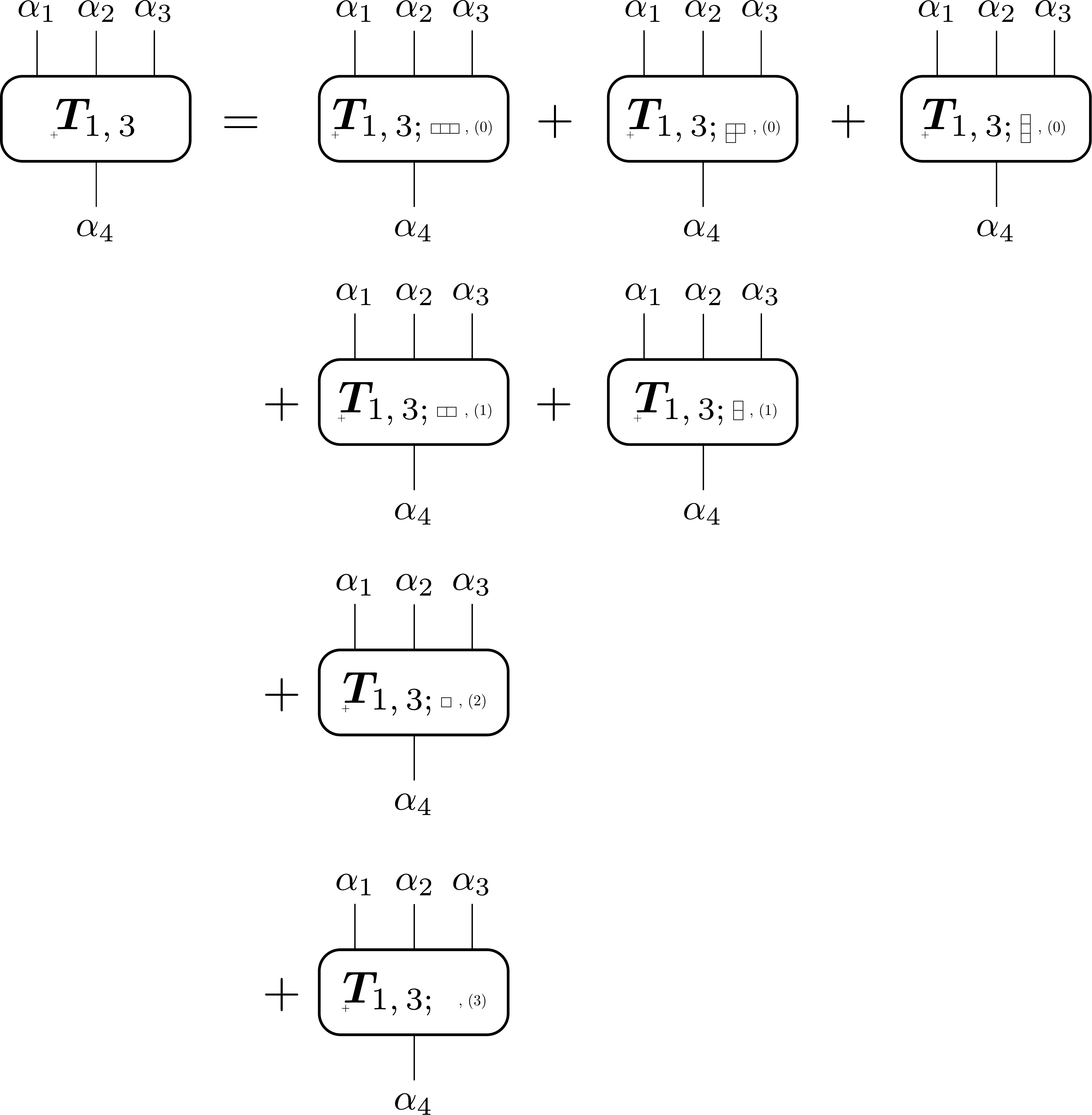}
    \caption{Contributions to $\bm{T}$ matrix block $\bm{T}_{1,\,3}$}
    \label{fig:t_1_3_contributions}
\end{figure}

\begin{figure}[!htbp]
    \centering
    \begin{subfigure}{0.45\textwidth}
        \centering
        \includegraphics[width=\textwidth]{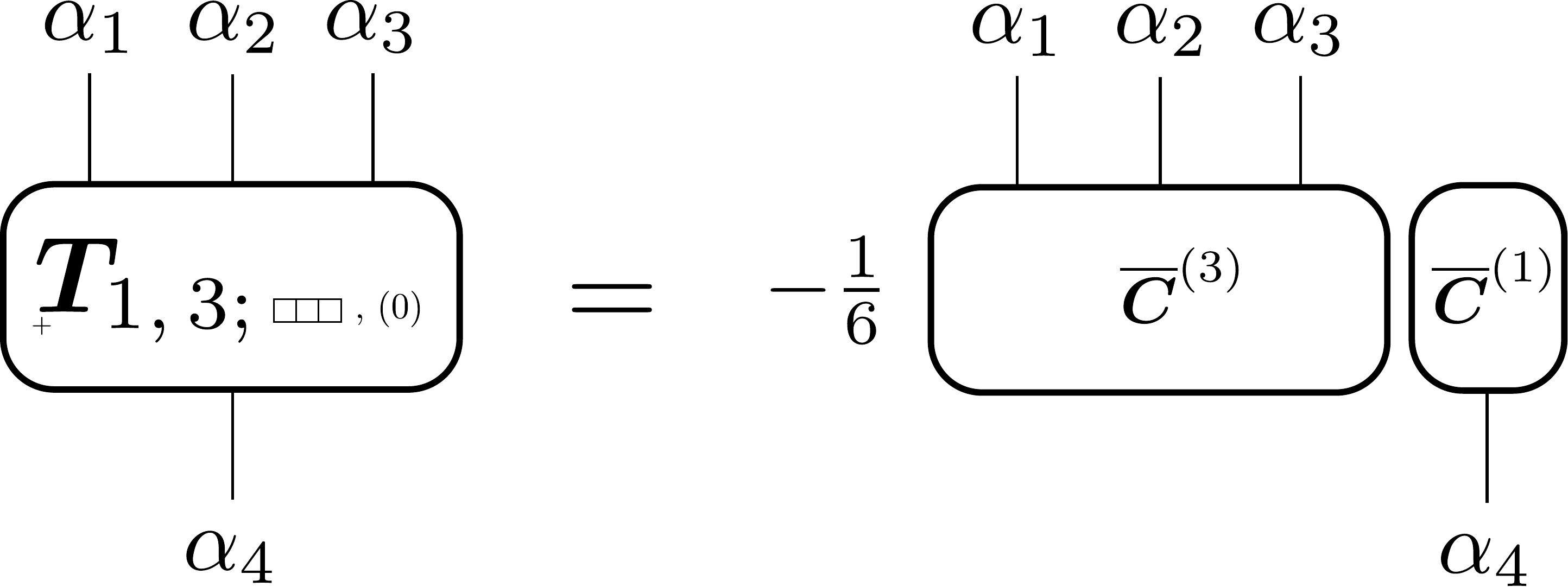}
        \caption{Contribution 1}
    \end{subfigure}
    \hspace*{0.08\textwidth}
    \begin{subfigure}{0.45\textwidth}
        \centering
        \includegraphics[width=0.9\textwidth]{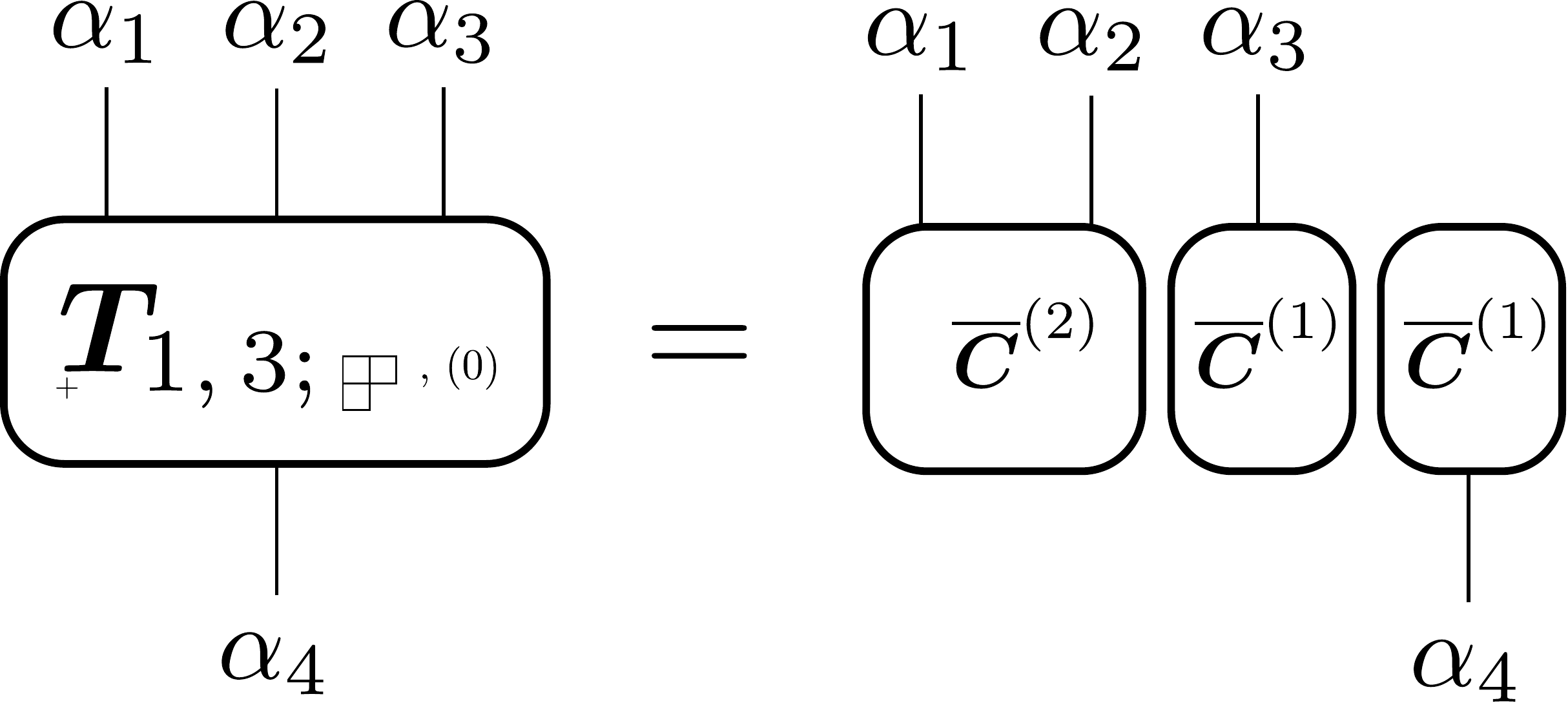}
        \caption{Contribution 2}
    \end{subfigure}\\\vspace*{10pt}
    \begin{subfigure}{0.45\textwidth}
        \centering
        \includegraphics[width=\textwidth]{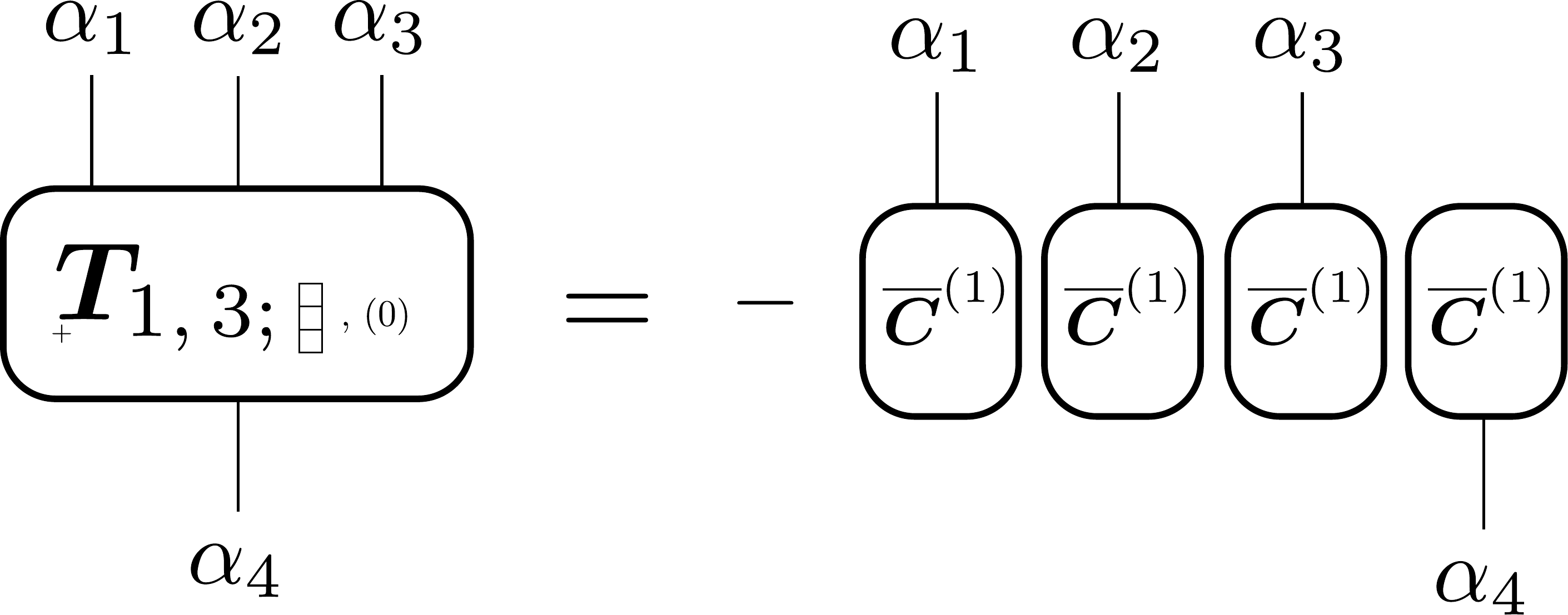}
        \caption{Contribution 3}
    \end{subfigure}
    \hspace*{0.08\textwidth}
    \begin{subfigure}{0.45\textwidth}
        \centering
        \includegraphics[width=0.9\textwidth]{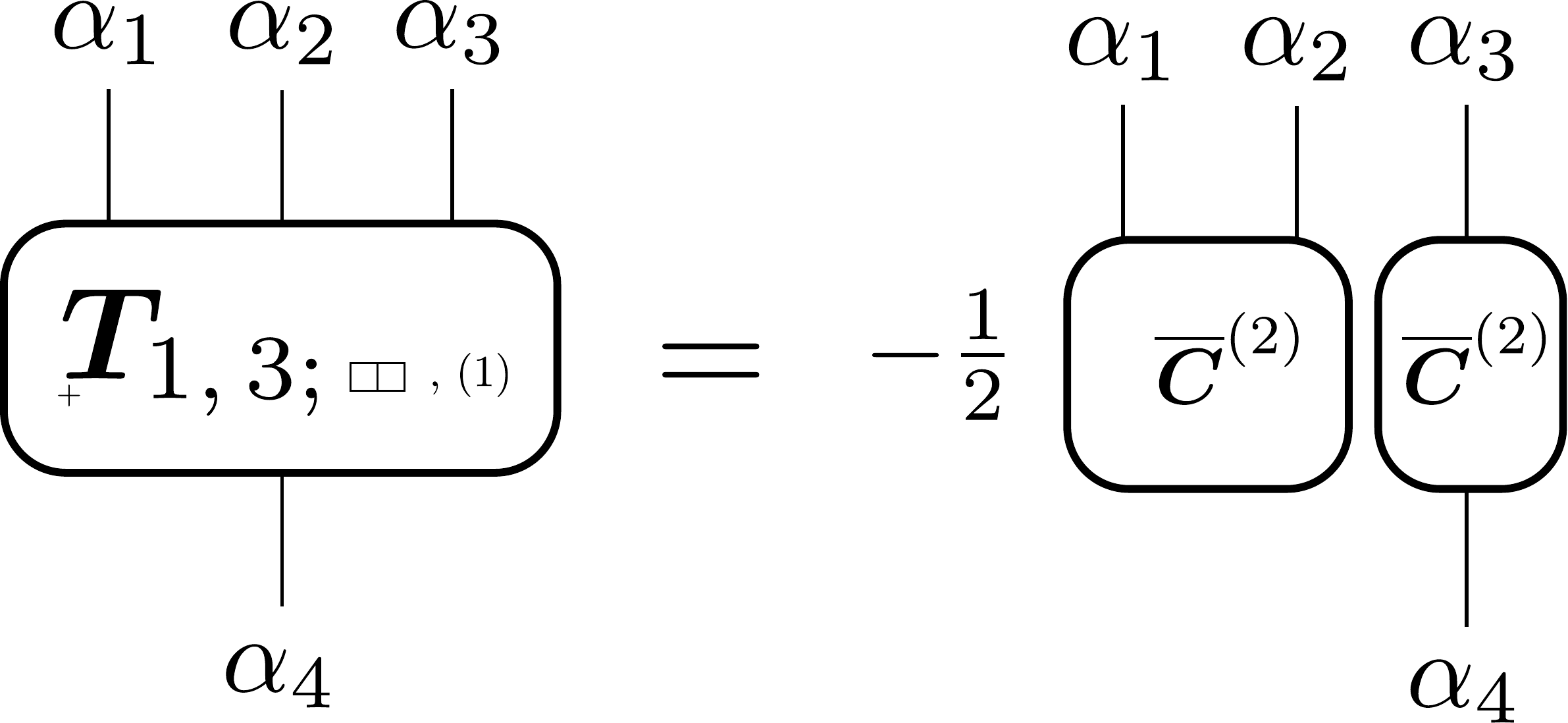}
        \caption{Contribution 4}
    \end{subfigure}\\\vspace*{10pt}
    \begin{subfigure}{0.45\textwidth}
        \centering
        \includegraphics[width=0.85\textwidth]{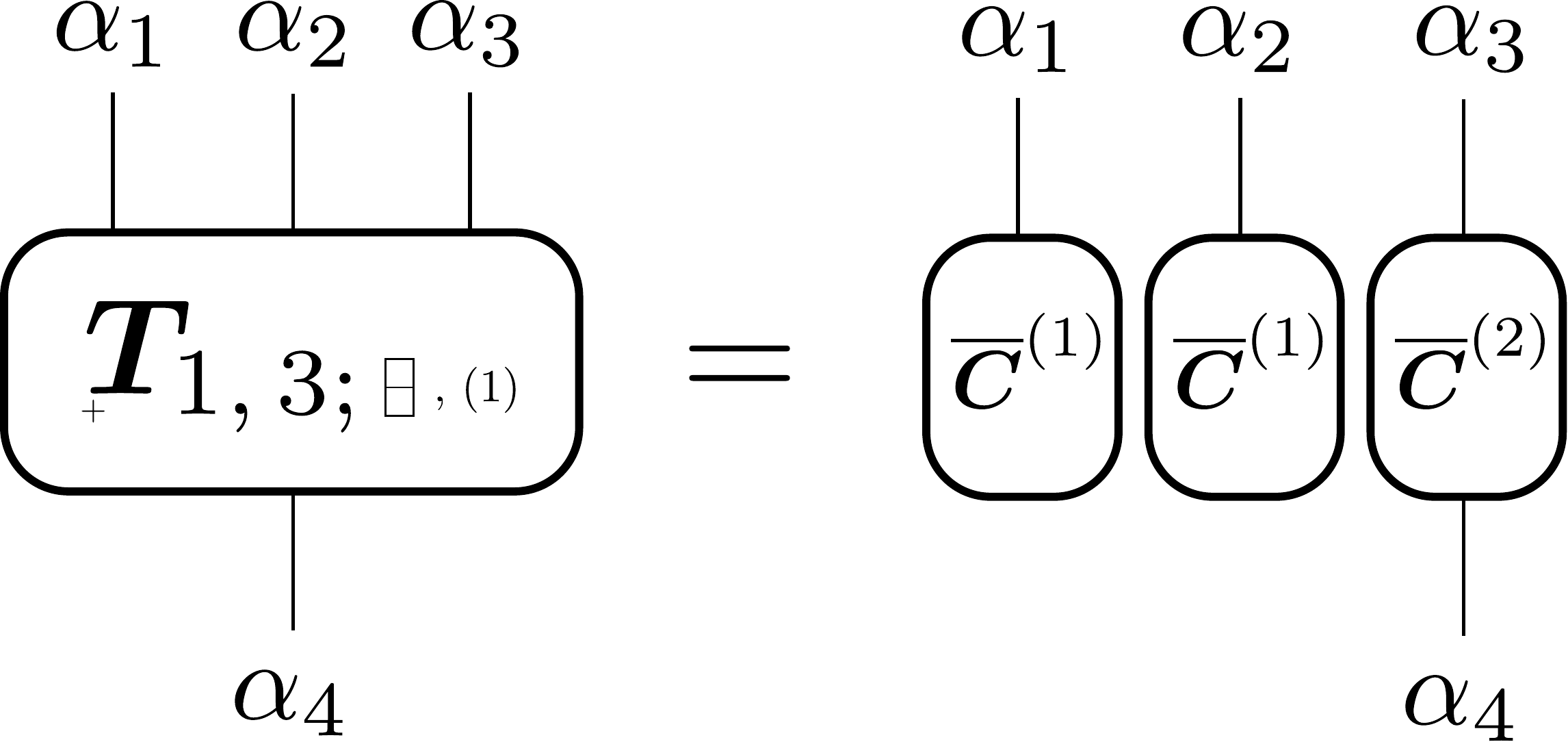}
        \caption{Contribution 5}
    \end{subfigure}
    \hspace*{0.08\textwidth}
    \begin{subfigure}{0.45\textwidth}
        \centering
        \includegraphics[width=0.85\textwidth]{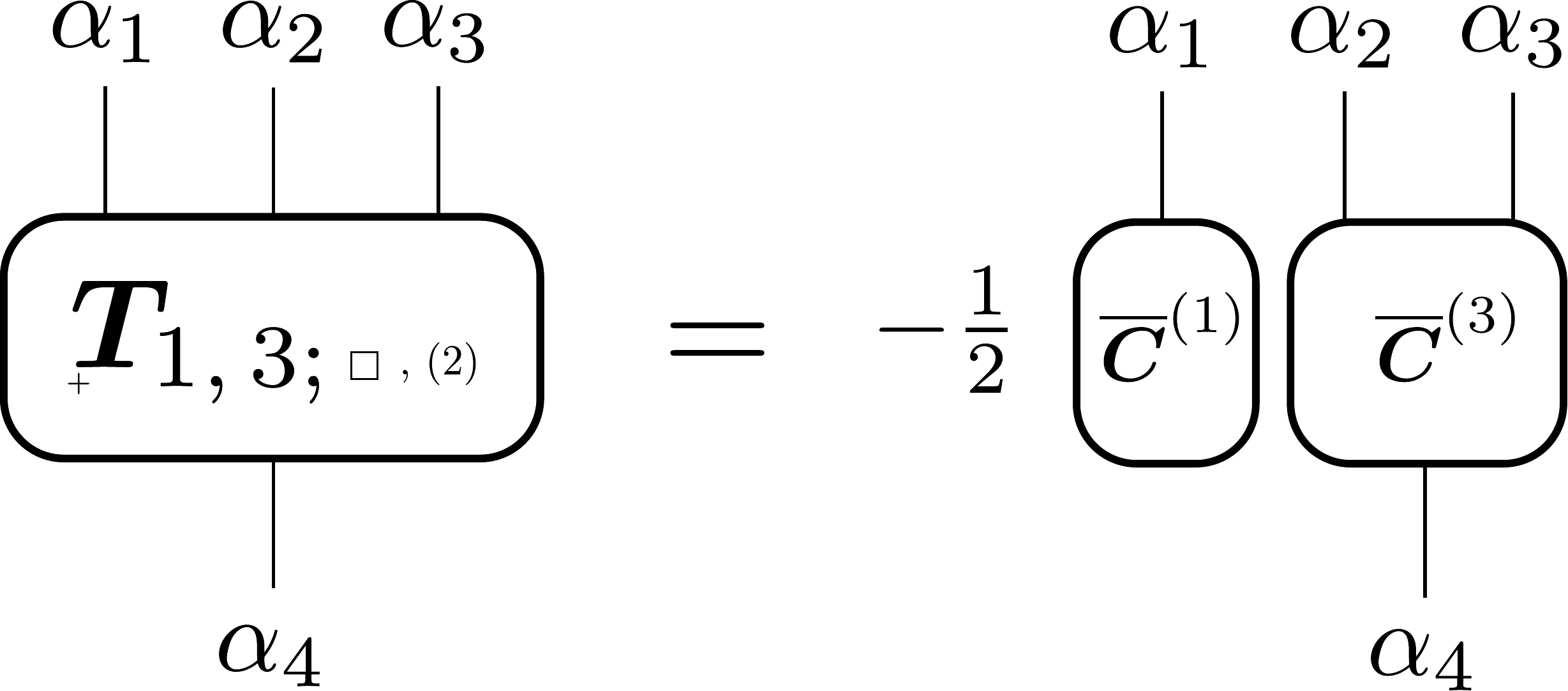}
        \caption{Contribution 6}
    \end{subfigure}\\\vspace*{10pt}
    \begin{subfigure}{0.45\textwidth}
        \centering
        \includegraphics[width=0.85\textwidth]{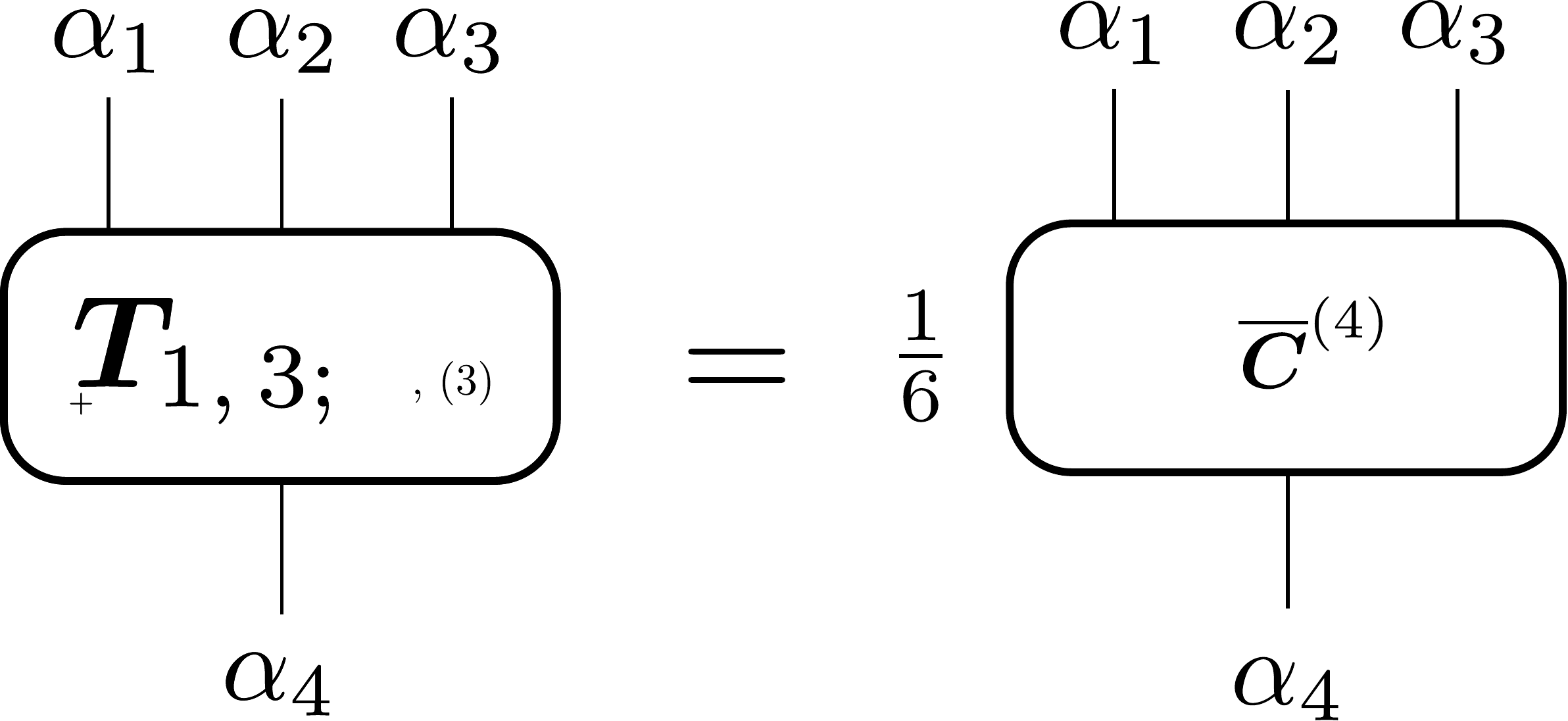}
        \caption{Contribution 7}
    \end{subfigure}
    \caption{Expression of each contribution to $\bm{T}_{1,\,2}$ in terms of noninteracting correlation tensors}
    \label{fig:t_1_3_contribution_by_contribution}
\end{figure}

The single $\bm{T}$ now has the following non-negligible contributions:
\begin{align}
    \left[\bm{T}\overline{\bm{\Theta}^*}\right]_1 & = \bm{T}_{1, 1}\left[\overline{\bm{\Theta}^*}\right]_1 + \bm{T}_{1, 2}\left[\overline{\bm{\Theta}^*}\right]_2 + \bm{T}_{1, 3}\left[\overline{\bm{\Theta}^*}\right]_3 + \mathcal{O}\left(\lambda^9\right)\\
    & = \bm{T}_{1,\,1}\overline{\bm{\theta}^*} + \bm{T}_{1,\,2}\overline{\bm{\theta}^*}^{\otimes 2} + \bm{T}_{1,\,3}\overline{\bm{\theta}^*}^{\otimes 3} + \mathcal{O}\left(\lambda^9\right).
\end{align}
The only contribution of order exactly $7$ in this equation is:
\begin{align}
    \bm{T}_{1,\,3}\left[\overline{\bm{\Theta}^*}\right]_3.
\end{align}
We therefore need the expression of $\bm{T}_{1, 3}$, shown as a tensor network on figures \ref{fig:t_1_3_contributions} and \ref{fig:t_1_3_contribution_by_contribution}

\begin{figure}[!htbp]
    \centering
    \includegraphics[width=0.8\textwidth]{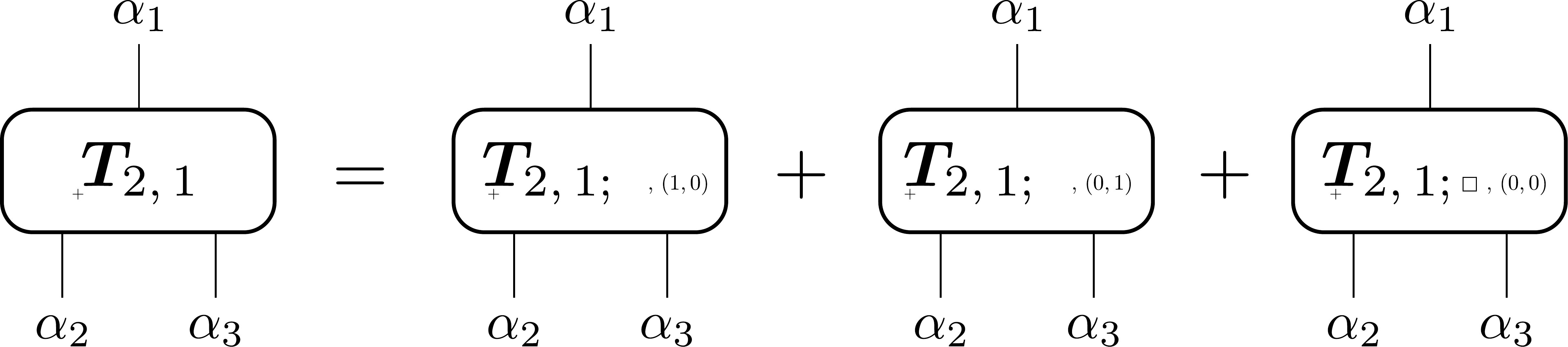}
    \caption{Contributions to $\bm{T}$ matrix block $\bm{T}_{2,\,1}$}
    \label{fig:t_2_1_contributions}
\end{figure}

\begin{figure}[!htbp]
    \centering
    \begin{subfigure}{0.45\textwidth}
        \includegraphics[width=0.75\textwidth]{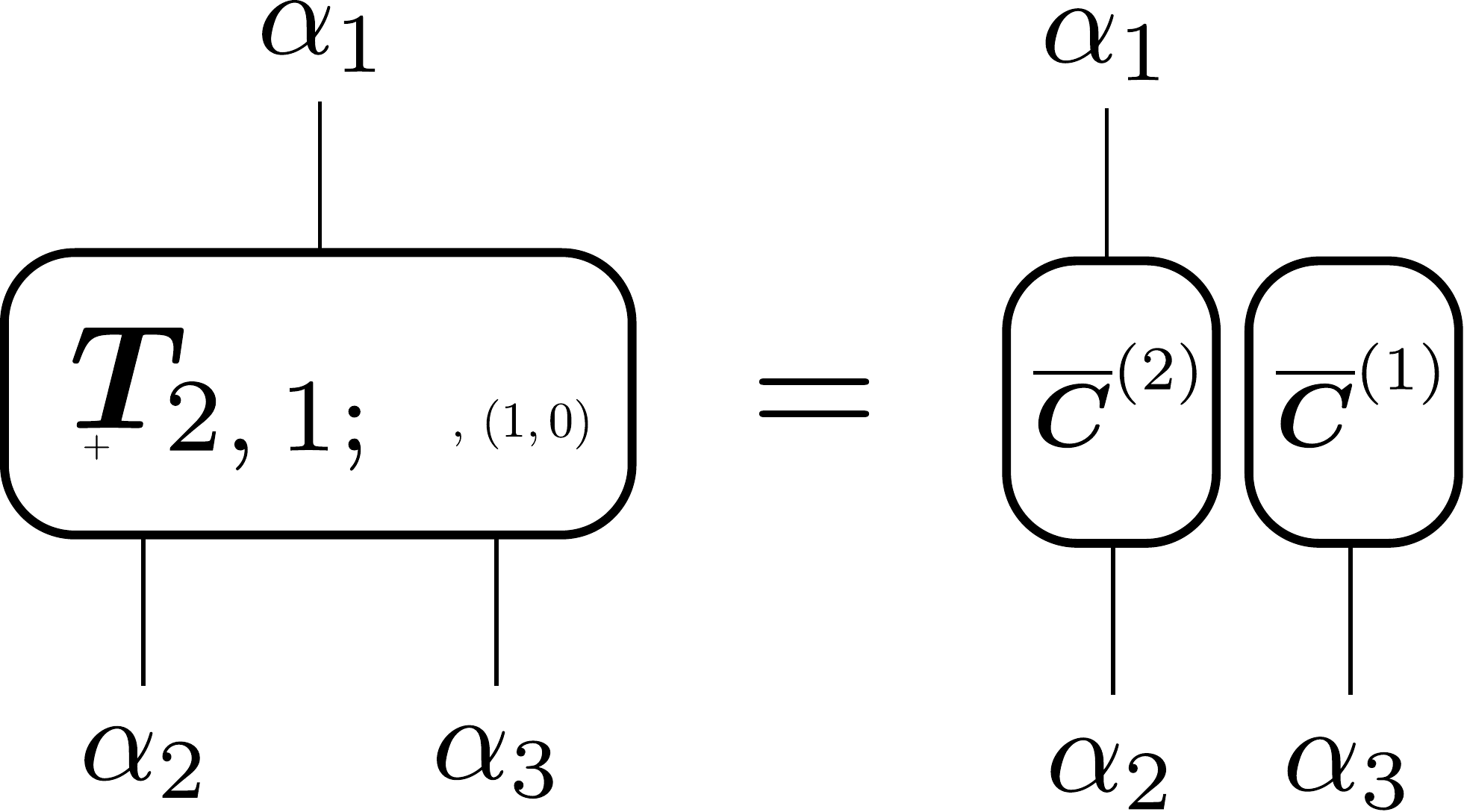}
        \caption{Contribution 1}
    \end{subfigure}
    \hspace*{0.08\textwidth}
    \begin{subfigure}{0.45\textwidth}
        \includegraphics[width=0.75\textwidth]{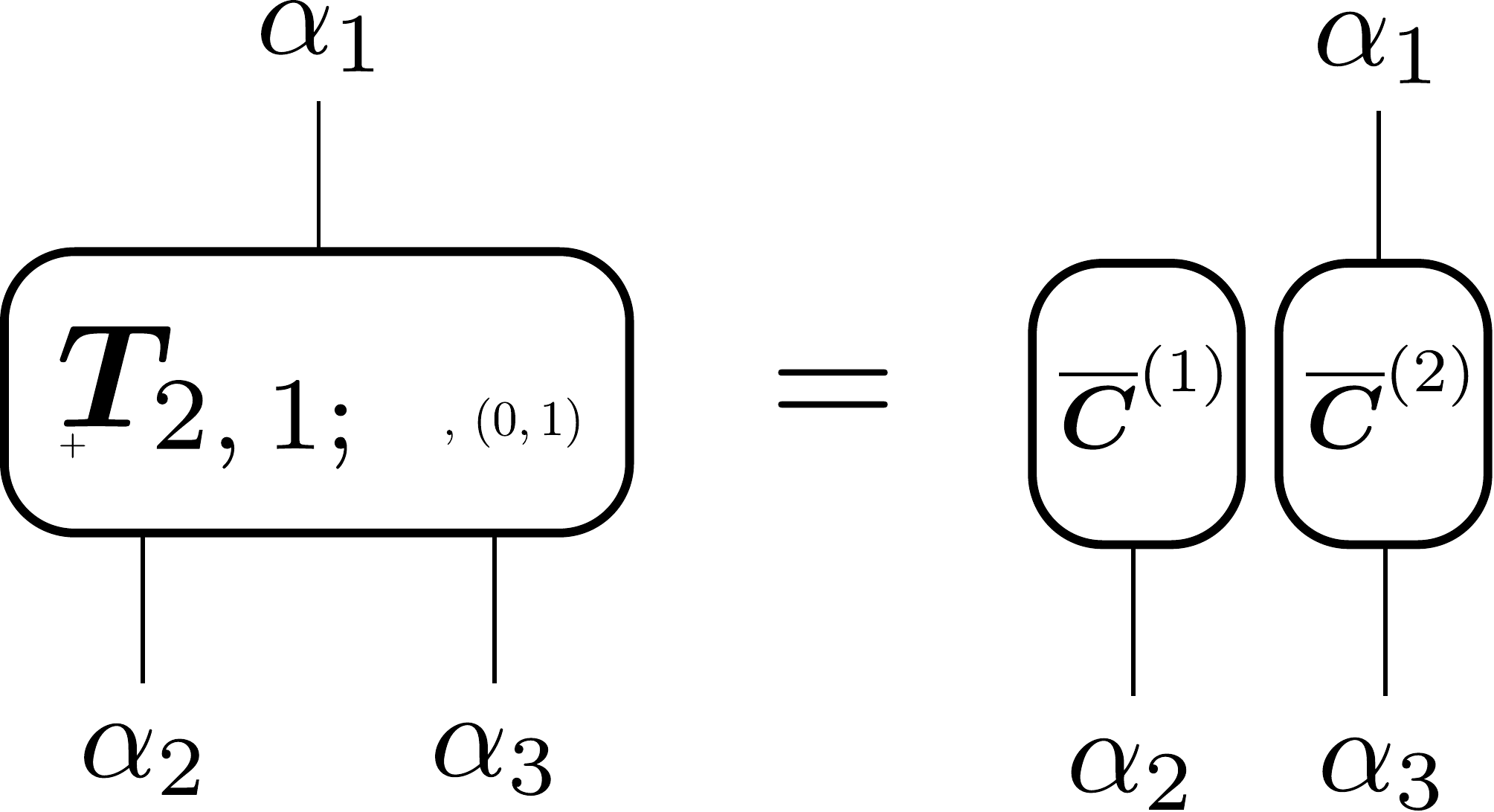}
        \caption{Contribution 2}
    \end{subfigure}\\\vspace*{10pt}
    \begin{subfigure}{0.45\textwidth}
        \includegraphics[width=\textwidth]{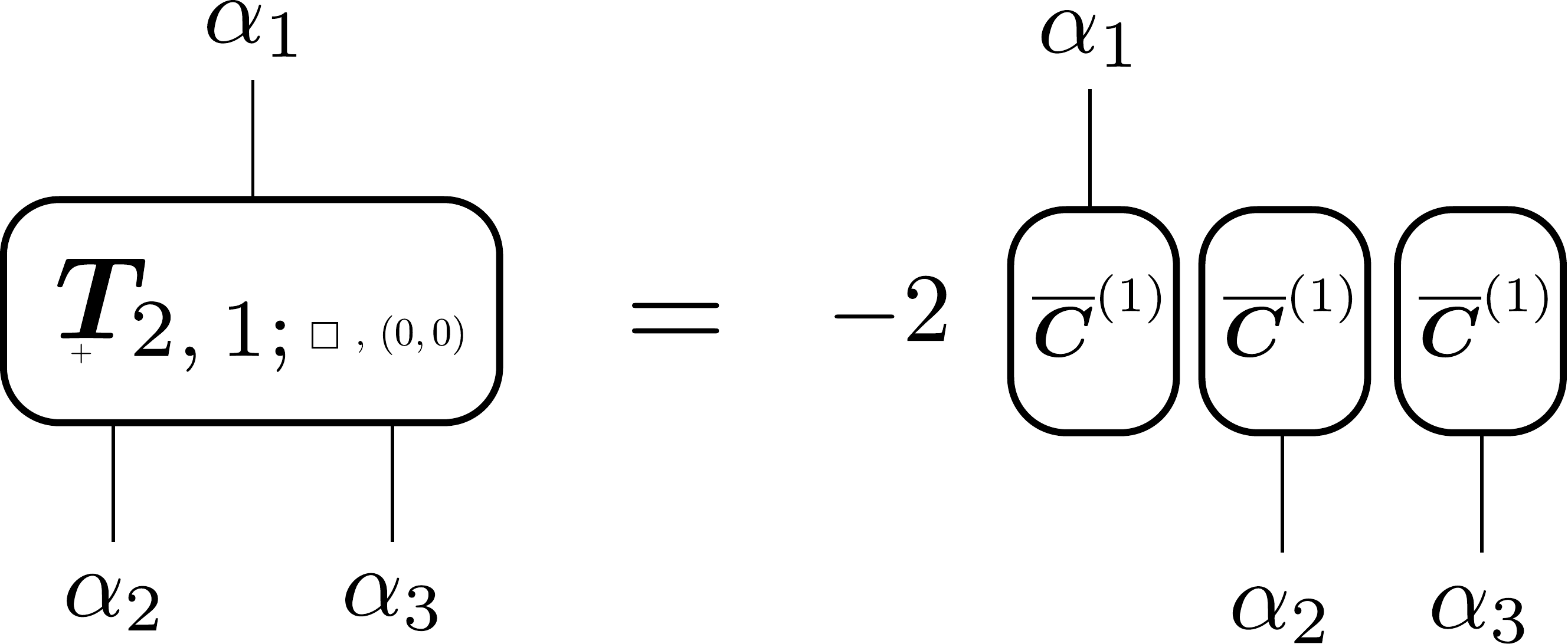}
        \caption{Contribution 3}
    \end{subfigure}
    \caption{Expression of each contribution to $\bm{T}_{2,\,1}$ in terms of noninteracting correlation tensors}
    \label{fig:t_2_1_contribution_by_contribution}
\end{figure}

This concludes the description of contribution $\left[\bm{T}\overline{\bm{\Theta}^*}\right]_1$, resulting from applying a single power of $\bm{T}$, to the saddle point $\bm{\theta}^*$. We now look at the contributions resulting from applying two powers of $\bm{T}$:
\begin{align}
    \left[\bm{T}^2\overline{\bm{\Theta}^*}\right]_1 & = \bm{T}_{1,\,1}\bm{T}_{1,\,1}\left[\overline{\bm{\Theta}^*}\right]_1 + \bm{T}_{1,\,2}\bm{T}_{2,\,1}\left[\overline{\bm{\Theta}^*}\right]_1 + \bm{T}_{1,\,1}\bm{T}_{1,\,2}\left[\overline{\bm{\Theta}^*}\right]_2 + \mathcal{O}\left(\lambda^9\right)\\
    & = \bm{T}_{1,\,1}\bm{T}_{1,\,1}\overline{\bm{\theta}^*} + \bm{T}_{1,\,2}\bm{T}_{2,\,1}\overline{\bm{\theta}^*} + \bm{T}_{1,\,1}\bm{T}_{1,\,2}\overline{\bm{\theta}^*}^{\otimes 2} + \mathcal{O}\left(\lambda^9\right).
\end{align}
More specifically, the only terms of order exactly $\lambda^7$ are:
\begin{align}
    \bm{T}_{1,\,2}\bm{T}_{2,\,1}\overline{\bm{\theta}^*} + \bm{T}_{1,\,1}\bm{T}_{1,\,2}\overline{\bm{\theta}^*}^{\otimes 2}
\end{align}
Apart from already computed $\bm{T}_{1,\,1}, \bm{T}_{1,\,2}$ (see figures \ref{fig:t_1_1_contributions}, \ref{fig:t_1_1_contribution_by_contribution}, \ref{fig:t_1_2_contributions}, \ref{fig:t_1_2_contribution_by_contribution}), this matrix-vector product involves block $\bm{T}_{2,\,1}$, expressed as tensor network diagrams in figures \ref{fig:t_2_1_contributions} and \ref{fig:t_2_1_contribution_by_contribution}.

We have now described contributions from applying one or two powers of $\bm{T}$ to $\overline{\bm{\Theta}^*}$. Let us then consider the contribution of the cubic power of $\bm{T}$:
\begin{align}
    \left[\bm{T}^3\overline{\bm{\Theta}^*}\right]_1 & = \bm{T}_{1,\,1}\bm{T}_{1,\,1}\bm{T}_{1,\,1}\left[\overline{\bm{\Theta}^*}\right]_1\\
    & = \bm{T}_{1,\,1}\bm{T}_{1,\,1}\bm{T}_{1,\,1}\overline{\bm{\theta}^*}.
\end{align}
This has a unique term, which is indeed of order exactly $7$:
\begin{align}
    \bm{T}_{1,\,1}\bm{T}_{1,\,1}\bm{T}_{1,\,1}\overline{\bm{\theta}^*}.
\end{align}

\begin{figure}[!htbp]
    \centering
    \begin{subfigure}{0.32\textwidth}
        \centering
        \includegraphics[width=\textwidth]{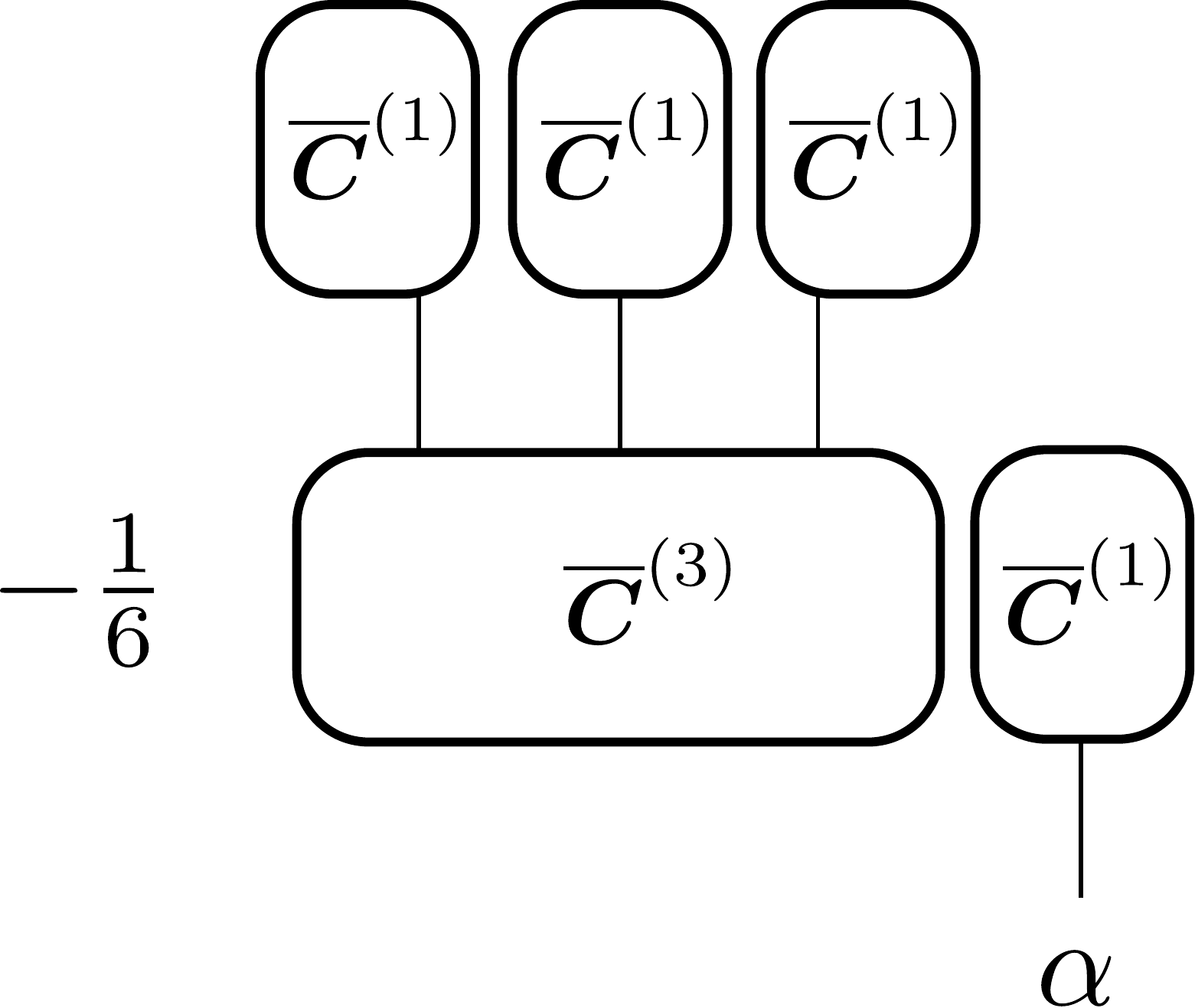}
        \caption{Contribution 1}
    \end{subfigure}
    \begin{subfigure}{0.3\textwidth}
        \centering
        \includegraphics[width=0.8\textwidth]{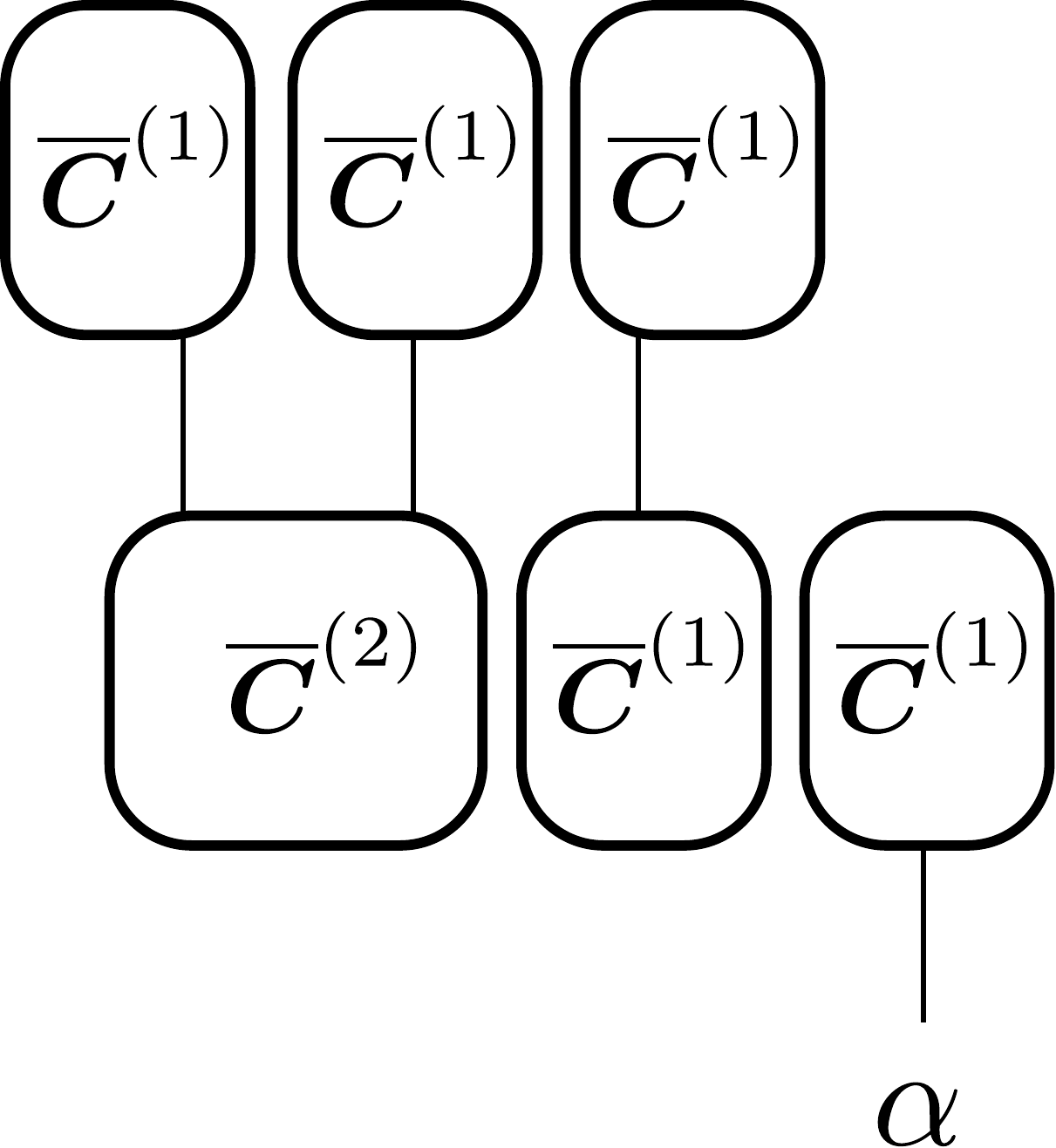}
        \caption{Contribution 2}
    \end{subfigure}
    \begin{subfigure}{0.35\textwidth}
        \centering
        \includegraphics[width=0.9\textwidth]{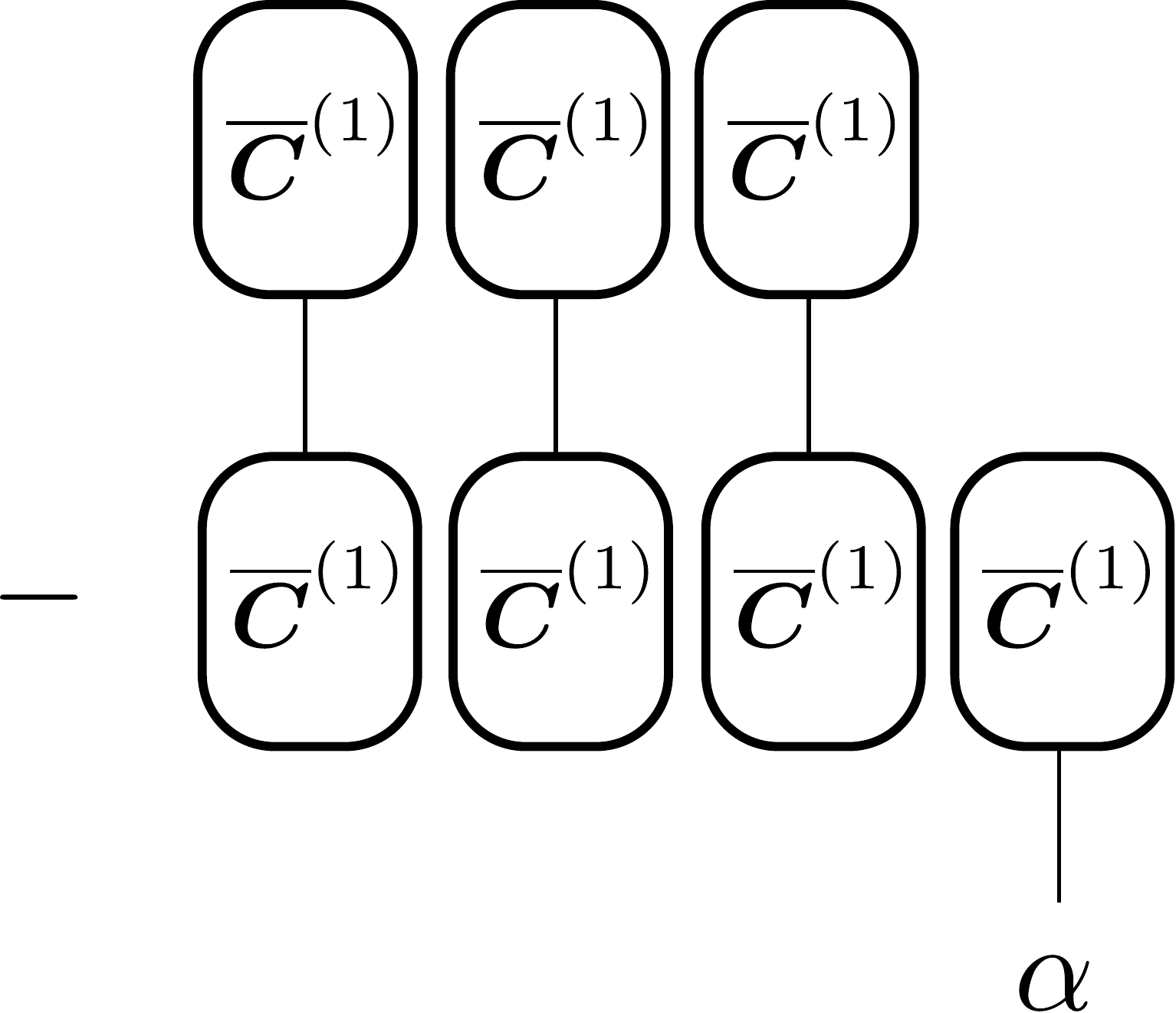}
        \caption{Contribution 3}
    \end{subfigure}\\
    \vspace*{20px}
    \begin{subfigure}{0.32\textwidth}
        \centering
        \includegraphics[width=0.8\textwidth]{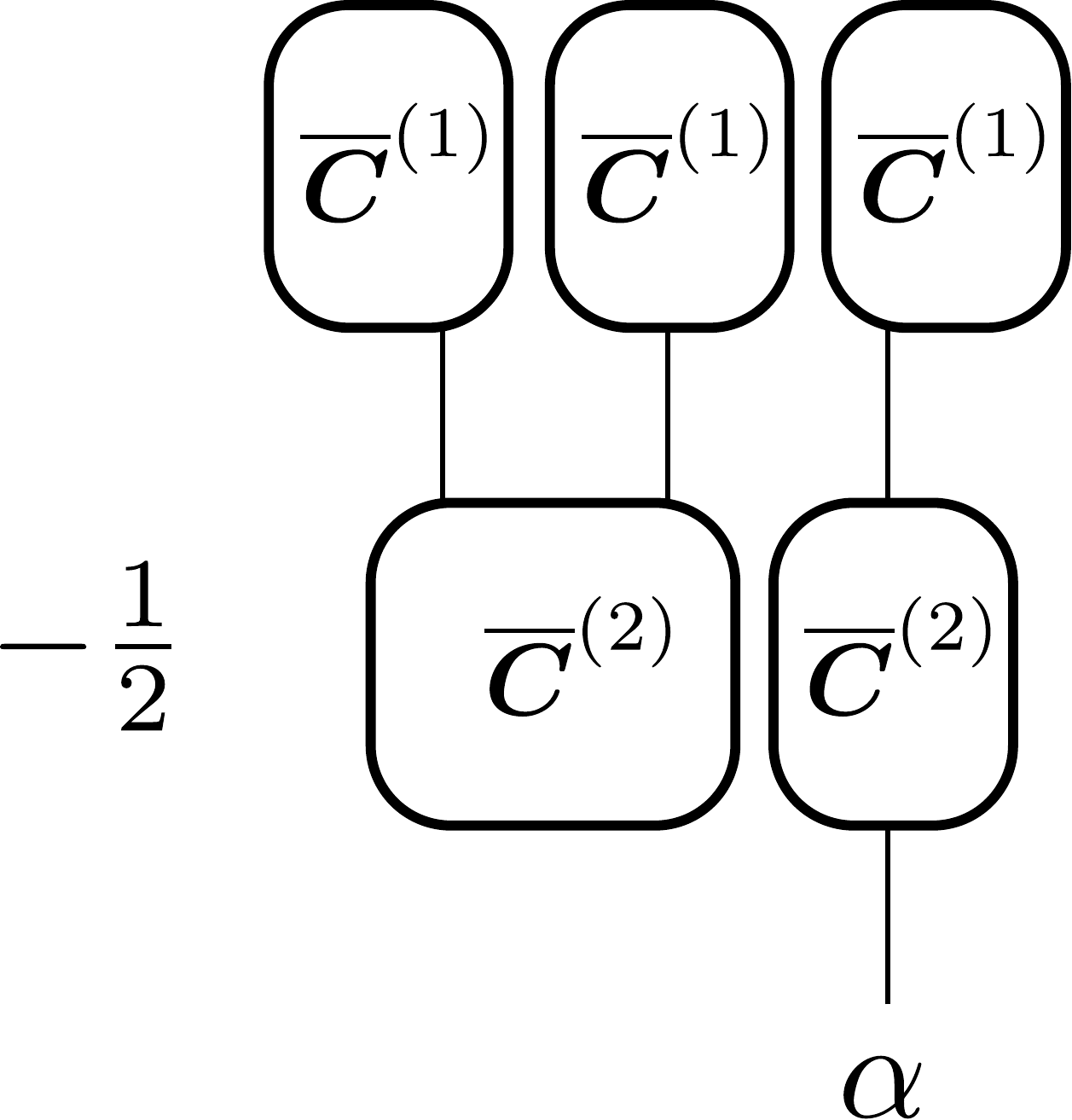}
        \caption{Contribution 4}
    \end{subfigure}
    \begin{subfigure}{0.35\textwidth}
        \centering
        \includegraphics[width=0.54\textwidth]{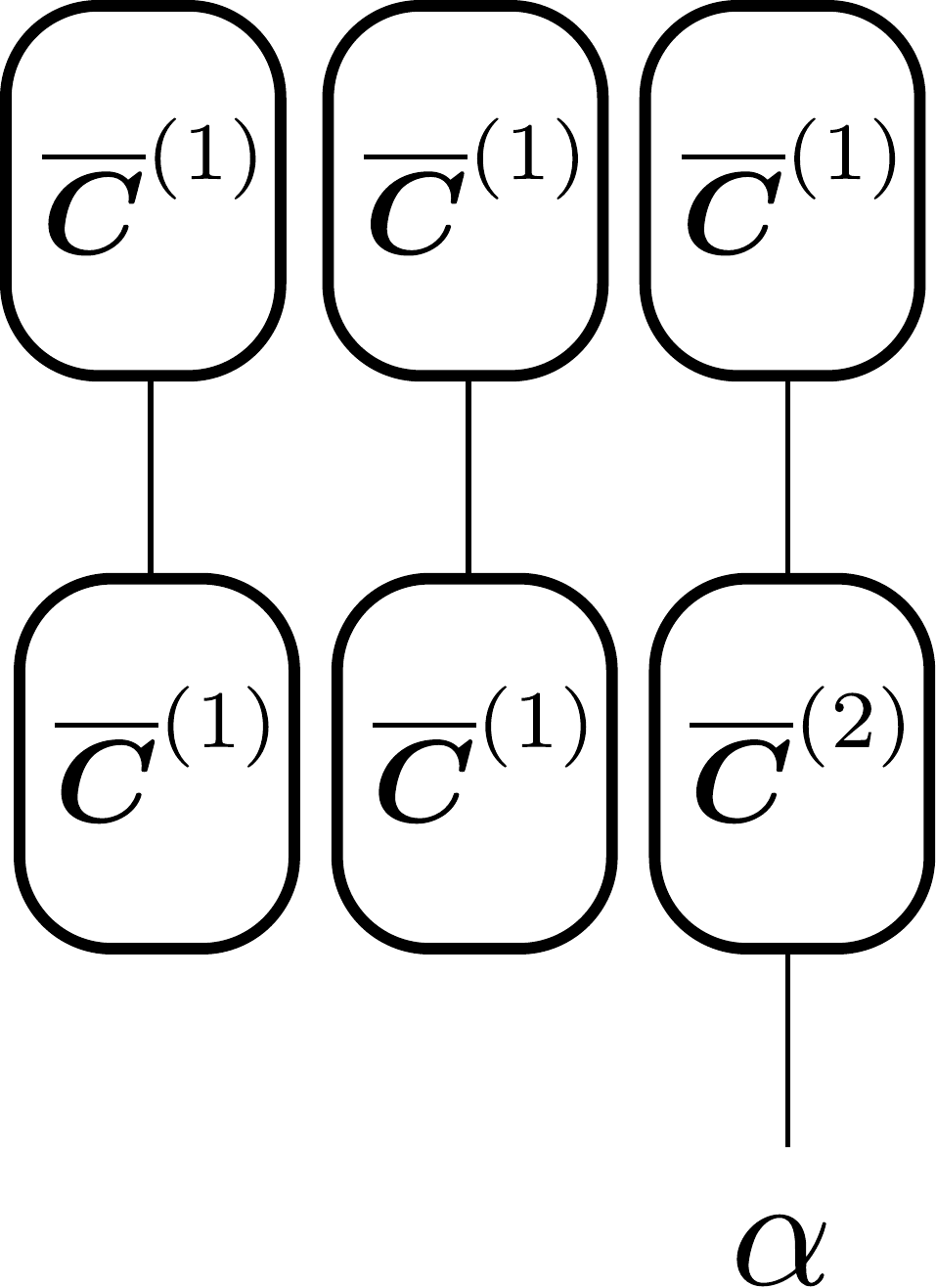}
        \caption{Contribution 5}
    \end{subfigure}
    \begin{subfigure}{0.3\textwidth}
        \centering
        \includegraphics[width=0.85\textwidth]{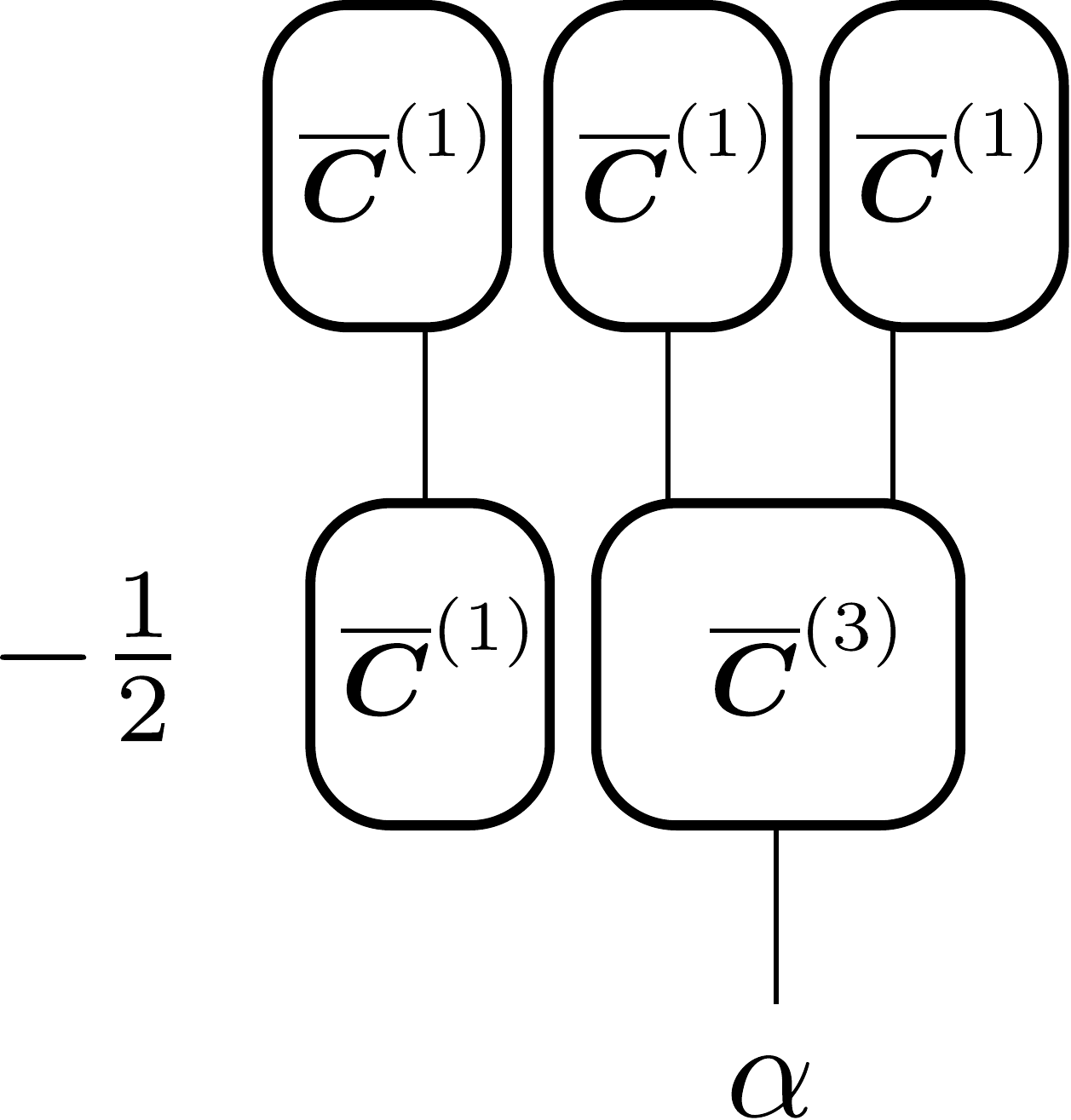}
        \caption{Contribution 6}
    \end{subfigure}\\
    \vspace*{20px}
    \begin{subfigure}{0.3\textwidth}
        \centering
        \includegraphics[width=0.8\textwidth]{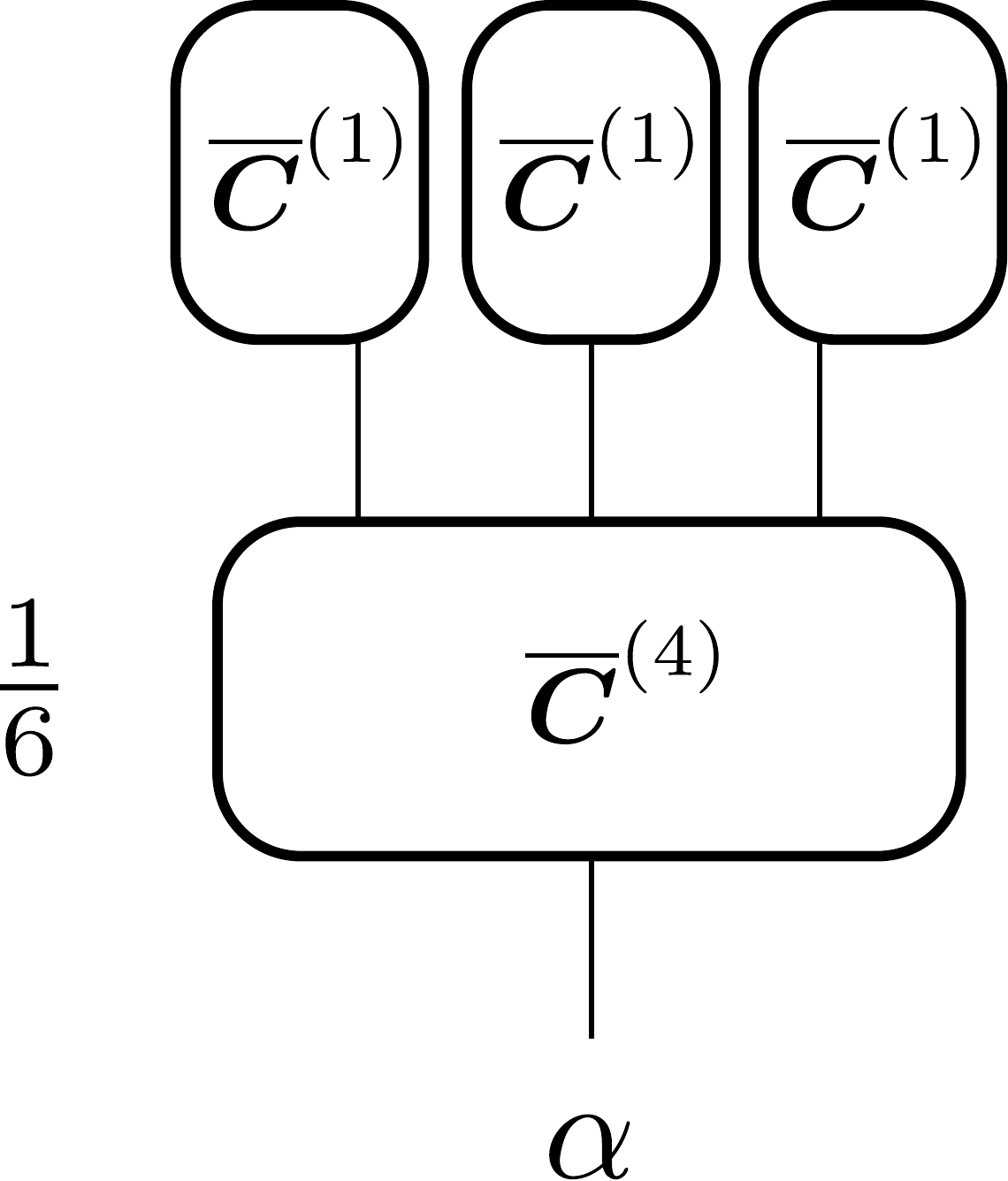}
        \caption{Contribution 7}
    \end{subfigure}
    \caption{Order 7 contributions to saddle point $\bm{\theta}^* = \left(\theta^*_{\alpha}\right)$ from linear $\bm{T}$ term $\bm{T}\overline{\bm{\Theta}^*}$}
    \label{fig:order_7_t_contributions}
\end{figure}

Multiplying $\bm{T}$ matrix blocks together, one may obtain the diagrammatic representation of order 7 contributions. Similar to the order 5 case, we organize these according to the power of $\bm{T}$ they arise from in the series expansion of $\left(\bm{I} - \bm{T}\right)^{-1}$. The linear, quadratic, and cubic contributions are given on figure \ref{fig:order_7_t_contributions}, \ref{fig:order_7_t_square_contributions} and \ref{fig:order_7_t_cube_contributions} respectively.

\begin{figure}[!htbp]
    \centering
    \begin{subfigure}{0.32\textwidth}
        \centering
        \includegraphics[width=0.7\textwidth]{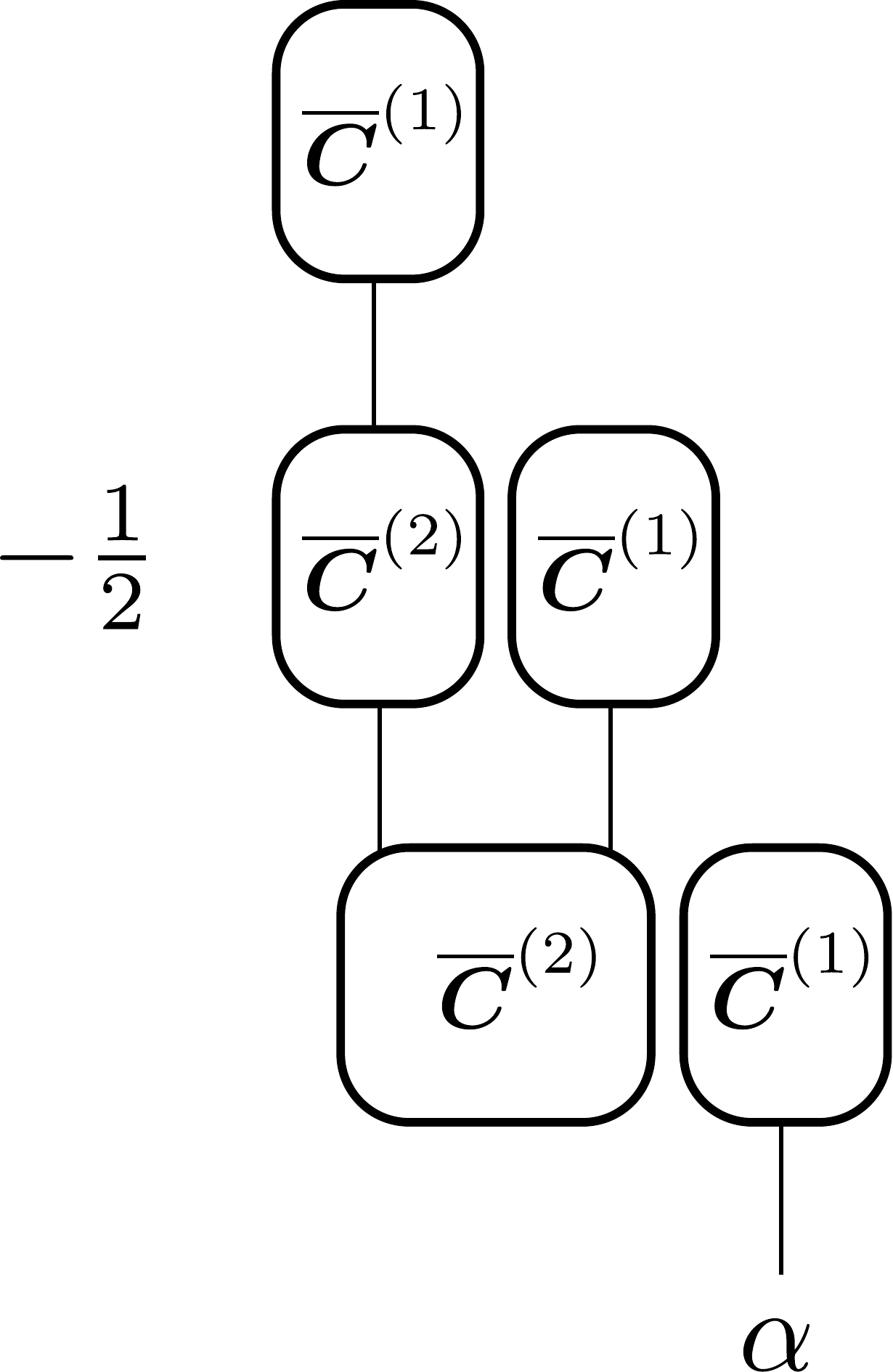}
        \caption{Contribution 1}
    \end{subfigure}
    \begin{subfigure}{0.32\textwidth}
        \centering
        \includegraphics[width=0.55\textwidth]{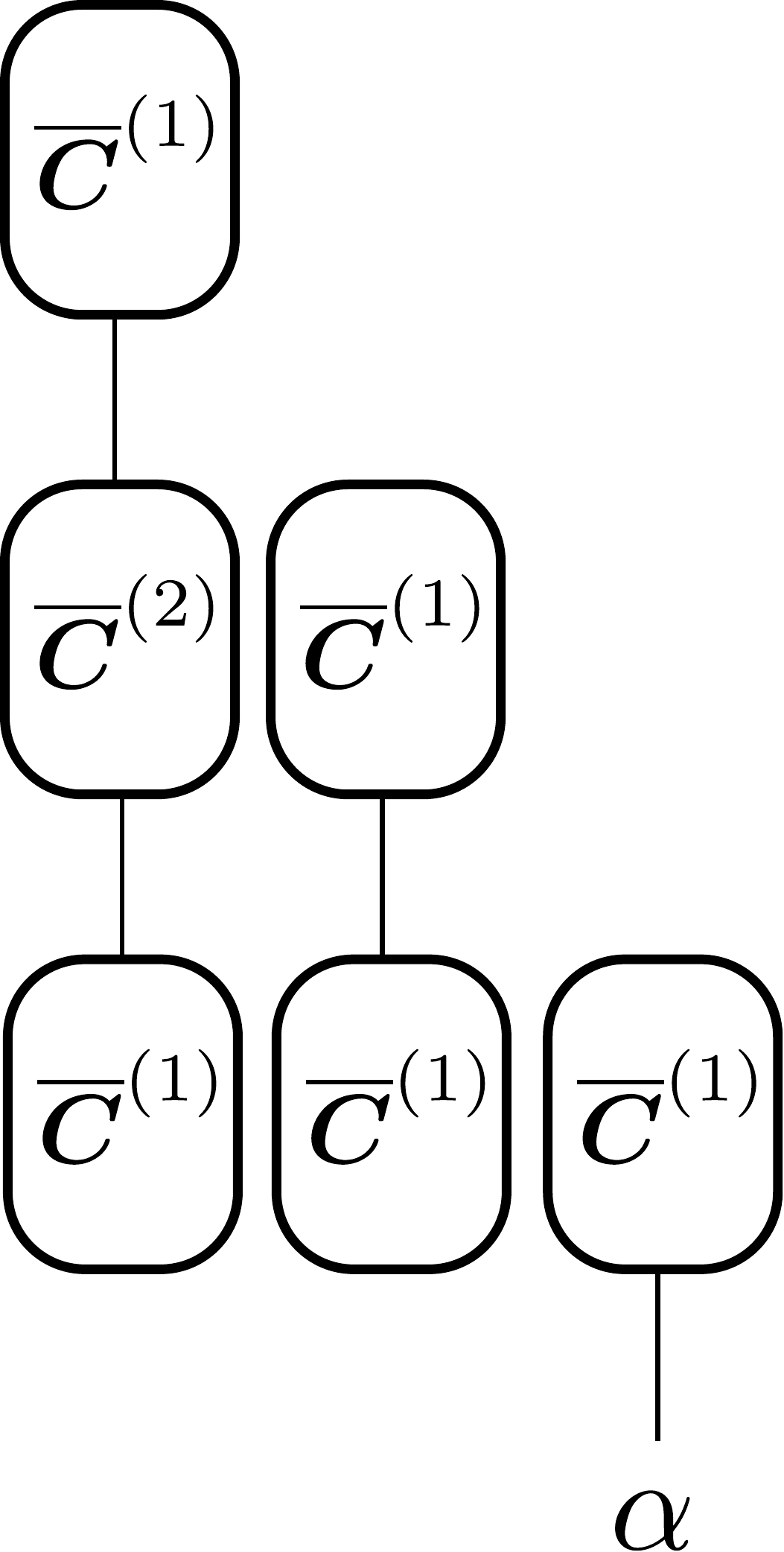}
        \caption{Contribution 2}
    \end{subfigure}
    \begin{subfigure}{0.32\textwidth}
        \centering
        \includegraphics[width=0.5\textwidth]{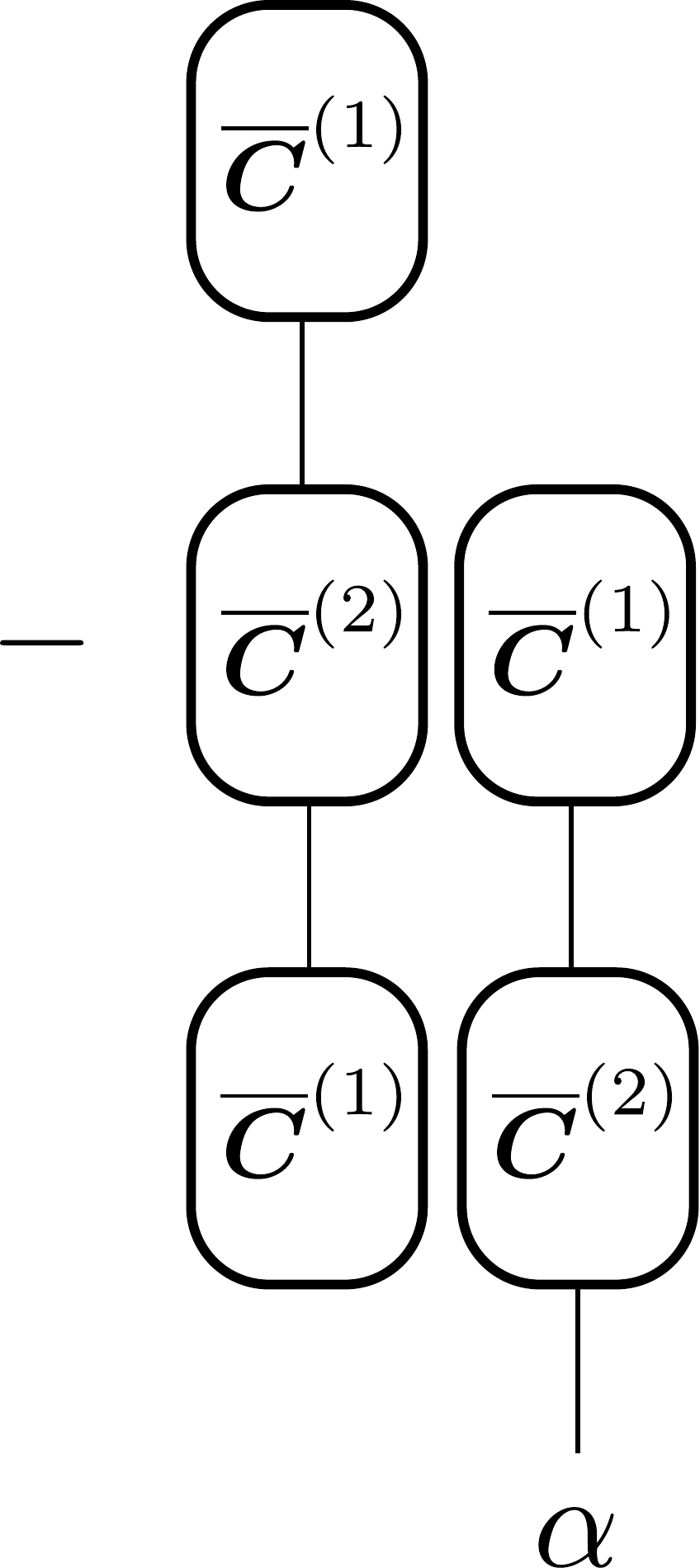}
        \caption{Contribution 3}
    \end{subfigure}\\
    \vspace*{20px}
    \begin{subfigure}{0.32\textwidth}
        \centering
        \includegraphics[width=0.5\textwidth]{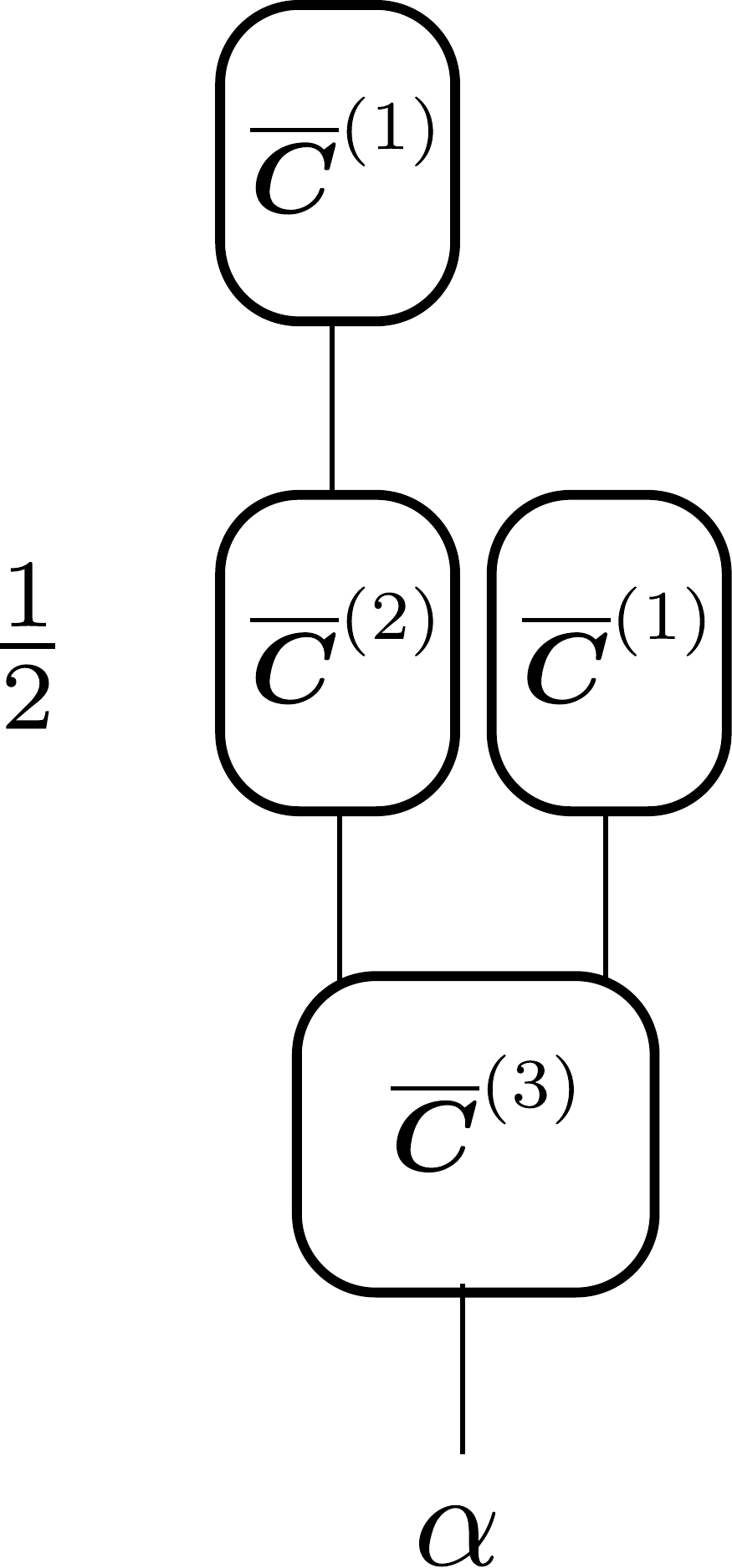}
        \caption{Contribution 4}
    \end{subfigure}
    \begin{subfigure}{0.32\textwidth}
        \centering
        \includegraphics[width=0.7\textwidth]{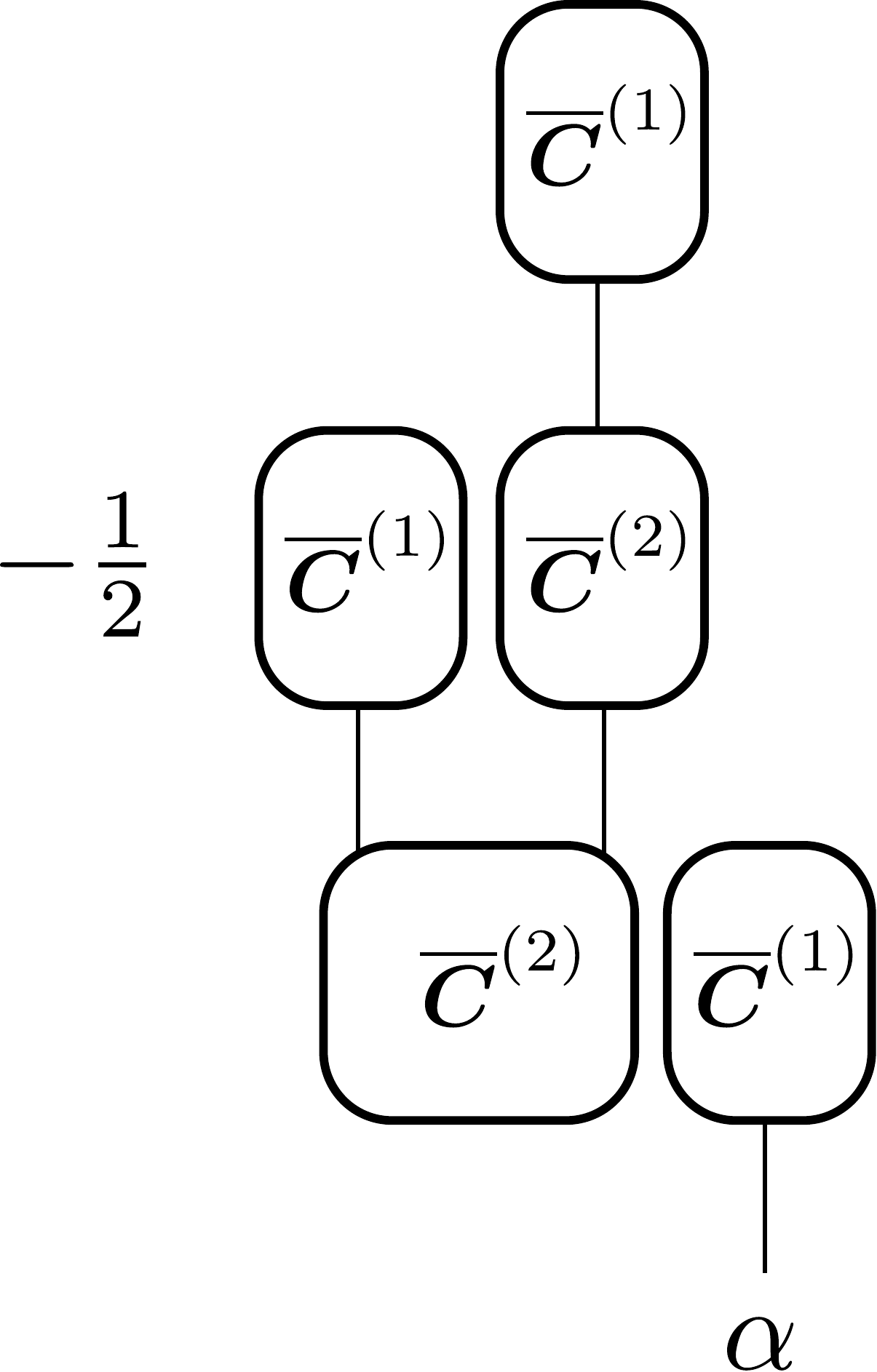}
        \caption{Contribution 5}
    \end{subfigure}
    \begin{subfigure}{0.32\textwidth}
        \centering
        \includegraphics[width=0.55\textwidth]{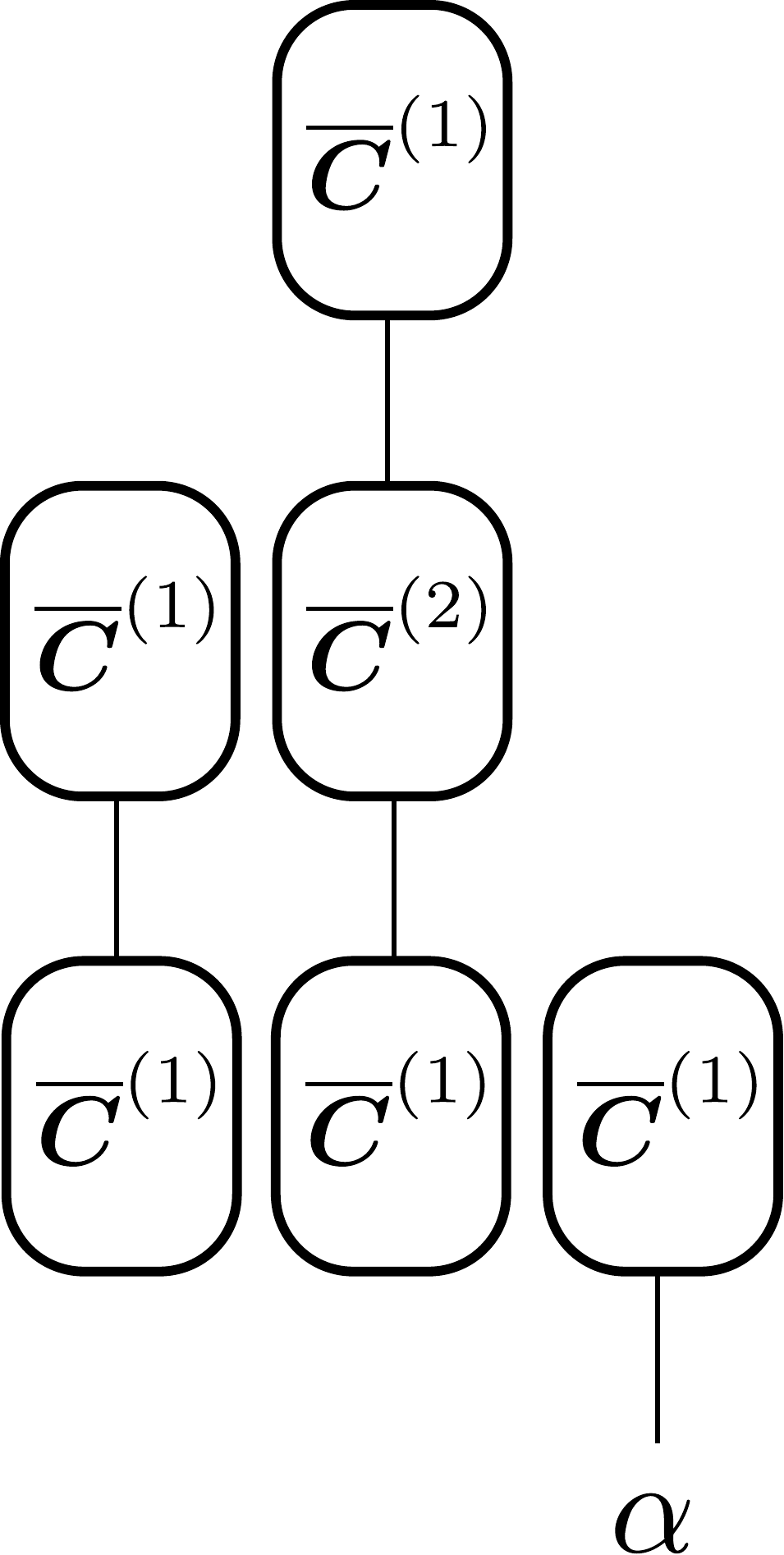}
        \caption{Contribution 6}
    \end{subfigure}\\
    \vspace*{20px}
    \begin{subfigure}{0.32\textwidth}
        \centering
        \includegraphics[width=0.5\textwidth]{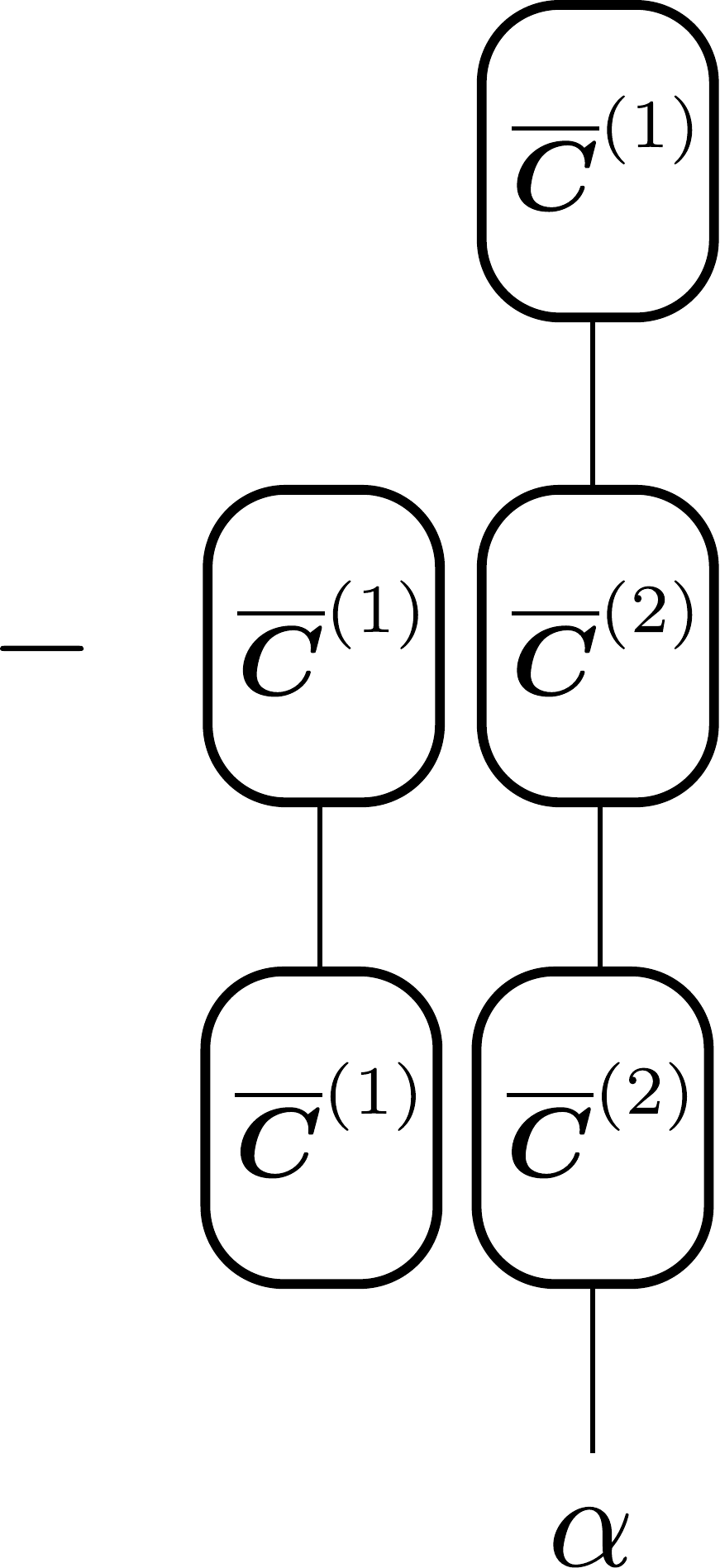}
        \caption{Contribution 7}
    \end{subfigure}
    \begin{subfigure}{0.32\textwidth}
        \centering
        \includegraphics[width=0.5\textwidth]{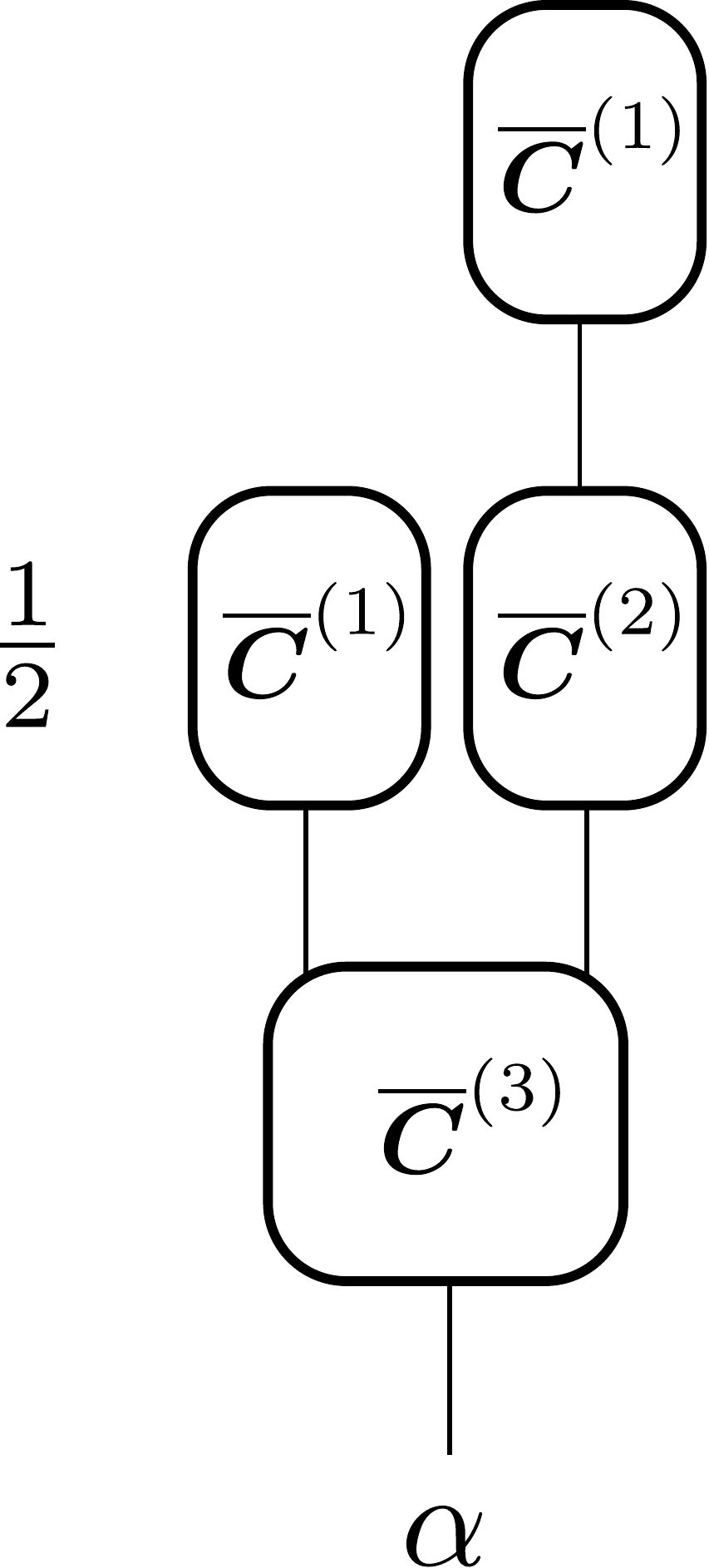}
        \caption{Contribution 8}
    \end{subfigure}
    \begin{subfigure}{0.32\textwidth}
        \centering
        \includegraphics[width=0.7\textwidth]{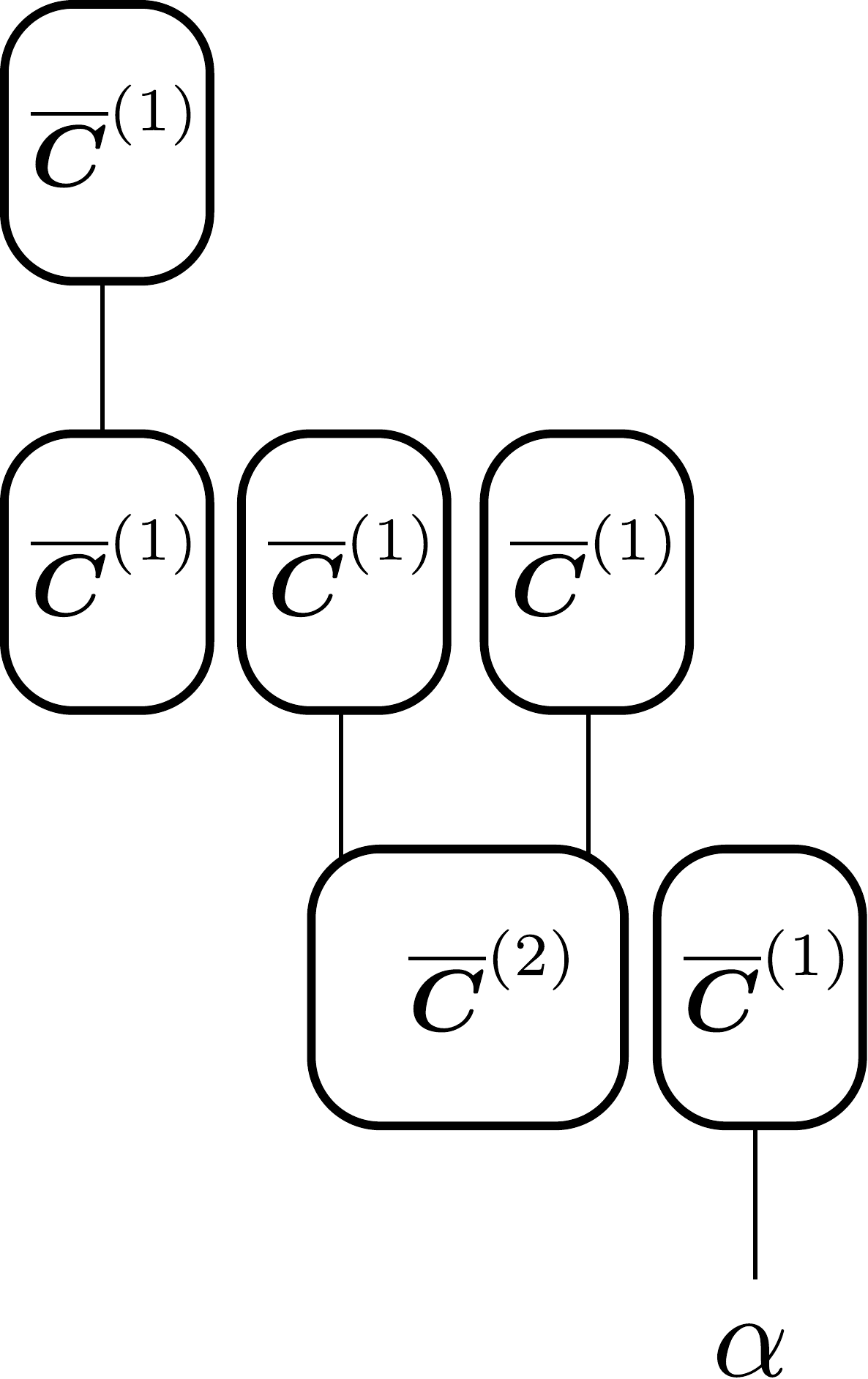}
        \caption{Contribution 9}
    \end{subfigure}
\end{figure}
\begin{figure}
    \ContinuedFloat
    \begin{subfigure}{0.35\textwidth}
        \centering
        \includegraphics[width=0.85\textwidth]{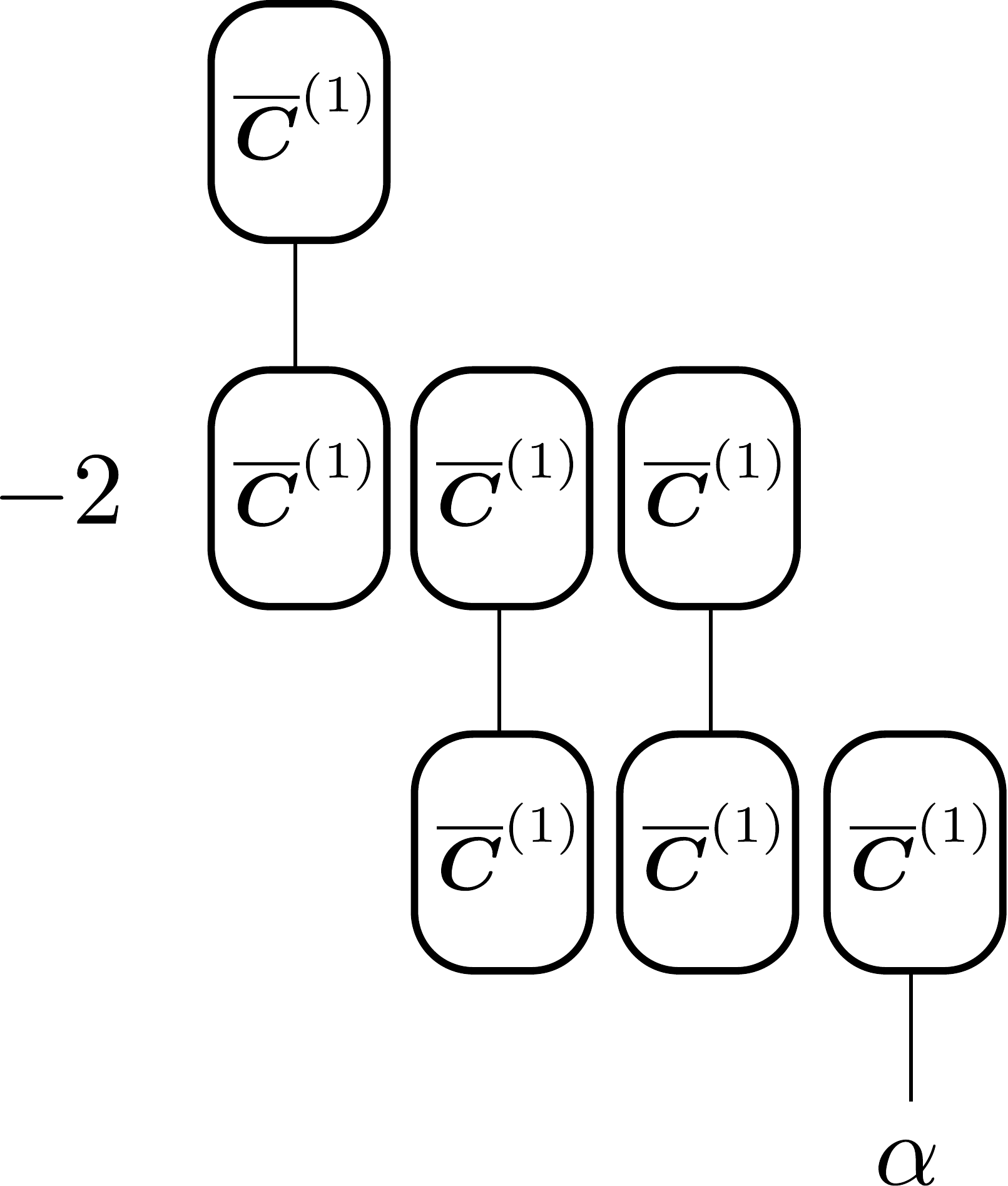}
        \caption{Contribution 10}
    \end{subfigure}
    \begin{subfigure}{0.31\textwidth}
        \centering
        \includegraphics[width=0.7\textwidth]{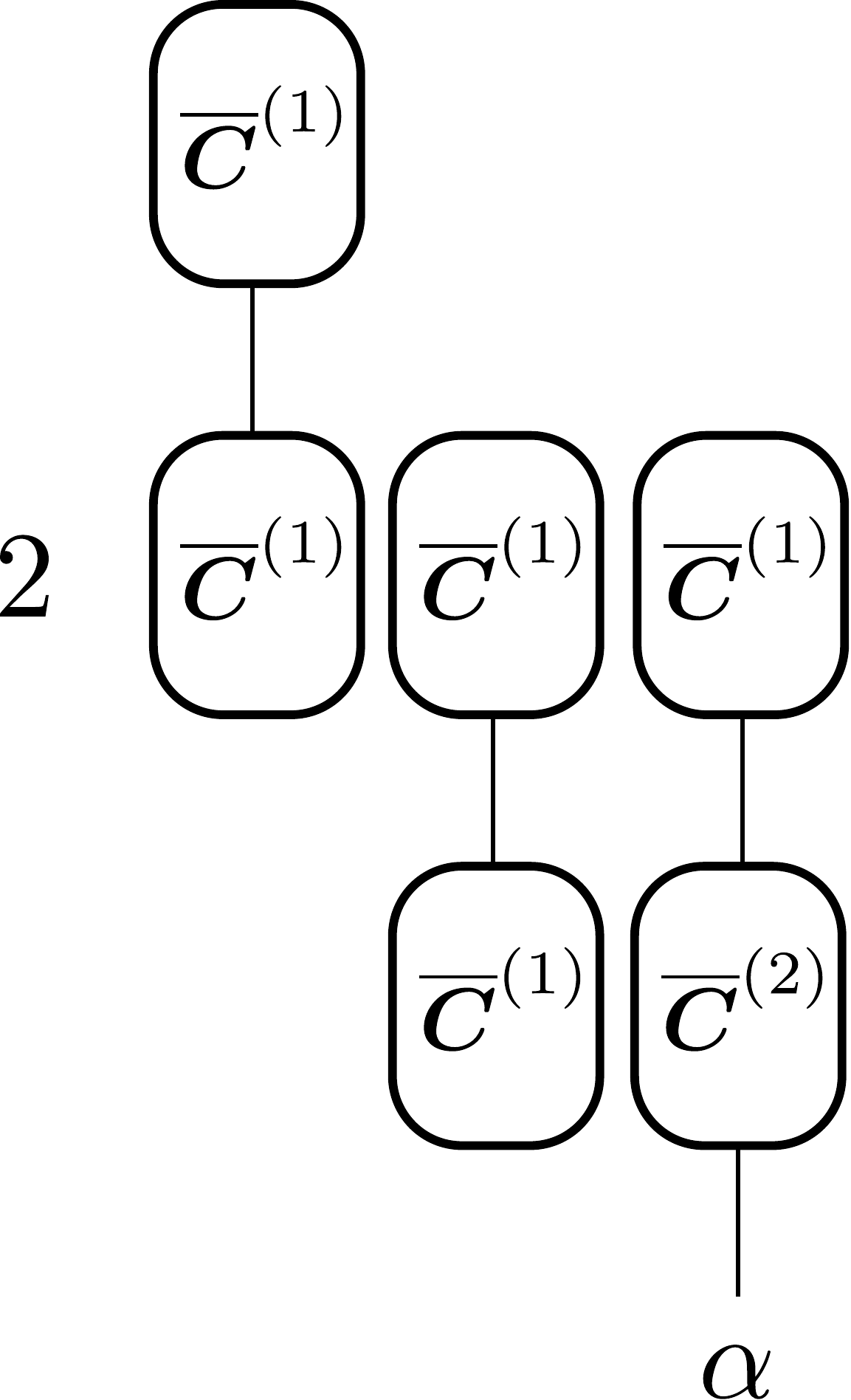}
        \caption{Contribution 11}
    \end{subfigure}
    \begin{subfigure}{0.31\textwidth}
        \centering
        \includegraphics[width=0.73\textwidth]{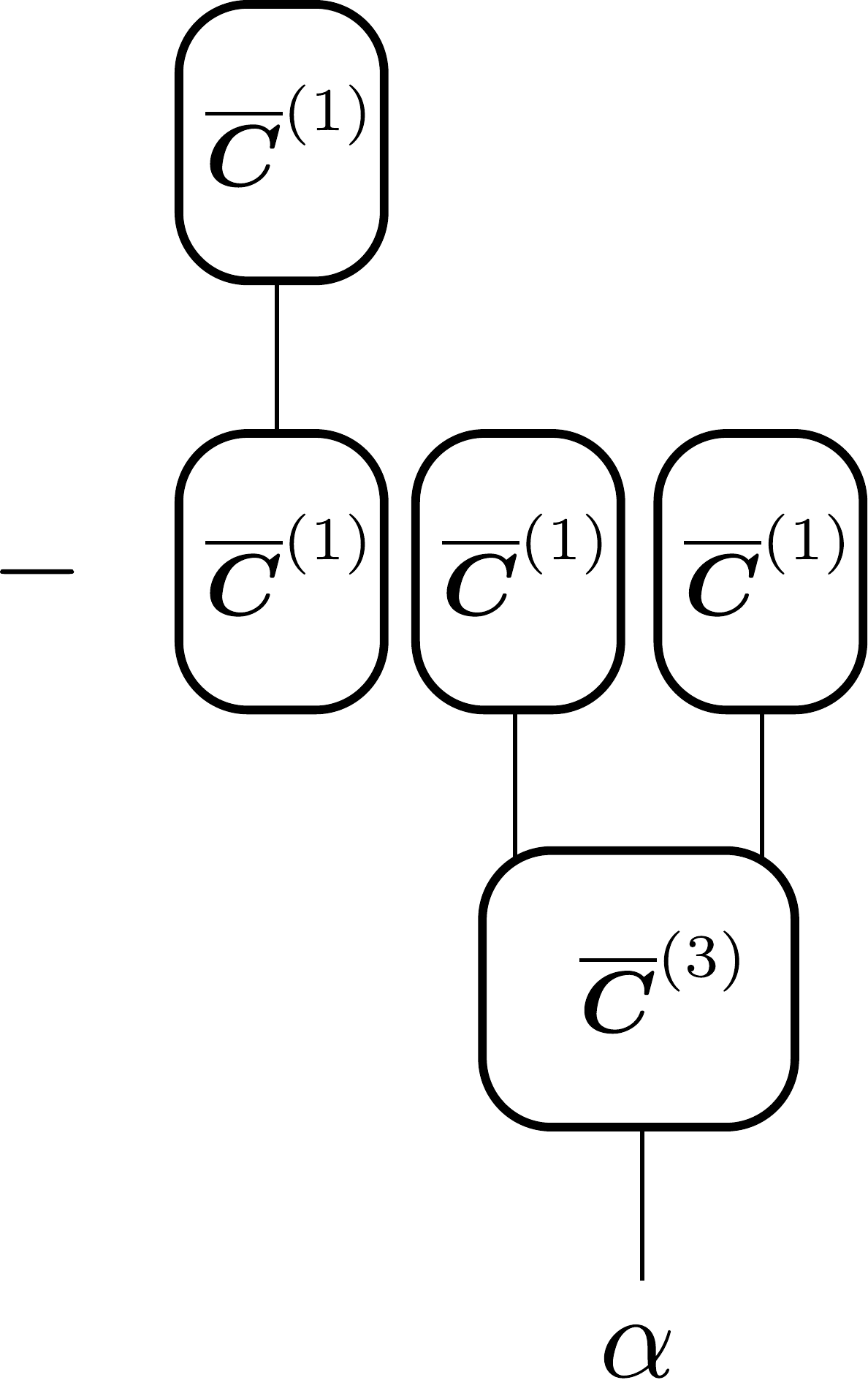}
        \caption{Contribution 12}
    \end{subfigure}\\
    \vspace*{20px}
    \begin{subfigure}{0.32\textwidth}
        \centering
        \includegraphics[width=0.8\textwidth]{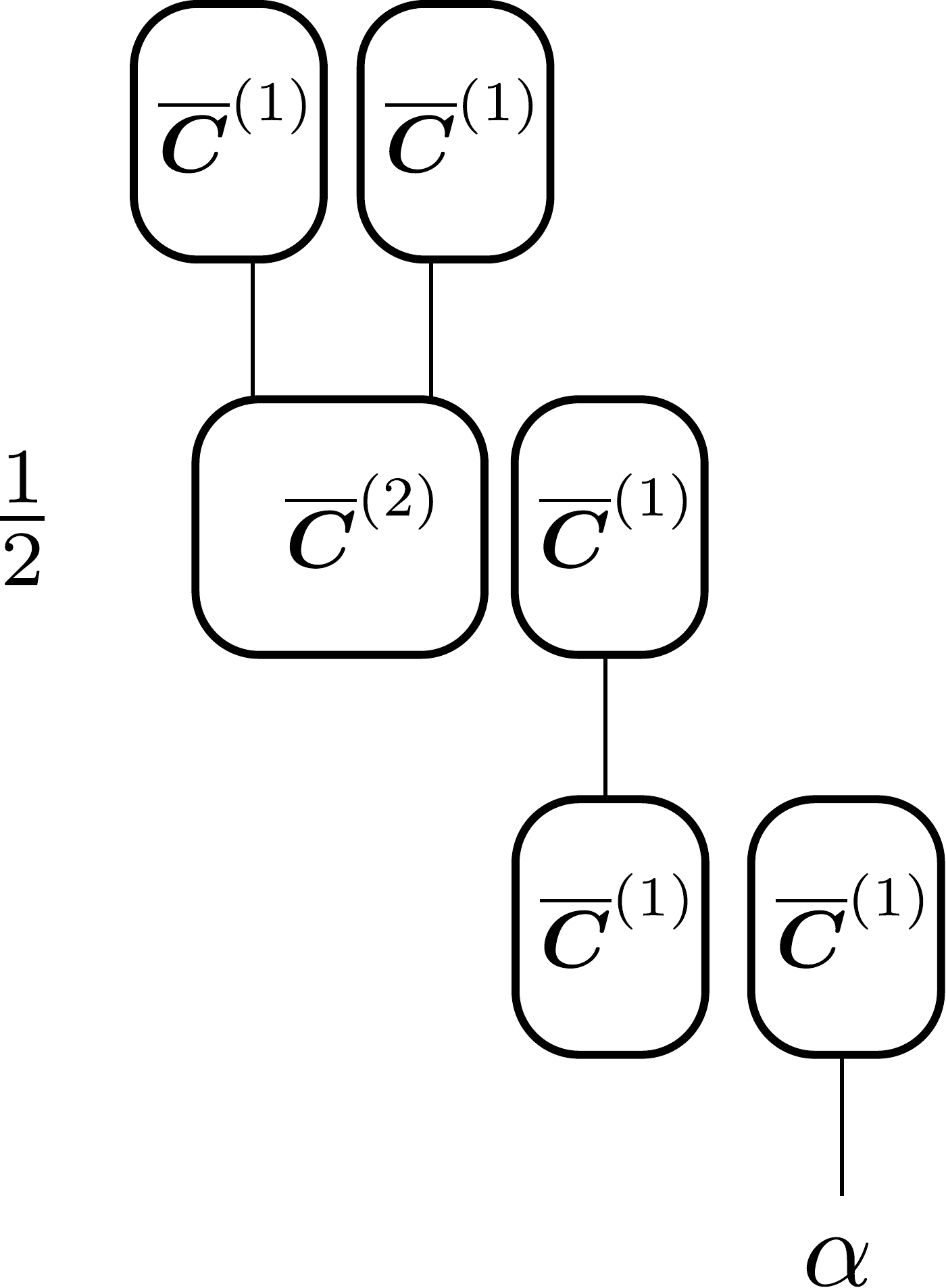}
        \caption{Contribution 13}
    \end{subfigure}
    \begin{subfigure}{0.32\textwidth}
        \centering
        \includegraphics[width=0.67\textwidth]{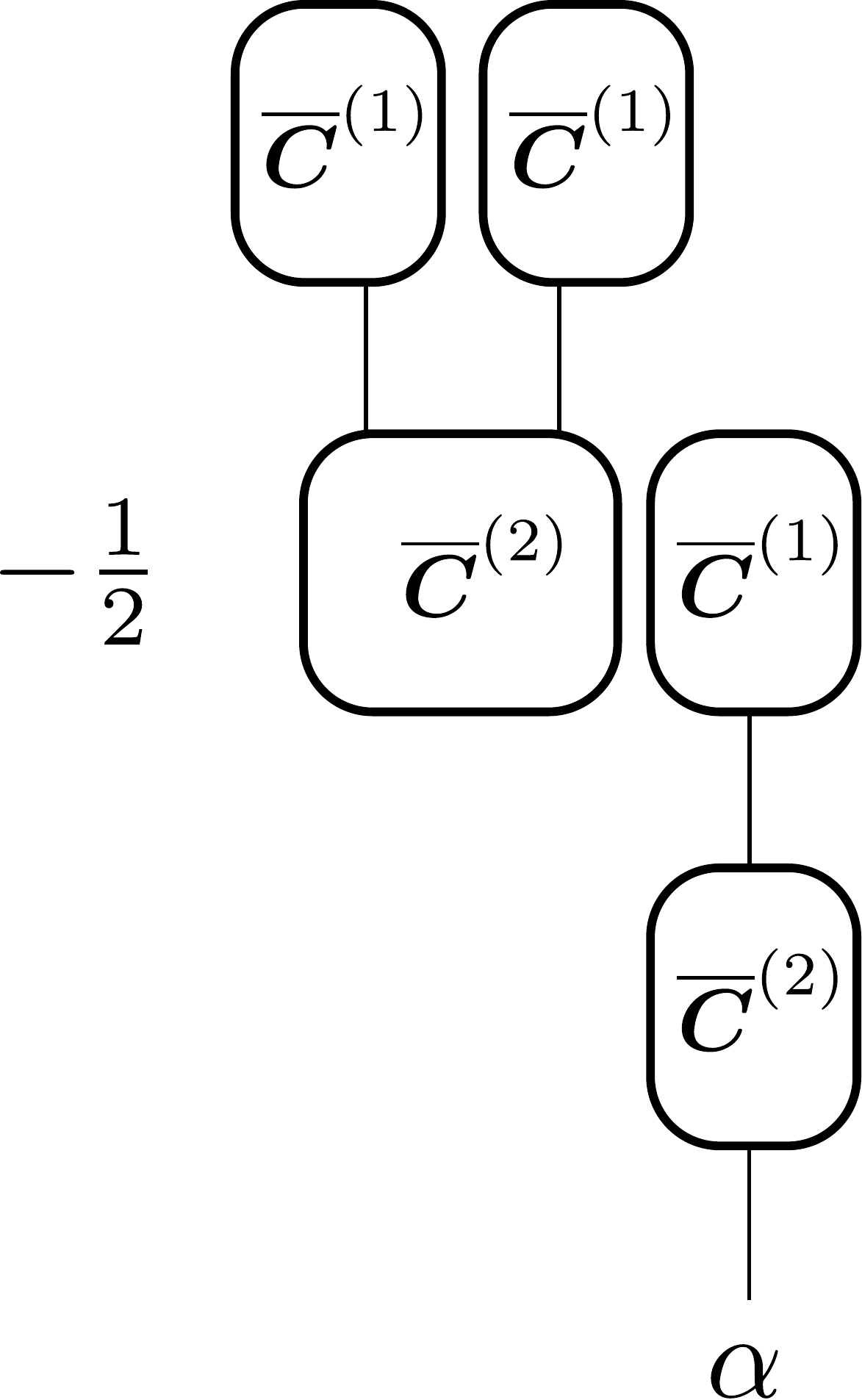}
        \caption{Contribution 14}
    \end{subfigure}
    \begin{subfigure}{0.32\textwidth}
        \centering
        \includegraphics[width=0.9\textwidth]{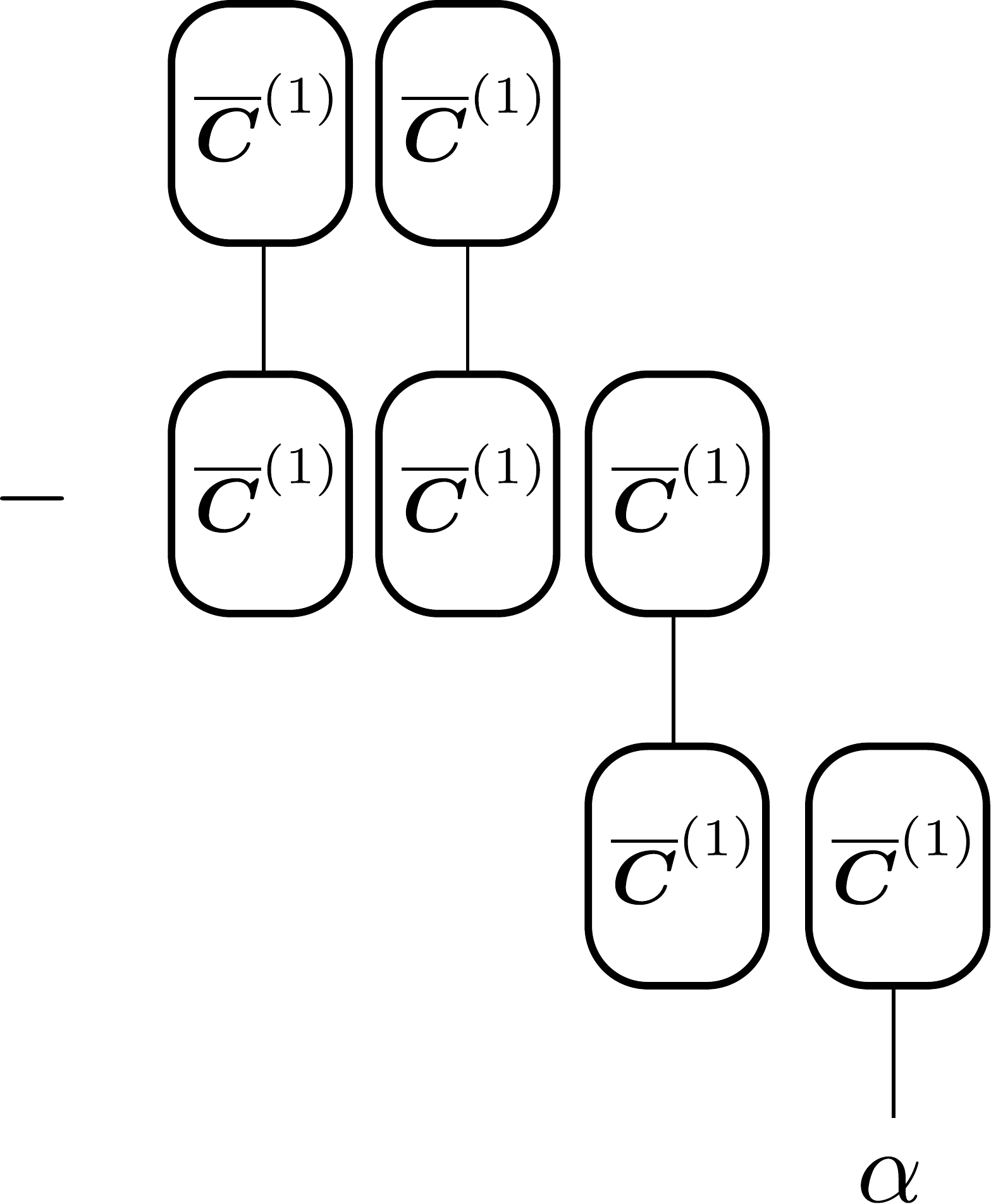}
        \caption{Contribution 15}
    \end{subfigure}\\
    \vspace*{20px}
    \begin{subfigure}{0.32\textwidth}
        \centering
        \includegraphics[width=0.55\textwidth]{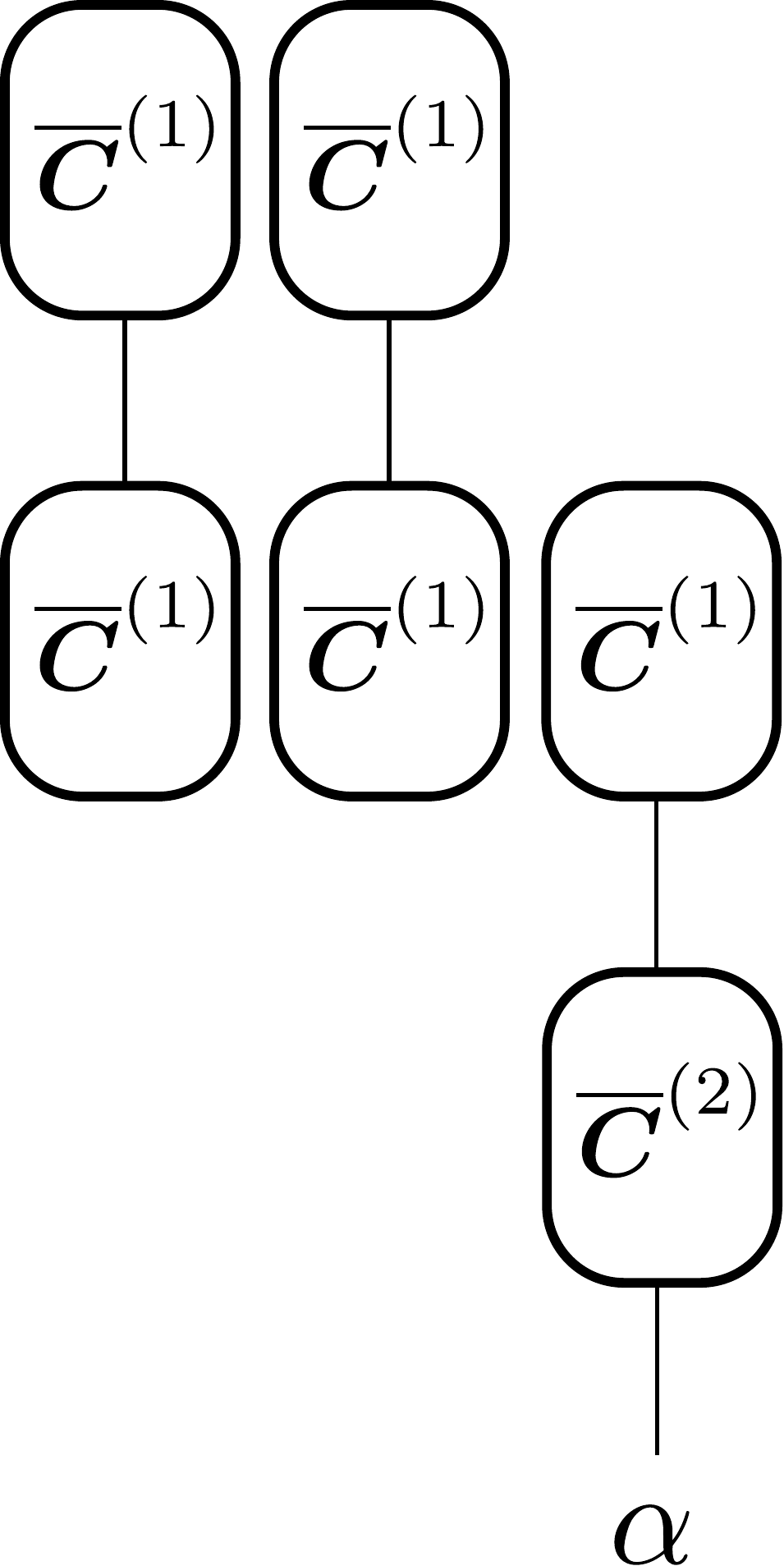}
        \caption{Contribution 16}
    \end{subfigure}
    \begin{subfigure}{0.32\textwidth}
        \centering
        \includegraphics[width=0.55\textwidth]{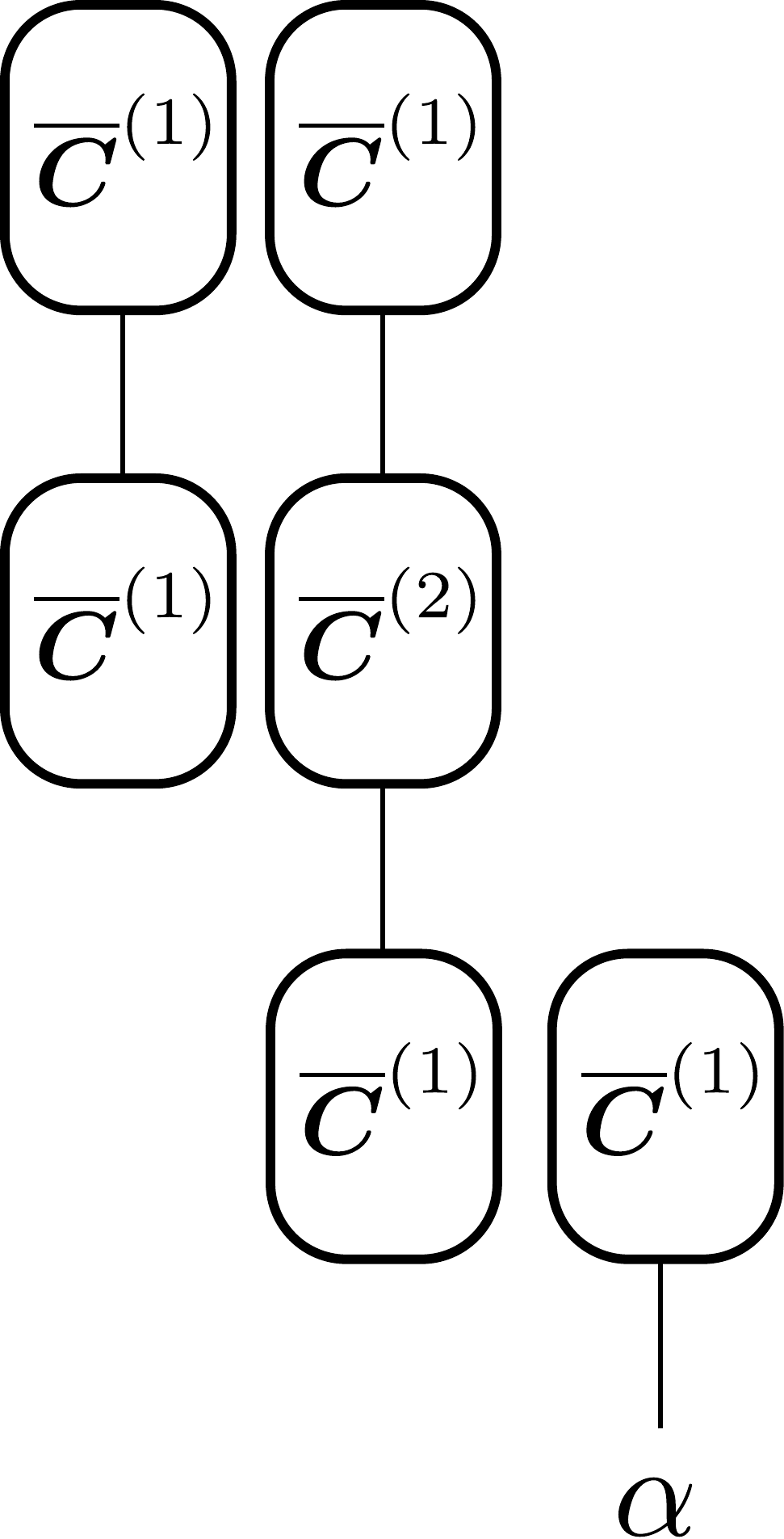}
        \caption{Contribution 17}
    \end{subfigure}
    \begin{subfigure}{0.32\textwidth}
        \centering
        \includegraphics[width=0.5\textwidth]{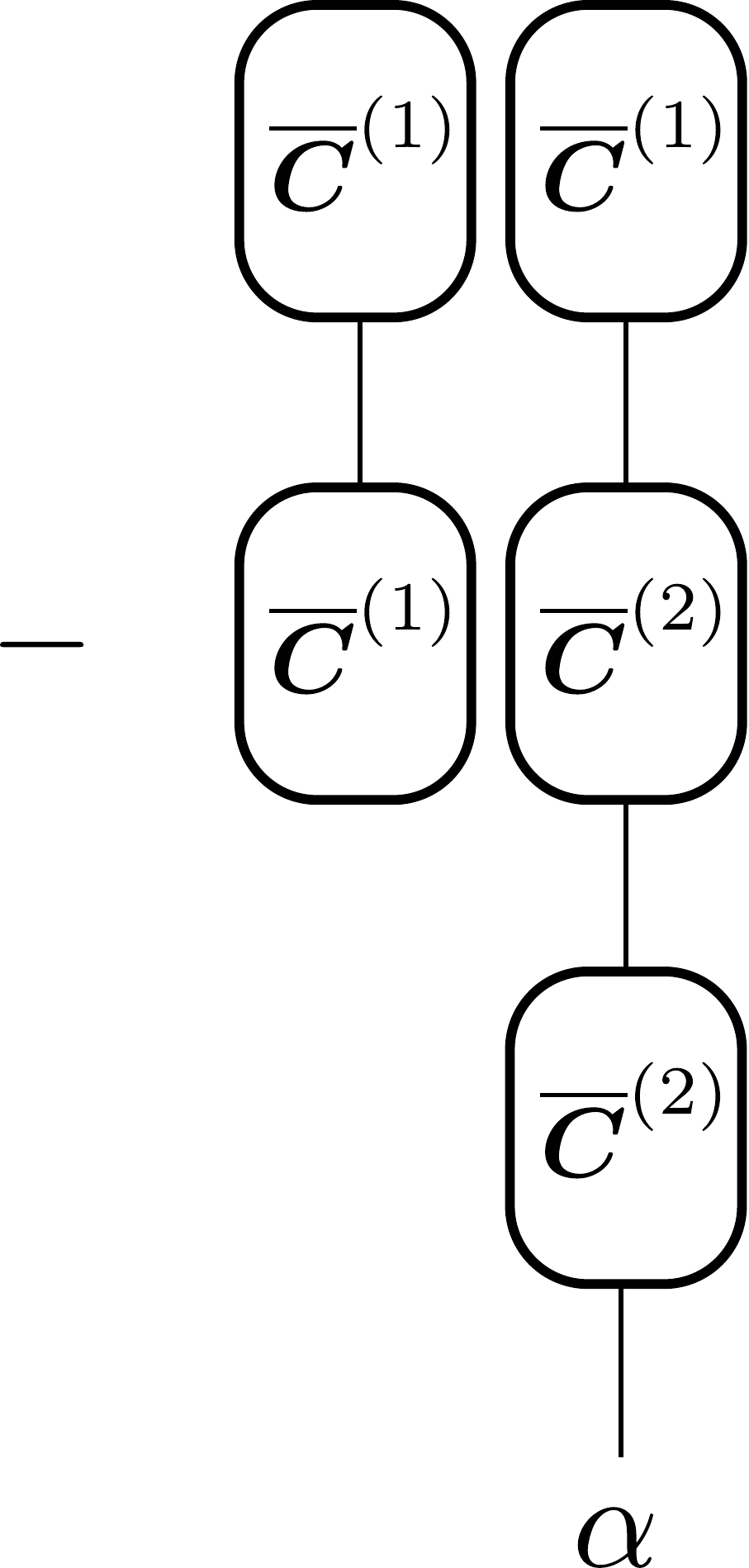}
        \caption{Contribution 18}
    \end{subfigure}
\end{figure}
\begin{figure}
    \ContinuedFloat
    \centering
    \begin{subfigure}{0.35\textwidth}
        \centering
        \includegraphics[width=0.6\textwidth]{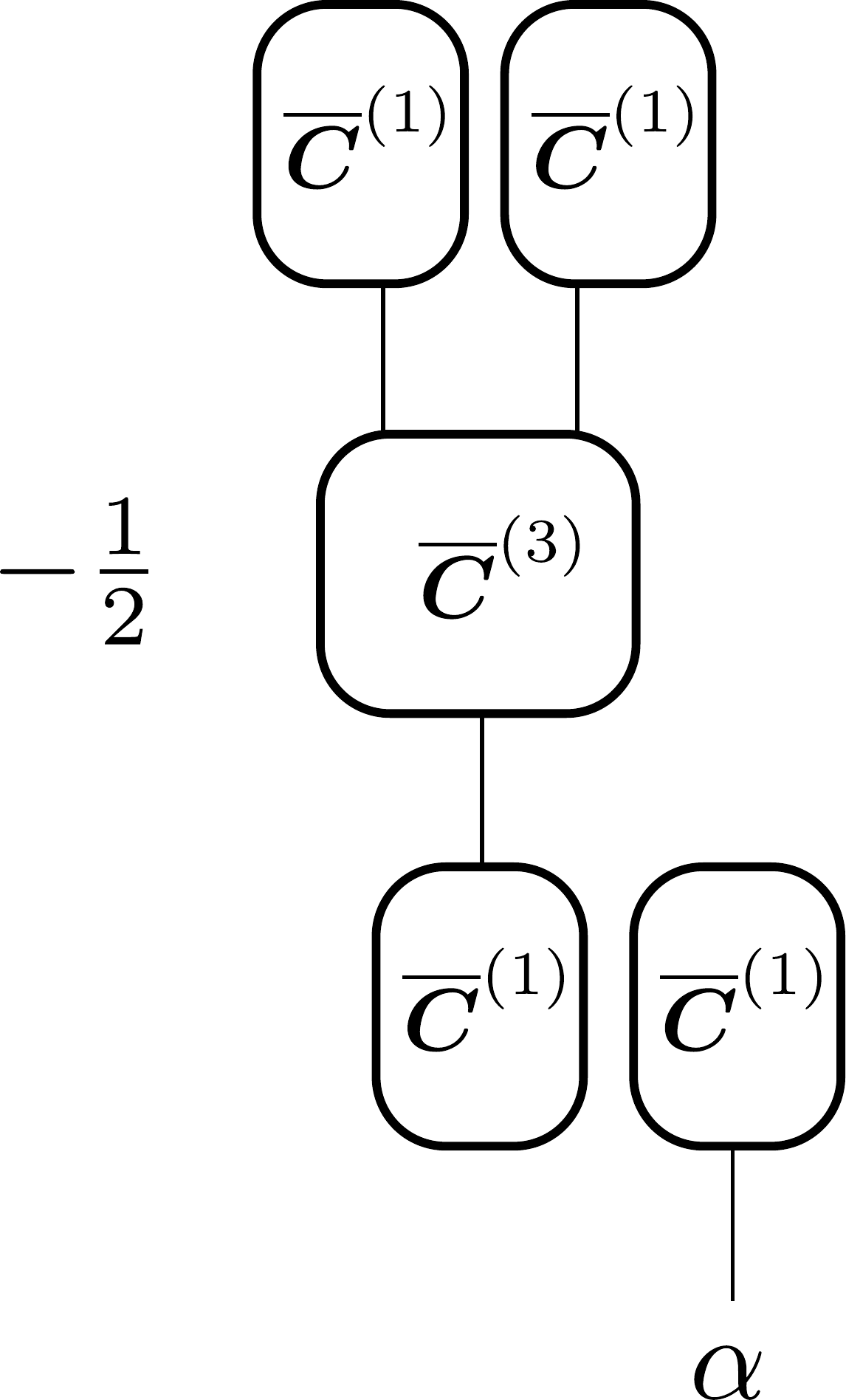}
        \caption{Contribution 19}
    \end{subfigure}
    \begin{subfigure}{0.35\textwidth}
        \centering
        \includegraphics[width=0.45\textwidth]{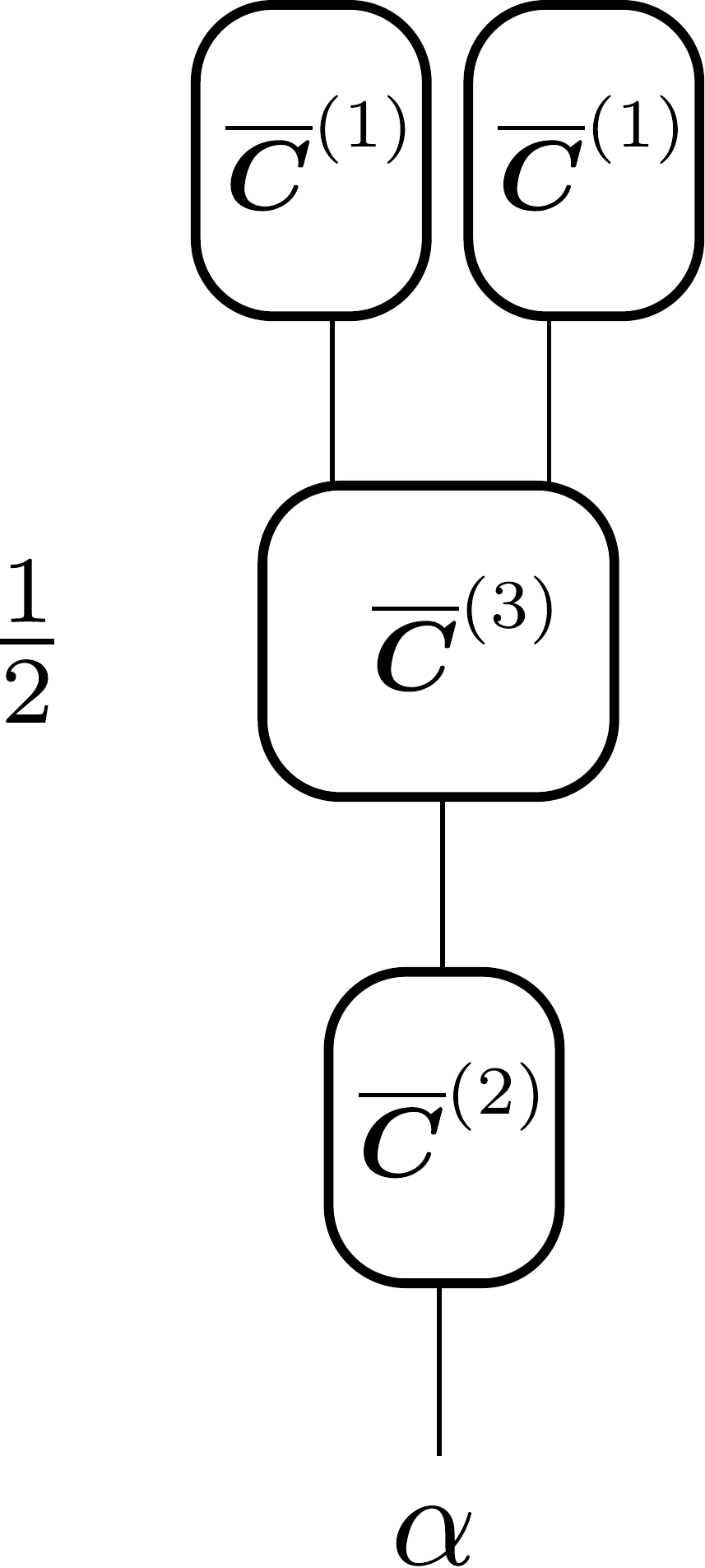}
        \caption{Contribution 20}
    \end{subfigure}
    \caption{Order 7 contributions to saddle point $\bm{\theta}^* = \left(\theta^*_{\alpha}\right)$ from quadratic $\bm{T}$ term $\bm{T}^2\overline{\bm{\Theta}^*}$. More specifically, contributions $1$ to $12$ arise from blocks product $\bm{T}_{1,\,2}\bm{T}_{2,\,1}\overline{\bm{\theta}^*}$, while contributions $13$ to $20$ come from blocks product $\bm{T}_{1,\,1}\bm{T}_{1,\,2}\overline{\bm{\theta}^*}^{\otimes 2}$.}
    \label{fig:order_7_t_square_contributions}
\end{figure}

\begin{figure}[!htbp]
    \centering
    \begin{subfigure}{0.34\textwidth}
        \centering
        \includegraphics[width=\textwidth]{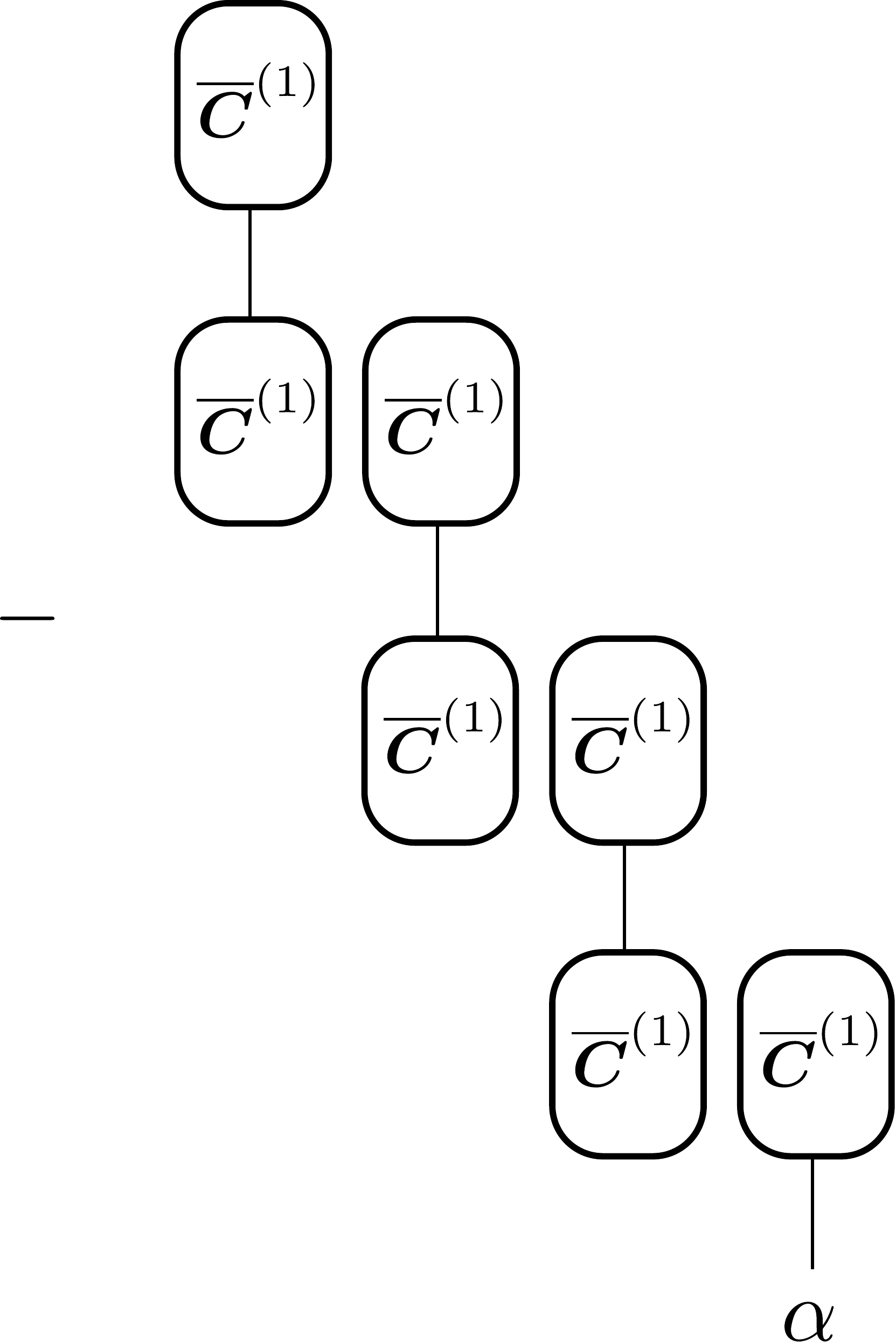}
        \caption{Contribution 1}
    \end{subfigure}
    \begin{subfigure}{0.32\textwidth}
        \centering
        \includegraphics[width=0.65\textwidth]{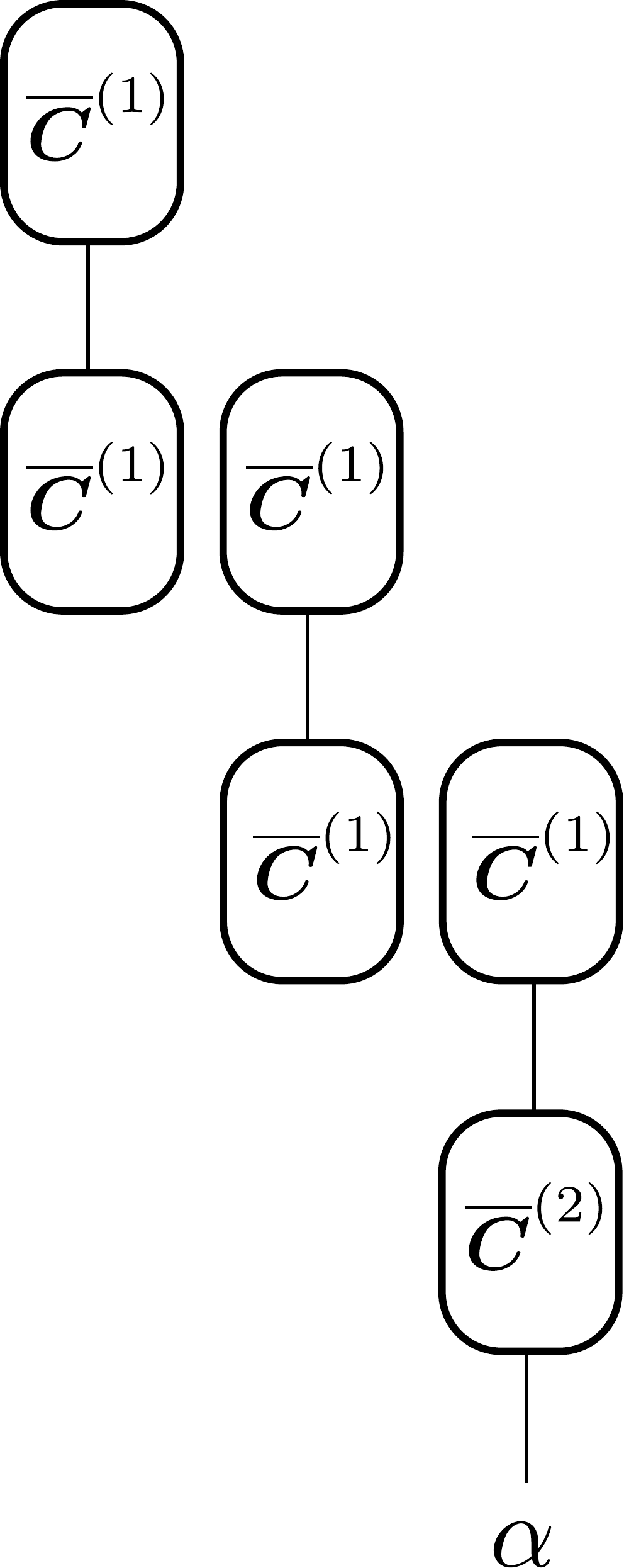}
        \caption{Contribution 2}
    \end{subfigure}
    \begin{subfigure}{0.32\textwidth}
        \centering
        \includegraphics[width=0.65\textwidth]{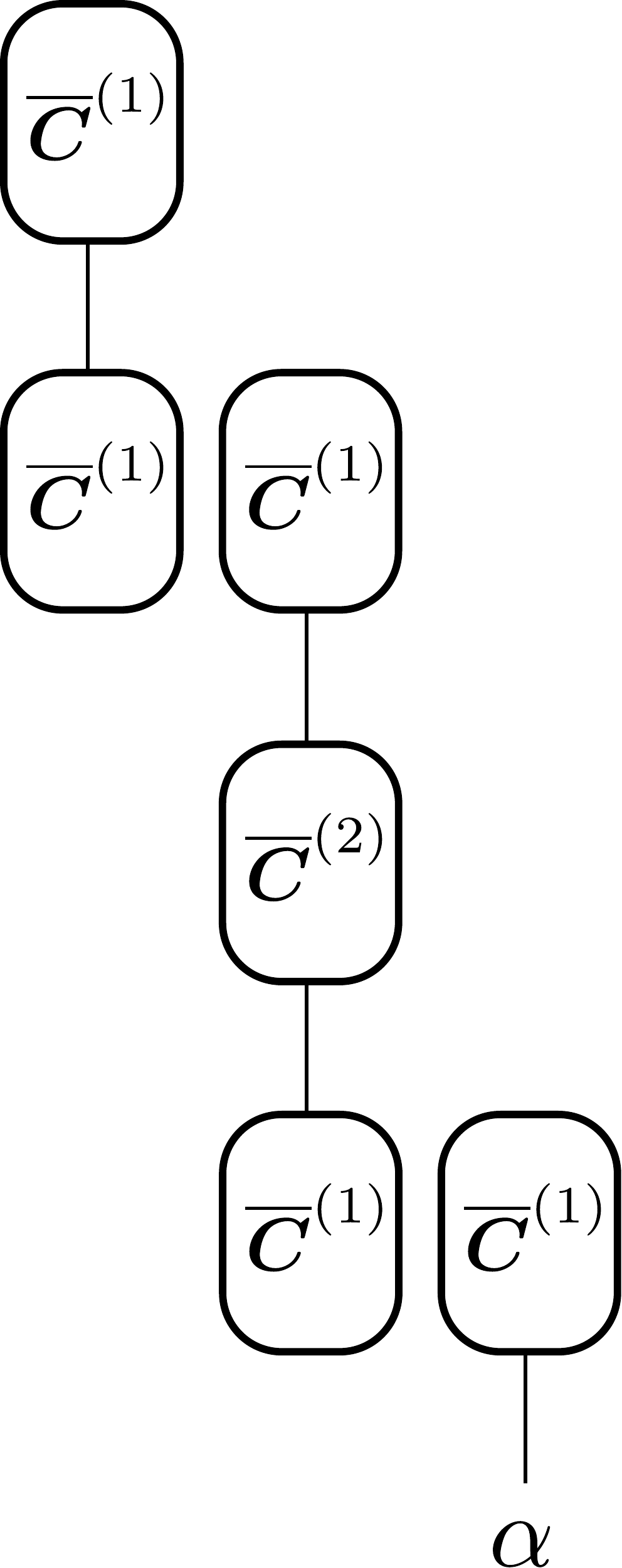}
        \caption{Contribution 3}
    \end{subfigure}\\
    \vspace*{20px}
    \begin{subfigure}{0.32\textwidth}
        \centering
        \includegraphics[width=0.65\textwidth]{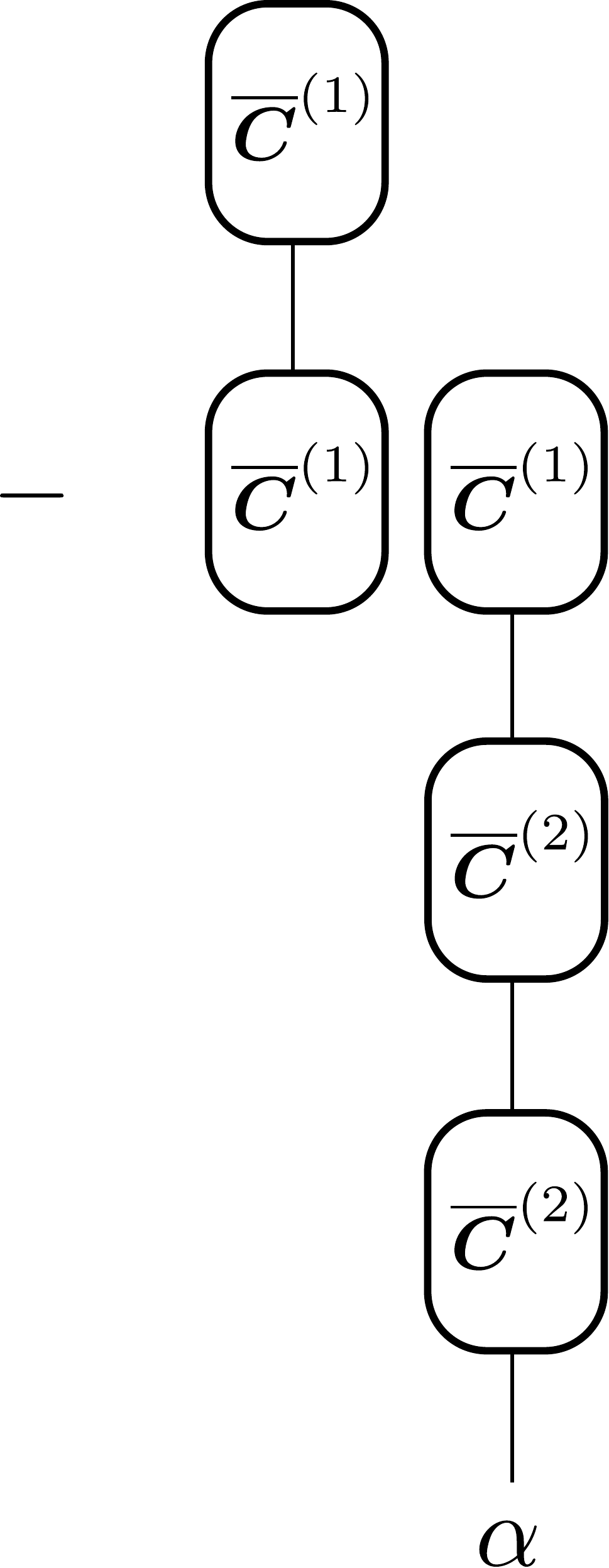}
        \caption{Contribution 4}
    \end{subfigure}
    \begin{subfigure}{0.33\textwidth}
        \centering
        \includegraphics[width=0.65\textwidth]{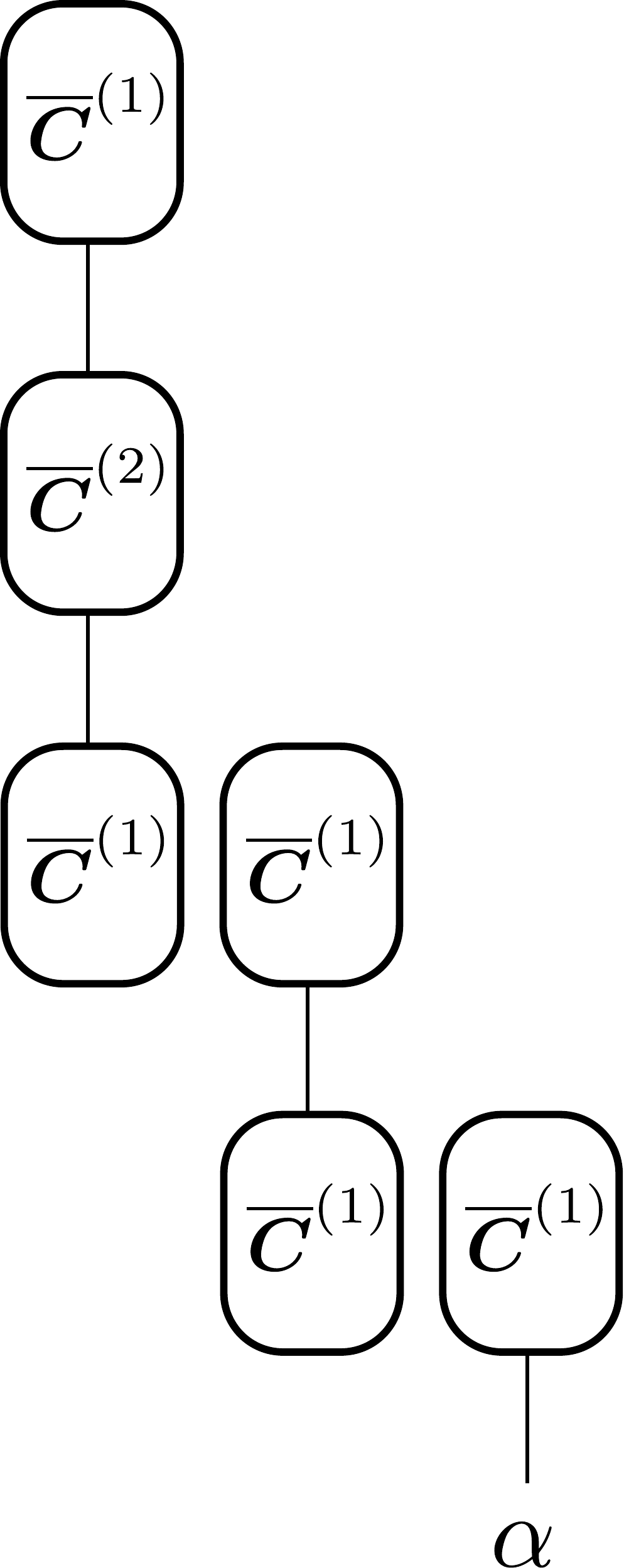}
        \caption{Contribution 5}
    \end{subfigure}
    \begin{subfigure}{0.32\textwidth}
        \centering
        \includegraphics[width=0.65\textwidth]{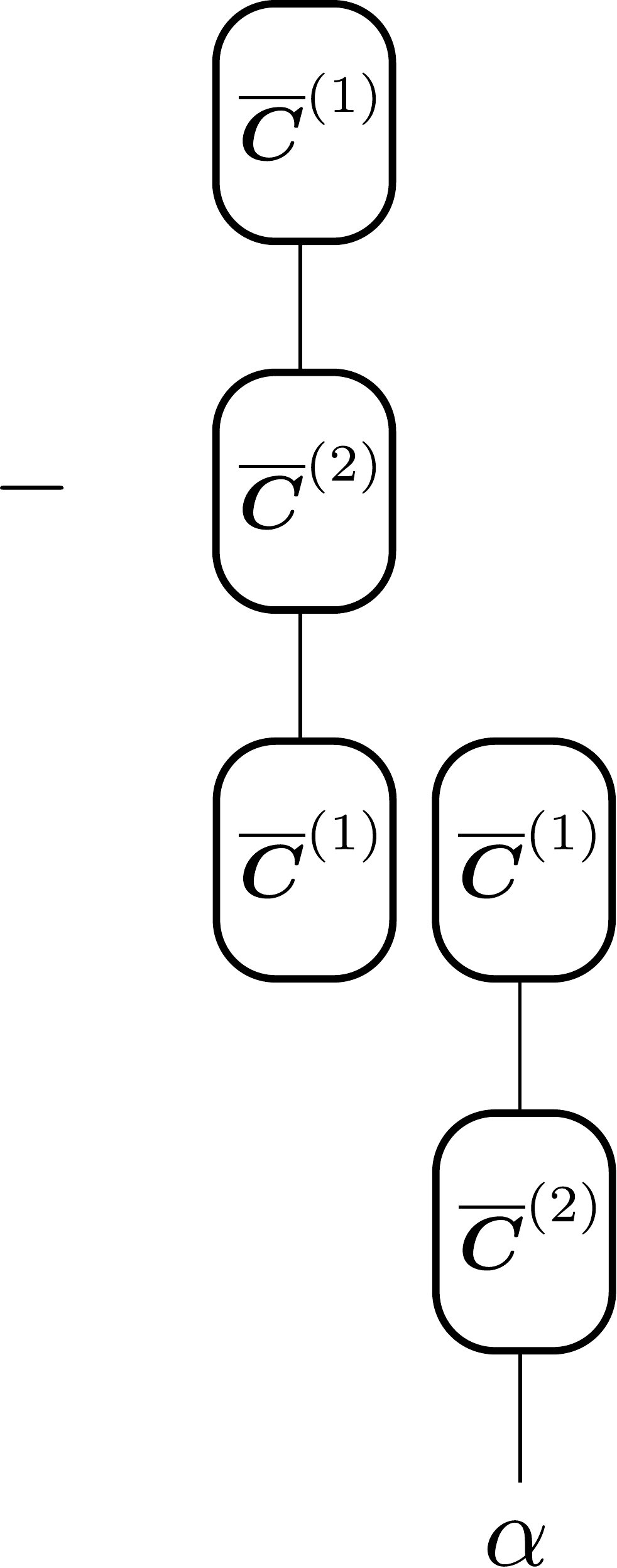}
        \caption{Contribution 6}
    \end{subfigure}
\end{figure}
\begin{figure}
    \ContinuedFloat
    \centering
    \begin{subfigure}{0.35\textwidth}
        \centering
        \includegraphics[width=0.57\textwidth]{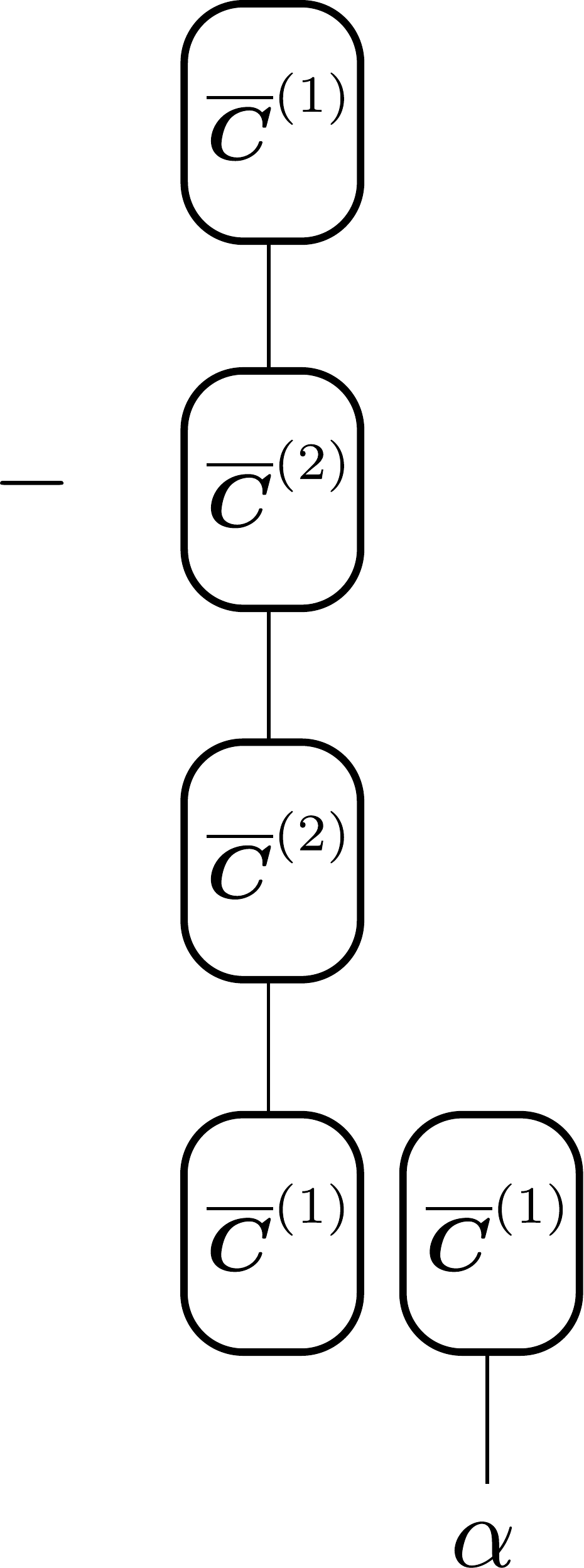}
        \caption{Contribution 7}
    \end{subfigure}
    \begin{subfigure}{0.35\textwidth}
        \centering
        \includegraphics[width=0.18\textwidth]{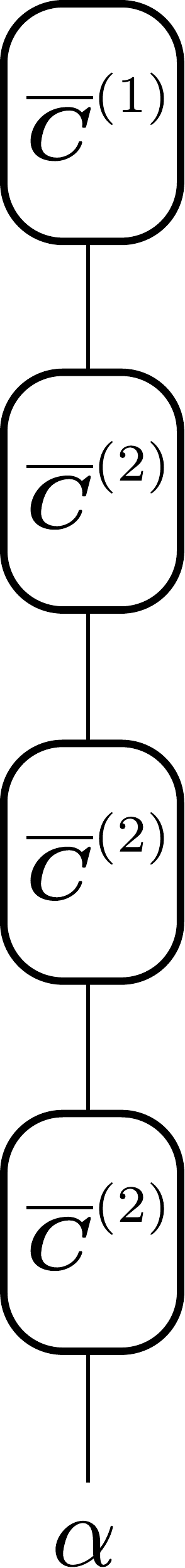}
        \caption{Contribution 8}
    \end{subfigure}
    \caption{Order 7 contributions to saddle point $\bm{\theta}^* = \left(\theta^*_{\alpha}\right)_{\alpha \in \mathcal{A}}$ from cubic $\bm{T}$ term $\bm{T}^3\overline{\bm{\Theta}^*}$.}
    \label{fig:order_7_t_cube_contributions}
\end{figure}

\clearpage

\end{document}